\documentclass[11pt, twoside, reqno]{amsart}
\usepackage[caption=false, font=normalsize]{subfig}
\usepackage[capposition=top]{floatrow}
\usepackage[thm_section]{macros}
\usepackage{makecell}
\usepackage[colorinlistoftodos]{todonotes}

\begin{document}
\begin{titlepage}
\begin{spacing}{1}
\title{\textbf{\Large
Using Multiple Outcomes to \\\vspace{0.5em} Adjust Standard Errors for Spatial Correlation
}
}
\author{
\begin{tabular}[t]{c@{\extracolsep{4em}}c} 
\large{Stefano DellaVigna} &  \large{Guido Imbens}\vspace{-0.7em}\\ \vspace{-1em}
\small{UC Berkeley and NBER} & \small{Stanford University} \\ \vspace{-0.7em} \\
\large{Woojin Kim} &  \large{David M. Ritzwoller}\vspace{-0.7em}\\ \vspace{-1em}
\small{Stanford University} & \small{Harvard University} \\ \vspace{-0.7em}
\end{tabular}%
\\
}
\date{%
\today\\ Email: sdellavi@econ.berkeley.edu, imbens@stanford.edu, wjnkim@stanford.edu, david.ritzwoller@cmsa.fas.harvard.edu.\\
We thank audiences at the University of Zurich, Munich, Mannheim, Bonn, USC, and the California Econometrics Conference, as well as Sasha O. Becker, Bruno Caprettini, Jiafeng Chen, Patrick Kline, Ulrich M{\"u}ller,  Michael Pollmann, Liyang Sun, and Mark Watson, for helpful comments and conversations. We are grateful to Malek Hassouneh, Rohan Jha, Chenxi Jiang, Afras Sial, and Yi Yu for exceptional research assistance. We thank Desmond Ang for sharing data used in one of the applications. Imbens acknowledges support from the Office of Naval Research under grant numbers N00014-17-1-2131 and N00014-19-1-2468 and a gift from Amazon. Kim acknowledges support from the National Institute on Aging to the National Bureau of Economic Research through grant number T32-AG000186. Ritzwoller acknowledges support from the National Science Foundation through grant DGE-1656518. Generative AI products from Anthropic, OpenAI, and Refine were used in the preparation of this manuscript. Code implementing the method developed in this paper is available at the link \url{https://github.com/wjnkim/tmo}.}
                      
\begin{abstract}
\smalltonormalsize{
A researcher interested in a particular outcome variable often observes dozens of other outcomes for the same units. If these auxiliary outcomes exhibit the same cross-sectional correlation as the outcome of interest, they can be used to estimate that correlation directly. We develop a procedure that uses estimates of this form to adjust standard errors in a given regression problem for cross-sectional dependence. We show, theoretically and with simulations, that the resulting standard errors can be substantially more accurate. Re-analyzing nine papers, we find that the proposed correction can make a meaningful difference in practice.
}
\end{abstract}
\end{spacing}
\end{titlepage}
\maketitle
\thispagestyle{empty}
\setcounter{page}{0}
\newpage
\begin{spacing}{1.4}

\section{Introduction\label{sec: introduction}} 

Economic outcomes are often linked across locations. Productivity shocks to one region may propagate through trade linkages and labor market flows \citep{greenstone2010identifying,bustos2016agricultural,giroud2024propagation}, fiscal policies in one municipality may affect the design of policies in other locations \citep{case1993budget,dellavigna2022policy}, and political events in one jurisdiction may determine voter behavior in others \citep{besley1995incumbent,madestam2013political}. Accordingly, data that exhibit spatial dependence are ubiquitous in economics. In 2023 alone, nearly half of all empirical, observational papers published in five leading general-interest journals primarily consider data indexed by locations in space.

Estimates of uncertainty can be substantially distorted when dependent observations are treated as independent. Existing  methods for correcting standard errors for cross-sectional dependence fall into two broad categories. The first consists of methods that cluster standard errors based on a partition of space. A prototypical instance of this approach allows for dependence between U.S.\ counties in the same state (for discussion, see \citealt{moulton1986random, moulton1990illustration}). The second set of methods bounds dependence by a decreasing function of geographic distance \citep{conley1999gmm, bester2016fixed, muller2022spatial, muller2023spatial}. 

Both approaches impose strong, ex ante restrictions on how spatial dependence varies with geography, in one case with sharp boundaries, and in the other with functions of distance. However, the correlation between economic outcomes at two locations need not be determined by their geographic distance alone. For instance, similarities in economic or political behavior could instead arise from factors like urbanization, income, or education levels, some of which may be unobserved.

\begin{figure}[p]
    \centering
    \caption{Correlation with Two California Counties\label{fig:SFModoc}}
    \medskip
    \begin{tabular}{l | l}
        \textbf{Correlation w/ San Francisco County}
        & \textbf{Correlation w/ Modoc County} \\
        \hline
        \textit{California} &\\
            \includegraphics[width=0.35\linewidth]{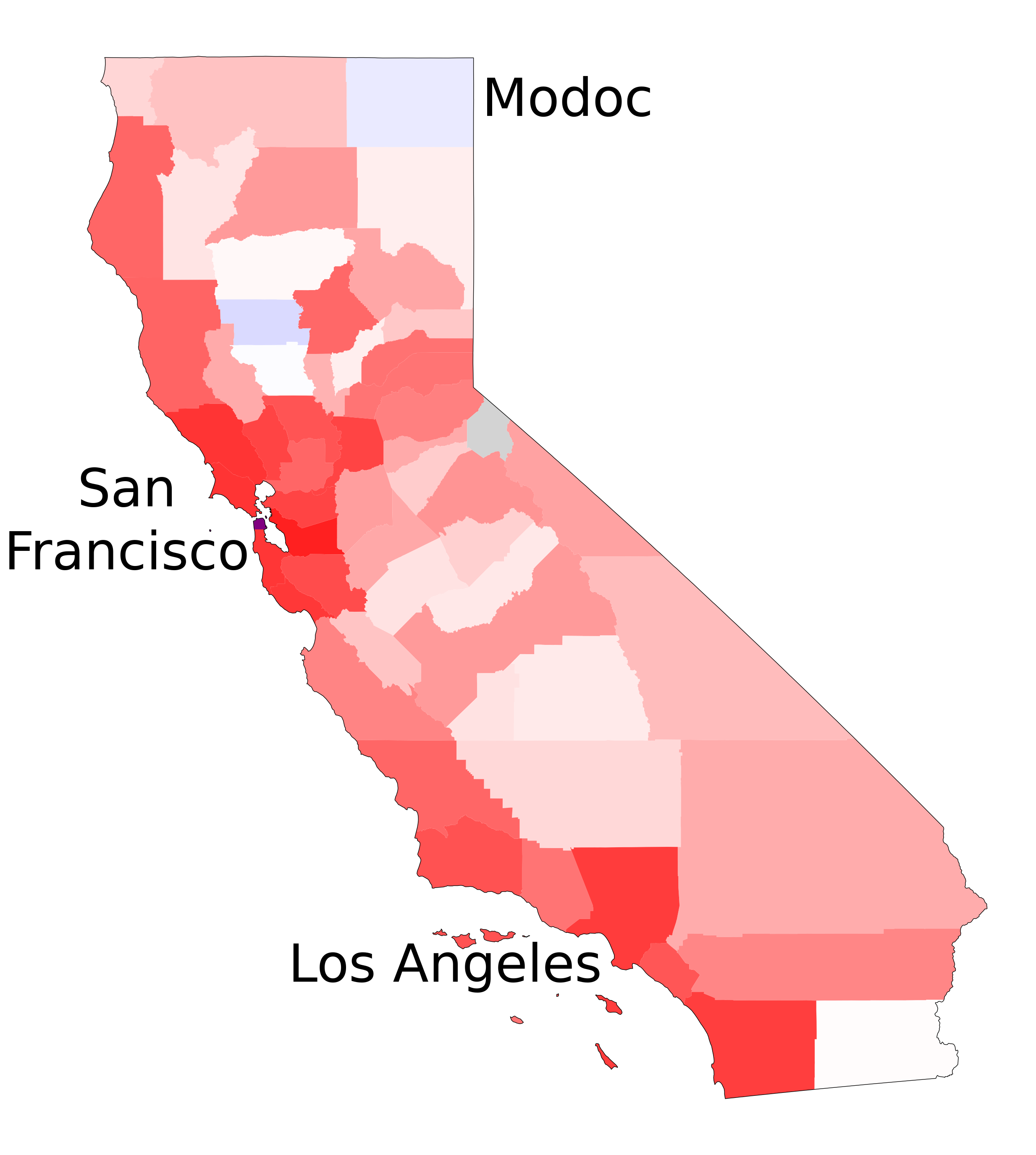}
            & \includegraphics[width=0.35\linewidth]{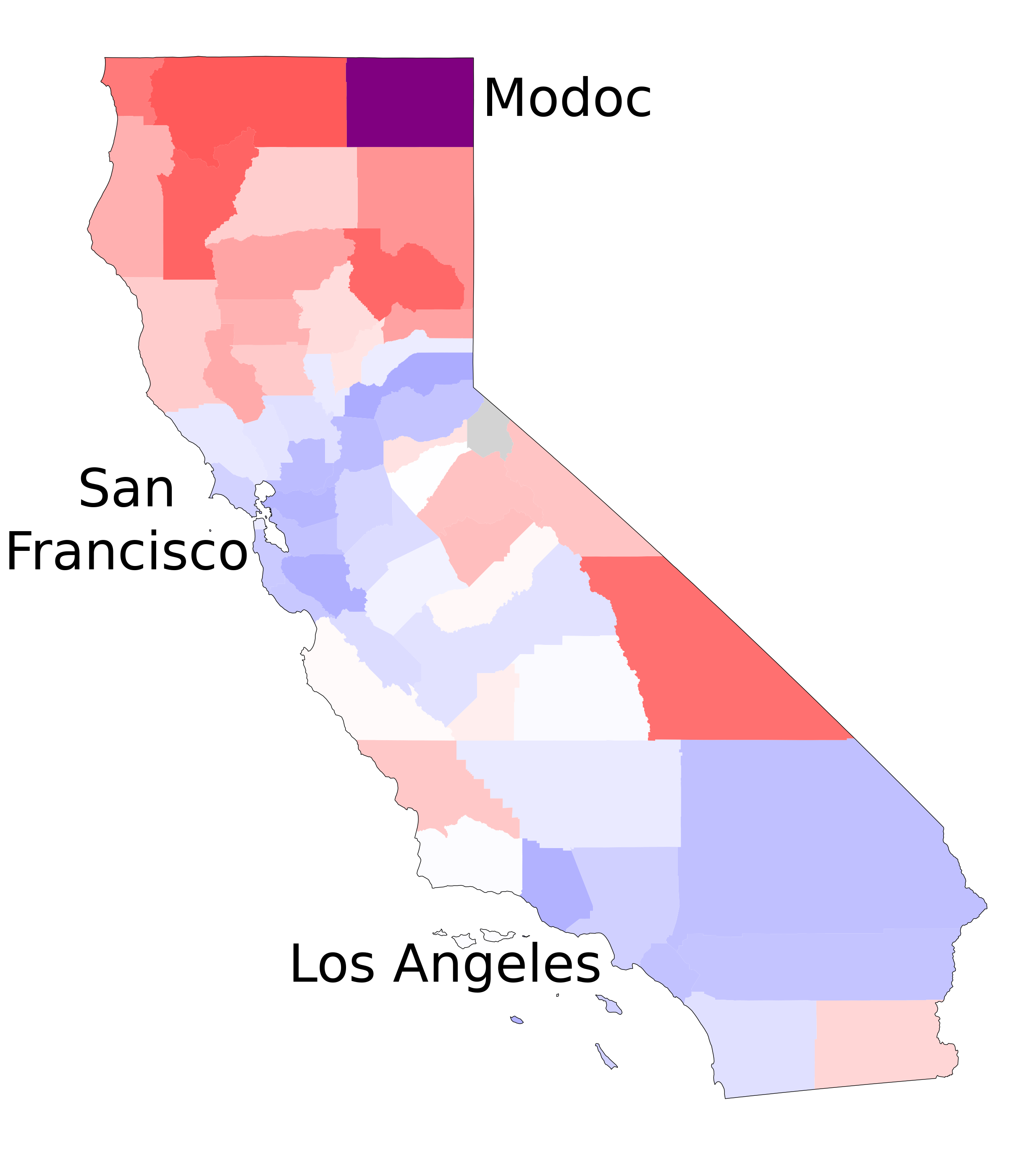} \\
        
        \textit{North Dakota} & \\
            \includegraphics[width=0.35\linewidth]{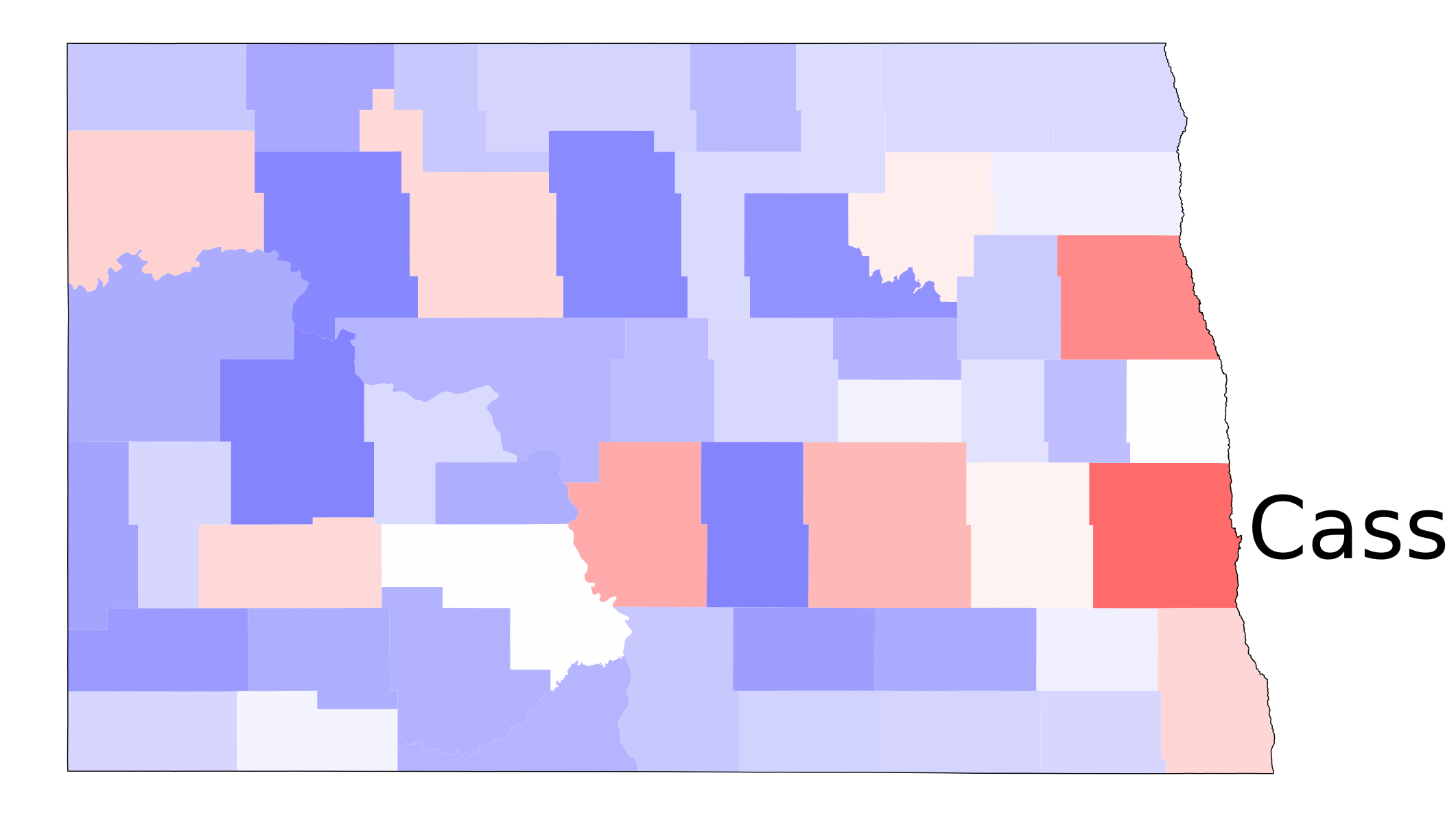}
            & \includegraphics[width=0.35\linewidth]{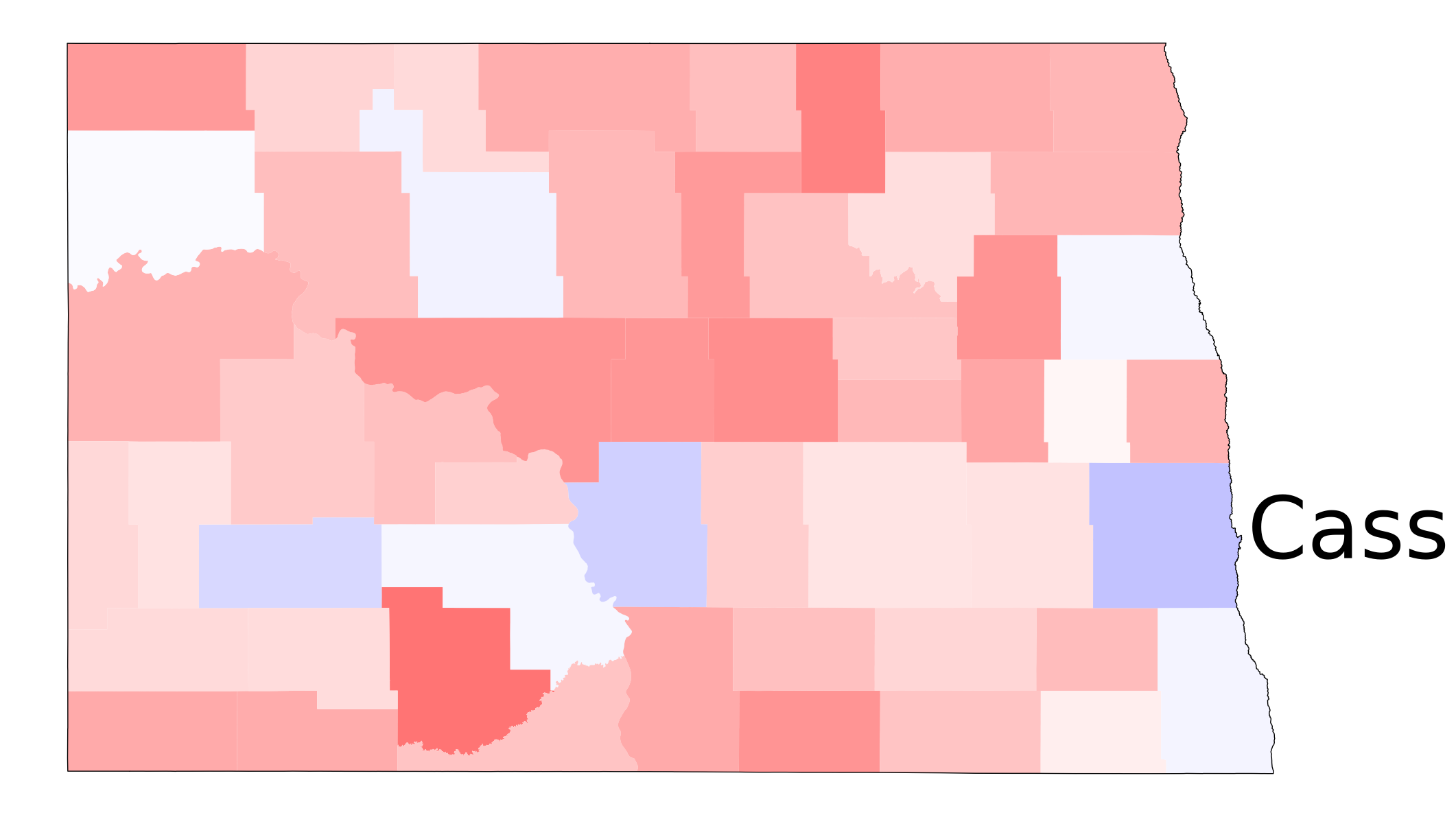} \\
        
        \textit{New York} & \\
            \includegraphics[width=0.35\linewidth]{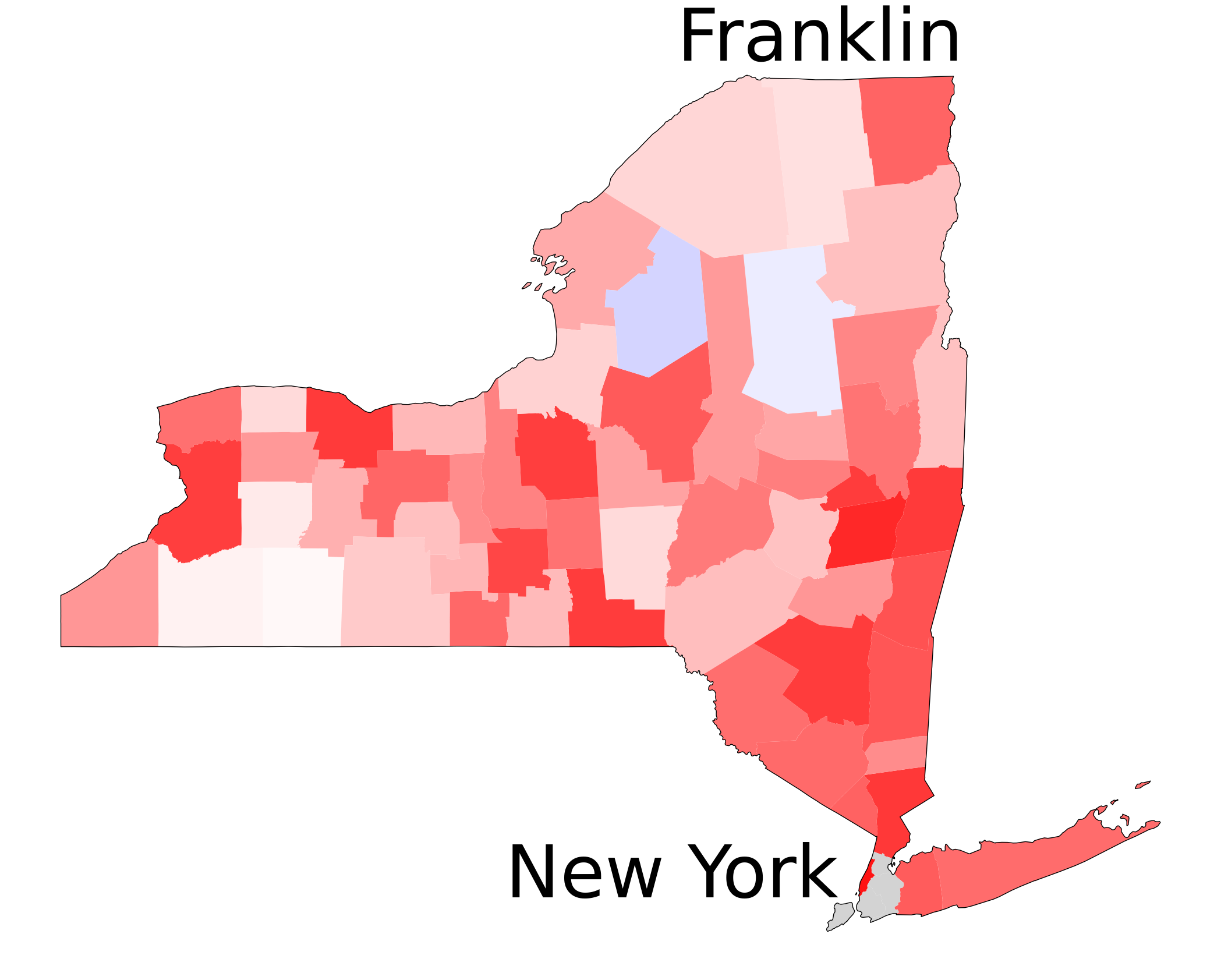}
            & \includegraphics[width=0.35\linewidth]{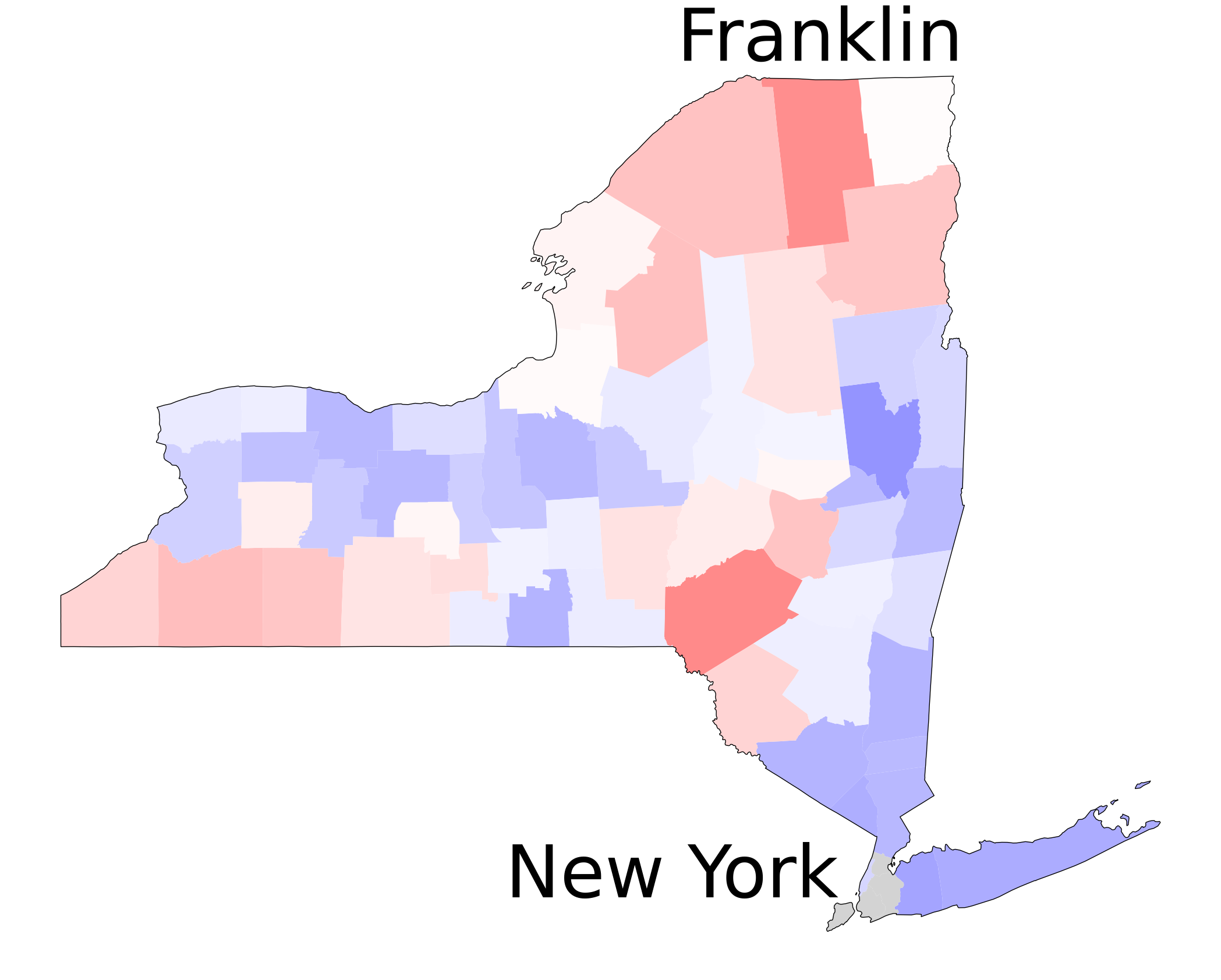} \\
        
        \multicolumn{2}{c}{\includegraphics[width=0.7\linewidth]{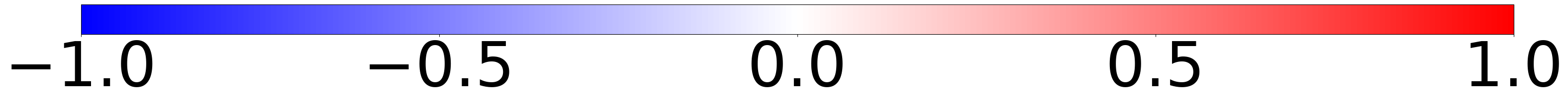}} \\  
    \end{tabular}
    
    \medskip
    \justifying
    {\noindent \footnotesize
    Notes: \cref{fig:SFModoc} displays heat-maps giving the correlation of a collection of economic outcomes between pairs of U.S.\ counties. These outcomes are listed in \cref{sec: simulation county data} and have been normalized to have mean zero and unit variance. The first column gives the correlation between outcomes in San Francisco County, California, and outcomes in each county in the states of California, North Dakota, and New York. The second column is analogous, but gives correlations relative to Modoc County, California.
    \par}
\end{figure}

To illustrate this point, we collect a set of 50 U.S. county-level economic outcomes (including the unemployment rate, the average income, the poverty rate, etc.) for 3,060 U.S. counties.\footnote{We list these outcomes, and give further details concerning their construction, primarily from the U.S.\ Census, in \cref{sec: simulation county data}.} We normalize each outcome to have mean zero and unit variance. The left column of \cref{fig:SFModoc} displays the correlation between the outcomes of San Francisco County, California, and the outcomes of every other county in the states of California, North Dakota, and New York. As one might expect, San Francisco is more correlated with other densely populated areas, such as Los Angeles County and New York County, than with the less populated areas in the rural state of North Dakota. Conversely, the right column of \cref{fig:SFModoc} indicates that Modoc County, the third least populous county in the state of California, with an economy driven by the agricultural sector \citep{modoc2023}, is more correlated with very distant less populated areas in North Dakota and New York than with closer, but more populated, areas in California.\footnote{These results generalize. \cref{fig:CANYCountyCorr_app} displays the correlation in outcomes between each pair of counties in California, North Dakota, and New York, sorting by state and population. Both population and geography are important determinants of the correlation between two counties. \cref{fig: county correlation} gives an analogous visualization for each pair of counties in the United States. The same pattern holds.}

Methods based on the assumption that cross-sectional dependence is determined by geographic proximity cannot capture many of the features in \cref{fig:SFModoc}, and so may not perform well in regression problems constructed using data that exhibit dependence with this structure. Indeed, in \cref{sec: simulation}, we calibrate a simulation to the estimates of the correlation between pairs of U.S.\ counties introduced above. We find that many standard adjustments for spatial dependence yield standard errors that are badly biased.

In this paper, we introduce a method that uses observations of multiple outcome variables to adjust standard errors for spatial dependence. We detail the proposal in \cref{sec: proposal}. We give a condition under which collections of outcomes, of the type considered above, can be used to estimate dependence across units. In particular, rather than directly restricting spatial dependence to be determined by, say, geographic proximity, we assume that the structure of spatial correlation is shared across outcomes. This is a highly stylized restriction---so too are alternative assumptions that relate spatial dependence to geography. The main appeal is that, in settings where the main features of spatial correlation are plausibly captured by a collection of available outcomes, spatial dependence can be estimated directly from the data.

We propose a procedure, which we call Thresholding Multiple Outcomes (TMO), for using estimates of spatial dependence to adjust standard errors in a given regression setting. Our procedure is premised on the hypothesis that pairs of locations can be effectively categorized as either ``null'' or ``non-null.'' Outcomes at non-null location pairs can exhibit meaningful correlation, whereas outcomes at null pairs cannot. The method consists of three steps. First, we estimate the correlation across locations using a collection of auxiliary outcomes. Second, we use these estimates to determine a threshold: pairs of locations whose correlation exceeds this threshold are treated as ``non-null,'' while pairs below the threshold are treated as ``null.'' Third, we construct standard errors by allowing for dependence among location pairs treated as ``non-null.'' This approach is analogous to the construction of clustered standard errors. There, pairs of locations in the same cluster are assumed to correlate \emph{a priori}. Here, we use the auxiliary outcomes to \emph{determine} the pairs of locations that correlate.  Revisiting the calibrated simulation from \cref{sec: simulation}, we find that this procedure exhibits substantially smaller bias, and more accurate rejection rates, than existing methods.

The fundamental statistical challenge, then, is to choose a threshold that best distinguishes between null and non-null pairs of units. In \cref{sec: risk approximation sec} we develop a closed-form approximation to the bias and root-mean-squared error of a simplified version of the TMO estimator. We show that, in realistic settings where null and non-null pairs of locations cannot be differentiated without non-negligible error, the (potentially non-vanishing) absolute bias of the estimator determines its root-mean-squared error to leading order. Thus, in \cref{sec: threshold choice}, we propose a method for choosing the threshold, based on optimizing a regularized estimate of the absolute bias. The absolute bias and root-mean-squared error achieved by this feasible choice differ from their respective infeasible minima by terms that converge to zero.

In \cref{sec: applications}, we document that 48\% of empirical, observational papers published in five leading general-interest economics journals in 2023 consider data indexed by locations, with the most common unit being the U.S.\ county. We re-analyze seven of these papers. For each paper, we use a set of county-level outcomes, taken from each paper's replication packages as well as from external sources, to construct the TMO estimator. We find that the TMO estimator can substantively change the estimated standard errors. Comparing TMO to leading alternatives, we find that in some cases, it leads to a similar adjustment, while in other cases, the resultant standard errors are qualitatively different. We also re-analyze two additional papers---one that studies commuting zones and another that studies countries. 

We conclude in \cref{sec: conclusion} by detailing our recommendations for practice. Additional Figures and Tables are in \cref{sec: app fig tab}, with further details and extensions in \cref{sec: app simulation}. Details concerning our empirical applications are in \cref{app: empirical}. Proofs for all results stated in the main text are in \cref{sec: proofs}.\footnote{Code implementing the method developed in this paper is available at the link \url{https://github.com/wjnkim/tmo}.}

\subsubsection*{Related Literature}

We contribute to an extensive econometric literature concerning quantification of uncertainty in data with cross-sectional dependence. Much of this research focuses on methods that partition units into clusters according to an observable characteristic, allowing for correlation among observations within the same cluster. \cite{liang1986longitudinal} introduce the first heteroskedasticity-consistent clustered standard error (see also \citealt{mackinnon1985some}, \citealt{bell2002bias}, and \citealt{imbens2016robust} for alternatives that exhibit better finite-sample performance). \cite{bertrand2004much} provide a demonstration of the importance of these methods in panel data. \cite{cameron2015practitioner} give a comprehensive review. As in this approach, we assume that only a small proportion of pairs of units correlate. By contrast, we do not assume that these pairs are known \textit{ex ante}, but rather, can be learned from observations of auxiliary outcomes.  

A subset of this literature develops methods premised on the assumption that dependence can be bounded by a function of the distance between units. This line of work was initiated by \cite{conley1999gmm}, who emphasizes that dependence should be modeled by measures of ``economic distance'' (see also \citealt{kelejian2007hac}, \citealt{kim2011spatial}, \citealt{bester2016fixed}, \cite{colella2023acreg} and \citealt{muller2022spatial,muller2023spatial} for alternative estimators based on this program). Despite this, most applications of this framework parameterize dependence in terms of geographic proximity alone. Our contribution can be framed as giving a general procedure for \emph{estimating} economic distance on the basis of auxiliary outcomes, as well as a method for choosing a suitable bandwidth that accommodates non-negligible sampling error in these estimates.

Separately, \cite{barrios2012clustering}, \cite{muller2024spatial}, and \cite{conley2025standard} demonstrate that applying standard variance estimators to spatially correlated data often yields spurious findings. The simulation experiments in \cref{sec: simulation} generate results consistent with this view. The core of our proposal is that, due to the widespread availability of additional outcome data, some of the cross-sectional correlation that is common across measured outcomes can be recovered. 

A related literature uses repeated panel observations to learn features of  cross-sectional relationships. For instance, \cite{manresa2016estimating} and \cite{de2025identifying}, among others, apply this idea to identify network interactions. More closely related to our setting, \cite{bailey2016two,bailey2019multiple} use repeated panel observations to estimate cross-sectional covariance or correlation matrices. By contrast, our interest is in using estimates of a cross-sectional covariance matrix as an input into a variance estimator. Our framework is closely related to the method developed in \cite{bai2024standard}, who construct a variance estimator by using the time dimension of a panel to estimate and threshold pairwise long-run covariances of regression scores. Their consistency results require conditions that imply that errors in distinguishing null and non-null pairs of units have a negligible effect on the variance estimator. We instead focus on settings in which these errors remain important and develop a method for choosing a threshold to minimize the resulting bias.

Finally, we contribute to a growing literature that considers how standard statistical procedures can be improved by incorporating multiple outcomes. For example, \cite{ludwig2019machine}, \cite{sun2023using}, and \cite{abadie2024doubly} consider how collections of outcomes can be used to improve estimates of causal effects. Our primary focus, by contrast, is on obtaining more accurate estimates of uncertainty in regression problems.

\section{Standard Adjustments for Cross-Sectional Dependence Are Biased\label{sec: simulation}}

In this section, we use a simulation to measure the performance of several methods for adjusting standard errors for cross-sectional dependence. We consider settings with the following structure. A researcher observes the data $(Y_i, W_i,X_i)$ for each unit $i$ in $1,\ldots,n$, where $Y_i$ is a scalar outcome, $W_i$ is a scalar treatment, and $X_i$ is a vector of covariates. For the sake of concreteness, $Y_i$ could measure the change in the income per capita in U.S.\ county $i$ over a given time period, $W_i$ could denote the change in the percent of the population in county $i$ that has completed a college degree, and $X_i$ could collect a set of state fixed effects. The researcher assumes that the data arise from the linear model 
\begin{equation} \label{eq: linear model}
    Y_i = \alpha + \tau W_i + \theta^\top X_i + \varepsilon_i,\quad\mathbb{E}[\varepsilon_i|W,X]=0,\quad i = 1,\ldots,n~,
\end{equation}
where $W=(W_i)_{i=1}^n$ and $X=(X_i)_{i=1}^n$ collect the treatment and covariate measurements over the full sample of units. Let $\hat{\tau}$ denote the ordinary least squares estimate of the parameter $\tau$. Our interest is in estimating an appropriate standard error for $\hat{\tau}$. 

Often, it is unreasonable to assume that the components of the residual vector $\varepsilon = (\varepsilon_i)^n_{i=1}$ are uncorrelated. Factors that determine outcomes in San Francisco, say, plausibly also determine outcomes in Berkeley. There are two common ways of adjusting standard errors for cross-sectional dependence. One set of methods is based on clustering standard errors based on a partition of the space, thereby imposing geographic borders on the scope of dependence \citep{moulton1986random, moulton1990illustration}. A second set of methods bounds dependence in terms of geographic distance \citep{conley1999gmm, bester2016fixed, muller2022spatial, muller2023spatial}. 

How well do these methods address the types of cross-sectional dependence exhibited by the data considered in applied economics? In \cref{sec: measuring}, we measure the main features of the spatial dependence exhibited by the types of U.S. county-level outcomes considered in applied economics. In \cref{sec: simulation design}, we use these measurements to calibrate a simulation and document the performance of several widely applied approaches to adjusting standard errors for spatial dependence.

\subsection{Measuring the Structure of Cross-Sectional Dependence\label{sec: measuring}}

Our premise is that non-trivial correlation structures can be inferred from measurements of multiple outcomes. To operationalize this idea, we collect 50 U.S. county-level outcomes as in \cref{sec: introduction}, including measurements of income, poverty, employment, and health. We give further details in \cref{sec: simulation county data}.

Given a ``representative'' outcome $Y_i$ and treatment $W_i$, and a pair of counties $i$ and $i^\prime$, should we expect the residuals $\varepsilon_i$ and $\varepsilon_{i^\prime}$ to correlate? We argue that an answer to this question can be recovered by averaging the product of the empirical versions of these residuals across the outcomes in our sample. Formally, let $Y^{(1)}_i, \ldots, Y^{(d)}_i$ denote the $d$ outcomes measured for county $i$. For each outcome $Y_i^{(j)}$, let $\hat{\varepsilon}^{(j)}_i$ denote the empirical residual from the least squares regression of $Y^{(j)}_i$ on a vector of state fixed effects. Let $\tilde{\varepsilon}^{(j)}_i$ denote a version of $\hat{\varepsilon}^{(j)}_i$ that has been standardized to have  variance one across units, given by
\begin{equation}
\tilde\varepsilon_{i}^{(j)}
=
\hat{\varepsilon}^{(j)}_i 
\Biggl/\sqrt{\frac{1}{n} \sum_{i=1}^n (\hat{\varepsilon}^{(j)}_i)^2}~.
\end{equation}
The covariance and correlation between the residuals for counties $i$ and $i^\prime$ \emph{across outcomes} are given by
\begin{equation}\label{eq: outcome correlation}
\widehat{\textrm{cov}}_{i,i^{\prime}} = \frac{1}{d} \sum_{j=1}^d (\tilde{\varepsilon}^{(j)}_i - \bar{\varepsilon}_i) (\tilde{\varepsilon}^{(j)}_{i^{\prime}} - \bar{\varepsilon}_{i^{\prime}})
\quad
\text{and}
\quad
\widehat{\textrm{corr}}_{i,i^\prime} = \frac{\widehat{\textrm{cov}}_{i,i^{\prime}}}{\sqrt{\widehat{\textrm{cov}}_{i,i} \widehat{\textrm{cov}}_{i^\prime,i^{\prime}}}}~,
\end{equation}
respectively, where $\bar{\varepsilon}_i=\frac{1}{d} \sum_{j=1}^d \tilde{\varepsilon}^{(j)}_i$ is the average of the normalized residuals across outcomes. The idea is that the statistic $\widehat{\textrm{corr}}_{i,i^\prime}$ measures the tendency for residuals associated with the pair $i,i'$ to be correlated. 

\begin{figure}[t]
\begin{centering}
\caption{Characteristics of Highly Correlated County Pairs\label{fig: predictors of correlation}}
\medskip{}
\begin{tabular}{c}
\includegraphics[width=\textwidth]{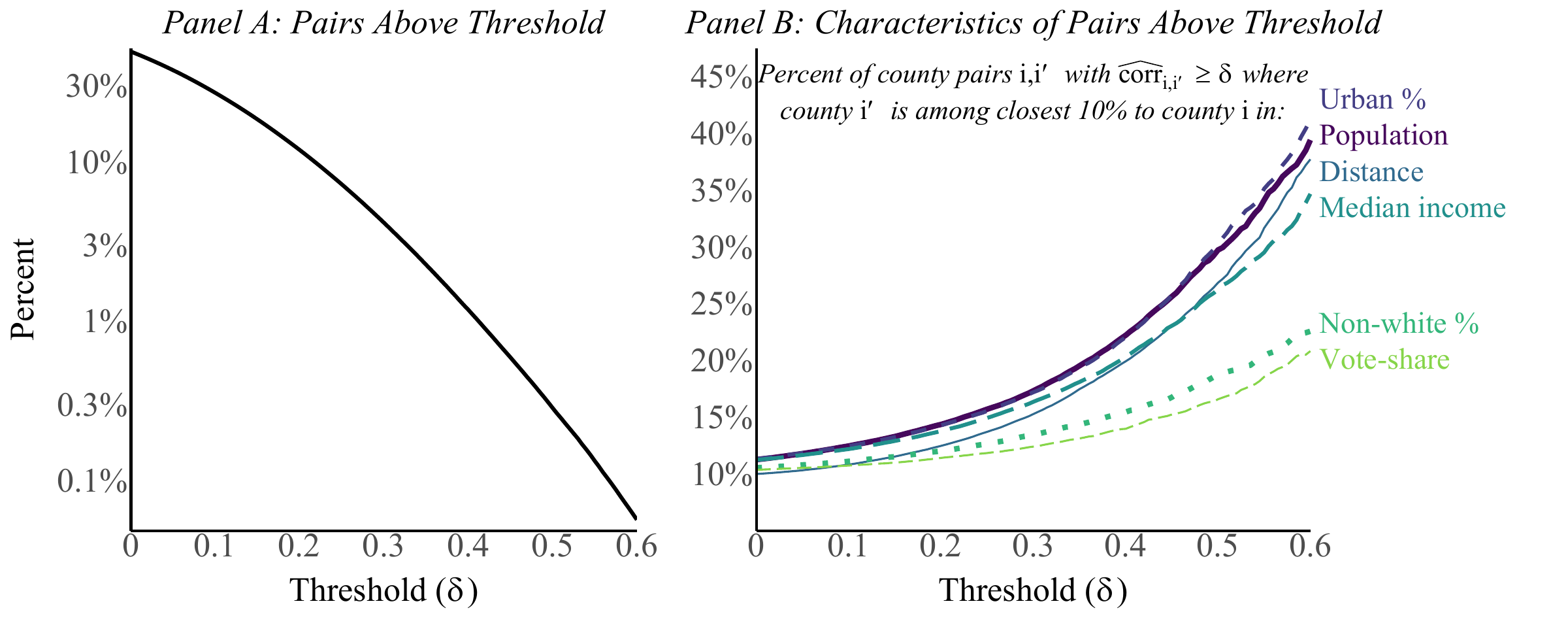}
\tabularnewline
\end{tabular}
\par\end{centering}
\medskip{}
\justifying
{\noindent\footnotesize{}Notes: \cref{fig: predictors of correlation} displays the characteristics of highly correlated pairs of counties. Panel A shows the percent of county pairs $i,i^\prime$ with $\widehat{\textrm{corr}}_{i,i^\prime}\geq\delta$. The y-axis is displayed on a logarithmic scale. Panel B shows, among these pairs, the percent for which county $i^\prime$ is among the 10\% of counties closest to county $i$ on each characteristic.}{\footnotesize\par}
\noindent\hrulefill
\end{figure}

These measurements behave intuitively. Panel A of \cref{fig: predictors of correlation} plots, as a function of the positive threshold $\delta$, the proportion of county pairs $(i,i^\prime)$ for which the correlation estimate $\widehat{\textrm{corr}}_{i,i^\prime}$ is at least $\delta$. Panel B plots, among these pairs, the proportion for which county $i^\prime$ is among the 10\% of counties closest to county $i$ along each of several observable characteristics. Proximity on observable characteristics is very predictive of having highly correlated residuals. Geographic proximity does not tell the whole story, however. Demographic and socioeconomic proximity, e.g., income, population, and urbanization, are similarly predictive.\footnote{\cref{tab: closeness overlap} shows that the characteristics identify largely distinct sets of county pairs. There is little overlap between the pairs of counties that are close in one dimension and those that are close in another.} 

\subsection{Calibrated Simulation\label{sec: simulation design}}

We now assess, through a calibrated simulation, whether ignoring the correlations measured in \cref{fig: predictors of correlation} leads to biased standard errors. We use the estimates $\widehat{\textrm{corr}}_{i,i^\prime}$ to parameterize a covariance matrix $\Sigma$ for the residual vector $\varepsilon$ in the specification \eqref{eq: linear model}. There are two desiderata. First, the non-zero off-diagonal elements of $\Sigma$ should primarily consist of pairs of counties whose correlation estimate $\widehat{\textrm{corr}}_{i,i^\prime}$ is likely to reflect real dependence. Second, the covariance matrix $\Sigma$ must be a proper, positive-definite covariance matrix, so that it can be used to draw simulation replicates.  To achieve these goals, we partition the counties into 99 clusters of roughly equal size using a deterministic rule, described in \cref{sec: app design}. We first choose 99 seed counties with large positive correlations with many other counties, and then grow the clusters sequentially by adding an unassigned county with the highest average correlation to the current members of a cluster that is not yet full. The components of the matrix $\Sigma$ are given by $\widehat{\textrm{corr}}_{i,i^\prime}$ if $i,i'$ belong to the same cluster and are zero otherwise. We set the treatment $W$ as the change in the percentage of population in county $i$ with a completed college degree from 1980 to 2009. For each of 1,000 simulation replicates, we draw a random vector $\varepsilon$ from $\mathsf{N}(0,\Sigma)$, set $Y = \tau W + \varepsilon$, where $\tau = 0$, and apply various methods to estimate the standard error associated with the regression of $Y$ on $W$. 

\cref{tab: Bias Estimates} displays the results.  The first row displays the bias for heteroskedasticity-consistent standard errors (HC1). Across the 1,000 simulation draws, the mean estimated standard error is only 0.50 of the true standard error, leading to a rejection rate of 0.34. Conventional methods to address spatial correlation, for example, clustering by state (row 2), using \cite{conley1999gmm} standard errors with a 150-mile radius (row 3), or the Spatial Correlation Principal Components (SCPC) method of \cite{muller2022spatial, muller2023spatial}, show little improvement, with the average standard error still at roughly 0.55 of the true standard error.\footnote{\cite{conley1999gmm} standard errors are consistent only if the number of location pairs within the bandwidth is small relative to the squared sample size. A bandwidth of 150 miles includes 3.2\% of county pairs. A 300-mile bandwidth would include 11.4\% of pairs, moving beyond where we expect consistency to hold. The methods in \cite{bester2016fixed} and \cite{muller2022spatial, muller2023spatial}, by contrast, control size under various conditions in strongly correlated data. Accordingly, the SCPC method of \cite{muller2022spatial, muller2023spatial} exhibits a lower rejection rate, as it uses a larger critical value that accounts for the variability of the variance estimator.}

\begin{table}[t] 
\centering
\begin{centering}
\caption{Comparison of Methods for Adjusting for Spatial Dependence\label{tab: Bias Estimates}}
\medskip{}
\begin{tabular} {l c c} \toprule Method & Mean(Est. SE / True SE) & Rejection Rate \\ \hline 
(1) HC1 & 0.50 & 0.34 \\
(2) Cluster by state & 0.54 & 0.29 \\
(3) Uniform kernel (150mi) & 0.55 & 0.30 \\
(4) SCPC & 0.54 & 0.18 \\ \hline
(5) TMO & 0.91 & 0.09 \\
\bottomrule \end{tabular}

\par\end{centering}
\medskip{}
\justifying
{\noindent\footnotesize{}Notes: \cref{tab: Bias Estimates} reports the performance of various methods in the calibrated simulation experiment outlined in \cref{sec: simulation design}. The treatment of interest is the change in the percent of college graduates from 1980 to 2009. We use 1,000 replications. For each replication, we draw 50 independent auxiliary outcomes. \say{HC1} refers to HC1 heteroskedasticity-consistent standard errors \citep{hinkley1977jackknifing}. \say{Uniform kernel (150mi.)} refers to the standard error estimator proposed by \cite{conley1999gmm}, implemented with a uniform kernel and a bandwidth of 150 miles. \say{SCPC} refers to the Spatial Correlation Principal Components method proposed by \cite{muller2022spatial, muller2023spatial}. \say{TMO} refers to the Thresholding Multiple Outcomes estimator developed in \cref{sec: proposal,sec: risk approximation sec,sec: threshold choice}.}{\footnotesize\par}
\noindent\hrulefill
\end{table}

\section{Thresholding Multiple Outcomes\label{sec: proposal}}

The results in \cref{tab: Bias Estimates} are concerning. In a setting calibrated to approximate the cross-sectional dependence exhibited by outcomes frequently studied in economics, standard methods can be severely biased. In this section, we propose an alternative method. We specify the problem, introduce our two main assumptions, and then detail our method and illustrate its performance in the simulation introduced in \cref{sec: simulation design}.

\subsection{Problem}

As before, we consider the linear model
\begin{equation} \label{eq: linear model re appear}
    Y^{(0)}_i = \alpha + \tau^{(0)} W_i + \varepsilon^{(0)}_i,\quad\mathbb{E}[\varepsilon^{(0)}_i|W]=0,\quad\quad i = 1,\ldots,n~,
\end{equation}
where $Y^{(0)}_i$ and $W_i$ denote an outcome and treatment of interest. Extensions to settings with covariates, instruments, and panel data are treated in \cref{sec: extensions}. The statistic $\hat{\tau}^{(0)}$ denotes the ordinary least squares estimate of $\tau^{(0)}$ in the specification \eqref{eq: linear model re appear} and we are interested in estimating an appropriate standard error for $\hat{\tau}^{(0)}$. For the purpose of this paper, we take a model-based perspective, and interpret the problem as desiring an estimate of the conditional variance
\begin{equation} \label{eq: standard error expression}
    V_n = \Var(\hat{\tau}^{(0)} \mid W) = (\bar{W}^\top \bar{W})^{-1}\left( \bar{W}^\top \Sigma^{(0)} \bar{W}\right)(\bar{W}^\top \bar{W})^{-1}~,
\end{equation}
where
\begin{align}
\bar{W} & = (W_i - \frac{1}{n} \sum_{k=1}^n W_k)_{i=1}^n~,\nonumber\\
\Sigma^{(0)} & = (\sigma^{(0)}_{i,i^\prime})_{i,i^\prime=1}^n~,\quad\text{and}\quad\sigma^{(0)}_{i,i^\prime} = \Cov(\varepsilon^{(0)}_{i}, \varepsilon^{(0)}_{i^\prime} \mid W)~.\label{eq: Sigma 0 def}
\end{align}
As a consequence, estimation of the conditional variance $V_n$ reduces to estimation of the covariance matrix $\Sigma^{(0)}$. This perspective is shared by many papers in the literature on spatial dependence (see, e.g., \citealt{kelejian2007hac,colella2023acreg,muller2023spatial}).

\subsection{Multiple Outcomes and Proportionality\label{sec: multiple outcomes}}

In the absence of further conditions, the variance \eqref{eq: standard error expression} is not consistently estimable. As a consequence, researchers impose a variety of restrictions on the structure of cross-sectional correlation. A common assumption is that all the off-diagonal elements of the matrix $\Sigma^{(0)}$ are equal to zero \citep{white1980heteroskedasticity}. In other cases, researchers assume that units can be partitioned into clusters on the basis of some shared characteristic \citep{moulton1986random,moulton1990illustration,liang1986longitudinal}, or that cross-sectional correlation can be bounded above by a function of geographic proximity \citep{conley1999gmm, bester2016fixed, muller2022spatial, muller2023spatial}. 

We take an alternative route. Our approach is premised on the observation that, often, in a single study, researchers observe many outcome variables for the same units. For example, a researcher, interested in the correlation across U.S.\ counties between average income and the proportion of the population with a college education, also observes the outcomes considered in \cref{sec: simulation}. Formally, for each unit $i$, in addition to the outcome of interest $Y_i^{(0)}$, we assume that we also observe a collection $Y^{(1)}_i,\ldots,Y^{(d)}_i$ of $d$ auxiliary outcomes. Let $\varepsilon_{i}^{(j)}$ and $\hat{\varepsilon}_{i}^{(j)}$ denote the population and empirical residuals associated with unit $i$ when outcome $Y^{(j)}_i$ replaces $Y^{(0)}_i$ in the linear model \eqref{eq: linear model re appear}, respectively.

The basic idea is to use the auxiliary outcomes to estimate the residual correlation matrix associated with the outcome of interest. To rationalize this approach, we must impose some conditions. First, we impose a restriction that implies that the residual \emph{correlation} matrix is shared across outcomes. We call a collection of outcomes with this property \emph{proportional}.

\begin{assumption}[Proportionality]\label{assu: proportionality}$ $\\
There exist positive semi-definite matrices $\Lambda_n = (\lambda_{i,i'})_{i,i'=1}^n$ and $\Gamma_d = (\gamma_{j,j'})_{j,j'=0}^d$ such that
\begin{align}
    \Cov(\varepsilon^{(j)}_i,\varepsilon^{(j')}_{i'} \mid W) = \lambda_{i,i'} \gamma_{j,j'}\label{eq: proportionality}
\end{align}
for each pair of units $i,i'$ and outcomes $j,j'$, respectively. We impose the normalization $\gamma_{0,0} = 1$.
\end{assumption}
\noindent The restriction \eqref{eq: proportionality} requires that the covariance between the residuals associated with two units and a pair of, potentially distinct, outcomes depends only on the identity of the outcomes through a constant factor.\footnote{\cref{assu: proportionality} can be motivated by a factor model. For instance, suppose that we can write $\varepsilon_i^{(j)} = r^{-1/2} U_j^\top F_i$, where the factors $U_j$ and $F_i$ have dimension $r$. If the factors $U_j$ and $F_i$ are independent, and $\Cov(F_{i,k},F_{i',l}) = \lambda_{i,i'} \mathbb{I}\{k = l\}$ and $\Cov(U_{j,k},U_{j',l}) = \gamma_{j,j'} \mathbb{I}\{k = l\}$ both hold, then we can evaluate $\Cov(\varepsilon^{(j)}_i,\varepsilon^{(j')}_{i'}) = \lambda_{i,i'} \gamma_{j,j'}$.}

As a consequence, the empirical residuals for the auxiliary outcomes can be used to estimate the covariance matrix $\Sigma^{(0)}$. Our approach for operationalizing this idea has three steps. First, we normalize the residuals to have variance one across outcomes. That is, we compute
\begin{equation}\label{eq: tilde epsilon}
\tilde\varepsilon_{i}^{(j)}
=
\hat{\varepsilon}^{(j)}_i \hat{\gamma}_{j}^{-1/2}~,
\quad\text{where}\quad
\hat{\gamma}_{j} = \frac{1}{n} \sum_{i=1}^n (\hat{\varepsilon}^{(j)}_i)^2
\end{equation}
denotes the sample variance of the residuals for the $j$th outcome. Second, we orthogonalize these normalized residuals for the auxiliary outcomes relative to the normalized residuals for the outcome of interest. In particular, we solve the regression
\begin{align}
    \tilde{\psi}^{(j)} = \arg\min_{\psi \in \mathbb{R}} \left\{ \sum_{i=1}^n ( \tilde{\varepsilon}_i^{(j)} - \psi \tilde{\varepsilon}_i^{(0)})^2
    \right\}
    \quad
    \text{and compute}\quad
    \tilde{\varepsilon}^{(j;0)}_i = \tilde{\varepsilon}^{(j)}_i - \tilde{\psi}^{(j)} \tilde{\varepsilon}_i^{(0)}
    \label{eq: first stage regression}
\end{align}
for each auxiliary outcome $j$. We discuss the motivation for implementing this step in further detail in \cref{sec: bias}. Finally, we compute the covariance and correlation between these normalized and orthogonalized residuals for units $i$ and $i^\prime$, across outcomes, by
\begin{equation}\label{eq: lambda ii}
\tilde{\lambda}_{i,i^{\prime}} = \frac{1}{d} \sum_{j=1}^d \tilde{\varepsilon}^{(j;0)}_i\tilde{\varepsilon}^{(j;0)}_{i'}
\quad\text{and}\quad
\tilde{\rho}_{i,i^\prime} = \frac{\tilde{\lambda}_{i,i^{\prime}}}{\sqrt{\tilde{\lambda}_{i,i} \tilde{\lambda}_{i^{\prime},i^{\prime}}}}~.
\end{equation}
The statistic $\tilde{\rho}_{i,i^\prime}$ is closely related to the correlation estimates considered in \cref{sec: simulation}.\footnote{The values of the statistics $\tilde{\lambda}_{i,i^{\prime}}$ and $\tilde{\rho}_{i,i^\prime}$ are invariant to changes in the sign of any of the outcomes.}

Proportionality is a highly stylized condition and should be viewed as only ever holding approximately for any given collection of outcomes. The same holds, of course, for assumptions that decompose spatial correlation into clusters or bound spatial correlation in terms of geography. By imposing \cref{assu: proportionality}, we appeal to the intuition that the cross-sectional correlation exhibited by the auxiliary outcomes captures the main features of the  cross-sectional correlation of the outcome of interest. By doing so, we are ignoring aspects of spatial correlation that might be particular to any given outcome.\footnote{Assumptions equivalent to proportionality have been considered in other fields of applied statistics. In particular, a literature in the analysis of micro-array experiments, initiated by \cite{efron2010correlated}, develops procedures for multiple hypothesis testing in proportional data. See, e.g., \cite{muralidharan2010detecting}, \cite{allen2012inference}, and \cite{hoff2016limitations} for further discussion. More generally, \citet{dawid1981some} gives a classical discussion of applications of models satisfying \cref{assu: proportionality} to Bayesian inference. \cite{gupta2018matrix} gives a textbook treatment of related distributions.} 

\subsection{Sparsity}

Proportionality, however, is not enough. Even if a collection of auxiliary outcomes that satisfy condition \eqref{eq: proportionality} is available (and even if these outcomes were mutually independent), any estimator of the conditional variance \eqref{eq: standard error expression} can still be made arbitrarily inaccurate. (We formalize this claim in \cref{sec: minimax}.) 

We thus enforce the restriction that the residuals for \emph{most} pairs of units are \emph{not} meaningfully correlated. That is, pairs of units can be categorized as ``nulls'' and ``non-nulls.'' The components of the residual covariance matrix $\Sigma^{(0)}$ that correspond to the null pairs are equal to zero, the components that correspond to non-nulls can be large, and the nulls outnumber the non-nulls.

We express this condition as follows. Throughout, the quantities $c < 1$ and $C$ denote universal positive constants, whose values may change at each appearance.  We write $a\lesssim b$ if $a\leq Cb$ and $a\asymp b$ if the reverse inequality also holds.

\begin{assumption}[Sparsity and Balance]\label{assu: sparsity}$ $\\
The components of the matrix $\Lambda_n$ satisfy
\begin{align}
    \max_{i\in\{1,\ldots,n\}}
    \sum_{i'\neq i}
    \mathbb{I}\{\lambda_{i,i'}\neq0\}
    &\leq s
    \quad\text{and}\quad
    \max_{i\in\{1,\ldots,n\}}
    \frac{1}{n}
    \sum_{i'=1}^n|\lambda_{i,i'}|
    \lesssim \frac{1}{n^2} \sum_{i=1}^n\sum_{i'=1}^n \lambda_{i,i'}\label{eq: sparsity and balance}
\end{align}
for an integer $1\leq s<n-1$.
\end{assumption}

\noindent The first condition in \eqref{eq: sparsity and balance} requires that at most $s$ off-diagonal elements of any row of $\Lambda_n$ are not zero.\footnote{The results of this paper should hold without substantive differences if the indicator $\mathbb{I}\{\lambda_{i,i'}\neq0\}$ were replaced by $\vert \lambda_{i,i'}\vert^q$, for $q < 1$. Conditions like \cref{assu: sparsity} were first applied to the normal-means problem by \cite{abramovich2006special}, and have been applied to estimation of covariance matrices in independent data by \cite{bickel2008covariance} and \cite{cai2011adaptive}, among others.} In the asymptotic analysis that follows, we focus our attention on regimes in which, at a minimum, $s$ is negligible relative to the sample size $n$. Similar assumptions are required in the literature on estimation of clustered standard errors \citep{hansen2019asymptotic}.\footnote{Methods for constructing clustered standard errors with large clusters are either conservative \citep{ibragimov2010t,ibragimov2016inference},  require the joint limiting distribution of estimation errors from regressions fit separately within each cluster to be invariant to separate sign changes \citep{canay2017randomization}, or require the limiting covariance matrices of the explanatory variables to be proportional across clusters \citep{canay2021wild}.}  The second condition in \eqref{eq: sparsity and balance} requires that the average absolute covariance associated with any one unit is not disproportionately large relative to the average covariance across all pairs. In our asymptotic analysis, this will play the role of a Lindeberg condition. This also rules out, for instance, situations where there is extensive cancellation between positive and negative covariances.

\subsection{Thresholding Multiple Outcomes}

The sparsity encoded in \cref{assu: sparsity} is a generalization of the idea that underlies clustering. There, pairs of units in the same cluster are non-null \textit{a priori}. Here, our aim is to use the auxiliary outcomes to \emph{determine} the pairs of units that are non-nulls.  We propose to make this determination by choosing pairs of units whose empirical residual correlation $\tilde{\rho}_{i,i^\prime}$ is large in absolute value.\footnote{An alternative approach, not pursued in this paper, is to use the estimates $\tilde{\rho}_{i,i^\prime}$ to learn clusters of units (see, e.g., \citealt{cao2024inference}). This has the disadvantage that, necessarily, some geographically proximate units are determined to not correlate.} We estimate
\begin{equation} \label{eq: arbitrary delta}
    \tilde{\sigma}^{(0)}_{i,i^\prime}(\delta) 
    =
    \begin{cases}
    \hat{\varepsilon}^{(0)}_i \hat{\varepsilon}^{(0)}_i ~,& i = i^\prime, \\
    \hat{\varepsilon}^{(0)}_i \hat{\varepsilon}^{(0)}_{i^\prime}\text{ } 
    \mathbb{I}\{ \vert \tilde{\rho}_{i,i^\prime}\vert \geq \delta \}~, & i\neq	i^\prime~.
    \end{cases}
\end{equation}
Collect these estimates into the matrix $\tilde{\Sigma}^{(0)}_{n,d}(\delta)$ and plug this matrix into the expression \eqref{eq: standard error expression}, giving
\begin{equation}\label{eq: V hat delta general}
\tilde{V}_{n,d}(\delta) = \left(\bar{W}^\top \bar{W}\right)^{-1} \left( \bar{W}^\top \tilde{\Sigma}_{n,d}^{(0)}(\delta) \bar{W}\right)\left(\bar{W}^\top \bar{W}\right)^{-1}~.
\end{equation}
This construction is analogous to heteroskedasticity-robust or cluster-robust standard errors that use terms like $ \hat{\varepsilon}^{(0)}_i \hat{\varepsilon}^{(0)}_{i^\prime} $ to estimate non-zero elements of residual covariance matrices \citep{white1980heteroskedasticity,liang1986longitudinal}.\footnote{As the matrix $\tilde{\Sigma}^{(0)}_{n,d}(\delta)$ is not necessarily positive definite, the estimator \eqref{eq: V hat delta general} is not necessarily positive. Negative variance estimates are not observed in our simulations or applications.} The difference is that, here, the auxiliary outcomes are used to determine which elements of the residual covariance matrix will be non-zero. The estimator \eqref{eq: V hat delta general} is thus minimally sensitive to \cref{assu: proportionality}. Its value is affected by the auxiliary outcomes only through the choice of which covariances to zero out.

We refer to the estimator \eqref{eq: V hat delta general} as the ``TMO'' variance estimator, for ``Thresholding Multiple Outcomes,'' and summarize its construction in \cref{alg: tmo}. It remains to specify, and justify, a procedure for choosing the threshold $\delta$. This is the purpose of the formal analysis detailed in \cref{sec: risk approximation sec,sec: threshold choice}.

\begin{remark}
    In the formulation \eqref{eq: arbitrary delta}, the TMO estimator \eqref{eq: V hat delta general} does not threshold diagonal elements of the residual covariance matrix. That is, in effect, we augment the standard \cite{white1980heteroskedasticity} heteroskedasticity-consistent variance estimate. The same idea can be applied to wrap our approach around other existing variance estimators. For instance, an analogous estimator could be constructed where elements that correspond to pairs of units in the same cluster, e.g., state, are never thresholded to zero. The spatial standard errors proposed by \cite{conley1999gmm} and \cite{muller2022spatial, muller2023spatial} can be augmented analogously. See \cref{app: augmenting} for further discussion. We compare the performance of these alternatives in \cref{sec: applications}.\hfill$\blacksquare$
\end{remark}

\begin{remark}
    The TMO estimator \eqref{eq: V hat delta general} can be viewed as an instance of the variance estimator proposed in \cite{conley1999gmm}, where the ``distance'' between units $i$ and $i'$ is $\arccos(\vert \tilde{\rho}_{i,i'}\vert)$ and the kernel is uniform with bandwidth $\arccos(\delta)$. The essential difference is that this measure of distance is explicitly estimated and, thereby, stochastic, in the sense that $|\lambda_{i,i'}|$ is not bounded by a decreasing function of $\arccos(\vert \tilde{\rho}_{i,i'}\vert)$.\footnote{\citet{conley1999gmm}, \cite{conley2007spatial}, \citet{kelejian2007hac}, and \citet{kim2011spatial} give extensions to settings where distance is measured with error, by imposing conditions that imply that the scale of this error is negligible relative to the bandwidth used by the estimator. By contrast, we focus on settings where estimation error in $\tilde{\rho}_{i,i'}$ is non-negligible relative to the relevant thresholds $\delta$, and, in fact, makes a first-order contribution to the performance of the estimator.} \cite{conley1999gmm} emphasizes that dependence should be modeled with the ``economic distance'' between units. Nevertheless, in most applications of this method, economic distance is proxied by geographic distance. Our contribution can, thus, be framed as giving a general procedure for \emph{estimating} economic distance on the basis of auxiliary outcomes, as well as a method for choosing a suitable bandwidth.\hfill$\blacksquare$
\end{remark}

\begin{algorithm}[t]
\caption{Thresholding Multiple Outcomes (TMO)} \label{alg: tmo}
\setstretch{1.35}
\SetAlgoHangIndent{0pt}
\KwIn{Outcome $Y_i^{(0)}$ and treatment $W_i$ for units $i$ in $1,...,n$}

Identify a collection of auxiliary outcomes $Y_i^{(1)},\ldots,Y_i^{(d)}$

For each unit $i$ and outcome $j$, compute the residuals $\tilde\varepsilon_{i}^{(j;0)}$, defined in \eqref{eq: tilde epsilon} and \eqref{eq: first stage regression}

For each pair of units $i\neq i^\prime$, measure the correlation $\tilde{\rho}_{i,i^\prime}$, defined in \eqref{eq: lambda ii}

Choose the threshold $\hat{\delta}_{n,d}^\star$ by applying \cref{alg: threshold}

Collect the estimates
\begin{equation*}
    \tilde{\sigma}^{(0)}_{i,i^\prime}(\hat{\delta}_{n,d}^\star) 
    =
    \begin{cases}
    \hat{\varepsilon}^{(0)}_i \hat{\varepsilon}^{(0)}_i ~,& i = i^\prime, \\
    \hat{\varepsilon}^{(0)}_i \hat{\varepsilon}^{(0)}_{i^\prime}\text{ } 
    \mathbb{I}\{ \vert \tilde{\rho}_{i,i^\prime}\vert \geq \hat{\delta}_{n,d}^\star \}~, & i\neq	i^\prime~,
    \end{cases}
    \quad\text{into}\quad\tilde{\Sigma}_{n,d}^{(0)}(\hat{\delta}_{n,d}^\star)
\end{equation*}

\nonl \textbf{Return} the variance estimate
\begin{equation*}
\tilde{V}_{n,d}(\hat{\delta}_{n,d}^\star) = \left(\bar{W}^\top \bar{W}\right)^{-1} \left( \bar{W}^\top \tilde{\Sigma}_{n,d}^{(0)}(\hat{\delta}_{n,d}^\star) \bar{W}\right)\left(\bar{W}^\top \bar{W}\right)^{-1}
\end{equation*}

\nonl \hrulefill
\setstretch{1}

\nonl {\footnotesize{}Notes: \cref{alg: tmo} details the Thresholding Multiple Outcomes (TMO) estimator proposed in this paper. Extensions of this procedure to settings with covariates, instrumental variables, and panel data are given in \cref{sec: extensions}. }{\footnotesize\par}
\end{algorithm}
\smallskip

\subsection{Performance\label{sec: performance}}

Before proceeding with a more technical analysis of threshold choice, we return to the calibrated simulation considered in \cref{sec: simulation}. The fifth row of \cref{tab: Bias Estimates} displays estimates of the bias and rejection rate of the TMO estimator. In each replication, we generate 50 independent auxiliary outcomes from the same specification used to generate the outcome of interest. The bias of TMO is substantially smaller than that of standard methods, and its associated rejection rate is closer to the nominal Type I error rate. 

\begin{figure}[t]
\begin{centering}
\caption{Distributional Comparison of Performance Across Treatments\label{fig: simulation cdfs}}
\medskip{}
\begin{tabular}{cc}
\textit{Panel A: Bias} & \textit{Panel B: Rejection Rate}\tabularnewline
 \includegraphics[width=0.5\textwidth]{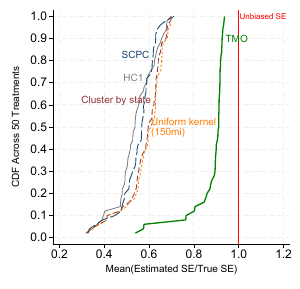}  & \includegraphics[width=0.5\textwidth]{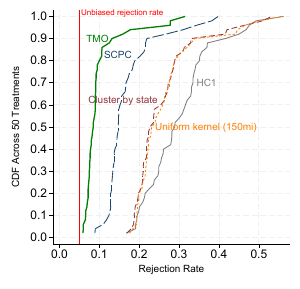}
\tabularnewline
\end{tabular}
\par\end{centering}
\medskip{}
\justifying
{\noindent\footnotesize{}Notes: \cref{fig: simulation cdfs} displays CDFs of the bias and rejection rates associated with various methods across several specifications of the calibrated simulation experiment outlined in \cref{sec: simulation design}. In particular, the distributions are taken across the specifications where each of the 50 outcome variables discussed in  \cref{sec: simulation} is set as the treatment variable. The vertical red lines display the hypothetical performance of an unbiased estimator and an unbiased test, respectively.}{\footnotesize\par}
\noindent\hrulefill
\end{figure}

A concern is that these results may be driven by the treatment considered in \cref{sec: simulation}, i.e., the change in the percentage of the population in county $i$ with a completed college degree. We thus replicate the Monte Carlo experiment, setting each of the 50 outcome variables as the treatment variable. \cref{fig: simulation cdfs} displays the distribution functions of the bias and rejection rate associated with each method across these specifications. The TMO standard errors have uniformly better bias and more accurate rejection rates. 

We note that this simulation is particularly well-suited to the TMO estimator, as the restrictions imposed by \cref{assu: proportionality} are well-specified and the spatial correlation structure used in the simulation is sparse. More diffuse, geographically structured correlation patterns may be better suited for alternative methods. We interpret these results as demonstrating that, \textit{when} \cref{assu: proportionality,assu: sparsity} are plausible, the TMO estimator can account for realistic forms of spatial correlation that are not captured by existing methods. 

\section{Bias, Risk, and Consistency\label{sec: risk approximation sec}}

In this section, we study a specialization of the setting above. There are two main results. First, we give closed-form approximations to the bias and root-mean-squared error of a simplified version of the TMO estimator. Second, we use these approximations to characterize the settings in which the root-mean-squared error of this estimator is determined, to leading order, by its absolute bias, and argue that this bias-dominated regime is empirically relevant. These results enable us to develop a procedure for choosing the threshold $\delta$, which we treat in the subsequent section.

\subsection{Simplifications}

For the sake of simplicity, we treat a specialization of the problem in \cref{sec: proposal}. We simplify the problem in three ways. First, we focus attention on inference for the mean. In this case, we can write 
\begin{align}
    \hat{\varepsilon}_i^{(j)} = Y_i^{(j)} - \bar{Y}^{(j)}~,
    \quad\text{where}\quad
    \bar{Y}^{(j)} = \frac{1}{n} \sum_{i=1}^n Y_i^{(j)}~,
\end{align}
and our interest is in estimating the parameter
\begin{align}
    V_n 
    = \Var(\bar{Y}^{(0)})= \frac{1}{n^2} \sum_{i=1}^n \sum_{i^\prime=1}^n \sigma_{i,i^\prime}^{(0)}~,
    \quad\text{where}\quad \Sigma^{(j)} = (\sigma_{i,i
    '}^{(j)})_{i,i'=1}^n\label{eq: objective}
\end{align}
denotes the covariance matrix of the random vector $Y^{(j)}=(Y^{(j)}_i)_{i=1}^n$. For each unit $i$, write $Y_i = (Y_i^{(j)})_{j=0}^d$ and let  $Y$ denote the matrix with rows $Y_i$.

Second, we assume that the matrices $\Lambda_n$ and $\Gamma_d$ introduced in \cref{assu: proportionality} are homoskedastic and well-conditioned, in the following sense.

\begin{assumption}[Homoskedasticity]\label{assu: homoskedasticity} Each of the diagonal components of $\Lambda_n$ and $\Gamma_d$ is equal to one.
\end{assumption}

\begin{assumption}[Well-Conditioned Covariance]\label{assu: well conditioned}$ $\\
The matrices $\Gamma_d$ and $\Lambda_{n}$ are uniformly well-conditioned, in the sense that
\begin{align}
    u^\top \Gamma_{d} u
        \asymp \| u \|_2^2
        \quad\text{and}\quad
       v^\top \Lambda_{n} v \asymp \| v \|_2^2
\end{align}
for every $u$ in $\mathbb{R}^{d+1}$, $v$ in $\mathbb{R}^n$, and each $n,d$.
\end{assumption}

\noindent \cref{assu: homoskedasticity,assu: well conditioned} enable us to consider the following simplified version of the TMO estimator. Define the statistic
\begin{align}
\hat{\lambda}_{i,i^{\prime}} &= \frac{1}{d} \sum_{j=1}^d 
\hat{\varepsilon}^{(j;0)}_i
\hat{\varepsilon}^{(j;0)}_{i'}~,
\quad\text{where}\label{eq: baseline auxiliary covariance statistic}\\
\hat{\varepsilon}^{(j;0)}_i &= \hat{\varepsilon}^{(j)}_i - \hat{\psi}^{(j)} \hat{\varepsilon}_i^{(0)}
\quad\text{and}
\quad
\hat{\psi}^{(j)} = \arg\min_{\psi} \sum_{i=1}^n ( \hat{\varepsilon}_i^{(j)} - \psi \hat{\varepsilon}_i^{(0)})^2~,\label{eq: baseline auxiliary covariance residuals}
\end{align}
as well as the simplified estimator
\begin{align}
    \hat{V}_{n,d}(\delta) = \frac{1}{n^2} \sum_{i=1}^n \sum_{i^\prime =1}^n\hat{\sigma}^{(0)}_{i,i'}(\delta) ~,
    \quad\text{where}\quad
    \hat{\sigma}^{(0)}_{i,i^\prime}(\delta) 
    =
    \hat{\varepsilon}^{(0)}_i \hat{\varepsilon}^{(0)}_{i^\prime}\text{ } 
    \mathbb{I}\{ \vert \hat{\lambda}_{i,i^\prime} \vert \geq \delta \}~.\label{eq: V hat}
\end{align}
Relative to the TMO estimator \eqref{eq: V hat delta general}, the simplified estimator \eqref{eq: V hat} uses the covariance \eqref{eq: baseline auxiliary covariance statistic} to discriminate between nulls and non-nulls and screens diagonal elements of the covariance matrix. Moreover, the covariance \eqref{eq: baseline auxiliary covariance statistic} is constructed with residuals that have not been normalized. These simplifications will not have a qualitative impact on our theoretical results, but substantially expedite the required calculations.

Finally, we assume that the data under consideration have a joint Gaussian distribution.

\begin{assumption}[Gaussianity]\label{assu: gaussianity} The joint distribution of the components of the matrix $Y$ is Gaussian.    
\end{assumption}

\noindent \cref{assu: gaussianity} will not allow us to dispense with asymptotics. For instance, many of our results will require the use of bounds on the discrepancy between the distribution of the covariance \eqref{eq: baseline auxiliary covariance statistic} and a Gaussian distribution, which shrink as the sample size $n$ and number of auxiliary outcomes $d$ increase. Nevertheless, our analysis relies on extensive use of the exact Gaussian structure imposed by \cref{assu: gaussianity}.

\subsection{Reparametrization}

To state the results that follow, it will be convenient to implement a reparameterization of the threshold $\delta$. To do this, we will need to introduce some further notation. In our analysis, the sequence
\begin{align}
    v_{0}^{(d)}
    &=\frac{1}{d^2}\|\bar\Gamma_d\|_{\mathrm F}^2~,
    \quad\text{where}\quad \bar\Gamma_d = (\gamma_{j,j'})_{j,j'=1}^d~,
\end{align}
approximates the sampling variability of the covariance \eqref{eq: baseline auxiliary covariance statistic} for a null pair of units. In turn, the function
\begin{align}
    \bar F_{i,i'}^+(\delta)
    &=P\{|\hat\lambda_{i,i'}|\geq\delta\}
    \label{eq: feasible null screening probability}
\end{align}
measures the probability that the pair $i,i'$ is included in the estimator \eqref{eq: V hat} when the threshold $\delta$ is used. 

Now, consider the set of thresholds
\begin{align}
    \Delta_{n,d}(\eta)
    &=\left\{
        \delta\geq0:
        \bar{\mathbf{F}}^{+}_{n,d}(\delta)
        \leq V_n^{\eta}
    \right\}~,
    \quad\text{where}\quad\bar{\mathbf{F}}^{+}_{n,d}(\delta)
    =\max\{\bar F_{i,i'}^+(\delta):\lambda_{i,i'}=0\}~,
    \label{eq: selected threshold set}
\end{align}
for each $\eta \geq 0$. The set $\Delta_{n,d}(\eta)$ collects thresholds where the probability of including a null pair of units in the estimator \eqref{eq: V hat} is at most $V_n^{\eta}$. Written differently,  for each $\delta$ in $\Delta_{n,d}(\eta)$, the expected proportion of the null pairs that are selected is smaller than $V_n^{\eta}$. We characterize this set in the following Lemma.

\begin{lem}[Reparametrization]
\label{lem: selected set characterization} Suppose that
\cref{assu: gaussianity,assu: proportionality,assu: homoskedasticity,assu: sparsity,assu: well conditioned} hold.

\noindent\textbf{(i)} The set $\Delta_{n,d}(\eta)$ is non-empty and takes the form
\begin{align}
    \Delta_{n,d}(\eta)
    &=[\underline\delta_{n,d}(\eta),\infty)~,
    \label{eq: selected threshold ray}
\end{align}
for all $n \geq 3$  and $d \geq 2$.

\noindent\textbf{(ii)} If $\eta>0$ and the limits
\begin{align}
    \frac{s}{n} \to 0~,
    \quad
    dV_n^2
    \log\left(V_n^{-1}\right)
    &\to0~,
    \quad\text{and}\quad
    \log\left(V_n^{-1}\right)
    =o(d)
\end{align}
hold, then
\begin{align}
    \underline\delta_{n,d}(\eta)
    &=
    \sqrt{
        2\eta v_{0}^{(d)}\log\left(V_n^{-1}\right)
    }(1+o(1))
    \label{eq: selected threshold endpoint}
\end{align}
as $n,d\to\infty$.
\end{lem}

\noindent In the results that follow, it will be helpful to interpret the threshold $\delta$ in terms of the reparameterization in $\eta$ suggested by the representation \eqref{eq: selected threshold endpoint}. That is, roughly speaking, by choosing the threshold $\delta = (2\eta v_{0}^{(d)}\log(V_n^{-1}))^{1/2}$, we ensure that the expected proportion of nulls that are retained in the estimator \eqref{eq: V hat} is bounded by $V_n^\eta$, where the maintained conditions imply that $1/n\lesssim V_n \lesssim s/n$. This representation is obtained by developing a moderate deviations approximation to the quadratic form \eqref{eq: baseline auxiliary covariance statistic}, based on an application of the G\"artner-Ellis Theorem (see, e.g., \citealp{dembozeitouni2010}, Theorem 2.3.6(c)). 

\subsection{Bias Approximation\label{sec: bias}}

We begin by giving a closed-form approximation to the bias of the TMO estimator \eqref{eq: V hat}, as the threshold parameter $\delta$ varies. Define the Gaussian tail probability
\begin{align}
        F^{+}(\delta,\mu, v) = P\{ \vert Z_{\mu,v} \vert  \geq \delta \}~,
        \quad\text{where}\quad
        Z_{\mu,v} \sim \mathsf{N}(\mu,v)
\end{align}
as well as the parameters
\begin{align}
    v^{(d)}_{i,i'} & = v_0^{(d)} (1 + \lambda^2_{i,i'})
    \quad\text{and}\quad 
    Q_{i,i'}(\delta)
    =F^{+}(\delta,\lambda_{i,i'},v^{(d)}_{i,i'})
    -F^{+}(\delta,0,v_{0}^{(d)})
\end{align}
for each pair of units $i,i'$. The parameter $v^{(d)}_{i,i'}$ is an approximation to the variance of the statistic \eqref{eq: baseline auxiliary covariance statistic}. Thus, the function $Q_{i,i'}(\delta)$ should be interpreted as a Gaussian approximation to the difference between the probability of retaining the pair $i,i'$ in the TMO estimator and the probability of retaining a null pair.

The following Lemma shows that the bias of the estimator \eqref{eq: V hat} is well-approximated by the function
\begin{align}
    B_{n,d}(\delta)
    &=\frac{1}{n^2}\sum_{i,i'=1}^n
    \lambda_{i,i'}\left(Q_{i,i'}(\delta)-1\right)\label{eq: B def}
\end{align}
for each $\delta \geq 0$. Throughout this section, we focus attention on regimes in which, for a fixed $\eta > 0$, the limit 
\begin{align}
    dV_n^{1+\eta}
    \log(V_n^{-1}) + \frac{s\sqrt d}{n}\to 0
    \label{eq: bias approximation conditions}
\end{align}
holds as $n,d\to \infty$.\footnote{The asymptotic regime \eqref{eq: bias approximation conditions} has the potentially counterintuitive feature that the number of auxiliary outcomes cannot grow too quickly. Roughly speaking, as $\hat{\lambda}_{i,i'}$ is constructed using empirical, rather than population, residuals, it has a bias for $\lambda_{i,i'}$ of order $s/n$, while its standard deviation is of order $d^{-1/2}$. The condition $s\sqrt{d}/n\to0$ ensures that this bias is negligible relative to the standard deviation, allowing us to center a Gaussian approximation at $\lambda_{i,i'}$. The regime considered here is realistic, as the number of auxiliary outcomes available to a researcher is typically much smaller than the number of units.}  

\begin{lem}[Bias Approximation]\label{lem: bias decomposition}$ $\\
Suppose that, for a fixed $\eta \in [0,1]$, \cref{assu: gaussianity,assu: proportionality,assu: homoskedasticity,assu: sparsity,assu: well conditioned} hold. If the condition \eqref{eq: bias approximation conditions} holds, then the bound
\begin{align}
    \sup_{\delta\in\Delta_{n,d}(\eta)}
    \left|
        \frac{
            \mathbb{E}[\hat V_{n,d}(\delta)-V_n]
            -B_{n,d}(\delta)
        }{V_n}
    \right|
    &\lesssim
    dV_n^{1+\eta} \log(V_n^{-1})
    +
    \frac{1}{\sqrt d}
    +\frac{s\sqrt d}{n}
    ~,\label{eq: bias approximation}
\end{align}
holds for all sufficiently large $n,d$.
\end{lem}

\begin{figure}[t]
\begin{centering}
\caption{Bias and Risk\label{fig: bias and risk}}
\medskip{}
\begin{tabular}{c}
\textit{Panel A: Relative Bias} \\
\includegraphics[width=\textwidth]{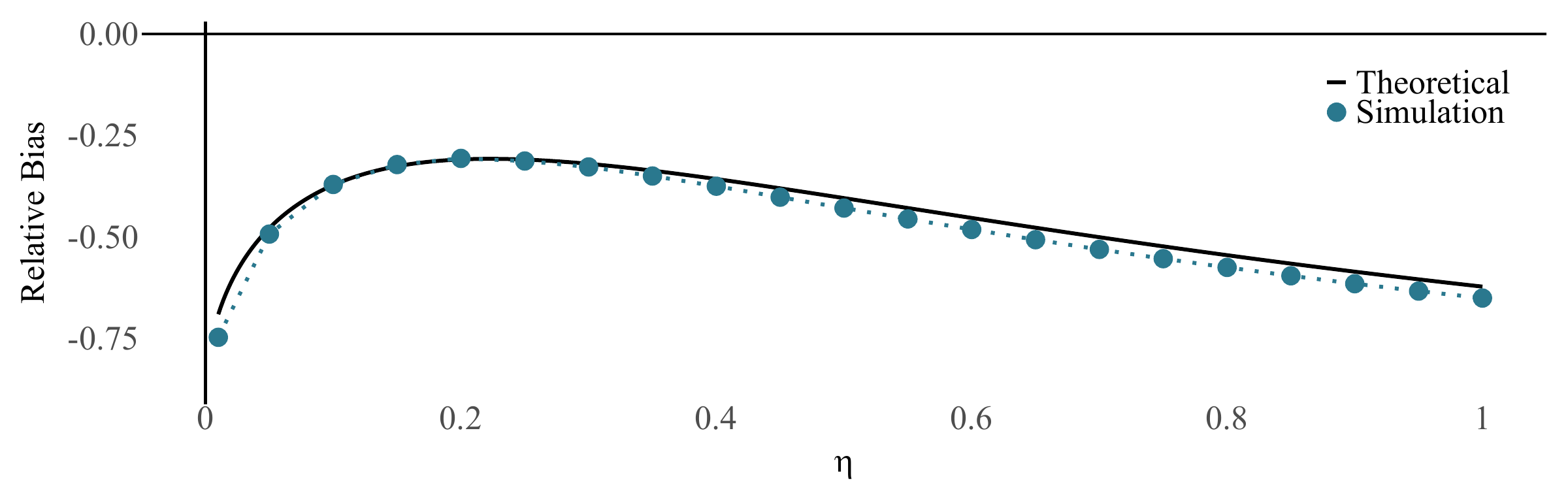} \\
\textit{Panel B: Relative Risk}\\
\includegraphics[width=\textwidth]{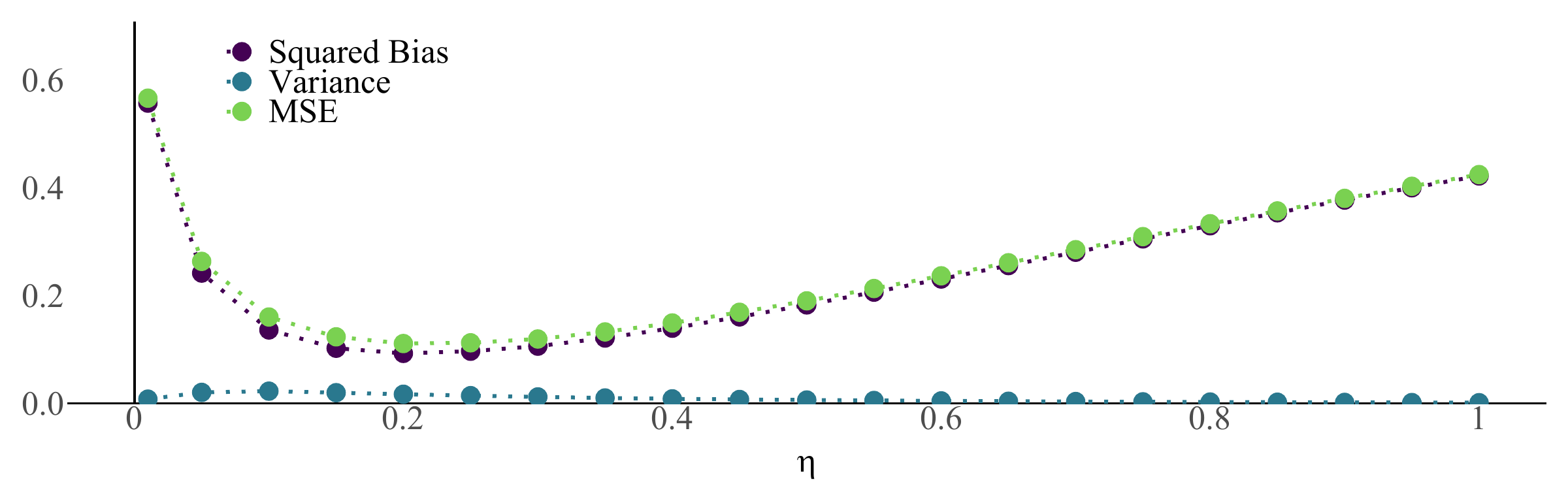}
\tabularnewline
\end{tabular}
\par\end{centering}
\medskip{}
\justifying
{\noindent\footnotesize{}Notes: \cref{fig: bias and risk} displays measures of performance of the estimator \eqref{eq: V hat} in a version of the calibrated simulation described in \cref{sec: simulation design}, where the target parameter is now the variance of the empirical mean of the outcome of interest. Panel A compares measurements of the relative bias to the Gaussian approximation $B_{n,d}(\delta)/V_n$ defined in \eqref{eq: B def}. Panel B displays the relative squared bias, variance, and mean-squared error. The x-axes of both panels are displayed in terms of the index $\eta$ through the reparametrization $\delta=(2\eta v_0^{(d)}\log(V_n^{-1}))^{1/2}$.}{\footnotesize\par}
\noindent\hrulefill
\end{figure}

\noindent Panel A of \cref{fig: bias and risk} displays the relative bias of the estimator \eqref{eq: V hat} in a simulation analogous to the environment considered in \cref{sec: performance}, where the target parameter is now the variance of the empirical mean of the outcome of interest. The relative bias is displayed over a range of values of $\eta$, using the reparameterization $\delta = (2\eta v_{0}^{(d)}\log(V_n^{-1}))^{1/2}$ introduced in the previous subsection. The true relative bias is overlaid with the approximation $B_{n,d}(\delta)/V_n$, which holds with a high level of accuracy.

\begin{remark}
    The bound \eqref{eq: bias approximation} is obtained by first showing that the bias admits the decomposition
    \begin{align}
        \mathbb{E}[\hat{V}_{n,d}(\delta) - V_n] \approx B_{n,d}(\delta) + 
        \frac{1}{n^2}\sum_{i,i'=1}^n 
        \Cov(\hat{\varepsilon}^{(0)}_i\hat{\varepsilon}^{(0)}_{i'},
        \mathbb{I}\{\vert \hat{\lambda}_{i,i'} \vert \geq \delta \} )~,\label{eq: bias selection remark}
    \end{align}
    where the error in this approximation contributes the second two terms in the bound \eqref{eq: bias approximation}. Roughly speaking, the approximation \eqref{eq: bias selection remark} follows by applying a suitable central limit theorem to show that the true selection probability $\bar{F}_{i,i'}^+(\delta)$ is close to the Gaussian approximation $F^{+}(\delta,\lambda_{i,i'},v^{(d)}_{i,i'})$.

    The second term in \eqref{eq: bias selection remark} can be interpreted as the influence of ``selection.'' \cref{lem: bias decomposition} follows by developing a suitable bound for this term. For this step, it is essential that we use the orthogonalized residuals $\hat{\varepsilon}^{(j;0)}$, rather than the residual $\hat{\varepsilon}^{(j)}$, to construct the statistic $\hat{\lambda}_{i,i'}$. In the latter case, it is possible to show that the selection term is not necessarily negligible relative to $V_n$, unless the $R^2$ of the population regression of the outcome of interest on the auxiliary outcomes converges to zero.
    The main idea underlying the use of orthogonalized residuals is that, as the data have a joint Gaussian distribution,  if the statistic $\hat{\lambda}_{i,i'}$ were constructed with the residuals from the \emph{population} regression of $\hat{\varepsilon}^{(j)}$ on $\hat{\varepsilon}^{(0)}$, then the vector $\hat{\varepsilon}^{(0)}$ and each $\hat{\lambda}_{i,i'}$ would be independent, and so it would follow that $\Cov(\hat{\varepsilon}^{(0)}_i\hat{\varepsilon}^{(0)}_{i'}, \mathbb{I}\{\vert \hat{\lambda}_{i,i'} \vert \geq \delta \} ) = 0$. The core difficulty then is that we are only able to implement the feasible version of this orthogonalization.\hfill$\blacksquare$
\end{remark}

\begin{remark}
    Both the bias and the bias approximation $B_{n,d}(\delta)$ are fixed at $-V_n$ at the endpoints $\delta=0$ and  $\delta=\infty$. To see this, observe that at $\delta = \infty$, no pairs are retained, and so the estimator is exactly zero. On the other hand, at $\delta = 0$, every pair is retained and so the estimator again reduces to $\hat{V}_{n,d}(\delta) = n^{-2}(\sum_{i=1}^n \hat{\varepsilon}_i^{(0)})^2=0$. Likewise, at both endpoints the function $Q_{i,i'}(\delta)=0$, and so $B_{n,d}(\delta)=-\frac{1}{n^2}\sum_{i,i'=1}^n \lambda_{i,i'}=-V_n$. 

    To interpret the behavior of the bias approximation away from the endpoints, observe that we can write
    \begin{align}
    \frac{B_{n,d}(\delta)}{V_n}
    &=-\frac{1}{V_n}
    \frac{1}{n^2}
    \sum_{i,i'=1}^n
    \lambda_{i,i'}
    (
    1-F^+(\delta,\lambda_{i,i'},v_{i,i'}^{(d)})
    )
    -F^+(\delta,0,v_0^{(d)})~.
    \label{eq: risk approximation relative bias representation}
    \end{align}
    If each $\lambda_{i,i'}$ is weakly positive, then the bias approximation \eqref{eq: risk approximation relative bias representation} is always weakly negative. In this case, the first term in \eqref{eq: risk approximation relative bias representation} decreases away from zero as the threshold increases, as more non-nulls are omitted. On the other hand, the second term in \eqref{eq: risk approximation relative bias representation} increases to zero as the threshold increases, as fewer nulls are retained. If one could set $\delta$ to perfectly discriminate between nulls and non-nulls, then the term $Q_{i,i'}(\delta)=1$ and so the bias approximation would reduce to zero.
    
    However, if the error in the estimator $\hat{\lambda}_{i,i'}$ is non-negligible relative to the value of $\lambda_{i,i'}$, then it is not possible to consistently identify that the pair $i,i'$ is non-null. If there is a sufficient mass of such pairs, then it is possible that no sequence of thresholds will ever discriminate between nulls and non-nulls without non-negligible error. Irreducible error of this sort is exhibited by \cref{fig: bias and risk}.\footnote{In \cref{app: examples}, we give an example where the relative bias approximation \eqref{eq: risk approximation relative bias representation} converges to a constant that can be made arbitrarily close to $-1$ for any sequence of thresholds.}\hfill$\blacksquare$
\end{remark}

\subsection{Risk Approximation\label{sec: risk approximation}}

In this section, we give an approximation to the relative root-mean-squared error
\begin{align}
    R_{n,d}(\delta)
    &=
    \frac{1}{V_n}
    \sqrt{\mathbb{E}\left[
    (
            \hat V_{n,d}(\delta)-V_n
    )^2
    \right]}\label{eq: rmse def}
\end{align}
of the estimator \eqref{eq: V hat} as the threshold $\delta$ varies. We show that the Gaussian approximation to the bias introduced in the previous subsection additionally provides an accurate approximation to the risk \eqref{eq: rmse def}. 

This result is obtained through an application of the following Lemma, which shows that, for sequences of thresholds $\delta\geq\underline\delta_{n,d}(\eta)$, the standard deviation of the estimator \eqref{eq: V hat} can be controlled by the sequence
\begin{align}
    \mathcal{V}_{n,d}(s,\eta) = V_n^{\eta/2}
    +\sqrt{
        dV_n^{1+\eta}
        \log(V_n^{-1})
    }
    +\frac{1}{\sqrt d}
    +\left(
        \frac{s\sqrt d}{n}
    \right)^{1/2}~.\label{eq: mathcal V}
\end{align}
Observe that, by contrast to the bound \eqref{eq: bias approximation} on the quality of the bias approximation, the sequence \eqref{eq: mathcal V} is only convergent if $\eta$ is strictly positive. That is, the following bound applies only if the expected proportion of nulls retained in the estimator can be bounded by a polynomial factor of $1/n \lesssim V_n \lesssim s/n$.

\begin{lem}[Standard Deviation Bound]\label{lem: variance bound}$ $\\
Suppose that \cref{assu: gaussianity,assu: proportionality,assu: homoskedasticity,assu: sparsity,assu: well conditioned} hold. If the condition \eqref{eq: bias approximation conditions} holds for $\eta \in (0,1]$, then
\begin{align}
    \sup_{\delta\in\Delta_{n,d}(\eta)}
    \frac{\sqrt{\Var(\hat{V}_{n,d}(\delta))}}
    {V_n}
    &\lesssim
    \mathcal{V}_{n,d}(s,\eta)\label{eq: std bound}
\end{align}
holds for all sufficiently large $n,d$.
\end{lem}

\noindent In \cref{app: examples}, we show that the term $V_n^{\eta/2}$ cannot be removed from the bound \eqref{eq: std bound}.\footnote{The proof of \cref{lem: variance bound} extended to the case $\eta > 1$ yields the rate $\mathcal{V}_{n,d}(s,1)$. Thus, we restrict attention to the case $\eta\in(0,1]$.}

By combining \cref{lem: bias decomposition} and \cref{lem: variance bound}, we obtain the following risk approximation.

\begin{theorem}[Risk Approximation]\label{thm: risk approximation}$ $\\
Suppose that \cref{assu: gaussianity,assu: proportionality,assu: homoskedasticity,assu: sparsity,assu: well conditioned} hold. If the condition \eqref{eq: bias approximation conditions} holds for $\eta\in(0,1]$, then
\begin{align}
    \sup_{\delta\in\Delta_{n,d}(\eta)}
    \left|
        R_{n,d}(\delta) -
        \frac{\left|B_{n,d}(\delta)\right|
        }{V_n}
    \right|
    &\lesssim
    \mathcal{V}_{n,d}(s,\eta)
\end{align}
holds for all sufficiently large $n,d$.
\end{theorem}

\begin{remark}
    \cref{thm: risk approximation} implies that the relative root-mean-square error admits the representation
    \begin{align}
        R_{n,d}(\underline{\delta}_{n,d}(\eta))
        =
        \frac{
            \left|B_{n,d}(\underline{\delta}_{n,d}(\eta))\right|
        }{V_n}
        +
        O\left(\mathcal{V}_{n,d}(s,\eta)\right),
        \quad\text{where}\quad
        V_n^\eta
        \lesssim
        \mathcal{V}_{n,d}(s,\eta)
        =
        o(1)~.
    \end{align}
    The sequence $\mathcal{V}_{n,d}(s,\eta)$ bounds the stochastic variability of the estimator, as well as the error in the bias approximation, and converges to zero. The relative absolute bias, by contrast, need not converge to zero and can remain of constant order for every sequence of thresholds. In this sense, in settings where null and non-null pairs cannot be differentiated without non-negligible error, the relative absolute bias determines the relative risk to first order. In Panel B of \cref{fig: bias and risk}, we display the relative squared bias, variance, and mean-squared error of the estimator \eqref{eq: V hat} in the same simulation considered in Panel A. The relative mean-squared error closely follows the relative squared bias, while the relative variance is substantially smaller.\hfill$\blacksquare$
\end{remark}

\begin{remark}
    The main difficulty in the proof of \cref{lem: variance bound} is in giving a suitable bound for the variance of a quantity analogous to the selection term in the approximation \eqref{eq: bias selection remark}. We do this by comparing this term to a more tractable quantity, where the orthogonalized residuals $\hat{\varepsilon}_{i}^{(j;0)}$ are replaced by the residuals from the population regression of $\hat{\varepsilon}^{(j)}$ on $\hat{\varepsilon}^{(0)}$. This comparison is enabled by the application of a method for evaluating non-differentiable functions of Gaussian vectors under perturbations to their covariance matrices, due to \citet{plackett1954} and \citet{price1958}. See \citet{voigtlaender2021} for a modern treatment.\hfill$\blacksquare$
\end{remark}

\subsection{Three Regimes\label{sec: minimax}}

We now delineate three asymptotic regimes in which the TMO estimator exhibits qualitatively different behavior. These regimes are differentiated by the mass of non-null pairs of units that are difficult to distinguish from nulls. In the first two regimes, the TMO estimator is consistent in relative root-mean-squared error, with its rate of convergence determined by either its variance or its absolute bias, respectively. Here, when the bias dominates the variance, the estimator achieves the minimax optimal rate of convergence, up to logarithmic factors. In the third regime, no estimator is uniformly consistent in relative root-mean-squared error, and the risk of the TMO estimator is determined, to leading order, by its non-vanishing absolute bias.

To express these results, let $\mathcal{P}_{n,d}(s)$ denote the class of distributions $P$ under which \cref{assu: gaussianity,assu: proportionality,assu: homoskedasticity,assu: sparsity,assu: well conditioned} hold.  Throughout this subsection, we append the subscript $P$ to make the dependence of various parameters and moments on $P$ explicit. We consider the minimax relative root-mean-squared error.

\begin{defn}\label{defn: minimax risks}
Define the relative root-mean-square error
\begin{align}
    R_{n,d}(T,P) = 
    \frac{1}{V_{n,P}}
        \sqrt{\mathbb{E}_P[(T-V_{n,P})^2]}~.
\end{align}
The minimax relative root-mean-square error is given by
\begin{align}
    \mathfrak{M}_{n,d}(s) 
    &= 
    \inf_{T}\sup_{P\in\mathcal P_{n,d}(s)}
    R_{n,d}(T,P)~,
\end{align}
where the infimum is taken over all measurable functions $T$ of the data $Y$.
\end{defn}

\noindent In the following Theorem, we give a generic lower bound on the minimax relative root-mean-square error for any estimator for $V_n$. We then show that, in regimes where its bias dominates its variance, the TMO estimator \eqref{eq: V hat} achieves this bound, up to logarithmic factors. 

\begin{theorem}[Minimax Optimality] \label{thm: constrained minimax lower bound} Suppose that $n,d,s\to\infty$.\\
\noindent\textbf{(i)} If it holds that $s\sqrt{d}/n \to 0$, then 
\begin{align}
    \mathfrak{M}_{n,d}(s)
    &\gtrsim
    \frac{1}{\log^2(s)}
    \min\left\{\frac{s}{\sqrt d},1\right\}~.
    \label{eq: constrained minimax all estimator lower bound}
\end{align}

\noindent\textbf{(ii)} Fix $\eta\in(0,1]$. If, additionally, it holds that
\begin{align}
    d\left(\frac{s}{n}\right)^{2}
    \log\left(\frac{n}{s}\right)
    &\to0
    \quad\text{and}\quad
    \log n=o(d)~,
    \label{eq: constrained minimax endpoint conditions}
\end{align}
then
\begin{align}
    \mathfrak{M}_{n,d}(s)
    & \gtrsim \frac{1}{\sqrt{\log n}\,\log^2(s)}
    \sup_{P\in\mathcal P_{n,d}(s)}
    \left\{
            \frac{
                |B_{n,d,P}(\underline\delta_{n,d,P}(\eta)) |
            }{
                V_{n,P}
            }
            -C\cdot\mathcal{V}_{n,d,P}(s,\eta)
    \right\}~.
    \label{eq: constrained minimax endpoint bias lower bound}
\end{align}
\end{theorem}

\begin{remark}
    The proof of \cref{thm: constrained minimax lower bound} follows from an application of Le Cam's method of two fuzzy hypotheses, discussed in  Chapter 2 of \cite{tsybakov-2009}. We use a construction based on moment matching. Related constructions have been used by \citet{cai-low-2011}, who consider estimation of a non-smooth functional in the normal-means model, and \citet{wu-yang-2016,wu-yang-2019}, who consider estimation of entropy and support size in discrete distributions. See \cite{wu-yang-2020} for a review.\hfill$\blacksquare$
\end{remark}

\cref{thm: constrained minimax lower bound}  implies that the behavior of the TMO estimator \eqref{eq: V hat} can be classified into three regimes, distinguished by the size of the covariance mass
    \begin{align}\label{eq: regime determinant}
    \kappa_{n,d}(\eta) = \frac{1}{V_n}\frac{1}{n^2}
    \sum_{i,i'=1}^n
    \lambda_{i,i'}
    \left(
        1-
        F^+\left(
            \underline{\delta}_{n,d}(\eta),
            \lambda_{i,i'},
            v_{i,i'}^{(d)}
        \right)
    \right)
    \end{align}
carried by non-null pairs of units weighted by the probability that they are included in the estimator. 

Suppose, first, that the sequence $\kappa_{n,d}(\eta)$ converges to zero. Observe that this sequence is the first term in the representation \eqref{eq: risk approximation relative bias representation} for the bias approximation given in \cref{lem: bias decomposition}, and that the second term in that representation is necessarily smaller than $\mathcal{V}_{n,d}(s,\eta)$. Thus, in this case, \cref{thm: risk approximation} implies that the TMO estimator is consistent in relative root-mean-squared error. Here, there are two regimes, distinguished according to whether $\kappa_{n,d}(\eta)$ is smaller or larger than $\mathcal{V}_{n,d}(s,\eta)$. In the latter case, Part (ii) of \cref{thm: constrained minimax lower bound} implies that the TMO estimator is not only consistent but also achieves the minimax rate of convergence, up to logarithmic factors. In the former case, where the estimator is consistent, but potentially suboptimal, the mass of non-nulls that are difficult to identify is necessarily (arguably, unrealistically) very small. 

Suppose, on the other hand, that the sequence $\kappa_{n,d}(\eta)$ does not converge to zero. In this case, \cref{lem: selected set characterization}, together with \cref{assu: sparsity}, implies that the condition $s\sqrt{\log(n)}/\sqrt{d} \to 0$ cannot hold. Thus, Part (i) of \cref{thm: constrained minimax lower bound} implies that, up to logarithmic factors, no variance estimator, including the TMO variance estimator, will be uniformly consistent in relative root-mean-squared error.\footnote{\cite{bai2024standard} consider a related thresholding estimator, using repeated observations over time to estimate cross-sectional covariances. Their consistency theory imposes an approximate-sparsity condition that, in effect, implies $\kappa_{n,d}(\eta)\to0$. We instead allow $\kappa_{n,d}(\eta)$ to remain bounded away from zero and develop methods for optimizing the resulting non-vanishing bias.} Here, \cref{thm: risk approximation} implies that the risk of the TMO estimator is determined, to leading order, by its relative absolute bias. 

\begin{figure}[t]
\begin{centering}
\caption{The Distribution of Correlations Between Pairs of U.S.\ Counties\label{fig: county histogram}}
\medskip{}
\begin{tabular}{c}
\includegraphics[width=\textwidth]{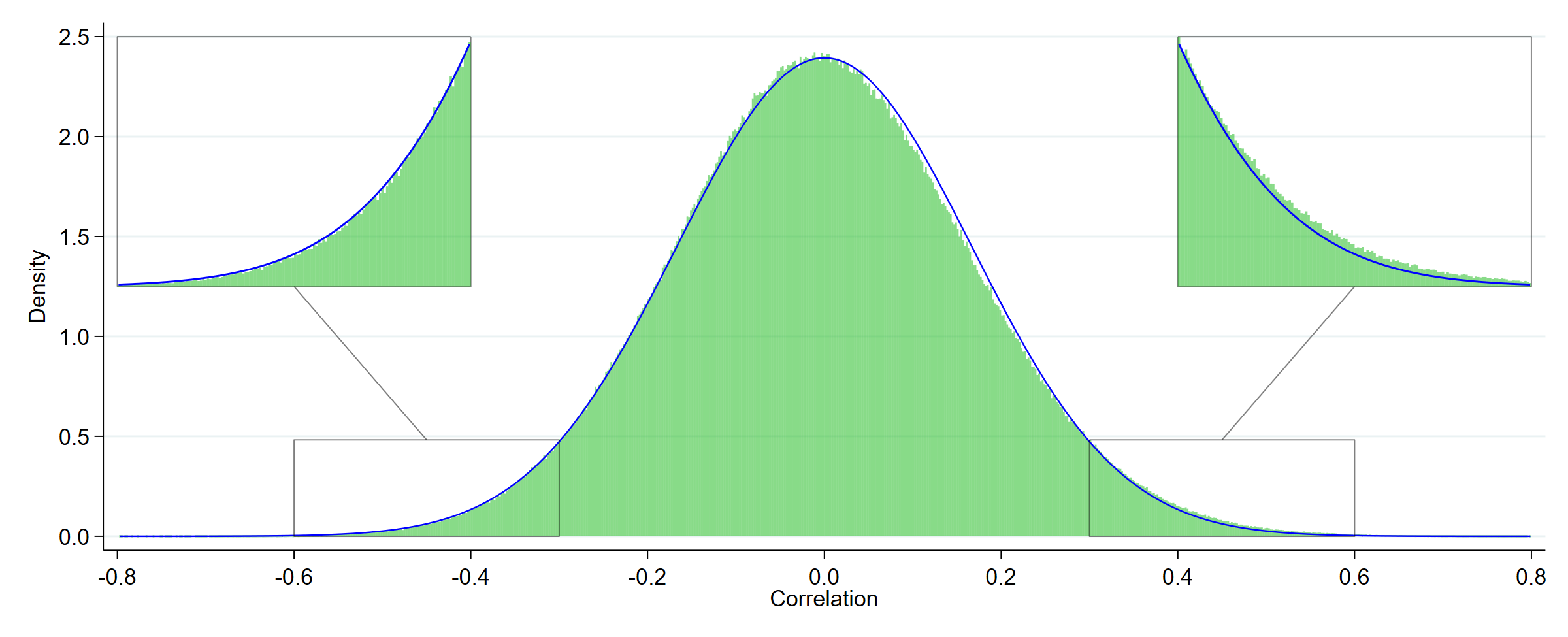}
\end{tabular}
\par\end{centering}
\medskip{}
\justifying
{\noindent\footnotesize{}Notes: \cref{fig: county histogram} displays a histogram of the estimates $\tilde{\rho}_{i,i^\prime}$ over each pair of distinct U.S.\ counties. A mean-zero, Gaussian density function is overlaid in blue. The scaling of this density function has been chosen to best fit the corresponding quartiles of the distribution of the estimates $\tilde{\rho}_{i,i^\prime}$. The left and right tails of the distribution are displayed as enlarged insets.}{\footnotesize\par}
\end{figure}

In our view, this third regime is more realistic. In \cref{fig: county histogram}, we display a histogram of the correlation estimates $\tilde{\rho}_{i,i^\prime}$ used in the example considered in \cref{sec: simulation}. The center of this distribution is well-approximated by a Gaussian density centered at zero and scaled so that the 95th percentile is at 0.27. If most pairs of units are nulls, then this is a good approximation to the null distribution of the correlation estimates. Only pairs of units whose outcomes have a true correlation very close to -1 or 1 can be differentiated from this distribution with negligible error. Thus, we develop a method for choosing a threshold $\delta$ to minimize the absolute bias, without requiring that the minimum attainable bias converge to zero.

\section{Threshold Choice\label{sec: threshold choice}}

In this section, we propose a method for choosing the threshold parameter $\delta$ based on the minimization of a regularized estimate of the absolute bias of the TMO estimator \eqref{eq: V hat}. We motivate both this criterion and the necessity for regularization, before introducing the method. We then give pointwise bounds on the regret associated with this choice, in terms of both the relative absolute bias and root-mean-square error.

\subsection{Absolute Bias Minimization}

In \cref{thm: risk approximation}, we show that in the realistic regime where null and non-null pairs of units cannot be differentiated without non-negligible error, the relative root-mean-squared error of the TMO estimator \eqref{eq: V hat} is determined, to first order, by its potentially non-vanishing, relative absolute bias. Thus, we propose to choose the threshold $\delta$ by minimizing (an estimate of) the absolute bias. Formally, fix an integer $g\geq 2$ and a grid $0 < \eta_1 < \eta_2 < \cdots < \eta_g \leq 1$.\footnote{We do not consider values of $\eta$ larger than one, as the bound on the relative standard deviation obtained by the proof of \cref{lem: variance bound} does not improve in this regime.} Our objective is to set the threshold $\delta$ at the (maximal) point $\underline{\delta}_{n,d}(\eta_k)$ associated with the minimal absolute bias. That is, we aim to choose the index
\begin{align}
    k^\star
    &= \underset{k\in[g]}{\max\arg\min}~
      \vert \operatorname{Bias}_{n,d}(k)\vert ~,
    \quad\text{where}\quad
    \operatorname{Bias}_{n,d}(k)
    =\mathbb E\left[
        \hat V_{n,d}(\underline{\delta}_{n,d}(\eta_k))-V_n
      \right]~,
    \label{eq: true bias and optimizer}
\end{align}
and $[g]=\{1,\ldots,g\}$. Methods for choosing bandwidths or tuning parameters based on minimizing the leading order term in asymptotic approximations to measures of risk are common. For instance, there is a well-developed literature that applies this principle to bandwidth selection for HAC covariance estimators \citep{andrews1991heteroskedasticity,newey1994automatic,kim2011spatial,kim2013spatiotemporal}. The distinctive feature of our application is that, in the regime of interest, the leading order term in our approximation to the root-mean-squared error is composed of the absolute bias alone, rather than a composition of the absolute bias and the standard deviation.

In the simulation considered in \cref{sec: risk approximation sec}, the bias of the TMO estimator \eqref{eq: V hat} is always non-positive. For the sake of simplicity, we impose a condition that implies that this property holds generically.

\begin{assumption}[Non-Positive Bias]\label{assu: nonpositive}
It holds that
\begin{align}
    0 \leq \frac{1}{n^2}\sum_{i,i'=1}^n \lambda_{i,i'}\mathbb{I}\{ \vert \lambda_{i,i'}\vert  \geq t\} \leq V_n
\end{align}
for each $t \geq 0$. 
\end{assumption}

\noindent \cref{assu: nonpositive} stipulates that there is more weight associated with positive non-nulls than negative non-nulls in each tail set $\vert \lambda_{i,i'} \vert \geq t$. In settings where there is aggregate positive dependence, in the sense that $1/n < V_n$, this is not particularly stringent. Settings with aggregate negative dependence, where the reverse holds, are less plausible in practice and are less concerning, as unadjusted standard errors are already conservative. 

Under \cref{assu: nonpositive}, the obvious approach for minimizing the absolute bias of the estimator \eqref{eq: V hat} is to simply choose the index $k$ associated with the largest variance estimator. That is, one might choose 
\begin{align}
    \check{k} = \underset{k\in[g]}{\max\arg\max} \, \hat{V}_{n,d}(\underline{\delta}_{n,d}(\eta_k))~,
    \label{eq: check k}
\end{align}
where, for the sake of simplicity (and for the time being), we assume that the thresholds $\underline{\delta}_{n,d}(\eta_k)$ are known. The premise is that, as the bias is always negative, maximizing the estimator should minimize the bias. 

Unfortunately, this method exhibits an undesirable pathology. To illustrate, suppose that all of the off-diagonal elements of $\Lambda_n$ are non-negative. In this case, for any two grid points $\eta_1 < \eta_2$, the definition \eqref{eq: selected threshold set} and the representation \eqref{eq: risk approximation relative bias representation} imply that
\begin{align}
    & V_n^{-1}(B_{n,d}(\underline{\delta}_{n,d}(\eta_2))
    - B_{n,d}(\underline{\delta}_{n,d}(\eta_1))) \nonumber\\
    &\quad\leq F^+(\underline{\delta}_{n,d}(\eta_1),0,v_0^{(d)})
    - F^+(\underline{\delta}_{n,d}(\eta_2),0,v_0^{(d)})
    \leq V_n^{\eta_1+o(1)}~. 
\end{align}
Thus, up to the approximation error in \cref{lem: bias decomposition}, as we increase $\eta$ above $\eta_1$, the relative bias can improve by at most $V_n^{\eta_1+o(1)}$. By contrast, \cref{lem: variance bound} indicates that the relative standard deviation of each estimator $\hat{V}_{n,d}(\underline{\delta}_{n,d}(\eta_k))$ can be (at least) of order $V_n^{\eta_k/2}$. Suppose then that, heuristically, the noise about the expectation $B_{n,d}(\underline{\delta}_{n,d}(\eta_k)) + V_n$ of the estimator $\hat{V}_{n,d}(\underline{\delta}_{n,d}(\eta_k))$ is symmetric with standard deviation $V_n^{\eta_k/2}$. In this case, as the standard deviation of the estimator at $k = 1$ is of a strictly larger order than both its standard deviation at larger values of $k$ and the change in its expectation as $k$ increases, the procedure will tend to choose $\check{k}=1$ when the noise at $k=1$ is positive, regardless of the value of $\eta$ that minimizes the true bias. That is, when $\eta_1$ is suboptimal, the procedure chooses $\check{k}=1$ with probability approaching one half. We give an example that exhibits this behavior in \cref{app: examples}.

\subsection{Regularization and Regret}

The procedure \eqref{eq: check k} rewards positive sampling error. When coupled with the fact that smaller values of $\eta$ are potentially associated with sampling error with a substantially larger scale, this encourages spuriously small threshold choices. In this section, we develop a method for regularizing the objective function used by the rule \eqref{eq: check k} to account for differences in stochastic variability across thresholds. 

To introduce this method, we require some further notation. In particular, let $\hat{V}^{(j)}_{n,d}(\delta)$ be defined analogously to the TMO estimator \eqref{eq: V hat}, except that outcome $j$ is treated as the outcome of interest and the outcome $0$ is treated as an auxiliary outcome. Define the statistics
\begin{align}
    \hat{\gamma}_{j}= \frac{1}{n}\sum_{i=1}^n (\hat{\varepsilon}^{(j)}_i)^2~,
    \quad
    \hat{\gamma}_{0,j}= \frac{1}{n}\sum_{i=1}^n \hat{\varepsilon}^{(0)}_i \hat{\varepsilon}^{(j)}_i~,
    \quad\text{and}\quad
    \hat{\omega}_{0,j}
    =
    \frac{ \hat{\gamma}_{0,j}}
    {\sqrt{\hat{\gamma}_0 \hat{\gamma}_j }}~.
\end{align}
We consider the choice
\begin{align}
        \hat{k} = \underset{k \in [g]}{\max\arg\max} \left\{ \hat{V}_{n,d}(\underline{\delta}_{n,d}(\eta_k)) - \hat{\xi}_{n,d}(\underline{\delta}_{n,d}(\eta_k)) \right\}~,\label{eq: hat k}
\end{align}
where
\begin{align}
        \hat{\xi}^2_{n,d}(\delta)
    = \frac{1}{d}
    \sum_{j=1}^d 
    \frac{(\hat{V}_{n,d}^{(j)}(\delta) - \hat{V}_{n,d}^{(0)}(\delta))^2}{1 - \hat{\omega}^2_{0,j}}\label{eq: xi def}
\end{align}
and again we assume that the thresholds $\underline{\delta}_{n,d}(\eta_k)$ are known. The selector \eqref{eq: hat k} is analogous to the naive selector \eqref{eq: check k}, except that the raw estimates $\hat{V}_{n,d}(\underline{\delta}_{n,d}(\eta_k))$ have been regularized by the statistics \eqref{eq: xi def}. Roughly speaking, the statistics \eqref{eq: xi def} measure the scale of the stochastic variability of the estimator $\hat{V}_{n,d}^{(0)}(\delta)$. \cite{chernozhukov2025policy} apply selection rules with a broadly related structure to policy choice.

The following Theorem gives high-probability bounds on the bias and root-mean-square error associated with the procedure that uses the selector \eqref{eq: hat k}, relative to the index \eqref{eq: true bias and optimizer} that minimizes the true absolute bias or the index that minimizes the true root-mean-squared error. 

\begin{theorem}[Consistent Regret Minimization]
\label{thm: relative bias regret}$ $\\
Suppose that  $g$ is fixed, \cref{assu: gaussianity,assu: proportionality,assu: homoskedasticity,assu: sparsity,assu: well conditioned,assu: nonpositive} hold, and that
\begin{align}
    \mathcal{R}_{n,d}(s) = \sqrt{dV_n}\log^{3/2}(V_n^{-1}) + \frac{\log(V_n^{-1})}{\sqrt{d}} + \left(\frac{s\sqrt{d}}{n}\right)^{1/2}
    &\to0~.
    \label{eq: regret rate condition}
\end{align}
It holds that
\begin{align}
    \frac{1}{V_n}\left(
    \vert \operatorname{Bias}_{n,d}(\hat k)\vert 
    -
    \vert \operatorname{Bias}_{n,d}(k^\star) \vert 
        \right)
    =O_p\left(
        V_n^{\eta_{k^\star}/2}
        +\mathcal{R}_{n,d}(s)
    \right)
    \label{eq: relative bias regret bound}
\end{align}
and that, moreover,
\begin{align}
    R_{n,d}(\underline\delta_{n,d}(\eta_{\hat k}))
    - \min_{k \in [g]} R_{n,d}(\underline\delta_{n,d}(\eta_k))
    =O_p\left(
        V_n^{\eta_{k^\star}/2}
        +\mathcal{R}_{n,d}(s)
    \right)~.\label{eq: relative risk regret bound}
\end{align}
\end{theorem}

\begin{figure}[t]
\begin{centering}
\caption{Distribution of Regularized Threshold Choices \label{fig: threshold choice distributions}}
\medskip{}
\begin{tabular}{c}
\includegraphics[width=\textwidth]{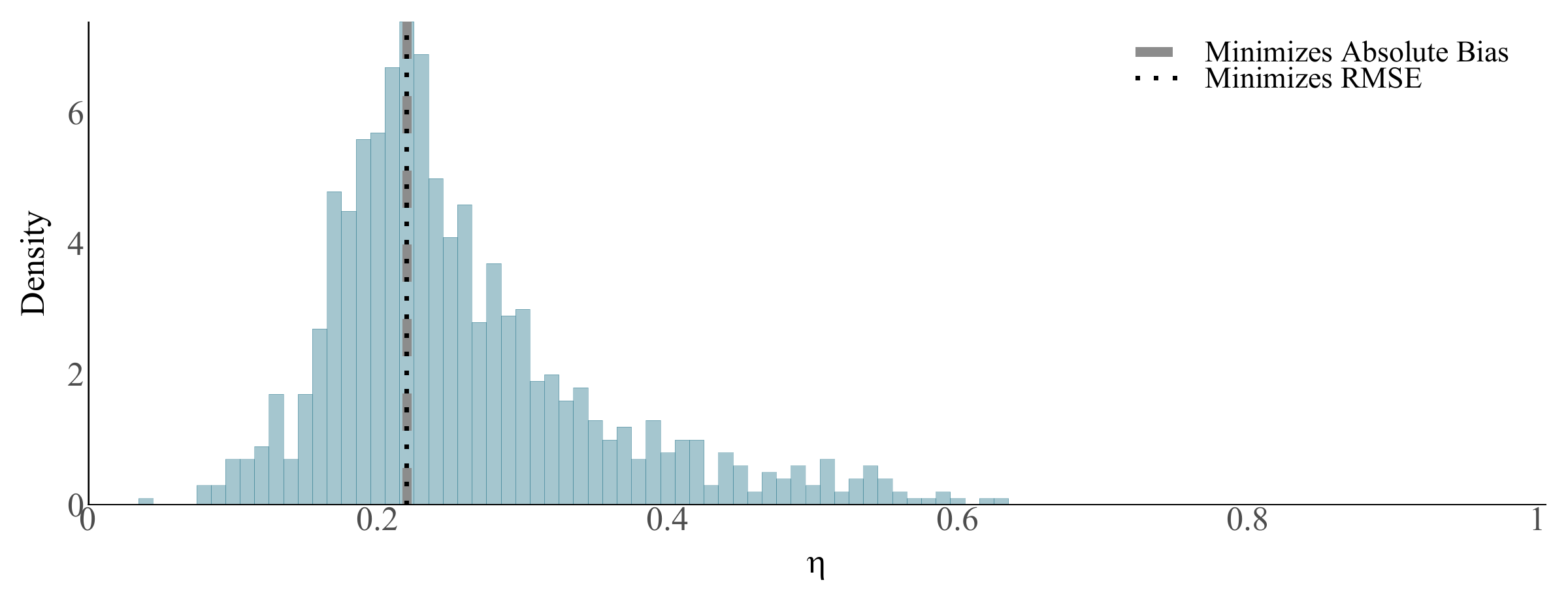}
\tabularnewline
\end{tabular}
\par\end{centering}
\medskip{}
\justifying
{\noindent\footnotesize{}Notes: \cref{fig: threshold choice distributions} displays the distribution of the grid point $\eta_{\hat{k}}$ selected by the regularized procedure \eqref{eq: hat k} over 1,000 replications of the calibrated simulation used in \cref{fig: bias and risk}. In each replication, the procedure searches over the grid $\eta \in \{0.01,0.02,\ldots,1\}$. The dashed and dotted vertical lines denote the values of $\eta$ whose corresponding threshold $\underline{\delta}_{n,d}(\eta)$ minimizes the absolute bias and root-mean-squared error of the TMO estimator \eqref{eq: V hat}, respectively.}{\footnotesize\par}
\noindent\hrulefill
\end{figure}

\noindent \cref{fig: threshold choice distributions} displays a histogram of the threshold $\eta$ chosen by the procedure \eqref{eq: hat k} in the simulation considered in \cref{sec: risk approximation sec}. Vertical dashed and dotted lines are placed at the values of $\eta$ that minimize the absolute bias and root-mean-square error, respectively. The selected values of $\eta$ cluster around these values.

\begin{remark}
    The essential feature of the bounds \eqref{eq: relative bias regret bound} and \eqref{eq: relative risk regret bound} obtained in \cref{thm: relative bias regret} is that they depend on the parameter $\eta$ only through the rate $V_n^{\eta_{k^\star}/2}$ associated with the \emph{oracle} threshold $\eta_{k^\star}$ that minimizes the absolute bias. In particular, \cref{lem: bias decomposition} and \cref{lem: variance bound} imply that an identical bound, where the oracle quantity $V_n^{\eta_{k^\star}/2}$ is replaced by the larger rate $V_n^{\eta_{1}/2}$, holds for the rule that always chooses $k=1$. The same, slower, rate will also hold for the unregularized rule \eqref{eq: check k}. In \cref{app: examples}, we give examples that illustrate that the rates obtained in \cref{thm: relative bias regret} are non-vacuous, in the sense that the bounds \eqref{eq: relative bias regret bound} and \eqref{eq: relative risk regret bound} are not obtained by the rule that always chooses $k=1$, and are not satisfied by the unregularized rule \eqref{eq: check k}.\hfill$\blacksquare$
\end{remark}

\begin{remark}
    The regularized selector \eqref{eq: hat k} can be motivated by considering an oracle inequality. In particular, the definition \eqref{eq: hat k} implies that the bias at the selected index $\hat{k}$ satisfies the inequality
    \begin{align}
    &\operatorname{Bias}_{n,d}(k^\star)
      -\operatorname{Bias}_{n,d}(\hat k)
    \nonumber\\
    &\leq
    \underbrace{
      |\hat V_{n,d}(k^\star)
        -\mathbb E[\hat V_{n,d}(k^\star)]|
      +\hat\xi_{n,d}(k^\star)}_{\text{Oracle Variability}}
    +\max_{k\in[g]}
      \underbrace{~[\hat V_{n,d}(k)
        -\hat\xi_{n,d}(k)
        -\mathbb E[\hat V_{n,d}(k)]]_+}_{\text{Optimism}}~,
    \label{eq: regret calibration reduction main}
    \end{align}
    where $[\cdot]_{+}=\max\{\cdot,0\}$ and we have replaced the threshold $\underline{\delta}_{n,d}(\eta_k)$ by its index $k$ with an abuse of notation (see e.g., \citealp{barron1999risk} and \citealp{goldenshluger2011bandwidth} for canonical applications of bounds with this structure to penalized model selection). The bound \eqref{eq: regret calibration reduction main} suggests that we would like the statistic $\hat\xi_{n,d}(k)$ to ensure that the terms that contribute ``Optimism'' are rarely positive, without introducing too much variability at the oracle index $k^\star$.
    
    To illustrate the main mechanism that motivates our construction, suppose, heuristically, that
    \begin{align}
        \hat{V}^{(j)}_{n,d}(k) = V_n + B_n(k) + Z_k^{(j)}~,
        \quad\text{where}\quad
        Z_k^{(j)}\overset{\text{i.i.d.}}\sim \mathsf{N}(0, \nu_k)~,
    \end{align}
    where we again use the index $k$ in the place of the threshold $\underline{\delta}_{n,d}(\eta_k)$ with an abuse of notation. In this case, conditional on the quantity $Z_k^{(0)}$, we should expect that
    \begin{align}
        \hat{\xi}^2_{n,d}(k)
    = \frac{1}{d}
    \sum_{j=1}^d 
    \frac{(\hat{V}_{n,d}^{(j)}(k) - \hat{V}_{n,d}^{(0)}(k))^2}{1 - \hat{\omega}^2_{0,j}}
    \overset{p}{\to} (Z_k^{(0)})^2 + \nu_k
    \end{align}
    and thereby 
    \begin{align}
        \hat V_{n,d}(k)
        -\hat\xi_{n,d}(k)
        -\mathbb E[\hat V_{n,d}(k)]
        \overset{p}{\to}  Z_k^{(0)} - \sqrt{(Z_k^{(0)})^2 + \nu_k} < 0~.
    \end{align}
    Hence, in this stylized example, the terms that contribute to the second term in the inequality \eqref{eq: regret calibration reduction main} should rarely be positive, and the regret will be determined by the scale of the variability of the objective function at the oracle index $k^\star$. \cref{thm: relative bias regret} shows that this intuition holds under our maintained conditions.\hfill$\blacksquare$
\end{remark}

\subsection{Endpoint Estimation}

The regularized threshold selector developed in the previous subsection is infeasible, in the sense that we have assumed that the values of the thresholds
\begin{align}
    \underline{\delta}_{n,d}(\eta_k) = \sqrt{2\eta_k v_0^{(d)}\log(V_n^{-1})}(1+o(1))\label{eq: endpoint asymptotic represent}
\end{align}
were known. To conclude the section, we close the procedure by giving a consistent estimator of these values.

There are two unknown quantities in the representation \eqref{eq: endpoint asymptotic represent}: the sequence $v_0^{(d)}$ and the target parameter $V_n$ itself. To estimate the former quantity, we fit the interquartile range of a Gaussian distribution to the empirical distribution of the covariances $\hat\lambda_{i,i'}$. Formally, we use the estimator
\begin{align}
    \hat{v}_0^{(d)} = \frac{1}{4}\frac{1}{\Phi^{-1}(3/4)^2}
    \left(
    \hat{q}_{0}(3/4) -  \hat{q}_{0}(1/4)
    \right)^2~,
\end{align}
where $\Phi(\cdot)$ denotes the standard Gaussian distribution function and
\begin{align}
    \hat{q}_{0}(z) 
    = \inf\bigg\{
            t\in\mathbb R:
            \frac{1}{n^2}\sum_{i,i'=1}^n
            \mathbb I\{\hat\lambda_{i,i'}\leq t\}
            \geq z
        \bigg\}
\end{align}
denotes the empirical quantile of the covariances $\hat\lambda_{i,i'}$. The idea is that, as the number of non-null pairs is negligible relative to the number of null pairs, the center of the distribution of the covariances should largely be determined by the distribution of the null pairs. A similar approach is used in a multiple testing application in \cite{efron2007}. In \cref{fig: county histogram}, we show that a Gaussian distribution, fit in this way, provides a high-quality approximation to the distribution of the estimates $\tilde{\rho}_{i,i'}$ used in \cref{sec: simulation}.

Finally, for the purpose of estimating the endpoint \eqref{eq: endpoint asymptotic represent}, we assume that we have access to a ``pilot'' estimator $\check{V}_n$ of the variance $V_n$. We do not require that this estimator is consistent. Instead, we only require that it identifies the correct rate of convergence. For positive random sequences $A_{n,d}$ and $B_{n,d}$, we write $A_{n,d}=\Theta_p(B_{n,d})$ if $P\{cB_{n,d}\leq A_{n,d}\leq CB_{n,d}\}\to1$ as $n,d\to\infty$. 

\begin{assumption}[Pilot Variance Estimator]
\label{assu: pilot variance estimator} The pilot estimator $\check V_n$ satisfies $\check V_n =\Theta_p(V_n)$.
\end{assumption}

\noindent In practice, we suggest using a standard spatial variance estimators as a pilot, such as clustered variance estimators or spatial HAC estimators. Here, all that is required is that the cross-sectional dependence captured by these estimators is non-negligible relative to total cross-sectional dependence.

In the following theorem, we show that the empirical threshold
\begin{align}
    \hat{\underline\delta}_{n,d}(\eta)
    &=
    \sqrt{
        2\eta\hat v_0^{(d)}
        \log\left(\check V_n^{-1}\right)
    }
    \label{eq: estimated threshold endpoint}
\end{align}
is consistent for the endpoint $\underline\delta_{n,d}(\eta)$.\footnote{This construction is not sensitive to failures of \cref{assu: pilot variance estimator}. In particular, it is straightforward to show that if $\check V_n=\Theta_p(V_n^{\kappa})$ for some $\kappa>0$, then $\hat{\underline\delta}_{n,d}(\eta)$ is consistent for $\underline\delta_{n,d}(\kappa\eta)$.} 

\begin{theorem}[Endpoint Estimation Consistency]\label{thm: threshold choice}$ $\\
Suppose that, for fixed $\eta\in(0,1]$, \cref{assu: gaussianity,assu: proportionality,assu: homoskedasticity,%
assu: sparsity,assu: well conditioned,assu: pilot variance estimator}
hold. If
\begin{align}
    \frac{s}{n}
    &\to0~,
    \quad
    dV_n^2\log\left(V_n^{-1}\right)
    \to0~,
    \quad\text{and}\quad
    \log\left(V_n^{-1}\right)
    =o(d)~,
\end{align}
then
\begin{align}
    \left| \hat{\underline\delta}_{n,d}(\eta) / \underline\delta_{n,d}(\eta)
        -1
    \right|
    &\overset{p}{\to}0
\end{align}
as $n,d\to\infty$.
\end{theorem}

\noindent \cref{alg: threshold} summarizes the proposed method for threshold choice. For ease of reference, the algorithm displays a generalization of the procedure developed in this Section, applicable to the general regression problem considered in \cref{sec: proposal,sec: simulation}, and used in our applications.\footnote{In \cref{alg: threshold}, we replace $\check V_n$ with the scale-normalized estimator $\hat{\gamma}^{-1}_0\hat{\sigma}_W^2\check V_n$, where $\hat{\sigma}_W^2$ denotes the sample variance of the treatments. This ensures that the procedure is invariant to changes in the units of either the outcome or treatment.}

\begin{algorithm}[t]
\caption{Threshold Choice} \label{alg: threshold}
\setstretch{1.35}
\SetAlgoHangIndent{0pt}
\KwIn{Outcomes $Y_i^{(0)},\ldots,Y_i^{(d)}$ and treatment $W_i$ for each $i$ in $[n]$. Grid $0 < \eta_1 < \cdots <\eta_g = 1$.}

Construct a pilot variance estimate $\check{V}_n$

For each pair of units $i\neq i^\prime$, measure the correlation $\tilde{\rho}_{i,i^\prime}$, defined in \eqref{eq: lambda ii}

For each outcome $j$ in $1,\ldots,d$, compute the statistics
\begin{align}
    \hat{\omega}_{0,j}
    &=
    \frac{ \hat{\gamma}_{0,j}}
    {\sqrt{\hat{\gamma}_0 \hat{\gamma}_j }}
    \quad\text{and}\quad
    \hat{\varphi}_{0,j}=
    \hat{\gamma}_0 / \hat{\gamma}_j~,
    \quad\text{where}\quad
    \hat{\gamma}_{j}= \frac{1}{n}\sum_{i=1}^n (\hat{\varepsilon}^{(j)}_i)^2
    \quad\text{and}\quad
\hat{\gamma}_{0,j}= \frac{1}{n}\sum_{i=1}^n \hat{\varepsilon}^{(0)}_i \hat{\varepsilon}^{(j)}_i\nonumber
\end{align}

Compute the estimator
\begin{align}
    \hat{v}_{0}^{(d)}
    = \frac{1}{4}\frac{1}{\Phi^{-1}(3/4)^2}
    \left(
    \hat{q}_{0}(3/4) -  \hat{q}_{0}(1/4)
    \right)^2~,
    \quad\text{where}\quad
    \hat{q}_{0}(z) 
    = \inf_t \bigg\{
            \frac{1}{n^2}\sum_{i,i'=1}^n
            \mathbb I\{\tilde\rho_{i,i'}\leq t\}
            \geq z
        \bigg\} \nonumber 
\end{align}

For each $k$ in $1,\ldots,g$, compute the grid-point
\begin{align}
    \hat{\underline\delta}_{n,d}(\eta_k)
    &=
    \sqrt{
        2\eta_k \hat v_0^{(d)}
        \log\left( \hat{\gamma}_0 \hat{\sigma}_{W}^{-2}\check{V}_n^{-1}\right)
    }~, 
    \quad\text{where}\quad \hat{\sigma}_{W}^2 = \frac{1}{n} \sum_{i=1}^n \left(W_i - \frac{1}{n} \sum_{j=1}^n W_j\right)^2
    \nonumber
\end{align}

\ForEach{$\delta$ in $\hat{\underline\delta}_{n,d}(\eta_1), \ldots, \hat{\underline\delta}_{n,d}(\eta_g)$}{

    For each $j$ in $0,\ldots,d$, compute $\tilde{V}_{n,d}^{(j)}(\delta)$ using \eqref{eq: V hat delta general}, with $Y^{(j)}$ treated as the  outcome of interest
        
    Construct the statistic
    \begin{align*}
        \hat{\xi}^2_{n,d}(\delta)
        &=
        \frac{1}{d}
        \sum_{j=1}^d
        \frac{
            \left(
                \hat{\varphi}_{0,j}
                \tilde{V}_{n,d}^{(j)}(\delta)
                -
                \tilde{V}_{n,d}^{(0)}(\delta)
            \right)^2
        }{
            1-\hat{\omega}_{0,j}^2
        }
    \end{align*}
}
\textbf{Select}
\[
    \hat k
    =
    \underset{k\in[g]}{\max\arg\max}
    \left\{
        \tilde V_{n,d}^{(0)}(
        \hat{\underline\delta}_{n,d}(\eta_k))
        -\hat\xi_{n,d}
        (\hat{\underline\delta}_{n,d}(\eta_k))
    \right\}
    \quad\text{and return}
    \quad \hat{\delta}_{n,d}^{\star}
    =\hat{\underline\delta}_{n,d}(\eta_{\hat{k}})
\]

\nonl \hrulefill
\setstretch{1}

\nonl {\footnotesize{}Notes:  \cref{alg: threshold} details a method for choosing the threshold $\delta$ to use with the TMO estimator defined in \cref{alg: tmo}. In practice, we use the subset of the grid $\eta_1 = 0.1, \eta_2 = 0.15, \ldots, \eta_g = 1$ such that at most 5\% of pairs of units are included in the TMO estimator.}{\footnotesize\par}
\end{algorithm}
\smallskip

\section{Applications\label{sec: applications}}

We now apply the TMO variance estimator to a sample of recently published papers.

\subsection{Spatial Correlation in Applied Economics\label{sec: survey overview}}

We begin by documenting the potential for spatial correlation in a set of recently published papers. We review all 370 papers published in 2023 in the \textit{American Economic Review}, \textit{Econometrica}, the \textit{Journal of Political Economy}, the \textit{Review of Economic Studies}, and the \textit{Quarterly Journal of Economics}. We hand-code each paper as belonging to four (overlapping) categories: Econometrics (26 papers), Economic Theory (108 papers), Macroeconomics (66 papers), and Empirical work (198 papers). We take an expansive view of the latter category, including macroeconomic, econometric, or theoretical papers with empirical applications. Within the 198 papers with empirical content, we classify 17 as using laboratory experiments, 29 as using data from randomized field experiments (RCTs), 23 as descriptive (e.g., measures of inequality), and 131 as observational (with a couple papers using multiple methods). The observational category includes non-randomized studies of how variation in a variable $W_i$ affects an outcome variable $Y_i$. Within these 131 papers, we code whether a main specification can be affected by spatial correlation.

Spatial correlation plays a role in a striking number of cases: 63 of 131 observational papers, or 48 percent. The most common geographic unit at which the correlation is likely to occur is the U.S.\ county, with 15 papers. Among the 33 papers with U.S.\ geographic units and potential spatial correlation, the next most common geographic units are states (7 papers), census blocks (2 papers), zip codes (2 papers), and commuting zones (2 papers). Among the 30 papers with non-U.S.\ geographic units and potential spatial correlation, the most common geographic unit is the country (7 papers), followed by Chinese counties (2 papers) and a variety of other local data units (e.g., French municipalities, Austrian regions, and Ukrainian districts). Given that the most common spatial setting in this sample is the county, accounting for 8 percent of all empirical papers, we focus on these papers, systematically revisiting the estimates in the 15 county-level papers.

\subsection{Results}

For the 15 papers with U.S.\ county-level observations, we query replication packages and, where necessary, authors to access the data. For eight of the 15 papers, we cannot reproduce a main specification because a key variable is confidential. In the remaining seven cases, we reproduce a main empirical specification and construct a set of auxiliary outcomes by combining the variables from each paper's replication package with publicly available data, matching the geographic units and approximate measurement periods. Further details are given in \cref{app: empirical}.

\begin{table}[t]
\centering
\caption{TMO Standard Errors\label{tab:papers_summary}}
{%
\renewcommand{\arraystretch}{1.35}
\begin{adjustbox}{max width=\linewidth}
\begin{tabular}{l c c c c c c} \toprule
Paper & SE Ratio & Units \(n\) & Outcomes \(d\) & Periods & $\hat{\eta}$ & \% $\geq\hat{\delta}$ \\
(1) & (2) & (3) & (4) & (5) & (6) & (7) \\
\midrule
\multicolumn{7}{l}{\textbf{Panel A. County-level papers from 2023 survey}} \\
\cite{cook2023evolution} & \textcolor{blue}{\textbf{2.06}} & 3104 & 50 & 12 & 0.30 & 3.64 \\
\cite{bazzi2023other} & \textcolor{blue}{\textbf{1.57}} & 1886 & 139 & 1 & 0.30 & 4.48 \\
\cite{caprettini2023new} & \textcolor{blue}{\textbf{1.52}} & 3022 & 50 & 1 & 0.35 & 2.27 \\
\cite{bernini2023race} & \textcolor{blue}{\textbf{1.39}} & 971 & 168 & 1 & 0.50 & 2.10 \\
\cite{calderon2023racial} & \textcolor{blue}{\textbf{1.37}} & 1263 & 70 & 3 & 0.50 & 1.44 \\
\cite{esposito2023reconciliation} & \textcolor{blue}{\textbf{1.26}} & 786 & 248 & 132 & 0.30 & 4.92 \\
\cite{moscona2023does} & \textcolor{blue}{\textbf{1.18}} & 3000 & 37 & 2 & 0.45 & 0.65 \\
\midrule
\multicolumn{7}{l}{\textbf{Panel B. Non-county examples}} \\
\cite{chetty2014land} & \textcolor{blue}{\textbf{1.60}} & 693 & 129 & 1 & 0.40 & 1.33 \\
\cite{acemoglu2019democracy} & \textcolor{blue}{\textbf{1.04}} & 175 & 57 & 47 & 0.70 & 0.51 \\
\bottomrule \end{tabular}

\end{adjustbox}
}

\medskip
\justifying
\begin{spacing}{0.9}
{\noindent\footnotesize
Notes: Panel A of \cref{tab:papers_summary} shows the results from applying TMO to seven county-level papers. Panel B shows the results for two non-county examples. Column 2 reports the ratio of the TMO standard error to a baseline standard error constructed without any adjustment for cross-sectional correlation. In cross-sectional applications, this baseline is the usual HC1 heteroskedasticity-robust standard error \citep{white1980heteroskedasticity}. In panel applications, the baseline clusters by locations \citep{bertrand2004much}.  Columns 3--5 report the numbers of units, auxiliary outcomes, and periods, respectively. Column 6 gives the selected value of $\eta$ and Column 7 gives the percentage of eligible location pairs whose estimated correlation $\tilde{\rho}_{i,i'}$ exceeds the selected threshold in absolute value. In each application, we search over the grid $\eta$ in $0.1,0.15,\ldots,1$, subject to the restriction that this percentage is at most 5\%.}
\end{spacing}

\noindent\hrulefill
\end{table}

We apply the procedure in \cref{alg: tmo} and choose the threshold using \cref{alg: threshold}. In particular, we search over each $\eta_k$ in $0.1,0.15,\ldots,1$ to maximize the objective \eqref{eq: hat k}, subject to the restriction that the proportion of pairs included in the estimator is at most 5\%.\footnote{\cref{fig: application objective paths} displays the values of $\tilde{V}_{n,d}(\hat{\underline\delta}_{n,d}(\eta_k))$ and $\hat\xi_{n,d}(\hat{\underline\delta}_{n,d}(\eta_k))$, as well as their difference, for each application over the range of $\eta_k$ under consideration.}  Column 2 in Panel A of \cref{tab:papers_summary} gives the ratio of the TMO standard error to a baseline standard error constructed without any adjustment for cross-sectional correlation.\footnote{In cross-sectional applications, this baseline is the usual HC1 heteroskedasticity-robust standard error \citep{white1980heteroskedasticity}. In panel applications, the baseline clusters by locations \citep{bertrand2004much}. \cref{tab:papers_summary_original} gives analogous results, where the baseline is the estimate given in the published paper.} TMO increases the standard error in all cases, with a median increase of 39\%. To illustrate that TMO is applicable to geographic units beyond U.S.\ counties, we revisit also \cite{chetty2014land}, which considers U.S.\ commuting-zone-level data, and \cite{acemoglu2019democracy}, which studies country-level data. The results are recorded in Panel B of \cref{tab:papers_summary} and are consistent.

In \cref{tab: comparison main}, we compare TMO to standard corrections for spatial dependence. Each entry is the ratio of a standard error to the same baseline used in \cref{tab:papers_summary}. Column 1 gives this ratio for standard errors clustered at the state level. Column 2 gives the analogous ratio where TMO augments state-level clusters, i.e., pairs of counties in the same state are never thresholded. In most cases, the standard error increases.

\begin{table}[t] 
\centering 
\caption{Augmenting with Alternative Spatial Standard Errors\label{tab: comparison main}}
\resizebox{\textwidth}{!}{%
\begin{tabular}{l c c c c c c} \toprule
& \multicolumn{2}{c}{State Clusters} & \multicolumn{2}{c}{Uniform Kernel} & \multicolumn{2}{c}{SCPC} \\
\cmidrule(lr){2-3} \cmidrule(lr){4-5} \cmidrule(lr){6-7}
& & +TMO & & +TMO & & +TMO \\
& (1) & (2) & (3) & (4) & (5) & (6) \\
\midrule
\cite{cook2023evolution} & 1.63 [3.1\%] & 2.45 [7.9\%] & 0.98 [3.2\%] & 1.93 [6.5\%] & 0.64 & 1.61 \\
\cite{bazzi2023other} & 1.60 [3.9\%] & 1.85 [6.8\%] & 1.69 [4.5\%] & 1.90 [7.3\%] & 1.39 & 1.78 \\
\cite{caprettini2023new} & 1.87 [3.1\%] & 2.10 [6.7\%] & 2.16 [3.2\%] & 2.33 [5.3\%] & 2.29 & 2.56 \\
\cite{bernini2023race}\textsuperscript{\textdaggerdbl} & 1.24 [0.9\%] & 1.49 [2.4\%] & 1.74 [9.0\%] & 1.84 [9.8\%] & 1.97 & 2.02 \\
\cite{calderon2023racial}\textsuperscript{\textdagger} & 1.21 [4.6\%] & 1.51 [8.2\%] & 1.58 [5.2\%] & 1.77 [8.4\%] & 1.90 & 1.91 \\
\cite{esposito2023reconciliation} & 1.33 [3.6\%] & 1.29 [3.9\%] & 1.53 [3.9\%] & 1.63 [8.6\%] & 2.31 & 2.32 \\
\cite{moscona2023does} & 1.84 [3.1\%] & 1.87 [3.4\%] & 1.76 [3.2\%] & 1.79 [3.5\%] & 1.72 & 1.94 \\
\cite{chetty2014land} & 1.97 [2.9\%] & 2.22 [3.7\%] & 1.91 [2.6\%] & 2.16 [3.4\%] & 3.00 & 3.18 \\
\cite{acemoglu2019democracy} & -- & -- & 1.07 [3.1\%] & 1.09 [3.5\%] & 1.01 & 0.98 \\
\bottomrule \end{tabular}

}

\medskip
\justifying
\begin{spacing}{0.9}
{\noindent\footnotesize
Notes: \cref{tab: comparison main} compares several methods for adjusting standard errors for spatial correlation. Each entry is the ratio of the standard error produced by the method in that column to the same baseline standard error used in \cref{tab:papers_summary}. Wherever applicable, the percentage of location pairs allowed to correlate is displayed in brackets. Column 1 reports state-clustered standard errors. Column 2 adds TMO to state clusters (these columns do not apply to the country-level \cite{acemoglu2019democracy} paper). Column 3 reports \cite{conley1999gmm} standard errors using a uniform kernel. Column 4 combines TMO with that estimator. The bandwidth is 150 miles, except for the country-level \cite{acemoglu2019democracy} application, where it is 650 miles and covers 3\% of country pairs. Column 5 reports the \cite{muller2022spatial,muller2023spatial} SCPC standard error. Column 6 integrates TMO with SCPC. Columns 5 and 6 do not report a percentage of pairs because SCPC does not directly ``zero out'' terms as in \eqref{eq: V hat delta general}. $^\dagger$\cite{calderon2023racial} conduct a weighted regression, which is not supported by the current SCPC Stata package. These estimates show the adjustment for the unweighted specification.\textsuperscript{\textdaggerdbl}For \cite{bernini2023race}, Columns 1 and 2 use the paper’s judicial-division clusters in place of the state clusters, since there are only 11 states in the sample.}
\end{spacing}

\noindent\hrulefill
\end{table}

Columns 3 and 5 display analogous results for \cite{conley1999gmm} standard errors constructed with a uniform kernel and \cite{muller2022spatial,muller2023spatial} SCPC standard errors. Columns 4 and 6 display analogous ratios, where now TMO has been applied to augment \cite{conley1999gmm} and SCPC, respectively. Combining the two forms of correction tends to lead to further increases in the standard error. This is intuitive---distance-based methods are better suited to capture more diffuse, geographically driven dependencies. The TMO standard errors are better suited to capture strong dependencies that are not driven by geographic proximity.

These applications exhibit spatial correlation predicted well by both geographic and non-geographic factors. In an exercise similar to \cref{fig: predictors of correlation}, \cref{tab: predicting correlations in papers} displays, for each of the U.S.\ county- or commuting-zone-level papers, the proportion of highly correlated pairs of locations that are similar along observable characteristics. Pairs of locations with large correlations in absolute value tend to be close (within the top 10\%) in at least one of these observable characteristics. While geographic proximity is typically the first or second most predictive dimension, a substantial proportion of highly correlated pairs are not geographically close.

\section{Conclusion\label{sec: conclusion}}

Economic outcomes are often connected across locations. Estimates of uncertainty that do not account for these connections can exhibit substantial biases. This paper introduces a method for adjusting standard errors for spatial correlation. We call this method ``Thresholding Multiple Outcomes'' (TMO). The essential input is a collection of auxiliary outcome variables. The main assumption is that the spatial correlation in the residuals for a regression problem of interest is shared by the analogous residuals constructed using these auxiliary outcomes. Under this assumption, the auxiliary outcomes can be used to estimate spatial correlation. We propose to use these estimates to determine pairs of locations that are very correlated. Standard errors in the original regression problem can then be adjusted by allowing for correlation among these pairs.

We view the TMO procedure as a complement to standard error corrections based on geographic distance. Indeed, in the examples above, combining the TMO estimator with clustered, spatial HAC, or SCPC standard errors often impacts the resultant estimate. Combinations of this form are straightforward to implement and are likely to offer the greatest robustness to different forms of cross-sectional dependence.

\begin{figure}[t]
    \begin{centering}
    \caption{TMO Adjustment for Different Sets of Auxiliary Outcomes\label{fig: threshold example outcomes}}
    \medskip{}
    \begin{tabular}{c}
    \includegraphics[width=\textwidth]{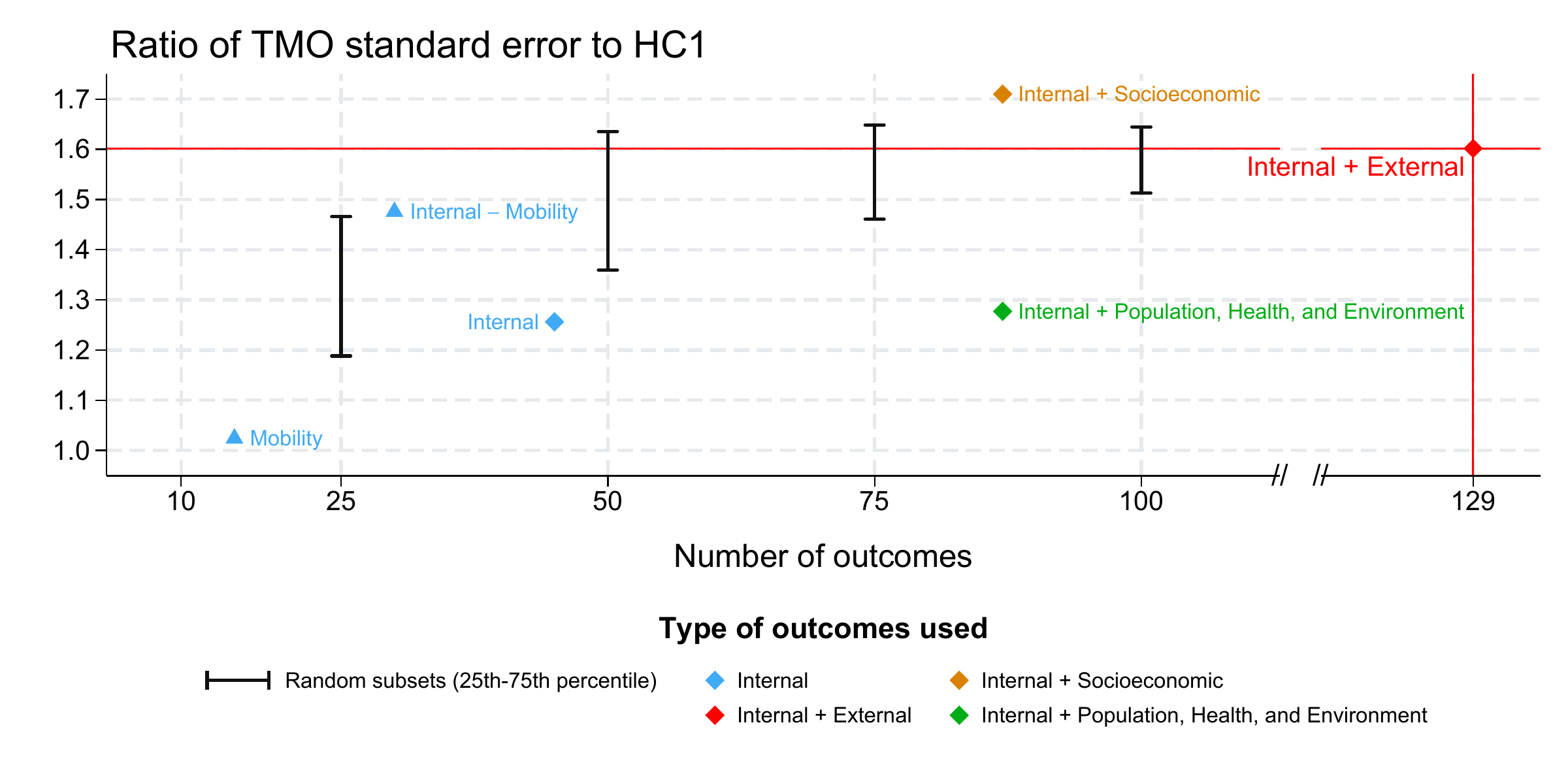}
    \tabularnewline
    \end{tabular}
    \par\end{centering}
    \medskip
    \justifying
   {\noindent\footnotesize{}Notes: \cref{fig: threshold example outcomes} shows the ratio of the TMO standard error to the heteroskedasticity-robust standard error for \cite{chetty2014land}, for different auxiliary outcomes. The \textit{Internal} estimate uses 45 outcomes from the paper's replication package. The \textit{External} set contains 84 county-level outcomes described in \cref{app: empirical}, aggregated to commuting zones. The vertical intervals show the inter-quartile range of the TMO adjustment across 20 randomly drawn subsets of the set of 129 Internal and External outcomes. The number of outcomes in each subset is shown on the $x$-axis. For example, the interval at $x=50$ summarizes the adjustments obtained from 20 rounds of drawing 50 outcomes and applying TMO. The \textit{Mobility} set contains the 15 mobility-related variables among the Internal outcomes, with the remaining 30 Internal outcomes in \textit{Internal--Mobility}. We divide the External outcomes into two groups, the \textit{Socioeconomic} group with 42 economic, educational, housing, and government and social-program outcomes, and the \textit{Population, Health, and Environment} group with the remaining 42 External outcomes. The red lines mark the adjustment and number of outcomes obtained using the full Internal and External datasets.}{\footnotesize\par}
   \noindent\hrulefill
\end{figure}

We conclude with some observations concerning the construction of relevant sets of auxiliary outcomes,  using \cite{chetty2014land} as a running example. \cref{fig: threshold example outcomes} displays the ratio of the TMO standard error to a heteroskedasticity-robust standard error across different choices of the auxiliary outcomes. First, we compute the TMO standard error using random subsets of the 129 auxiliary outcomes, comprising 45 outcomes from the paper’s replication package and 84 of the county-level outcomes described in \cref{app: empirical}, aggregated to commuting zones. The vertical intervals show the 25th and 75th percentiles of the adjustment, for subsets of 25, 50, 75, and 100 outcomes. The size of the adjustment increases with the number of outcomes. With fewer outcomes, estimates of the correlation between outcomes at two locations are noisier, reducing our ability to distinguish null pairs from non-null pairs. Nevertheless, with a sufficiently large number of outcomes, the procedure is relatively insensitive to changes in the composition of auxiliary outcomes.

Second, we examine the impact of characteristics of the auxiliary outcomes. In \cite{chetty2014land}, the outcome of interest is absolute upward mobility. The adjustment associated with using the 45 ``Internal'' outcomes sourced from the paper's replication package is moderate. Further, if the TMO method is implemented using only the 15 such variables  related to alternative measures of mobility, the adjustment is negligible. Conversely, if these mobility outcomes are removed, the TMO adjustment increases. Intuitively, these outcomes are highly correlated with the outcome of interest and much of their variation is removed when they are residualized on the outcome of interest in Step 2 of \cref{alg: tmo}.

On the other hand, if outcome variables are too different from the outcome of interest, the structure of their spatial correlation might tend to be different. To see this, we group the ``External'' outcomes, i.e., those introduced in \cref{app: empirical}, into two groups: ``Socioeconomic'' (e.g., median household income), which are likely more related to mobility, and ``Population, Health, and Environment'' (e.g., particulate-matter exposure). When added to the Internal outcome variables, the Socioeconomic variables induce a larger change in the standard error, even though the two groups contain the same number of auxiliary outcomes. These observations point to a ``sweet spot,'' where the most useful auxiliary outcomes are substantively, but not too mechanically, related to the outcome of interest.

Of course, in the application to \cite{chetty2014land}, we do not know what the standard error should be, i.e., whether larger adjustments are in fact indicative of better performance. \cref{fig: simulated outcome selection} displays the results of a set of analogous experiments in the context of the simulation exercise in \cref{sec: simulation}. Reassuringly, similar patterns emerge. The TMO standard error moves toward the true standard error as the number of relevant auxiliary outcomes increases. Holding the number of auxiliary outcomes fixed, using recent changes rather than recent levels or historical changes also produces an estimate closer to the truth.

Finally, we stress that we see more avenues for further research. For example, although our estimator can be applied to panel settings, as we have done in some of the examples discussed, data sets with panel structures have special features which merit their own analysis. More generally, the information on spatial correlation obtained from multiple outcomes could be useful for other purposes, such as constructing measures of exposure to shocks for estimation of spillover effects or for informing an experimental design.

\end{spacing}
\begin{spacing}{1.18}
\bibliographystyle{apalike}
\bibliography{references.bib}
\end{spacing}
\newpage

\begin{appendix}
\renewcommand\thefigure{\thesection.\arabic{figure}}
\renewcommand\thetable{\thesection.\arabic{table}}
\setcounter{figure}{0}
\setcounter{table}{0}

\begin{center}
\large{\it Supplemental Appendix to:}
\vskip0.2cm
\begin{spacing}{1}
\textbf{\Large Using Multiple Outcomes to \\\vspace{0.5em} Adjust Standard Errors for Spatial Correlation\daggerfootnote{\textit{Date}: \today}} \\\vspace{1em}
\begin{tabular}[t]{c@{\extracolsep{4em}}c} 
\large{Stefano DellaVigna} &  \large{Guido Imbens}\vspace{-0.4em}\\ \vspace{-0.7em}
\small{UC Berkeley and NBER} & \small{Stanford University} \\ \vspace{-0.7em} \\
\large{Woojin Kim} &  \large{David M. Ritzwoller}\vspace{-0.4em}\\ \vspace{-0.7em}
\small{Stanford University} & \small{Harvard University} \\ \vspace{-0.4em}
\end{tabular}%
\end{spacing}
\end{center}
\begin{spacing}{1.13}
\DoToC
\end{spacing}
\thispagestyle{empty}
\setcounter{page}{0}
\setcounter{figure}{0}   
\newpage

\begin{spacing}{1.4}
\normalsize 

\section{Auxiliary Figures and Tables\label{sec: app fig tab}}
\subsection{Figures}

\begin{figure}[h]
    \caption{Spatial Dependence Across Counties in Selected States \label{fig:CANYCountyCorr_app}}
    \medskip{}
    \begin{centering}
    \begin{tabular}{c}
    \includegraphics[width=\textwidth]{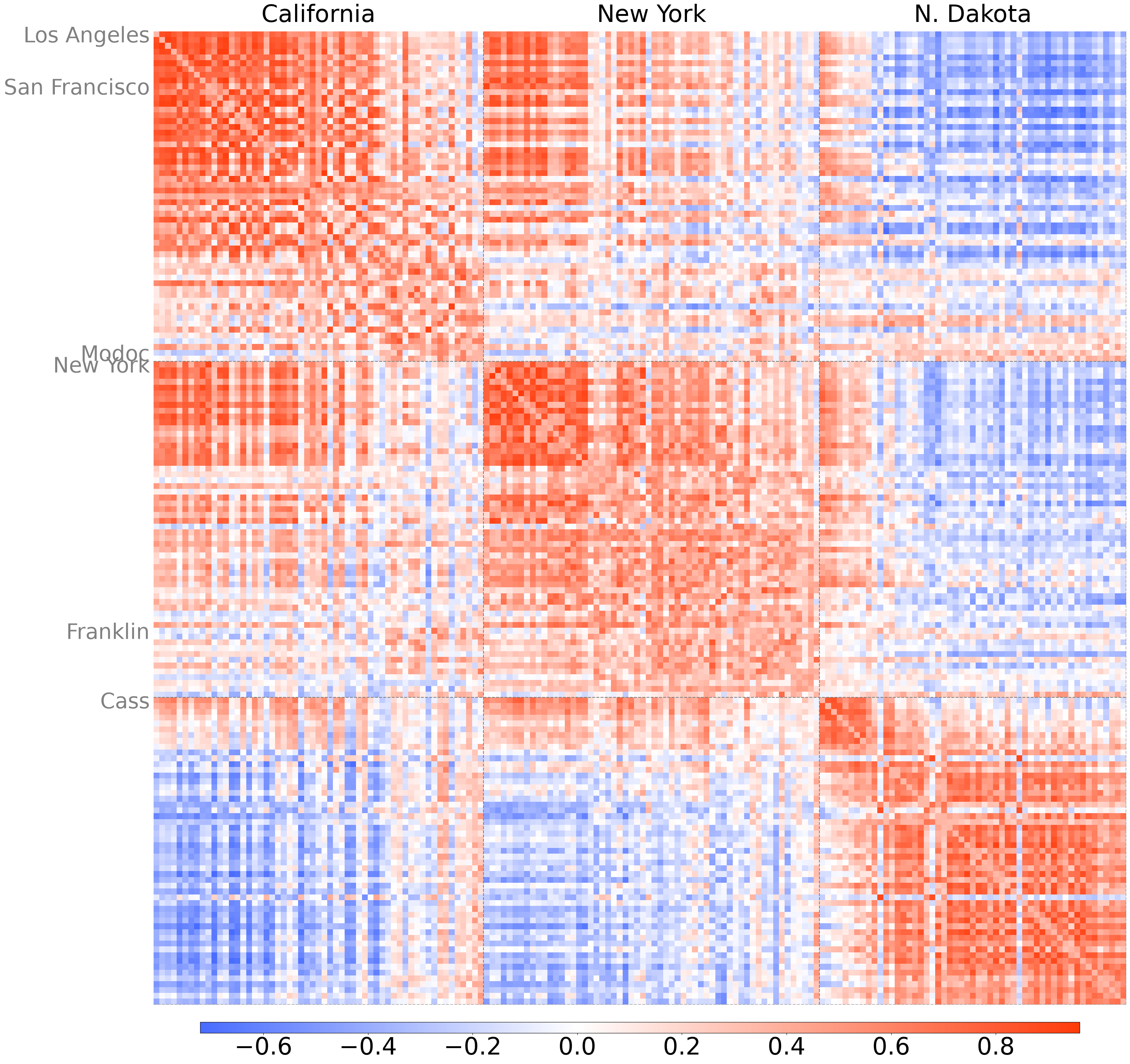}\\
    \end{tabular}
    \par\end{centering}
    
\medskip{}
\justifying
{\noindent \footnotesize{}Notes: \cref{fig:CANYCountyCorr_app} displays correlations across the standardized levels of the 50 outcomes for each pair of counties in California, North Dakota, and New York State, using the same measure as in \cref{fig:SFModoc}. Counties are sorted by population within each state.}{\footnotesize\par}
\end{figure}

\begin{figure}[h]
\begin{centering}
\caption{Spatial Dependence Across U.S. Counties}
\label{fig: county correlation}
\medskip{}
\begin{tabular}{c}
\includegraphics[width=\textwidth]{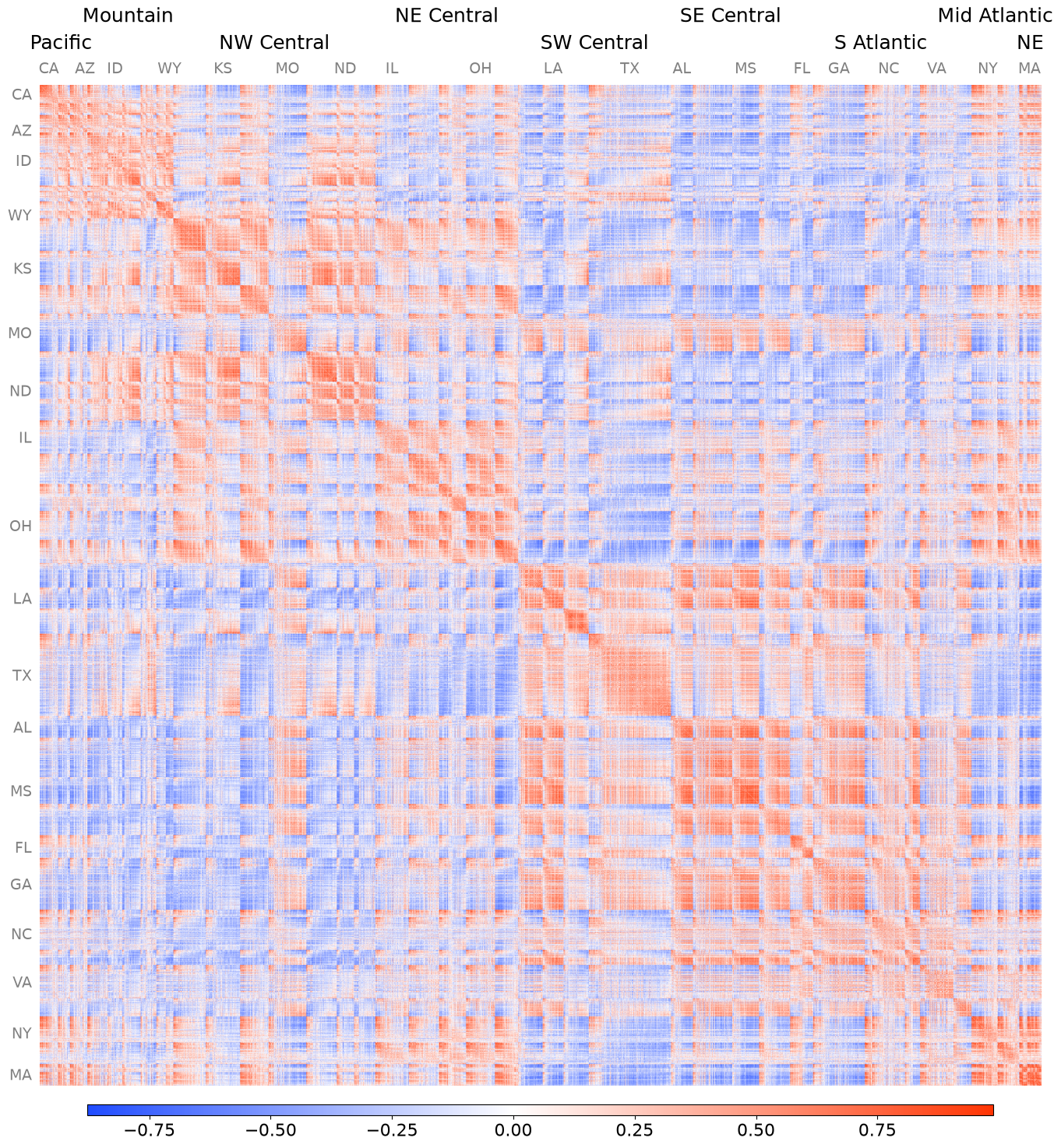}
\end{tabular}
\par\end{centering}
\medskip{}
\justifying
{\noindent \footnotesize{}Notes: \cref{fig: county correlation} displays correlations across the standardized levels of the 50 outcomes for each pair of counties, using the same measure as in \cref{fig:SFModoc}. Counties are sorted by region, state, and population within each state.}{\footnotesize\par}
\end{figure}

\end{spacing}

\clearpage
\begin{landscape}
\begin{figure}[t]
\begin{centering}
\caption{Threshold Selection in Nine Applications\label{fig: application objective paths}}
\medskip{}
\includegraphics[width=\linewidth]{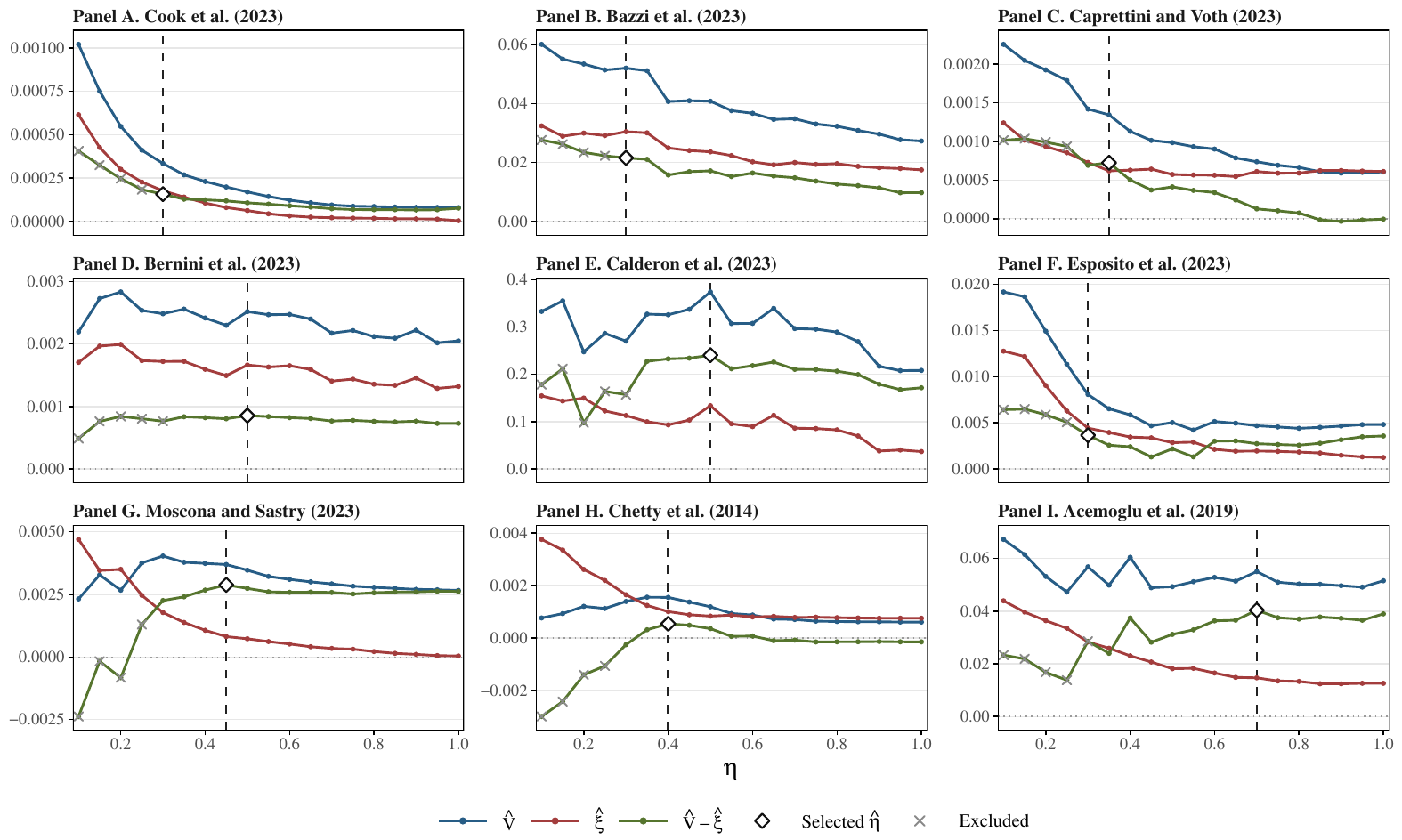}
\par\end{centering}
\medskip{}
\justifying
{\noindent\footnotesize{}Notes: \cref{fig: application objective paths}
displays the threshold-selection criterion for the nine applications considered in \cref{tab:papers_summary}. At each value of $\eta$, the blue and red lines report the variance estimate $\hat V_{n,d}$ and regularization term $\hat\xi_{n,d}$. The green line reports their difference. The gray crosses mark values of $\eta$ excluded because more than 5\% of eligible location pairs have an estimated correlation that exceeds the corresponding threshold in absolute value. The open diamond and dashed vertical line mark the value $\eta_{\hat{k}}$ that maximizes the difference over the remaining grid points. The vertical scale differs across panels.}{\footnotesize\par}
\end{figure}
\end{landscape}
\clearpage

\begin{figure}[t]
\begin{centering}
\caption{Recommendations for Outcome Selection\label{fig: simulated outcome selection}}
\medskip{}
\begin{tabular}{c}
\includegraphics[width=\textwidth]{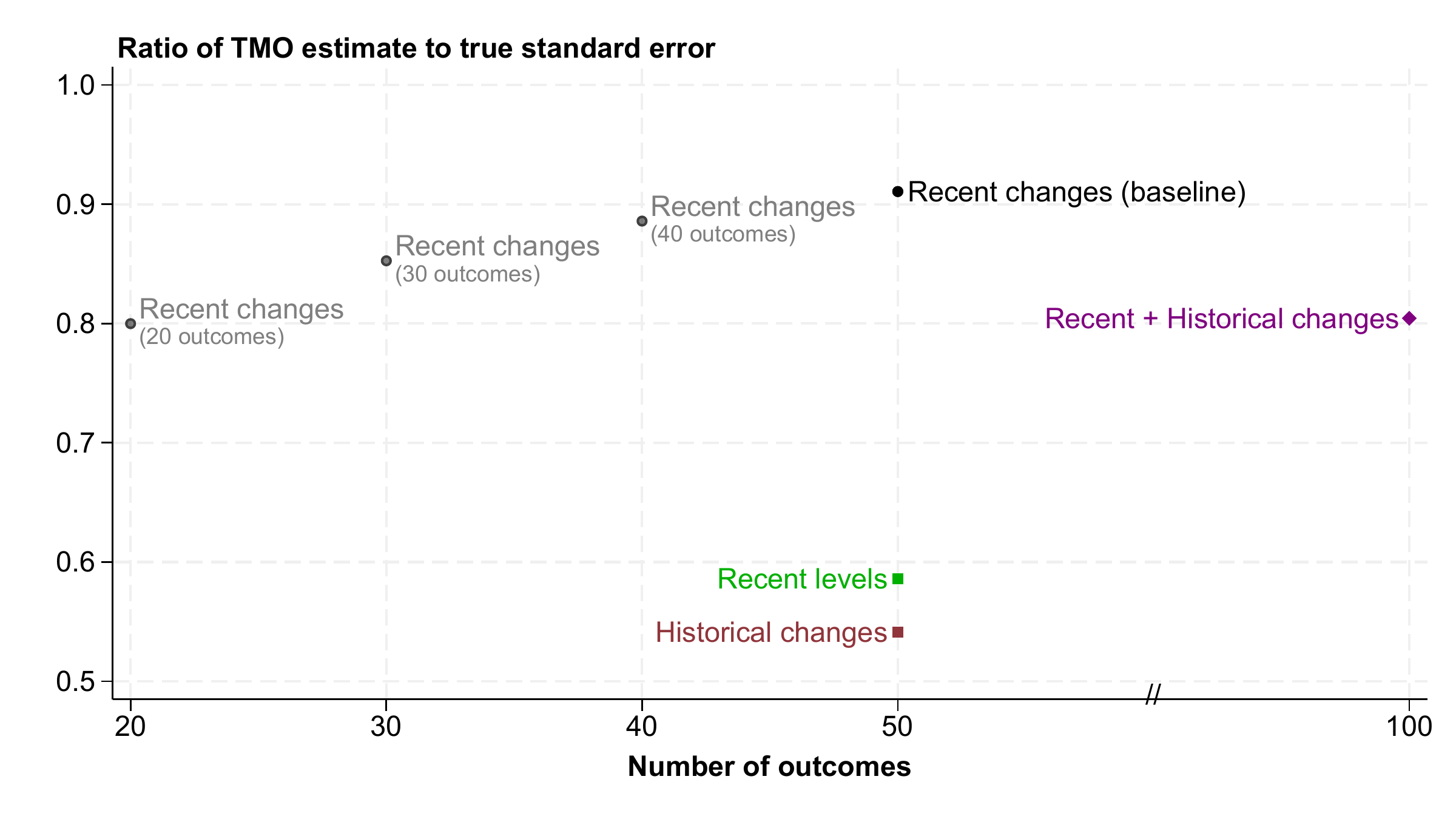}
\tabularnewline
\end{tabular}
\par\end{centering}
\medskip{}
\justifying
{\noindent\footnotesize{}Notes: \cref{fig: simulated outcome selection} reports the performance of the TMO estimator across seven sets of auxiliary outcomes in the calibrated simulation described in \cref{sec: simulation design}. Each point is the average, over 1,000 simulation replications, of the ratio of the TMO standard error to the true standard error. The outcome of interest is generated from the covariance structure calibrated using recent changes in county-level outcomes. The \textit{Recent changes} specifications use 20, 30, 40, or 50 auxiliary outcomes generated from this same covariance structure. The \textit{Recent levels} and \textit{Historical changes} specifications each use 50 auxiliary outcomes generated from separate covariance structures calibrated using, respectively, recent levels and historical changes in county-level outcomes. The \textit{Recent + historical changes} specification combines 50 auxiliary outcomes from the recent-change calibration with 50 from the historical-change calibration.}{\footnotesize\par}
\end{figure}
\clearpage

\subsection{Tables}

\begin{table}[h]
\centering
\caption{Overlap Between the Characteristics in \cref{fig: predictors of correlation}\label{tab: closeness overlap}}
\medskip
\small
\setlength{\tabcolsep}{4pt}
\renewcommand{\arraystretch}{1.12}
\begin{tabularx}{\textwidth}{@{}p{0.29\textwidth}*{6}{>{\centering\arraybackslash}X}>{\centering\arraybackslash}p{0.105\textwidth}@{}} \toprule
& \multicolumn{7}{c}{\% of those county-pairs where $i^\prime$ is \textbf{also} in the top 10\% closest to $i$ in:} \\
\cmidrule(lr){2-8}
\makecell[tl]{County-pairs $(i,i^\prime)$ where $i^\prime$ is in\\the top 10\% closest to $i$ in:} & \makecell[tc]{Pop.} & \makecell[tc]{Urban\\\%} & \makecell[tc]{Dist.} & \makecell[tc]{Median\\income} & \makecell[tc]{Non-\\white \%} & \makecell[tc]{Vote-\\share} & \makecell[tc]{Any of the\\other five} \\ \midrule
\hspace{1em}Population & \textcolor[gray]{0.55}{100} & 23 & 13 & 12 & 12 & 13 & 52 \\
\hspace{1em}Urban \% & 23 & \textcolor[gray]{0.55}{100} & 12 & 12 & 13 & 12 & 50 \\
\hspace{1em}Distance & 13 & 12 & \textcolor[gray]{0.55}{100} & 13 & 15 & 14 & 48 \\
\hspace{1em}Median income & 12 & 12 & 13 & \textcolor[gray]{0.55}{100} & 12 & 11 & 43 \\
\hspace{1em}Non-white \% & 12 & 13 & 15 & 12 & \textcolor[gray]{0.55}{100} & 12 & 46 \\
\hspace{1em}Vote-share & 13 & 12 & 14 & 11 & 12 & \textcolor[gray]{0.55}{100} & 44 \\
\bottomrule \end{tabularx}

\medskip
\justifying
{\noindent\footnotesize Notes: Each row takes the ordered county pairs $(i,i^\prime)$ in which county $i^\prime$ is among the 10\% of counties closest to county $i$ on the row's characteristic. Each of the first six columns gives the percent of those pairs in which $i^\prime$ is also among the 10\% closest to $i$ on the column's characteristic. The last column gives the percentage of those pairs in which $i'$ is also among the closest 10\% on at least one of the other five characteristics.}
\end{table}

\begin{table}[h]
\centering
\caption{Survey of Papers Published in Leading Economics Journals\label{tab:survey}}
\setlength\extrarowheight{-3pt}
\begin{tabular}{l c c c c c c}
\toprule
& All & \textit{AER} & \textit{ECTA} & \textit{JPE} & \textit{RES} & \textit{QJE} \\
\hline 

All papers ($N$) & 370 & 93 & 66 & 71 & 92 & 48 \\

\multicolumn{7}{l}{\textbf{Fields (not mutually exclusive)}} \\
Economic Theory & 108 & 18 & 23 & 34 & 27 & 6 \\
Econometrics & 26 & 4 & 12 & 2 & 6 & 2 \\
Macroeconomics & 66 & 15 & 12 & 15 & 17 & 7 \\
Empirical & 198 & 67 & 20 & 30 & 48 & 33 \\

\multicolumn{7}{l}{\textbf{\hspace{1em}Method (not mutually exclusive)}} \\
\hspace{1em}Descriptive & 23 & 5 & 1 & 3 & 10 & 4 \\
\hspace{1em}Randomized control trial & 29 & 11 & 4 & 6 & 4 & 4 \\
\hspace{1em}Lab experiment & 17 & 8 & 2 & 2 & 2 & 3 \\
\hspace{1em}Observational & 131 & 44 & 13 & 19 & 33 & 22 \\

\multicolumn{7}{l}{\textbf{\hspace{2em}Potential for spatial correlation}} \\
\hspace{2em}No & 68 & 24 & 10 & 12 & 12 & 10 \\
\hspace{2em}Yes & 63 & 20 & 3 & 7 & 21 & 12 \\

\hspace{3em}U.S.\ unit of observation & 33 & 11 & 2 & 2 & 11 & 7 \\
\hspace{4em}\textit{Most common:} County & 15 & 4 & 1 & 1 & 3 & 6 \\
\hspace{4em}\textit{Second most common:} State & 7 & 2 & 1 & 1 & 2 & 1 \\

\hspace{3em}Non-U.S.\ unit of observation & 30 & 9 & 1 & 5 & 10 & 5 \\
\hspace{4em}\textit{Most common:} Country & 7 & 3 & 1 & 2 & 0 & 1 \\
\hspace{4em}\textit{Second most common:} Chinese county & 2 & 1 & 0 & 0 & 0 & 1 \\

\bottomrule
\end{tabular}

\medskip
\justifying
{\noindent\footnotesize Notes: \cref{tab:survey} gives the results of a survey of papers published in the \textit{American Economic Review}, \textit{Econometrica}, \textit{Journal of Political Economy}, \textit{Review of Economic Studies}, and \textit{Quarterly Journal of Economics} in 2023. Each paper is categorized by field. Empirical papers are categorized by their methodology. Observational papers are categorized according to whether their analyses have the potential to be affected by spatial correlation.}
\end{table}

\newgeometry{left=0.5in, right=0.5in, top=1in, bottom=1in}
\begin{table}[H]
\centering
\caption{Ratio of TMO to Original Standard Errors in Nine Recent Papers\label{tab:papers_summary_original}}
{
\footnotesize
\setlength{\tabcolsep}{3pt}
\renewcommand{\arraystretch}{1.15}
\begin{tabular*}{\linewidth}{@{\extracolsep{\fill}} c c c c c c c c c c} \toprule
SE Ratio & TMO & Original SE & \shortstack{Original \\ clustering} & Coefficient & Units \(n\) & Periods & Outcomes \(d\) & $\hat{\eta}$ & \shortstack{Eligible \% \\ $\geq\hat{\delta}$} \\
(1) & (2) & (3) & (4) & (5) & (6) & (7) & (8) & (9) & (10) \\
\midrule
\multicolumn{10}{@{}l@{}}{\textbf{Panel A. County-level papers from 2023 survey}} \\
\multicolumn{10}{@{}l@{}}{\rlap{\hspace{1em}\cite{cook2023evolution}: Effect of White casualties in WWII on Green Book establishments (Table 3 Column 4)}} \\
\hspace{2em}\textcolor{blue}{\textbf{2.06}} & 0.018 & 0.009 & County & 0.065 & 3104 & 12 & 50 & 0.30 & 3.64 \\
\multicolumn{10}{@{}l@{}}{\rlap{\hspace{1em}\cite{bazzi2023other}: Effect of Southern Whites \% on 2016 Trump vote-share (Table 2 Column 4)}} \\
\hspace{2em}\textcolor{blue}{\textbf{1.39}} & 0.233 & 0.167 & \(60\times60\) mile grid cells & 1.026 & 1886 & 1 & 139 & 0.35 & 3.39 \\
\multicolumn{10}{@{}l@{}}{\rlap{\hspace{1em}\cite{caprettini2023new}: Effect of New Deal grants on purchase of WWII bonds (Table 2 Column 1)}} \\
\hspace{2em}\textcolor{blue}{\textbf{1.52}} & 0.037 & 0.024 & None (HC1) & 0.191 & 3022 & 1 & 50 & 0.35 & 2.27 \\
\multicolumn{10}{@{}l@{}}{\rlap{\hspace{1em}\cite{bernini2023race}: Effect of Black \% $\times$ Voting Rights Act on share of black elected officials (Table 2 Column 4)}} \\
\hspace{2em}\textcolor{blue}{\textbf{1.20}} & 0.054 & 0.045 & Judicial division & 0.103 & 971 & 1 & 168 & 0.60 & 1.50 \\
\multicolumn{10}{@{}l@{}}{\rlap{\hspace{1em}\cite{calderon2023racial}: Effect of Black population on Democratic vote-share (Table 2 Column 6)}} \\
\hspace{2em}\textcolor{blue}{\textbf{1.37}} & 0.612 & 0.446 & County & 1.885 & 1263 & 3 & 70 & 0.50 & 1.44 \\
\multicolumn{10}{@{}l@{}}{\rlap{\hspace{1em}\cite{esposito2023reconciliation}: Effect of The Birth of a Nation screening on patriotic vs. divisive language in newspapers (Table 3 Column 3)}} \\
\hspace{2em}\textcolor{blue}{\textbf{1.26}} & 0.090 & 0.071 & County & 1.092 & 786 & 132 & 248 & 0.30 & 4.92 \\
\multicolumn{10}{@{}l@{}}{\rlap{\hspace{1em}\cite{moscona2023does}: Effect of extreme temperature exposure $\times$ innovation on price per agricultural land acre (Table 3 Column 1)}} \\
\hspace{2em}\textcolor{blue}{\textbf{1.02}} & 0.096 & 0.095 & State & 0.249 & 3000 & 2 & 37 & 0.60 & 0.34 \\
\midrule
\multicolumn{10}{@{}l@{}}{\textbf{Panel B. Non-county examples}} \\
\multicolumn{10}{@{}l@{}}{\rlap{\hspace{1em}\cite{chetty2014land}: Correlation between share of African American and upward mobility (Figure 8 Row 1)}} \\
\hspace{2em}\textcolor{blue}{\textbf{1.13}} & 0.055 & 0.048 & State & -0.356 & 693 & 1 & 129 & 0.55 & 0.82 \\
\multicolumn{10}{@{}l@{}}{\rlap{\hspace{1em}\cite{acemoglu2019democracy}: Effect of democracy on log GDP per capita (Table 2 Column 3)}} \\
\hspace{2em}\textcolor{blue}{\textbf{1.04}} & 0.235 & 0.226 & Country & 0.787 & 175 & 47 & 57 & 0.70 & 0.51 \\
\bottomrule \end{tabular*}

}

\medskip
\justifying
\begin{spacing}{0.9}
{\noindent\footnotesize
Notes: Panel A of \cref{tab:papers_summary_original} shows the results from applying TMO to seven recent county-level papers surveyed in \cref{tab:survey}. Panel B shows the results for two non-county examples, \cite{chetty2014land} at the commuting-zone level and \cite{acemoglu2019democracy} at the country level. Column 1 reports the ratio of the TMO standard error in Column 2 to the original standard error in Column 3. The TMO standard error augments the original level of clustering, reported in Column 4. HC1 denotes a heteroskedasticity-consistent standard error without clustering. Column 5 reports the coefficient on the main regressor. Columns 6 and 7 report the numbers of locations and periods. Column 8 reports the number of auxiliary outcomes. Column 9 reports the selected value of $\eta$. Column 10 reports the percentage of eligible location pairs whose estimated correlation exceeds the selected threshold in absolute value. Eligible pairs are distinct location pairs outside the set protected by the paper's original clustering.} 
\end{spacing}
\end{table}

\clearpage
\restoregeometry

\begin{landscape}
\begin{table}[H]
\centering
\caption{Observable Similarity Among Selected Pairs in U.S.\ Applications\label{tab: predicting correlations in papers}}
{
\renewcommand{\arraystretch}{1.4}
\begin{tabular}{l c c c c c c c c c} \toprule
& & & \multicolumn{7}{c}{\textit{Among eligible pairs $(i,i^\prime)$ with $\lvert\tilde{\rho}_{i,i^\prime}\rvert\geq\hat{\underline\delta}_{n,d}(\eta_{\hat{k}})$, percentage in which}} \\
& & & \multicolumn{7}{c}{\textit{at least one unit ranks among the 10\% closest to the other in:}} \\
\cmidrule(lr){4-10}
& $\hat{\underline\delta}_{n,d}(\eta_{\hat{k}})$ & Eligible \% selected & Distance & Population & Urban \% & Median income & Non-white \% & Vote-share & Any of $\leftarrow$ \\
& (1) & (2) & (3) & (4) & (5) & (6) & (7) & (8) & (9) \\
\hline
\cite{cook2023evolution} & 0.53 & 3.64 & 33 & 18 & 17 & 16 & 17 & 14 & 65 \\
\cite{bazzi2023other} & 0.36 & 3.39 & 24 & 19 & 17 & 18 & 15 & 15 & 62 \\
\cite{caprettini2023new} & 0.46 & 2.27 & 25 & 15 & 15 & 14 & 15 & 13 & 59 \\
\cite{bernini2023race} & 0.26 & 1.50 & 23 & 16 & 15 & 17 & 17 & 15 & 61 \\
\cite{calderon2023racial} & 0.55 & 1.44 & 34 & 19 & 19 & 14 & 16 & 17 & 66 \\
\cite{esposito2023reconciliation} & 0.26 & 4.92 & 16 & 14 & 13 & 13 & 13 & 12 & 53 \\
\cite{moscona2023does} & 0.65 & 0.34 & 27 & 20 & 20 & 21 & 20 & 16 & 67 \\
\cite{chetty2014land} & 0.58 & 0.82 & 19 & 29 & 26 & 25 & 21 & 19 & 67 \\
\bottomrule \end{tabular}

}

\medskip
\justifying
\begin{spacing}{1}
{\noindent\footnotesize
Notes: \cref{tab: predicting correlations in papers} reports, for each U.S.\ application in \cref{tab:papers_summary}, observable similarity among the pairs selected by TMO. Column 1 reports the selected threshold $\hat{\underline\delta}_{n,d}(\eta_{\hat{k}})$. Column 2 reports the percentage of eligible pairs---distinct pairs outside the protected set---for which $\lvert\tilde{\rho}_{i,i^\prime}\rvert\geq\hat{\underline\delta}_{n,d}(\eta_{\hat{k}})$. Columns 3--8 report, among these selected pairs, the percentage for which at least one unit ranks among the 10\% closest to the other in the characteristic listed in the column header. Column 9 reports the percentage satisfying this criterion for at least one of the six characteristics. Columns 3--9 use pairs with complete covariate and distance information.}
\end{spacing}
\end{table}

\end{landscape}

\begin{spacing}{1.4}
\clearpage

\section{Details and Extensions\label{sec: app simulation}}

\subsection{Outcome Selection and Cleaning\label{sec: simulation county data}}
We use 50 county-level outcome variables from the following sources:
\begin{itemize}
\item The Social Determinants of Health Database from the Agency for Healthcare Research and Quality (AHRQ). \scriptsize{\url{https://www.ahrq.gov/sdoh/data-analytics/sdoh-data.html#download}}\normalsize
\item U.S. Census Bureau. (2011). ``USA Counties.'' Statistical Compendia. \scriptsize{\url{https://www.census.gov/library/publications/2011/compendia/usa-counties-2011.html}}\normalsize
\end{itemize}
We retain outcomes from USA Counties or AHRQ that are observed for every county in this sample and measure nonnegative quantities. We also drop one of two near-duplicate child-poverty rates. The resulting 50 outcomes comprise 33 from USA Counties and 17 from AHRQ.

The variables and their measurement years are listed in Tables \ref{tab:simoutcomes1}--\ref{tab:simoutcomes3}, displayed in \cref{app: panel tables}. They cover economics (e.g., income and unemployment), health (e.g., death rates and lack of health insurance), demographics (e.g., population density and the Black population), housing (e.g., home values and rents), and environment (e.g., water use and air pollution).

We normalize variables that scale with population, such as the number of Black residents, private nonfarm establishments, and physicians, by county population. For each variable, we take $\ln(x)$ if all observed values are positive and $\ln(1+x)$ if some are zero. For USA Counties, we use the first and last usable years with observations for at least 3,000 counties; for AHRQ, we use 2009 and 2018, requiring the same coverage in both years. Before differencing, we winsorize a variable within a year at the 0.1 and 99.9 percentiles if any county's value is more than eight standard deviations from the mean. We then take the difference between the two endpoint values and apply the same conditional winsorization rule to the long difference.

\subsection{Simulation Design\label{sec: app design}}
In this Appendix, we detail how we construct the covariance matrix $\Sigma$ for the simulation considered in \cref{sec: simulation design}. Our aim is to ensure that its non-zero off-diagonal elements primarily correspond to ``non-null'' pairs of counties, while maintaining positive definiteness. We take as input the estimates $\widehat{\textrm{corr}}_{i,i^\prime}$ defined in \eqref{eq: outcome correlation}, computed from the 50 outcomes above after removing state fixed effects and standardizing each residual series. We partition the counties into 99 clusters: 90 containing 31 counties and nine containing 30 counties.

We form the clusters in two steps. First, we choose 99 seed counties sequentially. At each step, we score each eligible county by the sum of its up to 30 largest positive correlations with other eligible counties. We select the county with the largest score, then exclude it and up to 30 of its most correlated eligible counties from subsequent seed selection. Only the selected seeds are assigned to clusters at this stage.

Second, starting with each seed as a one-county cluster, we repeatedly make the single best placement. Among all pairs of an unassigned county and a cluster that has not reached its current capacity, we choose the pair with the highest average correlation between the county and the cluster's existing members, and add the county to that cluster. We initially set every cluster's capacity to 30. Once all clusters contain 30 counties, we raise their capacities to 31 and assign the remaining 90 counties by the same rule. 

We set $\Sigma_{i,i^\prime}=\widehat{\textrm{corr}}_{i,i^\prime}$ when counties $i$ and $i^\prime$ belong to the same cluster, and set it to zero otherwise. Each of the resulting within-cluster correlation matrices is positive definite, so the block-diagonal matrix $\Sigma$ is positive definite.

\subsection{Alternative Estimators and Data Structures\label{sec: extensions}}

In this Appendix, we detail several extensions to the method proposed in \cref{sec: proposal}.

\subsubsection{Covariates}

In the main text, for ease of exposition, we treat the case where there are no covariates. The extension to the case with covariates is straightforward, by the Frisch-Waugh-Lovell Theorem. In particular, let $\tilde W_i$ and $\tilde Y_i^{(j)}$ denote the residuals from regressing $W_i$ and $Y_i^{(j)}$ on a covariate vector $X_i$, respectively. A standard error for $\hat{\tau}^{(0)}$ can then be constructed as before, by using these data.

\subsubsection{Instrumental Variables Regression}

For each unit $i$, suppose that we observe the instrumental variable $Z_i$. For the outcome of interest and each auxiliary outcome, we regress the outcome on $W_i$ and the covariates by two-stage least squares, using $Z_i$ as the instrument for $W_i$. We compute residuals using the estimated coefficients and the observed treatment $W_i$. We use these residuals in \cref{alg: tmo}. In the variance formula \eqref{eq: V hat delta general}, we replace $\bar W$ by the vector of first-stage fitted treatment values, after removing covariates as above. We caution that this approach is not robust to weak identification. The construction of identification-robust standard error estimates for settings with spatial correlation is an interesting direction for future research. See \cite{andrews2019weak} for a recent review of identification-robust  standard errors for linear instrumental variables regression in non-homoskedastic data.

\subsubsection{Panel Regressions}

Suppose that we observe the collection of outcomes across $t$ time periods. That is, we observe $(Y_{i,s}^{(0)}, Y^{(1)}_{i,s},\ldots,Y^{(d)}_{i,s})$ for each unit $i$ in $1,\ldots,n$ and period $s$ in $1,\ldots,t$. Here, we are often interested in constructing standard errors for coefficients in regressions of the form
\begin{equation} \label{eq: linear model panel}
    Y^{(0)}_{i,s} = \alpha + \tau^{(0)} W_{i,s} + \theta^\top X_{i,s} +\varepsilon^{(0)}_{i,s}~,
\end{equation}
where, in many cases, the covariates $X_{i,s}$ contain various fixed effects. To adjust standard errors for correlation across units, there are at least two options. First, we might account for the correlation between the units $i$ and $i^\prime$ only \textit{within} each time period. That is, the residuals $(\varepsilon^{(j)}_{i,s},\varepsilon^{(j)}_{i^{\prime},s})$ are allowed to be correlated, but the residuals $(\varepsilon^{(j)}_{i,s},\varepsilon^{(j)}_{i',s'})$, for $s{\neq}s^{\prime}$, are not. An alternative option would be to allow for correlation \textit{across} time periods as well. For the purposes of this paper, we adopt the latter approach.  In particular, we estimate the covariance between the summed scores of units $i$ and $i^\prime$ with the estimator
\begin{align}
\tilde{\lambda}_{i,i^{\prime}} = \frac{1}{d} \sum_{j=1}^d
 \tilde{w}^{(j;0)}_{i}\tilde{w}^{(j;0)}_{i'}~,
\quad\text{where}\quad
w^{(j)}_{i} = \sum_{s=1}^t \tilde{W}_{i,s}\hat{\varepsilon}^{(j)}_{i,s}~.
\end{align}
That is, in effect, we apply the procedure of the main text with the scores $w^{(j)}_{i}$ in place of $\hat{\varepsilon}^{(j)}_{i}$. Standard errors are then adjusted, as before, allowing for correlation across units and across time periods.

\subsection{Worked Examples\label{app: examples}}

In this Appendix, we specify several examples that substantiate claims made in the main text. 

\subsubsection{Non-Vanishing Bias}

We begin by giving an example where the relative bias approximation \eqref{eq: risk approximation relative bias representation} converges to a constant that can be made arbitrarily close to $-1$ for any sequence of thresholds.

\begin{lem}\label{lem: bias irreducible}
    Fix $a>0$. For each integer $k$, let $d=k$, $n = k^2(k^2+1)$, $s=k^2$, and 
    \begin{align}
        \Lambda_n = I_{k^2} \otimes 
        \left( 
            \left(1 - \frac{a}{k^2}\right)I_{k^2 + 1} + \frac{a}{k^2} \mathbf{1}_{k^2+1} \mathbf{1}_{k^2+1}^{\top}
        \right)
        \quad\text{and}\quad
        \Gamma_d = I_{d+1}~.
    \end{align}
    This sequence satisfies the conditions of \cref{lem: bias decomposition} and, for any sequence of thresholds $\delta_k$, it holds that
    \begin{align}
        \underset{k\to\infty}{\lim\inf}
        \frac{\vert B_{n,d}(\delta_k)\vert}{V_n}
        \geq \frac{a}{1+a}~.
    \end{align}
\end{lem}

\subsubsection{Sharpness of \cref{lem: variance bound}} Next, we show that the term $V_n^{\eta/2}$ cannot be removed from the bound obtained in \cref{lem: variance bound}.

\begin{lem}\label{lem: variance sharpness}
    Fix $0<\eta<1/2$ and choose $\eta <\alpha < 1-\eta$. Let $d=\lfloor n^{\alpha} \rfloor$, $s=1$, $\Lambda_n = I_n$, and $\Gamma_{d}=I_{d+1}$. It holds that
    \begin{align}
        \frac{\sqrt{\Var(\hat{V}_{n,d}(\underline{\delta}_{n,d}(\eta))}}{V_n}
        =V_n^{\eta/2+o(1)}.
    \end{align}
\end{lem}

\subsubsection{Failure of the Unregularized Selector} In turn, we exhibit a pathology in the behavior of the unregularized selector $\check{k}$ defined in \eqref{eq: check k}. Fix $g=2$ and assume that $0 < 2\eta_1 < \eta_2 <1/4$. 

\begin{lem}\label{lem: pathology}
    Choose $2\eta_2 <\alpha < 1-2\eta_1$. Let $d=\lfloor n^{\alpha} \rfloor$, $s=1$, $\Lambda_n = I_n$, and $\Gamma_{d}=I_{d+1}$. It holds that
    \begin{align}
        \frac{\operatorname{Bias}_{n,d}(k)}{V_n} = -V_{n}^{\eta_k+o(1)},\quad k \in \{1,2\}~.
    \end{align}
    However, it also holds that $P\{\check{k} = 1\} \to 1/2$ and, moreover, that
    \begin{align}
    P\left\{ 
    \frac{1}{V_n}\left(
    \vert \operatorname{Bias}_{n,d}(\check k)\vert 
    -
    \vert \operatorname{Bias}_{n,d}(k^\star) \vert 
        \right)
    > C (V_n^{\eta_{k^\star}/2}
        +\mathcal{R}_{n,d}(s))
    \right\}
    \to 1/2
    \label{eq: relative bias regret failure}
\end{align}
for every fixed $C > 0$.
\end{lem}

\subsection{Augmenting Distance-Based Estimators\label{app: augmenting}}

The TMO estimator can be used to augment existing spatial standard errors that are based on modeling dependence in terms of geographic distance. Recall that, for the method detailed in \cref{sec: proposal}, the covariance $\sigma^{(0)}_{i,i^\prime}$ is estimated by the quantity
\begin{equation}\label{eq: app diagonal}
    \hat{\sigma}^{(0)}_{i,i^\prime}(\delta) 
    =
    \begin{cases}
    \hat{\varepsilon}^{(0)}_i \hat{\varepsilon}^{(0)}_i ~,& i = i^\prime, \\
    \hat{\varepsilon}^{(0)}_i \hat{\varepsilon}^{(0)}_{i^\prime}\text{ } 
    \mathbb{I}\{ \vert \tilde{\rho}_{i,i^\prime}\vert \geq \delta \}~, & i\neq	i^\prime~.
    \end{cases}
\end{equation}
That is, the TMO estimator does not threshold diagonal elements of the residual covariance matrix---in effect, augmenting the standard \cite{white1980heteroskedasticity} heteroskedasticity-consistent variance estimate.

The same idea can be applied to augment alternative standard error estimates. Let $\mathcal{C} \subseteq [n]^2$ denote a subset of the pairs of units $(i,i^\prime)$ that includes all pairs of the form $(i,i)$. These might be pairs of units in the same cluster, or pairs of units whose geographic distance is smaller than some threshold. Suppose that, moreover, we have at our disposal an alternative estimator of the covariance $\sigma^{(0)}_{i,i^\prime}$ for each pair $(i,i^\prime)$ in $\mathcal{C}$. Denote this estimator by $\tilde{\sigma}^{(0)}_{i,i^\prime}$. This estimator could be simply $\tilde{\sigma}^{(0)}_{i,i^\prime}=\hat{\varepsilon}^{(0)}_i \hat{\varepsilon}^{(0)}_{i^\prime}$ or it could be, e.g., the estimator associated with a \cite{conley1999gmm} standard error with a chosen kernel and bandwidth. We can construct the covariance matrix estimate by collecting the estimates
\begin{equation}\label{eq: aug}
    \hat{\sigma}^{(0)}_{i,i^\prime}(\delta)
    =
    \begin{cases}
    \tilde{\sigma}^{(0)}_{i,i^\prime} ~,& (i, i^\prime)\in\mathcal{C}, \\
    \hat{\varepsilon}^{(0)}_i \hat{\varepsilon}^{(0)}_{i^\prime}\text{ } 
    \mathbb{I}\{ \vert \tilde{\rho}_{i,i^\prime}\vert \geq \delta \}~, & (i, i^\prime)\not\in\mathcal{C}~.
    \end{cases}
\end{equation}
In this case, the optimal threshold $\hat{\delta}_{n,d}^{\star}$ should be estimated by considering the distribution of only those pairs of units $(i, i^\prime)$ not in $\mathcal{C}$. 

A similar approach can be taken to use TMO to augment the \cite{muller2022spatial,muller2023spatial} SCPC standard error estimator. \cite{muller2022spatial,muller2023spatial} estimate the variance with an estimator of the form
\begin{equation}
    \hat{V}_{\mathrm{SCPC}}(q)
    =
    \frac{(\bar{W}^\top\bar{W})^{-2}}{q}
    \sum_{j=1}^q
    r_j^\top
    \mathsf{diag}(\bar{W})
    \hat{\varepsilon}^{(0)}
    (\hat{\varepsilon}^{(0)})^\top
    \mathsf{diag}(\bar{W})
    r_j~,
\end{equation}
where $r_j$ is an approximation to the $j$th principal component of $\mathsf{diag}(\bar W)\Var(\varepsilon^{(0)})\mathsf{diag}(\bar W)$. This estimator can, thus, be augmented with the TMO standard error by taking
\begin{equation}
    \hat{V}_{\mathrm{SCPC}+\mathrm{TMO}}(\delta,q)
    =
    \frac{(\bar{W}^\top\bar{W})^{-2}}{q}
    \sum_{j=1}^q
    r_j^\top
    \mathsf{diag}(\bar{W})
    \hat{\Sigma}_{\mathrm{SCPC}}(\delta)
    \mathsf{diag}(\bar{W})
    r_j
    +
    (\bar{W}^\top\bar{W})^{-2}
    \bar{W}^\top\hat{\Sigma}(\delta)\bar{W},
\end{equation}
where $\hat{\Sigma}_{\mathrm{SCPC}}(\delta)$ collects
\begin{equation}
    \hat{\sigma}^{\mathrm{SCPC}}_{i,i^\prime}(\delta)
    =
    \begin{cases}
    \hat{\varepsilon}^{(0)}_i
    \hat{\varepsilon}^{(0)}_{i^\prime},
    & (i,i^\prime)\notin\mathcal C
      \text{ and }
      \vert\tilde{\rho}_{i,i^\prime}\vert<\delta,\\
    0, & \text{otherwise},
    \end{cases}
\end{equation}
and $\hat{\Sigma}(\delta)$ is defined in \eqref{eq: aug}.

\section{Further Details for Empirical Applications\label{app: empirical}}

For each application, we begin with candidate auxiliary outcomes from the original replication package and supplement them with external data. We select external sources at the appropriate geographic level and construct levels or changes to match the form of the main outcome, aligning time periods where feasible. To the combined set of variables, we apply the rules in \cref{tab: outcome rules} in two steps. First, we provide Claude Opus 5.5 with the context of the paper, the specification of interest, and the paper's replication code and ask it to filter out variables according to the criteria in Panel A; the model flags uncertain classifications for human review. We then apply the statistical rules in Panel B in code. In settings where some outcomes are missing for some locations, we estimate the correlation between residuals at two locations using only the auxiliary outcomes observed at both locations.

For some applications, we assemble a shared county dataset from USA Counties, the historical County and City Data Book (hereafter, County Data Book), the Agency for Healthcare Research and Quality's Social Determinants of Health Database, and supplemental natality and life-expectancy data.\footnote{Sources: USA Counties, \url{https://www.census.gov/library/publications/2011/compendia/usa-counties-2011.html}; County and City Data Book, \url{https://doi.org/10.3886/ICPSR07736.v2}; AHRQ, \url{https://www.ahrq.gov/sdoh/data-analytics/sdoh-data.html}; natality data compiled by Bailey and coauthors, \url{https://doi.org/10.3886/E100229V4}; and Institute for Health Metrics and Evaluation life tables, \url{https://doi.org/10.6069/3WQ2-TG23}.} For others, in different time periods, we find alternate sources of data detailed below.

\begin{table}[t]
\centering
\caption{Construction of the Auxiliary Outcomes\label{tab: outcome rules}}
\medskip
\small
\renewcommand{\arraystretch}{1.15}
\begin{tabularx}{\textwidth}{@{}l X@{}} \toprule
\multicolumn{2}{@{}l}{\textbf{Panel A. Filters for auxiliary outcomes}} \\
\multicolumn{2}{@{}l}{\textit{Drop the following:}} \\
1 & \textit{Treatment-related variables}: the treatment variable itself, instruments, components (e.g., that sum to the treatment), alternate measures of the treatment, and variables used to assign or predict the treatment \\
2 & \textit{Identifiers} (e.g., FIPS code), \textit{time, period, sample, weights, and metadata fields} \\
3 & \textit{Fixed geographic features} (e.g., coordinates, land area, elevation, and long-run climate) \\
4 & \textit{Variables derived from other variables} (e.g., quantiles or logs) if those variables are included as auxiliary outcomes \\
\midrule
\multicolumn{2}{@{}l}{\textbf{Panel B. Statistical screens on the remaining candidates}} \\
\multicolumn{2}{@{}l}{\textit{Drop variables for which:}} \\
1 & The residual is missing for more than 50\% of the locations \\
2 & There is no residual variation or, in panels, no variation in location-level scores \\
3 & $|\hat{\rho}|>0.8$ or $|\hat{\rho}_S|>0.95$ with the treatment, a control, or an instrument \\
4 & $|\hat{\rho}|>0.8$ or $|\hat{\rho}_S|>0.95$ with the outcome of interest (in panels, also $|\hat{\rho}|>0.8$ between location-level scores) \\
5 & $|\hat{\rho}_S|>0.95$ or $|\hat{\rho}|>\sqrt{0.95}$ (i.e., $R^2>0.95$) with another auxiliary outcome\textsuperscript{\textdagger} \\
6 & It reconstructs the outcome residual in combination with other auxiliary residuals\textsuperscript{\textdagger} \\
\bottomrule
\end{tabularx}

\medskip
\justifying
\begin{spacing}{1}
{\noindent\footnotesize
Notes: The statistic $\hat{\rho}$ is the Pearson correlation \textit{after} partialing out from both variables any fixed effects included in the specification. The statistic $\hat{\rho}_S$ denotes the rank correlation of the raw variables. \textsuperscript{\textdagger}When one variable must be dropped from a group (e.g., two variables that are monotonic transformations of one another), drop the variable with more missing raw observations. Break ties by greater absolute skewness. For B6, drop the variable whose removal most reduces the $R^2$ for reconstructing the outcome.}
\end{spacing}
\end{table}

\subsection{\cite{bazzi2023other}}

\cite{bazzi2023other} examine the effects of the twentieth-century migration of Southern whites within the United States in a sample of 1,886 counties. The main finding is that the Southern white migration in the early twentieth century is associated with significantly higher Republican vote shares in the twenty-first century. \cite{bazzi2023other} consider the model
\begin{equation}
	\text{Vote}_c =\tau \cdot \text{Southern Whites}_{c,1940} + \gamma\cdot X_{c} + \alpha_s + \varepsilon_{c}~,
\end{equation}
where $c$ indexes counties and $s$ indexes states. The main outcome, $\text{Vote}_c$, is the vote share received by Donald Trump in 2016. The regressor of interest, $\text{Southern Whites}_{c,1940}$, is the Southern-born white population share in 1940. The regression is instrumented by a shift-share variable constructed with predetermined Southern white migration networks as of 1900 and predicted aggregate migration flows out of the South for each decade from 1900 to 1940. The control variables $X_{c}$ are historical economic factors (population density, manufacturing employment, average farm values), ideological factors (Union Army enlistment, mortality rates from the U.S. Civil War), and the vote share for Woodrow Wilson in 1912. The variable $\alpha_s$ denotes a state fixed effect. Standard errors are clustered across counties in $60 \times 60$ mile grid cells.

The main result is displayed in their Table 2, Panel A, Column (4).\footnote{The authors do not explicitly indicate the preferred specification. However, on page 1591, they write ``Our main estimating equation takes the following form." Thus, we infer that Table 2 displays their ``main'' results. On Page 1595, they write "Secondary specifications control for additional potential sorting correlates." Hence columns 5 and 7 are not the main specifications. Therefore, we use Panel A column 4 as the main specification.} The authors implement several adjustments for spatial correlation. They report alternative standard errors, constructed with the methods proposed by \cite{conley1999gmm}, \cite{colella2023acreg}, and \cite{adao2019shift}, as well as through a wild bootstrap with clustering by state. The corrections for the main specification are shown in their Column (4) of their Table A.3. 

We combine replication-package variables with the shared county outcome dataset. These external outcomes enter as standardized levels averaged over their available years. Retained auxiliaries include Republican presidential vote shares, public-school enrollment, and remaining life expectancy at ages 65 and over.

\subsection{\cite{bernini2023race}}

\cite{bernini2023race} study the impact of the Voting Rights Act (VRA) on the racial makeup of local governments in the U.S.\ South. They consider cross-sectional data consisting of 971 counties in 11 states. They find that federal scrutiny of Southern states after the VRA had a sizable impact on the extent to which enfranchisement led to Black office holding. \cite{bernini2023race} consider the model
\begin{align*}
    \Delta \text{Share Black Elected}_{cs} 
    &=\gamma\text{Percent Black}_{1960} + \tau \text{Percent Black}_{1960}*\text{Covered}_{cs} \\
    &+ \delta_1 X_{cs}+ \delta_2 X_{cs}*\text{Covered}_{cs} + I_s + \epsilon_{cs}~,
\end{align*}
where $c$ denotes a county and $s$ denotes state. The main outcome, $\Delta \text{Share Black Elected}_{cs}$, is the change in the share of Black elected officials between 1964 and 1980. The regressor of interest, $\text{Percent Black}_{1960}*\text{Covered}_{cs}$, is the interaction of the Black share in 1960 and a dummy for federal scrutiny. Control variables $X_{cs}$ are pre-VRA county characteristics (unemployment rate, the percent of families below the poverty line, the percent of unskilled workers, agricultural productivity,  population (ln), percent urban population, cotton share, pro- and anti-Black activism,  Republican share). The variables $I_s$ denote state fixed-effects. Standard errors are clustered by judicial division. Results are shown in their Table 2 Column 4.\footnote{On pages 1019-1020, \cite{bernini2023race} write ``Our preferred specification in column 4 of table 2."}

We combine replication-package variables with long differences from the shared county outcome dataset. Retained auxiliaries include changes in gubernatorial-election turnout, NAACP branches, and remaining life expectancy at ages 65 and over.

\subsection{\cite{caprettini2023new}}

\cite{caprettini2023new} study the complementarity between patriotism in World War II and public-good provision in the New Deal. They consider cross-sectional data covering 3022 counties. They find that higher government spending at  the county level in the 1930s is positively correlated with more patriotic actions during World War II. \cite{caprettini2023new} consider the model
\begin{equation*}
	\text{WWII Patriotism}_i = \tau \text{New Deal Grants}_i + \gamma \text{WWI Patriotism}_i+\delta X_i +\xi_s +u_i~,
\end{equation*}
where $i$ denotes a county. The main outcome, $\text{WWII Patriotism}_i$, denotes the average purchases of war bonds per capita in 1944. The regressor of interest, $\text{New Deal Grants}_i$, denotes the New Deal grants per 1930 population. Control variables $\text{WWI Patriotism}_i$ and $X_i$ are WWI volunteering rate, WWI medals per 1000 inhabitants, socioeconomic characteristics in 1930 (log of population, unemployment rate, share of veterans, share of African Americans, and share of people born in major Axis countries), share of people living on a farm in 1930, 1929 farm income, an urban dummy, 1898-1928 average Democratic vote share, value of WWII war contracts per capita, average 1939 wage of employees, and 1930 share of men. The variables $\xi_s$ denote state fixed effects. Heteroskedasticity-robust standard errors are used.

Results are shown in their Table 2 Panel A Column 1.\footnote{On Page 481, \cite{caprettini2023new} write ``Figure I  summarizes our main result". Similarly, the phrase ``to go beyond the graphical evidence"  on Page 483 suggests that the results in this section (Table 2) are the main results. The first three columns report results for three equally important measures of patriotism, so we choose the first as the main specification.} \cite{caprettini2023new} discuss spatial error correlation. They report \cite{conley1999gmm} standard errors in Table 2 Panel B and Table 3.

We supplement replication-package variables with county-level measures from the historical County Data Book. We linearly interpolate these measures between observed years and select their 1944 values. Retained auxiliaries include Second World War volunteers per 100 residents, the 1940 FDR vote share, and manufacturing employment.

\subsection{\cite{moscona2023does}}

\cite{moscona2023does} study how innovation shapes the economic impact of climate change. The specification reproduced here uses 3,000 counties in two decade observations, labeled 1950 and 2010. They find that counties' exposure to climate-change-induced innovation significantly decreases the local economic damage from extreme temperatures. \cite{moscona2023does} consider the regression model
\begin{align*}
		\text{log Agr Land Price}_{i,t} 
        &= \delta_i + \alpha_{s(i),t}+\tau \text{Extreme Exposure}_{i,t}+\gamma \text{Innovation Exposure}_{i,t} \\
	&+\phi (\text{Extreme Exposure}_{i,t}* \text{Innovation Exposure}_{i,t}) + \epsilon_{i,t}
\end{align*}
where $i$ denotes county, $s$ denotes state, and $t$ indexes the 1950 and 2010 observations (with land values measured in the 1959 and 2017 agricultural censuses, respectively). The main outcome, $\text{log Agr}$ $\text{Land Price}_{i,t}$, is the price per acre of cultivated land. The regressor of interest, $\text{Extreme Exposure}_{i,t}*\text{Innovation Exposure}_{i,t}$, is the interaction of measures of a county's exposures to extreme temperature and innovation, respectively. The variables $\delta_i$ denote county fixed effects. The variables $\alpha_{s(i),t}$ denote state-by-year fixed effects. The authors report standard errors double clustered at the county and state-by-decade levels in parentheses and clustered by state in brackets. Results are shown in Table 3 Column 1 in the paper. In Table A.3, we augment the standard errors clustered by state.\footnote{The authors do not explicitly indicate the preferred specification. However, on page 645, they write ``Sections IV and V present our main results." Thus, we conclude that Table 3 displays the paper's main county level results. Within Table 3, we pick Column 1 as the main specification, because the authors choose this specification to perform downstream analysis. For example, on Page 680, they write ``Figure VI reports the marginal effect of exposure to extreme  heat (y-axis) for quantiles of the innovation exposure distribution (x-axis), using the specification from column (1)".} \cite{moscona2023does} report \cite{conley1999gmm} standard errors in their Table A21. 

We supplement replication-package variables with outcomes from the United States Censuses of Agriculture compiled by \cite{hainesetal2018icpsr}, and measures available in both the historical County Data Book and USA Counties. For observations labeled 1950 and 2010, the agricultural data come from the 1950 and 2012 censuses, respectively. The other measures use 1950 County Data Book values and the available USA Counties observations closest to 2010. Retained auxiliaries include measures of the number of farms, irrigated acreage, and agricultural employment.

\subsection{\cite{esposito2023reconciliation}}

\cite{esposito2023reconciliation} study the effect of the Lost Cause narrative---a revisionist and racist  retelling of the U.S.\ Civil War---on national reunification and racial discrimination. They consider a panel dataset consisting of 786 counties between 1910 and 1920. They find that screenings of \textit{The Birth of a Nation} shifted public discourse toward more patriotic and less divisive language, increased military enlistment, and fostered cultural convergence. This paper considers an instrumental variables regression model
\begin{equation*}
	\text{Reconciliation}_{ct}=\tau \text{BON}_{ct} + \alpha_c +\alpha_t+\pi_{p(ct)}+\epsilon_{ct}
\end{equation*}
where $c$ is county and $t$ is year-month. The main outcome, $\text{Reconciliation}_{ct}$, represents the relative log frequencies of patriotic keywords to divisive ones in local newspapers. The regressor of interest, $\text{BON}_{ct}$, is equal to 1 after the screening of \textit{The Birth of a Nation}. The regression is instrumented with the screening of another movie, \textit{The Million Dollar Mystery}. The terms $\alpha_c$ and $\alpha_t$ denote county and month-year fixed effects; $\pi_{p(ct)}$ denotes fixed effects for percentiles of the number of newspaper pages available in each county-month. Standard errors are clustered at the county level. Results are shown in Table 3 Panel A Column 3.\footnote{On page 1479, the authors write ``In our baseline analysis..." Likewise, on page 1479, they write ``we compute our main dependent variable, $\text{Reconciliation}_{ct}$." On page 1472, \cite{esposito2023reconciliation} write ``our preferred solution is to instrument the treatment in a two-stage least squares (2SLS) version of equation (1)."} The paper clusters the standard errors at the state level and also applies the estimator proposed in \cite{colella2023acreg} in Table B34 Panel C Column 1.

We combine replication-package variables with county-level economic and demographic measures constructed from U.S. Census data obtained through IPUMS USA.\footnote{\url{https://usa.ipums.org/usa/}} We aggregate using household or person weights, linearly interpolate between Census years, and match the annual measures to the county-month panel for 1910--1920. Retained auxiliaries include a county-month indicator for whether any newspaper page contains ``White only'' together with ``help'' and ``wanted'', and the county homeownership rate.

\subsection{\cite{calderon2023racial}}

\cite{calderon2023racial} study the political effects of the migration of four million African Americans from the South to the North. They analyze a panel of 1,263 non-Southern counties, stacking changes over three decades from 1940 to 1970. \cite{calderon2023racial} find that the Great Migration increased support for the Democratic Party, increased Congress members' propensity to promote civil rights legislation, and encouraged pro-civil rights activism outside the U.S.\ South. This paper considers the instrumental variables regression model
\begin{equation*}
	\Delta \text{demsh}_{c \tau}=\delta_{s \tau} + \tau \Delta \text{Bl}_{c \tau} + \gamma X_{c \tau} +u_{c \tau}~,
\end{equation*}
where $c$ is county and $\tau$ is decade. The main outcome, $\Delta \text{demsh}_{c \tau}$, is the change in the Democratic vote share during decade $\tau$. The regressor of interest, $\Delta \text{Bl}_{c \tau}$, is the change in the Black population share. The regression is instrumented with a shift-share predictor of Black inflow. Control variables $X_{c \tau}$ are interactions between decade dummies and 1940 county characteristics (Black population share and Democratic incumbency in Congressional elections). The variable $\delta_{s \tau}$ collects interactions between decade and state dummies. The regression is weighted by 1940 county population. Standard errors are clustered at the county level. Results are displayed in Table 2 Column 6.\footnote{The authors note that this is the paper's preferred specification. On page 179, \cite{calderon2023racial} write "especially for our preferred specification (Column 6)"} \cite{calderon2023racial} additionally clusters their standard errors at the commuting zone level and reports the confidence interval constructed according to \cite{adao2019shift}. These results are displayed in Tables D.21 and D.22.

We combine replication-package variables with changes in county-level measures from the historical County Data Book. We linearly interpolate the underlying series and calculate changes over each observation's decade: 1940--1950, 1950--1960, or 1960--1970. Retained auxiliaries include changes in congressional-election turnout and manufacturing employment.

\subsection{\cite{cook2023evolution}}

\cite{cook2023evolution} study factors correlated with the provision of nondiscriminatory services. They consider a panel of 3,104 counties observed in 12 years between 1939 and 1955. They find that declines in the white population led to increases in the number of nondiscriminatory businesses. This paper considers the linear regression model
\begin{equation*}
\text{Asinh}(N_{0ct})=\tau \text{Asinh}(\text{casualties}_c) \cdot \text{postWWII}_t +\phi_c +\xi_t +\epsilon_{ct}	
\end{equation*}
where $c$ denotes county and $t$ denotes year. The main outcome, $N_{0ct}$, is the number of Green Book establishments in county $c$ at time $t$. The regressor of interest, $\text{Asinh}(\text{casualties}_c) * \text{postWWII}_t$, is the interaction of the number of white casualties in World War II and a post-war indicator. The variables $\phi_c$ and $\xi_t$ denote county and year fixed effects. Standard errors are clustered at the county level. Results are shown in Table 3 Panel A Column 4.\footnote{On page 77, the authors write ``our preferred specification uses county and year fixed effects,'' which reflects this specification.}

We combine replication-package variables with measures from the historical County Data Book. We linearly interpolate these measures between observed years and match them to each county-year observation. Retained auxiliaries include measures of Green Book hotels, service-sector employment, and retail sales.

\subsection{\cite{chetty2014land}}

\cite{chetty2014land} study features of inter-generational mobility in the United States. This paper finds that the absolute upward mobility (the expected income rank of children from families at the 25 percentile of the national parent income distribution) is negatively correlated with the proportion of the population in a commuting zone that is African American. This paper considers the linear regression model
\begin{equation*}
\text{absmob}_c=\tau \text{frac black}_c +\delta_s + \epsilon_{c}	
\end{equation*}
where $s$ denotes state and $c$ denotes commuting zone. The main outcome, $\text{absmob}_c$, is the measure of absolute upward mobility in commuting zone $c$. The regressor of interest, $\text{frac black}_c$, is the share of African American people in commuting zone $c$. Standard errors are clustered at the state level. Results are shown in the first row of Figure VIII in the paper.\footnote{We pick this specification because the authors write ``Perhaps the most obvious pattern from the maps in Figure VI is that inter-generational mobility is lower in areas with larger African American populations, such as the Southeast" on Page 1605.}

We combine replication-package variables with commuting-zone aggregates of the shared county outcome bank. We form population-weighted averages of standardized county levels using 2000 county populations and match them by commuting-zone identifier. Retained auxiliaries include the Gini coefficient, labor-force participation, public-school enrollment, and divorce rates.

\subsection{\cite{acemoglu2019democracy}} 

\cite{acemoglu2019democracy} estimate the impact of democracy on economic growth. They consider panel data of 175 countries between 1960 and 2010. They find that democratization increases GDP  per capita in the long run. This paper considers the linear regression model
\begin{equation*}
	\text{lgdppc}_{c,t} = \tau \text{demo}_{c,t}+\sum_{j=1}^4 \gamma_j \text{lgdppc}_{c,t-j}+\alpha_c+\delta_t+\epsilon_{c,t}
\end{equation*}
where $c$ denotes country and $t$ denotes year. Results are shown in Table 2 Column 3.\footnote{On page 59, \cite{acemoglu2019democracy} write ``Column 3,  which is our preferred specification, includes four lags of GDP per capita."} The main outcome, $\text{lgdppc}_{c,t}$, is the log of GDP per capita. The regressor of interest, $\text{demo}_{c,t}$, is a dichotomous measure of democracy. Control variables are the lags of log GDP per capita. The quantities $\delta_t$ and $\alpha_c$ are year and country fixed effects. Standard errors are clustered at the country level.\footnote{When lagged outcomes are included among the control variables, population residuals need not have mean zero conditional on the regressors from all periods, as required by the framework developed in \cref{sec: proposal}. Under suitable conditions, the usual clustered variance estimator, which has the same sandwich form as TMO, remains consistent in panels with lagged outcomes as both the number of units and the number of periods increase \citep{galvao2014estimation}. Here, we augment the country-clustered standard errors used in \cite{acemoglu2019democracy} to allow for residual correlation across countries, without providing a complete theoretical justification for this extension.}

We match replication-package variables to an external country-year panel by harmonized country code and calendar year. The panel includes World Development Indicators and Database of Political Institutions measures.\footnote{\url{https://databank.worldbank.org/source/world-development-indicators}; \url{https://datacatalog.worldbank.org/search/dataset/0039819/database-of-political-institutions}.} Retained auxiliaries include educational attainment, market reforms, the Gini index, central-government debt as a share of GDP, federalism, and executive ideology.

\section{Proofs for Results Stated in the Main Text\label{sec: proofs}}

We allow the constants $c$ and $C$ to change values from line to line and to depend arbitrarily on the scalar $\eta$. Throughout the proofs given in this appendix, we make use of the results collected in \cref{app: auxiliary lemmas}. Proofs for Lemmas stated in this Appendix are collected in \cref{app: selected set characterization proof} though \cref{app: gaussian calculus}.

\subsection{Proof of \cref{lem: selected set characterization}}

\subsubsection{Part (i)}

The result relies on the application of the following Lemma.

\begin{lem}
\label{lem: feasible screening continuity}$ $\\
Suppose that
\cref{assu: gaussianity,assu: proportionality,assu: homoskedasticity,assu: well conditioned} hold. If $n\geq3$ and $d\geq2$, then the distribution of $\hat\lambda_{i,i'}$ is non-atomic for every $i,i'\in[n]$.
\end{lem}

\noindent As $s<n-1$, \cref{assu: sparsity} ensures that the set of pairs $i,i'$ satisfying $\lambda_{i,i'}=0$ is non-empty, and so $\bar{\mathbf{F}}_{n,d}^+(\delta)$ is well-defined. \cref{lem: feasible screening continuity} implies that every function $\bar{F}_{i,i'}^+(\cdot)$ is continuous, non-increasing, and converges to zero as its argument diverges. Moreover, \cref{assu: sparsity,assu: proportionality} and the normalization $\gamma_{0,0}=1$ imply that $V_n>0$. Consequently, the function $\bar{\mathbf{F}}^{+}_{n,d}(\cdot)$ is the maximum of a non-empty finite collection of continuous, non-increasing functions that converge to zero. Hence, the set $\Delta_{n,d}(\eta)$ is non-empty and upward closed, and therefore takes the form in \eqref{eq: selected threshold ray}.\hfill\qed

\subsubsection{Part (ii)}

The result is obtained through the application of the following moderate deviation approximation.

\begin{lem}
\label{lem: feasible null moderate deviation}$ $\\
Suppose that, for a fixed $\eta\in(0,1]$,
\cref{assu: gaussianity,assu: proportionality,assu: homoskedasticity,assu: sparsity,assu: well conditioned} hold.
If $s/n\to0$,
\begin{align}\label{eq: mod dev high level rate}
    dV_n^2
    \log(V_n^{-1})
    &\to0~,
    \quad\text{and}\quad
    \log(V_n^{-1})
    =o(d)~,
\end{align}
then, for each fixed $\kappa>0$, it holds that
\begin{align}
    \sup_{\substack{i,i'\in[n]\\\lambda_{i,i'}=0}}
    \left|
        \frac{
            \log P\left\{
                |\hat\lambda_{i,i'}|
                \geq
                \kappa\sqrt{v_{0}^{(d)}}
                \sqrt{
                    2\eta\log\left(
                        V_n^{-1}
                    \right)
                }
            \right\}
        }{
            \log\left(1/V_n\right)
        }
        +\kappa^2\eta
    \right|
    &\to0
    \label{eq: feasible null moderate deviation}
\end{align}
as $n,d\to\infty$.
\end{lem}

\noindent Define the sequence $\delta_{n,d}^{\star}(\eta)
    =\sqrt{
        2\eta v_{0}^{(d)}
        \log(V_n^{-1})
    }$
and set $\epsilon\in(0,1)$. Applying \cref{lem: feasible null moderate deviation} with $\kappa=1+\epsilon$ and $\kappa=1-\epsilon$, and then taking the maximum over the null pairs, gives
\begin{align}
    \log\bar{\mathbf{F}}^{+}_{n,d}\left(
        (1+\epsilon)\delta_{n,d}^{\star}(\eta)
    \right)
    &=-(1+\epsilon)^2\eta
    \log(V_n^{-1})
    +o\left(\log(V_n^{-1})\right)
    \quad\text{and}
    \nonumber\\
    \log\bar{\mathbf{F}}^{+}_{n,d}\left(
        (1-\epsilon)\delta_{n,d}^{\star}(\eta)
    \right)
    &=-(1-\epsilon)^2\eta
    \log(V_n^{-1})
    +o\left(\log(V_n^{-1})\right)~.
    \label{eq: selected threshold endpoint brackets}
\end{align}
Combining \cref{lem: basic inequalities n}, display \eqref{eq: var bounds}, and \eqref{eq: selected threshold endpoint brackets}, we obtain
\begin{align}
    \log\left(
        \frac{
            \bar{\mathbf{F}}^{+}_{n,d}\left(
                (1+\epsilon)\delta_{n,d}^{\star}(\eta)
            \right)
        }{V_n^\eta}
    \right)
    &=-\eta(2\epsilon+\epsilon^2)
    \log(V_n^{-1})
    +o\left(\log(V_n^{-1})\right)
    \to-\infty\quad\text{and}
    \nonumber\\
    \log\left(
        \frac{
            \bar{\mathbf{F}}^{+}_{n,d}\left(
                (1-\epsilon)\delta_{n,d}^{\star}(\eta)
            \right)
        }{V_n^\eta}
    \right)
    &=\eta(2\epsilon-\epsilon^2)
    \log(V_n^{-1})
    +o\left(\log(V_n^{-1})\right)
    \to\infty~.
    \label{eq: selected threshold probability brackets}
\end{align}
Since $s/n\to0$, Part (i) applies for all sufficiently large $n$.
Consequently, \eqref{eq: selected threshold probability brackets} and
the representation \eqref{eq: selected threshold ray} imply that
\begin{align}
    (1-\epsilon)\delta_{n,d}^{\star}(\eta)
    &<
        \underline\delta_{n,d}(\eta)
    \leq(1+\epsilon)\delta_{n,d}^{\star}(\eta)
    \label{eq: eps sandwich}
\end{align}
for all sufficiently large $n,d$. Letting $\epsilon\downarrow0$ in
\eqref{eq: eps sandwich} gives \eqref{eq: selected threshold endpoint}.\hfill\qed

\subsection{Proof of \cref{lem: bias decomposition}}

Define the infeasible statistics
\begin{align}
    \tilde\varepsilon_{0,i}^{(j)}
    &=\hat\varepsilon_i^{(j)}
    -\gamma_{0,j}\hat\varepsilon_i^{(0)}
    \quad\text{and}\quad
    \tilde\lambda_{0,i,i'}
    =\frac{1}{d}\sum_{j=1}^d
    \tilde\varepsilon_{0,i}^{(j)}
    \tilde\varepsilon_{0,i'}^{(j)}~.
\end{align}
The result is obtained by considering the components of the following exact decomposition of the bias.

\begin{lem}\label{lem: exact bias decomposition}$ $\\
If \cref{assu: gaussianity,assu: proportionality} hold, then 
\begin{align}
    \mathbb{E}[\hat V_{n,d}(\delta)-V_n]
    &=\frac{1}{n^2}\sum_{i,i'=1}^n
    \left(
        \mu_{i,i'}\bar F^+_{0,i,i'}(\delta)-\lambda_{i,i'}
    \right)
    +\frac{1}{n^2}\sum_{i,i'=1}^n\bar S_{i,i'}(\delta)~,
    \label{eq: bias decomposition state}
\end{align}
where
\begin{align}
    \mu_{i,i'}
    &=\lambda_{i,i'}
    -\frac{1}{n}\sum_{k=1}^n
    (\lambda_{i,k}+\lambda_{i',k})
    +\frac{1}{n^2}\sum_{k,l=1}^n\lambda_{k,l}~,
    \quad
    \bar F^+_{0,i,i'}(\delta)
    =P\{|\tilde\lambda_{0,i,i'}|\geq\delta\}~,\quad\text{and}\label{eq: centered unit covariance}\\
    \bar S_{i,i'}(\delta)
    &=\mathbb{E}\left[
        \hat\varepsilon_i^{(0)}\hat\varepsilon_{i'}^{(0)}
        \left(
            \mathbb{I}\{|\hat\lambda_{i,i'}|\geq\delta\}
            -\mathbb{I}\{|\tilde\lambda_{0,i,i'}|\geq\delta\}
        \right)
    \right]~,\nonumber
\end{align}
for each pair of units $i,i'$. 
\end{lem}

\noindent In particular, we apply the parts of the following Lemma to approximate the terms appearing in \eqref{eq: bias decomposition state}.

\begin{lem}\label{lem: decomposition approximation}$ $\\
Define the quantity
\begin{align}
    \bar Q_{i,i'}(\delta)
    &=\bar F^+_{0,i,i'}(\delta)
    -F^+(\delta,0,v^{(d)}_0)
    \label{eq: oracle centered screening probability}
\end{align}
for each pair $i,i'$. Suppose that, for a fixed $\eta \in [0,1]$, the conditions of \cref{lem: bias decomposition} hold and $s/n\leq c$ for a sufficiently small constant $c$.

\noindent \textbf{(i)} It holds that
\begin{align}
    \sup_{i,i'\in[n]}\sup_{\delta\geq0}
    |\bar Q_{i,i'}(\delta)-Q_{i,i'}(\delta)|
    &\lesssim
    \frac{1}{\sqrt d}+ V_n\sqrt{d}~.
    \label{eq: oracle screening approximation}
\end{align}

\noindent\textbf{(ii)} If, additionally, it holds that
\begin{align}
    V_n
    \left(
        \sqrt{
            d\log(V_n^{-1})
        }
        +\log(V_n^{-1})
    \right)
    & \leq c
    \label{eq: feasible residualization lemma condition}
\end{align}
for a sufficiently small constant $c$, then
\begin{align}
    \sup_{\delta\in\Delta_{n,d}(\eta)}
    \left|
        \frac{1}{n^2}\sum_{i,i'=1}^n\bar S_{i,i'}(\delta)
    \right|
    &\lesssim
    \frac{V_n}{\sqrt d}
    +\frac{s}{n}V_n\sqrt d
    +
    V_n^{2+\eta}
    \left(
        d\log(V_n^{-1})
        +\log^2\left(V_n^{-1}\right)
    \right)~.
    \label{eq: feasible residualization bound}
\end{align}
\end{lem}

\noindent 
To this end, observe that \cref{lem: exact bias decomposition} and the fact that $\sum_{i,i'} \mu_{i,i'} = 0$ imply that
\begin{align}
    \mathbb{E}[\hat V_{n,d}(\delta)-V_n]
    &=\frac{1}{n^2}\sum_{i,i'=1}^n
    (\mu_{i,i'}\bar Q_{i,i'}(\delta)-\lambda_{i,i'})
    +\frac{1}{n^2}\sum_{i,i'=1}^n\bar S_{i,i'}(\delta)~.
    \label{eq: bias decomposition apply 1}
\end{align}
Now, for each pair $i,i'$, observe that we can write
\begin{align}
    \mu_{i,i'}\bar Q_{i,i'}(\delta)-\lambda_{i,i'}
    &=\lambda_{i,i'}(Q_{i,i'}(\delta)-1)
    +(\mu_{i,i'}-\lambda_{i,i'})Q_{i,i'}(\delta)
    +\mu_{i,i'}
    (\bar Q_{i,i'}(\delta)-Q_{i,i'}(\delta))~.
    \label{eq: screening decomposition}
\end{align}
Observe that $Q_{i,i'}(\delta)=0$ whenever $\lambda_{i,i'}=0$, and satisfies the bound $|Q_{i,i'}(\delta)|\leq1$ otherwise. As a consequence, \cref{assu: sparsity} and \cref{lem: basic inequalities n}, display \eqref{eq: n bound 3}, imply that
\begin{align}
    \sup_{\delta\in\Delta_{n,d}(\eta)}
    \left|
        \frac{1}{n^2}\sum_{i,i'=1}^n
        (\mu_{i,i'}-\lambda_{i,i'})Q_{i,i'}(\delta)
    \right|
    &\lesssim\frac{s}{n}V_n~.
    \label{eq: weighted centering bound}
\end{align}
Part (i) of \cref{lem: decomposition approximation} and \cref{lem: basic inequalities n}, display \eqref{eq: n bound 3}, imply that
\begin{align}
    \sup_{\delta\in\Delta_{n,d}(\eta)}
    \left|
        \frac{1}{n^2}\sum_{i,i'=1}^n
        \mu_{i,i'}
        (\bar Q_{i,i'}(\delta)-Q_{i,i'}(\delta))
    \right|
    &\lesssim
    V_n
    \left(
        \frac{1}{\sqrt d}
        +V_n\sqrt d
    \right)~.
    \label{eq: tail stability aggregate}
\end{align}
Thus, by plugging \eqref{eq: screening decomposition}, \eqref{eq: weighted centering bound}, and \eqref{eq: tail stability aggregate} into \eqref{eq: bias decomposition apply 1}, we find that
\begin{align}
    & \sup_{\delta\in\Delta_{n,d}(\eta)}
    \left|
\mathbb{E}[\hat V_{n,d}(\delta)-V_n]
- B_{n,d}(\delta) 
-\frac{1}{n^2}\sum_{i,i'=1}^n\bar S_{i,i'}(\delta)
    \right| \label{eq: bias pre S bound}\\
    & 
    \qquad \qquad \lesssim 
    V_n\left(\frac{s}{n} + 
        \frac{1}{\sqrt d}
        +V_n\sqrt d 
    \right) 
    \lesssim 
    V_n\left(\frac{s\sqrt{d}}{n} + 
        \frac{1}{\sqrt d} 
    \right) \nonumber
\end{align}
for all sufficiently large $n,d$.

Now, Part (ii) of \cref{lem: decomposition approximation} and \eqref{eq: bias pre S bound} give
\begin{align}
    \sup_{\delta\in\Delta_{n,d}(\eta)}
    \left|
        \frac{\mathbb{E}[\hat V_{n,d}(\delta)-V_n]
        -B_{n,d}(\delta)}{V_n}
    \right|
    \lesssim
    \frac{1}{\sqrt d}
    +\frac{s\sqrt d}{n}
    +
    V_n^{1+\eta}
    \left(
        d\log(V_n^{-1})
        +\log^2\left(V_n^{-1}\right)
    \right)     \label{eq: relative bias remainder bound}
\end{align}
for all sufficiently large $n,d$. To absorb the final term in \eqref{eq: relative bias remainder bound}, observe that
\begin{align}
    V_n^{1+\eta}\log^2\left(V_n^{-1}\right)
    &\leq
    \begin{cases}
        dV_n^{1+\eta}\log\left(V_n^{-1}\right)~,
        &\log\left(V_n^{-1}\right)\leq d~,\\
        d^2e^{-d}~,
        &\log\left(V_n^{-1}\right)>d~,
    \end{cases}
    \lesssim
    dV_n^{1+\eta}\log\left(V_n^{-1}\right)
    +\frac{1}{\sqrt d}~.
    \label{eq: bias logarithmic square absorption}
\end{align}
Substituting \eqref{eq: bias logarithmic square absorption} into \eqref{eq: relative bias remainder bound} gives
\begin{align}
    \sup_{\delta\in\Delta_{n,d}(\eta)}
    \left|
        \frac{\mathbb{E}[\hat V_{n,d}(\delta)-V_n]
        -B_{n,d}(\delta)}{V_n}
    \right|
    &\lesssim
    V_n^{1+\eta}d\log\left(V_n^{-1}\right)
    +\frac{1}{\sqrt d}
    +\frac{s\sqrt d}{n}~,
\end{align}
for all sufficiently large $n,d$.\hfill\qed

\subsection{Proof of \cref{lem: variance bound}}

Define the infeasible statistics
\begin{align}
    \varepsilon_{0,i}^{(j)}
    &=Y_i^{(j)}-\tau^{(j)}
    -\gamma_{0,j}\left(Y_i^{(0)}-\tau^{(0)}\right)~,\nonumber\\
    \check{\lambda}_{0,i,i'}
    & =\frac{1}{d}\sum_{j=1}^d
    \varepsilon_{0,i}^{(j)}
    \varepsilon_{0,i'}^{(j)}~,\quad\text{and}\quad\check V_{n,d}^{(0)}(\delta)
    =\frac{1}{n^2}\sum_{i,i'=1}^n
    \hat\varepsilon_i^{(0)}\hat\varepsilon_{i'}^{(0)}
    \mathbb{I}\left\{
        |\check{\lambda}_{0,i,i'}|\geq\delta
    \right\}
\end{align}
as well as the infeasible and feasible conditional expectations
\begin{align}
    G_{0,n,d}(\delta)
    &=\mathbb{E}\left[
        \check V_{n,d}^{(0)}(\delta)
        \mid\hat\varepsilon^{(0)}
    \right]
    \quad\text{and}\quad
    G_{n,d}(\delta)
    =\mathbb{E}\left[
        \hat V_{n,d}(\delta)
        \mid\hat\varepsilon^{(0)}
    \right]~,
    \label{eq: variance conditional means}
\end{align}
respectively.

Observe that the law of total variance implies that
\begin{align}
    \Var\left(\hat V_{n,d}(\delta)\right)
    &=\mathbb{E}\left[
        \Var\left(
            \hat V_{n,d}(\delta)
            \mid\hat\varepsilon^{(0)}
        \right)
    \right]
    +\Var\left(G_{n,d}(\delta)\right)~.
    \label{eq: feasible total variance decomposition}
\end{align}
The triangle inequality, the fact that centering cannot increase a second moment, and a second application of the law of total variance give
\begin{align}
    \sqrt{\Var(G_{n,d}(\delta))}
    &\leq
    \mathbb{E}\left[
        \left(G_{n,d}(\delta)-G_{0,n,d}(\delta)\right)^2
    \right]^{1/2}
    +\sqrt{\Var(\check V_{n,d}^{(0)}(\delta))}~.
\end{align}
Combining this inequality with \eqref{eq: feasible total variance decomposition} yields
\begin{align}
    \Var\left(\hat V_{n,d}(\delta)\right)^{1/2}
    &\leq
    \Var\left(\check V_{n,d}^{(0)}(\delta)\right)^{1/2}
    \nonumber\\
    &
    +\mathbb{E}\left[\left(G_{n,d}(\delta)-G_{0,n,d}(\delta)\right)^2\right]^{1/2}
    +
    \mathbb{E}\left[
            \Var\left(
                \hat V_{n,d}(\delta)
                \mid\hat\varepsilon^{(0)}
                \right)
            \right]^{1/2}~.
    \label{eq: variance component reduction}
\end{align}
We apply the following Lemma to bound the three terms on the right-hand side of \eqref{eq: variance component reduction}.

\begin{lem}
\label{lem: variance approximation components}$ $\\
Suppose that, for a fixed $\eta\in(0,1]$,
\cref{assu: gaussianity,assu: proportionality,assu: homoskedasticity,%
assu: sparsity,assu: well conditioned} hold, $d\geq2$, and $s/n\leq c$ for
$c>0$ sufficiently small. If
\begin{align}
    V_n
    \left(
        \sqrt{
            d\log(V_n^{-1})
        }
        +\log(V_n^{-1})
    \right)
    &\leq c~,
    \label{eq: variance component local condition}
\end{align}
then the following conclusions hold.

\noindent\textbf{(i)} It holds that
\begin{align}
    \sup_{\delta\in\Delta_{n,d}(\eta)}
    \Var\left(\check V_{n,d}^{(0)}(\delta)\right)
    &\leq CV_n^2
    \left(
        \frac{s+1}{n}+V_n^\eta
    \right)~.
    \label{eq: population oracle variance component}
\end{align}

\noindent\textbf{(ii)} It holds that
\begin{align}
    &\sup_{\delta\in\Delta_{n,d}(\eta)}
    \mathbb{E}\left[\left(G_{n,d}(\delta)-G_{0,n,d}(\delta)\right)^2\right]^{1/2}
    \nonumber\\
    &\quad\leq CV_n
    \Bigg(
        \frac{1}{\sqrt d}
        +\frac{s\sqrt d}{n}
        +V_n\sqrt d +V_n^{1+\eta}
        \left(
            d\log(V_n^{-1})
            +\log^2\left(V_n^{-1}\right)
        \right)
    \Bigg)~.
    \label{eq: conditional mean variance component}
\end{align}

\noindent\textbf{(iii)} It holds that
\begin{align}
    \sup_{\delta\in\Delta_{n,d}(\eta)}
    \mathbb{E}\left[
        \Var\left(
            \hat V_{n,d}(\delta)
            \mid\hat\varepsilon^{(0)}
        \right)
    \right]
    &\leq CV_n^2
    \left(
        \frac{s+1}{n}
        +V_n^\eta
        +\frac{s\sqrt d}{n}
        +dV_n^{1+\eta}
        \log(V_n^{-1})
    \right)~.
    \label{eq: conditional covariance variance component}
\end{align}
\end{lem}

\noindent To apply \cref{lem: variance approximation components}, observe that $s/n\leq c$ and $d \geq 2$ for all sufficiently large $n,d$ under the conditions of the Lemma. In turn, \cref{lem: basic inequalities n}, display \eqref{eq: var bounds}, implies that
\begin{align}
    V_n
    &\lesssim\frac{s+1}{n}
    \leq\frac{2s}{n}\to0
    \quad\text{and}\quad
    V_n\sqrt d
    \lesssim\frac{(s+1)\sqrt d}{n}
    \leq\frac{2s\sqrt d}{n}\to0~.
    \label{eq: variance target dimension consequence}
\end{align}
Thus, since $\eta\in(0,1]$, we have that
\begin{align}
    V_n
    \sqrt{
        d\log(V_n^{-1})
    }
    \leq
    \left(
        dV_n^{1+\eta}
        \log(V_n^{-1})
    \right)^{1/2}
    &\to0~,
    \nonumber\\
    V_n
    \log(V_n^{-1})
    \to0~,
    \quad\text{and}\quad
    V_n^\eta
    \log^2\left(V_n^{-1}\right)
    &\to0~.
    \label{eq: variance local and logarithmic consequences}
\end{align}
Hence, the conditions of \cref{lem: variance approximation components} apply for all sufficiently large $n,d$. We now bound the three terms in \eqref{eq: variance component reduction} explicitly. Part (i) of \cref{lem: variance approximation components}, \cref{lem: basic inequalities n}, display \eqref{eq: var bounds}, and $s\geq1$, $d\geq2$ imply that
\begin{align}
    \sup_{\delta\in\Delta_{n,d}(\eta)}
    \frac{
        \sqrt{\Var\left(\check V_{n,d}^{(0)}(\delta)\right)}
    }{
        V_n
    }
    \lesssim
    \left(
        \left(\frac{s\sqrt d}{n}\right)^{1/2}
        +V_n^{\eta/2}
    \right)~.
    \label{eq: normalized oracle variance bound}
\end{align}
The bound $x\log^2(x^{-1})\lesssim1$ for $x\in(0,1]$ gives
\begin{align}
    V_n^{1+\eta}\log^2(V_n^{-1})
    &\lesssim V_n^\eta~.
    \label{eq: variance logarithmic square absorption}
\end{align}
Part (ii) of \cref{lem: variance approximation components}, \cref{lem: basic inequalities n}, display \eqref{eq: var bounds}, \eqref{eq: variance target dimension consequence}, and \eqref{eq: variance logarithmic square absorption} give
\begin{align}
   & \sup_{\delta\in\Delta_{n,d}(\eta)}
    \frac{
        \mathbb{E}\left[
            \left(
                G_{n,d}(\delta)-G_{0,n,d}(\delta)
            \right)^2
        \right]^{1/2}
    }{
        V_n
    } \nonumber \\
    &\lesssim
    \frac{1}{\sqrt d}
        +V_n^{\eta/2} +
        \left(\frac{s\sqrt d}{n}\right)^{1/2}
        +\left(
            dV_n^{1+\eta}\log(V_n^{-1})
        \right)^{1/2}~.
    \label{eq: normalized conditional mean difference bound}
\end{align}
Part (iii) of \cref{lem: variance approximation components} and \cref{lem: basic inequalities n}, display \eqref{eq: var bounds}, likewise give
\begin{align}
    \sup_{\delta\in\Delta_{n,d}(\eta)}
    \frac{
        \mathbb{E}\left[
            \Var\left(
                \hat V_{n,d}(\delta)
                \mid\hat\varepsilon^{(0)}
            \right)
        \right]^{1/2}
    }{
        V_n
    }
    &\lesssim
    V_n^{\eta/2}
        +\left(\frac{s\sqrt d}{n}\right)^{1/2}
    +
        \left(
            dV_n^{1+\eta}\log(V_n^{-1})
        \right)^{1/2}~.
    \label{eq: normalized conditional covariance bound}
\end{align}
Substituting \eqref{eq: normalized oracle variance bound}, \eqref{eq: normalized conditional mean difference bound}, and \eqref{eq: normalized conditional covariance bound} into \eqref{eq: variance component reduction} yields
\begin{align}
    \sup_{\delta\in\Delta_{n,d}(\eta)}
    \frac{\sqrt{\Var(\hat V_{n,d}(\delta))}}{V_n}
    &\lesssim
    V_n^{\eta/2}
    +\left(
        dV_n^{1+\eta}\log(V_n^{-1})
    \right)^{1/2}
    +\frac{1}{\sqrt d}
    +\left(
        \frac{s\sqrt d}{n}
    \right)^{1/2}~,
\end{align}
as required.\hfill\qed

\subsection{Proof of \cref{thm: risk approximation}}

Consider the bias-variance decomposition
\begin{align}
    \mathbb{E}[(\hat V_{n,d}(\delta)-V_n)^2]
    &=\mathbb{E}\left[
        \hat V_{n,d}(\delta)-V_n
    \right]^2
    +\Var\left(\hat V_{n,d}(\delta)\right)~.
    \label{eq: risk decomposition}
\end{align}
Since \eqref{eq: risk decomposition} implies that the mean-square-error is no smaller than the absolute bias, the triangle inequality and the bound $\sqrt{x^2+y}\leq |x|+\sqrt y$ for $y\geq0$ give
\begin{align}
    &\left|
        \mathbb{E}[(\hat V_{n,d}(\delta)-V_n)^2]^{1/2}-\left|B_{n,d}(\delta)\right|
    \right| \nonumber \\
    &\leq
    \mathbb{E}[(\hat V_{n,d}(\delta)-V_n)^2]^{1/2}
    -\left|
        \mathbb{E}\left[
            \hat V_{n,d}(\delta)-V_n
        \right]
    \right|
    +
    \left|
        \mathbb{E}\left[
            \hat V_{n,d}(\delta)-V_n
        \right]
        -B_{n,d}(\delta)
    \right|
    \nonumber\\
    &\leq
    \sqrt{\Var\left(\hat V_{n,d}(\delta)\right)}
    +\left|
        \mathbb{E}\left[
            \hat V_{n,d}(\delta)-V_n
        \right]
        -B_{n,d}(\delta)
    \right|~.
    \label{eq: risk approximation reduction}
\end{align}
Dividing \eqref{eq: risk approximation reduction} by $V_n$, taking the supremum over $\Delta_{n,d}(\eta)$, and applying \cref{lem: bias decomposition} and \cref{lem: variance bound} yield
\begin{align}
    \sup_{\delta\in\Delta_{n,d}(\eta)}
    \left|R_{n,d}(\delta)-
        \frac{
            \left|B_{n,d}(\delta)\right|
        }{V_n}
    \right|
    \lesssim
    V_n^{\eta/2}
    +\left(
        dV_n^{1+\eta}\log\left(V_n^{-1}\right)
    \right)^{1/2}
    +\frac{1}{\sqrt d}
    +\left(
        \frac{s\sqrt d}{n}
    \right)^{1/2}~,
\end{align}
as required.\hfill\qed

\subsection{Proof of \cref{thm: constrained minimax lower bound}}

\subsubsection{Part (i)}

The result is obtained by applying the following information-theoretic lower bound. 

\begin{lem}[\citealp{tsybakov-2009}, Theorem 2.15 (i)]
\label{lem: two prior lower bound}$ $\\
Let $\Theta$ be a topological parameter space equipped with its Borel $\sigma$-field, let $F:\Theta\to\mathbb R$ be continuous, and let $P_\theta$ denote the distribution of the data under each $\theta\in\Theta$ in the family. Suppose that $\theta\mapsto P_\theta(A)$ is Borel measurable for every measurable event $A$. Let $\pi_0$ and $\pi_1$ be probability
measures on
$\Theta$, and let
\begin{align}
    \mathbb Q_h(A)
    &=\int_\Theta P_\theta(A)\,d\pi_h(\theta),
    \qquad h\in\{0,1\}~,
    \label{eq: two prior induced laws}
\end{align}
denote the induced data distributions. If there exist
$\zeta\in\mathbb R$, $\Delta>0$, and $\beta_0,\beta_1\in[0,1)$ such that
\begin{align}
    \pi_0\{\theta:F(\theta)\leq\zeta-\Delta\}
    &\geq1-\beta_0
    \quad\text{and}\quad
    \pi_1\{\theta:F(\theta)\geq\zeta+\Delta\}
    \geq1-\beta_1~,
    \label{eq: two prior separation conditions}
\end{align}
then
\begin{align}
    \inf_{\widehat F}
    \sup_{\theta\in\Theta}
    P_\theta\{|\widehat F-F(\theta)|\geq\Delta\}
    &\geq
    \frac12\left(
        1-\operatorname{TV}(\mathbb Q_1,\mathbb Q_0)
        -\beta_0-\beta_1
    \right)~,
    \label{eq: two prior probability lower bound}
\end{align}
where the infimum is over all measurable estimators of the functional.
\end{lem}

\noindent We also use of the following standard facts concerning information distances and Gaussian distributions.

\begin{lem}[\citealp{wainwright-2019}, Chapter 15]\label{lem: standard information distance bounds}$ $\\
\noindent \textbf{(i)} Let $P$ and $Q$ be probability measures, with $P$ absolutely continuous with respect to $Q$. Then
\begin{align}
    \operatorname{TV}(P,Q)
    &\leq
        \sqrt{\frac12D_{\mathrm{KL}}(P\|Q)}~.
    \label{eq: cited Pinsker inequality}
\end{align}

\noindent \textbf{(ii)} Let $P$ and $Q$ be probability measures, with $P$ absolutely continuous with respect to $Q$. Then, for every positive integer $m$,
\begin{align}
    D_{\mathrm{KL}}\left(
        P^{\otimes m}
        \,\middle\|\,
        Q^{\otimes m}
    \right)
    &=
    mD_{\mathrm{KL}}(P\|Q)~.
    \label{eq: cited KL product identity}
\end{align}

\noindent \textbf{(iii)} Let $\Sigma_0$ and $\Sigma_1$ be positive definite $q$ by $q$ matrices. Then
\begin{align}
    D_{\mathrm{KL}}\left(
        \mathsf N(0,\Sigma_1)
        \,\middle\|\,
        \mathsf N(0,\Sigma_0)
    \right)
    &=
    \frac12\left(
        \operatorname{tr}(\Sigma_0^{-1}\Sigma_1)-q
        -\log\det(\Sigma_0^{-1}\Sigma_1)
    \right)~.
    \label{eq: cited Gaussian KL identity}
\end{align}
\end{lem}

We begin by constructing two candidate priors, toward the application of \cref{lem: two prior lower bound}. Our construction is based on moment matching, and relies on the following Lemma. Similar constructions have been used in, e.g., \citet{cai-low-2011} and \citet{wu-yang-2016,wu-yang-2019,wu-yang-2020}.

\begin{lem}
\label{lem: positive moment pair}$ $\\
For every integer $L\geq2$, there exist probability measures $\varpi_0$ and $\varpi_1$ on $[0,1]$ such that the equalities
\begin{align}
    \int x^r\,d\varpi_0(x)
    &=
    \int x^r\,d\varpi_1(x),
    \qquad r=2,\ldots,L
    \label{eq: positive moment matching}
\end{align}
and
\begin{align}
    \left|
        \int x\,d\varpi_1(x)
        -
        \int x\,d\varpi_0(x)
    \right|
    &=
    \frac{1}{L^2}
    \label{eq: positive first moment gap}
\end{align}
hold.
\end{lem}

\noindent For the sake of simplicity, we restrict attention to the case that $n$ is even. Odd values of $n$ follow by an analogous argument. Fix a constant $C_1>2/\log(2)$ and choose
\begin{align}
    L
    &=\left\lceil C_1\log(2+s)\right\rceil~.
    \label{eq: moment matching order}
\end{align}
Then $L\geq2$ for all sufficiently large sequence indices. Let $M_1,\ldots,M_s$ be independent uniform perfect matchings of $[n]$, i.e., partitions of $[n]$ into $n/2$ disjoint unordered pairs. Let $A_{M_\ell}$ denote the adjacency matrix induced by $M_\ell$. Under prior $h\in\{0,1\}$, draw $x_1,\ldots,x_s$ independently from $\varpi_h$ and define
\begin{align}
    \Lambda_n
    &=
    I_n+
    \vartheta\sum_{\ell=1}^s x_\ell A_{M_\ell}
    \quad\text{and}\quad
    \vartheta
    =
    c_0\min\left\{\frac{1}{\sqrt d},\frac{1}{s}\right\}~,
    \label{eq: positive weighted matching covariance}
\end{align}
where $c_0>0$ is a sufficiently small to ensure that
\begin{align}
    (d+1)\vartheta^2
    &\leq2c_0^2
    \leq\kappa~.
    \label{eq: matching amplitude likelihood condition}
\end{align}
Let $\varrho_h$ denote the resulting probability measure over the matchings and weights. 

Set $\Gamma_d=I_{d+1}$ for both priors. Place independent $\mathsf N(0,\varphi_n^2)$ priors on the common means of the $d+1$ outcomes, independently of the matchings and weights, where
\begin{align}
    n\varphi_n^2
    &=(d+1)^2~.
    \label{eq: diffuse common mean prior}
\end{align}
Let $\pi_h$ be the product of $\varrho_h$ and these common-mean priors, and let $\overline{\mathbb Q}_h$ denote the data distribution induced by the corresponding product prior. Write $\theta=(M_1,\ldots,M_s,x_1,\ldots,x_s,\mu_0,\ldots,\mu_d)$, let $P_\theta$ denote the corresponding law, and let $\Theta$ be the union of the two prior supports. Give the matching coordinates the discrete topology and the weights and means their Euclidean topology.

We now verify that both priors are supported on $\mathcal P_{n,d}(s)$. Since $\|A_{M_\ell}\|_{\mathrm{op}}=1$ and $0\leq x_\ell\leq1$, we can evaluate
\begin{align}
    \left\|
        \vartheta\sum_{\ell=1}^s x_\ell A_{M_\ell}
    \right\|_{\mathrm{op}}
    &\leq
    \vartheta\sum_{\ell=1}^s x_\ell
    \leq c_0~.
    \label{eq: matching covariance conditioning}
\end{align}
It follows that
\begin{align}
    (1-c_0)I_n
    &\preceq\Lambda_n\preceq(1+c_0)I_n~,
    \qquad
    \lambda_{i,i}=1~,
    \quad\text{and}
    \quad
    \max_{i\in[n]}
    \sum_{i'\neq i}\mathbb I\{\lambda_{i,i'}\neq0\}
    \leq s~.
    \label{eq: matching covariance class conditions}
\end{align}
Since every row sum equals $1+\vartheta\sum_{\ell=1}^s x_\ell$, we have
\begin{align}
    V_{n,P}
    &=
    \frac{1+\vartheta\sum_{\ell=1}^s x_\ell}{n}
    \quad\text{and}\quad
    \max_{i\in[n]}\sum_{i'=1}^n|\lambda_{i,i',P}|
    =nV_{n,P}~.
    \label{eq: matching covariance target}
\end{align}
The first identity in \eqref{eq: matching covariance target} shows that $\theta \mapsto V_{n,P_\theta}$ is continuous because it is affine in the weights and does not depend on the remaining coordinates. Consequently, the two conditions in \cref{assu: sparsity} hold. The constructed Gaussian laws with common outcome means, together with $\Gamma_d=I_{d+1}$ and \eqref{eq: matching covariance conditioning}, also satisfy \cref{assu: gaussianity,assu: proportionality,assu: homoskedasticity,assu: well conditioned}. Hence, every law in the support of either prior belongs to $\mathcal P_{n,d}(s)$.

We now evaluate the separation of the target parameter $V_{n,P_\theta}$ under the two priors. Define
\begin{align}
    \bar x_h
    &=\int x\,d\varpi_h(x),
    \qquad h\in\{0,1\}~.
    \label{eq: matching weight means}
\end{align}
After interchanging the indices if necessary, \cref{lem: positive moment pair} implies that $\bar x_1-\bar x_0=L^{-2}$. Since the weights are independent and take values between zero and one, we have that
\begin{align}
    \Var_{\varrho_h}\left(\sum_{\ell=1}^s x_\ell\right)
    &=\sum_{\ell=1}^s\Var_{\varpi_h}(x_\ell)
    \leq s~.
    \label{eq: matching weight sum variance}
\end{align}
Thus, Chebyshev's inequality and \eqref{eq: matching weight sum variance} give
\begin{align}
    \varrho_h\left\{
        \left|
            \sum_{\ell=1}^s x_\ell-s\bar x_h
        \right|
        >
        \frac{s}{4L^2}
    \right\}
    &\leq
    \frac{
        \sum_{\ell=1}^s\Var_{\varpi_h}(x_\ell)
    }{
        s^2/(16L^4)
    }
    \leq
    \frac{16L^4}{s}~.
    \label{eq: matching target concentration}
\end{align}
Define the parameters
\begin{align}
    \zeta
    &=
    \frac{1}{n}
    +\frac{\vartheta s}{2n}(\bar x_0+\bar x_1)
    \quad\text{and}\quad
    \Delta
    =
    \frac{\vartheta s}{4nL^2}~.
    \label{eq: matching target midpoint and radius}
\end{align}
On the complementary event in
\eqref{eq: matching target concentration}, the identity
\eqref{eq: matching covariance target} gives
\begin{align}
    \left|
        V_{n,P_\theta}
        -\frac{1+\vartheta s\bar x_h}{n}
    \right|
    &=
    \frac{\vartheta}{n}
    \left|
        \sum_{\ell=1}^s x_\ell-s\bar x_h
    \right|
    \leq
    \frac{\vartheta s}{4nL^2}
    =
    \Delta~.
    \label{eq: matching target good event}
\end{align}
Since $\bar x_1-\bar x_0=L^{-2}$, we can write
\begin{align}
    \frac{1+\vartheta s\bar x_0}{n}
    &=
    \zeta-2\Delta
    \quad\text{and}\quad
    \frac{1+\vartheta s\bar x_1}{n}
    =
    \zeta+2\Delta~.
    \label{eq: matching target center locations}
\end{align}
Consequently, the bound \eqref{eq: matching target good event} implies that $V_{n,P_\theta} \leq \zeta-\Delta$ under prior 0 and $V_{n,P_\theta} \geq \zeta+\Delta$ under prior 1 on the corresponding complementary events in \eqref{eq: matching target concentration}. Since $\pi_h$ is the product of $\varrho_h$ and the common-mean prior, while $V_{n,P_\theta}$ does not depend on the common means, we obtain
\begin{align}
    \pi_0\{\theta:V_{n,P_\theta}\leq\zeta-\Delta\}
    &\geq
    1-\frac{16L^4}{s}
    \quad\text{and}\quad
    \pi_1\{\theta:V_{n,P_\theta}\geq\zeta+\Delta\}
    \geq
    1-\frac{16L^4}{s}~,
    \label{eq: matching target prior separation}
\end{align}
respectively.

Thus, it remains to bound the total variation distance between the two induced data distributions. Let $U$ have orthonormal columns spanning the orthogonal complement of the constant vector, and define
\begin{align}
    \widetilde A_M
    &=U^\top A_MU~.
    \label{eq: residual matching involution}
\end{align}
We make use of the the following Lemma. Similar bounds for moment-matched Gaussian mixtures are used by, e.g., \citet{cai-low-2011}. 

\begin{lem}
\label{lem: one layer likelihood bound}$ $\\
Suppose that $n$ is even, $L\geq2$, $0<c_0\leq1/4$, $\vartheta\geq0$, and $(d+1)\vartheta^2\leq\kappa$ for a sufficiently small universal constant $\kappa$. Let $\varpi_0$ and $\varpi_1$ be the measures constructed in \cref{lem: positive moment pair}. Let $M$ be a uniform perfect matching of $[n]$, let $A_M$ be its adjacency matrix, let $U$ have orthonormal columns spanning $\bm{1}^{\perp}$, and set $\widetilde A_M=U^\top A_MU$. Suppose also that a positive definite matrix $K$ satisfies
\begin{align}
    (1-c_0)I_{n-1}
    &\preceq K
    \preceq(1+c_0)I_{n-1}~.
    \label{eq: one layer background bounds}
\end{align}
For $h\in\{0,1\}$, let $\mathbb Q_{h,K}$ be the mixture of $d+1$ independent centered Gaussian vectors with covariance
\begin{equation}
    K+\vartheta x\widetilde A_M~,
    \label{eq: one layer conditional covariance}
\end{equation}
where $M$ and $x$ are independent and $x$ has law $\varpi_h$ under the corresponding hypothesis. There exists a universal constant $C>0$, independent of $n,d,L$, and $K$, such
that
\begin{align}
    \left\|
        \frac{d\mathbb Q_{1,K}-d\mathbb Q_{0,K}}{dP_K}
    \right\|_{L^2(P_K)}^2
    &\leq
    \frac{C(d+1)\vartheta^2}{(n-1)L^4}
    +C2^{-2L}~,
    \label{eq: one layer signed likelihood bound}
\end{align}
where $P_K$ is the law of $d+1$ independent centered Gaussian vectors with common covariance matrix $K$.
\end{lem}

\noindent We proceed with a telescoping argument. In particular, let $\mathbb Q^{(r)}$ denote the distribution of the residual coordinates when the first $r$ weights have law $\varpi_1$ and the remaining weights have law $\varpi_0$. To compare the distributions $\mathbb Q^{(r)}$ and $\mathbb Q^{(r-1)}$, begin by conditioning on every matching and weight except the pair $(M_r,x_r)$. The common conditional background covariance is
\begin{align}
    K_r
    &=
    I_{n-1}
    +\vartheta\sum_{\ell\neq r}x_\ell\widetilde A_{M_\ell}~.
    \label{eq: common matching layer background}
\end{align}
The bound \eqref{eq: matching covariance conditioning} implies that $K_r$ satisfies the condition \eqref{eq: one layer background bounds}. Observe that Cauchy--Schwarz implies that
\begin{align}
    \operatorname{TV}(\mathbb Q_{1,K_r},\mathbb Q_{0,K_r})
    &=
    \frac12\int
    \left|
        \frac{d\mathbb Q_{1,K_r}-d\mathbb Q_{0,K_r}}{dP_{K_r}}
    \right|dP_{K_r}
    \leq
    \frac12
    \left\|
        \frac{d\mathbb Q_{1,K_r}-d\mathbb Q_{0,K_r}}{dP_{K_r}}
    \right\|_{L^2(P_{K_r})}~,
    \label{eq: one matching L2 to TV}
\end{align}
where we use the notation $\mathbb Q_{h,K}$ introduced in \cref{lem: one layer likelihood bound}. In turn, for probability kernels $R_z$ and $S_z$ and a probability measure $\nu$, the definition of total variation gives
\begin{align}
    \operatorname{TV}\left(
        \int R_z\,d\nu(z),
        \int S_z\,d\nu(z)
    \right)
    &=
    \sup_A\left|
        \int(R_z(A)-S_z(A))\,d\nu(z)
    \right|
    \nonumber\\
    &\leq
    \int\sup_A|R_z(A)-S_z(A)|\,d\nu(z)
    =
    \int\operatorname{TV}(R_z,S_z)\,d\nu(z)~,
    \label{eq: common mixture TV inequality}
\end{align}
where the suprema range over the measurable events. Applying \eqref{eq: common mixture TV inequality} to the common distribution of the conditioned variables gives
\begin{align}
    \operatorname{TV}(\mathbb Q^{(r)},\mathbb Q^{(r-1)})
    &\leq
    \mathbb E\left[
        \operatorname{TV}(\mathbb Q_{1,K_r},\mathbb Q_{0,K_r})
    \right]
    \leq
    C\left(
        \frac{\sqrt{d+1}\vartheta}{\sqrt{n-1}L^2}
        +2^{-L}
    \right)~,
    \label{eq: one matching layer TV bound}
\end{align}
where the second inequality applies \cref{lem: one layer likelihood bound} and \eqref{eq: one matching L2 to TV}. Hence, we can conclude that
\begin{align}
    \operatorname{TV}(\mathbb Q^{(s)},\mathbb Q^{(0)})
    &\leq
    \sum_{r=1}^s
    \operatorname{TV}(\mathbb Q^{(r)},\mathbb Q^{(r-1)})
    \leq
    C\left(
        \frac{s\sqrt{d+1}\vartheta}{\sqrt{n-1}L^2}
        +s2^{-L}
    \right)
    \label{eq: all matching layers TV bound}
\end{align}
by the triangle inequality.

Now, observe that we can write
\begin{align}
    \frac{1}{n}\bm{1}^{\top}\Lambda_n\bm{1}
    &=
    1+\vartheta\sum_{\ell=1}^s x_\ell
    \quad\text{and}\quad
    \frac{1}{\sqrt n}U^\top\Lambda_n\bm{1}
    =0~.
    \label{eq: constant residual covariance zero}
\end{align}
Consequently, \cref{assu: gaussianity,assu: proportionality} and the choice $\Gamma_d=I_{d+1}$,  imply that
\begin{align}
    \left(
        \begin{pmatrix}
            n^{-1/2}\bm{1}^{\top}Y^{(j)}\\
            U^\top Y^{(j)}
        \end{pmatrix}
    \right)_{j=0}^d
    &\quad\sim
    \prod_{j=0}^d
    \left(
        \mathsf N\left(
            \sqrt n\,\mu_j,
            1+\vartheta\sum_{\ell=1}^s x_\ell
        \right)
        \otimes
        \mathsf N(0,U^\top\Lambda_nU)
    \right)
    \label{eq: conditional constant residual factorization}
\end{align}
conditional on the matchings and weights. Integrating the mean $\tau^{(j)}\sim\mathsf N(0,\varphi_n^2)$ gives
\begin{align}
    \int_{\mathbb R}
    \mathsf N\left(
        \sqrt n\,\tau,
        1+\vartheta\sum_{\ell=1}^s x_\ell
    \right)
    d\mathsf N(0,\varphi_n^2)(\tau)
    &=
    \mathsf N\left(
        0,
        (d+1)^2+1+\vartheta\sum_{\ell=1}^s x_\ell
    \right)~,
    \label{eq: integrated constant coordinate law}
\end{align}
Thus, conditional on the matchings and weights, the $d+1$ variables $\frac{1}{\sqrt n}\sum_{i=1}^nY_i^{(j)}$ have the joint distribution
\begin{align}
    \mathbb G(x_1,\ldots,x_s)
    &=
    \prod_{j=0}^d
    \mathsf N\left(
        0,
        (d+1)^2+1+\vartheta\sum_{\ell=1}^s x_\ell
    \right)~.
    \label{eq: conditional constant coordinate law}
\end{align}
Define the reference distribution
\begin{align}
    \mathbb G_{\mathrm{ref}}
    & =
    \prod_{j=0}^d
    \mathsf N(0,(d+1)^2)~.
    \label{eq: constant coordinate product laws}
\end{align}
Parts (ii) and (iii) of
\cref{lem: standard information distance bounds}, applied with
$m=d+1$, $q=1$,
\begin{equation}
    \Sigma_1
    =
    (d+1)^2+1+\vartheta\sum_{\ell=1}^s x_\ell~,
    \quad\text{and}\quad
    \Sigma_0=(d+1)^2~,
\end{equation}
give
\begin{align}
    D_{\mathrm{KL}}\left(
        \mathbb G(x_1,\ldots,x_s)
        \,\middle\|\,
        \mathbb G_{\mathrm{ref}}
    \right)
    &=
    \frac{d+1}{2}
    \left(
        \frac{
            1+\vartheta\sum_{\ell=1}^s x_\ell
        }{
            (d+1)^2
        }
        -
        \log\left(
            1+
            \frac{
                1+\vartheta\sum_{\ell=1}^s x_\ell
            }{
                (d+1)^2
            }
        \right)
    \right)
    \nonumber\\
    &\leq
    \frac{d+1}{4}
    \left(
        \frac{
            1+\vartheta\sum_{\ell=1}^s x_\ell
        }{
            (d+1)^2
        }
    \right)^2
    \leq
    \frac{C}{(d+1)^3}~,
    \label{eq: masked coordinate Gaussian divergence}
\end{align}
where the second inequality uses the elementary bound 
\begin{align}
    t-\log(1+t)
    &=
    \int_0^t\frac{u}{1+u}\,du
    \leq
    \frac{t^2}{2}
    \label{eq: logarithm quadratic bound}
\end{align}
and the final inequality uses $0\leq\vartheta\sum_{\ell=1}^s x_\ell\leq c_0$. Hence, we find that
\begin{align}
    \operatorname{TV}\left(
        \mathbb G(x_1,\ldots,x_s),
        \mathbb G_{\mathrm{ref}}
    \right)
    &\leq
    \sqrt{
        \frac12
        D_{\mathrm{KL}}\left(
            \mathbb G(x_1,\ldots,x_s)
            \,\middle\|\,
            \mathbb G_{\mathrm{ref}}
        \right)
    }
    \leq
    C(d+1)^{-3/2}~.
    \label{eq: masked coordinate TV bound}
\end{align}
by applying Part (i) of \cref{lem: standard information distance bounds}.

With this in place, apply the orthogonal transformation
\begin{align}
    \left(Y^{(j)}\right)_{j=0}^d
    &\longmapsto
    \left(
        \frac{1}{\sqrt n}\bm{1}^{\top}Y^{(j)},
        U^\top Y^{(j)}
    \right)_{j=0}^d~.
    \label{eq: constant residual orthogonal transformation}
\end{align}
Since this transformation is a measurable bijection, total variation distance is unchanged. For fixed matching and weights $M_1,\ldots,M_s,x_1,\ldots,x_s$, define the measure 
\begin{align}
    \mathbb H(M_1,\ldots,M_s,x_1,\ldots,x_s)
    &=
    \prod_{j=0}^d
    \mathsf N(0,U^\top\Lambda_nU)~.
    \label{eq: conditional transformed residual law}
\end{align}
The representations \eqref{eq: conditional constant residual factorization} and \eqref{eq: integrated constant coordinate law} imply that we can write
\begin{align}
    \overline{\mathbb Q}_h
    &=
    \int
    \left(
        \mathbb H(M_1,\ldots,M_s,x_1,\ldots,x_s)
        \otimes
        \mathbb G(x_1,\ldots,x_s)
    \right)
    d\varrho_h,
    \qquad h\in\{0,1\}~,
    \label{eq: full mixture conditional representation}
\end{align}
and that moreover
\begin{align}
    \mathbb Q^{(s)}\otimes\mathbb G_{\mathrm{ref}}
    &=
    \int
    \left(
        \mathbb H(M_1,\ldots,M_s,x_1,\ldots,x_s)
        \otimes
        \mathbb G_{\mathrm{ref}}
    \right)
    d\varrho_1\quad\text{and}
    \nonumber\\
    \mathbb Q^{(0)}\otimes\mathbb G_{\mathrm{ref}}
    &=
    \int
    \left(
        \mathbb H(M_1,\ldots,M_s,x_1,\ldots,x_s)
        \otimes
        \mathbb G_{\mathrm{ref}}
    \right)
    d\varrho_0~.
    \label{eq: reference full mixture representations}
\end{align}
Hence, \eqref{eq: common mixture TV inequality}, \eqref{eq: masked coordinate TV bound}, \eqref{eq: full mixture conditional representation}, \eqref{eq: reference full mixture representations},  imply that
\begin{align}
    &\operatorname{TV}\left(
        \overline{\mathbb Q}_1,
        \mathbb Q^{(s)}\otimes\mathbb G_{\mathrm{ref}}
    \right)
    \nonumber\\
    &\quad\leq
    \int
    \operatorname{TV}\left(
        \mathbb H(M_1,\ldots,M_s,x_1,\ldots,x_s)
        \otimes
        \mathbb G(x_1,\ldots,x_s),
        \mathbb H(M_1,\ldots,M_s,x_1,\ldots,x_s)
        \otimes
        \mathbb G_{\mathrm{ref}}
    \right)
    d\varrho_1
    \nonumber\\
    &\quad=
    \int
    \operatorname{TV}\left(
        \mathbb G(x_1,\ldots,x_s),
        \mathbb G_{\mathrm{ref}}
    \right)
    d\varrho_1
    \leq
    C(d+1)^{-3/2}
    \label{eq: prior one full mixture replacement}
\end{align}
and analogously
\begin{align}
    \operatorname{TV}\left(
        \overline{\mathbb Q}_0,
        \mathbb Q^{(0)}\otimes\mathbb G_{\mathrm{ref}}
    \right)
    &\leq
    C(d+1)^{-3/2}~,
    \label{eq: full mixture constant coordinate replacement}
\end{align}
where we have used the fact that
\begin{align}
    \operatorname{TV}(H\otimes P,H\otimes Q) = \operatorname{TV}(P,Q) = \operatorname{TV}(P\otimes H,Q\otimes H) \label{eq: tv invariance}
\end{align}
for probability measures $P$, $Q$, and $H$. Hence, the triangle inequality implies that
\begin{align}
    & \operatorname{TV}(
        \overline{\mathbb Q}_1,
        \overline{\mathbb Q}_0
    ) \nonumber \\
    &\leq
    \operatorname{TV}\left(
        \overline{\mathbb Q}_1,
        \mathbb Q^{(s)}\otimes\mathbb G_{\mathrm{ref}}
    \right)
    +
    \operatorname{TV}\left(
        \mathbb Q^{(s)}\otimes\mathbb G_{\mathrm{ref}},
        \mathbb Q^{(0)}\otimes\mathbb G_{\mathrm{ref}}
    \right)
    +
    \operatorname{TV}\left(
        \mathbb Q^{(0)}\otimes\mathbb G_{\mathrm{ref}},
        \overline{\mathbb Q}_0
    \right)
    \nonumber\\
    &\leq
    \operatorname{TV}\left(
        \mathbb Q^{(s)},
        \mathbb Q^{(0)}
    \right)
    +2C(d+1)^{-3/2}
    \leq
    C\left(
        \frac{
            s\sqrt{d+1}\vartheta
        }{
            \sqrt{n-1}L^2
        }
        +s2^{-L}
        +(d+1)^{-3/2}
    \right)~,
    \label{eq: full data matching mixture TV bound}
\end{align}
where the second inequality uses \eqref{eq: prior one full mixture replacement}, \eqref{eq: full mixture constant coordinate replacement}, and \eqref{eq: tv invariance} and the final inequality uses \eqref{eq: all matching layers TV bound}. Observe that the choice \eqref{eq: moment matching order} implies that
\begin{align}
    \frac{L^4}{s}
    &\lesssim\frac{\log^4(2+s)}{s}
    \quad\text{and}\quad
    s2^{-L}
    \leq
    s(2+s)^{-C_1\log 2}
    \label{eq: matching prior exceptional probability convergence}
\end{align}
and that the choice \eqref{eq: positive weighted matching covariance} implies that
\begin{align}
    s\sqrt d\,\vartheta
    &=c_0\min\{s,\sqrt d\}
    \label{eq: matching TV amplitude simplification}
\end{align}
and thereby
\begin{align}
    \left(
        \frac{s\sqrt{d+1}\vartheta}{\sqrt{n-1}L^2}
    \right)^2
    &\lesssim
    \frac{\min\{s,\sqrt d\}^2}{nL^4}
    \leq
    \frac{s\sqrt d}{n}~,
    \label{eq: maintained condition controls TV}
\end{align}
where the final inequality uses $L\geq1$. By plugging these bounds into \eqref{eq: full data matching mixture TV bound}, we find that
\begin{align}
    \operatorname{TV}(\overline{\mathbb Q}_1,
        \overline{\mathbb Q}_0)
    &
    \leq C
    \left(
    \left(\frac{s\sqrt d}{n}\right)^{1/2}
    + 
    s(2+s)^{-C_1\log 2}
    +(d+1)^{-3/2}
    \right) \to 0
    \label{eq: matching mixture TV convergence}
\end{align}
as $n,s,d\to\infty$.

It remains to apply \cref{lem: two prior lower bound}.  Take $\zeta$ and $\Delta$ from \eqref{eq: matching target midpoint and radius}. The bounds in \eqref{eq: matching target prior separation} verify \eqref{eq: two prior separation conditions} with $\beta_0=\beta_1=16L^4/s$, respectively. For every estimator $T$ and every $\theta$ supports of both prior distributions, we have that
\begin{align}
    \frac{\mathbb E_{P_\theta}[(T-F(\theta))^2]}{F(\theta)^2}
    &\geq
    \frac{n^2\Delta^2}{(1+c_0)^2}
    P_\theta\{|T-F(\theta)|\geq\Delta\}~.
    \label{eq: probability to relative risk conversion}
\end{align}
Combining \cref{lem: two prior lower bound}, display \eqref{eq: two prior probability lower bound}, \eqref{eq: matching target prior separation}, and \eqref{eq: probability to relative risk conversion}, we obtain
\begin{align}
    \inf_T
    \sup_{\theta\in\Theta}
    \frac{\mathbb E_{P_\theta}[(T-F(\theta))^2]}{F(\theta)^2}
    &\geq
    \frac{c\vartheta^2s^2}{L^4}
    \left(
        1-
        \operatorname{TV}(\overline{\mathbb Q}_1,
            \overline{\mathbb Q}_0)
        -\frac{32L^4}{s}
    \right)~.
    \label{eq: fuzzy matching lower bound}
\end{align}
By the choices
\begin{align}
    \vartheta^2s^2
    &=
    c_0^2\min\left\{\frac{s^2}{d},1\right\}
    \quad\text{and}\quad
    L\asymp\log(2+s)~,
    \label{eq: matching target separation rate}
\end{align}
the preceding display and the bound \eqref{eq: matching mixture TV convergence} imply that
\begin{align}
    \inf_T
    \sup_{P\in\mathcal P_{n,d}(s)}
    \frac{\mathbb E_P[(T-V_{n,P})^2]}{V_{n,P}^2}
    &\gtrsim
    \frac{1}{\log^4(2+s)}
    \min\left\{\frac{s^2}{d},1\right\}~,
    \label{eq: constrained minimax squared lower bound}
\end{align}
as required.\hfill\qed

\subsubsection{Part (ii)}

Observe that \cref{lem: basic inequalities n}, display \eqref{eq: var bounds}, gives
\begin{align}
    \frac{c}{n}
    &\leq V_{n,P}
    \leq\frac{C(s+1)}{n}
    \label{eq: constrained minimax objective range}
\end{align}
uniformly over $P\in\mathcal P_{n,d}(s)$. Together with
\eqref{eq: constrained minimax endpoint conditions}, these bounds verify the conditions of Part (ii) of \cref{lem: selected set characterization} uniformly over $P\in\mathcal P_{n,d}(s)$. Thus, Part (ii) of \cref{lem: selected set characterization} implies that
\begin{align}
    \sup_{P\in\mathcal P_{n,d}(s)}
    \left|
        \frac{
            \underline\delta_{n,d,P}(\eta)
        }{
            \sqrt{
                2\eta v_{0,P}^{(d)}
                \log(1/V_{n,P})
            }
        }
        -1
    \right|
    &\to0~.
    \label{eq: constrained minimax uniform endpoint characterization}
\end{align}
Moreover, \cref{lem: auxiliary outcome covariance bounds}, display \eqref{eq: auxiliary outcome covariance bounds}, and \eqref{eq: constrained minimax uniform endpoint characterization} imply that
\begin{align}
    \underline\delta_{n,d,P}(\eta)^2
    &\lesssim
    \frac{1}{d}\log\left(\frac{1}{V_{n,P}}\right)
    \label{eq: constrained minimax uniform endpoint scale}
\end{align}
uniformly over $P\in\mathcal P_{n,d}(s)$.

Consider the decomposition
\begin{align}
    \frac{B_{n,d,P}(\delta)}{V_{n,P}}
    &=-F^+(\delta,0,v_{0,P}^{(d)})
    -\frac{1}{n^2V_{n,P}}
    \sum_{i,i'=1}^n
    \lambda_{i,i',P}
    \left(
        1-F^+(\delta,\lambda_{i,i',P},v_{i,i',P}^{(d)})
    \right)~.
    \label{eq: constrained exact B decomposition}
\end{align}
We begin by bounding the first term in \eqref{eq: constrained exact B decomposition} at $\underline\delta_{n,d,P}(\eta)$. Observe that the  Gaussian tail bound
\begin{align}
    F^+(\delta,0,v)
    &\leq2\exp\left(-\frac{\delta^2}{2v}\right)
    \label{eq: constrained gaussian ideal null bound}
\end{align}
and \eqref{eq: constrained minimax uniform endpoint scale} give
\begin{align}
    F^+(\underline\delta_{n,d,P}(\eta),0,v_{0,P}^{(d)})
    &\leq V_{n,P}^{\eta/2}
    \leq\mathcal V_{n,d,P}(s,\eta)
    \label{eq: constrained ideal null standard deviation bound}
\end{align}
uniformly over $P\in\mathcal P_{n,d}(s)$. Next, we bound the second term in \eqref{eq: constrained exact B decomposition}. \cref{lem: auxiliary outcome covariance bounds}, display \eqref{eq: auxiliary outcome covariance bounds}, gives
\begin{align}
    \max_{i,i'\in[n]}v_{i,i',P}^{(d)}
    &\lesssim\frac1d
    \label{eq: constrained covariance and variance bounds}
\end{align}
uniformly over $P\in\mathcal P_{n,d}(s)$. Fix a pair $i \neq i'$. If $|\lambda_{i,i',P}|\leq 2\underline\delta_{n,d,P}(\eta)$, then it holds that
\begin{align}
    &|\lambda_{i,i',P}|
    P\left\{
        |Z_{\lambda_{i,i',P},v_{i,i',P}^{(d)}}|
        <\underline\delta_{n,d,P}(\eta)
    \right\}
    \leq
    2\underline\delta_{n,d,P}(\eta)~.
    \label{eq: constrained weak edge false negative mass}
\end{align}
On the other hand, if $|\lambda_{i,i',P}|> 2\underline\delta_{n,d,P}(\eta)$, then the triangle inequality and \eqref{eq: constrained covariance and variance bounds} give
\begin{align}
    |\lambda_{i,i',P}|P\left\{
        |Z_{\lambda_{i,i',P},v_{i,i',P}^{(d)}}|
        <\underline\delta_{n,d,P}(\eta)
    \right\}
    &\leq
    |\lambda_{i,i',P}|P\left\{
        |Z_{\lambda_{i,i',P},v_{i,i',P}^{(d)}}
        -\lambda_{i,i',P}|
        >\frac{|\lambda_{i,i',P}|}{2}
    \right\}
    \nonumber\\
    &\leq
    2|\lambda_{i,i',P}|\exp\left(
        -\frac{\lambda_{i,i',P}^2}{8v_{i,i',P}^{(d)}}
    \right) \nonumber \\
    & \leq
    2|\lambda_{i,i',P}|\exp(-cd\lambda_{i,i',P}^2)
    \lesssim \frac{1}{\sqrt{d}}~,
    \label{eq: constrained strong edge false negative probability}
\end{align}
where the second inequality applies \eqref{eq: constrained gaussian ideal null bound} to the centered Gaussian variable with threshold $|\lambda_{i,i',P}|/2$, and the third inequality uses \eqref{eq: constrained covariance and variance bounds}, and the final inequality uses 
\begin{align}
    \sup_{x\geq0}x\exp(-cdx^2)
    &=\frac{1}{\sqrt{2cde}}
    \lesssim\frac1{\sqrt d}~.
\end{align}
Hence, we find that
\begin{align}
    |\lambda_{i,i',P}|
    P\left\{
        |Z_{\lambda_{i,i',P},v_{i,i',P}^{(d)}}|
        <\underline\delta_{n,d,P}(\eta)
    \right\}
    \lesssim
    \underline\delta_{n,d,P}(\eta)+\frac1{\sqrt d}~.
    \label{eq: constrained per edge false negative mass}
\end{align}
by combining the two cases. For the case that $i = i^\prime$, observe that \eqref{eq: constrained minimax objective range}, \eqref{eq: constrained minimax endpoint conditions}, and \eqref{eq: constrained minimax uniform endpoint scale} imply that
\begin{align}
    \sup_{P\in\mathcal P_{n,d}(s)}
    \underline\delta_{n,d,P}(\eta)^2
    &\lesssim
    \frac{\log n}{d}
    \to0~.
    \label{eq: constrained diagonal endpoint convergence}
\end{align}
Thus, for all sufficiently large $n,d,s$, the triangle inequality, \eqref{eq: constrained covariance and variance bounds}, and \eqref{eq: constrained diagonal endpoint convergence} give
\begin{align}
    1-F^+(\underline\delta_{n,d,P}(\eta),1,v_{i,i,P}^{(d)})
    &\leq
    P\left\{
        |Z_{1,v_{i,i,P}^{(d)}}-1|>\frac12
    \right\}
    \leq
    2\exp(-cd)
    \label{eq: constrained diagonal false negative mass}
\end{align}
uniformly over $P\in\mathcal P_{n,d}(s)$, where the final inequality applies \eqref{eq: constrained gaussian ideal null bound} and \eqref{eq: constrained covariance and variance bounds}.

By \cref{assu: sparsity}, there are at most $ns$ nonzero off-diagonal entries. For every $i,i'$, we have
\begin{align}
    1-F^+\left(
        \underline\delta_{n,d,P}(\eta),
        \lambda_{i,i',P},
        v_{i,i',P}^{(d)}
    \right)
    &=
    P\left\{
        \left|
            Z_{\lambda_{i,i',P},v_{i,i',P}^{(d)}}
        \right|
        <
        \underline\delta_{n,d,P}(\eta)
    \right\}~.
    \label{eq: constrained false negative probability identity}
\end{align}
Therefore, \eqref{eq: constrained exact B decomposition} gives
\begin{align}
    &\left|
        \frac{
            B_{n,d,P}(\underline\delta_{n,d,P}(\eta))
        }{
            V_{n,P}
        }
        +
        F^+\left(
            \underline\delta_{n,d,P}(\eta),
            0,
            v_{0,P}^{(d)}
        \right)
    \right|
    \nonumber\\
    &\qquad=
    \frac{1}{n^2V_{n,P}}
    \left|
        \sum_{i=1}^n\sum_{i'\neq i}
        \lambda_{i,i',P}
        P\left\{
            \left|
                Z_{\lambda_{i,i',P},v_{i,i',P}^{(d)}}
            \right|
            <
            \underline\delta_{n,d,P}(\eta)
        \right\}
    \right.
    \nonumber\\
    &\hspace{62mm}\left.
        {}+
        \sum_{i=1}^n
        P\left\{
            \left|
                Z_{1,v_{i,i,P}^{(d)}}
            \right|
            <
            \underline\delta_{n,d,P}(\eta)
        \right\}
    \right|
    \nonumber\\
    &\qquad\leq
    \frac{1}{n^2V_{n,P}}
    \sum_{i=1}^n\sum_{i'\neq i}
    |\lambda_{i,i',P}|
    P\left\{
        \left|
            Z_{\lambda_{i,i',P},v_{i,i',P}^{(d)}}
        \right|
        <
        \underline\delta_{n,d,P}(\eta)
    \right\}
    \nonumber\\
    &\qquad\quad+
    \frac{1}{n^2V_{n,P}}
    \sum_{i=1}^n
    P\left\{
        \left|
            Z_{1,v_{i,i,P}^{(d)}}
        \right|
        <
        \underline\delta_{n,d,P}(\eta)
    \right\}~.
    \label{eq: constrained endpoint off diagonal diagonal decomposition}
\end{align}
For the off-diagonal sum, \eqref{eq: constrained per edge false negative mass} implies that
\begin{align}
    &\sum_{i=1}^n\sum_{i'\neq i}
    |\lambda_{i,i',P}|
    P\left\{
        \left|
            Z_{\lambda_{i,i',P},v_{i,i',P}^{(d)}}
        \right|
        <
        \underline\delta_{n,d,P}(\eta)
    \right\}
    \nonumber\\
    &\qquad=
    \sum_{i=1}^n\sum_{i'\neq i}
    \mathbb I\{\lambda_{i,i',P}\neq0\}
    |\lambda_{i,i',P}|
    P\left\{
        \left|
            Z_{\lambda_{i,i',P},v_{i,i',P}^{(d)}}
        \right|
        <
        \underline\delta_{n,d,P}(\eta)
    \right\}
    \nonumber\\
    &\qquad\lesssim
    \left(
        \underline\delta_{n,d,P}(\eta)
        +\frac{1}{\sqrt d}
    \right)
    \sum_{i=1}^n\sum_{i'\neq i}
    \mathbb I\{\lambda_{i,i',P}\neq0\}
    \leq
    ns\left(
        \underline\delta_{n,d,P}(\eta)
        +\frac{1}{\sqrt d}
    \right)~.
    \label{eq: constrained off diagonal false negative aggregation}
\end{align}
For the diagonal sum, \eqref{eq: constrained diagonal false negative mass} gives
\begin{align}
    &\sum_{i=1}^n
    P\left\{
        \left|
            Z_{1,v_{i,i,P}^{(d)}}
        \right|
        <
        \underline\delta_{n,d,P}(\eta)
    \right\}
    =
    \sum_{i=1}^n
    \left(
        1-
        F^+\left(
            \underline\delta_{n,d,P}(\eta),
            1,
            v_{i,i,P}^{(d)}
        \right)
    \right)
    \leq
    2n\exp(-cd)~.
    \label{eq: constrained diagonal false negative aggregation}
\end{align}
Substituting \eqref{eq: constrained off diagonal false negative aggregation} and \eqref{eq: constrained diagonal false negative aggregation} into \eqref{eq: constrained endpoint off diagonal diagonal decomposition} gives
\begin{align}
    &\left|
        \frac{
            B_{n,d,P}(\underline\delta_{n,d,P}(\eta))
        }{
            V_{n,P}
        }
        +
        F^+\left(
            \underline\delta_{n,d,P}(\eta),
            0,
            v_{0,P}^{(d)}
        \right)
    \right|
    \nonumber\\
    &\qquad\lesssim
    \frac{1}{n^2V_{n,P}}
    \left\{
        ns\left(
            \underline\delta_{n,d,P}(\eta)
            +\frac{1}{\sqrt d}
        \right)
        +n\exp(-cd)
    \right\}
    \nonumber\\
    &\qquad=
    \frac{s}{nV_{n,P}}
    \left(
        \underline\delta_{n,d,P}(\eta)
        +\frac{1}{\sqrt d}
    \right)
    +
    \frac{\exp(-cd)}{nV_{n,P}}~.
    \label{eq: constrained endpoint aggregation}
\end{align}
Additionally, $0\leq1-F^+(\delta,\lambda,v)\leq1$ and \cref{lem: basic inequalities n}, display \eqref{eq: n bound 1}, imply that
\begin{align}
    &\left|
        \frac{B_{n,d,P}(\underline\delta_{n,d,P}(\eta))}{V_{n,P}}
        +F^+(\underline\delta_{n,d,P}(\eta),0,v_{0,P}^{(d)})
    \right|
    \leq
    \frac{1}{n^2V_{n,P}}
    \sum_{i,i'=1}^n|\lambda_{i,i',P}|
    \lesssim1~.
    \label{eq: constrained endpoint crude bound}
\end{align}
As a consequence, by combining \eqref{eq: constrained minimax objective range}, \eqref{eq: constrained minimax uniform endpoint scale}, \eqref{eq: constrained endpoint aggregation}, and \eqref{eq: constrained endpoint crude bound}, we obtain
\begin{align}
    &\left(
        \frac{B_{n,d,P}(\underline\delta_{n,d,P}(\eta))}{V_{n,P}}
        +F^+(\underline\delta_{n,d,P}(\eta),0,v_{0,P}^{(d)})
    \right)^2
    \nonumber\\
    &\qquad\lesssim
    \min\left\{
        \frac{s^2}{dn^2V_{n,P}^2}
        \left(
            1+\log\left(\frac{1}{V_{n,P}}\right)
        \right),
        1
    \right\}
    +\exp(-cd)~.
    \label{eq: constrained pointwise endpoint envelope}
\end{align}
Observe that 
\begin{align}
    \exp(-cd)
    &=o\left(\frac{\log n}{d}\right)
    \quad\text{and}\quad
    \frac{\log n}{d}
    \leq
    \min\left\{\frac{s^2\log n}{d},1\right\}
    \label{eq: constrained exponential remainder absorption}
\end{align}
for all sufficiently large $n,d,s$, where the second inequality uses $\log n=o(d)$. Hence, we can conclude that
\begin{align}
    &\sup_{P\in\mathcal P_{n,d}(s)}
    \left(
        \frac{B_{n,d,P}(\underline\delta_{n,d,P}(\eta))}{V_{n,P}}
        +F^+(\underline\delta_{n,d,P}(\eta),0,v_{0,P}^{(d)})
    \right)^2
    \lesssim
    \min\left\{\frac{s^2\log n}{d},1\right\}~.
    \label{eq: constrained endpoint envelope}
\end{align}
by plugging \eqref{eq: constrained exponential remainder absorption} into \eqref{eq: constrained pointwise endpoint envelope}.

To conclude the proof, observe that the elementary inequality
\begin{align}
    \frac{
        \sqrt{\min\{s^2\log n/d,1\}}
    }{
        \sqrt{\log n}
    }
    &\leq
    \min\left\{\frac{s}{\sqrt d},1\right\}
    \label{eq: constrained endpoint envelope rate comparison}
\end{align}
and \eqref{eq: constrained endpoint envelope} imply that
\begin{align}
    &\frac{1}{\sqrt{\log n}\,\log^2(2+s)}
    \sup_{P\in\mathcal P_{n,d}(s)}
    \left|
        \frac{B_{n,d,P}(\underline\delta_{n,d,P}(\eta))}{V_{n,P}}
        +F^+(\underline\delta_{n,d,P}(\eta),0,v_{0,P}^{(d)})
    \right|
    \nonumber\\
    &\qquad\lesssim
    \frac{1}{\log^2(2+s)}
    \min\left\{\frac{s}{\sqrt d},1\right\}
    \lesssim
    \mathfrak{M}_{n,d}(s)~,
    \label{eq: constrained structural root mean square comparison}
\end{align}
where the second inequality follows from Part (i) of the Lemma. In turn, the triangle inequality implies that
\begin{align}
    \frac{|B_{n,d,P}(\underline\delta_{n,d,P}(\eta))|}{V_{n,P}}
    &\leq
    \left|
        \frac{B_{n,d,P}(\underline\delta_{n,d,P}(\eta))}{V_{n,P}}
        +F^+(\underline\delta_{n,d,P}(\eta),0,v_{0,P}^{(d)})
    \right|
    +F^+(\underline\delta_{n,d,P}(\eta),0,v_{0,P}^{(d)})~.
    \label{eq: constrained literal B linear decomposition bound}
\end{align}
By combining \eqref{eq: constrained literal B linear decomposition bound} and \eqref{eq: constrained ideal null standard deviation bound}, we obtain
\begin{align}
    \frac{|B_{n,d,P}(\underline\delta_{n,d,P}(\eta))|}{V_{n,P}}
    -C\mathcal V_{n,d,P}(s,\eta)
    &\leq
    \left|
        \frac{B_{n,d,P}(\underline\delta_{n,d,P}(\eta))}{V_{n,P}}
        +F^+(\underline\delta_{n,d,P}(\eta),0,v_{0,P}^{(d)})
    \right|~.
    \label{eq: constrained literal B linear envelope bound}
\end{align}
The proof is then complete by taking the supremum over $P\in\mathcal P_{n,d}(s)$ in \eqref{eq: constrained literal B linear envelope bound}, dividing by $\sqrt{\log n}\,\log^2(2+s)$, and applying the inequality \eqref{eq: constrained structural root mean square comparison}.
\hfill\qed

\subsection{Proof of \cref{thm: relative bias regret}}

Throughout the proofs supporting this Lemma, we use the shorthand
\begin{align}
    \hat V_{n,d}(\underline{\delta}_{n,d}(\eta_{k}))
    = \hat V_{n,d}(k)
    \quad\text{and}\quad
    \hat\xi_{n,d}(\underline{\delta}_{n,d}(\eta_{k}))
    = \hat\xi_{n,d}(k)
\end{align}
with an abuse of notation. 

We begin by showing that \cref{assu: nonpositive} implies that $B_{n,d}(\delta)$ is always non-positive. Observe that
\begin{align}
    Q(\delta,x) = F^{+}(\delta,x,v_{0}^{(d)}(1+x^2))-F^{+}(\delta,0,v_{0}^{(d)})
\end{align}
is non-decreasing in $x$ and satisfies $Q(\delta,0)=0$ and $0 \leq Q(\delta,x) \leq 1$. Let $ t_1 <\ldots <t_m \leq 1$ denote the distinct positive values of $\vert \lambda_{i,i'} \vert $ and set $t_0 = 0$. Thus, we can write
\begin{align}
    Q(\delta,\vert \lambda_{i,i'} \vert) =
    \sum_{r =1}^m (Q(\delta,t_r)-Q(\delta,t_{r-1}))\mathbb{I}\{ \vert \lambda_{i,i'} \vert \geq t_r \}
\end{align}
and thereby
\begin{align}
    \frac{1}{n^2}\sum_{i,i'=1}^n \lambda_{i,i'}Q_{i,i'}(\delta)
    = \sum_{r=1}^m(Q(\delta,t_r)-Q(\delta,t_{r-1})) 
    \left( \frac{1}{n^2}\sum_{i,i'=1}^n \lambda_{i,i'} \mathbb{I}\{ \vert \lambda_{i,i'} \vert \geq t_r\}\right)~.
\end{align}
Hence, \cref{assu: nonpositive} and the fact that $ Q(\delta,x)$ is non-decreasing imply that
\begin{align}
    0 \leq \frac{1}{n^2}\sum_{i,i'=1}^n \lambda_{i,i'}Q_{i,i'}(\delta) 
    \leq V_n
    \sum_{r=1}^m(Q(\delta,t_r)-Q(\delta,t_{r-1})) = V_n Q(\delta,t_m) \leq V_n
\end{align}
and thereby 
\begin{align}
    -V_n \leq B_{n,d}(\delta)= \frac{1}{n^2}\sum_{i,i'=1}^n \lambda_{i,i'}Q_{i,i'}(\delta)-V_n \leq 0~,\label{eq: negative bias approx}
\end{align}
as required.

With this in place, the bound \eqref{eq: relative bias regret bound} follows from the application of the following Lemma.

\begin{lem}
\label{lem: fixed multiplier calibration}$ $\\
Suppose that the conditions of \cref{thm: relative bias regret} hold.

\noindent \textbf{(i)} It holds that
\begin{align}
    &\max_{k\in[g]}
    \frac{[\hat V_{n,d}(k)-\hat\xi_{n,d}(k)
      -\mathbb E[\hat V_{n,d}(k)]]_+}{V_n}
    =O_p\left(
      \frac{\log(V_n^{-1})}{\sqrt d}
      +\sqrt{dV_n}\log^{3/2}(V_n^{-1})
      +\frac{s\sqrt d}{n}
    \right)~,
    \label{eq: selected side calibration}
\end{align}
where $[\cdot]_{+}=\max\{0,\cdot\}$.

\noindent \textbf{(ii)} For each $k \in [g]$, it holds that
\begin{align}
    &\frac{|\hat V_{n,d}(k)
      -\mathbb E[\hat V_{n,d}(k)]|+\hat\xi_{n,d}(k)}{V_n}
    \nonumber\\
    &\quad=O_p\left(
      V_n^{\eta_k/2}+\frac{\log(V_n^{-1})}{\sqrt d}
      +\sqrt{dV_n}\log^{3/2}(V_n^{-1})
      +\frac{s\sqrt d}{n}
    \right)~.
    \label{eq: optimizer side calibration}
\end{align}
\end{lem}

\noindent By \eqref{eq: negative bias approx},
\eqref{eq: bias approximation}, \eqref{eq: regret rate condition}, and the
fact that $g$ is fixed,
\begin{align}
    \max_{k\in[g]}[\operatorname{Bias}_{n,d}(k)]_+
    &\leq
      \max_{k\in[g]}
      \left|
        \operatorname{Bias}_{n,d}(k)
        -B_{n,d}(\underline{\delta}_{n,d}(\eta_k))
      \right|
    \lesssim
      V_n\mathcal{R}_{n,d}(s)~.
    \label{eq: positive bias approximation bound}
\end{align}
As a consequence, it holds that
\begin{align}
    &\left|\operatorname{Bias}_{n,d}(\hat k)\right|
      -\left|\operatorname{Bias}_{n,d}(k^\star)\right|
    \nonumber\\
    &\quad\leq
      \operatorname{Bias}_{n,d}(k^\star)
      -\operatorname{Bias}_{n,d}(\hat k)
      +2[\operatorname{Bias}_{n,d}(\hat k)]_+
    \nonumber\\
    &\quad=
      \mathbb E[\hat V_{n,d}(k^\star)]
      -\hat V_{n,d}(k^\star)
      +\hat\xi_{n,d}(k^\star)
    \nonumber\\
    &\quad+
      \left(
        \hat V_{n,d}(k^\star)
        -\hat\xi_{n,d}(k^\star)
        -\hat V_{n,d}(\hat k)
        +\hat\xi_{n,d}(\hat k)
      \right)
    +
      \left(
        \hat V_{n,d}(\hat k)
        -\hat\xi_{n,d}(\hat k)
        -\mathbb E[\hat V_{n,d}(\hat k)]
      \right)
      +2[\operatorname{Bias}_{n,d}(\hat k)]_+
    \nonumber\\
    &\quad\leq
      \left|
        \hat V_{n,d}(k^\star)
        -\mathbb E[\hat V_{n,d}(k^\star)]
      \right|
      +\hat\xi_{n,d}(k^\star)
    \nonumber\\
    &\quad+
      \max_{k\in[g]}
      \left[
        \hat V_{n,d}(k)
        -\hat\xi_{n,d}(k)
        -\mathbb E[\hat V_{n,d}(k)]
      \right]_+
      +2\max_{k\in[g]}[\operatorname{Bias}_{n,d}(k)]_+~.
    \label{eq: regret calibration reduction}
\end{align}
The first inequality follows from the identity
$|x|=-x+2[x]_+$ and the inequality $-|y|\leq y$. The final inequality follows
from the definition of $\hat k$. The proof of
\eqref{eq: relative bias regret bound} is complete by combining
\eqref{eq: positive bias approximation bound} with Parts (i) and (ii) of
\cref{lem: fixed multiplier calibration} in
\eqref{eq: regret calibration reduction}.

Finally, we obtain the bound \eqref{eq: relative risk regret bound}. This bound follows from the application of the following Lemma. 
\begin{lem}
\label{lem: risk selection bound}
If the conditions of \cref{thm: relative bias regret} hold, then
\begin{align}
    \frac{
        \sqrt{
            \Var\left(
                \hat V_{n,d}
                (\underline\delta_{n,d}(\eta_\ell))
            \right)
        }
    }{V_n}
    \bigg|_{\ell=\hat k}
    &=
    O_p\left(
        V_n^{\eta_{k^\star}/2}
        +\mathcal R_{n,d}(s)
    \right)~.
    \label{eq: risk regret selected standard deviation}
\end{align}
\end{lem}
\noindent In particular, observe that the bias-variance decomposition gives
\begin{align}
    R_{n,d}(\underline\delta_{n,d}(\eta_k))
    &\geq
    \frac{|\operatorname{Bias}_{n,d}(k)|}{V_n}
    \geq
    \frac{|\operatorname{Bias}_{n,d}(k^\star)|}{V_n}
\end{align}
for every $k\in[g]$. Moreover, it holds that
\begin{align}
    R_{n,d}(\underline\delta_{n,d}(\eta_{\hat k}))
    &\leq
    \frac{|\operatorname{Bias}_{n,d}(\hat k)|}{V_n}
    +
    \frac{
        \sqrt{
            \Var\left(
                \hat V_{n,d}
                (\underline\delta_{n,d}(\eta_\ell))
            \right)
        }
    }{V_n}
    \bigg|_{\ell=\hat k}~.
\end{align}
Consequently, we can conclude that
\begin{align}
    &R_{n,d}(\underline\delta_{n,d}(\eta_{\hat k}))
      -\min_{k\in[g]}
       R_{n,d}(\underline\delta_{n,d}(\eta_k))
    \nonumber\\
    &\leq
      \frac{
        |\operatorname{Bias}_{n,d}(\hat k)|
        -|\operatorname{Bias}_{n,d}(k^\star)|
      }{V_n}
      +
      \frac{
        \sqrt{
          \Var\left(
            \hat V_{n,d}
            (\underline\delta_{n,d}(\eta_\ell))
          \right)
        }
      }{V_n}
      \Big|_{\ell=\hat k}
    =
      O_p\left(
        V_n^{\eta_{k^\star}/2}
        +\mathcal R_{n,d}(s)
      \right)~,
    \label{eq: risk regret final assembly}
\end{align}
by \eqref{eq: relative bias regret bound} and \cref{lem: risk selection bound}, as required.\hfill\qed

\subsection{Proof of \cref{thm: threshold choice}}

Throughout the proof of this Theorem, we use notation and apply several results developed in the proofs supporting \cref{thm: risk approximation}. Consider the infeasible quantities
\begin{align}
    \varepsilon_{0,i}^{(j)}
    &=
    Y_i^{(j)}-\tau^{(j)}
    -\gamma_{0,j}\left(Y_i^{(0)}-\tau^{(0)}\right)
    \quad\text{and}\quad
    \check\lambda_{0,i,i'}
    =
    \frac{1}{d}\sum_{j=1}^d
    \varepsilon_{0,i}^{(j)}
    \varepsilon_{0,i'}^{(j)}
    \label{eq: threshold choice oracle statistic}
\end{align}
respectively. For every pair satisfying $\lambda_{i,i'}=0$,
\eqref{eq: population residual covariance and independence} and
\cref{lem: rotation and regression} give
\begin{align}
    \check\lambda_{0,i,i'}
    &\overset{d}{=}L_{i,i'}(\Lambda_n)~,
    &m_{i,i'}(\Lambda_n)
    &=0~,\quad\text{and}
    &v_{i,i'}(\Lambda_n)
    &=\frac{\|\Omega_d\|_{\mathrm F}^2}{d^2}~,
\end{align}
where $L_{i,i'}(\Lambda_n)$ is the Gaussian product defined in \eqref{eq: gaussian product approximation representation}.  Moreover, \cref{lem: auxiliary outcome covariance bounds}, display \eqref{eq: auxiliary outcome weight and scale bounds}, gives $v_{i,i'}(\Lambda_n)/v_0^{(d)}=1+O(d^{-1})$. Therefore, \cref{lem: gaussian product normal approximation}, display \eqref{eq: gaussian product approximation cdf}, and \eqref{eq: gaussian location scale comparison} imply that
\begin{align}
    \sup_{\substack{i,i'\in[n]\\\lambda_{i,i'}=0}}
    \sup_{x\in\mathbb R}
    \left|
        P\left\{
            \check\lambda_{0,i,i'}
            \leq x\sqrt{v_0^{(d)}}
        \right\}
        -\Phi(x)
    \right|
    &\lesssim\frac{1}{\sqrt d}
    \label{eq: threshold choice oracle null cdf}
\end{align}
uniformly over null pairs.

By \cref{assu: sparsity,assu: homoskedasticity}, the number of pairs satisfying $i=i'$ or $\lambda_{i,i'}\neq0$ is at most $n(s+1)$. Hence, \eqref{eq: threshold choice oracle null cdf} gives
\begin{align}
    &\left|
        \mathbb E\left[
            \frac{1}{n^2}\sum_{i,i'=1}^n
            \mathbb I\left\{
                \check\lambda_{0,i,i'}
                \leq x\sqrt{v_0^{(d)}}
            \right\}
        \right]
        -\Phi(x)
    \right|
    \lesssim
    \frac{s+1}{n}
    +\frac{1}{\sqrt d}
    \to0
    \label{eq: threshold choice oracle cdf mean}
\end{align}
for every fixed $x\in\mathbb R$. By \eqref{eq: population residual covariance and independence}, the indicators associated with the pairs $(i,i')$ and $(k,k')$ are independent unless at least one of $\lambda_{i,k}$, $\lambda_{i,k'}$, $\lambda_{i',k}$, or $\lambda_{i',k'}$ is nonzero. As there at most $4n^3(s+1)$ such
quadruples, we have that
\begin{align}
    &\Var\left(
      \frac{1}{n^2}\sum_{i,i'=1}^n
      \mathbb I\left\{
        \check\lambda_{0,i,i'}\leq x\sqrt{v_0^{(d)}}
      \right\}
    \right)
    \leq\frac{4(s+1)}{n}
    \to0~.
    \label{eq: threshold choice oracle cdf variance}
\end{align}
Thus, we find that
\begin{align}
    \frac{1}{n^2}\sum_{i,i'=1}^n
    \mathbb I\left\{
        \check\lambda_{0,i,i'}
        \leq x\sqrt{v_0^{(d)}}
    \right\}
    &\overset{p}{\to}\Phi(x)
    \label{eq: threshold choice oracle empirical cdf}
\end{align}
for every fixed $x\in\mathbb R$.

We now replace the infeasible statistics by the feasible statistics. By \cref{lem: basic inequalities n}, display \eqref{eq: var bounds}, and $s/n\to0$, the conditions of \cref{rem: positivity} apply for all sufficiently large $n$, and therefore $\|\hat\varepsilon^{(0)}\|_2>0$ almost surely. The definitions of the centered residuals and the least-squares coefficient give
\begin{align}
    \hat\varepsilon^{(j)}
    &=
    \gamma_{0,j}\hat\varepsilon^{(0)}
    +H_n\varepsilon_0^{(j)}
    \quad\text{and}\quad
    \hat\psi^{(j)}
    =
    \gamma_{0,j}
    +
    \frac{
        (\hat\varepsilon^{(0)})^\top
        H_n\varepsilon_0^{(j)}
    }{
        \|\hat\varepsilon^{(0)}\|_2^2
    }
    \label{eq: threshold choice exact residualization coefficients}
\end{align}
for every $j\in[d]$. Since $H_n\hat\varepsilon^{(0)}=\hat\varepsilon^{(0)}$, substituting \eqref{eq: threshold choice exact residualization coefficients} into the expression for $\hat\varepsilon^{(j;0)}$ gives
\begin{align}
    \hat\varepsilon^{(j;0)}
    &=
    \left(
        H_n-
        \frac{
            \hat\varepsilon^{(0)}
            (\hat\varepsilon^{(0)})^\top
        }{
            \|\hat\varepsilon^{(0)}\|_2^2
        }
    \right)
    \varepsilon_0^{(j)}
    =H_{n,0}\varepsilon_0^{(j)}
    \label{eq: threshold choice exact residualization}
\end{align}
for every $j\in[d]$. Consequently, \eqref{eq: threshold choice oracle statistic} and \eqref{eq: threshold choice exact residualization} imply that
\begin{align}
    \left(
        \hat\lambda_{i,i'}
    \right)_{i,i'=1}^n
    &=
    H_{n,0}
    \left(
        \check\lambda_{0,i,i'}
    \right)_{i,i'=1}^n
    H_{n,0}~,
    \label{eq: threshold choice exact matrix residualization}
\end{align}
almost surely. In turn, the identity $\bm 1^\top\hat\varepsilon^{(0)}=0$ implies that
\begin{align}
    I_n-H_{n,0}
    &=
    \frac{1}{n}\bm 1\bm 1^\top
    +
    \frac{
        \hat\varepsilon^{(0)}
        (\hat\varepsilon^{(0)})^\top
    }{
        \|\hat\varepsilon^{(0)}\|_2^2
    }
    \label{eq: threshold choice residualization complement}
\end{align}
is an orthogonal projection of rank two. For any symmetric matrix $A$, expansion around this projection and contraction of the Frobenius norm give
\begin{align}
    H_{n,0}AH_{n,0}-A
    &=
    -(I_n-H_{n,0})A
    -A(I_n-H_{n,0})
    +(I_n-H_{n,0})A(I_n-H_{n,0})
    \quad\text{and}\quad
    \nonumber\\
    \|H_{n,0}AH_{n,0}-A\|_{\mathrm F}
    &\leq
    3\|(I_n-H_{n,0})A\|_{\mathrm F}~.
    \label{eq: threshold choice projection frobenius bound}
\end{align}
The independence in \eqref{eq: population residual covariance and independence}, Part (i) of \cref{lem: rotation and regression}, and \cref{lem: wick} thereby give
\begin{align}
    \mathbb E\left[
        \left(
            \left(
                \check\lambda_{0,i,i'}
            \right)_{i,i'=1}^n
        \right)^2
        \,\middle|\,
        \hat\varepsilon^{(0)}
    \right]
    &=
    \frac{1}{d^2}
    \sum_{j=1}^d
    \theta_{d,j}^2
    \left(
        \operatorname{tr}(\Lambda_n)\Lambda_n
        +2\Lambda_n^2
    \right)
    +
    \frac{1}{d^2}
    \sum_{\substack{j,j'=1\\j\neq j'}}^d
    \theta_{d,j}\theta_{d,j'}\Lambda_n^2
    \nonumber\\
    &\quad=
    \frac{
        \|\Omega_d\|_{\mathrm F}^2
        \operatorname{tr}(\Lambda_n)
    }{
        d^2
    }\Lambda_n
    +
    \frac{
        \operatorname{tr}(\Omega_d)^2
        +\|\Omega_d\|_{\mathrm F}^2
    }{
        d^2
    }\Lambda_n^2
    \label{eq: threshold choice oracle matrix second moment}
\end{align}
almost surely.
Since $I_n-H_{n,0}$ is a rank-two projection, positive
semi-definiteness of $\Lambda_n$ gives
\begin{align}
    \operatorname{tr}\left(
        (I_n-H_{n,0})\Lambda_n
    \right)
    &\leq2\|\Lambda_n\|_{\mathrm{op}}
    \quad\text{and}\quad
    \operatorname{tr}\left(
        (I_n-H_{n,0})\Lambda_n^2
    \right)
    \leq2\|\Lambda_n\|_{\mathrm{op}}^2~.
    \label{eq: threshold choice projected covariance traces}
\end{align}
Moreover, \cref{lem: basic inequalities n}, display \eqref{eq: n bound 1}, and symmetry of $\Lambda_n$ imply that $\|\Lambda_n\|_{\mathrm{op}}\lesssim nV_n$, while \cref{assu: homoskedasticity} implies that $\operatorname{tr}(\Lambda_n)=n$. Thus, \cref{lem: auxiliary outcome covariance bounds}, display \eqref{eq: auxiliary outcome covariance bounds}, \eqref{eq: threshold choice exact matrix residualization}, \eqref{eq: threshold choice projection frobenius bound}, \eqref{eq: threshold choice oracle matrix second moment}, and \eqref{eq: threshold choice projected covariance traces} imply that
\begin{align}
    &\mathbb E\left[
        \frac{1}{n^2}
        \sum_{i,i'=1}^n
        \left(
            \hat\lambda_{i,i'}
            -\check\lambda_{0,i,i'}
        \right)^2
    \right]
    \nonumber\\
    &\quad=
    \frac{1}{n^2}
    \mathbb E\left[
        \left\|
            H_{n,0}
            \left(
                \check\lambda_{0,i,i'}
            \right)_{i,i'=1}^n
            H_{n,0}
            -
            \left(
                \check\lambda_{0,i,i'}
            \right)_{i,i'=1}^n
        \right\|_{\mathrm F}^2
    \right]
    \nonumber\\
    &\quad\leq
    \frac{9}{n^2}
    \mathbb E\left[
        \operatorname{tr}\left(
            (I_n-H_{n,0})
            \left(
                \left(
                    \check\lambda_{0,i,i'}
                \right)_{i,i'=1}^n
            \right)^2
        \right)
    \right]
    \lesssim
    \frac{1}{n^2}
    \left(
        \frac{dn}{d^2}\,nV_n
        +n^2V_n^2
    \right)
    \lesssim
    \frac{V_n}{d}+V_n^2
    \label{eq: threshold choice feasible oracle mean square}
\end{align}
for all sufficiently large $n,d$. \cref{lem: auxiliary outcome covariance bounds}, display \eqref{eq: auxiliary outcome covariance bounds}, implies that $v_0^{(d)}\asymp d^{-1}$. Hence, \eqref{eq: threshold choice feasible oracle mean square} gives
\begin{align}
    \mathbb E\left[
        \frac{1}{n^2}
        \sum_{i,i'=1}^n
        \frac{
            \left(
                \hat\lambda_{i,i'}
                -\check\lambda_{0,i,i'}
            \right)^2
        }{
            v_0^{(d)}
        }
    \right]
    &\lesssim
    V_n+dV_n^2
    \to0~,
    \label{eq: threshold choice normalized feasible oracle mean square}
\end{align}
where the convergence follows from
\cref{lem: basic inequalities n}, display \eqref{eq: var bounds},
$\log(V_n^{-1})\to\infty$, and
$dV_n^2\log(V_n^{-1})\to0$.

With this in place, fix $x\in\mathbb R$ and $\epsilon>0$. The bound
\begin{align}
    &\left|
        \mathbb I\left\{
            \hat\lambda_{i,i'}
            \leq x\sqrt{v_0^{(d)}}
        \right\}
        -
        \mathbb I\left\{
            \check\lambda_{0,i,i'}
            \leq x\sqrt{v_0^{(d)}}
        \right\}
    \right|
    \nonumber\\
    &\quad\leq
    \mathbb I\left\{
        \left|
            \check\lambda_{0,i,i'}
            -x\sqrt{v_0^{(d)}}
        \right|
        \leq\epsilon\sqrt{v_0^{(d)}}
    \right\}
    +
    \frac{
        \left(
            \hat\lambda_{i,i'}
            -\check\lambda_{0,i,i'}
        \right)^2
    }{
        \epsilon^2v_0^{(d)}
    }
    \label{eq: threshold choice indicator comparison}
\end{align}
holds for every pair $i,i'$. The first term on the right-hand side satisfies
\begin{align}
    &\mathbb I\left\{
        \left|
            \check\lambda_{0,i,i'}
            -x\sqrt{v_0^{(d)}}
        \right|
        \leq\epsilon\sqrt{v_0^{(d)}}
    \right\}
    \nonumber\\
    &\quad\leq
    \mathbb I\left\{
        \check\lambda_{0,i,i'}
        \leq(x+\epsilon)\sqrt{v_0^{(d)}}
    \right\}
    -
    \mathbb I\left\{
        \check\lambda_{0,i,i'}
        \leq(x-2\epsilon)\sqrt{v_0^{(d)}}
    \right\}
    \label{eq: threshold choice oracle small interval}
\end{align}
for every pair $i,i'$. Taking expectations in \eqref{eq: threshold choice indicator comparison}, summing over the pairs, and applying \eqref{eq: threshold choice oracle cdf mean}, \eqref{eq: threshold choice normalized feasible oracle mean square}, and \eqref{eq: threshold choice oracle small interval}, we obtain
\begin{align}
    &\limsup_{n,d\to\infty}
    \mathbb E\Bigg[
        \frac{1}{n^2}\sum_{i,i'=1}^n
        \left|
            \mathbb I\left\{
                \hat\lambda_{i,i'}
                \leq x\sqrt{v_0^{(d)}}
            \right\}
            -
            \mathbb I\left\{
                \check\lambda_{0,i,i'}
                \leq x\sqrt{v_0^{(d)}}
            \right\}
        \right|
    \Bigg]
    \nonumber\\
    &\quad\leq
    \Phi(x+\epsilon)-\Phi(x-2\epsilon)
    \label{eq: threshold choice empirical cdf comparison}
\end{align}
for every fixed $\epsilon>0$.
Letting $\epsilon\downarrow0$ in
\eqref{eq: threshold choice empirical cdf comparison} and using the
continuity of $\Phi$ shows that its left-hand side is zero. Combining
this conclusion with
\eqref{eq: threshold choice oracle empirical cdf}, we obtain
\begin{align}
    \frac{1}{n^2}\sum_{i,i'=1}^n
    \mathbb I\left\{
        \hat\lambda_{i,i'}
        \leq x\sqrt{v_0^{(d)}}
    \right\}
    &\overset{p}{\to}\Phi(x)
    \label{eq: threshold choice feasible empirical cdf}
\end{align}
for every fixed $x\in\mathbb R$.

With this in place, fix $u\in\{1/4,3/4\}$ and $\epsilon>0$. Monotonicity of the empirical distribution function gives
\begin{align}
    &P\Bigg\{
        \frac{
            \inf\left\{
                t\in\mathbb R:
                \frac{1}{n^2}\sum_{i,i'=1}^n
                \mathbb I\{\hat\lambda_{i,i'}\leq t\}
                \geq u
            \right\}
        }{
            \sqrt{v_0^{(d)}}
        }
        \notin
        \left(
            \Phi^{-1}(u)-\epsilon,
            \Phi^{-1}(u)+\epsilon
        \right]
    \Bigg\}
    \nonumber\\
    &\quad\leq
    P\left\{
        \frac{1}{n^2}\sum_{i,i'=1}^n
        \mathbb I\left\{
            \hat\lambda_{i,i'}
            \leq
            \left(
                \Phi^{-1}(u)-\epsilon
            \right)
            \sqrt{v_0^{(d)}}
        \right\}
        \geq u
    \right\}
    \nonumber\\
    &\qquad+
    P\left\{
        \frac{1}{n^2}\sum_{i,i'=1}^n
        \mathbb I\left\{
            \hat\lambda_{i,i'}
            \leq
            \left(
                \Phi^{-1}(u)+\epsilon
            \right)
            \sqrt{v_0^{(d)}}
        \right\}
        <u
    \right\}
    \to0
    \label{eq: threshold choice empirical quantiles}
\end{align}
as $n,d\to\infty$. Thus, using $\Phi^{-1}(1/4)=-\Phi^{-1}(3/4)$ and substituting the two limits in \eqref{eq: threshold choice empirical quantiles} into the expression for $\hat v_0^{(d)}$, we find that
\begin{align}
\hat v_0^{(d)}/v_0^{(d)}
    &\overset{p}{\to}1
    \label{eq: threshold choice fitted variance consistency}
\end{align}
as $n,d\to\infty$.

Now, observe that \Cref{assu: pilot variance estimator} implies that there are constants $0<c<C<\infty$ such that
\begin{align}
    P\left\{
        c
        \leq
        \frac{\check V_n}{V_n}
        \leq C
    \right\}
    &\to1
    \label{eq: threshold choice pilot event}
\end{align}
as $n,d\to\infty$. On the event in \eqref{eq: threshold choice pilot event}, the bounds
\begin{align}
    1-\frac{\log C}{\log(V_n^{-1})}
    &\leq
    \frac{
        \log(\check V_n^{-1})
    }{
        \log(V_n^{-1})
    }
    \leq
    1
    -\frac{\log c}{\log(V_n^{-1})}
    \label{eq: threshold choice pilot logarithm bounds}
\end{align}
hold simultaneously. Since $\log(V_n^{-1})\to\infty$, combining \eqref{eq: threshold choice pilot event} and \eqref{eq: threshold choice pilot logarithm bounds} gives
\begin{align}
    \frac{
        \log(\check V_n^{-1})
    }{
        \log(V_n^{-1})
    }
    &\overset{p}{\to}1
    \label{eq: threshold choice pilot logarithm}
\end{align}
as $n,d\to\infty$. Part (ii) of \cref{lem: selected set characterization}, gives
\begin{align}
    \frac{
        \sqrt{
            2\eta v_0^{(d)}
            \log(V_n^{-1})
        }
    }{
        \underline\delta_{n,d}\left(\eta\right)
    }
    &\to1
    \label{eq: threshold choice population endpoint}
\end{align}
as $n,d\to\infty$. Hence, by combining \eqref{eq: threshold choice fitted variance consistency}, \eqref{eq: threshold choice pilot logarithm}, and \eqref{eq: threshold choice population endpoint}, we obtain
\begin{align}
    &\left|
        \frac{
            \hat{\underline\delta}_{n,d}(\eta)
        }{
            \underline\delta_{n,d}\left(\eta\right)
        }
        -1
    \right|
    =
    \left|
        \left(
            \frac{\hat v_0^{(d)}}{v_0^{(d)}}
        \right)^{1/2}
        \left(
            \frac{
                \log(\check V_n^{-1})
            }{
                \log(V_n^{-1})
            }
        \right)^{1/2}
        \frac{
            \sqrt{
                2\eta v_0^{(d)}
                \log(V_n^{-1})
            }
        }{
            \underline\delta_{n,d}\left(\eta\right)
        }
        -1
    \right|
    \overset{p}{\to}0
\end{align}
as $n,d\to\infty$.\hfill\qed

\end{spacing}
\newpage

\begin{center}
\thispagestyle{empty}
\large{\it Additional Material Supporting:}
\vskip0.2cm
\begin{spacing}{1}
\textbf{\Large Using Multiple Outcomes to \\\vspace{0.5em} Adjust Standard Errors for Spatial Correlation\daggerfootnote{\textit{Date}: \today}} \\\vspace{1em}
\begin{tabular}[t]{c@{\extracolsep{4em}}c} 
\large{Stefano DellaVigna} &  \large{Guido Imbens}\vspace{-0.4em}\\ \vspace{-0.7em}
\small{UC Berkeley and NBER} & \small{Stanford University} \\ \vspace{-0.7em} \\
\large{Woojin Kim} &  \large{David M. Ritzwoller}\vspace{-0.4em}\\ \vspace{-0.7em}
\small{Stanford University} & \small{Harvard University} \\ \vspace{-0.4em}
\end{tabular}%
\end{spacing}
\end{center}
\begin{spacing}{1.13}
\DoToC
\end{spacing}
\thispagestyle{empty}
\setcounter{page}{0}
\setcounter{figure}{0}   
\newpage

\begin{spacing}{1.3}

\begin{landscape}
\section{Panel of County Level Outcomes\label{app: panel tables}}

\renewcommand\thetable{\thesection.\arabic{table}a}
\begin{table}[htbp!] \centering \renewcommand{\arraystretch}{1.15} \begin{threeparttable} \newcolumntype{L}{>{\raggedright\arraybackslash\footnotesize}X} \newcolumntype{C}{>{\centering\arraybackslash}X} \newcolumntype{S}{>{\centering\arraybackslash}X} \caption{County outcomes used in simulation} \begin{tabularx}{\linewidth}{L l c c} \toprule Description & Source & First year & Last year \\ \hline 
1. Annual mean of Particulate Matter (PM2.5) concentration (ug/m3) & AHRQ & 2009 & 2018 \\
2. Average household size & AHRQ & 2009 & 2018 \\
3. Median distance in miles to the nearest emergency department, calculated using population weighted tract centroids in the county & AHRQ & 2009 & 2018 \\
4. Median distance in miles to the nearest obstetrics department, calculated using population weighted tract centroids in the county & AHRQ & 2009 & 2018 \\
5. Median distance in miles to the nearest pediatric ICU, calculated using population weighted tract centroids in the county & AHRQ & 2009 & 2018 \\
6. Median selected monthly owner costs for houses with a mortgage (dollars) & AHRQ & 2009 & 2018 \\
7. Percentage of families with children that are single-parent families & AHRQ & 2009 & 2018 \\
8. Percentage of households with same-sex unmarried partner & AHRQ & 2009 & 2018 \\
9. Percentage of limited English speaking households & AHRQ & 2009 & 2018 \\
10. Percentage of occupied housing units with utility gas heating & AHRQ & 2009 & 2018 \\
11. Percentage of population reporting American Indian and Alaska Native race alone & AHRQ & 2009 & 2018 \\
12. Percentage of population reporting multiple races & AHRQ & 2009 & 2018 \\
13. Percentage of population that speaks Spanish (ages 5 and over) & AHRQ & 2009 & 2018 \\
14. Total cardiovascular disease deaths per 100,000 population (ages 35 and over) & AHRQ & 2009 & 2018 \\
15. Total number of days with daily maximum heat index, absolute threshold: 90F & AHRQ & 2009 & 2018 \\
16. Total number of hospital beds per 1,000 population & AHRQ & 2009 & 2018 \\
17. Total number of hospitals per 1,000 population & AHRQ & 2009 & 2018 \\
18. All persons 18 to 64 years without health insurance, percent & U.S. Census Bureau & 2005 & 2007 \\
19. All persons under 18 years without health insurance, percent & U.S. Census Bureau & 2005 & 2007 \\
20. Average travel time to work for workers 16 years and over who did not work at home & U.S. Census Bureau & 1990 & 2009 \\
\bottomrule \end{tabularx} \label{tab:simoutcomes1} \begin{tablenotes} \footnotesize \item \textit{Sources:} AHRQ: \href{www.ahrq.gov/sdoh/data-analytics/sdoh-data.html#download}{Agency for Healthcare Research and Quality}, U.S. Census Bureau: \href{www.census.gov/library/publications/2011/compendia/usa-counties-2011.html}{USA Counties}. USA Counties variables based on complete counts are indicated in the description. \end{tablenotes} \end{threeparttable} \end{table}

\addtocounter{table}{-1}
\newpage
\renewcommand\thetable{\thesection.\arabic{table}b}
\begin{table}[htbp!] \centering \renewcommand{\arraystretch}{1.15} \begin{threeparttable} \newcolumntype{L}{>{\raggedright\arraybackslash\footnotesize}X} \newcolumntype{C}{>{\centering\arraybackslash}X} \newcolumntype{S}{>{\centering\arraybackslash}X} \caption{County outcomes used in simulation} \begin{tabularx}{\linewidth}{L l c c} \toprule Description & Source & First year & Last year \\ \hline 
21. Births per 1,000 population & U.S. Census Bureau & 1970 & 2007 \\
22. Black population (complete count) & U.S. Census Bureau & 1990 & 2010 \\
23. Civilian labor force & U.S. Census Bureau & 1990 & 2010 \\
24. Civilian labor force unemployment rate & U.S. Census Bureau & 1990 & 2010 \\
25. Deaths per 1,000 population & U.S. Census Bureau & 1970 & 2007 \\
26. Educational attainment - persons 25 years and over - percent bachelor's degree or higher & U.S. Census Bureau & 1980 & 2009 \\
27. Educational attainment - persons 25 years and over - percent high school graduate or higher & U.S. Census Bureau & 1980 & 2009 \\
28. Employment in government - state and local (NAICS) & U.S. Census Bureau & 2001 & 2007 \\
29. Infant deaths per 1,000 live births & U.S. Census Bureau & 1990 & 2007 \\
30. Median contract rent of specified renter-occupied housing units paying cash rent (complete count) & U.S. Census Bureau & 1980 & 2009 \\
31. Median household income & U.S. Census Bureau & 1995 & 2009 \\
32. Median selected monthly owner costs of specified owner-occupied housing units with a mortgage & U.S. Census Bureau & 1980 & 2000 \\
33. Median value of specified owner-occupied housing units (complete count) & U.S. Census Bureau & 1980 & 2009 \\
34. New private housing units authorized by building permits - total & U.S. Census Bureau & 2004 & 2010 \\
35. People of all ages in poverty - percent & U.S. Census Bureau & 1995 & 2009 \\
36. Per capita personal income & U.S. Census Bureau & 1969 & 2000 \\
37. Persons 16 to 19 years not enrolled in school and not a high school graduate & U.S. Census Bureau & 1990 & 2000 \\
38. Population per square mile & U.S. Census Bureau & 1980 & 2010 \\
39. Private nonfarm establishments & U.S. Census Bureau & 1990 & 2009 \\
40. Private nonfarm establishments - arts, entertainment and recreation (NAICS 71) & U.S. Census Bureau & 2002 & 2009 \\
\bottomrule \end{tabularx} \label{tab:simoutcomes2} \begin{tablenotes} \footnotesize \item \textit{Sources:} AHRQ: \href{www.ahrq.gov/sdoh/data-analytics/sdoh-data.html#download}{Agency for Healthcare Research and Quality}, U.S. Census Bureau: \href{www.census.gov/library/publications/2011/compendia/usa-counties-2011.html}{USA Counties}. USA Counties variables based on complete counts are indicated in the description. \end{tablenotes} \end{threeparttable} \end{table}

\addtocounter{table}{-1}
\newpage
\renewcommand\thetable{\thesection.\arabic{table}c}
\begin{table}[htbp!] \centering \renewcommand{\arraystretch}{1.15} \begin{threeparttable} \newcolumntype{L}{>{\raggedright\arraybackslash\footnotesize}X} \newcolumntype{C}{>{\centering\arraybackslash}X} \newcolumntype{S}{>{\centering\arraybackslash}X} \caption{County outcomes used in simulation} \begin{tabularx}{\linewidth}{L l c c} \toprule Description & Source & First year & Last year \\ \hline 
41. Related children age 5 to 17 in families in poverty - percent & U.S. Census Bureau & 1995 & 2009 \\
42. Renter-occupied housing units (complete count) & U.S. Census Bureau & 1990 & 2010 \\
43. Resident population under 18 years, percent & U.S. Census Bureau & 2000 & 2009 \\
44. Resident population: Median age (complete count) & U.S. Census Bureau & 1980 & 2010 \\
45. Resident population: total females, percent & U.S. Census Bureau & 2000 & 2009 \\
46. Total physicians & U.S. Census Bureau & 2004 & 2009 \\
47. Valuation of new private housing units authorized by building permits & U.S. Census Bureau & 2004 & 2010 \\
48. Vehicles available per occupied housing unit & U.S. Census Bureau & 1990 & 2000 \\
49. Vote cast for president- percent Republican & U.S. Census Bureau & 1980 & 2008 \\
50. Water use: per capita use & U.S. Census Bureau & 1990 & 2005 \\
\bottomrule \end{tabularx} \label{tab:simoutcomes3} \begin{tablenotes} \footnotesize \item \textit{Sources:} AHRQ: \href{www.ahrq.gov/sdoh/data-analytics/sdoh-data.html#download}{Agency for Healthcare Research and Quality}, U.S. Census Bureau: \href{www.census.gov/library/publications/2011/compendia/usa-counties-2011.html}{USA Counties}. USA Counties variables based on complete counts are indicated in the description. \end{tablenotes} \end{threeparttable} \end{table}

\end{landscape}
\clearpage
\renewcommand\thetable{\thesection.\arabic{table}}

\section{Auxiliary Lemmas\label{app: auxiliary lemmas}}

We begin by fixing some notation that will be used repeatedly over the course of the ensuing proofs. Throughout, denote the centering matrix and the covariance matrix of the centered residuals by
\begin{align}
    H_n
    &=I_n-\frac{1}{n}\bm{1}\bm{1}^{\top}
    \quad\text{and}\quad
    M_n
    =H_n\Lambda_nH_n = (\mu_{i,i'})_{i,i'=1}^n~,
    \label{eq: exact screening covariance notation}
\end{align}
respectively.  Write $\bar{\Gamma}_d=(\gamma_{j,j'})_{j,j'=1}^d$ and $\bar{\Gamma}_{0,d} = (\gamma_{0,j})_{j=1}^d$ and define
\begin{align}
    \Omega_d
    &=\bar\Gamma_d-
    \bar\Gamma_{0,d}^{\top}\bar\Gamma_{0,d} = (\omega_{j,j'})_{j,j'=1}^d
    \label{eq: exact screening outcome covariance}
\end{align}
and let $\theta_{d,1},\ldots,\theta_{d,d}$ denote its eigenvalues. Define the set
\begin{align}
    \mathcal S_n
    &=\left\{
        (i,i')\in[n]^2:i=i'\ \text{or}\ \lambda_{i,i'}\neq0
    \right\}
\end{align}
of non-null unit pairs.

For any positive semi-definite matrix $\Xi$, define the quantities
\begin{align}
    \psi_{i,i'}(\Xi)
    &=\left(\Xi_{i,i}\Xi_{i',i'}\right)^{1/2}
    \quad\text{and}\quad
    \rho_{i,i'}(\Xi)
    =\frac{\Xi_{i,i'}}{\psi_{i,i'}(\Xi)}~,
    \label{eq: exact screening pair parameters}
\end{align}
as well as
\begin{align}
    m_{i,i'}(\Xi)
    &=\frac{\operatorname{tr}(\Omega_d)}{d}\Xi_{i,i'}
    \quad\text{and}\quad
    v_{i,i'}(\Xi)
    =\frac{\|\Omega_d\|_{\mathrm F}^2}{d^2}
    \left(
        \Xi_{i,i}\Xi_{i',i'}+\Xi_{i,i'}^2
    \right)~,
    \label{eq: feasible normal parameters}
\end{align}
where we adopt the convention that $\rho_{i,i'}(\Xi)=0$ if $\psi_{i,i'}(\Xi)=0$. We also adopt the convention that $F^+(\delta,m_{i,i'}(\Xi),0) =\mathbb{I}\{|m_{i,i'}(\Xi)|\geq\delta\}$ when $v_{i,i'}(\Xi)=0$.

For $a\geq0$, $\rho\in[-1,1]$, and $\delta\geq0$, define the function
\begin{align}
    H_{\delta,d}(a,\rho)
    &=P\left\{
        \left|
            \frac{a}{d}\sum_{j=1}^d
            \theta_{d,j}Z_j
            \left(\rho Z_j+\sqrt{1-\rho^2}W_j\right)
        \right|\geq\delta
    \right\},
    \label{eq: feasible screening tail}
\end{align}
where $Z_1,\ldots,Z_d,W_1,\ldots,W_d$ are independent standard Gaussian
variables. Finally, define the sequence
\begin{align}
    \beta_{\delta,d}
    &=\left.
    \frac{\mathrm{d}}{\mathrm{d} a}H_{\delta,d}(a,\rho)
    \right|_{(a,\rho)=(1,0)}
\end{align}
for each $d \geq 1$.

\subsection{Auxiliary Lemmas\label{sec: aux B}}

We state several basic Lemmas that will be used repeatedly throughout the proofs. 

\begin{lem}
\label{lem: basic inequalities n}$ $\\
Suppose that \cref{assu: proportionality,assu: sparsity,assu: homoskedasticity,assu: well conditioned} hold. Then
\begin{align}
    \operatorname{rank}(M_n)
    &=n-1
    \quad\text{and}\quad
    \min_{i\in[n]}\mu_{i,i}
    \gtrsim1~.
    \label{eq: n rank and diagonal}
\end{align}
Moreover, there exists a constant $c^\star < 1/4$ such that, if $s/n \leq c^\star$, then the bounds
\begin{align}
        \frac{1}{n}& \lesssim V_n\lesssim \frac{s+1}{n} \leq 2 c^{\star}\label{eq: var bounds} 
\end{align}
and
\begin{align}
    \max_{i \in [n]} \sum_{i'=1}^n \vert \lambda_{i,i'}\vert
    & \lesssim nV_n~,
    & \sum_{i,i'=1}^n \vert \lambda_{i,i'}\vert
    & \lesssim n^2V_n~,\label{eq: n bound 1}\\
    \mu_{i,i} &\asymp 1~,
    & \sup_{i,i'} \vert \mu_{i,i'}\vert 
    &\lesssim 1~,\label{eq: n bound 2}\\
    \max_{i,i'\in[n]} \vert \mu_{i,i'} - \lambda_{i,i'}\vert 
    & \lesssim V_n~,
    & \sum_{i,i'=1}^n \vert \mu_{i,i'}\vert
    & \lesssim n^2V_n~,\label{eq: n bound 3}\\
    \| M_n \|_{\mathrm op}
    & \lesssim n V_n~,
    & \operatorname{tr}(M_n) 
    & = n(1 - V_n) \asymp n\label{eq: n bound 4}
\end{align}
hold.
\end{lem}

\begin{lem}
\label{lem: auxiliary outcome covariance bounds}$ $\\
Suppose that \cref{assu: proportionality,assu: homoskedasticity,assu: well conditioned} hold and $d\geq2$. It holds that
\begin{align}
    \| \bar{\Gamma}_{0,d}\|_2^2 \lesssim 1~.
\end{align}
Moreover, the matrix $\Omega_d$ is uniformly well-conditioned, in the sense that
\begin{align}
    u^\top\Omega_du
    &\asymp\|u\|_2^2
    \label{eq: auxiliary outcome uniform conditioning}
\end{align}
for every $u\in\mathbb R^d$, and satisfies
\begin{align}
    \|\Omega_d\|_{\mathrm{op}}
    &\lesssim1~,
    &\|\Omega_d\|_{\mathrm F}^2
    &\asymp d~,
    &\operatorname{tr}(\Omega_d) 
    & = d + O(1)~,
    &\text{and}\quad
    \|\Omega_d\|_{\mathrm F}^2
    =\|\bar\Gamma_d\|_{\mathrm F}^2+O(1)~.
    \label{eq: auxiliary outcome covariance bounds}
\end{align}
as well as
\begin{align}
    \frac{\max_{j\in[d]}\theta_{d,j}}
    {(\sum_{j=1}^d\theta_{d,j}^2)^{1/2}}
    &\lesssim\frac{1}{\sqrt d}
    \quad\text{and}\quad
    \frac{\|\Omega_d\|_{\mathrm F}}{d}
    =\sqrt{v_{0}^{(d)}}\left(1+O(d^{-1})\right)~.
    \label{eq: auxiliary outcome weight and scale bounds}
\end{align}
\end{lem}

In turn, at various points throughout the proofs, we make use of the following facts concerning collections of Gaussian random variables. The first result follows from repeated application of well-known facts concerning Gaussian ensembles. See, for reference, \citet{dawid1981some} and \citet[Theorem~1.2.11]{muirhead1982}.

\begin{lem}
\label{lem: rotation and regression}$ $\\
Let $X=(X^{(j)})_{j=0}^m$ be a centered Gaussian random matrix that satisfies
the condition
\begin{align}
    \Cov(X_i^{(j)},X_{i'}^{(j')})
    &=A_{i,i'}B_{j,j'}
\end{align}
for positive semi-definite matrices $A$ and $B$. Define the sub-matrices $\bar B=(B_{j,j'})_{j,j'=1}^m$, $\bar B_0=(B_{0,j})_{j=1}^m$, and $\bar X=(X^{(j)})_{j=1}^m$, respectively. 

\noindent\textbf{(i)} Suppose that the matrix $B$ admits the decomposition
\begin{align}
    B
    &=U\mathsf{diag}(\theta_0,\ldots,\theta_m)U^\top
\end{align}
for an orthogonal matrix $U$. There exist independent Gaussian vectors
$G^{(0)},\ldots,G^{(m)}$, each distributed as $\mathsf{N}(0,A)$, such that the
distributional equality
\begin{align}
    XU
    &\overset{d}{=}
    (\sqrt{\theta_0}G^{(0)},\ldots,\sqrt{\theta_m}G^{(m)})
\end{align}
holds.

\noindent\textbf{(ii)} If $B_{0,0} = 1$ and 
\begin{align}
    \widetilde X_0 =\bar X-X^{(0)}\bar B_0~,
\end{align}
then it holds that
\begin{align}
    \widetilde X_0
    &\ci X^{(0)}
    \quad\text{and}\quad
    \Cov\left(
        \widetilde X_{0,i}^{(j)},
        \widetilde X_{0,i'}^{(j')}
    \right)
    =A_{i,i'}
    \left(
        \bar B-\bar B_0^{\top}\bar B_0
    \right)_{j,j'}~.
\end{align}

\noindent\textbf{(iii)} If $B_{0,0} = 1$, then the representation
\begin{align}
    X^{(0)}
    &=\bar X\bar B^\dagger\bar B_0^{\top}+\xi
\end{align}
holds, where the centered Gaussian vector $\xi$ satisfies
\begin{align}
    \xi
    &\ci\bar X
    \quad\text{and}\quad
    \Cov(\xi)
    =\left(
        1-\bar B_0\bar B^\dagger\bar B_0^{\top}
    \right)A~.
\end{align}
\end{lem}

\noindent The second result is often referred to as ``Wick's formula.'' See, e.g., \cite{isserlis1918formula}.

\begin{lem}
\label{lem: wick}$ $\\
If $(W_1,\ldots,W_{2m})$ is a centered Gaussian vector, then the equality
\begin{align}
    \mathbb{E}\left[\prod_{j=1}^{2m}W_j\right]
    &=\sum_{\pi\in\Pi_{2m}}
    \prod_{(k,l)\in\pi}\Cov(W_k,W_l)
\end{align}
holds, where $\Pi_{2m}$ denotes the set of pairings of $[2m]$, i.e., partitions of $[2m]$ into sets of cardinality $2$.
\end{lem}

\begin{remark}\label{rem: positivity}
Observe that, if $n \geq 2$, the vector $\hat\varepsilon^{(0)}$ is distributed as $\mathsf{N}(0,M_n)$ by \cref{assu: gaussianity}. Let $a_{n,1},\ldots,a_{n,n}$ denote the eigenvalues of $M_n$ and let $Z$ be distributed as $\mathsf{N}(0,I_n)$. Since $M_n$ is positive semi-definite, $M_n^{1/2}Z$ is centered Gaussian with covariance $M_n$. Thus, by applying a standard representation for Gaussian quadratic forms (see e.g., \citealp{muirhead1982}, Proof of Theorem 1.4.2), we obtain
\begin{align}
    \|\hat\varepsilon^{(0)}\|_2^2
    &\overset{d}{=}Z^{\top}M_nZ
    \overset{d}{=}\sum_{j=1}^na_{n,j}W_j^2~,
    \label{eq: projection stability target norm representation}
\end{align}
for independent standard Gaussian variables $W_1,\ldots,W_n$. By \cref{lem: basic inequalities n}, exactly $n-1 \geq 1$ eigenvalues $a_{n,j}$ are positive. Hence, we can conclude that
\begin{align}
    \|\hat\varepsilon^{(0)}\|_2^2>0
\end{align}
holds almost surely.\hfill$\blacksquare$
\end{remark}

\subsection{Proof of \cref{lem: basic inequalities n}}

By \cref{assu: well conditioned}, for every $v\in\mathbb R^n$, it holds that
\begin{align}
    v^\top M_nv
    &=(H_nv)^\top\Lambda_n(H_nv)
    \asymp\|H_nv\|_2^2~.
\end{align}
Consequently, $\ker(M_n)=\ker(H_n)=\operatorname{span}\{\bm{1}\}$, and therefore $\operatorname{rank}(M_n)=n-1$. \cref{assu: well conditioned}, applied to $\Lambda_n$, gives
\begin{align}
    \mu_{i,i}
    &=(H_ne_i)^\top\Lambda_n(H_ne_i)
    \gtrsim\|H_ne_i\|_2^2
    =1-\frac{1}{n}
    \gtrsim1
\end{align}
for every $i\in[n]$, where the final bound follows from $1\leq s<n$. Thus, \eqref{eq: n rank and diagonal} holds.

\cref{assu: proportionality,assu: homoskedasticity} imply that
\begin{align}
    V_n
    &=
    \frac{1}{n^2}\bm{1}^{\top}\Lambda_n\bm{1}
    =
    \frac{1}{n^2}\sum_{i,i'=1}^n\lambda_{i,i'}
    \quad\text{and}\quad
    \lambda_{i,i}=1
    \label{eq: basic objective identity}
\end{align}
for every $i\in[n]$. Moreover, every two-dimensional principal submatrix of $\Lambda_n$ is positive semi-definite, so its determinant is nonnegative and
\begin{align}
    |\lambda_{i,i'}|^2
    &\leq
    \lambda_{i,i}\lambda_{i',i'}
    =1
\end{align}
for every $i,i'\in[n]$. Hence, \cref{assu: well conditioned}, applied to $\Lambda_n$, \cref{assu: sparsity}, and $s/n\leq c^\star$ give
\begin{align}
    \frac{1}{n}
    &\lesssim V_n
    \leq
    \frac{1}{n^2}\sum_{i,i'=1}^n|\lambda_{i,i'}|
    \leq
    \frac{s+1}{n}
    \leq
    \frac{2s}{n}
    \leq
    2c^\star~.
    \label{eq: basic objective bounds}
\end{align}
This establishes \eqref{eq: var bounds}.

\cref{assu: sparsity} also gives
\begin{align}
    \max_{i\in[n]}\sum_{i'=1}^n|\lambda_{i,i'}|
    &\lesssim
    nV_n~,\quad\text{and}\quad
    \sum_{i,i'=1}^n|\lambda_{i,i'}|
    &\leq
    n\max_{i\in[n]}\sum_{i'=1}^n|\lambda_{i,i'}|
    \lesssim
    n^2V_n~.
\end{align}
Thus, \eqref{eq: n bound 1} holds. Expanding the definition $M_n=H_n\Lambda_nH_n$ and applying \eqref{eq: basic objective identity}, we obtain
\begin{align}
    \mu_{i,i'}-\lambda_{i,i'}
    &=
    -\frac{1}{n}\sum_{k=1}^n\lambda_{i,k}
    -\frac{1}{n}\sum_{k=1}^n\lambda_{k,i'}
    +\frac{1}{n^2}\sum_{k,l=1}^n\lambda_{k,l}
    \nonumber\\
    &=
    -\frac{1}{n}\sum_{k=1}^n\lambda_{i,k}
    -\frac{1}{n}\sum_{k=1}^n\lambda_{k,i'}
    +V_n~.
    \label{eq: basic centering identity}
\end{align}
Since $\Lambda_n$ is symmetric, \cref{assu: sparsity} and \eqref{eq: basic centering identity} imply that
\begin{align}
    \sup_{i,i'\in[n]}
    |\mu_{i,i'}-\lambda_{i,i'}|
    &\leq
    \frac{2}{n}
    \max_{i\in[n]}\sum_{k=1}^n|\lambda_{i,k}|
    +V_n
    \lesssim
    V_n~.
    \label{eq: basic centering deviation}
\end{align}
Summing this bound and applying \eqref{eq: n bound 1}, we also obtain
\begin{align}
    \sum_{i,i'=1}^n|\mu_{i,i'}|
    &\leq
    \sum_{i,i'=1}^n|\lambda_{i,i'}|
    +n^2\sup_{i,i'\in[n]}
    |\mu_{i,i'}-\lambda_{i,i'}|
    \lesssim
    n^2V_n~.
\end{align}
Together with \eqref{eq: basic centering deviation}, this proves \eqref{eq: n bound 3}.

The bound \eqref{eq: n rank and diagonal}, \eqref{eq: basic centering deviation}, $\lambda_{i,i}=1$, and \eqref{eq: basic objective bounds} imply that $\mu_{i,i}\asymp1$ uniformly over $i\in[n]$. Moreover, \cref{assu: proportionality} and the definition $M_n=H_n\Lambda_nH_n$ imply that $M_n$ is positive semi-definite. Thus, every two-dimensional principal submatrix of $M_n$ is positive semi-definite, so its determinant is nonnegative and
\begin{align}
    |\mu_{i,i'}|^2
    &\leq
    \mu_{i,i}\mu_{i',i'}
    \lesssim1
\end{align}
for every $i,i'\in[n]$. These bounds establish \eqref{eq: n bound 2}.

For every $u\in\mathbb{R}^n$, symmetry of $\Lambda_n$ and the bound $2|u_i u_{i'}|\leq u_i^2+u_{i'}^2$ give
\begin{align}
    |u^\top\Lambda_nu|
    &\leq
    \sum_{i,i'=1}^n
    |\lambda_{i,i'}||u_i||u_{i'}|
    \nonumber\\
    &\leq
    \frac{1}{2}\sum_{i,i'=1}^n
    |\lambda_{i,i'}|(u_i^2+u_{i'}^2)
    =
    \sum_{i=1}^nu_i^2
    \sum_{i'=1}^n|\lambda_{i,i'}|
    \leq
    \left(
        \max_{i\in[n]}\sum_{i'=1}^n|\lambda_{i,i'}|
    \right)\|u\|_2^2~.
\end{align}
Taking the supremum over unit vectors, using \cref{assu: sparsity}, and observing that $H_n$ is an orthogonal projection, we find that
\begin{align}
    \|M_n\|_{\mathrm{op}}
    &=
    \|H_n\Lambda_nH_n\|_{\mathrm{op}}
    \leq
    \|\Lambda_n\|_{\mathrm{op}}
    \leq
    \max_{i\in[n]}\sum_{i'=1}^n|\lambda_{i,i'}|
    \lesssim
    nV_n~.
    \label{eq: basic operator bounds}
\end{align}
Finally, cyclicity of the trace, $H_n^2=H_n$, \eqref{eq: basic objective identity}, and $\operatorname{tr}(\Lambda_n)=n$ yield
\begin{align}
    \operatorname{tr}(M_n)
    &=
    \operatorname{tr}(H_n\Lambda_nH_n)
    =
    \operatorname{tr}(\Lambda_nH_n)
    \nonumber\\
    &=
    \operatorname{tr}(\Lambda_n)
    -\frac{1}{n}\bm{1}^\top\Lambda_n\bm{1}
    =
    n\left(1-V_n\right)~.
    \label{eq: basic trace identity}
\end{align}
The bound \eqref{eq: basic objective bounds} and $c^\star<1/4$ imply that
\begin{align}
    \frac{n}{2}
    &\leq
    \operatorname{tr}(M_n)
    \leq
    n~.
\end{align}
Together with \eqref{eq: basic operator bounds} and \eqref{eq: basic trace identity}, this proves \eqref{eq: n bound 4}.\hfill\qed

\subsection{Proof of \cref{lem: auxiliary outcome covariance bounds}}

\cref{assu: well conditioned}, applied to $\Gamma_d$, gives
\begin{align}
    \|\bar\Gamma_{0,d}\|_2
    &\leq
    \|(\gamma_{0,j})_{j=0}^d\|_2
    \leq
    \|\Gamma_d\|_{\mathrm{op}}
    \lesssim1~.
    \label{eq: auxiliary outcome row bound}
\end{align}
For every $u\in\mathbb R^d$, the normalization $\gamma_{0,0}=1$ in
\cref{assu: proportionality} implies that
\begin{align}
    u^{\top}\Omega_du
    &=
    \begin{pmatrix}
        -\bar\Gamma_{0,d}u\\
        u
    \end{pmatrix}^{\top}
    \Gamma_d
    \begin{pmatrix}
        -\bar\Gamma_{0,d}u\\
        u
    \end{pmatrix}~.
    \label{eq: auxiliary outcome schur complement}
\end{align}
\cref{assu: well conditioned}, applied to $\Gamma_d$, \eqref{eq: auxiliary outcome row bound}, and \eqref{eq: auxiliary outcome schur complement} therefore imply that
\begin{align}
    \|u\|_2^2
    &\lesssim
    u^\top\Omega_du
    \lesssim
    \|u\|_2^2+|\bar\Gamma_{0,d}u|^2
    \lesssim
    \|u\|_2^2~.
    \label{eq: auxiliary outcome schur complement bounds}
\end{align}
This proves \eqref{eq: auxiliary outcome uniform conditioning}. In turn, it holds that
\begin{align}
    \|\Omega_d\|_{\mathrm{op}}
    &\lesssim1~\quad\text{and}\quad
    \|\Omega_d\|_{\mathrm F}^2
    =\sum_{j=1}^d\theta_{d,j}^2\asymp d
    \label{eq: auxiliary outcome operator and frobenius bounds}
\end{align}
as well as
\begin{align}
    \operatorname{tr}(\Omega_d)
    &=
    \operatorname{tr}(\bar\Gamma_d)
    -\|\bar\Gamma_{0,d}\|_2^2
    =
    d-\|\bar\Gamma_{0,d}\|_2^2
    =
    d+O(1)~,
    \label{eq: auxiliary outcome trace identity}
\end{align}
where the final equality follows from \eqref{eq: auxiliary outcome row bound}.

Direct expansion implies that
\begin{align}
    \|\Omega_d\|_{\mathrm F}^2
    &=\|\bar\Gamma_d\|_{\mathrm F}^2
    -2\bar\Gamma_{0,d}\bar\Gamma_d\bar\Gamma_{0,d}^{\top}
    +\|\bar\Gamma_{0,d}\|_2^4~.
    \label{eq: auxiliary outcome frobenius expansion}
\end{align}
\cref{assu: well conditioned}, applied to $\Gamma_d$, together with
\eqref{eq: auxiliary outcome row bound}, implies that
\begin{align}
    \left|
        -2\bar\Gamma_{0,d}\bar\Gamma_d
        \bar\Gamma_{0,d}^{\top}
        +\|\bar\Gamma_{0,d}\|_2^4
    \right|
    &\leq
    2\|\bar\Gamma_d\|_{\mathrm{op}}
    \|\bar\Gamma_{0,d}\|_2^2
    +\|\bar\Gamma_{0,d}\|_2^4
    \lesssim1~.
    \label{eq: auxiliary outcome frobenius correction bound}
\end{align}
Combining \eqref{eq: auxiliary outcome frobenius expansion} and
\eqref{eq: auxiliary outcome frobenius correction bound}, we obtain
\begin{align}
    \|\Omega_d\|_{\mathrm F}^2
    &=
    \|\bar\Gamma_d\|_{\mathrm F}^2+O(1)~.
    \label{eq: auxiliary outcome frobenius comparison}
\end{align}
Combining \eqref{eq: auxiliary outcome operator and frobenius bounds}, \eqref{eq: auxiliary outcome trace identity}, and \eqref{eq: auxiliary outcome frobenius comparison} proves \eqref{eq: auxiliary outcome covariance bounds}.

Since $\Omega_d$ is positive semi-definite, its eigenvalues satisfy
\begin{align}
    \max_{j\in[d]}\theta_{d,j}
    &=\|\Omega_d\|_{\mathrm{op}}\quad\text{and}\quad
    \left(\sum_{j=1}^d\theta_{d,j}^2\right)^{1/2}
    =\|\Omega_d\|_{\mathrm F}~.
\end{align}
Consequently, \eqref{eq: auxiliary outcome operator and frobenius bounds} gives
\begin{align}
    \frac{\max_{j\in[d]}\theta_{d,j}}
    {(\sum_{j=1}^d\theta_{d,j}^2)^{1/2}}
    &=
    \frac{\|\Omega_d\|_{\mathrm{op}}}
    {\|\Omega_d\|_{\mathrm F}}
    \lesssim
    \frac{1}{\sqrt d}~.
    \label{eq: auxiliary outcome spectral ratio bound}
\end{align}
Finally, \cref{assu: well conditioned}, applied to the principal submatrix $\bar\Gamma_d$ of $\Gamma_d$, implies that
\begin{align}
    \|\bar\Gamma_d\|_{\mathrm F}^2
    &=
    \operatorname{tr}(\bar\Gamma_d^2)
    \asymp d~.
\end{align}
This bound, together with \eqref{eq: auxiliary outcome frobenius comparison}, gives
\begin{align}
    \left|
        \frac{\|\Omega_d\|_{\mathrm F}}
        {\|\bar\Gamma_d\|_{\mathrm F}}
        -1
    \right|
    &=
    \frac{
        \left|
            \|\Omega_d\|_{\mathrm F}^2
            -\|\bar\Gamma_d\|_{\mathrm F}^2
        \right|
    }{
        \|\bar\Gamma_d\|_{\mathrm F}
        \left(
            \|\Omega_d\|_{\mathrm F}
            +\|\bar\Gamma_d\|_{\mathrm F}
        \right)
    }
    \lesssim
    \frac{1}{d}~.
\end{align}
Together with \eqref{eq: auxiliary outcome spectral ratio bound}, the
definition of $v_{0}^{(d)}$ gives
\begin{align}
    \frac{\|\Omega_d\|_{\mathrm F}}{d}
    &=
    \frac{\|\bar\Gamma_d\|_{\mathrm F}}{d}
    \left(1+O(d^{-1})\right)
    =
    \sqrt{v_{0}^{(d)}}
    \left(1+O(d^{-1})\right)~,
    \label{eq: auxiliary outcome scale comparison}
\end{align}
as required.\hfill\qed

\subsection{Proof of \cref{lem: rotation and regression}}

Since a deterministic linear transformation of a Gaussian vector is Gaussian, the vectors $((XU)^{(k)})_{k=0}^m$ are jointly Gaussian. We can evaluate
\begin{align}
    \Cov((XU)_i^{(k)},(XU)_{i'}^{(l)})
    &=\sum_{j,j'=0}^m
    U_{j,k}U_{j',l}
    \Cov(X_i^{(j)},X_{i'}^{(j')})
    \nonumber\\
    &=A_{i,i'}
    \sum_{j,j'=0}^m
    U_{j,k}B_{j,j'}U_{j',l}
    =A_{i,i'}(U^{\top} BU)_{k,l}
    =A_{i,i'}\theta_k\mathbb{I}\{k=l\}~.
\end{align}
Thus, for arbitrary vectors $t_0,\ldots,t_m\in\mathbb{R}^n$, the Gaussian characteristic function gives
\begin{align}
    \mathbb{E}\left[
    \exp\left(
    \mathrm{i}\sum_{k=0}^m t_k^{\top}(XU)^{(k)}
    \right)
    \right]
    &=\exp\left(
    -\frac12\sum_{k,l=0}^m
    t_k^{\top}
    \Cov((XU)^{(k)},(XU)^{(l)})t_l
    \right)
    \nonumber\\
    &=\prod_{k=0}^m
    \exp\left(-\frac{\theta_k}{2}t_k^{\top} At_k\right)~.
    \label{eq: rotated gaussian characteristic function}
\end{align}
The right-hand side of
\eqref{eq: rotated gaussian characteristic function} is the joint
characteristic function of $(\sqrt{\theta_k}G^{(k)})_{k=0}^m$ when $G^{(0)},\ldots,G^{(m)}$ are independent Gaussian. Uniqueness of characteristic functions therefore implies that
\begin{align}
    XU
    &\overset{d}{=}
    (\sqrt{\theta_0}G^{(0)},\ldots,
    \sqrt{\theta_m}G^{(m)})~,
\end{align}
as required.\hfill\qed

\subsubsection{Part (ii)}

Observe that we can evaluate
\begin{align}
    \Cov\left(
        \widetilde X_{0,i}^{(j)},
        X_{i'}^{(0)}
    \right)
    &=
    \Cov\left(X_i^{(j)},X_{i'}^{(0)}\right)
    -
    B_{0,j}
    \Cov\left(X_i^{(0)},X_{i'}^{(0)}\right)
    \nonumber\\
    &=
    A_{i,i'}
    \left(
        B_{j,0}-B_{0,j}B_{0,0}
    \right)
    =
    0~,
    \label{eq: auxiliary on target zero covariance}
\end{align}
where the final equality follows from the symmetry of $B$ and the normalization $B_{0,0} = 1$. The vectors $X^{(0)},\widetilde X_0^{(1)},\ldots,\widetilde X_0^{(m)}$ are jointly Gaussian because they are linear transformations of the original jointly Gaussian vector. Choose arbitrary $t,u^{(1)},\ldots,u^{(m)}\in\mathbb{R}^n$ as test vectors. The condition \eqref{eq: auxiliary on target zero covariance} implies that
\begin{align}
    \Cov\left(
        t^{\top} X^{(0)},
        \sum_{j=1}^m
        (u^{(j)})^{\top}\widetilde X_0^{(j)}
    \right)
    &=
    \sum_{i,i'=1}^n\sum_{j=1}^m
    t_i u_{i'}^{(j)}
    \Cov\left(
        X_i^{(0)},
        \widetilde X_{0,i'}^{(j)}
    \right)
    =
    0~.
\end{align}
The Gaussian characteristic function therefore gives
\begin{align}
    \mathbb{E}\left[
        \exp\left(
            \mathrm{i}t^{\top} X^{(0)}
            +
            \mathrm{i}\sum_{j=1}^m
            (u^{(j)})^{\top}\widetilde X_0^{(j)}
        \right)
    \right]
    &=
    \exp\left(
        -\frac{1}{2}\Var(t^{\top} X^{(0)})
        -
        \frac{1}{2}
        \Var\left(
            \sum_{j=1}^m
            (u^{(j)})^{\top}\widetilde X_0^{(j)}
        \right)
    \right)
    \nonumber\\
    &=
    \mathbb{E}\left[
        \exp(\mathrm{i}t^{\top} X^{(0)})
    \right]
    \mathbb{E}\left[
        \exp\left(
            \mathrm{i}\sum_{j=1}^m
            (u^{(j)})^{\top}\widetilde X_0^{(j)}
        \right)
    \right]~.
\end{align}
Thus, the joint characteristic function factors for every choice of
$t,u^{(1)},\ldots,u^{(m)}$, which establishes the asserted
independence. The independence, moreover, implies that
\begin{align}
    \Cov\left(
        \widetilde X_{0,i}^{(j)},
        \widetilde X_{0,i'}^{(j')}
    \right)
    &=
    A_{i,i'}
    \left(
        B_{j,j'}
        -
        B_{j,0}B_{0,j'}
        -
        B_{0,j}B_{0,j'}
        +
        B_{0,j}B_{0,j'}B_{0,0}
    \right)
    \nonumber\\
    &=
    A_{i,i'}
    \left(
        B_{j,j'}-B_{j,0}B_{0,j'}
    \right)
    =
    A_{i,i'}
    \left(
        \bar B-\bar B_0^{\top}\bar B_0
    \right)_{j,j'}~,
\end{align}
where the second equality uses the normalization $B_{0,0} = 1$ and the third uses covariance symmetry, completing the proof.\hfill\qed

\subsubsection{Part (iii)}

We begin by establishing the range condition needed when $\bar B$ is
singular. Fix $u\in\ker(\bar B)$ arbitrarily. Positive semi-definiteness of $B$
implies that, for every $t\in\mathbb{R}$,
\begin{align}
    0
    &\leq
    \begin{pmatrix}
        t & u^{\top}
    \end{pmatrix}
    \begin{pmatrix}
        1 & \bar B_0\\
        \bar B_0^{\top} & \bar B
    \end{pmatrix}
    \begin{pmatrix}
        t\\u
    \end{pmatrix}
    =
    t^2+2t\bar B_0u+u^{\top}\bar Bu
    =
    t^2+2t\bar B_0u~.
    \label{eq: gaussian regression range quadratic}
\end{align}
Taking $t=-\bar B_0u$ in
\eqref{eq: gaussian regression range quadratic} gives
$-(\bar B_0u)^2\geq0$ for the chosen vector. Consequently, $\bar B_0$
vanishes throughout this kernel. Since
$I_m-\bar B^\dagger\bar B$ is the
orthogonal projection onto $\ker(\bar B)$, it follows that
\begin{align}
    \bar B_0
    (I_m-\bar B^\dagger\bar B)
    &=0,
    \quad\text{and hence}\quad
    \bar B_0
    =
    \bar B_0\bar B^\dagger\bar B~.
    \label{eq: gaussian regression range identity}
\end{align}
Now, define the residual
\begin{align}
    \xi
    &=X^{(0)}
    -
    \bar X\bar B^\dagger\bar B_0^{\top}~.
\end{align}
For every $i,i'$, the covariance between $\xi_i$
and the $i'$th row of $\bar X$ satisfies
\begin{align}
    \Cov\left(\xi_i,\bar X_{i',\cdot}\right)
    &=
    A_{i,i'}
    \left(
        \bar B_0
        -
        \bar B_0\bar B^\dagger\bar B
    \right)
    =
    0~,
    \label{eq: gaussian regression zero covariance}
\end{align}
where the final equality follows from
\eqref{eq: gaussian regression range identity}.

The vectors $\xi,X^{(1)},\ldots,X^{(m)}$ are jointly Gaussian because
they are linear transformations of the original jointly Gaussian vector.
Choose arbitrary $t,u^{(1)},\ldots,u^{(m)}\in\mathbb{R}^n$ as test
vectors.
The condition \eqref{eq: gaussian regression zero covariance} implies
\begin{align}
    \Cov\left(
        t^{\top}\xi,
        \sum_{j=1}^m (u^{(j)})^{\top} X^{(j)}
    \right)
    &=
    \sum_{i,i'=1}^n\sum_{j=1}^m
    t_i u_{i'}^{(j)}
    \Cov\left(\xi_i,X_{i'}^{(j)}\right)
    =
    0~.
\end{align}
The Gaussian characteristic function therefore gives
\begin{align}
    \mathbb{E}\left[
        \exp\left(
            \mathrm{i}t^{\top}\xi
            +
            \mathrm{i}\sum_{j=1}^m
            (u^{(j)})^{\top} X^{(j)}
        \right)
    \right]
    &=
    \exp\left(
        -\frac{1}{2}\Var(t^{\top}\xi)
        -
        \frac{1}{2}
        \Var\left(
            \sum_{j=1}^m (u^{(j)})^{\top} X^{(j)}
        \right)
    \right)
    \nonumber\\
    &=
    \mathbb{E}\left[\exp(\mathrm{i}t^{\top}\xi)\right]
    \mathbb{E}\left[
        \exp\left(
            \mathrm{i}\sum_{j=1}^m
            (u^{(j)})^{\top} X^{(j)}
        \right)
    \right]~.
\end{align}
Thus, the joint characteristic function factors for every choice of
$t,u^{(1)},\ldots,u^{(m)}$, which establishes the asserted
independence. The independence, moreover, implies that
\begin{align}
    \Cov(\xi_i,\xi_{i'})
    &=
    A_{i,i'}
    \left(
        1
        -
        2\bar B_0\bar B^\dagger\bar B_0^{\top}
        +
        \bar B_0
        \bar B^\dagger
        \bar B
        \bar B^\dagger
        \bar B_0^{\top}
    \right)
    \nonumber\\
    &=
    A_{i,i'}
    \left(
        1-\bar B_0\bar B^\dagger\bar B_0^{\top}
    \right)~,
\end{align}
where the final equality uses the Moore--Penrose identity
$\bar B^\dagger\bar B\bar B^\dagger=\bar B^\dagger$ for the generalized
inverse. Hence,
\begin{align}
    \Cov(\xi)
    &=
    \left(
        1-\bar B_0\bar B^\dagger\bar B_0^{\top}
    \right)A~,
\end{align}
as required.\hfill\qed

\section{Proofs for Lemmas Supporting \cref{lem: selected set characterization}\label{app: selected set characterization proof}}

\subsection{Proof of \cref{lem: feasible screening continuity}}

By \cref{rem: positivity}, the condition $n\geq3$ implies that $\|\hat\varepsilon^{(0)}\|_2^2>0$ almost surely. On this event, define
\begin{align}
    H_{n,0}
    &=H_n-
    \frac{
        \hat\varepsilon^{(0)}
        (\hat\varepsilon^{(0)})^{\top}
    }{\|\hat\varepsilon^{(0)}\|_2^2}
    \quad\text{and}\quad
    \widetilde M_n
    =H_{n,0}M_nH_{n,0}=(\tilde{\mu}_{i,i'})_{i,i'=1}^n~.
    \label{eq: exact screening projected covariance}
\end{align}
Since $H_nH_{n,0}=H_{n,0}$, \cref{assu: well conditioned}, applied to $\Lambda_n$, implies that
\begin{align}
    \tilde\mu_{i,i}
    &=
    (H_{n,0}e_i)^\top\Lambda_n(H_{n,0}e_i)
    \gtrsim
    \|H_{n,0}e_i\|_2^2~,
\end{align}
where $e_i$ is the $i$th basis vector.  In turn, it holds that $H_{n,0}e_i=0$ only if $\hat\varepsilon^{(0)}\in\operatorname{span}(H_ne_i)$, an event of probability zero. It follows that
\begin{align}
    \min_{i\in[n]}\tilde\mu_{i,i}
    >0
    \label{eq: continuity positive projected diagonal}
\end{align}
almost surely by a union bound. The result is obtained by invoking the following representation.

\begin{lem}
\label{lem: exact feasible screening representation}$ $\\
Suppose that \cref{assu: gaussianity,assu: proportionality,assu: well conditioned} hold and that $n \geq 2$.  Conditional on $\hat\varepsilon^{(0)}$, it holds that
\begin{align}
    (\hat\lambda_{i,i'})_{i,i'\in[n]}
    &\overset{d}{=}
    \left(
        \frac{1}{d}\sum_{j=1}^d
        \theta_{d,j}G_i^{(j)}G_{i'}^{(j)}
    \right)_{i,i'\in[n]}~,
    \label{eq: exact feasible screening joint representation}
\end{align}
where $G^{(1)},\ldots,G^{(d)}$ are independent with distribution $\mathsf N(0,\widetilde M_n)$. Consequently, for every $i,i'\in[n]$,
\begin{align}
    \hat\lambda_{i,i'}
    &\overset{d}{=}
    \frac{\psi_{i,i'}(\widetilde M_n)}{d}
    \sum_{j=1}^d\theta_{d,j}Z_j
    \left(
        \rho_{i,i'}(\widetilde M_n)Z_j
        +
        \sqrt{1-\rho_{i,i'}(\widetilde M_n)^2}W_j
    \right)~,
    \label{eq: exact conditional screening representation}
\end{align}
where $Z_1,\ldots,Z_d,W_1,\ldots,W_d$ are independent standard Gaussian
variables.
\end{lem}

\noindent  By \eqref{eq: continuity positive projected diagonal}, $\psi_{i,i'}(\widetilde M_n)>0$. Moreover, \cref{lem: auxiliary outcome covariance bounds}, display \eqref{eq: auxiliary outcome covariance bounds}, implies that $\Omega_d$ is positive semi-definite and nonzero. Hence, at least one eigenvalue $\theta_{d,j}$ is positive. The corresponding summand in \eqref{eq: exact conditional screening representation} is a nonzero scalar multiple of
\begin{align}
    Z\left(\rho Z+\sqrt{1-\rho^2}W\right)~,
    \label{eq: continuity gaussian product}
\end{align}
where $Z$ and $W$ are independent standard Gaussian variables. If $|\rho|<1$, then, conditional on $Z=z$ for any $z\neq0$, the variable in \eqref{eq: continuity gaussian product} has a nondegenerate Gaussian distribution and is thereby non-atomic. If $|\rho|=1$, it equals $\rho Z^2$ and is also non-atomic. This summand is independent of the remaining summands in \eqref{eq: exact conditional screening representation}, so their sum is non-atomic. Thus, the conditional distribution of $\hat\lambda_{i,i'}$ is non-atomic almost surely. Taking expectations proves the result for every $i,i'\in[n]$.\hfill\qed

\subsection{Proof of \cref{lem: feasible null moderate deviation}}

Fix $\kappa>0$. The result is obtained though a successive application of the following Lemmas. 

\begin{lem}
\label{lem: null pair projection stability}$ $\\
Suppose that \cref{assu: gaussianity,assu: proportionality,assu: homoskedasticity,assu: sparsity} hold and that $s/n\to0$. For every fixed $q>0$, there is a constant $C_q$ such that
\begin{align}
    \sup_{\substack{i,i'\in[n]\\\lambda_{i,i'}=0}}
    P\left\{
        \left|
            \psi_{i,i'}(\widetilde M_n)-1
        \right|
        +\left|
            \rho_{i,i'}(\widetilde M_n)
        \right|
        >C_qV_n
        \log(V_n^{-1})
    \right\}
    &=o\left(V_n^q\right)
    \label{eq: null pair projection stability}
\end{align}
as $n\to\infty$.
\end{lem}

\begin{lem}
\label{lem: weighted gaussian product moderate deviation}$ $\\
Let $\theta_{d,1},\ldots,\theta_{d,d}$ be nonnegative numbers satisfying
\begin{align}
    \sum_{j=1}^d\theta_{d,j}^2
    &>0
    \quad\text{and}\quad
    \frac{\max_{j\in[d]}\theta_{d,j}}
    {(\sum_{j=1}^d\theta_{d,j}^2)^{1/2}}
    \lesssim\frac{1}{\sqrt d}~.
    \label{eq: weighted gaussian product weight condition}
\end{align}
Let $Z_1,\ldots,Z_d,W_1,\ldots,W_d$ be independent standard Gaussian
variables. For $|\rho|\leq1$, define
\begin{align}
    S_{d,\rho}
    &=\frac{1}{\sqrt{1+\rho^2}}
    \sum_{j=1}^d
    \frac{\theta_{d,j}}
    {(\sum_{k=1}^d\theta_{d,k}^2)^{1/2}}
    \left(
        Z_j(\rho Z_j+\sqrt{1-\rho^2}W_j)-\rho
    \right)~.
\end{align}
If $u_d\to\infty$ and $u_d=o(\sqrt d)$, then
\begin{align}
    \log P\{S_{d,\rho}\geq u_d\}
    &=-\frac{u_d^2}{2}(1+o(1))
    \quad\text{and}\quad
    \log P\{S_{d,\rho}\leq-u_d\}
    =-\frac{u_d^2}{2}(1+o(1))~.
    \label{eq: weighted gaussian product moderate deviation}
\end{align}
The approximations are uniform over $|\rho|\leq1$ and over weight arrays
satisfying \eqref{eq: weighted gaussian product weight condition} with a
common implicit constant. They continue to hold when $u_d$ is replaced by
any $\widetilde u_d$ satisfying $\widetilde u_d/u_d\to1$.
\end{lem}

\noindent Set $q=\kappa^2\eta+1$, and let $C_q$ be the constant supplied by \cref{lem: null pair projection stability}. For each null pair, define the event
\begin{align}
    \mathcal E_{i,i'}
    &=\left\{
        \left|
            \psi_{i,i'}(\widetilde M_n)-1
        \right|
        +\left|
            \rho_{i,i'}(\widetilde M_n)
        \right|
        \leq C_qV_n
        \log(V_n^{-1})
    \right\}
    \label{eq: null moderate deviation projection event}
\end{align}
\cref{lem: null pair projection stability} implies that
\begin{align}
    \sup_{\substack{i,i'\in[n]\\\lambda_{i,i'}=0}}
    P\{\mathcal E_{i,i'}^c\}
    &=o\left(V_n^q\right)~.
    \label{eq: null moderate deviation projection probability}
\end{align}
Since $x\log(1/x)\to0$ as $x\downarrow0$, it holds that 
\begin{align}\label{eq: psi rho tilde bounds}
    \psi_{i,i'}(\widetilde M_n)=1+o(1)
    \quad\text{and}\quad
    \rho_{i,i'}(\widetilde M_n)=o(1)
\end{align}
uniformly on $\mathcal E_{i,i'}$ and over null pairs. In particular, $\psi_{i,i'}(\widetilde M_n)>0$ for all sufficiently large $n,d$. Hence, \cref{lem: exact feasible screening representation} implies
\begin{align}
    S_{d,\rho_{i,i'}(\widetilde M_n)}
    &\overset{d}{=} \frac{
        \hat\lambda_{i,i'} -
        \psi_{i,i'}(\widetilde M_n) \rho_{i,i'}(\widetilde M_n)\operatorname{tr}(\Omega_d)/d
    }{
        \psi_{i,i'}(\widetilde M_n)\sqrt{1+\rho_{i,i'}(\widetilde M_n)^2}\|\Omega_d\|_{\mathrm F}/d
    }
    \label{eq: null moderate deviation standardization}
\end{align}
conditional on $\hat\varepsilon^{(0)}$, where $S_{d,\rho}$ is defined in \cref{lem: weighted gaussian product moderate deviation}. 

By \cref{lem: auxiliary outcome covariance bounds}, display \eqref{eq: auxiliary outcome weight and scale bounds}, the conditional scale in \eqref{eq: null moderate deviation standardization} satisfies
\begin{align}
    \frac{
        \psi_{i,i'}(\widetilde M_n)\sqrt{1+\rho_{i,i'}(\widetilde M_n)^2}\|\Omega_d\|_{\mathrm F}/d
    }{\sqrt{v_{0}^{(d)}}}
    &=
    \psi_{i,i'}(\widetilde M_n)
    \sqrt{1+\rho_{i,i'}(\widetilde M_n)^2}
    \left(1+O(d^{-1})\right)
    =1+o(1)
    \label{eq: null moderate deviation conditional scale}
\end{align}
uniformly on $\mathcal E_{i,i'}$ and over null pairs. In turn, Cauchy--Schwarz and \cref{lem: auxiliary outcome covariance bounds}, display \eqref{eq: auxiliary outcome weight and scale bounds}, give
\begin{align}
    \frac{
        |\psi_{i,i'}(\widetilde M_n) \rho_{i,i'}(\widetilde M_n)|\operatorname{tr}(\Omega_d)/d
    }{
        \sqrt{v_{0}^{(d)}}
        \sqrt{\log(1/V_n)}
    }
    &\leq
    |\psi_{i,i'}(\widetilde M_n)\rho_{i,i'}(\widetilde M_n)|
    \sqrt{
        \frac{d}{\log(1/V_n)}
    }
    \left(1+O(d^{-1})\right)
    \nonumber\\
    &\lesssim
    V_n
    \sqrt{
        d\log(V_n^{-1})
    }
    \to0
    \label{eq: null moderate deviation conditional mean}
\end{align}
uniformly on $\mathcal E_{i,i'}$ and over null pairs, where the second inequality follows from the bound \eqref{eq: psi rho tilde bounds}, and the convergence follows from the first condition in \eqref{eq: mod dev high level rate}.

Now, define the threshold
\begin{align}
    t_{n,d}(\kappa)
    &=\kappa\sqrt{v_{0}^{(d)}}
    \sqrt{
        2\eta\log(V_n^{-1})
    }~.
\end{align}
Observe that \eqref{eq: null moderate deviation conditional scale} and \eqref{eq: null moderate deviation conditional mean} imply that
\begin{align}
    \frac{
        t_{n,d}(\kappa)
        \mathbin{\pm}\psi_{i,i'}(\widetilde M_n)\rho_{i,i'}(\widetilde M_n)\operatorname{tr}(\Omega_d)/d
    }{
        \psi_{i,i'}(\widetilde M_n)\sqrt{1+\rho_{i,i'}(\widetilde M_n)^2}\|\Omega_d\|_{\mathrm F}/d
    }
    &=\kappa
    \sqrt{
        2\eta\log(V_n^{-1})
    }(1+o(1))
    \label{eq: null moderate deviation conditional thresholds}
\end{align}
uniformly on $\mathcal E_{i,i'}$ and over null pairs. By \cref{lem: basic inequalities n}, display \eqref{eq: var bounds}, $V_n\to0$, and so $t_{n,d}(\kappa)/\sqrt{v_{0}^{(d)}}\to\infty$. In turn, the second condition in \eqref{eq: mod dev high level rate} implies that $t_{n,d}(\kappa)/\sqrt{v_{0}^{(d)}}=o(\sqrt d)$. Moreover, positive semi-definiteness of $\widetilde M_n$ implies that $|\rho_{i,i'}(\widetilde M_n)|\leq1$. In turn, \cref{lem: auxiliary outcome covariance bounds} implies that $\Omega_d$ is positive semi-definite, so $\theta_{d,j}\geq0$ for each $j\in[d]$. By \cref{lem: auxiliary outcome covariance bounds}, display \eqref{eq: auxiliary outcome covariance bounds}, it holds that $\sum_{j=1}^d\theta_{d,j}^2=\|\Omega_d\|_{\mathrm F}^2>0$. Finally, by \cref{lem: auxiliary outcome covariance bounds}, display \eqref{eq: auxiliary outcome weight and scale bounds}, the remaining condition in
\eqref{eq: weighted gaussian product weight condition} holds. Thus, the conditions of \cref{lem: weighted gaussian product moderate deviation} hold with $u_d=t_{n,d}(\kappa)/\sqrt{v_{0}^{(d)}}$.

By the representation \eqref{eq: null moderate deviation standardization}, we can write
\begin{align}
    P\left\{
        \hat\lambda_{i,i'}
        \geq t_{n,d}(\kappa)
        \mid\hat\varepsilon^{(0)}
    \right\}
    &=
    P\left\{
        S_{d,\rho_{i,i'}(\widetilde M_n)}
        \geq
        \frac{
            t_{n,d}(\kappa)
            -\psi_{i,i'}(\widetilde M_n)
            \rho_{i,i'}(\widetilde M_n)
            \operatorname{tr}(\Omega_d)/d
        }{
            \psi_{i,i'}(\widetilde M_n)
            \sqrt{1+\rho_{i,i'}(\widetilde M_n)^2}
            \|\Omega_d\|_{\mathrm F}/d
        }
    \right\}~.
\end{align}
By  \eqref{eq: null moderate deviation conditional thresholds}, the threshold on the right-hand side is uniformly equal to $u_d(1+o(1))$ on $\mathcal E_{i,i'}$ and over null pairs. Hence, \cref{lem: weighted gaussian product moderate deviation} implies that
\begin{align}
    \log P\left\{
        \hat\lambda_{i,i'}
        \geq t_{n,d}(\kappa)
        \mid\hat\varepsilon^{(0)}
    \right\}
    &=-\frac{u_d^2}{2}(1+o(1))
    \nonumber\\
    &=-\kappa^2\eta
    \log(V_n^{-1})
    +o\left(
        \log(V_n^{-1})
    \right)~,
\end{align}
uniformly on $\mathcal E_{i,i'}$ and over null pairs. Applying both conclusions of\cref{lem: weighted gaussian product moderate deviation} through \eqref{eq: null moderate deviation standardization} and \eqref{eq: null moderate deviation conditional thresholds} gives
\begin{align}
    \log P\left\{
        |\hat\lambda_{i,i'}|
        \geq t_{n,d}(\kappa)
        \mid\hat\varepsilon^{(0)}
    \right\}
    &=-\kappa^2\eta
    \log(V_n^{-1})
    +o\left(
        \log(V_n^{-1})
    \right)
    \label{eq: null moderate deviation conditional conclusion}
\end{align}
uniformly on $\mathcal E_{i,i'}$ and over null pairs. To remove the conditioning, the uniformity in \eqref{eq: null moderate deviation conditional conclusion} implies that there exists a nonnegative sequence $r_{n,d}\to0$ such that
\begin{align}
    &\exp\left(
        -(\kappa^2\eta+r_{n,d})
        \log(V_n^{-1})
    \right)P\{\mathcal E_{i,i'}\}
    \nonumber\\
    &\leq
    P\left\{
        |\hat\lambda_{i,i'}|
        \geq t_{n,d}(\kappa)
    \right\}
    \nonumber\\
    &\leq
    \exp\left(
        -(\kappa^2\eta-r_{n,d})
        \log(V_n^{-1})
    \right)
    +P\{\mathcal E_{i,i'}^c\}~.
    \label{eq: null moderate deviation unconditional sandwich}
\end{align}
Since $q>\kappa^2\eta$, the bound \eqref{eq: null moderate deviation projection probability} implies that the final term in \eqref{eq: null moderate deviation unconditional sandwich} is negligible relative to $V_n^{\kappa^2\eta}$. Taking logarithms in that display and using $P\{\mathcal E_{i,i'}\}\to1$ uniformly over null pairs yields \eqref{eq: feasible null moderate deviation}.\hfill\qed

\subsection{Proof of \cref{lem: exact feasible screening representation}}

Define the oracle residual vector
\begin{align}
    \tilde\varepsilon_0^{(j)}
    &=\hat\varepsilon^{(j)}
    -\gamma_{0,j}\hat\varepsilon^{(0)}~.
    \label{eq: exact screening oracle residual}
\end{align}
By \cref{assu: gaussianity,assu: proportionality}, the array
$(\hat\varepsilon^{(j)})_{j=0}^d$ is centered and jointly Gaussian, with
covariance
\begin{align}
    \Cov(\hat\varepsilon_k^{(j)},\hat\varepsilon_l^{(j')})
    &=\mu_{k,l}\gamma_{j,j'}~.
\end{align}
Thus, by identifying $X=(\hat\varepsilon^{(j)})_{j=0}^d$, $A=M_n$, and
$B=\Gamma_d$, Part (ii) of \cref{lem: rotation and regression} gives
\begin{align}
    (\tilde\varepsilon_0^{(j)})_{j=1}^d
    &\ci\hat\varepsilon^{(0)}
    \quad\text{and}\quad
    \Cov\left(
        \tilde\varepsilon_{0,k}^{(j)},
        \tilde\varepsilon_{0,l}^{(j')}
    \right)
    =\mu_{k,l}\omega_{j,j'}
    \label{eq: exact screening oracle covariance}
\end{align}
because $\gamma_{0,0}=1$ under \cref{assu: proportionality}. Observe that we
can write
\begin{align}
    \hat\psi^{(j)}
    &=\gamma_{0,j}
    +\frac{
        (\hat\varepsilon^{(0)})^{\top}
        \tilde\varepsilon_0^{(j)}
    }{\|\hat\varepsilon^{(0)}\|_2^2}
    \quad\text{and thereby}\quad
    \hat\varepsilon^{(j;0)}
    =H_{n,0}\tilde\varepsilon_0^{(j)}~.
\end{align}
Thus, conditional on $\hat\varepsilon^{(0)}$, the components of the feasible
residual array $(\hat\varepsilon^{(j;0)})_{j=1}^d$ are
centered, jointly Gaussian, and satisfy
\begin{align}
    \Cov\left(
        \hat\varepsilon_k^{(j;0)},
        \hat\varepsilon_l^{(j';0)}
        \mid\hat\varepsilon^{(0)}
    \right)
    &=\tilde{\mu}_{k,l}\omega_{j,j'}~.
    \label{eq: exact screening feasible covariance}
\end{align}
Conditional on $\hat\varepsilon^{(0)}$, Part
(i) of \cref{lem: rotation and regression}, applied with
$X=(\hat\varepsilon^{(j;0)})_{j=1}^d$, $A=\widetilde M_n$, and $B=\Omega_d$, implies that we
can write
\begin{align}
    \hat\lambda_{i,i'}
    &\overset{d}{=}
    \frac{1}{d}\sum_{j=1}^d
    \theta_{d,j}G_i^{(j)}G_{i'}^{(j)}~,
    \quad\text{where}\quad
    G^{(1)},\ldots,G^{(d)}
    \overset{\mathrm{ind}}{\sim}
    \mathsf{N}(0,\widetilde M_n)
    \label{eq: exact screening rotated representation}
\end{align}
conditional on $\hat\varepsilon^{(0)}$. If
$\psi_{i,i'}(\widetilde M_n)>0$, each Gaussian pair in
\eqref{eq: exact screening rotated representation} admits the representation
\begin{align}
    (G_i^{(j)},G_{i'}^{(j)})
    &\overset{d}{=}
    \left(
        \tilde{\mu}_{i,i}^{1/2}Z_j,
        \tilde{\mu}_{i',i'}^{1/2}
        \left(
            \rho_{i,i'}(\widetilde M_n)Z_j
            +\sqrt{1-\rho_{i,i'}(\widetilde M_n)^2}W_j
        \right)
    \right)~.
\end{align}
Indeed, the vector on the right of this display is centered Gaussian and has
the same covariance matrix as the vector on the left. If
$\psi_{i,i'}(\widetilde M_n)=0$, at least one of $G_i^{(j)}$ and
$G_{i'}^{(j)}$ has variance zero, so
$G_i^{(j)}G_{i'}^{(j)}=0$ almost surely. Substitution into
\eqref{eq: exact screening rotated representation} therefore gives
\eqref{eq: exact conditional screening representation} in both cases.\hfill\qed

\subsection{Proof of \cref{lem: null pair projection stability}}

Let $a_{n,1},\ldots,a_{n,n}$ be the eigenvalues of $M_n$. Direct integration of the standard Gaussian density gives
\begin{align}
    \mathbb{E}[\exp(-ta_{n,j}W_j^2)]
    &=\frac{1}{\sqrt{2\pi}}
    \int_{\mathbb{R}}
    \exp\left(-\frac{1+2ta_{n,j}}{2}w^2\right)\,dw
    =(1+2ta_{n,j})^{-1/2}~,
\end{align}
for each $t\geq0$ and $j\in[n]$. Choose $c>0$ sufficiently small that $cn\leq\operatorname{tr}(M_n)/2$ for all sufficiently large $n$ and $2c\leq1/2$, and set $t=c/\|M_n\|_{\mathrm{op}}$. Since $a_{n,j}\leq\|M_n\|_{\mathrm{op}}$, it follows that $2ta_{n,j}\leq1/2$ for every $j\in[n]$. Markov's inequality, \eqref{eq: projection stability target norm representation}, and the preceding identity give
\begin{align}
    P\{\|\hat\varepsilon^{(0)}\|_2^2\leq cn\}
    &\leq
    \exp\left(
        \frac{t}{2}\operatorname{tr}(M_n)
    \right)
    \prod_{j=1}^n(1+2ta_{n,j})^{-1/2}
    \nonumber\\
    &\leq
    \exp\left(
        -\frac{t}{2}\operatorname{tr}(M_n)
        +Ct^2\operatorname{tr}(M_n^2)
    \right)
    \leq
    \exp\left(
        -c\frac{\operatorname{tr}(M_n)}
        {\|M_n\|_{\mathrm{op}}}
    \right)
    \leq\exp\left(-\frac{c}{V_n}\right)~.
    \label{eq: projection stability target norm probability}
\end{align}
The first inequality also uses $cn\leq\operatorname{tr}(M_n)/2$. The second uses $\log(1+x)\geq x-Cx^2$ for all sufficiently small $x\geq0$. The third follows from $\operatorname{tr}(M_n^2)\leq \|M_n\|_{\mathrm{op}}\operatorname{tr}(M_n)$ and the sufficiently small choice of $c$. The final inequality follows from \cref{lem: basic inequalities n}, display \eqref{eq: n bound 4}.

Since $\hat\varepsilon^{(0)}$ has covariance $M_n$, \cref{lem: basic inequalities n}, displays \eqref{eq: n bound 2} and \eqref{eq: n bound 4}, imply that
\begin{align}
    \Var(\hat\varepsilon_k^{(0)})
    &=\mu_{k,k}\lesssim1\quad\text{and}
    \nonumber\\
    \Var((M_n\hat\varepsilon^{(0)})_k)
    &=e_k^{\top}M_n^3e_k
    \leq\|M_n\|_{\mathrm{op}}^2\mu_{k,k}
    \lesssim n^2V_n^2~.
    \label{eq: projection stability coordinate variances}
\end{align}
For a constant $K>0$, define
\begin{align}
    \mathcal G_{i,i'}(K)
    &=\left\{
        \|\hat\varepsilon^{(0)}\|_2^2\geq cn
    \right\}
    \cap\left\{
        \max_{k\in\{i,i'\}}
        |\hat\varepsilon_k^{(0)}|
        \leq K\sqrt{
            \log(V_n^{-1})
        }
    \right\}
    \nonumber\\
    &\quad\cap\left\{
        \max_{k\in\{i,i'\}}
        |(M_n\hat\varepsilon^{(0)})_k|
        \leq KnV_n
        \sqrt{
            \log(V_n^{-1})
        }
    \right\}~.
    \label{eq: projection stability good event}
\end{align}
Integrating the standard Gaussian density gives $P\{|Z|\geq x\}\leq2e^{-x^2/2}$. Hence, \eqref{eq: projection stability target norm probability} and \eqref{eq: projection stability coordinate variances} yield
\begin{align}
    \sup_{\substack{i,i'\in[n]\\\lambda_{i,i'}=0}}
    P\{\mathcal G_{i,i'}(K)^c\}
    &\lesssim
    \exp\left(
        -cK^2\log(V_n^{-1})
    \right)
    +\exp\left(-\frac{c}{V_n}\right)~.
    \label{eq: projection stability good event probability}
\end{align}
Expanding \eqref{eq: exact screening projected covariance}, we obtain
\begin{align}
    \tilde\mu_{k,l}-\mu_{k,l}
    &=-\frac{
        \hat\varepsilon_k^{(0)}
        (M_n\hat\varepsilon^{(0)})_l
        +(M_n\hat\varepsilon^{(0)})_k
        \hat\varepsilon_l^{(0)}
    }{\|\hat\varepsilon^{(0)}\|_2^2}
    +
    \frac{
        (\hat\varepsilon^{(0)})^{\top}
        M_n\hat\varepsilon^{(0)}
    }{\|\hat\varepsilon^{(0)}\|_2^4}
    \hat\varepsilon_k^{(0)}
    \hat\varepsilon_l^{(0)}~.
    \label{eq: projection stability exact perturbation}
\end{align}
On $\mathcal G_{i,i'}(K)$, the first fraction in \eqref{eq: projection stability exact perturbation} is bounded in absolute value by $CK^2V_n\log(1/V_n)$. Moreover, $(\hat\varepsilon^{(0)})^{\top}M_n\hat\varepsilon^{(0)} \leq\|M_n\|_{\mathrm{op}}\|\hat\varepsilon^{(0)}\|_2^2$, so \cref{lem: basic inequalities n}, display \eqref{eq: n bound 4}, gives the same bound for the second fraction. Consequently,
\begin{align}
    \max_{k,l\in\{i,i'\}}
    |\tilde\mu_{k,l}-\mu_{k,l}|
    &\lesssim
    K^2V_n
    \log(V_n^{-1})
    \label{eq: projection stability feasible entries}
\end{align}
on $\mathcal G_{i,i'}(K)$. Combining \cref{lem: basic inequalities n}, display \eqref{eq: n bound 3}, with \eqref{eq: projection stability feasible entries}, and using $V_n\log(1/V_n)\to0$, we find that
\begin{align}
    |\psi_{i,i'}(\widetilde M_n)-1|
    +|\rho_{i,i'}(\widetilde M_n)|
    &\lesssim
    K^2V_n
    \log(V_n^{-1})
    \label{eq: projection stability pair parameters}
\end{align}
on this event. Choose $K$ so that $cK^2>q$, and choose $C_q$ to dominate the implicit constant in \eqref{eq: projection stability pair parameters}. Then \eqref{eq: projection stability good event probability} is $o(V_n^q)$, so \eqref{eq: null pair projection stability} follows.
\hfill\qed

\subsection{Proof of \cref{lem: weighted gaussian product moderate deviation}}

Define the standardized eigenvalue
\begin{align}
    \tilde{\theta}_{d,j}
    &=\frac{\theta_{d,j}}
    {(\sum_{k=1}^d\theta_{d,k}^2)^{1/2}}~.
\end{align}
The definition of $S_{d,\rho}$ and the weight condition \eqref{eq: weighted gaussian product weight condition} imply that
\begin{align}
    \sum_{j=1}^d\tilde{\theta}_{d,j}^2
    &=1
    \quad\text{and}\quad
    \max_{j\in[d]}\tilde{\theta}_{d,j}
    \lesssim\frac{1}{\sqrt d}~.
    \label{eq: weighted gaussian product normalized weights}
\end{align}
We begin by approximating the log moment generating function of each summand. For $|\rho|\leq1$, define
\begin{align}
    \ell_\rho(t)
    &=\log\mathbb{E}\left[
        \exp\left(
            t\frac{
                Z_j(\rho Z_j+\sqrt{1-\rho^2}W_j)-\rho
            }{\sqrt{1+\rho^2}}
        \right)
    \right]~.
\end{align}
Conditioning on $Z_j$ and evaluating the standard Gaussian integral gives
\begin{align}
    \ell_\rho(t)
    &=-\frac{\rho t}{\sqrt{1+\rho^2}}
    -\frac{1}{2}\log\left(
        1-\frac{2\rho t}{\sqrt{1+\rho^2}}
        -\frac{(1-\rho^2)t^2}{1+\rho^2}
    \right)~,
    \label{eq: weighted gaussian product exact log mgf}
\end{align}
whenever the argument of the logarithm is positive. That argument is uniformly bounded away from zero for $|t|\leq1/4$ and $|\rho|\leq1$. Hence, the third derivative of $\ell_\rho$ is uniformly bounded on this set. Since $\ell_\rho(0)=\ell_\rho'(0)=0$ and $\ell_\rho''(0)=1$, Taylor's theorem gives
\begin{align}
    \left|\ell_\rho(t)-\frac{t^2}{2}\right|
    &\leq C|t|^3~,
    \label{eq: weighted gaussian product log mgf expansion}
\end{align}
for every $|t|\leq1/4$ and $|\rho|\leq1$.

Next, we apply the following scalar consequence of the G\"artner--Ellis theorem.

\begin{lem}[\citealp{dembozeitouni2010}, Theorem 2.3.6(c)]
\label{lem: gartner ellis theorem}$ $\\
Let $X_m$ be a sequence of random variables and let $v_m\to\infty$. Suppose that, for each $t\in\mathbb{R}$, the limit
\begin{align}
    \mathcal L(t)
    &=\lim_{m\to\infty}
    \frac{1}{v_m}
    \log\mathbb{E}[\exp(v_mtX_m)]
    \label{eq: gartner ellis limiting log mgf}
\end{align}
exists and is finite, and that $\mathcal L$ is differentiable on $\mathbb{R}$. Define
\begin{align}
    \mathcal L^*(x)
    &=\sup_{t\in\mathbb{R}}\{tx-\mathcal L(t)\}~.
\end{align}
Then, for every Borel set $D\subseteq\mathbb{R}$, the sequence of bounds
\begin{align}
    -\inf_{x\in D^{\circ}}\mathcal L^*(x)
    &\leq
    \liminf_{m\to\infty}
    \frac{1}{v_m}\log P\{X_m\in D\}
    \nonumber\\
    &\leq
    \limsup_{m\to\infty}
    \frac{1}{v_m}\log P\{X_m\in D\}
    \leq
    -\inf_{x\in\overline D}\mathcal L^*(x)
    \label{eq: gartner ellis large deviation principle}
\end{align}
hold, where $D^{\circ}$ and $\overline D$ denote the interior and closure of $D$, respectively.
\end{lem}

\noindent Fix an arbitrary sequence $\rho_d\in[-1,1]$ and an arbitrary sequence of weight arrays satisfying \eqref{eq: weighted gaussian product weight condition} with a common implicit constant. For each fixed $t\in\mathbb{R}$, the condition $u_d=o(\sqrt d)$ and \eqref{eq: weighted gaussian product normalized weights} imply that $|t|u_d\max_j\tilde{\theta}_{d,j}\to0$. Thus, independence and \eqref{eq: weighted gaussian product log mgf expansion} give
\begin{align}
    \left|
        \frac{1}{u_d^2}
        \log\mathbb{E}\left[
            \exp\left(tu_dS_{d,\rho_d}\right)
        \right]
        -\frac{t^2}{2}
    \right|
    &\leq
    C|t|^3u_d\sum_{j=1}^d\tilde{\theta}_{d,j}^3
    \leq
    C|t|^3u_d\max_{j\in[d]}\tilde{\theta}_{d,j}
    \to0~.
    \label{eq: weighted gaussian product scaled log mgf}
\end{align}
The second inequality uses the nonnegativity of the weights and $\sum_j\tilde{\theta}_{d,j}^2=1$. Apply \cref{lem: gartner ellis theorem} with
\begin{align}
    X_d
    &=\frac{S_{d,\rho_d}}{u_d}~,
    \quad
    v_d
    =u_d^2~,
    \quad
    \mathcal L(t)
    =\frac{t^2}{2}~,
    \quad\text{and}\quad
    \mathcal L^*(x)
    =\frac{x^2}{2}~.
\end{align}
The expansion \eqref{eq: weighted gaussian product scaled log mgf} verifies the condition in \eqref{eq: gartner ellis limiting log mgf}. Since the infimum of $x^2/2$ over the interior and closure of each of $[1,\infty)$ and $(-\infty,-1]$ equals $1/2$, the bounds \eqref{eq: gartner ellis large deviation principle} yield
\begin{align}
    \frac{1}{u_d^2}\log P\{S_{d,\rho_d}\geq u_d\}
    &\to-\frac{1}{2}
    \quad\text{and}\quad
    \frac{1}{u_d^2}\log P\{S_{d,\rho_d}\leq-u_d\}
    \to-\frac{1}{2}~,
    \label{eq: weighted gaussian product gartner ellis tails}
\end{align}
along the arbitrary sequence. This proves uniformity, since a failure of
uniform convergence would produce a sequence contradicting
\eqref{eq: weighted gaussian product gartner ellis tails}. Finally,
$\widetilde u_d/u_d\to1$ implies that $\widetilde u_d\to\infty$ and
$\widetilde u_d=o(\sqrt d)$, so the same argument proves the
perturbed-threshold conclusion.
\hfill\qed

\section{Proofs for Lemmas Supporting \cref{lem: bias decomposition}}

\subsection{Proof of \cref{lem: exact bias decomposition}}

\cref{assu: proportionality} implies that the residual matrix is centered. Moreover, \cref{assu: proportionality} and \eqref{eq: exact screening covariance notation} give
\begin{align}
    \Cov(\hat\varepsilon_i^{(j)},\hat\varepsilon_{i'}^{(j')})
    &=\gamma_{j,j'}(H_n\Lambda_nH_n)_{i,i'}
    =\gamma_{j,j'}\mu_{i,i'}
    \quad\text{and}\quad
    \mathbb{E}[\hat\varepsilon_i^{(0)}\hat\varepsilon_{i'}^{(0)}]
    =\mu_{i,i'}~,
    \label{eq: residual covariance}
\end{align}
for all $i,i'\in[n]$ and $j,j'\in\{0,\ldots,d\}$, where the expectation identity also uses the normalization $\gamma_{0,0}=1$. By \cref{assu: gaussianity}, the residual matrix is jointly Gaussian. We may therefore apply Part (ii) of \cref{lem: rotation and regression} to \eqref{eq: residual covariance}, identifying $X=(\hat\varepsilon^{(j)})_{j=0}^d$, $A=M_n$, and $B=\Gamma_d$. The normalization $B_{0,0}=\gamma_{0,0}=1$ follows from \cref{assu: proportionality}. Thus, we find that
\begin{align}
    \hat\varepsilon^{(0)}
    &\ci
    (\tilde\varepsilon_0^{(1)},\ldots,
    \tilde\varepsilon_0^{(d)})~.
\end{align}
Consequently, the product $\hat\varepsilon_i^{(0)}\hat\varepsilon_{i'}^{(0)}$ is independent of $\mathbb{I}\{|\tilde\lambda_{0,i,i'}|\geq\delta\}$, and so we obtain
\begin{align}
    \mathbb{E}\left[
        \hat\varepsilon_i^{(0)}\hat\varepsilon_{i'}^{(0)}
        \mathbb{I}\{|\tilde\lambda_{0,i,i'}|\geq\delta\}
    \right]
    &=\mu_{i,i'}\bar F^+_{0,i,i'}(\delta)
    \label{eq: oracle selected target product}
\end{align}
for every pair $i,i'$ and threshold $\delta\geq0$. Hence, we can conclude that
\begin{align}
    \mathbb{E}[\hat V_{n,d}(\delta)-V_n]
    &=\frac{1}{n^2}\sum_{i,i'=1}^n
    \mathbb{E}\left[
        \hat\varepsilon_i^{(0)}\hat\varepsilon_{i'}^{(0)}
        \mathbb{I}\{|\tilde\lambda_{0,i,i'}|\geq\delta\}
    \right]
    -\frac{1}{n^2}\sum_{i,i'=1}^n\lambda_{i,i'}
    +
    \frac{1}{n^2}\sum_{i,i'=1}^n\bar S_{i,i'}(\delta)
    \nonumber\\
    &=\frac{1}{n^2}\sum_{i,i'=1}^n
    \left(
        \mu_{i,i'}\bar F^+_{0,i,i'}(\delta)-\lambda_{i,i'}
    \right)
    +\frac{1}{n^2}\sum_{i,i'=1}^n\bar S_{i,i'}(\delta)~,
\end{align}
as required.\hfill\qed

\subsection{Proof of \cref{lem: decomposition approximation}}

Take the constant $c$ in the statement of the result no larger than the constant $c^\star$ in \cref{lem: basic inequalities n}. The identity \eqref{eq: residual covariance} and Part (ii) of \cref{lem: rotation and regression} give
\begin{align}
    (\tilde\varepsilon_0^{(j)})_{j=1}^d
    &\ci\hat\varepsilon^{(0)}
    \quad\text{and}\quad
    \Cov\left(
        \tilde\varepsilon_{0,i}^{(j)},
        \tilde\varepsilon_{0,i'}^{(j')}
    \right)
    =\mu_{i,i'}\omega_{j,j'}~,
    \label{eq: oracle residual regression conclusion}
\end{align}
for all $i,i'\in[n]$ and $j,j'\in[d]$.

Next, \cref{lem: basic inequalities n}, display \eqref{eq: n bound 2}, implies that $\psi_{i,i'}(M_n)>0$ for every pair $i,i'$. Choose an orthogonal matrix $U_d$ satisfying $\Omega_d =U_d\mathsf{Diag}(\theta_{d,1},\ldots,\theta_{d,d})U_d^\top$, and write $\widetilde{\varepsilon}_0 =(\tilde\varepsilon_0^{(1)},\ldots,\tilde\varepsilon_0^{(d)})$ for the oracle residual matrix. By \eqref{eq: oracle residual regression conclusion}, Part (i) of \cref{lem: rotation and regression}, applied with $A=M_n$ and $B=\Omega_d$, gives independent vectors $G^{(1)},\ldots,G^{(d)}$, each distributed as $\mathsf N(0,M_n)$, for which
\begin{align}
    \widetilde{\varepsilon}_0 U_d
    &\overset{d}{=}
    \left(
        \sqrt{\theta_{d,1}}G^{(1)},\ldots,
        \sqrt{\theta_{d,d}}G^{(d)}
    \right)
\end{align}
holds. There exist independent standard Gaussian variables $Z_1,\ldots,Z_d,W_1,\ldots,W_d$ such that
\begin{align}
    \left(G_i^{(j)},G_{i'}^{(j)}\right)_{j=1}^d
    &\overset{d}{=}
    \left(
        \sqrt{\mu_{i,i}}Z_j,
        \sqrt{\mu_{i',i'}}
        \left(
            \rho_{i,i'}(M_n)Z_j
            +\sqrt{1-\rho_{i,i'}(M_n)^2}W_j
        \right)
    \right)_{j=1}^d~.
\end{align}
Thus, we find that
\begin{align}
    \tilde\lambda_{0,i,i'}
    &=\frac{1}{d}
    \left(\widetilde \varepsilon_0\widetilde \varepsilon_0^\top\right)_{i,i'}
    \nonumber\\
    &=\frac{1}{d}
    \left(
        (\widetilde \varepsilon_0U_d)(\widetilde \varepsilon_0U_d)^\top
    \right)_{i,i'}
    \nonumber\\
    &\overset{d}{=}
    \frac{1}{d}\sum_{j=1}^d
    \theta_{d,j}G_i^{(j)}G_{i'}^{(j)}
    \overset{d}{=}
    \frac{\psi_{i,i'}(M_n)}{d}
    \sum_{j=1}^d\theta_{d,j}Z_j
    \left(
        \rho_{i,i'}(M_n)Z_j
        +\sqrt{1-\rho_{i,i'}(M_n)^2}W_j
    \right)~,
\end{align}
for every pair $i,i'$. Consequently, we obtain
\begin{align}
    \bar F^+_{0,i,i'}(\delta)
    &=H_{\delta,d}\left(
        \psi_{i,i'}(M_n),\rho_{i,i'}(M_n)
    \right)~,
    \label{eq: oracle screening tail representation}
\end{align}
for every pair $i,i'$ and threshold $\delta\geq0$.

\subsubsection{Part (i)}

Observe that 
\begin{align}
    \sup_{\delta\geq0}
    |\bar Q_{i,i'}(\delta)-Q_{i,i'}(\delta)|&
    \leq 
    \sup_{\delta\geq0}
    \left|
        \bar F^+_{0,i,i'}(\delta)
        -F^+\left(
            \delta,m_{i,i'}(M_n),v_{i,i'}(M_n)
        \right)
    \right| \nonumber \\
    & + \sup_{\delta\geq0}
    \left|
        F^+\left(
            \delta,m_{i,i'}(M_n),v_{i,i'}(M_n)
        \right)
        -
        F^+\left(
            \delta,\lambda_{i,i'},v_{i,i'}^{(d)}
        \right)
    \right|~.\label{eq: part i break}
\end{align}
We bound each of the terms in \eqref{eq: part i break} in succession. 

To bound the first term, we apply the following Gaussian approximation.

\begin{lem}
\label{lem: gaussian product normal approximation}$ $\\
Suppose that
\cref{assu: proportionality,assu: homoskedasticity,assu: well conditioned}
hold and $d\geq1$. If $\Xi$ is a positive semi-definite matrix satisfying
\begin{align}
    \Xi_{i,i},\Xi_{i',i'}
    &\in[\underline a,\overline a]
\end{align}
for fixed constants $0<\underline a<\overline a<\infty$, then
\begin{align}
    \sup_{\delta\geq0}
    \left|
        H_{\delta,d}\left(
            \psi_{i,i'}(\Xi),\rho_{i,i'}(\Xi)
        \right)
        -F^+\left(
            \delta,m_{i,i'}(\Xi),v_{i,i'}(\Xi)
        \right)
    \right|
    &\lesssim\frac{1}{\sqrt d}~,
    \label{eq: gaussian product normal approximation}
\end{align}
where the implicit constant may depend on $\underline a$ and
$\overline a$~.
\end{lem}

\noindent If $d=1$, the conclusion follows because the absolute difference between two probabilities is bounded by one. We therefore maintain $d\geq2$. Since $M_n$ is positive semi-definite, \cref{lem: basic inequalities n}, display \eqref{eq: n bound 2}, implies that its diagonal entries lie in a fixed compact subset of $(0,\infty)$. Hence, \cref{lem: gaussian product normal approximation} and \eqref{eq: oracle screening tail representation} give
\begin{align}
    \sup_{\delta\geq0}
    \left|
        \bar F^+_{0,i,i'}(\delta)
        -F^+\left(
            \delta,m_{i,i'}(M_n),v_{i,i'}(M_n)
        \right)
    \right|
    &\lesssim\frac{1}{\sqrt d}
    \label{eq: oracle component two tail approximation}
\end{align}
uniformly over pairs $i,i'\in[n]$.

To bound the second term, observe that the bound $|\lambda_{i,i'}|\leq1$, \cref{lem: basic inequalities n}, display \eqref{eq: n bound 3}, and \cref{lem: auxiliary outcome covariance bounds}, display \eqref{eq: auxiliary outcome covariance bounds}, imply that
\begin{align}
    |m_{i,i'}(M_n)-\lambda_{i,i'}|
    &\leq
    \frac{\operatorname{tr}(\Omega_d)}{d}
    |\mu_{i,i'}-\lambda_{i,i'}|
    +
    \left|
        \frac{\operatorname{tr}(\Omega_d)}{d}-1
    \right|
    |\lambda_{i,i'}|
    \lesssim
    V_n+\frac{1}{d}~.
    \label{eq: oracle normal mean comparison}
\end{align}
The identity $x^2-y^2=(x-y)(x+y)$ and \cref{lem: basic inequalities n}, displays \eqref{eq: n bound 2} and \eqref{eq: n bound 3}, give
\begin{align}
    &\left|
        \mu_{i,i}\mu_{i',i'}+\mu_{i,i'}^2
        -(1+\lambda_{i,i'}^2)
    \right|
    \nonumber\\
    &\quad\leq
    |\mu_{i,i}-1||\mu_{i',i'}|
    +|\mu_{i',i'}-1|
    +|\mu_{i,i'}-\lambda_{i,i'}|
    \left(
        |\mu_{i,i'}|+|\lambda_{i,i'}|
    \right)
    \lesssim V_n~.
    \label{eq: oracle normal variance factor comparison}
\end{align}
Moreover, we can write
\begin{align}
    \frac{v_{i,i'}(M_n)}{v_{i,i'}^{(d)}}
    &=
    \frac{\|\Omega_d\|_{\mathrm F}^2}
    {\|\bar\Gamma_d\|_{\mathrm F}^2}
    \frac{
        \mu_{i,i}\mu_{i',i'}+\mu_{i,i'}^2
    }{
        1+\lambda_{i,i'}^2
    }~.
\end{align}
Since $1\leq1+\lambda_{i,i'}^2\leq2$, \eqref{eq: oracle normal variance factor comparison} controls the second factor by $1+O(V_n)$. In turn, \cref{lem: auxiliary outcome covariance bounds}, display \eqref{eq: auxiliary outcome covariance bounds}, controls the first factor by $1+O(d^{-1})$ and gives $\|\bar\Gamma_d\|_{\mathrm F}^2\asymp d$. Consequently, we find that
\begin{align}
    v_{i,i'}(M_n)
    &\asymp
    v_{i,i'}^{(d)}
    \asymp\frac{1}{d}
    \quad\text{and}\quad
    \left|
        \frac{v_{i,i'}(M_n)}{v_{i,i'}^{(d)}}-1
    \right|
    \lesssim
    V_n+\frac{1}{d}~,
    \label{eq: oracle normal variance comparison}
\end{align}
uniformly over pairs $i,i'$. The first relation in \eqref{eq: oracle normal variance comparison} places the variance ratio in a fixed compact subset of $(0,\infty)$. Hence, the mean value theorem applied to $x\mapsto\frac{1}{2}\log x$, together with \eqref{eq: oracle normal mean comparison} and \eqref{eq: oracle normal variance comparison}, gives
\begin{align}
    &\frac{
        |m_{i,i'}(M_n)-\lambda_{i,i'}|
    }{
        \sqrt{v_{i,i'}(M_n)}
    }
    +
    \left|
        \log\left(
            \sqrt{\frac{v_{i,i'}(M_n)}{v_{i,i'}^{(d)}}}
        \right)
    \right|
    \lesssim
    \sqrt d\left(
        V_n+\frac{1}{d}
    \right)
    +V_n+\frac{1}{d}
    \lesssim
    \frac{1}{\sqrt d}
    +V_n\sqrt d~.
\end{align}
To complete the proof, we apply the elementary inequality
\begin{align}
    \sup_{x\in\mathbb R}
    \left|
        \Phi\left(\frac{x-u}{\sigma}\right)
        -\Phi\left(\frac{x-\widetilde u}{\widetilde\sigma}\right)
    \right|
    &\lesssim
    \frac{|u-\widetilde u|}{\sigma}
    +\left|
        \log\left(\frac{\sigma}{\widetilde\sigma}\right)
    \right|
    \label{eq: gaussian location scale comparison}
\end{align}
for $\sigma,\widetilde\sigma>0$ and
$u,\widetilde u\in\mathbb R$. Observe that applying \eqref{eq: gaussian location scale comparison} twice gives the same bound, up to a universal constant, for two-sided tail probabilities. Applying this two-sided bound with the $M_n$-based normal as the first distribution and the $\lambda_{i,i'}$-based normal as the second gives
\begin{align}
    \sup_{\delta\geq0}
    \left|
        F^+\left(
            \delta,m_{i,i'}(M_n),v_{i,i'}(M_n)
        \right)
        -
        F^+\left(
            \delta,\lambda_{i,i'},v_{i,i'}^{(d)}
        \right)
    \right|
    &\lesssim
    \frac{1}{\sqrt d}
    +V_n\sqrt d~.
    \label{eq: oracle tail stability cdf comparison}
\end{align}
The proof is complete by plugging \eqref{eq: oracle component two tail approximation} and \eqref{eq: oracle tail stability cdf comparison} into \eqref{eq: part i break}.\hfill\qed

\subsubsection{Part (ii)}

Define the quantities
\begin{align}
    \Delta\psi_{i,i'}
    & =\psi_{i,i'}(\widetilde M_n)-\psi_{i,i'}(M_n)
    \quad\text{and}\nonumber \\
    \Delta H_{i,i'}(\delta)
    &=H_{\delta,d}\left(
        \psi_{i,i'}(\widetilde M_n),
        \rho_{i,i'}(\widetilde M_n)
    \right)
    -H_{\delta,d}\left(
        \psi_{i,i'}(M_n),
        \rho_{i,i'}(M_n)
    \right)~.\nonumber 
\end{align}
By \cref{lem: exact feasible screening representation} and
\eqref{eq: oracle screening tail representation}, we have
\begin{align}
    P\left\{
        |\hat\lambda_{i,i'}|\geq\delta
        \mid\hat\varepsilon^{(0)}
    \right\}
    &=H_{\delta,d}\left(
        \psi_{i,i'}(\widetilde M_n),
        \rho_{i,i'}(\widetilde M_n)
    \right)\quad\text{and}
    \nonumber\\
    P\{|\tilde\lambda_{0,i,i'}|\geq\delta\}
    &=H_{\delta,d}\left(
        \psi_{i,i'}(M_n),
        \rho_{i,i'}(M_n)
    \right)
    \label{eq: feasible conditional screening representations}
\end{align}
respectively. The result is based on the consideration of each of the components of the decomposition
\begin{align}
    \frac{1}{n^2}\sum_{i,i'=1}^n\bar S_{i,i'}(\delta)
    & = \frac{1}{n^2}\sum_{i,i'=1}^n\mathbb{E}\left[
        \hat\varepsilon_i^{(0)}
        \hat\varepsilon_{i'}^{(0)}
        \Delta H_{i,i'}(\delta)
    \right] \nonumber \\
    &=\frac{\beta_{\delta,d}}{n^2}
    \sum_{(i,i')\notin\mathcal S_n}
    \mathbb{E}\left[
        \hat\varepsilon_i^{(0)}
        \hat\varepsilon_{i'}^{(0)}
        \Delta\psi_{i,i'}
    \right]
    +
    \frac{1}{n^2}\sum_{(i,i')\in\mathcal S_n}
    \mathbb{E}\left[
        \hat\varepsilon_i^{(0)}
        \hat\varepsilon_{i'}^{(0)}
        \Delta H_{i,i'}(\delta)
    \right]
    \nonumber\\
    &+
    \frac{1}{n^2}\sum_{(i,i')\notin\mathcal S_n}
    \mathbb{E}\left[
        \hat\varepsilon_i^{(0)}
        \hat\varepsilon_{i'}^{(0)}
        \left(
            \Delta H_{i,i'}(\delta)
            -\beta_{\delta,d}\Delta\psi_{i,i'}
        \right)
    \right]~,
    \label{eq: feasible residualization exact aggregate decomposition}
\end{align}
where the equality follows from \eqref{eq: oracle residual regression conclusion} and \eqref{eq: feasible conditional screening representations}. We control each of the terms in \eqref{eq: feasible residualization exact aggregate decomposition} through the application of the following Lemma.

\begin{lem}
\label{lem: feasible residualization aggregates}$ $\\
Suppose that \cref{assu: gaussianity,assu: proportionality,assu: homoskedasticity,assu: sparsity,assu: well conditioned} hold, $d\geq2$, and $s/n\leq c$, where $c>0$ is a sufficiently small constant.

\noindent\textbf{(i)} It holds that
\begin{align}
    \sup_{\delta\geq0}
    \left|
        \frac{\beta_{\delta,d}}{n^2}
        \sum_{(i,i')\notin\mathcal S_n}
        \mathbb{E}\left[
            \hat\varepsilon_i^{(0)}
            \hat\varepsilon_{i'}^{(0)}
            \Delta\psi_{i,i'}
        \right]
    \right|
    &\lesssim
    \frac{s}{n}V_n\sqrt d~.
    \label{eq: feasible null scale aggregate}
\end{align}

\noindent\textbf{(ii)} It holds that
\begin{align}
    \sup_{\delta\geq0}
    \frac{1}{n^2}\sum_{(i,i')\in\mathcal S_n}
    \mathbb{E}\left[
        |\hat\varepsilon_i^{(0)}\hat\varepsilon_{i'}^{(0)}|
        |\Delta H_{i,i'}(\delta)|
    \right]
    &\lesssim
    \frac{V_n}{\sqrt d}
    +\frac{s}{n}V_n\sqrt d~.
    \label{eq: feasible sparse pair aggregate}
\end{align}

\noindent\textbf{(iii)} Fix $\eta\in[0,1]$. There exist constants
$c>0$ and $C<\infty$, such that, if
\begin{align}
    V_n
    \left(
        \sqrt{
            d\log(V_n^{-1})
        }
        +\log(V_n^{-1})
    \right)
    &\leq c~,
    \label{eq: feasible aggregate local condition}
\end{align}
then
\begin{align}
    &\sup_{\delta\in\Delta_{n,d}(\eta)}
    \frac{1}{n^2}\sum_{(i,i')\notin\mathcal S_n}
    \mathbb{E}\left[
        |\hat\varepsilon_i^{(0)}\hat\varepsilon_{i'}^{(0)}|
        |\Delta H_{i,i'}(\delta)
        -\beta_{\delta,d}\Delta\psi_{i,i'}|
    \right]
    \nonumber\\
    &\leq C
    V_n^{2+\eta}
    \left(
        d\log(V_n^{-1})
        +\log^2\left(V_n^{-1}\right)
    \right)~.
    \label{eq: feasible null pair aggregate}
\end{align}
\end{lem}

\noindent By Plugging Parts (i)-(iii) of \cref{lem: feasible residualization aggregates} into \eqref{eq: feasible residualization exact aggregate decomposition}, we obtain
\begin{align}
    \sup_{\delta\in\Delta_{n,d}(\eta)}
    \left|
        \frac{1}{n^2}\sum_{i,i'=1}^n\bar S_{i,i'}(\delta)
    \right|
    &\lesssim
    \frac{V_n}{\sqrt d}
    +\frac{s}{n}V_n\sqrt d
    +
    V_n^{2+\eta}
    \left(
        d\log(V_n^{-1})
        +\log^2\left(V_n^{-1}\right)
    \right)~,\label{eq: combine three parts}
\end{align}
as required.\hfill\qed

\subsection{Proof of \cref{lem: gaussian product normal approximation}}

Since the conclusion is immediate when $d=1$, we take $d\geq2$. By \cref{lem: auxiliary outcome covariance bounds}, the matrix $\Omega_d$ is positive semi-definite, and display \eqref{eq: auxiliary outcome covariance bounds} gives $\|\Omega_d\|_{\mathrm F}^2\asymp d$. Hence, its eigenvalues are nonnegative and have a positive squared sum. Now, let $G^{(1)},\ldots,G^{(d)}$ be independently distributed $\mathsf N(0,\Xi)$ and define
\begin{align}
    L_{i,i'}(\Xi)
    &=\frac{1}{d}\sum_{j=1}^d
    \theta_{d,j}G_i^{(j)}G_{i'}^{(j)}~.
\end{align}
Observe that we can write
\begin{align}
    H_{\delta,d}\left(
        \psi_{i,i'}(\Xi),\rho_{i,i'}(\Xi)
    \right)
    &=P\{|L_{i,i'}(\Xi)|\geq\delta\}~.
    \label{eq: gaussian product approximation representation}
\end{align}
Moreover, \cref{lem: wick} implies that
\begin{align}
    \mathbb E[G_i^{(j)}G_{i'}^{(j)}]
    &=\Xi_{i,i'}
    \quad\text{and}\quad
    \Var(G_i^{(j)}G_{i'}^{(j)})
    =\Xi_{i,i}\Xi_{i',i'}+\Xi_{i,i'}^2
    \label{eq: gaussian product approximation moments}
\end{align}
and so $L_{i,i'}(\Xi)$ has mean $m_{i,i'}(\Xi)$ and variance $v_{i,i'}(\Xi)>0$, where positivity follows from the lower bound on the diagonal entries of $\Xi$ and the positive squared sum established above.

The result is obtained through the application of the following standard Berry--Esseen inequality.

\begin{lem}[Theorem XVI.5.2, \citealp{feller1991introduction}]
\label{lem: berry esseen}$ $\\
Let $X_1,\ldots,X_m$ be independent variables such that
\begin{align}
    \mathbb{E}[X_k] = 0~,
    \quad
    \mathbb{E}[|X_k|^2] = \sigma_k^2~,
    \quad\text{and}\quad
    \mathbb{E}[|X_k|^3] = \zeta_k~.
\end{align}
If $\sum_{j=1}^m\sigma_j^2>0$, then the inequality
\begin{align}
    \sup_{x\in\mathbb R}
    \left|
        P\left\{
            \frac{\sum_{j=1}^mX_j}
            {(\sum_{j=1}^m\sigma_j^2)^{1/2}}
            \leq x
        \right\}
        -\Phi(x)
    \right|
    &\lesssim
    \frac{\sum_{j=1}^m\zeta_j}
    {(\sum_{j=1}^m\sigma_j^2)^{3/2}}
\end{align}
holds.
\end{lem}

\noindent To apply this result, take
\begin{align}
    X_j
    &=
    \frac{\theta_{d,j}}{d}
    \left(
        G_i^{(j)}G_{i'}^{(j)}-\Xi_{i,i'}
    \right)~.
\end{align}
The variables $X_1,\ldots,X_d$ are independent, mean zero, and satisfy the equality
\begin{align}
    L_{i,i'}(\Xi)
    &=
    m_{i,i'}(\Xi)+\sum_{j=1}^dX_j~.
\end{align}
Consequently, \cref{lem: berry esseen} implies that
\begin{align}
    \sup_{x\in\mathbb R}
    \left|
        P\{L_{i,i'}(\Xi)\leq x\}
        -
        P\{Z_{m_{i,i'}(\Xi),v_{i,i'}(\Xi)}\leq x\}
    \right|
    &\lesssim
    \frac{\sum_{j=1}^d\mathbb E[|X_j|^3]}
    {\left(\sum_{j=1}^d\mathbb E[X_j^2]\right)^{3/2}}~.
    \label{eq: gaussian product approximation berry esseen}
\end{align}
Observe that we can evaluate
\begin{align}
    \sum_{j=1}^d\mathbb E[X_j^2]
    &=
    \frac{1}{d^2}
    \left(
        \sum_{j=1}^d\theta_{d,j}^2
    \right)
    \left(
        \Xi_{i,i}\Xi_{i',i'}+\Xi_{i,i'}^2
    \right)
    =v_{i,i'}(\Xi)~,
    \label{eq: gaussian product approximation second moment}
\end{align}
by \eqref{eq: gaussian product approximation moments}. In turn, the Cauchy--Schwarz inequality implies that
\begin{align}
    \mathbb E\left[
        |G_i^{(j)}G_{i'}^{(j)}|^3
    \right]
    &\leq
    \left(
        \mathbb E[|G_i^{(j)}|^6]
        \mathbb E[|G_{i'}^{(j)}|^6]
    \right)^{1/2}
    \lesssim
    (\Xi_{i,i}\Xi_{i',i'})^{3/2}~.
\end{align}
Thus, using $|x-y|^3 \lesssim |x|^3+|y|^3$ together with
$|\Xi_{i,i'}|\leq (\Xi_{i,i}\Xi_{i',i'})^{1/2}$, we find that
\begin{align}
    \mathbb E\left[
        \left|
            G_i^{(j)}G_{i'}^{(j)}-\Xi_{i,i'}
        \right|^3
    \right]
    &\lesssim
    \left(
        \Xi_{i,i}\Xi_{i',i'}+\Xi_{i,i'}^2
    \right)^{3/2}
    \label{eq: gaussian product approximation third moment}
\end{align}
and that thereby
\begin{align}
    \mathbb E[|X_j|^3]
    &\lesssim
    \frac{\theta_{d,j}^3}{d^3}
    \left(
        \Xi_{i,i}\Xi_{i',i'}+\Xi_{i,i'}^2
    \right)^{3/2}~.
\end{align}
Combining the preceding bound with
\eqref{eq: gaussian product approximation second moment}, we obtain
\begin{align}
    \frac{\sum_{j=1}^d\mathbb E[|X_j|^3]}
    {\left(\sum_{j=1}^d\mathbb E[X_j^2]\right)^{3/2}}
    &\lesssim
    \frac{\sum_{j=1}^d\theta_{d,j}^3}
    {\left(\sum_{j=1}^d\theta_{d,j}^2\right)^{3/2}}
    \leq
    \frac{\max_{j\in[d]}\theta_{d,j}}
    {\left(\sum_{j=1}^d\theta_{d,j}^2\right)^{1/2}}
    \lesssim\frac{1}{\sqrt d}~,
    \label{eq: gaussian product approximation spectral ratio}
\end{align}
where the final inequality follows from \cref{lem: auxiliary outcome covariance bounds}, display \eqref{eq: auxiliary outcome weight and scale bounds}. Substituting \eqref{eq: gaussian product approximation spectral ratio} into \eqref{eq: gaussian product approximation berry esseen} gives
\begin{align}
    \sup_{x\in\mathbb R}
    \left|
        P\{L_{i,i'}(\Xi)\leq x\}
        -
        P\{Z_{m_{i,i'}(\Xi),v_{i,i'}(\Xi)}\leq x\}
    \right|
    &\lesssim\frac{1}{\sqrt d}~.
    \label{eq: gaussian product approximation cdf}
\end{align}
Taking left limits in \eqref{eq: gaussian product approximation cdf} gives the same bound for strict distribution functions. For $\delta>0$, we can write
\begin{align}
    P\{|L_{i,i'}(\Xi)|\geq\delta\}
    &=P\{L_{i,i'}(\Xi)\leq-\delta\}
    +1-P\{L_{i,i'}(\Xi)<\delta\}~.
\end{align}
Applying the weak and strict distribution-function bounds at $-\delta$ and $\delta$, respectively, and using \eqref{eq: gaussian product approximation representation} proves the required bound uniformly over $\delta>0$. At $\delta=0$, both probabilities in \eqref{eq: gaussian product normal approximation} equal one.
\hfill\qed

\subsection{Proof of \cref{lem: feasible residualization aggregates}}

We begin by collecting several facts and Lemmas that will be used to prove more than one part of the Lemma. Take the constant $c$ in the statement no larger than the constant $c^\star$ in \cref{lem: basic inequalities n}. Define
\begin{align}
    R_{i,i'}
    &=\max\left\{
        |\tilde\mu_{i,i}-\mu_{i,i}|,\,
        |\tilde\mu_{i',i'}-\mu_{i',i'}|,\,
        |\tilde\mu_{i,i'}-\mu_{i,i'}|
    \right\}
    \label{eq: feasible covariance perturbation magnitude}
\end{align}
for each pair $i,i'$. 

\begin{lem}
\label{lem: projection residualization bounds}$ $\\
Suppose that \cref{assu: gaussianity,assu: proportionality,assu: homoskedasticity,assu: sparsity,assu: well conditioned} hold. For every positive integer $m$, it holds that
\begin{align}
    \|\tilde\mu_{i,i'}-\mu_{i,i'}\|_{2m}
    &\lesssim mV_n~,
    \label{eq: feasible projection higher moments}
\end{align}
uniformly over $i,i'\in[n]$ and $m$. Moreover, it holds that
\begin{align}
    \|\tilde\mu_{i,i'}-\mu_{i,i'}\|_4
    +\mathbb E\left[
        |\hat\varepsilon_k^{(0)}\hat\varepsilon_{k'}^{(0)}|
        |\tilde\mu_{i,i'}-\mu_{i,i'}|
    \right]
    &\lesssim V_n
    \label{eq: feasible projection first moments}
\end{align}
and
\begin{align}
    \mathbb E\left[
        |\hat\varepsilon_k^{(0)}\hat\varepsilon_{k'}^{(0)}|
        |\tilde\mu_{i,i'}-\mu_{i,i'}|^2
    \right]
    &\lesssim V_n^2
    \label{eq: feasible projection second moments}
\end{align}
uniformly over $i,i',k,k'\in[n]$.

\end{lem}

\begin{lem}
\label{lem: screening derivative bounds}$ $\\
Suppose that \cref{assu: proportionality,assu: well conditioned} hold. Fix
$0<\underline a<1<\overline a<\infty$ and $\bar\rho\in(0,1)$. The map $(a,\rho)\mapsto H_{\delta,d}(a,\rho)$ is twice continuously differentiable on $[\underline a,\overline a]\times[-\bar\rho,\bar\rho]$. If
$p=H_{\delta,d}(a,\rho)>0$, then the bounds
\begin{align}
    \|\nabla H_{\delta,d}(a,\rho)\|_2
    &\lesssim
    p\left(
        \sqrt{d(1+\log(p^{-1}))}
        +1+\log(p^{-1})
    \right)\quad\text{and}
    \nonumber\\
    \|\nabla^2H_{\delta,d}(a,\rho)\|_{\mathrm{op}}
    &\lesssim
    p\left(
        d(1+\log(p^{-1}))
        +(1+\log(p^{-1}))^2
    \right)
    \label{eq: screening derivative probability bounds}
\end{align}
hold. If $p=0$, both derivatives vanish. Consequently, the bounds
\begin{align}
    \sup_{\delta\geq0}
    \sup_{\substack{
        a\in[\underline a,\overline a]\\
        |\rho|\leq\bar\rho
    }}
    \|\nabla H_{\delta,d}(a,\rho)\|_2
    &\lesssim\sqrt d
    \quad\text{and}\quad
    \sup_{\delta\geq0}|\beta_{\delta,d}|
    \lesssim\sqrt d
    \label{eq: uniform screening gradient}
\end{align}
hold, where the implicit constants may depend on $\underline a,\overline a$, and $\bar\rho$.
\end{lem}

\noindent By \cref{lem: basic inequalities n}, display
\eqref{eq: n bound 2}, there exist fixed constants
$0<\underline a<1<\overline a<\infty$ such that
\begin{align}
    \mu_{i,i}
    &\in[\underline a,\overline a]
    \label{eq: feasible centered diagonal rectangle}
\end{align}
uniformly over $i\in[n]$.

\subsubsection{Part (i)}

For $u,v\in[\underline a,\overline a]$ and $u+x,v+y\geq0$, Taylor's theorem when $|x|+|y|\leq\underline a/2$ and the arithmetic--geometric mean inequality when $|x|+|y|>\underline a/2$ give
\begin{align}
    \left|
        \sqrt{(u+x)(v+y)}
        -\sqrt{uv}
        -\frac{x+y}{2}
    \right|
    &\lesssim
    (|u-1|+|v-1|)(|x|+|y|)
    +x^2+y^2~.
\end{align}
Since $M_n$ and $\widetilde M_n$ are positive semi-definite, applying this inequality to their diagonal entries and using
\cref{lem: basic inequalities n}, display \eqref{eq: n bound 3}, gives
\begin{align}
    &\left|
        \Delta\psi_{i,i'}
        -\frac{1}{2}\left(
            \tilde\mu_{i,i}-\mu_{i,i}
            +\tilde\mu_{i',i'}-\mu_{i',i'}
        \right)
    \right|
    \nonumber\\
    &\quad\lesssim
    V_n
    \left(
        |\tilde\mu_{i,i}-\mu_{i,i}|
        +|\tilde\mu_{i',i'}-\mu_{i',i'}|
    \right)
    +|\tilde\mu_{i,i}-\mu_{i,i}|^2
    +|\tilde\mu_{i',i'}-\mu_{i',i'}|^2~.
    \label{eq: feasible scale perturbation bound}
\end{align}
Multiplying \eqref{eq: feasible scale perturbation bound} by $|\hat\varepsilon_i^{(0)}\hat\varepsilon_{i'}^{(0)}|$, taking expectations, and summing over $i,i'$ gives
\begin{align}
    &\frac{1}{n^2}\sum_{i,i'=1}^n
    \mathbb E\left[
        |\hat\varepsilon_i^{(0)}\hat\varepsilon_{i'}^{(0)}|
        \left|
            \Delta\psi_{i,i'}
            -\frac{1}{2}\left(
                \tilde\mu_{i,i}-\mu_{i,i}
                +\tilde\mu_{i',i'}-\mu_{i',i'}
            \right)
        \right|
    \right]
    \lesssim V_n^2~,
    \label{eq: feasible aggregate scale remainder}
\end{align}
where the final inequality follows from \cref{lem: projection residualization bounds}, displays \eqref{eq: feasible projection first moments} and \eqref{eq: feasible projection second moments}.

Now, observe that
\begin{align}
    &\sum_{i,i'=1}^n
    \hat\varepsilon_i^{(0)}\hat\varepsilon_{i'}^{(0)}
    \left(
        \tilde\mu_{i,i}-\mu_{i,i}
        +\tilde\mu_{i',i'}-\mu_{i',i'}
    \right)
    =
    2
    \left(
        \sum_{i=1}^n
        \hat\varepsilon_i^{(0)}
        \left(
            \tilde\mu_{i,i}-\mu_{i,i}
        \right)
    \right)
    \left(
        \sum_{i'=1}^n
        \hat\varepsilon_{i'}^{(0)}
    \right)
    =0~,
    \label{eq: centering cancellation}
\end{align}
by the equality $\sum_{i=1}^n\hat\varepsilon_i^{(0)}=0$. Hence, we have that
\begin{align}
    &\sum_{(i,i')\notin\mathcal S_n}
    \hat\varepsilon_i^{(0)}\hat\varepsilon_{i'}^{(0)}
    \left(
        \tilde\mu_{i,i}-\mu_{i,i}
        +\tilde\mu_{i',i'}-\mu_{i',i'}
    \right)
    =
    -\sum_{(i,i')\in\mathcal S_n}
    \hat\varepsilon_i^{(0)}\hat\varepsilon_{i'}^{(0)}
    \left(
        \tilde\mu_{i,i}-\mu_{i,i}
        +\tilde\mu_{i',i'}-\mu_{i',i'}
    \right)~.
    \label{eq: complement centering cancellation}
\end{align}
Adding and subtracting the linear term in \eqref{eq: feasible scale perturbation bound}, applying \eqref{eq: complement centering cancellation}, and using the triangle inequality gives
\begin{align}
    &\frac{1}{n^2}
    \left|
        \sum_{(i,i')\notin\mathcal S_n}
        \mathbb E\left[
            \hat\varepsilon_i^{(0)}
            \hat\varepsilon_{i'}^{(0)}
            \Delta\psi_{i,i'}
        \right]
    \right|
    \nonumber\\
    &\quad\leq
    \frac{1}{2n^2}
    \sum_{(i,i')\in\mathcal S_n}
    \mathbb E\left[
        |\hat\varepsilon_i^{(0)}\hat\varepsilon_{i'}^{(0)}|
        \left(
            |\tilde\mu_{i,i}-\mu_{i,i}|
            +|\tilde\mu_{i',i'}-\mu_{i',i'}|
        \right)
    \right]
    \nonumber\\
    &\quad+
    \frac{1}{n^2}
    \sum_{(i,i')\notin\mathcal S_n}
    \mathbb E\left[
        |\hat\varepsilon_i^{(0)}\hat\varepsilon_{i'}^{(0)}|
        \left|
            \Delta\psi_{i,i'}
            -\frac{1}{2}\left(
                \tilde\mu_{i,i}-\mu_{i,i}
                +\tilde\mu_{i',i'}-\mu_{i',i'}
            \right)
        \right|
    \right]
    \nonumber\\
    &\quad\lesssim
    \frac{|\mathcal S_n|}{n^2}
    V_n
    +V_n^2
    \lesssim
    \frac{s+1}{n}V_n
    \lesssim
    \frac{s}{n}V_n~,
\end{align}
where the first $\lesssim$ inequality follows from \cref{lem: projection residualization bounds}, display \eqref{eq: feasible projection first moments}, and \eqref{eq: feasible aggregate scale remainder}, the second follows from \cref{assu: sparsity} and \cref{lem: basic inequalities n}, display \eqref{eq: var bounds}, and the third uses $s\geq1$. Therefore, \cref{lem: screening derivative bounds}, display \eqref{eq: uniform screening gradient}, implies that
\begin{align}
    &\sup_{\delta\geq0}
    \bigg|
        \frac{\beta_{\delta,d}}{n^2}
        \sum_{(i,i')\notin\mathcal S_n}
        \mathbb E\left[
            \hat\varepsilon_i^{(0)}
            \hat\varepsilon_{i'}^{(0)}
            \Delta\psi_{i,i'}
        \right]
    \bigg|
    \lesssim
    \frac{s}{n}V_n\sqrt d~,
\end{align}
as required.\hfill\qed

\subsubsection{Part (ii)}

Fix $\rho_\star\in(0,1)$ and divide the set $\mathcal S_n$ into
\begin{align}
    \mathcal S_n^{\mathrm w}
    &=
    \left\{
        (i,i')\in\mathcal S_n:
        i\neq i',\ 
        |\lambda_{i,i'}|\leq\rho_\star
    \right\}
    \quad\text{and}\quad
    \mathcal S_n^{\mathrm s}
    =
    \mathcal S_n\setminus\mathcal S_n^{\mathrm w}~.
    \label{eq: feasible weak strong pairs}
\end{align}
The argument treats the two sets differently. For the weak pairs, the derivative bounds from \cref{lem: screening derivative bounds} apply. For the strong pairs, we instead use a Gaussian approximation.

We first record a parameter perturbation bound that applies to both sets. By \eqref{eq: feasible centered diagonal rectangle}, after choosing $c$ sufficiently small, on the event $\{R_{i,i'}\leq c\}$ the diagonal entries of $M_n$, $\widetilde M_n$, and every matrix on the line segment joining them lie in a fixed compact subset of $(0,\infty)$. Positive semi-definiteness also uniformly bounds the corresponding off-diagonal entries along this line segment. Hence, the derivatives of the maps defining $\psi_{i,i'}(\cdot)$ and $\rho_{i,i'}(\cdot)$ are uniformly bounded there. The mean value theorem therefore gives
\begin{align}
    |\Delta\psi_{i,i'}|
    +
    \left|
        \rho_{i,i'}(\widetilde M_n)
        -\rho_{i,i'}(M_n)
    \right|
    &\lesssim R_{i,i'}
    \label{eq: feasible parameter perturbation}
\end{align}
on $\{R_{i,i'}\leq c\}$, uniformly over $i,i'$.

We first consider the weak pairs. For a sufficiently small $\rho_\star$, the bounds \eqref{eq: basic centering deviation} and \eqref{eq: feasible centered diagonal rectangle} imply that
\begin{align}
    \psi_{i,i'}(M_n)
    &\in[\underline a,\overline a]
    \quad\text{and}\quad
    |\rho_{i,i'}(M_n)|
    \leq\bar\rho
    \label{eq: feasible weak parameter rectangle}
\end{align}
for every $(i,i')\in\mathcal S_n^{\mathrm w}$ and some fixed $\bar\rho<1$. We choose the constants in this rectangle with sufficient slack and then decrease $c$ if necessary. It follows from \eqref{eq: feasible parameter perturbation} that, on $\{R_{i,i'}\leq c\}$, the parameter pairs associated with $M_n$, $\widetilde M_n$, and every matrix between them remain in a fixed rectangle on which \cref{lem: screening derivative bounds} applies. The mean value theorem, \eqref{eq: feasible parameter perturbation}, and \cref{lem: screening derivative bounds} therefore give
\begin{align}
    &|\Delta H_{i,i'}(\delta)|
    \mathbb{I}\{R_{i,i'}\leq c\}
    \nonumber\\
    &\quad\lesssim
    \sqrt d
    \left(
        |\Delta\psi_{i,i'}|
        +
        \left|
            \rho_{i,i'}(\widetilde M_n)
            -\rho_{i,i'}(M_n)
        \right|
    \right)
    \mathbb{I}\{R_{i,i'}\leq c\}
    \lesssim
    \sqrt d\,R_{i,i'}
    \mathbb{I}\{R_{i,i'}\leq c\}~.
\end{align}
On the complementary event $\{R_{i,i'} > c\}$, we use
\begin{align}
    |\Delta H_{i,i'}(\delta)|
    &\leq1
    \quad\text{and}\quad
    \mathbb{I}\{R_{i,i'}>c\}
    \lesssim R_{i,i'}^2~.
\end{align}
Splitting the expectation over the two events thus gives
\begin{align}
    &\sup_{\delta\geq0}
    \mathbb E\left[
        |\hat\varepsilon_i^{(0)}\hat\varepsilon_{i'}^{(0)}|
        |\Delta H_{i,i'}(\delta)|
    \right]
    \nonumber\\
    &\quad\lesssim
    \sqrt d\,
    \mathbb E\left[
        |\hat\varepsilon_i^{(0)}\hat\varepsilon_{i'}^{(0)}|
        R_{i,i'}
    \right]
    +
    \mathbb E\left[
        |\hat\varepsilon_i^{(0)}\hat\varepsilon_{i'}^{(0)}|
        R_{i,i'}^2
    \right]
    \lesssim
    V_n\sqrt d
    +
    V_n^2
    \lesssim
    V_n\sqrt d~,
    \label{eq: feasible weak pair bound}
\end{align}
where the second inequality follows from \cref{lem: projection residualization bounds}, displays \eqref{eq: feasible projection first moments} and \eqref{eq: feasible projection second moments}, while the final inequality follows from \cref{lem: basic inequalities n}, display \eqref{eq: var bounds}. Hence, we find that
\begin{align}
    &\sup_{\delta\geq0}
    \frac{1}{n^2}
    \sum_{(i,i')\in\mathcal S_n^{\mathrm w}}
    \mathbb E\left[
        |\hat\varepsilon_i^{(0)}\hat\varepsilon_{i'}^{(0)}|
        |\Delta H_{i,i'}(\delta)|
    \right]
    \lesssim
    \frac{s}{n}
    V_n\sqrt d~,
    \label{eq: feasible weak pair aggregate}
\end{align}
since \cref{assu: sparsity} implies that
$|\mathcal S_n^{\mathrm w}|\lesssim ns$.

We next consider the strong pairs. Since $\lambda_{i,i}=1>\rho_\star$, the bound \eqref{eq: feasible weak strong pairs} gives
\begin{align}
    \mathcal S_n^{\mathrm s}
    &=\left\{
        (i,i')\in[n]^2:
        |\lambda_{i,i'}|>\rho_\star
    \right\}
    \quad\text{and hence}
    \nonumber\\
    |\mathcal S_n^{\mathrm s}|
    &\leq
    \frac{1}{\rho_\star}
    \sum_{i,i'=1}^n|\lambda_{i,i'}|
    \lesssim
    n^2V_n~,
    \label{eq: feasible strong pair cardinality}
\end{align}
where the final inequality follows from  \cref{lem: basic inequalities n}, display \eqref{eq: n bound 1}. On $\{R_{i,i'}\leq c\}$, the diagonal entries of $M_n$ and $\widetilde M_n$ lie in a fixed compact subset of $(0,\infty)$, so \cref{lem: gaussian product normal approximation} applies to both matrices.

We first compare the means and variances of the two normal approximations. By \cref{lem: auxiliary outcome covariance bounds}, display \eqref{eq: auxiliary outcome covariance bounds}, $\|\Omega_d\|_{\mathrm F}>0$. Moreover, Cauchy--Schwarz gives $|\operatorname{tr}(\Omega_d)|\leq \sqrt d\|\Omega_d\|_{\mathrm F}$. Thus, \eqref{eq: feasible normal parameters}, \eqref{eq: feasible centered diagonal rectangle}, and the definition of $R_{i,i'}$ give, on $\{R_{i,i'}\leq c\}$, that
\begin{align}
    \frac{
        |m_{i,i'}(\widetilde M_n)-m_{i,i'}(M_n)|
    }{
        \sqrt{v_{i,i'}(M_n)}
    }
    &=
    \frac{|\operatorname{tr}(\Omega_d)|}
    {\|\Omega_d\|_{\mathrm F}}
    \frac{
        |\tilde\mu_{i,i'}-\mu_{i,i'}|
    }{
        \left(
            \mu_{i,i}\mu_{i',i'}+\mu_{i,i'}^2
        \right)^{1/2}
    }
    \lesssim\sqrt d\,R_{i,i'}~,
    \label{eq: feasible strong normal perturbation}
\end{align}
and, by the mean value theorem,
\begin{align}
    \frac{1}{2}
    \left|
        \log\left(
            \frac{
                \tilde\mu_{i,i}\tilde\mu_{i',i'}
                +\tilde\mu_{i,i'}^2
            }{
                \mu_{i,i}\mu_{i',i'}+\mu_{i,i'}^2
            }
        \right)
    \right|
    \lesssim R_{i,i'}~.
\end{align}
Now, on $\{R_{i,i'}\leq c\}$, the triangle inequality gives
\begin{align}
    |\Delta H_{i,i'}(\delta)|
    &\leq
    \left|
        H_{\delta,d}\left(
            \psi_{i,i'}(\widetilde M_n),
            \rho_{i,i'}(\widetilde M_n)
        \right)
        -
        F^+\left(
            \delta,
            m_{i,i'}(\widetilde M_n),
            v_{i,i'}(\widetilde M_n)
        \right)
    \right|
    \nonumber\\
    &\quad+
    \left|
        F^+\left(
            \delta,
            m_{i,i'}(\widetilde M_n),
            v_{i,i'}(\widetilde M_n)
        \right)
        -
        F^+\left(
            \delta,
            m_{i,i'}(M_n),
            v_{i,i'}(M_n)
        \right)
    \right|
    \nonumber\\
    &\quad+
    \left|
        F^+\left(
            \delta,
            m_{i,i'}(M_n),
            v_{i,i'}(M_n)
        \right)
        -
        H_{\delta,d}\left(
            \psi_{i,i'}(M_n),
            \rho_{i,i'}(M_n)
        \right)
    \right|~.
    \label{eq: feasible strong gaussian decomposition}
\end{align}
For the first and third terms, \cref{lem: gaussian product normal approximation}, display \eqref{eq: gaussian product normal approximation}, gives
\begin{align}
    \sup_{\delta\geq0}
    \left|
        H_{\delta,d}\left(
            \psi_{i,i'}(\Xi),
            \rho_{i,i'}(\Xi)
        \right)
        -
        F^+\left(
            \delta,
            m_{i,i'}(\Xi),
            v_{i,i'}(\Xi)
        \right)
    \right|
    &\lesssim\frac{1}{\sqrt d}~,
    \qquad
    \Xi\in\{M_n,\widetilde M_n\}~.
    \label{eq: feasible strong approximating tail}
\end{align}
For the middle term, applying \eqref{eq: gaussian location scale comparison} at the two endpoints $-\delta$ and $\delta$ gives
\begin{align}
    &\sup_{\delta\geq0}
    \left|
        F^+\left(
            \delta,
            m_{i,i'}(\widetilde M_n),
            v_{i,i'}(\widetilde M_n)
        \right)
        -
        F^+\left(
            \delta,
            m_{i,i'}(M_n),
            v_{i,i'}(M_n)
        \right)
    \right|
    \nonumber\\
    &\quad\lesssim
    \frac{
        |m_{i,i'}(\widetilde M_n)-m_{i,i'}(M_n)|
    }{
        \sqrt{v_{i,i'}(M_n)}
    }
    +
    \left|
        \log\left(
            \sqrt{
                \frac{v_{i,i'}(\widetilde M_n)}
                {v_{i,i'}(M_n)}
            }
        \right)
    \right|
    \lesssim
    \sqrt d\,R_{i,i'}~,
    \label{eq: feasible strong gaussian comparison}
\end{align}
where the final inequality follows from \eqref{eq: feasible strong normal perturbation}. Combining \eqref{eq: feasible strong gaussian decomposition}, \eqref{eq: feasible strong approximating tail}, and \eqref{eq: feasible strong gaussian comparison} gives
\begin{align}
    |\Delta H_{i,i'}(\delta)|
    \mathbb{I}\{R_{i,i'}\leq c\}
    &\lesssim
    \left(
        \frac{1}{\sqrt d}
        +\sqrt d\,R_{i,i'}
    \right)
    \mathbb{I}\{R_{i,i'}\leq c\}
    \label{eq: feasible strong pair event bound}
\end{align}
uniformly over $\delta\geq0$. 

On the complementary event $\{R_{i,i'} > c\}$, we again use $|\Delta H_{i,i'}(\delta)|\leq1$ and $\mathbb{I}\{R_{i,i'}>c\}\lesssim R_{i,i'}^2$. It follows that
\begin{align}
    &\sup_{\delta\geq0}
    \mathbb E\left[
        |\hat\varepsilon_i^{(0)}\hat\varepsilon_{i'}^{(0)}|
        |\Delta H_{i,i'}(\delta)|
    \right]
    \nonumber\\
    &\quad\lesssim
    \frac{1}{\sqrt d}
    \mathbb E\left[
        |\hat\varepsilon_i^{(0)}\hat\varepsilon_{i'}^{(0)}|
    \right]
    +
    \sqrt d\,
    \mathbb E\left[
        |\hat\varepsilon_i^{(0)}\hat\varepsilon_{i'}^{(0)}|
        R_{i,i'}
    \right]
    +
    \mathbb E\left[
        |\hat\varepsilon_i^{(0)}\hat\varepsilon_{i'}^{(0)}|
        R_{i,i'}^2
    \right]
    \nonumber\\
    &\quad\lesssim
    \frac{1}{\sqrt d}
    +V_n\sqrt d
    +V_n^2~,
    \label{eq: feasible strong pair bound}
\end{align}
where the final inequality follows from Cauchy--Schwarz, \eqref{eq: feasible centered diagonal rectangle}, and \cref{lem: projection residualization bounds}, displays \eqref{eq: feasible projection first moments} and \eqref{eq: feasible projection second moments}. Combining \eqref{eq: feasible strong pair bound} with \eqref{eq: feasible strong pair cardinality} yields
\begin{align}
    &\sup_{\delta\geq0}
    \frac{1}{n^2}
    \sum_{(i,i')\in\mathcal S_n^{\mathrm s}}
    \mathbb E\left[
        |\hat\varepsilon_i^{(0)}\hat\varepsilon_{i'}^{(0)}|
        |\Delta H_{i,i'}(\delta)|
    \right]
    \nonumber\\
    &\quad\lesssim
    V_n
    \left(
        \frac{1}{\sqrt d}
        +V_n\sqrt d
        +V_n^2
    \right)
    \nonumber\\
    &\quad=
    \frac{V_n}{\sqrt d}
    +V_n^2\sqrt d
    +V_n^3
    \lesssim
    \frac{V_n}{\sqrt d}
    +
    \frac{s}{n}
    V_n\sqrt d~.
    \label{eq: feasible strong pair aggregate}
\end{align}
The final inequality follows from \cref{lem: basic inequalities n}, display \eqref{eq: var bounds}, $s\geq1$, and $d\geq2$. Finally, \eqref{eq: feasible weak strong pairs} partitions $\mathcal S_n$ into the weak and strong pairs. Combining \eqref{eq: feasible weak pair aggregate} and \eqref{eq: feasible strong pair aggregate} therefore completes the proof.\hfill\qed

\subsubsection{Part (iii)}

The bound is obtained through the application of the following Lemma.

\begin{lem}
\label{lem: rare null pair screening}$ $\\
Suppose that \cref{assu: gaussianity,assu: proportionality,assu: homoskedasticity,assu: sparsity,assu: well conditioned} hold and $d\geq2$. Fix $\eta\in[0,1]$. There exist constants $c,C>0$ such that, if $s/n\leq c$ and
\begin{align}
    V_n
    \left(
        \sqrt{
            d\log(V_n^{-1})
        }
        +\log(V_n^{-1})
    \right)
    &\leq c~,
    \label{eq: rare null local condition}
\end{align}
then the following conclusions hold.

\noindent\textbf{(i)} Uniformly over
$\delta\in\Delta_{n,d}(\eta)$, it holds that
\begin{align}
    H_{\delta,d}(1,0)
    &\leq CV_n^\eta,
    \label{eq: rare null probability envelope}
\end{align}
and
\begin{align}
    |\beta_{\delta,d}|
    &\leq C
    V_n^\eta
    \left(
        \sqrt{
            d\log(V_n^{-1})
        }
        +\log(V_n^{-1})
    \right)~.
\end{align}

\noindent\textbf{(ii)} Uniformly over
$\delta\in\Delta_{n,d}(\eta)$ and pairs satisfying
$\lambda_{i,i'}=0$, on the event $\{R_{i,i'}\leq c\}$, it holds that
\begin{align}
    &|\Delta H_{i,i'}(\delta)
    -\beta_{\delta,d}\Delta\psi_{i,i'}|
    \nonumber\\
    &\quad\leq C
    V_n^\eta
    \left(
        d\log(V_n^{-1})
        +\log^2\left(V_n^{-1}\right)
    \right)
    \nonumber\\
    &\qquad\cdot
    \exp\left(
        C\left(
            \sqrt{
                d\log(V_n^{-1})
            }
            +\log(V_n^{-1})
        \right)R_{i,i'}
    \right)
    \left(
        V_n^2+R_{i,i'}^2
    \right)~.
    \label{eq: rare null likelihood expansion}
\end{align}

\noindent\textbf{(iii)} Uniformly over $i,i'\in[n]$, it holds that
\begin{align}
    & \mathbb E\bigg[
        |\hat\varepsilon_i^{(0)}\hat\varepsilon_{i'}^{(0)}|
        \left(V_n^2+R_{i,i'}^2\right)
        \\
        & \qquad\cdot
        \exp\left(
            C\left(
                \sqrt{
                    d\log(V_n^{-1})
                }
                +\log(V_n^{-1})
            \right)R_{i,i'}
        \right)
    \bigg]
    \leq CV_n^2~.
    \label{eq: feasible projection exponential moment}
\end{align}
\end{lem}

\noindent Fix $\delta\in\Delta_{n,d}(\eta)$. For every
$(i,i')\notin\mathcal S_n$, \cref{lem: projection residualization bounds}, \cref{lem: wick}, and positive semi-definiteness give
\begin{align}
    \|R_{i,i'}\|_4
    &\lesssim V_n~,
    &
    \|\hat\varepsilon_i^{(0)}\hat\varepsilon_{i'}^{(0)}\|_4
    &\lesssim1~,
    &
    |\Delta\psi_{i,i'}|\mathbb I\{R_{i,i'}>c\}
    &\lesssim R_{i,i'}~.
\end{align} 
Since $\lambda_{i,i'}=0$, Part (i) of \cref{lem: rare null pair screening}, Markov's inequality, and H\"older's inequality yield
\begin{align}
    &\mathbb E\left[
        |\hat\varepsilon_i^{(0)}\hat\varepsilon_{i'}^{(0)}|
        \mathbb{I}\{R_{i,i'}>c\}
        |\Delta H_{i,i'}(\delta)
        -\beta_{\delta,d}\Delta\psi_{i,i'}|
    \right]
    \nonumber\\
    &\quad\leq
    \|\hat\varepsilon_i^{(0)}\hat\varepsilon_{i'}^{(0)}\|_4
    P\{R_{i,i'}>c\}^{3/4}
    +
    |\beta_{\delta,d}|
    \|\hat\varepsilon_i^{(0)}\hat\varepsilon_{i'}^{(0)}\|_4
    \|R_{i,i'}\|_4P\{R_{i,i'}>c\}^{1/2}
    \nonumber\\
    &\quad\lesssim
    V_n^3
    \left(
        1+V_n^\eta
        \left(
            \sqrt{
                d\log(V_n^{-1})
            }
            +\log(V_n^{-1})
        \right)
    \right)
    \lesssim
    V_n^{2+\eta}
    \left(
        d\log(V_n^{-1})
        +\log^2\left(V_n^{-1}\right)
    \right)~.
\end{align}
On $\{R_{i,i'}\leq c\}$, Parts (ii) and (iii) of \cref{lem: rare null pair screening} and non-negativity give
\begin{align}
    &\mathbb E\left[
        |\hat\varepsilon_i^{(0)}\hat\varepsilon_{i'}^{(0)}|
        \mathbb{I}\{R_{i,i'}\leq c\}
        |\Delta H_{i,i'}(\delta)
        -\beta_{\delta,d}\Delta\psi_{i,i'}|
    \right]
    \nonumber\\
    &\quad\lesssim
    V_n^\eta
    \left(
        d\log(V_n^{-1})
        +\log^2\left(V_n^{-1}\right)
    \right)
    \nonumber\\
    &\qquad\cdot
    \mathbb E\left[
        |\hat\varepsilon_i^{(0)}\hat\varepsilon_{i'}^{(0)}|
        \left(V_n^2+R_{i,i'}^2\right)
        \exp\left(
            C\left(
                \sqrt{
                    d\log(V_n^{-1})
                }
                +\log(V_n^{-1})
            \right)R_{i,i'}
        \right)
    \right]
    \nonumber\\
    &\quad\lesssim
    V_n^{2+\eta}
    \left(
        d\log(V_n^{-1})
        +\log^2\left(V_n^{-1}\right)
    \right)~.
\end{align}
Combining the preceding two bounds, summing over at most $n^2$ null
pairs, and taking the supremum over $\delta\in\Delta_{n,d}(\eta)$ proves
\eqref{eq: feasible null pair aggregate}.\hfill\qed

\subsection{Proof of \cref{lem: projection residualization bounds}}

\cref{assu: well conditioned}, applied to $\Lambda_n$, implies that
\begin{align}
    u^\top M_nu
    &=
    (H_nu)^\top\Lambda_n(H_nu)
    \gtrsim
    \|H_nu\|_2^2~,
    \label{eq: projection residualization lower bound}
\end{align}
for every $u\in\mathbb R^n$. \Cref{lem: basic inequalities n}, display
\eqref{eq: n rank and diagonal}, and \cref{rem: positivity} imply that $\|\hat\varepsilon^{(0)}\|_2>0$ almost surely. Moreover, we have that
\begin{align}
    \|M_n^{1/2}e_i\|_2^2
    &=\mu_{i,i}
    \lesssim1
    \quad\text{and}\quad
    \|M_n^{3/2}e_i\|_2^2
    =(M_n^3)_{i,i}
    \leq
    \|M_n\|_{\mathrm{op}}^2\mu_{i,i}
    \lesssim
    n^2V_n^2~.
    \label{eq: projection residualization coordinate bounds}
\end{align}
by positive semi-definiteness of $M_n$ and \cref{lem: basic inequalities n}, displays \eqref{eq: n bound 2} and \eqref{eq: n bound 4}.

Now, let $Z\sim\mathsf N(0,I_n)$. Since $H_n$ is an orthogonal
projection of rank $n-1$, \citet[Theorem~1.5.6]{muirhead1982} implies
that $H_nZ/\|H_nZ\|_2$ is uniform on the unit sphere in the range of $H_n$. Moreover, \citet[Theorem~1.5.7(ii)]{muirhead1982}, applied to the rank-one projection onto the span of $H_nv$, gives
\begin{align}
    \frac{
        (v^\top H_nZ)^2
    }{
        \|H_nv\|_2^2\|H_nZ\|_2^2
    }
    &\sim
    \operatorname{Beta}\left(
        \frac{1}{2},
        \frac{n-2}{2}
    \right)
\end{align}
when $H_nv\neq0$. The conclusion is immediate when $H_nv=0$. Consequently, the moments of the beta distribution give
\begin{align}
    \mathbb E\left[
        \left|
            v^\top\frac{H_nZ}{\|H_nZ\|_2}
        \right|^{2q}
    \right]
    &=
    \|H_nv\|_2^{2q}
    \frac{
        \Gamma(q+1/2)\Gamma((n-1)/2)
    }{
        \Gamma(1/2)\Gamma(q+(n-1)/2)
    }
    \nonumber\\
    &=
    \|H_nv\|_2^{2q}
    \prod_{l=0}^{q-1}
    \frac{2l+1}{n-1+2l}
    \leq
    \left(
        \frac{2q}{n-1}
    \right)^q
    \|H_nv\|_2^{2q}\label{eq: projection gaussian direction product}
\end{align}
for every deterministic $v,w\in\mathbb R^n$ and positive integer $q$.

Observe that $H_nM_n=M_nH_n=M_n$ and that we can write
\begin{align}
    \tilde\mu_{i,i'}-\mu_{i,i'}
    &=
    -\frac{
        \hat\varepsilon_i^{(0)}
        (M_n\hat\varepsilon^{(0)})_{i'}
        +(M_n\hat\varepsilon^{(0)})_i
        \hat\varepsilon_{i'}^{(0)}
    }{
        \|\hat\varepsilon^{(0)}\|_2^2
    }
    +
    \frac{
        (\hat\varepsilon^{(0)})^\top
        M_n\hat\varepsilon^{(0)}
    }{
        \|\hat\varepsilon^{(0)}\|_2^4
    }
    \hat\varepsilon_i^{(0)}
    \hat\varepsilon_{i'}^{(0)}~.
    \label{eq: feasible covariance perturbation}
\end{align}
The Gaussian representation in \cref{rem: positivity}, the identity
$M_n^{1/2}H_n=M_n^{1/2}$, and
\eqref{eq: projection residualization lower bound} imply that
\begin{align}
    \left\|
        \frac{
            \hat\varepsilon_i^{(0)}
            (M_n\hat\varepsilon^{(0)})_{i'}
        }{
            \|\hat\varepsilon^{(0)}\|_2^2
        }
    \right\|_m
    &\lesssim
    \frac{m}{n}
    \|M_n^{1/2}e_i\|_2
    \|M_n^{3/2}e_{i'}\|_2
    \lesssim
    mV_n~,
    \label{eq: projection residualization linear term}
\end{align}
where the final inequality follows from
\eqref{eq: projection residualization coordinate bounds}. The same
bound holds for the second term in
\eqref{eq: feasible covariance perturbation}. Moreover, positive
semi-definiteness of $M_n$ gives
\begin{align}
    (\hat\varepsilon^{(0)})^\top
    M_n\hat\varepsilon^{(0)}
    &\leq
    \|M_n\|_{\mathrm{op}}
    \|\hat\varepsilon^{(0)}\|_2^2~.
\end{align}
Thus, another application of
\eqref{eq: projection gaussian direction product}, followed by
\cref{lem: basic inequalities n}, display \eqref{eq: n bound 4}, and
\eqref{eq: projection residualization coordinate bounds}, gives
\begin{align}
    &\left\|
        \frac{
            (\hat\varepsilon^{(0)})^\top
            M_n\hat\varepsilon^{(0)}
        }{
            \|\hat\varepsilon^{(0)}\|_2^4
        }
        \hat\varepsilon_i^{(0)}
        \hat\varepsilon_{i'}^{(0)}
    \right\|_m
    \nonumber\\
    &\quad\leq
    \|M_n\|_{\mathrm{op}}
    \left\|
        \frac{
            \hat\varepsilon_i^{(0)}
            \hat\varepsilon_{i'}^{(0)}
        }{
            \|\hat\varepsilon^{(0)}\|_2^2
        }
    \right\|_m
    \lesssim
    nV_n
    \frac{m}{n}
    \|M_n^{1/2}e_i\|_2
    \|M_n^{1/2}e_{i'}\|_2
    \lesssim
    mV_n~.
    \label{eq: projection residualization quadratic term}
\end{align}
Combining \eqref{eq: feasible covariance perturbation},
\eqref{eq: projection residualization linear term}, and
\eqref{eq: projection residualization quadratic term}, with $2m$ in
place of $m$, we conclude that
\begin{align}
    \|\tilde\mu_{i,i'}-\mu_{i,i'}\|_{2m}
    &\lesssim
    mV_n~.
    \label{eq: projection residualization all moments}
\end{align}
This bound holds uniformly over $i,i'\in[n]$ and the positive integers
$m$.

Finally, \cref{assu: gaussianity}, \cref{lem: wick},
\cref{lem: basic inequalities n}, display \eqref{eq: n bound 2}, and
H\"older's inequality imply that
\begin{align}
    \|\hat\varepsilon_k^{(0)}
    \hat\varepsilon_{k'}^{(0)}\|_2
    &\leq
    \|\hat\varepsilon_k^{(0)}\|_4
    \|\hat\varepsilon_{k'}^{(0)}\|_4
    \lesssim1~.
\end{align}
Cauchy--Schwarz and
\eqref{eq: projection residualization all moments}, applied with $m=1$
and $m=2$, respectively, therefore give
\begin{align}
    \mathbb E\left[
        |\hat\varepsilon_k^{(0)}\hat\varepsilon_{k'}^{(0)}|
        |\tilde\mu_{i,i'}-\mu_{i,i'}|
    \right]
    &\lesssim V_n
    \quad\text{and}\quad
    \nonumber\\
    \mathbb E\left[
        |\hat\varepsilon_k^{(0)}\hat\varepsilon_{k'}^{(0)}|
        |\tilde\mu_{i,i'}-\mu_{i,i'}|^2
    \right]
    &\lesssim V_n^2~.
\end{align}
Combining these bounds with
\eqref{eq: projection residualization all moments} proves
\eqref{eq: feasible projection higher moments},
\eqref{eq: feasible projection first moments} and
\eqref{eq: feasible projection second moments} uniformly over the
stated indices and positive integers.\hfill\qed

\subsection{Proof of \cref{lem: screening derivative bounds}}

Fix $\delta\geq0$. For $(a,\rho)$ in the stated rectangle, let $P_{a,\rho}$ denote the law under which the pairs $(G_j,\widetilde G_j)$, $j\in[d]$, are independent and each has a centered Gaussian distribution with covariance matrix
\begin{align}
    \Sigma_{a,\rho}
    &=
    a
    \begin{pmatrix}
        1&\rho\\
        \rho&1
    \end{pmatrix}~.
\end{align}
The density of one such pair is
\begin{align}
    f_{a,\rho}(g,\widetilde g)
    &=
    \frac{1}{
        2\pi a\sqrt{1-\rho^2}
    }
    \exp\left(
        -\frac{
            g^2-2\rho g\widetilde g+\widetilde g^2
        }{
            2a(1-\rho^2)
        }
    \right)~.
    \label{eq: screening gaussian pair density}
\end{align}
Moreover, we can write
\begin{align}
    H_{\delta,d}(a,\rho)
    &=P_{a,\rho}\{\mathcal A_{\delta,d}\}~,
    \quad\text{where}\quad
    \mathcal A_{\delta,d}
    =
    \left\{
        \left|
            \frac{1}{d}\sum_{j=1}^d
            \theta_{d,j}G_j\widetilde G_j
        \right|
        \geq\delta
    \right\}~.
    \label{eq: screening likelihood event}
\end{align}
In particular, the event $\mathcal A_{\delta,d}$ does not depend on $(a,\rho)$.

For $r\in\{a,\rho\}$, define the total likelihood score by
\begin{align}
    \mathcal U_r
    &=
    \frac{\mathrm d}{\mathrm d r}
    \log\left(
        \prod_{j=1}^d
        f_{a,\rho}(G_j,\widetilde G_j)
    \right)~.
\end{align}
On an open neighborhood of the parameter rectangle, the eigenvalues of $\Sigma_{a,\rho}$ are uniformly bounded above and away from zero. Consequently, each parameter derivative through order two of the joint density is bounded by a fixed polynomial in the observations times an integrable Gaussian density. Differentiation under the integral sign therefore gives
\begin{align}
    \frac{\mathrm d}{\mathrm d r}H_{\delta,d}(a,\rho)
    &=
    \mathbb E_{a,\rho}\left[
        \mathbb{I}\{\mathcal A_{\delta,d}\}\mathcal U_r
    \right]\quad\text{and}
    \nonumber\\
    \frac{\mathrm d^2}{\mathrm d t\,\mathrm d r}
    H_{\delta,d}(a,\rho)
    &=
    \mathbb E_{a,\rho}\left[
        \mathbb{I}\{\mathcal A_{\delta,d}\}
        \left(
            \mathcal U_r\mathcal U_t
            +\frac{\mathrm d \mathcal U_r}{\mathrm d t}
        \right)
    \right]
    \label{eq: screening likelihood differentiation}
\end{align}
for $r,t\in\{a,\rho\}$. These derivatives are continuous on the parameter rectangle by the same argument.

We next bound the random variables in \eqref{eq: screening likelihood differentiation}. Direct differentiation of \eqref{eq: screening gaussian pair density} gives
\begin{align}
    \mathcal U_a
    &=-\frac{d}{a}
    +\frac{1}{2a^2(1-\rho^2)}
    \sum_{j=1}^d
    \left(
        G_j^2-2\rho G_j\widetilde G_j+\widetilde G_j^2
    \right)
    \quad\text{and}\\
    \mathcal U_\rho
    &=\frac{d\rho}{1-\rho^2}
    +\frac{1}{a(1-\rho^2)^2}
    \sum_{j=1}^d
    \left(
        (1+\rho^2)G_j\widetilde G_j
        -\rho(G_j^2+\widetilde G_j^2)
    \right)~.
    \label{eq: screening likelihood scores}
\end{align}
Under $P_{a,\rho}$, we may write
\begin{align}
    G_j
    &=\sqrt a\,Z_j~,
    \quad
    \widetilde G_j
    =
    \sqrt a\left(
        \rho Z_j+\sqrt{1-\rho^2}W_j
    \right)~,
\end{align}
where $Z_1,\ldots,Z_d,W_1,\ldots,W_d$ are independent standard Gaussian variables. Substitution into \eqref{eq: screening likelihood scores} gives
\begin{align}
    \mathcal U_a
    &=\frac{1}{2a}
    \sum_{j=1}^d
    \left(
        Z_j^2+W_j^2-2
    \right)\quad\text{and}\quad
    \mathcal U_\rho
    =\sum_{j=1}^d
    \left(
        \frac{Z_jW_j}{\sqrt{1-\rho^2}}
        -\frac{\rho}{1-\rho^2}(W_j^2-1)
    \right)~.
    \label{eq: screening standardized likelihood scores}
\end{align}
Thus, each likelihood score is a centered Gaussian quadratic form. Differentiating \eqref{eq: screening likelihood scores} and substituting the same representation shows that $\frac{\mathrm d \mathcal U_r}{\mathrm d t}-\mathbb E_{a,\rho}[\frac{\mathrm d \mathcal U_r}{\mathrm d t}]$ is also a centered Gaussian quadratic form for $r,t\in\{a,\rho\}$. The matrices defining these quadratic forms are block diagonal and satisfy
\begin{align}
    \|\mathcal Q\|_{\mathrm{op}}
    &\lesssim1
    \quad\text{and}\quad
    \|\mathcal Q\|_{\mathrm F}^2
    \lesssim d~.
    \label{eq: screening score matrix bounds}
\end{align}
Direct differentiation also show that
\begin{align}
    \left|
        \mathbb E_{a,\rho}\left[
            \frac{\mathrm d \mathcal U_r}{\mathrm d t}
        \right]
    \right|
    &\lesssim d~.
    \label{eq: screening likelihood derivative mean}
\end{align}
Applying \cref{lem: gaussian quadratic form tail}, display
\eqref{eq: gaussian derivative quadratic tail}, to the centered quadratic
forms above and using \eqref{eq: screening score matrix bounds} gives
\begin{align}
    P_{a,\rho}\{|\mathcal U_r|>x\}
    &\leq
    C\exp\left(
        -c\min\left\{
            \frac{x^2}{d},x
        \right\}
    \right)\quad\text{and}
    \nonumber\\
    P_{a,\rho}\left\{
        \left|
            \frac{\mathrm d \mathcal U_r}{\mathrm d t}
            -\mathbb E_{a,\rho}\left[
                \frac{\mathrm d \mathcal U_r}{\mathrm d t}
            \right]
        \right|>x
    \right\}
    &\leq
    C\exp\left(
        -c\min\left\{
            \frac{x^2}{d},x
        \right\}
    \right)
    \label{eq: screening likelihood score tails}
\end{align}
uniformly over the parameter rectangle and $x\geq0$.

Let $\mathcal A$ be an event satisfying $P_{a,\rho}\{\mathcal A\}=p>0$. For every $x\geq0$, tail integration gives
\begin{align}
    \mathbb E_{a,\rho}\left[
        \mathbb{I}\{\mathcal A\}|\mathcal U_r|
    \right]
    &\leq
    px+
    \int_x^\infty
    P_{a,\rho}\{|\mathcal U_r|>u\}\,\mathrm du
    \quad\text{and}
    \nonumber\\
    \mathbb E_{a,\rho}\left[
        \mathbb{I}\{\mathcal A\}\mathcal U_r^2
    \right]
    &\leq
    px^2+
    2\int_x^\infty
    uP_{a,\rho}\{|\mathcal U_r|>u\}\,\mathrm du~.
\end{align}
Taking
\begin{align}
    x
    &=C\left(
        \sqrt{d(1+\log(p^{-1}))}
        +1+\log(p^{-1})
    \right)
\end{align}
and applying \eqref{eq: screening likelihood score tails} therefore gives
\begin{align}
    \mathbb E_{a,\rho}\left[
        \mathbb{I}\{\mathcal A\}|\mathcal U_r|
    \right]
    &\lesssim
    p\left(
        \sqrt{d(1+\log(p^{-1}))}
        +1+\log(p^{-1})
    \right)\quad\text{and}
    \nonumber\\
    \mathbb E_{a,\rho}\left[
        \mathbb{I}\{\mathcal A\}\mathcal U_r^2
    \right]
    &\lesssim
    p\left(
        d(1+\log(p^{-1}))
        +(1+\log(p^{-1}))^2
    \right)~.
    \label{eq: screening likelihood event score moments}
\end{align}
The second bound in \eqref{eq: screening likelihood score tails} and \eqref{eq: screening likelihood derivative mean} similarly give
\begin{align}
    \mathbb E_{a,\rho}\left[
        \mathbb{I}\{\mathcal A\}
        \left|
            \frac{\mathrm d \mathcal U_r}{\mathrm d t}
        \right|
    \right]
    &\lesssim
    p\left(
        d(1+\log(p^{-1}))
        +(1+\log(p^{-1}))^2
    \right)~.
    \label{eq: screening likelihood event derivative moment}
\end{align}
Suppose now that $p=H_{\delta,d}(a,\rho)>0$. Applying \eqref{eq: screening likelihood event score moments} and \eqref{eq: screening likelihood event derivative moment} with $\mathcal A=\mathcal A_{\delta,d}$ in \eqref{eq: screening likelihood differentiation}, together with $|\mathcal U_r\mathcal U_t|\leq\frac{1}{2}\left(\mathcal U_r^2+\mathcal U_t^2\right)$ proves \eqref{eq: screening derivative probability bounds}. If $p=0$, the strictly positive Gaussian densities in \eqref{eq: screening gaussian pair density} imply that $\mathcal A_{\delta,d}$ has probability zero at every point in the parameter rectangle. The map $H_{\delta,d}$ is then identically zero on the rectangle, and both derivatives vanish. 

To complete the proof, observe that right-hand sides in \eqref{eq: screening derivative probability bounds} are uniformly bounded over $p\in(0,1]$. Maximizing the first bound in \eqref{eq: screening derivative probability bounds} over $p\in(0,1]$ therefore gives the first conclusion in \eqref{eq: uniform screening gradient}. Since $(1,0)$ belongs to the parameter rectangle and $\beta_{\delta,d}$ is the $a$-coordinate derivative at that point, the second conclusion follows as well.
\hfill\qed

\subsection{Proof of \cref{lem: rare null pair screening}}

Fix a rectangle \([\underline a,\overline a]\times[-\bar\rho,\bar\rho]\) with \((1,0)\) in its interior. If \(\lambda_{i,i'}=0\), then \cref{lem: basic inequalities n}, displays \eqref{eq: n bound 2} and \eqref{eq: n bound 3}, and the mean value theorem give
\begin{align}
    \left\|
        \left(
            \psi_{i,i'}(M_n)-1,
            \rho_{i,i'}(M_n)
        \right)
    \right\|_2
    &\lesssim V_n~.
    \label{eq: rare oracle parameter bound}
\end{align}
Moreover, on \(\{R_{i,i'}\leq c\}\), \eqref{eq: feasible parameter perturbation} gives
\begin{align}
    \left\|
        \left(
            \psi_{i,i'}(\widetilde M_n)-\psi_{i,i'}(M_n),
            \rho_{i,i'}(\widetilde M_n)-\rho_{i,i'}(M_n)
        \right)
    \right\|_2
    &\lesssim R_{i,i'}~.
    \label{eq: rare feasible parameter perturbation}
\end{align}
By taking $c$ sufficiently small, \cref{lem: basic inequalities n}, displays \eqref{eq: var bounds} and \eqref{eq: n bound 2}, and the preceding two displays ensure that $\log(1/V_n)\geq1$ and that the parameter pairs associated with \(M_n\) and \(\widetilde M_n\), as well as the segments joining them to \((1,0)\), lie in the rectangle \([\underline a,\overline a]\times[-\bar\rho,\bar\rho]\) on \(\{R_{i,i'}\leq c\}\).

\subsubsection{Part (i)}

Since $s/n\leq c$ for a sufficiently small $c$, \cref{assu: sparsity} ensures that there exists a pair satisfying \(\lambda_{i,i'}=0\). Fix such a pair. Fix \(\delta\in\Delta_{n,d}(\eta)\). If $\eta=0$, then
\begin{align}
    H_{\delta,d}(1,0)
    &\leq1
    =V_n^\eta~.
\end{align}
Moreover, by taking $c$ sufficiently small, \cref{lem: basic inequalities n}, display \eqref{eq: var bounds}, gives $\log(1/V_n)\geq1$. Hence, \cref{lem: screening derivative bounds}, display \eqref{eq: uniform screening gradient}, gives
\begin{align}
    |\beta_{\delta,d}|
    &\lesssim\sqrt d
    V_n^\eta
    \left(
        \sqrt{
            d\log(V_n^{-1})
        }
        +\log(V_n^{-1})
    \right)~.
\end{align}
Thus, Part (i) holds when $\eta=0$. We therefore maintain that $\eta\in(0,1]$. If \(H_{\delta,d}(1,0)=0\), then \cref{lem: screening derivative bounds} gives \(\beta_{\delta,d}=0\), and the conclusion follows. We therefore maintain that \(H_{\delta,d}(1,0)>0\). The strictly positive density in \eqref{eq: screening gaussian pair density} implies that the Gaussian measures associated with all parameter pairs in the rectangle \([\underline a,\overline a]\times[-\bar\rho,\bar\rho]\) are mutually absolutely continuous. Hence, \(H_{\delta,d}\) is strictly positive throughout this rectangle.

Minkowski's inequality, monotonicity of $L_p$ norms, and \cref{lem: projection residualization bounds}, display \eqref{eq: feasible projection first moments}, give
\begin{align}
    \|R_{i,i'}\|_2
    &\leq
    \|\tilde\mu_{i,i}-\mu_{i,i}\|_2
    +\|\tilde\mu_{i',i'}-\mu_{i',i'}\|_2
    +\|\tilde\mu_{i,i'}-\mu_{i,i'}\|_2
    \lesssim V_n~.
    \label{eq: rare feasible parameter bounds}
\end{align}
Moreover, \eqref{eq: feasible conditional screening representations} and $\delta\in\Delta_{n,d}(\eta)$ give
\begin{align}
    \mathbb E\left[
        H_{\delta,d}\left(
            \psi_{i,i'}(\widetilde M_n),
            \rho_{i,i'}(\widetilde M_n)
        \right)
    \right]
    &=
    \bar F_{i,i'}^+(\delta)
    \leq
        V_n^\eta~.
    \label{eq: rare feasible probability envelope}
\end{align}
The result is obtained through the application of the following Lemma.

\begin{lem}
\label{lem: rare screening probability transport}$ $\\
Let \((a_0,\rho_0)\) and \((a_1,\rho_1)\) be points in the rectangle \([\underline a,\overline a]\times[-\bar\rho,\bar\rho]\), and suppose that \(H_{\delta,d}\) is strictly positive throughout the rectangle. There exists a constant \(c>0\), depending only on the rectangle, such that, if
\begin{align}
    &\|(a_1-a_0,\rho_1-\rho_0)\|_2
    \left(
        \sqrt{
            d\left(
                1+\log\left(
                    H_{\delta,d}(a_0,\rho_0)^{-1}
                \right)
            \right)
        }
        +1+\log\left(
            H_{\delta,d}(a_0,\rho_0)^{-1}
        \right)
    \right)
    \leq c~,
    \label{eq: rare screening transport condition}
\end{align}
then
\begin{align}
    \left|
        \log\left(
            \frac{H_{\delta,d}(a_1,\rho_1)}
            {H_{\delta,d}(a_0,\rho_0)}
        \right)
    \right|
    &\lesssim
    \|(a_1-a_0,\rho_1-\rho_0)\|_2
    \nonumber\\
    &\quad\cdot
    \left(
        \sqrt{
            d\left(
                1+\log\left(
                    H_{\delta,d}(a_0,\rho_0)^{-1}
                \right)
            \right)
        }
        +1+\log\left(
            H_{\delta,d}(a_0,\rho_0)^{-1}
        \right)
    \right)
    \lesssim1~,
\end{align}
where the implicit constant depends only on the rectangle.
\end{lem}

\noindent Define the event
\begin{align}
    \mathcal E_{i,i'}
    &=
    \left\{
        R_{i,i'}
        \leq CV_n
    \right\}~.\label{eq: rare screening good event}
\end{align}
By choosing $C$ sufficiently large, Markov's inequality and \eqref{eq: rare feasible parameter bounds} imply that
\begin{align}
    P\{\mathcal E_{i,i'}^c\}
    &\leq
    \frac{
        \mathbb E\left[R_{i,i'}^2\right]
    }{
        C^2V_n^2
    }
    \leq\frac{1}{2}~.
    \label{eq: rare screening good event probability}
\end{align}
Now, suppose, toward a contradiction, that
\begin{align}
    H_{\delta,d}(1,0)
    &>CV_n^\eta~.
    \label{eq: rare reference probability contradiction}
\end{align}
In this case, we can evaluate
\begin{align}
    1+\log\left(H_{\delta,d}(1,0)^{-1}\right)
    &<
    1+\eta\log\left(
        V_n^{-1}
    \right)
    -\log(C)
    \lesssim
    \log\left(
        V_n^{-1}
    \right)~.
    \label{eq: rare reference logarithm bound}
\end{align}
Moreover, on $\mathcal E_{i,i'}$, we have $R_{i,i'}\leq CV_n$. After decreasing the constant in the hypotheses, \eqref{eq: rare null local condition} therefore places the parameter pair $\psi_{i,i'}(\widetilde M_n),\rho_{i,i'}(\widetilde M_n)$ in the rectangle \([\underline a,\overline a]\times[-\bar\rho,\bar\rho]\) and makes \eqref{eq: rare feasible parameter perturbation} applicable. The triangle inequality, \eqref{eq: rare oracle parameter bound}, and \eqref{eq: rare feasible parameter perturbation} then give
\begin{align}
    \left\|
        \left(
            \psi_{i,i'}(\widetilde M_n)-1,
            \rho_{i,i'}(\widetilde M_n)
        \right)
    \right\|_2
    &\lesssim
    V_n+R_{i,i'}
    \lesssim V_n~.
    \label{eq: rare feasible reference distance}
\end{align}
After decreasing $c$, \eqref{eq: rare reference logarithm bound}, \eqref{eq: rare feasible reference distance}, and \eqref{eq: rare null local condition} verify the condition \eqref{eq: rare screening transport condition} with $(a_0,\rho_0)=(1,0)$ and $(a_1,\rho_1)=(\psi_{i,i'}(\widetilde M_n), \rho_{i,i'}(\widetilde M_n))$ on $\mathcal E_{i,i'}$. Thus, \cref{lem: rare screening probability transport} gives
\begin{align}
    H_{\delta,d}\left(
        \psi_{i,i'}(\widetilde M_n),
        \rho_{i,i'}(\widetilde M_n)
    \right)
    &\gtrsim H_{\delta,d}(1,0)
\end{align}
on \(\mathcal E_{i,i'}\). Consequently, \eqref{eq: rare feasible probability envelope} and \eqref{eq: rare screening good event probability} imply that
\begin{align}
    V_n^\eta
    &\geq
    \mathbb E\left[
        H_{\delta,d}\left(
            \psi_{i,i'}(\widetilde M_n),
            \rho_{i,i'}(\widetilde M_n)
        \right)
    \right]
    \nonumber\\
    &\geq
    \mathbb E\left[
        H_{\delta,d}\left(
            \psi_{i,i'}(\widetilde M_n),
            \rho_{i,i'}(\widetilde M_n)
        \right)
        \mathbb{I}\{\mathcal E_{i,i'}\}
    \right]
    \gtrsim
    H_{\delta,d}(1,0)P\{\mathcal E_{i,i'}\}
    \gtrsim
    CV_n^\eta~,
\end{align}
which contradicts the choice of $C$. Hence, we find that
\begin{align}
    H_{\delta,d}(1,0)
    &\leq
    CV_n^\eta
    \label{eq: rare transported probability envelope}
\end{align}
by contradiction.

For \(r\in\{1/2,1\}\) and \(p\in(0,1]\), we can evaluate
\begin{align}
    \frac{\mathrm d}{\mathrm d p}
    \left(
        p\left(1+\log(p^{-1})\right)^r
    \right)
    &=
    \left(1+\log(p^{-1})\right)^{r-1}
    \left(
        1+\log(p^{-1})-r
    \right)
    \geq0~.
\end{align}
Thus, \eqref{eq: rare transported probability envelope} therefore implies that, for \(r\in\{1/2,1\}\),
\begin{align}
    &H_{\delta,d}(1,0)
    \left(
        1+\log\left(H_{\delta,d}(1,0)^{-1}\right)
    \right)^r
    \nonumber\\
    &\quad\leq
    CV_n^\eta
    \left(
        1+\log\left(
            \left(
                CV_n^\eta
            \right)^{-1}
        \right)
    \right)^r
    \quad\lesssim
    V_n^\eta
    \log^r\left(
        V_n^{-1}
    \right)~.
\end{align}
Hence, \cref{lem: screening derivative bounds}, display \eqref{eq: screening derivative probability bounds}, gives
\begin{align}
    |\beta_{\delta,d}|
    &\leq
    \|\nabla H_{\delta,d}(1,0)\|_2
    \lesssim
    V_n^\eta
    \left(
        \sqrt{
            d\log\left(
                V_n^{-1}
            \right)
        }
        +\log\left(
            V_n^{-1}
        \right)
    \right)~,
\end{align}
by choosing $c$ sufficiently small to ensure that $CV_n^\eta\leq1$, as required.\hfill\qed

\subsubsection{Part (ii)}

Fix $\delta\in\Delta_{n,d}(\eta)$ and a pair satisfying $\lambda_{i,i'}=0$. If $H_{\delta,d}(1,0)=0$, mutual absolute continuity on the parameter rectangle \([\underline a,\overline a]\times[-\bar\rho,\bar\rho]\) implies that $H_{\delta,d}$ vanishes throughout the rectangle. Thus, \cref{lem: screening derivative bounds} gives $\beta_{\delta,d}=0$. The two parameter pairs in $\Delta H_{i,i'}(\delta)$ lie in the rectangle on $\{R_{i,i'}\leq c\}$ by the choice of $c$ at the beginning of the proof. Hence, $\Delta H_{i,i'}(\delta)=0$, and the conclusion follows. We therefore maintain that $H_{\delta,d}(1,0)>0$, so mutual absolute continuity implies that $H_{\delta,d}$ is strictly positive throughout the rectangle.

Observe that
\begin{align}
    H_{\delta,d}(a,\rho)
    &=H_{\delta,d}(a,-\rho)
    \quad\text{and thereby}\quad
    \left.
    \frac{\mathrm d}{\mathrm d \rho}H_{\delta,d}(1,\rho)
    \right|_{\rho=0}
    =0~.
    \label{eq: screening zero correlation derivative}
\end{align}
If $\eta=0$, then, for $q\in\{1,2\}$, direct differentiation and the limit as $p\downarrow0$ give
\begin{align}
    \sup_{0<p\leq1}
    p\left(1+\log(p^{-1})\right)^q
    &=
    \begin{cases}
        1,&q=1~,\\
        4/e,&q=2~.
    \end{cases}
    \label{eq: rare logarithmic probability bounds}
\end{align}
Consequently, \cref{lem: screening derivative bounds}, display \eqref{eq: screening derivative probability bounds}, implies that
\begin{align}
    \sup_{\substack{
        a\in[\underline a,\overline a]\\
        |\rho|\leq\bar\rho
    }}
    \left\|\nabla^2H_{\delta,d}(a,\rho)\right\|_{\mathrm{op}}
    &\lesssim d~.
    \label{eq: rare zero exponent hessian bound}
\end{align}
We therefore maintain that $\eta\in(0,1]$. Moreover, Part (i) of the Lemma, \cref{lem: basic inequalities n}, display \eqref{eq: var bounds}, and \eqref{eq: rare null local condition} imply that
\begin{align}
    \log(V_n^{-1})
    &\geq1,\quad\text{and}\quad
    H_{\delta,d}(1,0)
    \leq
    CV_n^\eta
    \leq e^{-1}~.
    \label{eq: rare screening logarithm range}
\end{align}
by setting $c$ sufficiently small.

Now, fix $(a,\rho)$ such that the segment joining $(1,0)$ and $(a,\rho)$ lies in the parameter rectangle \([\underline a,\overline a]\times[-\bar\rho,\bar\rho]\), and write
\begin{align}
    p(u)
    &=H_{\delta,d}\left(1+u(a-1),u\rho\right),
    \qquad u\in[0,1]~.
\end{align}
If $p(u)>CV_n^\eta$, let $u_0$ be the largest $v\in[0,u]$ such that $p(v)=CV_n^\eta$. This point exists by the bound in Part (i) and continuity, and $p(v)\geq CV_n^\eta$ for $v\in[u_0,u]$. Hence, \eqref{eq: rare screening logarithm range} gives
\begin{align}
    1+\log(p(v)^{-1})
    &\lesssim
    \log(V_n^{-1})~.
\end{align}
The chain rule, the Cauchy--Schwarz inequality, and \cref{lem: screening derivative bounds}, display \eqref{eq: screening derivative probability bounds}, imply that
\begin{align}
    \left|
        \frac{\mathrm d}{\mathrm d v}\log p(v)
    \right|
    &\lesssim
    \|(a-1,\rho)\|_2
    \left(
        \sqrt{
            d\log(V_n^{-1})
        }
        +\log(V_n^{-1})
    \right)
\end{align}
for $v\in[u_0,u]$. Integrating from $u_0$ to $u$ implies that
\begin{align}
    p(u)
    &\leq
    CV_n^\eta
    \exp\left(
        C\|(a-1,\rho)\|_2
        \left(
            \sqrt{
                d\log(V_n^{-1})
            }
            +\log(V_n^{-1})
        \right)
    \right)
    \label{eq: rare screening probability growth}
\end{align}
whenever $p(u)>CV_n^\eta$. The same bound holds under the reverse inequality because the exponential factor is at least one.

For $q\in\{1,2\}$ and $0<x\leq e^{-1}$, we can evaluate
\begin{align}
    \frac{\mathrm d}{\mathrm d x}
    \left(
        x\left(1+\log(x^{-1})\right)^q
    \right)
    &={}
    \left(1+\log(x^{-1})\right)^{q-1}
    \left(1+\log(x^{-1})-q\right)
    \geq0~.
\end{align}
If $p(u)\leq CV_n^\eta$, this monotonicity and \eqref{eq: rare screening logarithm range} bound $p(u)(1+\log(p(u)^{-1}))^q$ by a constant multiple of $V_n^\eta \log^q(1/V_n)$ in this case. If the reverse inequality holds, \eqref{eq: rare screening probability growth} gives the same bound multiplied by its exponential factor. Thus, \cref{lem: screening derivative bounds}, display \eqref{eq: screening derivative probability bounds}, implies that
\begin{align}
    &\left\|
        \nabla^2H_{\delta,d}\left(1+u(a-1),u\rho\right)
    \right\|_{\mathrm{op}}
    \nonumber\\
    &\quad\lesssim
    V_n^\eta
    \left(
        d\log(V_n^{-1})
        +\log^2\left(V_n^{-1}\right)
    \right)
    \nonumber\\
    &\qquad\cdot
    \exp\left(
        C\|(a-1,\rho)\|_2
        \left(
            \sqrt{
                d\log(V_n^{-1})
            }
            +\log(V_n^{-1})
        \right)
    \right)
    \label{eq: rare screening transported hessian}
\end{align}
uniformly over $u\in[0,1]$.

Taylor's theorem gives the exact identity
\begin{align}
    &H_{\delta,d}(a,\rho)-H_{\delta,d}(1,0)
    -\left\langle
        \nabla H_{\delta,d}(1,0),(a-1,\rho)
    \right\rangle
    \nonumber\\
    &\quad=
    \int_0^1(1-u)(a-1,\rho)
    \nabla^2H_{\delta,d}\left(1+u(a-1),u\rho\right)
    \begin{pmatrix}
        a-1\\
        \rho
    \end{pmatrix}
    \,\mathrm du~.
\end{align}
By \eqref{eq: screening zero correlation derivative}, the inner product on the left-hand side equals $\beta_{\delta,d}(a-1)$. Combining the exact identity with \eqref{eq: rare screening transported hessian} gives
\begin{align}
    &\left|
        H_{\delta,d}(a,\rho)-H_{\delta,d}(1,0)
        -\beta_{\delta,d}(a-1)
    \right|
    \nonumber\\*
    &\quad\lesssim
    V_n^\eta
    \left(
        d\log(V_n^{-1})
        +\log^2\left(V_n^{-1}\right)
    \right)
    \nonumber\\
    &\qquad\cdot
    \exp\left(
        C\left(
            \sqrt{
                d\log(V_n^{-1})
            }
            +\log(V_n^{-1})
        \right)
        \|(a-1,\rho)\|_2
    \right)
    \left((a-1)^2+\rho^2\right)~.
    \label{eq: rare screening taylor remainder}
\end{align}
On $\{R_{i,i'}\leq c\}$, \eqref{eq: rare oracle parameter bound} and
\eqref{eq: rare feasible parameter perturbation} give
\begin{align}
    \left\|
        \left(
            \psi_{i,i'}(M_n)-1,
            \rho_{i,i'}(M_n)
        \right)
    \right\|_2
    &\lesssim V_n
    \quad\text{and}
    \nonumber\\
    \left\|
        \left(
            \psi_{i,i'}(\widetilde M_n)-1,
            \rho_{i,i'}(\widetilde M_n)
        \right)
    \right\|_2
    &\lesssim V_n+R_{i,i'}~.
\end{align}
Applying \eqref{eq: rare screening taylor remainder} to these two parameter pairs, whose segments lie in the fixed rectangle by the choice of $c$ at the beginning of the proof, we thereby obtain
\begin{align}
    &|\Delta H_{i,i'}(\delta)
    -\beta_{\delta,d}\Delta\psi_{i,i'}|
    \nonumber\\
    &\quad\leq
    \left|
        H_{\delta,d}\left(
            \psi_{i,i'}(\widetilde M_n),
            \rho_{i,i'}(\widetilde M_n)
        \right)
        -H_{\delta,d}(1,0)
        -\beta_{\delta,d}
        \left(\psi_{i,i'}(\widetilde M_n)-1\right)
    \right|
    \nonumber\\
    &\qquad+
    \left|
        H_{\delta,d}\left(
            \psi_{i,i'}(M_n),
            \rho_{i,i'}(M_n)
        \right)
        -H_{\delta,d}(1,0)
        -\beta_{\delta,d}
        \left(\psi_{i,i'}(M_n)-1\right)
    \right|
    \nonumber\\
    &\quad\lesssim
    V_n^\eta
    \left(
        d\log(V_n^{-1})
        +\log^2\left(V_n^{-1}\right)
    \right)
    \nonumber\\
    &\qquad\cdot
    \exp\left(
        C\left(
            \sqrt{
                d\log(V_n^{-1})
            }
            +\log(V_n^{-1})
        \right)R_{i,i'}
    \right)
    \left(V_n^2+R_{i,i'}^2\right)~,
\end{align}
as required, where we have applied \eqref{eq: rare null local condition} to absorb the exponential term involving $V_n$.\hfill\qed

\subsubsection{Part (iii)}

Define the sequence
\begin{align}
    h_{n,d}
    &=C\left(
        \sqrt{
            d\log(V_n^{-1})
        }
        +\log(V_n^{-1})
    \right)~.
\end{align}
For every positive integer $m$, Minkowski's inequality and \cref{lem: projection residualization bounds}, display \eqref{eq: feasible projection higher moments}, give
\begin{align}
    \|R_{i,i'}\|_{2m}
    &\leq
    \|\tilde\mu_{i,i}-\mu_{i,i}\|_{2m}
    +\|\tilde\mu_{i',i'}-\mu_{i',i'}\|_{2m}
    +\|\tilde\mu_{i,i'}-\mu_{i,i'}\|_{2m}
    \lesssim
    mV_n~.
\end{align}
Moreover, \cref{assu: gaussianity}, \cref{lem: wick}, \cref{lem: basic inequalities n}, display \eqref{eq: n bound 2}, and H\"older's inequality imply that
\begin{align}
    \|\hat\varepsilon_i^{(0)}
    \hat\varepsilon_{i'}^{(0)}\|_2
    &\lesssim1~.
\end{align}
Consequently, Cauchy--Schwarz gives
\begin{align}
    \mathbb E\left[
        |\hat\varepsilon_i^{(0)}
        \hat\varepsilon_{i'}^{(0)}|
        R_{i,i'}^m
    \right]
    &\leq
    \|\hat\varepsilon_i^{(0)}
    \hat\varepsilon_{i'}^{(0)}\|_2
    \|R_{i,i'}\|_{2m}^m
    \leq
    \left(CmV_n\right)^m
    \label{eq: projection weighted high moments}
\end{align}
uniformly over $i,i'\in[n]$ and the positive integers $m$. Observe that
\begin{align}
    \mathbb E\left[
        |\hat\varepsilon_i^{(0)}
        \hat\varepsilon_{i'}^{(0)}|
    \right]
    &\leq
    \mu_{i,i}^{1/2}\mu_{i',i'}^{1/2}
    \lesssim1
    \label{eq: projection weighted zeroth moment}
\end{align}
by Cauchy--Schwarz and \cref{lem: basic inequalities n}, display \eqref{eq: n bound 2}. Observe also that, \cref{lem: basic inequalities n}, display \eqref{eq: var bounds}, and
\eqref{eq: rare null local condition} ensure that
\begin{align}
    \log(V_n^{-1})
    &\geq1
    \quad\text{and}\quad
    Ch_{n,d}V_n
    \leq\frac{1}{2}~.
    \label{eq: projection exponential series condition}
\end{align}
by setting $c$ sufficiently small.

With this in place, observe that, as $R_{i,i'}\geq0$, Tonelli's theorem gives the exact expansion
\begin{align}
    &\mathbb E\left[
        |\hat\varepsilon_i^{(0)}\hat\varepsilon_{i'}^{(0)}|
        \left(V_n^2+R_{i,i'}^2\right)
        \exp\left(h_{n,d}R_{i,i'}\right)
    \right]
    \nonumber\\
    &\quad=
    V_n^2
    \mathbb E\left[
        |\hat\varepsilon_i^{(0)}\hat\varepsilon_{i'}^{(0)}|
    \right] \nonumber \\
    & \quad +
    V_n^2
    \sum_{m=1}^{\infty}\frac{h_{n,d}^m}{m!}
    \mathbb E\left[
        |\hat\varepsilon_i^{(0)}\hat\varepsilon_{i'}^{(0)}|
        R_{i,i'}^m
    \right]
    +
    \sum_{m=0}^{\infty}\frac{h_{n,d}^m}{m!}
    \mathbb E\left[
        |\hat\varepsilon_i^{(0)}\hat\varepsilon_{i'}^{(0)}|
        R_{i,i'}^{m+2}
    \right]
    \nonumber\\
    &\quad\leq
    CV_n^2
    +V_n^2
    \sum_{m=1}^{\infty}
        \frac{h_{n,d}^m}{m!}
        \left(CmV_n\right)^m
    +
    \sum_{m=0}^{\infty}
        \frac{h_{n,d}^m}{m!}
        \left(C(m+2)V_n\right)^{m+2}
    \nonumber\\
    &\quad\leq
    CV_n^2
    \left(
        1+
        \sum_{m=1}^{\infty}
            \left(Ch_{n,d}V_n\right)^m
    \right)
    +
    CV_n^2
    \sum_{m=0}^{\infty}
        (m+1)(m+2)
        \left(Ch_{n,d}V_n\right)^m
    \lesssim V_n^2~,
\end{align}
as required, where the first inequality uses \eqref{eq: projection weighted high moments}
and \eqref{eq: projection weighted zeroth moment} and the remaining inequalities use the geometric series formula and \eqref{eq: projection exponential series condition}.\hfill\qed

\subsection{Proof of \cref{lem: rare screening probability transport}}

The chain rule gives
\begin{align}
    &\frac{\mathrm d}{\mathrm d u}
    H_{\delta,d}\left(
        a_0+u(a_1-a_0),
        \rho_0+u(\rho_1-\rho_0)
    \right)
    \nonumber\\
    &\quad=
    \left\langle
        \nabla H_{\delta,d}\left(
            a_0+u(a_1-a_0),
            \rho_0+u(\rho_1-\rho_0)
        \right),
        (a_1-a_0,\rho_1-\rho_0)
    \right\rangle~.
\end{align}
The Cauchy--Schwarz inequality and the first bound in
\eqref{eq: screening derivative probability bounds} consequently give
\begin{align}
    &\left|
        \frac{\mathrm d}{\mathrm d u}
        H_{\delta,d}\left(
            a_0+u(a_1-a_0),
            \rho_0+u(\rho_1-\rho_0)
        \right)
    \right|
    \nonumber\\
    &\quad\lesssim
    \|(a_1-a_0,\rho_1-\rho_0)\|_2
    H_{\delta,d}\left(
        a_0+u(a_1-a_0),
        \rho_0+u(\rho_1-\rho_0)
    \right)
    \nonumber\\
    &\qquad\cdot
    \left(
        \sqrt{
            d\left[
                1+\log\left(
                    H_{\delta,d}\left(
                        a_0+u(a_1-a_0),
                        \rho_0+u(\rho_1-\rho_0)
                    \right)^{-1}
                \right)
            \right]
        }
    \right.
    \nonumber\\
    &\hspace{40mm}\left.
        +1+\log\left(
            H_{\delta,d}\left(
                a_0+u(a_1-a_0),
                \rho_0+u(\rho_1-\rho_0)
            \right)^{-1}
        \right)
    \right)~.
    \label{eq: rare screening path derivative}
\end{align}
Because the probability in \eqref{eq: rare screening path derivative}
is positive, the chain rule also gives
\begin{align}
    &\frac{\mathrm d}{\mathrm d u}
    \left(
        1+\log\left(
            H_{\delta,d}\left(
                a_0+u(a_1-a_0),
                \rho_0+u(\rho_1-\rho_0)
            \right)^{-1}
        \right)
    \right)^{1/2}
    \nonumber\\
    &\quad=
    -\frac{
        \dfrac{\mathrm d}{\mathrm d u}
        H_{\delta,d}\left(
            a_0+u(a_1-a_0),
            \rho_0+u(\rho_1-\rho_0)
        \right)
    }{
        2H_{\delta,d}\left(
            a_0+u(a_1-a_0),
            \rho_0+u(\rho_1-\rho_0)
        \right)
    }
    \nonumber\\
    &\qquad\cdot
    \left[
        1+\log\left(
            H_{\delta,d}\left(
                a_0+u(a_1-a_0),
                \rho_0+u(\rho_1-\rho_0)
            \right)^{-1}
        \right)
    \right]^{-1/2}~.
    \label{eq: rare screening square root chain rule}
\end{align}
Taking absolute values and substituting in \eqref{eq: rare screening path derivative} gives
\begin{align}
    &\left|
        \frac{\mathrm d}{\mathrm d u}
        \left(
            1+\log\left(
                H_{\delta,d}\left(
                    a_0+u(a_1-a_0),
                    \rho_0+u(\rho_1-\rho_0)
                \right)^{-1}
            \right)
        \right)^{1/2}
    \right|    \label{eq: rare screening square root derivative}
    \\
    &\quad\lesssim
    \|(a_1-a_0,\rho_1-\rho_0)\|_2
    \left(
        \frac{
            \sqrt{
                d\left[
                    1+\log\left(
                        H_{\delta,d}\left(
                            a_0+u(a_1-a_0),
                            \rho_0+u(\rho_1-\rho_0)
                        \right)^{-1}
                    \right)
                \right]
            }
        }{
            \left[
                1+\log\left(
                    H_{\delta,d}\left(
                        a_0+u(a_1-a_0),
                        \rho_0+u(\rho_1-\rho_0)
                    \right)^{-1}
                \right)
            \right]^{1/2}
        }
    \right.
    \nonumber\\
    &\hspace{48mm}\left.
        +
        \frac{
            1+\log\left(
                H_{\delta,d}\left(
                    a_0+u(a_1-a_0),
                    \rho_0+u(\rho_1-\rho_0)
                \right)^{-1}
            \right)
        }{
            \left[
                1+\log\left(
                    H_{\delta,d}\left(
                        a_0+u(a_1-a_0),
                        \rho_0+u(\rho_1-\rho_0)
                    \right)^{-1}
                \right)
            \right]^{1/2}
        }
    \right)
    \nonumber\\
    &\quad\lesssim
    \|(a_1-a_0,\rho_1-\rho_0)\|_2
    \left(
        \sqrt d
        +
        \left[
            1+\log\left(
                H_{\delta,d}\left(
                    a_0+u(a_1-a_0),
                    \rho_0+u(\rho_1-\rho_0)
                \right)^{-1}
            \right)
        \right]^{1/2}
    \right)~.\nonumber 
\end{align}
Integrating \eqref{eq: rare screening square root derivative} from $0$ to $u$ gives
\begin{align}
    &\left(
        1+\log\left(
            H_{\delta,d}\left(
                a_0+u(a_1-a_0),
                \rho_0+u(\rho_1-\rho_0)
            \right)^{-1}
        \right)
    \right)^{1/2}
    \nonumber\\
    &\quad\leq
    \left(
        1+\log\left(
            H_{\delta,d}(a_0,\rho_0)^{-1}
        \right)
    \right)^{1/2}
    +C\sqrt d\,
    \|(a_1-a_0,\rho_1-\rho_0)\|_2
    \nonumber\\
    &\qquad+
    C\|(a_1-a_0,\rho_1-\rho_0)\|_2 \nonumber \\
    &\qquad \cdot\int_0^u
    \left(
        1+\log\left(
            H_{\delta,d}\left(
                a_0+v(a_1-a_0),
                \rho_0+v(\rho_1-\rho_0)
            \right)^{-1}
        \right)
    \right)^{1/2}
    \,\mathrm dv~.
    \label{eq: rare screening square root integral inequality}
\end{align}
Gronwall's inequality \cite[Chapter~III, \S1, Theorem~1.1, p.~24]{hartman2002ordinary},
applied to \eqref{eq: rare screening square root integral inequality}, yields
\begin{align}
    &\sup_{u\in[0,1]}
    \left(
        1+\log\left(
            H_{\delta,d}\left(
                a_0+u(a_1-a_0),
                \rho_0+u(\rho_1-\rho_0)
            \right)^{-1}
        \right)
    \right)^{1/2}
    \nonumber\\
    &\quad\leq
    \exp\left(
        C\|(a_1-a_0,\rho_1-\rho_0)\|_2
    \right)
    \left(
        \left[
            1+\log\left(
                H_{\delta,d}(a_0,\rho_0)^{-1}
            \right)
        \right]^{1/2}
    \right.
    \nonumber\\
    &\hspace{60mm}\left.
        +C\sqrt d\,
        \|(a_1-a_0,\rho_1-\rho_0)\|_2
    \right)~.
    \label{eq: rare screening logarithm transport}
\end{align}
Since $1+\log(H_{\delta,d}(a_0,\rho_0)^{-1})\geq1$, the condition
\eqref{eq: rare screening transport condition}
implies
\begin{align}
    \|(a_1-a_0,\rho_1-\rho_0)\|_2
    &\leq c
    \quad\text{and}
    \nonumber\\
    \sqrt d\,
    \|(a_1-a_0,\rho_1-\rho_0)\|_2
    &\leq
    \frac{c}{
        \left(
            1+\log\left(
                H_{\delta,d}(a_0,\rho_0)^{-1}
            \right)
        \right)^{1/2}
    }~.
\end{align}
Substitution into \eqref{eq: rare screening logarithm transport}
therefore gives
\begin{align}
    &\sup_{u\in[0,1]}
    \left(
        1+\log\left(
            H_{\delta,d}\left(
                a_0+u(a_1-a_0),
                \rho_0+u(\rho_1-\rho_0)
            \right)^{-1}
        \right)
    \right)^{1/2}
    \nonumber\\
    &\quad\lesssim
    \left(
        1+\log\left(
            H_{\delta,d}(a_0,\rho_0)^{-1}
        \right)
    \right)^{1/2}~.
    \label{eq: rare screening logarithm comparison}
\end{align}
The bound \eqref{eq: rare screening path derivative} also implies that
\begin{align}
    &\left|
        \frac{\mathrm d}{\mathrm d u}
        \log\left(
            H_{\delta,d}\left(
                a_0+u(a_1-a_0),
                \rho_0+u(\rho_1-\rho_0)
            \right)
        \right)
    \right|
        \label{eq: rare screening logarithm derivative}\\
    &\quad=
    \frac{
        \left|
            \dfrac{\mathrm d}{\mathrm d u}
            H_{\delta,d}\left(
                a_0+u(a_1-a_0),
                \rho_0+u(\rho_1-\rho_0)
            \right)
        \right|
    }{
        H_{\delta,d}\left(
            a_0+u(a_1-a_0),
            \rho_0+u(\rho_1-\rho_0)
        \right)
    }
    \nonumber\\
    &\quad\lesssim
    \|(a_1-a_0,\rho_1-\rho_0)\|_2
    \left(
        \sqrt{
            d\left[
                1+\log\left(
                    H_{\delta,d}\left(
                        a_0+u(a_1-a_0),
                        \rho_0+u(\rho_1-\rho_0)
                    \right)^{-1}
                \right)
            \right]
        }
    \right.
    \nonumber\\
    &\hspace{40mm}\left.
        +1+\log\left(
            H_{\delta,d}\left(
                a_0+u(a_1-a_0),
                \rho_0+u(\rho_1-\rho_0)
            \right)^{-1}
        \right)
    \right)~.
\end{align}
Integrating \eqref{eq: rare screening logarithm derivative} over
$u\in[0,1]$ and applying
\eqref{eq: rare screening logarithm comparison} yields
\begin{align}
    \left|
        \log\left(
            \frac{H_{\delta,d}(a_1,\rho_1)}
            {H_{\delta,d}(a_0,\rho_0)}
        \right)
    \right|
    &\lesssim
    \|(a_1-a_0,\rho_1-\rho_0)\|_2
    \nonumber\\
    &\quad\cdot
    \left(
        \sqrt{
            d\left(
                1+\log\left(
                    H_{\delta,d}(a_0,\rho_0)^{-1}
                \right)
            \right)
        }
    \right.
    \nonumber\\
    &\hspace{36mm}\left.
        +1+\log\left(
            H_{\delta,d}(a_0,\rho_0)^{-1}
        \right)
    \right)
    \lesssim1~,
\end{align}
completing the proof.\hfill\qed

\section{Proofs for Lemmas Supporting \cref{lem: variance bound}}

\subsection{Proof of \cref{lem: variance approximation components}}

Since $\hat\varepsilon^{(0)}=H_n(Y^{(0)}-\tau^{(0)}\bm{1})$, Part (ii) of \cref{lem: rotation and regression}, applied with $X=Y-\mathbb{E}[Y]$, $A=\Lambda_n$, and $B=\Gamma_d$, and \cref{assu: gaussianity,assu: proportionality} give
\begin{align}
    \left(\varepsilon_0^{(j)}\right)_{j=1}^d
    &\ci\hat\varepsilon^{(0)}~,
    \quad\text{and}\quad
    \Cov\left(
        \varepsilon_{0,i}^{(j)},
        \varepsilon_{0,i'}^{(j')}
    \right)
    =\lambda_{i,i'}\omega_{j,j'}
    \label{eq: population residual covariance and independence}
\end{align}
for all $i,i'\in[n]$ and $j,j'\in[d]$. Moreover, \cref{lem: auxiliary outcome covariance bounds} implies that $\Omega_d$ is positive semi-definite. Thus, Part (i) of \cref{lem: rotation and regression}, applied to $(\varepsilon_0^{(j)})_{j=1}^d$ with $A=\Lambda_n$ and $B=\Omega_d$, and \cref{assu: homoskedasticity} give
\begin{align}
    P\left\{
        \left|\check{\lambda}_{0,i,i'}\right|\geq\delta
    \right\}
    &=H_{\delta,d}(1,\lambda_{i,i'})
    \label{eq: population residualized screening representation}
\end{align}
for every pair $i,i'\in[n]$ and threshold $\delta\geq0$.

\subsubsection{Part (i)}

Recall from \eqref{eq: residual covariance} that $\Cov(\hat\varepsilon_i^{(0)},\hat\varepsilon_k^{(0)})=\mu_{i,k}$. Thus, the independence condition in \eqref{eq: population residual covariance and independence} and \cref{lem: wick} imply the decomposition
\begin{align}
    \Var\left(\check V_{n,d}^{(0)}(\delta)\right)
    &=\frac{1}{n^4}
    \sum_{i,i',k,k'=1}^n
    \left(
        \mu_{i,k}\mu_{i',k'}
        +\mu_{i,k'}\mu_{i',k}
    \right)
    P\left\{
        \left|\check{\lambda}_{0,i,i'}\right|\geq\delta,
        \left|\check{\lambda}_{0,k,k'}\right|\geq\delta
    \right\}
    \nonumber\\
    &+
    \frac{1}{n^4}
    \sum_{i,i',k,k'=1}^n
    \mu_{i,i'}\mu_{k,k'}
    \Cov\left(
        \mathbb{I}\left\{
            \left|\check{\lambda}_{0,i,i'}\right|\geq\delta
        \right\},
        \mathbb{I}\left\{
            \left|\check{\lambda}_{0,k,k'}\right|\geq\delta
        \right\}
    \right)~.
    \label{eq: population oracle exact variance decomposition}
\end{align}
We bound the two terms in this decomposition in succession.

First, \cref{lem: basic inequalities n}, displays
\eqref{eq: n bound 1} and \eqref{eq: n bound 3}, gives
\begin{align}
    \max_{i\in[n]}\sum_{i'=1}^n|\mu_{i,i'}|
    &\leq
    \max_{i\in[n]}\sum_{i'=1}^n|\lambda_{i,i'}|
    +n\max_{i,i'\in[n]}|\mu_{i,i'}-\lambda_{i,i'}|
    \lesssim nV_n~.
    \label{eq: population oracle centered row sums}
\end{align}
Thus, we can evaluate
\begin{align}
    \sum_{i,i',k,k'=1}^n
    \left|
        \mu_{i,k}\mu_{i',k'}
        +\mu_{i,k'}\mu_{i',k}
    \right|
    &\leq
    \sum_{i,i',k,k'=1}^n
    \left(
        |\mu_{i,k}||\mu_{i',k'}|
        +|\mu_{i,k'}||\mu_{i',k}|
    \right)
    \nonumber\\
    &=
    2\left(
        \sum_{i,k=1}^n|\mu_{i,k}|
    \right)^2
    \leq Cn^4V_n^2~.
    \label{eq: population oracle h sum}
\end{align}
In turn, we can evaluate
\begin{align}
    &\sum_{i,i',k,k'=1}^n
    \left|
        \mu_{i,k}\mu_{i',k'}
        +\mu_{i,k'}\mu_{i',k}
    \right|
    \mathbb I\left\{
        (i,i')\in\mathcal S_n
        \ \text{or}\ 
        (k,k')\in\mathcal S_n
    \right\}
    \nonumber\\
    &\quad\leq
    \sum_{(i,i')\in\mathcal S_n}\sum_{k,k'=1}^n
    \left|
        \mu_{i,k}\mu_{i',k'}
        +\mu_{i,k'}\mu_{i',k}
    \right|
    +
    \sum_{(k,k')\in\mathcal S_n}\sum_{i,i'=1}^n
    \left|
        \mu_{i,k}\mu_{i',k'}
        +\mu_{i,k'}\mu_{i',k}
    \right|
    \nonumber\\
    &\quad\leq
    4\sum_{(i,i')\in\mathcal S_n}
    \left(\sum_{k=1}^n|\mu_{i,k}|\right)
    \left(\sum_{k'=1}^n|\mu_{i',k'}|\right)
    \nonumber\\
    &\quad\leq
    4|\mathcal S_n|
    \left(
        \max_{u\in[n]}\sum_{v=1}^n|\mu_{u,v}|
    \right)^2
    \leq Cn^3(s+1)V_n^2~,
    \label{eq: population oracle support h sum}
\end{align}
by the union bound, the triangle inequality and the symmetry of $M_n$, the maximum row sum, and the bound $|\mathcal S_n|\leq n(s+1)$ from \cref{assu: sparsity} together with \eqref{eq: population oracle centered row sums}. Now, fix $(i,i'),(k,k')\notin\mathcal S_n$ and $\delta\in\Delta_{n,d}(\eta)$. We have that
\begin{align}
    P\left\{
        \left|\check{\lambda}_{0,i,i'}\right|\geq\delta,
        \left|\check{\lambda}_{0,k,k'}\right|\geq\delta
    \right\}
    &\leq
    P\left\{
        \left|\check{\lambda}_{0,i,i'}\right|\geq\delta
    \right\}^{1/2}
    P\left\{
        \left|\check{\lambda}_{0,k,k'}\right|\geq\delta
    \right\}^{1/2}
    \nonumber\\
    &\quad=
    H_{\delta,d}(1,0)
    \leq CV_n^\eta~,
\end{align}
by Cauchy--Schwarz, \eqref{eq: population residualized screening representation}, and \cref{lem: rare null pair screening}, Part (i). Thus, since every joint probability is at most one, we have that
\begin{align}
    &P\left\{
        \left|\check{\lambda}_{0,i,i'}\right|\geq\delta,
        \left|\check{\lambda}_{0,k,k'}\right|\geq\delta
    \right\}
    \leq
    \mathbb{I}\left\{
        (i,i')\in\mathcal S_n
        \ \text{or}\ 
        (k,k')\in\mathcal S_n
    \right\}
    +CV_n^\eta~.
    \label{eq: population oracle joint event bound}
\end{align}
Consequently, by plugging \eqref{eq: population oracle joint event bound} into the first term of \eqref{eq: population oracle exact variance decomposition}, we obtain
\begin{align}
    &\sup_{\delta\in\Delta_{n,d}(\eta)}
    \left|
        \frac{1}{n^4}
        \sum_{i,i',k,k'=1}^n
        \left(
            \mu_{i,k}\mu_{i',k'}
            +\mu_{i,k'}\mu_{i',k}
        \right)
        P\left\{
            \left|\check{\lambda}_{0,i,i'}\right|\geq\delta,
            \left|\check{\lambda}_{0,k,k'}\right|\geq\delta
        \right\}
    \right|
    \nonumber\\
    &\quad\leq
    \frac{1}{n^4}
    \sum_{i,i',k,k'=1}^n
    \left|
        \mu_{i,k}\mu_{i',k'}
        +\mu_{i,k'}\mu_{i',k}
    \right|
    \mathbb I\left\{
        (i,i')\in\mathcal S_n
        \ \text{or}\ 
        (k,k')\in\mathcal S_n
    \right\}
    \nonumber\\
    &\qquad+
    \frac{CV_n^\eta}{n^4}
    \sum_{i,i',k,k'=1}^n
    \left|
        \mu_{i,k}\mu_{i',k'}
        +\mu_{i,k'}\mu_{i',k}
    \right|
    \nonumber\\
    &\quad\leq
    \frac{C}{n^4}
    \left(
        n^3(s+1)V_n^2
        +n^4V_n^{2+\eta}
    \right)
    \leq
    C
    V_n^2
    \left(
        \frac{s+1}{n}
        +V_n^\eta
    \right)~,
    \label{eq: population oracle first variance bound}
\end{align}
where the second inequality uses \eqref{eq: population oracle h sum} and \eqref{eq: population oracle support h sum}.

We next bound the second term. When $\lambda_{i,k}=\lambda_{i,k'}=\lambda_{i',k}=\lambda_{i',k'}=0$, the condition \eqref{eq: population residual covariance and independence} implies that every component of $(\varepsilon_{0,i}^{(j)},\varepsilon_{0,i'}^{(j)})_{j=1}^d$ is uncorrelated with every component of $(\varepsilon_{0,k}^{(j)},\varepsilon_{0,k'}^{(j)})_{j=1}^d$. By \cref{assu: gaussianity}, these two collections are therefore independent. Hence, the corresponding indicator variables in \eqref{eq: population oracle exact variance decomposition} have zero covariance. Since the absolute covariance of two indicator variables is at most one, we have
\begin{align}
    &\left|
        \Cov\left(
            \mathbb{I}\left\{
                \left|\check{\lambda}_{0,i,i'}\right|\geq\delta
            \right\},
            \mathbb{I}\left\{
                \left|\check{\lambda}_{0,k,k'}\right|\geq\delta
            \right\}
        \right)
    \right| \nonumber \\
    & \leq
    \mathbb{I}\left\{
        \lambda_{i,k}\neq0
        \ \text{or}\ 
        \lambda_{i,k'}\neq0
        \ \text{or}\ 
        \lambda_{i',k}\neq0
        \ \text{or}\ 
        \lambda_{i',k'}\neq0
    \right\}~.
    \label{eq: population oracle covariance support bound}
\end{align}
By multiplying the indicator in \eqref{eq: population oracle covariance support bound} by $|\mu_{i,i'}\mu_{k,k'}|$ and summing over the four indices, we obtain
\begin{align}
    &\sum_{i,i',k,k'=1}^n
    |\mu_{i,i'}\mu_{k,k'}|
    \mathbb I\left\{
        \lambda_{i,k}\neq0
        \ \text{or}\ 
        \lambda_{i,k'}\neq0
        \ \text{or}\ 
        \lambda_{i',k}\neq0
        \ \text{or}\ 
        \lambda_{i',k'}\neq0
    \right\}
    \nonumber\\
    &\quad\leq
    \sum_{i,i',k,k'=1}^n
    |\mu_{i,i'}\mu_{k,k'}|
    \Big(
        \mathbb I\{\lambda_{i,k}\neq0\}
        +\mathbb I\{\lambda_{i,k'}\neq0\}
        +\mathbb I\{\lambda_{i',k}\neq0\}
        +\mathbb I\{\lambda_{i',k'}\neq0\}
    \Big)
    \nonumber\\
    &\quad=
    4\sum_{u,v=1}^n
    \mathbb{I}\{\lambda_{u,v}\neq0\}
    \left(\sum_{u'=1}^n|\mu_{u,u'}|\right)
    \left(\sum_{v'=1}^n|\mu_{v,v'}|\right)
    \nonumber\\
    &\quad\leq
    4n(s+1)
    \left(
        \max_{u\in[n]}\sum_{u'=1}^n|\mu_{u,u'}|
    \right)^2
    \leq Cn^3(s+1)V_n^2~,
    \label{eq: population oracle dependent coefficient sum}
\end{align}
where the first inequality follows from bounding the indicator of a union by the sum of its four indicators, the equality follows from relabeling the indices and the symmetry of $M_n$, the second inequality uses \cref{assu: sparsity}, which permits at most $n(s+1)$ nonzero entries of $\Lambda_n$, and the final inequality uses \eqref{eq: population oracle centered row sums}. Consequently, we can conclude that
\begin{align}
    &\sup_{\delta\geq0}
    \frac{1}{n^4}
    \left|
        \sum_{i,i',k,k'=1}^n
        \mu_{i,i'}\mu_{k,k'}
        \Cov\left(
            \mathbb{I}\left\{
                \left|\check{\lambda}_{0,i,i'}\right|\geq\delta
            \right\},
            \mathbb{I}\left\{
                \left|\check{\lambda}_{0,k,k'}\right|\geq\delta
            \right\}
        \right)
    \right|
    \lesssim
    V_n^2\frac{s+1}{n}~.
    \label{eq: population oracle second variance bound}
\end{align}
Combining \eqref{eq: population oracle exact variance decomposition}, \eqref{eq: population oracle first variance bound}, and \eqref{eq: population oracle second variance bound} completes the proof.\hfill\qed

\subsubsection{Part (ii)}

Observe that \eqref{eq: population residual covariance and independence} and \eqref{eq: population residualized screening representation}, together with the first equality in
\eqref{eq: feasible conditional screening representations}, imply the representation
\begin{align}
    & G_{n,d}(\delta)-G_{0,n,d}(\delta) \\
    &=\frac{\beta_{\delta,d}}{n^2}
    \sum_{(i,i')\notin\mathcal S_n}
    \hat\varepsilon_i^{(0)}
    \hat\varepsilon_{i'}^{(0)}
    \left(
        \psi_{i,i'}(\widetilde M_n)-1
    \right)
    \nonumber\\
    &\quad+
    \frac{1}{n^2}
    \sum_{(i,i')\in\mathcal S_n}
    \hat\varepsilon_i^{(0)}
    \hat\varepsilon_{i'}^{(0)}
    \left(
        H_{\delta,d}\left(
            \psi_{i,i'}(\widetilde M_n),
            \rho_{i,i'}(\widetilde M_n)
        \right)
        -H_{\delta,d}(1,\lambda_{i,i'})
    \right)
    \nonumber\\
    &\quad+
    \frac{1}{n^2}
    \sum_{(i,i')\notin\mathcal S_n}
    \hat\varepsilon_i^{(0)}
    \hat\varepsilon_{i'}^{(0)}
    \left(
        H_{\delta,d}\left(
            \psi_{i,i'}(\widetilde M_n),
            \rho_{i,i'}(\widetilde M_n)
        \right)
        -H_{\delta,d}(1,0)
        -\beta_{\delta,d}
        \left(
            \psi_{i,i'}(\widetilde M_n)-1
        \right)
    \right)~.
    \label{eq: conditional mean exact three term decomposition}
\end{align}
We bound these three terms with the following Lemma.

\begin{lem}
\label{lem: conditional mean residualization aggregates}$ $\\
Suppose that, for a fixed $\eta\in(0,1]$, \cref{assu: gaussianity,assu: proportionality,assu: homoskedasticity,assu: sparsity,assu: well conditioned} hold, $d\geq2$, and $s/n\leq c$, where $c>0$ is sufficiently small.

\noindent\textbf{(i)} It holds that
\begin{align}
    &\sup_{\delta\geq0}
    \frac{|\beta_{\delta,d}|}{n^2}
    \mathbb{E}\Bigg[
        \Bigg(
        \sum_{(i,i')\notin\mathcal S_n}
        \hat\varepsilon_i^{(0)}
        \hat\varepsilon_{i'}^{(0)}
        \left(
            \psi_{i,i'}(\widetilde M_n)-1
        \right)
        \Bigg)^2
    \Bigg]^{1/2}
    \lesssim
    V_n
    \left(
        \frac{s\sqrt d}{n}
        +V_n\sqrt d
    \right)~.
    \label{eq: conditional mean null scale aggregate}
\end{align}

\noindent\textbf{(ii)} It holds that
\begin{align}
    &\sup_{\delta\geq0}
    \frac{1}{n^2}
    \mathbb{E}\Bigg[
        \Bigg(
        \sum_{(i,i')\in\mathcal S_n}
        \hat\varepsilon_i^{(0)}
        \hat\varepsilon_{i'}^{(0)}
        \Bigg(
            H_{\delta,d}\left(
                \psi_{i,i'}(\widetilde M_n),
                \rho_{i,i'}(\widetilde M_n)
            \right)
            -H_{\delta,d}(1,\lambda_{i,i'})
        \Bigg)
        \Bigg)^2
    \Bigg]^{1/2}
    \nonumber\\
    &\quad\lesssim
    \frac{V_n}{\sqrt d}
    +\frac{s}{n}V_n\sqrt d
    +V_n^2\sqrt d~.
    \label{eq: conditional mean support aggregate}
\end{align}

\noindent\textbf{(iii)} If, additionally,
\eqref{eq: variance component local condition} holds, then
\begin{align}
    &\sup_{\delta\in\Delta_{n,d}(\eta)}
    \frac{1}{n^2}
    \mathbb{E}\Bigg[
        \Bigg(
        \sum_{(i,i')\notin\mathcal S_n}
        \hat\varepsilon_i^{(0)}
        \hat\varepsilon_{i'}^{(0)}
        \Bigg(
            H_{\delta,d}\left(
                \psi_{i,i'}(\widetilde M_n),
                \rho_{i,i'}(\widetilde M_n)
            \right)
            -H_{\delta,d}(1,0)
    \nonumber\\
    &\hspace{75mm}
            -\beta_{\delta,d}
            \left(
                \psi_{i,i'}(\widetilde M_n)-1
            \right)
        \Bigg)
        \Bigg)^2
    \Bigg]^{1/2}
    \nonumber\\
    &\quad\leq
    CV_n^{2+\eta}
    \left(
        d\log(V_n^{-1})
        +\log^2\left(V_n^{-1}\right)
    \right)~.
    \label{eq: conditional mean null remainder aggregate}
\end{align}
\end{lem}

\noindent In \eqref{eq: conditional mean support aggregate},
\cref{lem: basic inequalities n}, display \eqref{eq: var bounds}, absorbs the last term into the second.  Minkowski's inequality and Parts (i) through (iii) of \cref{lem: conditional mean residualization aggregates}, applied to \eqref{eq: conditional mean exact three term decomposition}, therefore give
\begin{align}
    &\sup_{\delta\in\Delta_{n,d}(\eta)}
    \mathbb{E}\left[
        \left(G_{n,d}(\delta)-G_{0,n,d}(\delta)\right)^2
    \right]^{1/2}
    \nonumber\\
    &\quad\leq
    CV_n
    \Bigg(
        \frac{1}{\sqrt d}
        +\frac{s\sqrt d}{n}
        +V_n^{1+\eta}
        \left(
            d\log(V_n^{-1})
            +\log^2\left(V_n^{-1}\right)
        \right)
    \Bigg)~,
\end{align}
as required.
\hfill\qed

\subsubsection{Part (iii)}

Fix $a=(i,i')$ and $b=(k,k')$, and let $T=\{i,i',k,k'\}$ be the set of
distinct indices in the two pairs. For any symmetric array $\Xi$ such that
$(\Xi_{u,v})_{u,v\in T}$ is positive semi-definite, let
$(G_u^{(1)})_{u\in T},\ldots,(G_u^{(d)})_{u\in T}$ be independent centered
Gaussian vectors with covariance $(\Xi_{u,v})_{u,v\in T}$. Define the quantity
\begin{align}
    \Psi_{a,b}(\Xi,\delta)
    &=\Cov\Bigg(
        \mathbb{I}\left\{
            \left|
                \frac{1}{d}\sum_{j=1}^d
                \theta_{d,j}G_i^{(j)}G_{i'}^{(j)}
            \right|\geq\delta
        \right\},
        \mathbb{I}\left\{
            \left|
                \frac{1}{d}\sum_{j=1}^d
                \theta_{d,j}G_k^{(j)}G_{k'}^{(j)}
            \right|\geq\delta
        \right\}
    \Bigg)~.
    \label{eq: generic product screening covariance}
\end{align}
Observe that we can write
\begin{align}
    &\Cov\left(
        \mathbb{I}\left\{
            |\hat\lambda_{i,i'}|\geq\delta
        \right\},
        \mathbb{I}\left\{
            |\hat\lambda_{k,k'}|\geq\delta
        \right\}
        \mid\hat\varepsilon^{(0)}
    \right)
    =\Psi_{a,b}(\widetilde M_n,\delta)~.
    \label{eq: feasible joint product screening covariance}
\end{align}
by \cref{lem: exact feasible screening representation}, display \eqref{eq: exact feasible screening joint representation}. Thus, by applying \eqref{eq: feasible joint product screening covariance}, we obtain
\begin{align}
    \mathbb{E}\left[
        \Var\left(
            \hat V_{n,d}(\delta)
            \mid\hat\varepsilon^{(0)}
        \right)
    \right]
    & =\frac{1}{n^4}\sum_{a,b\in[n]^2}
    \mathbb{E}\left[
        \hat\varepsilon_i^{(0)}
        \hat\varepsilon_{i'}^{(0)}
        \hat\varepsilon_k^{(0)}
        \hat\varepsilon_{k'}^{(0)}
        \Psi_{a,b}(\widetilde M_n,\delta)
    \right]\nonumber \\
    &=\frac{1}{n^4}\sum_{a,b\in[n]^2}
    \mathbb{E}\left[
        \hat\varepsilon_i^{(0)}
        \hat\varepsilon_{i'}^{(0)}
        \hat\varepsilon_k^{(0)}
        \hat\varepsilon_{k'}^{(0)}
    \right]\Psi_{a,b}(M_n,\delta)
    \nonumber\\
    &+\frac{1}{n^4}\sum_{a,b\in[n]^2}
    \mathbb{E}\Bigg[
        \hat\varepsilon_i^{(0)}
        \hat\varepsilon_{i'}^{(0)}
        \hat\varepsilon_k^{(0)}
        \hat\varepsilon_{k'}^{(0)}
        \Big(
            \Psi_{a,b}(\widetilde M_n,\delta)
            -\Psi_{a,b}(M_n,\delta)
        \Big)
    \Bigg]~.
    \label{eq: conditional covariance exact comparison}
\end{align}
The two parts of the following lemma bound, in order, the two terms on the
right-hand side of \eqref{eq: conditional covariance exact comparison}.

\begin{lem}
\label{lem: weighted product screening covariance bounds}$ $\\
Suppose that, for a fixed $\eta\in(0,1]$,
\cref{assu: gaussianity,assu: proportionality,assu: homoskedasticity,%
assu: sparsity,assu: well conditioned} hold, $d\geq2$, and $s/n\leq c$,
where $c>0$ is sufficiently small. If
\eqref{eq: variance component local condition} holds, then the following
conclusions hold.

\noindent\textbf{(i)} It holds that
\begin{align}
    &\sup_{\delta\in\Delta_{n,d}(\eta)}
    \frac{1}{n^4}\sum_{a,b\in[n]^2}
    \mathbb{E}\left[
        \hat\varepsilon_i^{(0)}\hat\varepsilon_{i'}^{(0)}
        \hat\varepsilon_k^{(0)}\hat\varepsilon_{k'}^{(0)}
    \right]\Psi_{a,b}(M_n,\delta)
    \lesssim CV_n^2
    \left(
        \frac{s+1}{n}
        +V_n^\eta
        +\frac{s\sqrt d}{n}
    \right)~.
    \label{eq: centered oracle conditional variance bound}
\end{align}

\noindent\textbf{(ii)} It holds that
\begin{align}
    &\sup_{\delta\in\Delta_{n,d}(\eta)}
    \left|
        \frac{1}{n^4}\sum_{a,b\in[n]^2}
        \mathbb{E}\Bigg[
            \hat\varepsilon_i^{(0)}\hat\varepsilon_{i'}^{(0)}
            \hat\varepsilon_k^{(0)}\hat\varepsilon_{k'}^{(0)}
            \Big(
                \Psi_{a,b}(\widetilde M_n,\delta)
                -\Psi_{a,b}(M_n,\delta)
            \Big)
        \Bigg]
    \right|
    \nonumber\\
    &\quad\lesssim V_n^2
    \Bigg(
        V_n^{\eta}
        +\frac{s\sqrt d}{n}
        +dV_n^{1+\eta}
        \log(V_n^{-1})
    \Bigg)~.
    \label{eq: centered product screening covariance perturbation}
\end{align}
\end{lem}

\noindent Parts (i) and (ii) of \cref{lem: weighted product screening covariance bounds}, applied to \eqref{eq: conditional covariance exact comparison}, give
\begin{align}
    \sup_{\delta\in\Delta_{n,d}(\eta)}
    \mathbb{E}\left[
        \Var\left(
            \hat V_{n,d}(\delta)
            \mid\hat\varepsilon^{(0)}
        \right)
    \right]
    &\lesssim V_n^2
    \left(
        \frac{s+1}{n}
        +V_n^\eta
        +\frac{s\sqrt d}{n}
        +dV_n^{1+\eta}
        \log(V_n^{-1})
    \right)~,
\end{align}
as required.
\hfill\qed

\subsection{Proof of \cref{lem: conditional mean residualization aggregates}}

We begin by recording several inequalities that will be used to prove more than one part of the Lemma. Observe that \cref{lem: basic inequalities n}, display \eqref{eq: n bound 3}, \Cref{assu: homoskedasticity}, and the triangle inequality give
\begin{align}
    \max_{k,l\in\{i,i'\}}
    |\tilde\mu_{k,l}-\lambda_{k,l}|
    &\lesssim
    V_n+R_{i,i'}
    \label{eq: conditional mean raw covariance perturbation}
\end{align}
uniformly over $i,i'\in[n]$. Minkowski's inequality and \cref{lem: projection residualization bounds}, display \eqref{eq: feasible projection higher moments}, imply that
\begin{align}
    \|R_{i,i'}\|_{2m}
    &\lesssim
    mV_n
    \label{eq: conditional mean perturbation moments}
\end{align}
for every positive integer $m$, uniformly over $i,i'\in[n]$ and $m$. Moreover, \cref{assu: gaussianity,assu: proportionality}, \cref{lem: wick}, and \cref{lem: basic inequalities n}, display \eqref{eq: n bound 2}, give
\begin{align}
    \|\hat\varepsilon_i^{(0)}
      \hat\varepsilon_{i'}^{(0)}\|_4
    &\lesssim1~.
    \label{eq: conditional mean target fourth moment}
\end{align}
Consequently, Cauchy--Schwarz gives
\begin{align}
    &\mathbb E\left[
        \left(
            \hat\varepsilon_i^{(0)}
            \hat\varepsilon_{i'}^{(0)}
        \right)^2
        R_{i,i'}^m
    \right]
    \leq
    \|\hat\varepsilon_i^{(0)}
      \hat\varepsilon_{i'}^{(0)}\|_4^2
    \|R_{i,i'}\|_{2m}^m
    \leq
    \left(
        CmV_n
    \right)^m~.
    \label{eq: conditional mean weighted perturbation moments}
\end{align}
for every positive integer $m$. Applying \eqref{eq: conditional mean weighted perturbation moments} with $m=4$ and using Minkowski's inequality, we obtain
\begin{align}
    \mathbb{E}\Bigg[
        \Bigg(
        \hat\varepsilon_i^{(0)}
        \hat\varepsilon_{i'}^{(0)}
        \left(
            V_n^2+R_{i,i'}^2
        \right)
        \Bigg)^2
    \Bigg]^{1/2}
    &\lesssim
    V_n^2~.
    \label{eq: conditional mean weighted quadratic moment}
\end{align}

\subsubsection{Part (i)}

Positive semi-definiteness of $\widetilde M_n$ implies that $\tilde\mu_{i,i},\tilde\mu_{i',i'}\geq0$. For nonnegative $x,y$, it holds that
\begin{align}
    \sqrt{xy}-1
    &=
    \frac{1}{2}\bigl((x-1)+(y-1)\bigr)
    -\frac{1}{2}(\sqrt{x}-\sqrt{y})^2~.
\end{align}
By applying this identity with $x=\tilde\mu_{i,i}$ and $y=\tilde\mu_{i',i'}$, we obtain
\begin{align}
    &\left|
        \psi_{i,i'}(\widetilde M_n)-1
        -\frac{1}{2}\left(
            \tilde\mu_{i,i}-1
            +\tilde\mu_{i',i'}-1
        \right)
    \right|
    \nonumber\\
    &\quad=
    \frac{1}{2}
    \left(
        \sqrt{\tilde\mu_{i,i}}
        -\sqrt{\tilde\mu_{i',i'}}
    \right)^2
    \nonumber\\
    &\quad\leq
    \left(
        \sqrt{\tilde\mu_{i,i}}-1
    \right)^2
    +
    \left(
        \sqrt{\tilde\mu_{i',i'}}-1
    \right)^2
    \leq
    \left(
        \tilde\mu_{i,i}-1
    \right)^2
    +
    \left(
        \tilde\mu_{i',i'}-1
    \right)^2
    \lesssim
    V_n^2+R_{i,i'}^2~,
    \label{eq: conditional mean scale expansion}
\end{align}
where the last inequality follows from \eqref{eq: conditional mean raw covariance perturbation} and \cref{assu: homoskedasticity}.

Now, observe that the equality $\sum_{i=1}^n\hat\varepsilon_i^{(0)}=0$ implies that
\begin{align}
    &\sum_{i,i'=1}^n
    \hat\varepsilon_i^{(0)}
    \hat\varepsilon_{i'}^{(0)}
    \left(
        \tilde\mu_{i,i}-1
        +\tilde\mu_{i',i'}-1
    \right)
    \nonumber\\
    &\quad=
    \left(
        \sum_{i=1}^n
        \hat\varepsilon_i^{(0)}
        (\tilde\mu_{i,i}-1)
    \right)
    \left(
        \sum_{i'=1}^n
        \hat\varepsilon_{i'}^{(0)}
    \right)
    +
    \left(
        \sum_{i=1}^n
        \hat\varepsilon_i^{(0)}
    \right)
    \left(
        \sum_{i'=1}^n
        \hat\varepsilon_{i'}^{(0)}
        (\tilde\mu_{i',i'}-1)
    \right)
    =0~.
    \label{eq: conditional mean scale linear cancellation}
\end{align}
Partitioning this zero sum according to whether $(i,i')\in\mathcal S_n$ implies the decomposition
\begin{align}
    \sum_{(i,i')\notin\mathcal S_n}
    \hat\varepsilon_i^{(0)}
    \hat\varepsilon_{i'}^{(0)}
    \left(
        \psi_{i,i'}(\widetilde M_n)-1
    \right)
    & =
    \sum_{(i,i')\notin\mathcal S_n}
    \hat\varepsilon_i^{(0)}
    \hat\varepsilon_{i'}^{(0)}
    \left(
        \psi_{i,i'}(\widetilde M_n)-1
        -\frac{1}{2}\left(
            \tilde\mu_{i,i}-1
            +\tilde\mu_{i',i'}-1
        \right)
    \right)
    \nonumber\\
    &-
    \frac{1}{2}
    \sum_{(i,i')\in\mathcal S_n}
    \hat\varepsilon_i^{(0)}
    \hat\varepsilon_{i'}^{(0)}
    \left(
        \tilde\mu_{i,i}-1
        +\tilde\mu_{i',i'}-1
    \right)~.
    \label{eq: conditional mean null scale exact decomposition}
\end{align}
Obseve that Minkowski's inequality, \eqref{eq: conditional mean scale expansion}, and \eqref{eq: conditional mean weighted quadratic moment}, imply that
\begin{align}
    &\frac{1}{n^2}
    \mathbb{E}\Bigg[
        \Bigg(
        \sum_{(i,i')\notin\mathcal S_n}
        \hat\varepsilon_i^{(0)}
        \hat\varepsilon_{i'}^{(0)}
        \left(
            \psi_{i,i'}(\widetilde M_n)-1
            -\frac{1}{2}\left(
                \tilde\mu_{i,i}-1
                +\tilde\mu_{i',i'}-1
            \right)
        \right)
        \Bigg)^2
    \Bigg]^{1/2}
    \nonumber\\
    &\quad\leq
    \frac{1}{n^2}
    \sum_{(i,i')\notin\mathcal S_n}
    \mathbb{E}\Bigg[
        \Bigg(
        \hat\varepsilon_i^{(0)}
        \hat\varepsilon_{i'}^{(0)}
        \left(
            \psi_{i,i'}(\widetilde M_n)-1
            -\frac{1}{2}\left(
                \tilde\mu_{i,i}-1
                +\tilde\mu_{i',i'}-1
            \right)
        \right)
        \Bigg)^2
    \Bigg]^{1/2}
    \nonumber\\
    &\quad\lesssim
    \frac{1}{n^2}
    \sum_{(i,i')\notin\mathcal S_n}
    \mathbb{E}\Bigg[
        \Bigg(
        \hat\varepsilon_i^{(0)}
        \hat\varepsilon_{i'}^{(0)}
        \left(
            V_n^2+R_{i,i'}^2
        \right)
        \Bigg)^2
    \Bigg]^{1/2}
    \lesssim
    \frac{1}{n^2}
    \sum_{(i,i')\notin\mathcal S_n}
    V_n^2
    \lesssim
    V_n^2~.
    \label{eq: conditional mean scale remainder sum}
\end{align}
In turn, for $k\in\{i,i'\}$, the bound \eqref{eq: conditional mean raw covariance perturbation} gives
\begin{align}
    \left|
        \tilde\mu_{k,k}-1
    \right|
    &\lesssim
    V_n+R_{i,i'}~.
\end{align}
Consequently, we have that
\begin{align}
    &\mathbb{E}\Bigg[
        \Bigg(
            \hat\varepsilon_i^{(0)}
            \hat\varepsilon_{i'}^{(0)}
            \left(
                \tilde\mu_{i,i}-1
                +\tilde\mu_{i',i'}-1
            \right)
        \Bigg)^2
    \Bigg]^{1/2}
    \nonumber\\
    &\quad\lesssim
    V_n
    \mathbb{E}\left[
        \left(
            \hat\varepsilon_i^{(0)}
            \hat\varepsilon_{i'}^{(0)}
        \right)^2
    \right]^{1/2}
    +
    \mathbb{E}\left[
        \left(
            \hat\varepsilon_i^{(0)}
            \hat\varepsilon_{i'}^{(0)}
        \right)^2
        R_{i,i'}^2
    \right]^{1/2}
    \nonumber\\
    &\quad\leq
    V_n
    \mathbb{E}\left[
        \left(
            \hat\varepsilon_i^{(0)}
            \hat\varepsilon_{i'}^{(0)}
        \right)^4
    \right]^{1/4}
    +
    \mathbb{E}\left[
        \left(
            \hat\varepsilon_i^{(0)}
            \hat\varepsilon_{i'}^{(0)}
        \right)^2
        R_{i,i'}^2
    \right]^{1/2}
    \lesssim
    V_n~,
    \label{eq: conditional mean supported linear pair bound}
\end{align}
where the final inequality uses \eqref{eq: conditional mean target fourth moment} and \eqref{eq: conditional mean weighted perturbation moments} with $m=2$. Minkowski's inequality, \eqref{eq: conditional mean supported linear pair bound}, and $|\mathcal S_n|\leq n(s+1)$ now give
\begin{align}
    &\frac{1}{2n^2}
    \mathbb{E}\Bigg[
        \Bigg(
        \sum_{(i,i')\in\mathcal S_n}
        \hat\varepsilon_i^{(0)}
        \hat\varepsilon_{i'}^{(0)}
        \left(
            \tilde\mu_{i,i}-1
            +\tilde\mu_{i',i'}-1
        \right)
        \Bigg)^2
    \Bigg]^{1/2}
    \nonumber\\
    &\quad\leq
    \frac{1}{2n^2}
    \sum_{(i,i')\in\mathcal S_n}
    \mathbb{E}\Bigg[
        \Bigg(
            \hat\varepsilon_i^{(0)}
            \hat\varepsilon_{i'}^{(0)}
            \left(
                \tilde\mu_{i,i}-1
                +\tilde\mu_{i',i'}-1
            \right)
        \Bigg)^2
    \Bigg]^{1/2}
    \lesssim
    \frac{|\mathcal S_n|}{n^2}
    V_n
    \lesssim
    \frac{s}{n}V_n~.
    \label{eq: conditional mean supported linear sum}
\end{align}
Finally, applying Minkowski's inequality to \eqref{eq: conditional mean null scale exact decomposition}, and then using \eqref{eq: conditional mean scale remainder sum} and \eqref{eq: conditional mean supported linear sum}, yields
\begin{align}
    &\frac{1}{n^2}
    \mathbb{E}\Bigg[
        \Bigg(
        \sum_{(i,i')\notin\mathcal S_n}
        \hat\varepsilon_i^{(0)}
        \hat\varepsilon_{i'}^{(0)}
        \left(
            \psi_{i,i'}(\widetilde M_n)-1
        \right)
        \Bigg)^2
    \Bigg]^{1/2}
    \lesssim
    V_n^2
    +
    \frac{s}{n}V_n~.
    \label{eq: conditional mean null scale l2}
\end{align}
Therefore, \cref{lem: screening derivative bounds}, display \eqref{eq: uniform screening gradient}, gives
\begin{align}
    &\sup_{\delta\geq0}
    \frac{|\beta_{\delta,d}|}{n^2}
    \mathbb{E}\Bigg[
        \Bigg(
        \sum_{(i,i')\notin\mathcal S_n}
        \hat\varepsilon_i^{(0)}
        \hat\varepsilon_{i'}^{(0)}
        \left(
            \psi_{i,i'}(\widetilde M_n)-1
        \right)
        \Bigg)^2
    \Bigg]^{1/2}
    \nonumber\\
    &\quad\lesssim
    \sqrt d\left(
        V_n^2
        +\frac{s}{n}V_n
    \right)
    =
    V_n
    \left(
        V_n\sqrt d
        +\frac{s\sqrt d}{n}
    \right)~,
\end{align}
as required.\hfill\qed

\subsubsection{Part (ii)}

Consider the decomposition
\begin{align}
    &\sum_{(i,i')\in\mathcal S_n}
    \hat\varepsilon_i^{(0)}
    \hat\varepsilon_{i'}^{(0)}
    \Bigg(
        H_{\delta,d}\left(
            \psi_{i,i'}(\widetilde M_n),
            \rho_{i,i'}(\widetilde M_n)
        \right)
        -H_{\delta,d}(1,\lambda_{i,i'})
    \Bigg)
    \nonumber\\
    &=
    \sum_{\substack{(i,i')\in\mathcal S_n\\i\neq i'}}
    \hat\varepsilon_i^{(0)}
    \hat\varepsilon_{i'}^{(0)}
    \Bigg(
        H_{\delta,d}\left(
            \psi_{i,i'}(\widetilde M_n),
            \rho_{i,i'}(\widetilde M_n)
        \right)
        -H_{\delta,d}(1,\lambda_{i,i'})
    \Bigg)
    \nonumber\\
    &+
    \sum_{i=1}^n
    (\hat\varepsilon_i^{(0)})^2
    \Bigg(
        H_{\delta,d}\left(
            \psi_{i,i}(\widetilde M_n),
            \rho_{i,i}(\widetilde M_n)
        \right)
        -H_{\delta,d}(1,1)
    \Bigg)
    \label{eq: conditional mean support exact decomposition}
\end{align}
where we have used \cref{assu: homoskedasticity}. 

We begin by considering the first term in \eqref{eq: conditional mean support exact decomposition}. Consider an off-diagonal pair $(i,i')\in\mathcal S_n$. The two-dimensional principal submatrix of $\Lambda_n$ indexed by $i,i'$ has eigenvalues $1+\lambda_{i,i'}$ and $1-\lambda_{i,i'}$. \cref{assu: well conditioned}, applied to $\Lambda_n$, therefore gives
\begin{align}
    1-|\lambda_{i,i'}|
    &\geq c~.
    \label{eq: conditional mean off diagonal correlation gap}
\end{align}
On the event $\{R_{i,i'}\leq c\}$, \eqref{eq: conditional mean raw covariance perturbation}, \eqref{eq: conditional mean off diagonal correlation gap}, and \eqref{eq: exact screening pair parameters} imply that
\begin{align}
    \left|
        \psi_{i,i'}(\widetilde M_n)-1
    \right|
    +
    \left|
        \rho_{i,i'}(\widetilde M_n)-\lambda_{i,i'}
    \right|
    &\leq
    C\left(
        V_n+R_{i,i'}
    \right)~,
    \nonumber\\
    \psi_{i,i'}(\widetilde M_n)&\in[1/2,3/2]~,\quad\text{and}\nonumber\\
    \left|
        \rho_{i,i'}(\widetilde M_n)
    \right|
    &\leq1-c~.
    \label{eq: conditional mean off diagonal parameter rectangle}
\end{align}
The mean value theorem and \eqref{eq: uniform screening gradient} therefore yield
\begin{align}
    &\sup_{\delta\geq0}
    \left|
        H_{\delta,d}\left(
            \psi_{i,i'}(\widetilde M_n),
            \rho_{i,i'}(\widetilde M_n)
        \right)
        -H_{\delta,d}(1,\lambda_{i,i'})
    \right|
    \mathbb I\{R_{i,i'}\leq c\}
    \nonumber\\
    &\quad\lesssim
    \sqrt d\left(
        V_n+R_{i,i'}
    \right)
    \mathbb I\{R_{i,i'}\leq c\}~.
    \label{eq: conditional mean off diagonal probability comparison}
\end{align}
On the other hand, observe that, on the event $R_{i,i'} > c$, it holds that
\begin{align}
    \mathbb{I}\{ R_{i,i'} > c \} \leq \frac{R_{i,i'}^4}{c^4}
    \quad\text{and}\quad
    \vert H_{\delta,d}\left(
            \psi_{i,i'}(\widetilde M_n),
            \rho_{i,i'}(\widetilde M_n)
        \right)
        -H_{\delta,d}(1,\lambda_{i,i'})\vert \leq 1\label{eq: R > c event}
\end{align}
and so therefore
\begin{align}
    & \mathbb E\left[
        \left(
            \hat\varepsilon_i^{(0)}
            \hat\varepsilon_{i'}^{(0)}
            \vert H_{\delta,d}\left(
            \psi_{i,i'}(\widetilde M_n),
            \rho_{i,i'}(\widetilde M_n)
        \right)
        -H_{\delta,d}(1,\lambda_{i,i'})\vert
        \right)^2
        \mathbb{I}\{ R_{i,i'} > c \}
    \right]^{1/2} \nonumber \\
    & \lesssim 
    \mathbb E\left[
        \left(
            \hat\varepsilon_i^{(0)}
            \hat\varepsilon_{i'}^{(0)}
        \right)^2
        R_{i,i'}^4
    \right]^{1/2}
    \lesssim V_n^2~,
\end{align}
where the second inequality follows from \eqref{eq: conditional mean weighted perturbation moments} with $m = 4$. Consequently, Minkowski's inequality gives
\begin{align}
    &\sup_{\delta\geq0}
    \mathbb E\Bigg[
        \Bigg(
            \hat\varepsilon_i^{(0)}
            \hat\varepsilon_{i'}^{(0)}
            \Bigg(
                H_{\delta,d}\left(
                    \psi_{i,i'}(\widetilde M_n),
                    \rho_{i,i'}(\widetilde M_n)
                \right)
                -H_{\delta,d}(1,\lambda_{i,i'})
            \Bigg)
        \Bigg)^2
    \Bigg]^{1/2}
    \nonumber\\
    &\quad\lesssim
    \sqrt d\,V_n
    \mathbb E\left[
        \left(
            \hat\varepsilon_i^{(0)}
            \hat\varepsilon_{i'}^{(0)}
        \right)^2
    \right]^{1/2}
    +
    \sqrt d\,
    \mathbb E\left[
        \left(
            \hat\varepsilon_i^{(0)}
            \hat\varepsilon_{i'}^{(0)}
        \right)^2
        R_{i,i'}^2
    \right]^{1/2}
    +
    V_n^2
    \nonumber\\
    &\quad\lesssim
        V_n\sqrt d
        +V_n^2~,
    \label{eq: conditional mean off diagonal l2 bound}
\end{align}
where final inequality follows from \eqref{eq: conditional mean weighted perturbation moments} with $m=2$. Hence, as there are at most $ns$ off-diagonal pairs in $\mathcal S_n$, Minkowski's inequality and \eqref{eq: conditional mean off diagonal l2 bound} give
\begin{align}
    &\sup_{\delta\geq0}
    \frac{1}{n^2}
    \mathbb E\Bigg[
        \Bigg(
        \sum_{\substack{(i,i')\in\mathcal S_n\\i\neq i'}}
        \hat\varepsilon_i^{(0)}
        \hat\varepsilon_{i'}^{(0)}
        \Bigg(
            H_{\delta,d}\left(
                \psi_{i,i'}(\widetilde M_n),
                \rho_{i,i'}(\widetilde M_n)
            \right)
            -H_{\delta,d}(1,\lambda_{i,i'})
        \Bigg)
        \Bigg)^2
    \Bigg]^{1/2}
    \nonumber\\
    &\quad\lesssim
    \frac{ns}{n^2}
    \left(
        V_n\sqrt d
        +V_n^2
    \right)
    \lesssim
    \frac{s}{n}V_n\sqrt d~,
    \label{eq: conditional mean off diagonal aggregate}
\end{align}
where the final inequality uses \cref{lem: basic inequalities n}, display \eqref{eq: var bounds}.

Next, we consider the second term in \eqref{eq: conditional mean support exact decomposition}. Consider a diagonal pair $(i,i)$. On the event $\{R_{i,i}\leq c\}$, \eqref{eq: conditional mean raw covariance perturbation} and \eqref{eq: exact screening pair parameters} give
\begin{align}
    \tilde\mu_{i,i}
    &\in[1/2,3/2],
    &
    \psi_{i,i}(\widetilde M_n)
    &=\tilde\mu_{i,i},\quad\text{and}
    &
    \rho_{i,i}(\widetilde M_n)
    &=1~.
\end{align}
Moreover, \cref{assu: homoskedasticity} and
\cref{lem: auxiliary outcome covariance bounds}, display
\eqref{eq: auxiliary outcome covariance bounds}, give
\begin{align}
    v_{i,i}(\Lambda_n)
    &=
    \frac{2\|\Omega_d\|_{\mathrm F}^2}{d^2}
    \asymp\frac{1}{d}>0~.
    \label{eq: conditional mean diagonal oracle variance}
\end{align}
The normal parameters in \eqref{eq: feasible normal parameters} satisfy
\begin{align}
    m_{i,i}(\widetilde M_n)-m_{i,i}(\Lambda_n)
    &=
    \frac{\operatorname{tr}(\Omega_d)}{d}
    (\tilde\mu_{i,i}-1)\quad\text{and}\quad
    \sqrt{
        \frac{v_{i,i}(\widetilde M_n)}
        {v_{i,i}(\Lambda_n)}
    }
    &=\tilde\mu_{i,i}~.
    \label{eq: conditional mean diagonal normal parameters}
\end{align}
Thus, on $\{R_{i,i}\leq c\}$, we can evaluate
\begin{align}
    &\frac{
        |m_{i,i}(\widetilde M_n)-m_{i,i}(\Lambda_n)|
    }{
        \sqrt{v_{i,i}(\Lambda_n)}
    }
    +
    \left|
        \log\left(
            \sqrt{
                \frac{v_{i,i}(\widetilde M_n)}
                {v_{i,i}(\Lambda_n)}
            }
        \right)
    \right|
    \leq
    C\sqrt d\left(
        V_n+R_{i,i}
    \right)~,
    \label{eq: conditional mean diagonal standardized parameters}
\end{align}
where we have used \eqref{eq: conditional mean raw covariance perturbation}, \eqref{eq: conditional mean diagonal oracle variance}, \cref{lem: auxiliary outcome covariance bounds}, display \eqref{eq: auxiliary outcome covariance bounds}, and the mean value theorem.

Now, the triangle inequality, \cref{lem: gaussian product normal approximation}, display \eqref{eq: gaussian product normal approximation}, \eqref{eq: gaussian location scale comparison}, and \eqref{eq: conditional mean diagonal standardized parameters} give
\begin{align}
    &\sup_{\delta\geq0}
    \left|
        H_{\delta,d}\left(
            \psi_{i,i}(\widetilde M_n),
            \rho_{i,i}(\widetilde M_n)
        \right)
        -H_{\delta,d}(1,1)
    \right|
    \mathbb I\{R_{i,i}\leq c\}
    \nonumber\\
    &\quad\leq
    C\left(
        \frac{1}{\sqrt d}
        +\sqrt d\left(
            V_n+R_{i,i}
        \right)
    \right)
    \mathbb I\{R_{i,i}\leq c\}~.
    \label{eq: conditional mean diagonal probability comparison}
\end{align}
Thus, by the bounds \eqref{eq: R > c event}, we can evaluate
\begin{align}
    &\sup_{\delta\geq0}
    \mathbb E\Bigg[
        \Bigg(
            (\hat\varepsilon_i^{(0)})^2
            \Bigg(
                H_{\delta,d}\left(
                    \psi_{i,i}(\widetilde M_n),
                    \rho_{i,i}(\widetilde M_n)
                \right)
                -H_{\delta,d}(1,1)
            \Bigg)
        \Bigg)^2
    \Bigg]^{1/2}
    \nonumber\\
    &\lesssim
    \frac{1}{\sqrt d}
    \mathbb E\left[
        (\hat\varepsilon_i^{(0)})^4
    \right]^{1/2}
    +
    \sqrt d\,V_n
    \mathbb E\left[
        (\hat\varepsilon_i^{(0)})^4
    \right]^{1/2} \nonumber \\
    & +
    \sqrt d\,
    \mathbb E\left[
        (\hat\varepsilon_i^{(0)})^4R_{i,i}^2
    \right]^{1/2}
    +
    \mathbb E\left[
        (\hat\varepsilon_i^{(0)})^4R_{i,i}^4
    \right]^{1/2}
    \lesssim
    \frac{1}{\sqrt d}
        +V_n\sqrt d
        +V_n^2~,
    \label{eq: conditional mean diagonal l2 bound}
\end{align}
where the last inequality follows from \eqref{eq: conditional mean target fourth moment} and \eqref{eq: conditional mean weighted perturbation moments} with $m=2$ and $m=4$. As there are $n$ diagonal pairs, Minkowski's inequality and \eqref{eq: conditional mean diagonal l2 bound} therefore give
\begin{align}
    &\sup_{\delta\geq0}
    \frac{1}{n^2}
    \mathbb E\Bigg[
        \Bigg(
        \sum_{i=1}^n
        (\hat\varepsilon_i^{(0)})^2
        \Bigg(
            H_{\delta,d}\left(
                \psi_{i,i}(\widetilde M_n),
                \rho_{i,i}(\widetilde M_n)
            \right)
            -H_{\delta,d}(1,1)
        \Bigg)
        \Bigg)^2
    \Bigg]^{1/2}
    \nonumber\\
    &\quad\lesssim
    \frac{1}{n}
    \left(
        \frac{1}{\sqrt d}
        +V_n\sqrt d
        +V_n^2
    \right)
    \lesssim
        \frac{V_n}{\sqrt d}
        +V_n^2\sqrt d~,
    \label{eq: conditional mean diagonal aggregate}
\end{align}
where the final inequality uses $n^{-1}\leq CV_n$ from \cref{lem: basic inequalities n}, display \eqref{eq: var bounds}, and $d\geq2$. To conclude the proof, observe that Minkowski's inequality applied to \eqref{eq: conditional mean support exact decomposition}, followed by \eqref{eq: conditional mean off diagonal aggregate} and \eqref{eq: conditional mean diagonal aggregate}, yields
\begin{align}
    &\sup_{\delta\geq0}
    \frac{1}{n^2}
    \mathbb E\Bigg[
        \Bigg(
        \sum_{(i,i')\in\mathcal S_n}
        \hat\varepsilon_i^{(0)}
        \hat\varepsilon_{i'}^{(0)}
        \Bigg(
            H_{\delta,d}\left(
                \psi_{i,i'}(\widetilde M_n),
                \rho_{i,i'}(\widetilde M_n)
            \right)
            -H_{\delta,d}(1,\lambda_{i,i'})
        \Bigg)
        \Bigg)^2
    \Bigg]^{1/2}
    \nonumber\\
    &\quad\lesssim
        \frac{V_n}{\sqrt d}
        +\frac{s}{n}V_n\sqrt d
        +V_n^2\sqrt d~,
\end{align}
as required.
\hfill\qed

\subsubsection{Part (iii)}

A power-series expansion of the exponential give
\begin{align}
    &\mathbb E\Bigg[
        \left(
            \hat\varepsilon_i^{(0)}
            \hat\varepsilon_{i'}^{(0)}
        \right)^2
        \left(
            V_n^2+R_{i,i'}^2
        \right)^2
        \exp\left(
            2C\left(
                \sqrt{
                    d\log\left(
                        V_n^{-1}
                    \right)
                }
                +\log\left(
                    V_n^{-1}
                \right)
            \right)R_{i,i'}
        \right)
    \Bigg]
    \nonumber\\
    &\lesssim
    V_n^4 +
    V_n^4
    \sum_{q=1}^{\infty}
    \frac{q^q}{q!}
    \Bigg(
        V_n
        \left(
            \sqrt{
                d\log\left(
                    V_n^{-1}
                \right)
            }
            +\log\left(
                V_n^{-1}
            \right)
        \right)
    \Bigg)^q
    \nonumber\\
    &+
    V_n^4
    \sum_{q=0}^{\infty}
    \frac{(q+4)^{q+4}}{q!}
    \Bigg(
        CV_n
        \left(
            \sqrt{
                d\log\left(
                    V_n^{-1}
                \right)
            }
            +\log\left(
                V_n^{-1}
            \right)
        \right)
    \Bigg)^q
    \nonumber\\
    &\leq
    V_n^4
    \sum_{q=0}^{\infty}
    (q+1)^4
    \Bigg(
        CV_n
        \left(
            \sqrt{
                d\log\left(
                    V_n^{-1}
                \right)
            }
            +\log\left(
                V_n^{-1}
            \right)
        \right)
    \Bigg)^q
    \leq
    V_n^4~,
    \label{eq: conditional mean squared exponential moment}
\end{align}
where the first inequality uses Tonelli's theorem, \eqref{eq: conditional mean target fourth moment}, and \eqref{eq: conditional mean weighted perturbation moments} with indices $q$ and $q+4$, the second inequality uses $q^q/q!\leq C^q$ and
$(q+4)^{q+4}/q!\leq C^q(q+1)^4$, and the final inequality follows from \eqref{eq: variance component local condition}, after decreasing its constant so that the base of the series is strictly smaller than one.

Now fix $(i,i')\notin\mathcal S_n$. On the event $\{R_{i,i'}\leq c\}$, the bound \eqref{eq: conditional mean raw covariance perturbation} implies that
\begin{align}
    \psi_{i,i'}(\widetilde M_n)
    &\in[1/2,3/2],\nonumber \\
    \left|
        \rho_{i,i'}(\widetilde M_n)
    \right|
    &\leq\frac{1}{2},
    \nonumber\\
    \left\|
        \left(
            \psi_{i,i'}(\widetilde M_n)-1,
            \rho_{i,i'}(\widetilde M_n)
        \right)
    \right\|_2
    &\leq
    C\left(
        V_n+R_{i,i'}
    \right)~.
    \label{eq: conditional mean raw null parameter rectangle}
\end{align}
Thus, the segment joining $\bigl(\psi_{i,i'}(\widetilde M_n), \rho_{i,i'}(\widetilde M_n)\bigr)$ to $(1,0)$ lies in the fixed rectangle on which the bound \eqref{eq: rare screening taylor remainder} holds. Combining that bound with \eqref{eq: conditional mean raw null parameter rectangle} and \eqref{eq: variance component local condition} gives
\begin{align}
    &\left|
        H_{\delta,d}\left(
            \psi_{i,i'}(\widetilde M_n),
            \rho_{i,i'}(\widetilde M_n)
        \right)
        -H_{\delta,d}(1,0)
        -\beta_{\delta,d}
        \left(
            \psi_{i,i'}(\widetilde M_n)-1
        \right)
    \right|
    \mathbb I\{R_{i,i'}\leq c\}
    \nonumber\\
    &\quad\leq
    CV_n^\eta
    \left(
        d\log\left(
            V_n^{-1}
        \right)
        +\log^2\left(
            V_n^{-1}
        \right)
    \right)
    \nonumber\\
    &\qquad\cdot
    \exp\left(
        C\left(
            \sqrt{
                d\log\left(
                    V_n^{-1}
                \right)
            }
            +\log\left(
                V_n^{-1}
            \right)
        \right)R_{i,i'}
    \right)
    \left(
        V_n^2+R_{i,i'}^2
    \right)~,
    \label{eq: conditional mean raw null expansion}
\end{align}
where \eqref{eq: variance component local condition} absorbs the factor involving $V_n$ into the constant. Hence, we find that
\begin{align}
    &\sup_{\delta\in\Delta_{n,d}(\eta)}
    \mathbb E\Bigg[
        \Bigg(
            \hat\varepsilon_i^{(0)}
            \hat\varepsilon_{i'}^{(0)}
            \Bigg(
                H_{\delta,d}\left(
                    \psi_{i,i'}(\widetilde M_n),
                    \rho_{i,i'}(\widetilde M_n)
                \right)
                -H_{\delta,d}(1,0) \nonumber \\
                & 
                \quad\quad\quad\quad\quad\quad\quad\quad\quad\quad\quad\quad\quad\quad\quad\quad\quad\quad-\beta_{\delta,d}
                \left(
                    \psi_{i,i'}(\widetilde M_n)-1
                \right)
            \Bigg)
            \mathbb I\{R_{i,i'}\leq c\}
        \Bigg)^2
    \Bigg]^{1/2}
    \nonumber\\
    &\quad\lesssim
    V_n^{2+\eta}
    \left(
        d\log\left(
            V_n^{-1}
        \right)
        +\log^2\left(
            V_n^{-1}
        \right)
    \right)~.
    \label{eq: conditional mean raw null good event}
\end{align}
by the bounds \eqref{eq: conditional mean raw null expansion} and \eqref{eq: conditional mean squared exponential moment}.

With this in place, observe that, for nonnegative $x,y$, $\sqrt{xy}$ lies between $x$ and $y$. Thus, positive semi-definiteness of $\widetilde M_n$ and \eqref{eq: conditional mean raw covariance perturbation} give
\begin{align}
    \left|
        \psi_{i,i'}(\widetilde M_n)-1
    \right|
    &\leq
    \max\left\{
        |\tilde\mu_{i,i}-1|,
        |\tilde\mu_{i',i'}-1|
    \right\}
    \lesssim
        V_n+R_{i,i'}~.
    \label{eq: conditional mean raw null scale envelope}
\end{align}
Moreover, \cref{lem: screening derivative bounds}, display \eqref{eq: uniform screening gradient} implies that
\begin{align}
    &\left|
        H_{\delta,d}\left(
            \psi_{i,i'}(\widetilde M_n),
            \rho_{i,i'}(\widetilde M_n)
        \right)
        -H_{\delta,d}(1,0)
        -\beta_{\delta,d}
        \left(
            \psi_{i,i'}(\widetilde M_n)-1
        \right)
    \right|
    \mathbb I\{R_{i,i'}>c\}
    \nonumber\\
    &\quad\leq
    C\left(
        1+\sqrt d\,R_{i,i'}
    \right)
    \mathbb I\{R_{i,i'}>c\}~.
    \label{eq: conditional mean raw null complement envelope}
\end{align}
In turn, on the event $\{R_{i,i'}>c\}$, it holds that$\mathbb I\{R_{i,i'}>c\}\leq c^{-8}R_{i,i'}^8$. Thus, \eqref{eq: conditional mean weighted perturbation moments}, applied with moment indices $8$ and $10$, gives
\begin{align}
    \mathbb E\left[
        \left(
            \hat\varepsilon_i^{(0)}
            \hat\varepsilon_{i'}^{(0)}
        \right)^2
        \mathbb I\{R_{i,i'}>c\}
    \right]
    &\leq
    c^{-8}
    \mathbb E\left[
        \left(
            \hat\varepsilon_i^{(0)}
            \hat\varepsilon_{i'}^{(0)}
        \right)^2
        R_{i,i'}^8
    \right]
    \leq
    CV_n^8\quad\text{and}
    \nonumber\\
    \mathbb E\left[
        \left(
            \hat\varepsilon_i^{(0)}
            \hat\varepsilon_{i'}^{(0)}
        \right)^2
        R_{i,i'}^2
        \mathbb I\{R_{i,i'}>c\}
    \right]
    &\leq
    c^{-8}
    \mathbb E\left[
        \left(
            \hat\varepsilon_i^{(0)}
            \hat\varepsilon_{i'}^{(0)}
        \right)^2
        R_{i,i'}^{10}
    \right]
    \leq
    CV_n^{10}~.
    \label{eq: conditional mean raw null complement fixed moments}
\end{align}
Hence, Minkowski's inequality, \eqref{eq: conditional mean raw null complement envelope}, and \eqref{eq: conditional mean raw null complement fixed moments} imply that
\begin{align}
    &\sup_{\delta\in\Delta_{n,d}(\eta)}
    \mathbb E\Bigg[
        \Bigg(
            \hat\varepsilon_i^{(0)}
            \hat\varepsilon_{i'}^{(0)}
            \Bigg(
                H_{\delta,d}\left(
                    \psi_{i,i'}(\widetilde M_n),
                    \rho_{i,i'}(\widetilde M_n)
                \right) \nonumber \\
                &\quad\quad\quad\quad\quad\quad\quad\quad\quad\quad
                -H_{\delta,d}(1,0)
                -\beta_{\delta,d}
                \left(
                    \psi_{i,i'}(\widetilde M_n)-1
                \right)
            \Bigg)
            \mathbb I\{R_{i,i'}>c\}
        \Bigg)^2
    \Bigg]^{1/2}
    \nonumber\\
    &\quad\lesssim
    V_n^4
    +\sqrt d\,V_n^5
    \lesssim
    V_n^4
    \leq
    V_n^{2+\eta}
    \left(
        d\log\left(
            V_n^{-1}
        \right)
        +\log^2\left(
            V_n^{-1}
        \right)
    \right)~,
    \label{eq: conditional mean raw null complement l2}
\end{align}
where the penultimate inequality follows from \eqref{eq: variance component local condition} and \eqref{eq: var bounds}.

To conclude the proof, observe that Minkowski's inequality, \eqref{eq: conditional mean raw null good event}, and \eqref{eq: conditional mean raw null complement l2} give
\begin{align}
    &\sup_{\delta\in\Delta_{n,d}(\eta)}
    \mathbb E\Bigg[
        \Bigg(
            \hat\varepsilon_i^{(0)}
            \hat\varepsilon_{i'}^{(0)}
            \Bigg(
                H_{\delta,d}\left(
                    \psi_{i,i'}(\widetilde M_n),
                    \rho_{i,i'}(\widetilde M_n)
                \right) \nonumber \\
                &\quad\quad\quad\quad\quad\quad\quad\quad\quad\quad\quad\quad\quad
                -H_{\delta,d}(1,0)
                -\beta_{\delta,d}
                \left(
                    \psi_{i,i'}(\widetilde M_n)-1
                \right)
            \Bigg)
        \Bigg)^2
    \Bigg]^{1/2}
    \nonumber\\
    &\quad\lesssim
    V_n^{2+\eta}
    \left(
        d\log\left(
            V_n^{-1}
        \right)
        +\log^2\left(
            V_n^{-1}
        \right)
    \right)~.
    \label{eq: conditional mean null pair root mean square bound}
\end{align}
Finally, Minkowski's inequality, \eqref{eq: conditional mean null pair root mean square bound}, and the bound $|[n]^2\setminus\mathcal S_n|\leq n^2$ yield
\begin{align}
    &\sup_{\delta\in\Delta_{n,d}(\eta)}
    \frac{1}{n^2}
    \mathbb E\Bigg[
        \Bigg(
            \sum_{(i,i')\notin\mathcal S_n}
            \hat\varepsilon_i^{(0)}
            \hat\varepsilon_{i'}^{(0)}
            \Bigg(
                H_{\delta,d}\left(
                    \psi_{i,i'}(\widetilde M_n),
                    \rho_{i,i'}(\widetilde M_n)
                \right)  \nonumber \\
                &\quad\quad\quad\quad\quad\quad\quad\quad\quad\quad\quad\quad\quad\quad\quad\quad
                -H_{\delta,d}(1,0)
                -\beta_{\delta,d}
                \left(
                    \psi_{i,i'}(\widetilde M_n)-1
                \right)
            \Bigg)
        \Bigg)^2
    \Bigg]^{1/2}
    \nonumber\\
    &\quad\leq
    \frac{1}{n^2}
    \sum_{(i,i')\notin\mathcal S_n}
    \sup_{\delta\in\Delta_{n,d}(\eta)}
    \mathbb E\Bigg[
        \Bigg(
            \hat\varepsilon_i^{(0)}
            \hat\varepsilon_{i'}^{(0)}
            \Bigg(
                H_{\delta,d}\left(
                    \psi_{i,i'}(\widetilde M_n),
                    \rho_{i,i'}(\widetilde M_n)
                \right) \nonumber \\
                &\quad\quad\quad\quad\quad\quad\quad\quad\quad\quad\quad\quad\quad\quad\quad\quad\quad\quad\quad
                -H_{\delta,d}(1,0)
                -\beta_{\delta,d}
                \left(
                    \psi_{i,i'}(\widetilde M_n)-1
                \right)
            \Bigg)
        \Bigg)^2
    \Bigg]^{1/2}
    \nonumber\\
    &\quad\lesssim
    V_n^{2+\eta}
    \left(
        d\log\left(
            V_n^{-1}
        \right)
        +\log^2\left(
            V_n^{-1}
        \right)
    \right)~,
\end{align}
as required.\hfill\qed

\subsection{Proof of \cref{lem: weighted product screening covariance bounds}}

\subsubsection{Part (i)}

Fix $\delta\geq0$. Consider the decomposition
\begin{align}
    &\frac{1}{n^4}\sum_{a,b\in[n]^2}
    \mathbb{E}\left[
        \hat\varepsilon_i^{(0)}\hat\varepsilon_{i'}^{(0)}
        \hat\varepsilon_k^{(0)}\hat\varepsilon_{k'}^{(0)}
    \right]\Psi_{a,b}(M_n,\delta)
    \nonumber\\
    &\quad=
    \frac{1}{n^4}\sum_{a,b\in[n]^2}
    \mathbb{E}\left[
        \hat\varepsilon_i^{(0)}\hat\varepsilon_{i'}^{(0)}
        \hat\varepsilon_k^{(0)}\hat\varepsilon_{k'}^{(0)}
    \right]
    \left(
        \Psi_{a,b}(M_n,\delta)
        -\Psi_{a,b}(\Lambda_n,\delta)
    \right)
    \nonumber\\
    &\quad+
    \frac{1}{n^4}\sum_{a,b\in[n]^2}
    \mathbb{E}\left[
        \hat\varepsilon_i^{(0)}\hat\varepsilon_{i'}^{(0)}
        \hat\varepsilon_k^{(0)}\hat\varepsilon_{k'}^{(0)}
    \right]\Psi_{a,b}(\Lambda_n,\delta)~.
    \label{eq: centered oracle exact comparison}
\end{align}
We begin by giving a bound for the first term in \eqref{eq: centered oracle exact comparison}. To do this, we require the following Lemma, which follows from an application of Theorem 1.1 of \citet{devroyemehrabianreddad2023}. 

\begin{lem}
\label{lem: finite gaussian covariance comparison}$ $\\
Let $m\leq4$ be a positive integer, and let $K_0$ and $K_1$ be
positive-definite $m\times m$ matrices. Suppose that, for some
$\kappa\in(0,1)$, it holds that
\begin{align}
    \kappa I_m
    &\preceq K_r\preceq\kappa^{-1}I_m,
    \qquad r\in\{0,1\}~.
    \label{eq: finite gaussian comparison domain}
\end{align}
For $r\in\{0,1\}$, let $P_r$ be the law of $d$ independent
$\mathsf{N}(0,K_r)$ random vectors. Then, for all Borel events
$E,F\subseteq(\mathbb{R}^m)^d$, the bound
\begin{align}
    \left|
        \Cov_{P_1}\left(\mathbb{I}\{E\},\mathbb{I}\{F\}\right)
        -\Cov_{P_0}\left(\mathbb{I}\{E\},\mathbb{I}\{F\}\right)
    \right|
    &\leq
    C_\kappa\sqrt d\,\|K_1-K_0\|_{\mathrm F}
    \label{eq: finite gaussian covariance transport}
\end{align}
holds.
\end{lem}

\noindent Fix $a=(i,i')$ and $b=(k,k')$, and let $T=\{i,i',k,k'\}$ be the set of
distinct indices in the tuple. \cref{assu: well conditioned}, applied to $\Lambda_n$, and \cref{assu: homoskedasticity} imply that
\begin{align}
    c\|x\|_2^2
    &\leq x^{\top}(\lambda_{u,v})_{u,v\in T}x
    \leq
    \operatorname{tr}\left(
        (\lambda_{u,v})_{u,v\in T}
    \right)\|x\|_2^2
    =|T|\|x\|_2^2
    \leq4\|x\|_2^2~.
    \label{eq: centered population block geometry}
\end{align}
\Cref{lem: basic inequalities n}, displays \eqref{eq: var bounds} and
\eqref{eq: n bound 3}, gives
\begin{align}
    \left\|
        (\mu_{u,v}-\lambda_{u,v})_{u,v\in T}
    \right\|_{\mathrm F}
    &\leq
    |T|\max_{u,v\in T}|\mu_{u,v}-\lambda_{u,v}|
    \lesssim V_n
    \lesssim\frac{s}{n}~.
    \label{eq: centered finite block perturbation}
\end{align}
Combining \eqref{eq: centered population block geometry} and \eqref{eq: centered finite block perturbation} gives a fixed $\kappa\in(0,1)$ such that
\begin{align}
    \kappa I_{|T|}
    &\preceq(\lambda_{u,v})_{u,v\in T}
    \preceq\kappa^{-1}I_{|T|}
    \quad\text{and}\quad
    \kappa I_{|T|}
    \preceq(\mu_{u,v})_{u,v\in T}
    \preceq\kappa^{-1}I_{|T|}~.
    \label{eq: centered local gaussian comparison geometry}
\end{align}
\cref{lem: finite gaussian covariance comparison}, combined with the bounds the bounds \eqref{eq: finite gaussian covariance transport} and \eqref{eq: centered finite block perturbation}, gives
\begin{align}
    \sup_{\delta\geq0}
    \left|
        \Psi_{a,b}(M_n,\delta)
        -\Psi_{a,b}(\Lambda_n,\delta)
    \right|
    &\lesssim\sqrt d\,V_n~.
    \label{eq: centered local gaussian covariance comparison}
\end{align}
\cref{lem: wick} and \cref{lem: basic inequalities n}, display \eqref{eq: n bound 3}, yields
\begin{align}
    &\frac{1}{n^4}\sum_{i,i',k,k'=1}^n
    \left|
        \mathbb{E}\left[
            \hat\varepsilon_i^{(0)}\hat\varepsilon_{i'}^{(0)}
            \hat\varepsilon_k^{(0)}\hat\varepsilon_{k'}^{(0)}
        \right]
    \right|
    \leq
    \frac{3}{n^4}
    \left(\sum_{u,v=1}^n|\mu_{u,v}|\right)^2
    \lesssim V_n^2~.
    \label{eq: centered four row target absolute sum}
\end{align}
Hence, we obtain
\begin{align}
    &\sup_{\delta\geq0}
    \left|
        \frac{1}{n^4}\sum_{a,b\in[n]^2}
        \mathbb{E}\left[
            \hat\varepsilon_i^{(0)}\hat\varepsilon_{i'}^{(0)}
            \hat\varepsilon_k^{(0)}\hat\varepsilon_{k'}^{(0)}
        \right]
        \left(
            \Psi_{a,b}(M_n,\delta)
            -\Psi_{a,b}(\Lambda_n,\delta)
        \right)
    \right|
    \lesssim
    V_n^3\sqrt d
    \lesssim
    V_n^2\frac{s\sqrt d}{n}~.
    \label{eq: centered covariance part one strong bound}
\end{align}
by combining \eqref{eq: centered local gaussian covariance comparison} and \eqref{eq: centered four row target absolute sum}, and then applying \cref{lem: basic inequalities n}, display \eqref{eq: var bounds}.

It remains to control the second term in \eqref{eq: centered oracle exact comparison}.  By \eqref{eq: population residual covariance and independence}, the residualized auxiliary outcomes are independent of $\hat\varepsilon^{(0)}$ and have row covariance $\Lambda_n$.  Hence, conditional on $\hat\varepsilon^{(0)}$, the covariance of $\mathbb{I}\{|\check\lambda_{0,i,i'}|\geq\delta\}$ and $\mathbb{I}\{|\check\lambda_{0,k,k'}|\geq\delta\}$ is $\Psi_{a,b}(\Lambda_n,\delta)$. Expanding the conditional variance of $\check V_{n,d}^{(0)}(\delta)$ and taking expectations therefore gives
\begin{align}
    &\frac{1}{n^4}\sum_{a,b\in[n]^2}
    \mathbb{E}\left[
        \hat\varepsilon_i^{(0)}\hat\varepsilon_{i'}^{(0)}
        \hat\varepsilon_k^{(0)}\hat\varepsilon_{k'}^{(0)}
    \right]\Psi_{a,b}(\Lambda_n,\delta)
    =
    \mathbb{E}\left[
        \Var\left(
            \check V_{n,d}^{(0)}(\delta)
            \mid\hat\varepsilon^{(0)}
        \right)
    \right]
    \leq\Var\left(\check V_{n,d}^{(0)}(\delta)\right)~,
    \label{eq: population conditional variance dominated}
\end{align}
where the inequality follows from the law of total variance. Part (i) of \cref{lem: variance approximation components} now gives
\begin{align}
    \sup_{\delta\in\Delta_{n,d}(\eta)}
    \frac{1}{n^4}\sum_{a,b\in[n]^2}
    \mathbb{E}\left[
        \hat\varepsilon_i^{(0)}\hat\varepsilon_{i'}^{(0)}
        \hat\varepsilon_k^{(0)}\hat\varepsilon_{k'}^{(0)}
    \right]\Psi_{a,b}(\Lambda_n,\delta)
    \lesssim
    V_n^2
    \left(
        \frac{s+1}{n}+V_n^\eta
    \right)~.
    \label{eq: centered population covariance reference bound}
\end{align}
Taking the supremum in \eqref{eq: centered oracle exact comparison} and substituting \eqref{eq: centered covariance part one strong bound} and \eqref{eq: centered population covariance reference bound}, we obtain
\begin{align}
    & \sup_{\delta\in\Delta_{n,d}(\eta)}
    \frac{1}{n^4}\sum_{a,b\in[n]^2}
    \mathbb{E}\left[
        \hat\varepsilon_i^{(0)}\hat\varepsilon_{i'}^{(0)}
        \hat\varepsilon_k^{(0)}\hat\varepsilon_{k'}^{(0)}
    \right]\Psi_{a,b}(M_n,\delta) 
    \lesssim
    V_n^2
    \left(
        \frac{s+1}{n}+V_n^\eta
        +\frac{s\sqrt d}{n}
    \right)~,
\end{align}
as required.\hfill\qed

\subsubsection{Part (ii)}

Define the set
\begin{align}
    \mathcal Q_n
    &=\left\{
        (i,i',k,k')\in[n]^4:
        |\{i,i',k,k'\}|=4
    \right\}~.
    \label{eq: four distinct index set}
\end{align}
as well as the array
\begin{align}
    \bar M_n=(\bar\mu_{u,v})_{u,v=1}^n~,
    \quad\text{where}\quad
    \bar\mu_{u,v}
    &=\mu_{u,v}
    +(\tilde\mu_{u,v}-\mu_{u,v})
    \mathbb{I}\left\{
        |\tilde\mu_{u,v}-\mu_{u,v}|\leq c
    \right\}
    \label{eq: clipped feasible covariance array}~.
\end{align}
For every $T\subseteq[n]$ with $|T|\leq4$, the definition of $\bar M_n$
gives
\begin{align}
    \left\|
        (\bar\mu_{u,v}-\mu_{u,v})_{u,v\in T}
    \right\|_{\mathrm{op}}
    &\leq4c~.
\end{align}
Together with \eqref{eq: centered population block geometry} and
\eqref{eq: centered finite block perturbation}, this implies that
\begin{align}
    cI_{|T|}
    &\preceq
    \left((1-t)\mu_{u,v}+t\bar\mu_{u,v}\right)_{u,v\in T}
    \preceq CI_{|T|}
    \label{eq: centered covariance clipped local geometry}
\end{align}
for each $t\in[0,1]$.

Now, consider the decomposition
\begin{align}
    &\frac{1}{n^4}\sum_{a,b\in[n]^2}
    \mathbb{E}\Bigg[
        \hat\varepsilon_i^{(0)}\hat\varepsilon_{i'}^{(0)}
        \hat\varepsilon_k^{(0)}\hat\varepsilon_{k'}^{(0)}
        \Big(
            \Psi_{a,b}(\widetilde M_n,\delta)
            -\Psi_{a,b}(M_n,\delta)
        \Big)
    \Bigg]
    \nonumber\\
    &\quad=
    \mathcal R_{1,n,d}(\delta)+\mathcal R_{2,n,d}(\delta)
    +\mathcal R_{3,n,d}(\delta)+\mathcal R_{4,n,d}(\delta)~,
    \label{eq: centered covariance exact four class decomposition}
\end{align}
where
\begin{align}
    \mathcal R_{1,n,d}(\delta)
    &=\frac{1}{n^4}\sum_{a,b\in[n]^2}
    \mathbb{E}\Bigg[
        \hat\varepsilon_i^{(0)}\hat\varepsilon_{i'}^{(0)}
        \hat\varepsilon_k^{(0)}\hat\varepsilon_{k'}^{(0)} 
        \Big(
            \Psi_{a,b}(\widetilde M_n,\delta)
            -\Psi_{a,b}(\bar M_n,\delta)
        \Big)
    \Bigg]~,
    \label{eq: feasible covariance first aggregate}\\
        \mathcal R_{2,n,d}(\delta)
    &=\frac{1}{n^4}
    \sum_{(i,i',k,k')\in[n]^4\setminus\mathcal Q_n}
    \mathbb{E}\Bigg[
        \hat\varepsilon_i^{(0)}\hat\varepsilon_{i'}^{(0)}
        \hat\varepsilon_k^{(0)}\hat\varepsilon_{k'}^{(0)} \bar{\Psi}_{(i,i'),(k,k')}(\delta)
    \Bigg]~,\label{eq: feasible covariance second aggregate}\\
        \mathcal R_{3,n,d}(\delta)
    &=\frac{1}{n^4}
    \sum_{\substack{(i,i',k,k')\in\mathcal Q_n\\
        (i,i')\in\mathcal S_n\ \mathrm{or}\
        (k,k')\in\mathcal S_n}}
    \mathbb{E}\Bigg[
        \hat\varepsilon_i^{(0)}\hat\varepsilon_{i'}^{(0)}
        \hat\varepsilon_k^{(0)}\hat\varepsilon_{k'}^{(0)} 
    \bar{\Psi}_{(i,i'),(k,k')}(\delta)
    \Bigg]~,\label{eq: feasible covariance third aggregate}\\
        \mathcal R_{4,n,d}(\delta)
    &=\frac{1}{n^4}
    \sum_{\substack{(i,i',k,k')\in\mathcal Q_n\\
        (i,i')\notin\mathcal S_n,\ (k,k')\notin\mathcal S_n}}
    \mathbb{E}\Bigg[
        \hat\varepsilon_i^{(0)}\hat\varepsilon_{i'}^{(0)}
        \hat\varepsilon_k^{(0)}\hat\varepsilon_{k'}^{(0)} \bar{\Psi}_{(i,i'),(k,k')}(\delta)
    \Bigg]~,\label{eq: feasible covariance fourth aggregate}
\end{align}
and
\begin{align}
    \bar{\Psi}_{(i,i'),(k,k')}(\delta) &= \Psi_{(i,i'),(k,k')}(\bar M_n,\delta)
            -\Psi_{(i,i'),(k,k')}(M_n,\delta)
\end{align}
We bound each of the terms in \eqref{eq: centered covariance exact four class decomposition} through the application of the following Lemma.

\begin{lem}
\label{lem: feasible product screening covariance classes}$ $\\
Suppose that, for a fixed $\eta\in(0,1]$,
\cref{assu: gaussianity,assu: proportionality,assu: homoskedasticity,%
assu: sparsity,assu: well conditioned} hold, $d\geq2$, and $s/n\leq c$,
where $c>0$ is sufficiently small. If, additionally,
\eqref{eq: variance component local condition} holds, then the following
conclusions hold.

\noindent\textbf{(i)} It holds that
\begin{align}
    \sup_{\delta\geq0}|\mathcal R_{1,n,d}(\delta)|
    &\leq CV_n^3~.
    \label{eq: feasible product screening first class bound}
\end{align}

\noindent\textbf{(ii)} It holds that
\begin{align}
    &\sup_{\delta\in\Delta_{n,d}(\eta)}
    |\mathcal R_{2,n,d}(\delta)|
    \lesssim V_n^2\Bigg(
        V_n^{\eta}
        +\frac{s\sqrt d}{n}
        +dV_n^{1+\eta}
        \log(V_n^{-1})
    \Bigg)~.
    \label{eq: feasible product screening second class bound}
\end{align}

\noindent\textbf{(iii)} It holds that
\begin{align}
    \sup_{\delta\in\Delta_{n,d}(\eta)}
    |\mathcal R_{3,n,d}(\delta)|
    &\lesssim V_n^2\frac{s\sqrt d}{n}~.
    \label{eq: feasible product screening third class bound}
\end{align}

\noindent\textbf{(iv)} It holds that
\begin{align}
    &\sup_{\delta\in\Delta_{n,d}(\eta)}
    |\mathcal R_{4,n,d}(\delta)|
    \lesssim 
    V_n^2\Bigg(
        V_n^{\eta}
        +\frac{s\sqrt d}{n}
        +dV_n^{1+\eta}
        \log(V_n^{-1})
    \Bigg)~.
    \label{eq: feasible product screening fourth class bound}
\end{align}
\end{lem}

\noindent Hence, the decomposition  \eqref{eq: centered covariance exact four class decomposition} and Parts (i) through (iv) of \cref{lem: feasible product screening covariance classes} imply that
\begin{align}
    &\sup_{\delta\in\Delta_{n,d}(\eta)}
        \Bigg|
            \frac{1}{n^4}\sum_{a,b\in[n]^2}
            \mathbb{E}\Bigg[
                \hat\varepsilon_i^{(0)}\hat\varepsilon_{i'}^{(0)}
                \hat\varepsilon_k^{(0)}\hat\varepsilon_{k'}^{(0)}
                {}\cdot\Big(
                    \Psi_{a,b}(\widetilde M_n,\delta)
                    -\Psi_{a,b}(M_n,\delta)
                \Big)
            \Bigg]
        \Bigg| \nonumber \\
    &\lesssim
    V_n^3
    +V_n^2\left(
        V_n^\eta
        +\frac{s\sqrt d}{n}
        +dV_n^{1+\eta}
        \log(V_n^{-1})
    \right)
    \nonumber\\
    &\lesssim
    V_n^2\left(
        V_n^\eta
        +\frac{s\sqrt d}{n}
        +dV_n^{1+\eta}
        \log(V_n^{-1})
    \right)~,
\end{align}
as required.\hfill\qed

\subsection{Proof of \cref{lem: finite gaussian covariance comparison}}

Put $B=K_0^{-1/2}K_1K_0^{-1/2}$, and let
$\theta_1,\ldots,\theta_m$ be its eigenvalues.  For the product covariances
$I_d\otimes K_0$ and $I_d\otimes K_1$, each $\theta_\ell-1$ occurs $d$
times.  Hence \citet[Theorem~1.1]{devroyemehrabianreddad2023} gives
\begin{align}
    \|P_1-P_0\|_{\mathrm{TV}}
    &\leq
    C\left(
        d\sum_{\ell=1}^m(\theta_\ell-1)^2
    \right)^{1/2}
    =C\sqrt d\|B-I_m\|_{\mathrm F}
    \leq C_\kappa\sqrt d\|K_1-K_0\|_{\mathrm F}~.
\end{align}
Since
$\Cov_{P_r}(\mathbb{I}\{E\},\mathbb{I}\{F\})
=P_r\{E\cap F\}-P_r\{E\}P_r\{F\}$, it follows that
\begin{align}
    \left|
        \Cov_{P_1}(\mathbb{I}\{E\},\mathbb{I}\{F\})
        -\Cov_{P_0}(\mathbb{I}\{E\},\mathbb{I}\{F\})
    \right|
    &\leq
    |P_1\{E\cap F\}-P_0\{E\cap F\}|
    +|P_1\{E\}-P_0\{E\}|
    \nonumber\\
    &\quad+|P_1\{F\}-P_0\{F\}|
    \nonumber\\
    &\leq3\|P_1-P_0\|_{\mathrm{TV}}
    \leq C_\kappa\sqrt d\|K_1-K_0\|_{\mathrm F}~,
\end{align}
as required.\hfill\qed

\section{Proofs for Lemmas Supporting \cref{thm: constrained minimax lower bound}}

\subsection{Proof of \cref{lem: positive moment pair}}

The proof uses Lagrange interpolation between Chebyshev--Lobatto points to construct $\varpi_0$ and $\varpi_1$, and verifies directly that the resultant measures satisfy \eqref{eq: positive moment matching} and \eqref{eq: positive first moment gap}. See \citet[Theorem~2.5]{rivlin-1990} and \citet[Section~2.2]{wu-yang-2020} for further discussion of the connection between moment matching and best uniform approximation. Consider the sequence
\begin{align}
    x_j
    &=
    \sin^2\left(\frac{j\pi}{2L}\right),
    \quad j=0,\ldots,L~,
    \label{eq: Chebyshev Lobatto points}
\end{align}
and define
\begin{align}
    p_L(x)
    &=
    \prod_{j=0}^L(x-x_j)
    \quad\text{and}\quad
    \chi_j
    =
    \frac{1}{p_L'(x_j)}~.
    \label{eq: Chebyshev barycentric weights}
\end{align}
For every polynomial $q$ of degree at most $L-1$, the coefficient of $x^L$ in its Lagrange interpolation formula is zero. Consequently, we have that
\begin{align}
    \sum_{j=0}^L
    \frac{q(x_j)}{p_L'(x_j)}
    &=0~.
    \label{eq: Lagrange leading coefficient identity}
\end{align}
Taking $q(x)=x^k$ in
\eqref{eq: Lagrange leading coefficient identity} gives
\begin{align}
    \sum_{j=0}^L\chi_jx_j^k
    &=0,
    \quad k=0,\ldots,L-1~.
    \label{eq: barycentric coefficient identities}
\end{align}
Define $w_j=\chi_j/x_j$ for $j=1,\ldots,L$ and $w_0=-\sum_{j=1}^Lw_j$ to balance the total mass. Then
\begin{align}
    \sum_{j=0}^Lw_j
    &=0,
    \quad\text{and}\quad
    \sum_{j=0}^Lw_jx_j^r
    =0~,
    \quad r=2,\ldots,L~.
    \label{eq: barycentric moment identities}
\end{align}
Indeed, $w_jx_j^r=\chi_jx_j^{r-1}$ for $j=1,\ldots,L$, while the term with $j=0$ is zero. The second equality therefore follows from \eqref{eq: barycentric coefficient identities} with $k=r-1$.

We next calculate the positive and negative masses of the signed measure
$\sum_{j=0}^Lw_j\delta_{x_j}$ constructed above. The factorization
\begin{align}
    p_L(x)
    &=c_Lx(x-1)U_{L-1}(1-2x)
    \label{eq: Lobatto polynomial factorization}
\end{align}
holds for a nonzero constant $c_L$, where $U_{L-1}(\cos(t))=\sin(Lt)/\sin(t)$. At $t_j=j\pi/L$, differentiation of this identity gives
\begin{align}
    U_{L-1}'(\cos(t_j))
    &=-\frac{L(-1)^j}{\sin^2(t_j)},
    \quad j=1,\ldots,L-1~.
    \label{eq: Chebyshev root derivatives}
\end{align}
Using $x_j(x_j-1)=-\sin^2(t_j)/4$ in \eqref{eq: Lobatto polynomial factorization}, we obtain
\begin{align}
    \frac{p_L'(0)}{c_L}
    &=-L,
    \quad
    \frac{p_L'(x_j)}{c_L}
    =\frac{L}{2}(-1)^{j+1},
    \quad
    \frac{p_L'(1)}{c_L}
    =L(-1)^{L-1}~.
    \label{eq: Lobatto polynomial root derivatives}
\end{align}
Taking ratios in \eqref{eq: Lobatto polynomial root derivatives} gives
\begin{align}
    \frac{\chi_j}{\chi_0}
    &=2(-1)^j,
    \quad j=1,\ldots,L-1,
    \quad\text{and}\quad
    \frac{\chi_L}{\chi_0}
    =(-1)^L~.
    \label{eq: Lobatto differentiation weight ratios}
\end{align}
The factorization of $z^m-1$ into its complex roots, with $z=e^{2it}$, gives the finite product identity
\begin{align}
    \sin(mt)
    &=c_m\prod_{k=0}^{m-1}
    \sin\left(t+\frac{k\pi}{m}\right)
    \label{eq: finite sine product}
\end{align}
for a nonzero constant $c_m$. Taking a logarithmic derivative in \eqref{eq: finite sine product} gives
\begin{align}
    m\cot(mt)
    &=\sum_{k=0}^{m-1}\cot\left(t+\frac{k\pi}{m}\right)
    \label{eq: cotangent decomposition}
\end{align}
and differentiating once more gives
\begin{align}
    m^2\csc^2(mt)
    &=\sum_{k=0}^{m-1}
    \csc^2\left(t+\frac{k\pi}{m}\right)~.
    \label{eq: differentiated cotangent decomposition}
\end{align}
Subtracting $\csc^2(t)$ and using $\csc^2(t)=t^{-2}+1/3+O(t^2)$ as $t\to0$ gives
\begin{align}
    \sum_{k=1}^{m-1}
    \csc^2\left(\frac{k\pi}{m}\right)
    &=\frac{m^2-1}{3}~.
    \label{eq: cosecant square sum}
\end{align}
Applying \eqref{eq: cosecant square sum} with $m=2L$ and using symmetry about $k=L$ gives
\begin{align}
    \sum_{j=1}^{L-1}
    \csc^2\left(\frac{j\pi}{2L}\right)
    &=\frac{2(L^2-1)}{3}~.
    \label{eq: half cosecant square sum}
\end{align}
Moreover, the even-index terms in the sum \eqref{eq: cosecant square sum} with $m=2L$ sum to $(L^2-1)/3$ by applying the same identity with $m=L$. Therefore,
\begin{align}
    2\sum_{j=1}^{L-1}(-1)^j
    \csc^2\left(\frac{j\pi}{2L}\right)
    +(-1)^L
    &=-\frac{2L^2+1}{3}~.
    \label{eq: alternating cosecant square sum}
\end{align}
Since $x_j^{-1}=\csc^2(j\pi/(2L))$, combining \eqref{eq: Lobatto differentiation weight ratios}, \eqref{eq: half cosecant square sum}, and \eqref{eq: alternating cosecant square sum} gives
\begin{align}
    \frac{w_0}{\chi_0}
    &=\frac{2L^2+1}{3}
    \quad\text{and}\quad
    \frac{1}{|\chi_0|}\sum_{j=1}^L|w_j|
    =\frac{4L^2-1}{3}~.
    \label{eq: Lobatto signed mass calculation}
\end{align}
The two identities in \eqref{eq: Lobatto signed mass calculation}, together with $\sum_{j=0}^Lw_j=0$, show that
\begin{align}
    \sum_{j:w_j>0}w_j
    &=-\sum_{j:w_j<0}w_j
    =L^2|\chi_0|~.
    \label{eq: Lobatto common sign mass}
\end{align}
Define
\begin{align}
    \varpi_1
    &=\frac{1}{L^2|\chi_0|}
    \sum_{j:w_j>0}w_j\delta_{x_j},
    \quad\text{and}\quad
    \varpi_0
    =\frac{1}{L^2|\chi_0|}
    \sum_{j:w_j<0}(-w_j)\delta_{x_j}~.
    \label{eq: positive moment pair measures}
\end{align}
These are probability measures by \eqref{eq: Lobatto common sign mass}. The moment identities in \eqref{eq: barycentric moment identities} prove \eqref{eq: positive moment matching}. Finally, we have that
\begin{align}
    \left|
        \int x\,d\varpi_1(x)
        -\int x\,d\varpi_0(x)
    \right|
    &=\frac{1}{L^2|\chi_0|}
    \left|\sum_{j=1}^L\chi_j\right|
    =\frac1{L^2}~,
    \label{eq: Lobatto first moment calculation}
\end{align}
where \eqref{eq: barycentric coefficient identities} with $k=0$ gives
$\sum_{j=1}^L\chi_j=-\chi_0$, as required.
\hfill\qed

\subsection{Proof of \cref{lem: one layer likelihood bound}}

The proof relies on the application of the following two Lemmas. Determinant representations of Gaussian likelihood second moments have been used in covariance-detection lower bounds, such as \citet{berthet-rigollet-2013}. \cref{lem: alternating cycle identity} is a specialization of a standard pairing-cycle enumeration. See, for instance, \citet{greenhill-janson-kim-wormald-2002}.

\begin{lem}
\label{lem: Gaussian covariance likelihood series}$ $\\
Let $q$ and $m$ be positive integers, let $A_\omega$ be a random symmetric $q$ by $q$ matrix independent of the reference Gaussian vectors, and suppose that $|u|\|A_\omega\|_{\mathrm{op}}\leq\rho<1$ almost surely. Let $P_0$ denote the law of $m$ independent $\mathsf N(0,I_q)$ vectors, and write $\mathbb E_0$ and $\|\cdot\|_{L^2(P_0)}$ for expectation and norm under this law. For each $x\in[0,1]$ and each realization of $A_\omega$, let $\ell_{x,\omega}$ denote the likelihood ratio, with respect to $P_0$, of the law of $m$ independent $\mathsf N(0,I_q+uxA_\omega)$ vectors. There exist nonnegative numbers $a_j$ such that, for any two probability measures $\varpi_0$ and $\varpi_1$ on $[0,1]$, it holds that
\begin{align}
    \left\|
        \int\mathbb E_\omega[\ell_{x,\omega}]
        \,d(\varpi_1-\varpi_0)(x)
    \right\|_{L^2(P_0)}^2
    &=
    \sum_{j=1}^{\infty}
    a_ju^{2j}
    \left(
        \int x^j\,d\varpi_1(x)
        -
        \int x^j\,d\varpi_0(x)
    \right)^2~.
    \label{eq: Gaussian covariance signed likelihood series}
\end{align}
Moreover, it holds that
\begin{align}
    a_1
    &=
    \frac{m}{2}
    \left\|
        \mathbb E_\omega[A_\omega]
    \right\|_{\mathrm F}^2~,
    \label{eq: Gaussian covariance first likelihood coefficient}
\end{align}
and, for an independent copy $A_{\omega'}$, that
\begin{align}
    \mathbb E_0[
        \ell_{1,\omega}\ell_{1,\omega'}
    ]
    &=
    \det(I_q-u^2A_\omega A_{\omega'})^{-m/2}~.
    \label{eq: Gaussian covariance cross likelihood}
\end{align}
\end{lem}

\begin{lem}
\label{lem: alternating cycle identity}$ $\\
Fix the integer $n=2N$. Let $M$ be a perfect matching and $M'$ be another perfect matching drawn uniformly at random. View $M\cup M'$ as a multigraph that retains one copy of an edge from each matching. Let $J_\ell$ be the number of alternating components of this multigraph containing $\ell$ edges from each matching. For a formal power series $g$ and a nonnegative integer $k$, write $[u^k]g(u)$ for the coefficient of $u^k$. Then, for nonnegative numbers $w_\ell$, it holds that
\begin{align}
    \mathbb E_{M'}\left[
        \prod_{\ell\geq1}w_\ell^{J_\ell}
    \right]
    &=
    \frac{
        [u^N]
        \exp\left(
            \frac12\sum_{\ell\geq1}
            \frac{w_\ell u^\ell}{\ell}
        \right)
    }{
        [u^N](1-u)^{-1/2}
    }~.
    \label{eq: alternating cycle coefficient identity}
\end{align}
\end{lem}

\noindent We being by proving the bound when $K=bI_{n-1}$. Fix $1-c_0\leq b\leq1+c_0$, put $t=\vartheta/b$, and choose $\kappa\leq9/64$. Since $c_0\leq1/4$, this choice gives
\begin{align}
    |t|
    &\leq\frac{\sqrt\kappa}{1-c_0}
    \leq\frac12~.
    \label{eq: one layer Hermite convergence condition}
\end{align}
The adjacency matrix $A_M$ satisfies $A_M\bm{1}=\bm{1}$ and $A_M^2=I_n$. Consequently, $A_M$ maps $\bm{1}^{\perp}$ to itself, and
\begin{align}
    \widetilde A_M^2
    &=I_{n-1}
    \quad\text{and}\quad
    \|\widetilde A_M\|_{\mathrm{op}}=1
    \label{eq: one layer matching operator bounds}
\end{align}
hold. Thus, \eqref{eq: one layer Hermite convergence condition} verifies the norm condition in \cref{lem: Gaussian covariance likelihood series} and ensures that the covariance matrices below are positive definite. After dividing each observation by $\sqrt b$, let $\ell_{x,M}$ denote the likelihood of $d+1$ independent Gaussian vectors with covariance $I_{n-1}+tx\widetilde A_M$ relative to the identity covariance. Then \cref{lem: Gaussian covariance likelihood series}, applied with $m=d+1$, $u=t$, and $A_\omega=\widetilde A_M$, gives
\begin{align}
    \left\|
        \frac{d\mathbb{Q}_{1,bI}-d\mathbb{Q}_{0,bI}}{dP_{bI}}
    \right\|_{L^2(P_{bI})}^2
    &=
    \sum_{j=1}^\infty
    a_jt^{2j}
    \left(
        \int x^j\,d\varpi_1(x)
        -
        \int x^j\,d\varpi_0(x)
    \right)^2~.
    \label{eq: exact one layer likelihood series}
\end{align}
For distinct $i$ and $i'$, symmetry gives $P\{\{i,i'\}\in M\}=1/(n-1)$. Hence, we have
\begin{align}
    \mathbb E_M[\widetilde A_M]
    &=U^\top
    \left(
        \frac{\bm{1}\bm{1}^\top-I_n}{n-1}
    \right)U
    =-\frac{1}{n-1}I_{n-1}
    \label{eq: one layer residual matching mean}
\end{align}
Consequently, \cref{lem: Gaussian covariance likelihood series} gives
\begin{align}
    a_1
    &=\frac{d+1}{2}
    \left\|\mathbb E_M[\widetilde A_M]\right\|_{\mathrm F}^2
    =
    \frac{d+1}{2(n-1)}~.
    \label{eq: first one layer likelihood coefficient}
\end{align}
and \eqref{eq: positive first moment gap} gives the exact
first-order contribution
\begin{align}
    a_1t^2
    \left(
        \int x\,d\varpi_1(x)-\int x\,d\varpi_0(x)
    \right)^2
    &=\frac{(d+1)t^2}{2(n-1)L^4}~.
    \label{eq: first one layer likelihood contribution}
\end{align}
The terms with $j=2,\ldots,L$ in \eqref{eq: exact one layer likelihood series} equal zero by \eqref{eq: positive moment matching}.

Now, let $M'$ be an independent uniform perfect matching. For fixed $M$ and $M'$, \cref{lem: Gaussian covariance likelihood series} gives
\begin{align}
    \mathbb{E}[
        \ell_{1,M}\ell_{1,M'}
    ]
    &=
    \det(I_{n-1}-t^2\widetilde A_M
    \widetilde A_{M'})^{-(d+1)/2}~.
    \label{eq: one layer Gaussian cross kernel}
\end{align}
For a nonnegative scalar $z$, superimpose $M$ and $M'$,  and let $J_\ell$ count the components containing $\ell$ edges from each matching. Each component produces two cycles of length $\ell$ in the permutation matrix $A_MA_{M'}$ on the full coordinate space. Hence, we obtain the representation
\begin{align}
    \det(I_n-zA_MA_{M'})
    &=\prod_{\ell\geq1}(1-z^\ell)^{2J_\ell}~.
    \label{eq: full matching product determinant}
\end{align}
The constant vector is an eigenvector with eigenvalue one. Therefore, we have
\begin{align}
    \det(I_{n-1}-z\widetilde A_M
        \widetilde A_{M'})
    &=(1-z)^{-1}
    \prod_{\ell\geq1}(1-z^\ell)^{2J_\ell}~.
    \label{eq: restricted matching product determinant}
\end{align}
Combining \eqref{eq: restricted matching product determinant} with \eqref{eq: one layer Gaussian cross kernel}, with $z=t^2$, gives
\begin{align}
    \det(I_{n-1}-z\widetilde A_M
        \widetilde A_{M'})^{-(d+1)/2}
    &=(1-z)^{(d+1)/2}
    \prod_{\ell\geq1}
    (1-z^\ell)^{-(d+1)J_\ell}~.
    \label{eq: one layer alternating cycle determinant}
\end{align}
Put $ w_\ell=(1-z^\ell)^{-(d+1)}$ and $r_j= [u^j](1-u)^{-1/2}$. Define
\begin{align}
    g(u)
    &=
    \exp\left(
        \frac12
        \sum_{\ell\geq1}
        \frac{(w_\ell-1)u^\ell}{\ell}
    \right)~.
    \label{eq: one layer analytic factor}
\end{align}
\cref{lem: alternating cycle identity} implies that
\begin{align}
    \mathbb{E}\left[
        \prod_{\ell\geq1}w_\ell^{J_\ell}
    \right]
    &=
    \frac{[u^{n/2}](1-u)^{-1/2}g(u)}{r_{n/2}}~.
    \label{eq: one layer coefficient identity}
\end{align}
The prefactor $(1-z)^{(d+1)/2}$ in
\eqref{eq: one layer alternating cycle determinant} is at most one.
Fix a universal constant $0<\kappa_0\leq1/4$ sufficiently small. If
$(d+1)z\leq\kappa_0$, then $z\leq1/4$ and
\begin{align}
    w_\ell-1
    &=\exp\left(
        (d+1)\log\left(\frac1{1-z^\ell}\right)
    \right)-1
    \leq C(d+1)z^\ell,
    \qquad \ell\geq1~,
    \label{eq: one layer weight bound}
\end{align}
where we used $-\log(1-y)\leq2y$ for $0\leq y\leq1/2$ and $e^v-1\leq Cv$ for $0\leq v\leq2\kappa_0$ throughout the stated range. Consequently, we find that
\begin{align}
    \log g(2)
    &\leq C(d+1)\sum_{\ell\geq1}
    \frac{(2z)^\ell}{\ell}
    \leq C\kappa_0~.
    \label{eq: one layer analytic factor bound}
\end{align}
Write $g(u)=\sum_{k\geq0}g_ku^k$ with coefficients indexed by the nonnegative integers. Since $0\leq z<1$ gives $w_\ell\geq1$, these coefficients are nonnegative. The generating function $(1-u)^{-1/2}$ satisfies
\begin{align}
    2(1-u)\frac{\mathrm d}{\mathrm d u}(1-u)^{-1/2}
    &=(1-u)^{-1/2}~.
    \label{eq: central binomial generating differential identity}
\end{align}
Comparing the coefficient of $u^j$ in
\eqref{eq: central binomial generating differential identity} gives
\begin{align}
    r_0
    &=1
    \quad\text{and}\quad
    r_{j+1}=\frac{2j+1}{2j+2}r_j,
    \qquad j\geq0~.
    \label{eq: central binomial coefficient recurrence}
\end{align}
Direct expansion gives
\begin{align}
    (2j+1)^2(4j+5)
    -4(j+1)^2(4j+1)
    &=1\quad\text{and}
    \nonumber\\
    4(j+1)^2(3j+1)
    -(2j+1)^2(3j+4)
    &=j~.
    \label{eq: central binomial induction inequalities}
\end{align}
These identities are equivalent to
\begin{align}
    \left(\frac{4j+1}{4j+5}\right)^{1/2}
    &\leq
    \frac{2j+1}{2j+2}
    \leq
    \left(\frac{3j+1}{3j+4}\right)^{1/2}~.
    \label{eq: central binomial induction ratios}
\end{align}
Since $r_0=1$, \eqref{eq: central binomial coefficient recurrence} and \eqref{eq: central binomial induction ratios} imply, by induction, that $(4j+1)^{-1/2}\leq r_j\leq(3j+1)^{-1/2}$. Consequently, we find that
\begin{align}
    r_j
    &=4^{-j}\binom{2j}{j}
    \asymp(j+1)^{-1/2},
    \qquad j\geq0~.
    \label{eq: central binomial coefficient bound}
\end{align}
For $0\leq k\leq n/4$, equation \eqref{eq: central binomial coefficient bound} gives $r_{n/2-k}/r_N\leq C$. For $n/4<k\leq n/2$, the same equation gives $r_{n/2-k}/r_{n/2}\leq C\sqrt{n/2}$. Hence, we find that
\begin{align}
    \sum_{k=0}^{\lfloor n/4\rfloor}
    g_k\frac{r_{n/2-k}}{r_{n/2}}
    &\leq Cg(1)\quad\text{and}\quad
    \sum_{k=\lfloor {n/4}\rfloor+1}^{n/2}
    g_k\frac{r_{n/2-k}}{r_{n/2}}
    \leq C\sqrt {n/2}\,2^{-n/4}g(2)~.
    \label{eq: one layer split coefficient bounds}
\end{align}
Since $g(1)\leq g(2)$ and $\sqrt{n/2}2^{-n/4}\lesssim1$, the bound in \eqref{eq: one layer analytic factor bound} implies that both right-hand sides in \eqref{eq: one layer split coefficient bounds} are bounded uniformly in the sample size. Thus, we find that
\begin{align}
    \sup_{0\leq z\leq\kappa_0/(d+1)}
    \mathbb{E}_{M,M'}\left[
        \det(I_{n-1}-z\widetilde A_M
        \widetilde A_{M'})^{-(d+1)/2}
    \right]
    &\leq C
    \label{eq: amplified one layer kernel bound}
\end{align}
by \eqref{eq: one layer alternating cycle determinant} and \eqref{eq: one layer coefficient identity}.

Choose the constant $\kappa$ in the statement so that
\begin{align}
    \kappa
    &\leq\frac{9}{64}\kappa_0~.
    \label{eq: one layer amplification constant}
\end{align}
Then $(d+1)(2t)^2\leq\kappa_0$ holds as required. Since $|\int x^j\,d\varpi_1(x)-\int x^j\,d\varpi_0(x)|\leq2$ and the coefficients $a_j$ are nonnegative, the kernel bound \eqref{eq: amplified one layer kernel bound}, applied with $z=(2t)^2$, and \cref{lem: Gaussian covariance likelihood series} give
\begin{align}
    1+\sum_{j=1}^{\infty}a_j(2t)^{2j}
    &=\mathbb{E}_{M,M'}\left[
        \det(I_{n-1}-(2t)^2\widetilde A_M
        \widetilde A_{M'})^{-(d+1)/2}
    \right]
    \leq C~.
    \label{eq: amplified one layer coefficient sum}
\end{align}
Therefore, we have that
\begin{align}
    &\sum_{j>L}a_jt^{2j}
    \left(
        \int x^j\,d\varpi_1(x)
        -\int x^j\,d\varpi_0(x)
    \right)^2
    \leq
    4\,2^{-2L}
    \sum_{j>L}a_j(2t)^{2j}
    \leq C2^{-2L}~.
    \label{eq: one layer amplified tail bound}
\end{align}
Thus, the term with $j=1$ is given by
\eqref{eq: first one layer likelihood contribution}, the terms with
$2\leq j\leq L$ vanish by \eqref{eq: positive moment matching}, and the
terms with $j>L$ satisfy \eqref{eq: one layer amplified tail bound}.
Since $b\geq1-c_0\geq3/4$, we also have
$t^2=\vartheta^2/b^2\leq16\vartheta^2/9$.
Together, these three ranges prove
\eqref{eq: one layer signed likelihood bound} when $K=bI_{n-1}$.

We now extend the bound to every matrix $K$ satisfying \eqref{eq: one layer background bounds}. Put $b_*=1-c_0$. For each of the $d+1$ vectors, let $X$ have the scalar experiment with covariance $b_*I_{n-1}+\vartheta x\widetilde A_M$, let $W$ be independent centered Gaussian with covariance $K-b_*I_{n-1}$, and take the observed output to be $Y=X+W$. Since $K-b_*I_{n-1}$ is positive semi-definite, this common kernel maps the scalar reference and mixtures to $P_K$, $\mathbb{Q}_{0,K}$, and $\mathbb{Q}_{1,K}$, respectively. Under the joint scalar reference law, the output likelihood satisfies
\begin{align}
    \frac{d\mathbb{Q}_{1,K}-d\mathbb{Q}_{0,K}}{dP_K}(Y)
    &=\mathbb{E}\left[
        \frac{
            d\mathbb{Q}_{1,b_*I}
            -
            d\mathbb{Q}_{0,b_*I}
        }{
            dP_{b_*I}
        }(X)
        \mid Y
    \right]~.
    \label{eq: one layer output likelihood identity}
\end{align}
Conditional Jensen's inequality applied to
\eqref{eq: one layer output likelihood identity} gives
\begin{align}
    \left\|
        \frac{d\mathbb{Q}_{1,K}-d\mathbb{Q}_{0,K}}{dP_K}
    \right\|_{L^2(P_K)}^2
    &\leq
    \left\|
        \frac{d\mathbb{Q}_{1,b_*I}-d\mathbb{Q}_{0,b_*I}}
        {dP_{b_*I}}
    \right\|_{L^2(P_{b_*I})}^2~.
    \label{eq: one layer likelihood contraction}
\end{align}
The proof is complete by taking $b=b_*$.
\hfill\qed

\subsection{Proof of \cref{lem: Gaussian covariance likelihood series}}

If $u=0$, define $a_1$ by \eqref{eq: Gaussian covariance first likelihood coefficient} and set $a_j=0$ for $j\geq2$. All three conclusions then hold. We thus assume that $u\neq0$. Let $H_r$ denote the normalized probabilists' Hermite polynomial determined by
\begin{align}
    \exp\left(tz-\frac{t^2}{2}\right)
    &=\sum_{r=0}^{\infty}H_r(z)\frac{t^r}{\sqrt{r!}}~.
    \label{eq: lower bound hermite generating function}
\end{align}
Taking $N=1$, $S=(t)$, and $X=Z/\sqrt{2}$ in the Kibble--Slepian formula of \citet[Theorem~1.1]{ismail-zhang-2016}, and expressing the resulting polynomials in this normalization, gives
\begin{align}
    \frac{1}{\sqrt{1+t}}
    \exp\left(
        \frac{tZ^2}{2(1+t)}
    \right)
    &=
    \sum_{r=0}^{\infty}
    \frac{\sqrt{(2r)!}\,t^r}{2^rr!}
    H_{2r}(Z),
    \qquad |t|<1~.
    \label{eq: even Hermite generating identity}
\end{align}
For $Z\sim\mathsf N(0,1)$, Hermite orthogonality gives
\begin{align}
    \mathbb E\left[
        H_r(Z)
        H_{r'}(Z)
    \right]
    &=
    \mathbb I\{r=r'\}~.
    \label{eq: probabilists Hermite orthogonality}
\end{align}
Consequently, we have that
\begin{align}
    \sum_{r=0}^{\infty}
    \left(
        \frac{\sqrt{(2r)!}\,t^r}{2^rr!}
    \right)^2
    \mathbb E\left[
        H_{2r}(Z)^2
    \right]
    &=
    \sum_{r=0}^{\infty}
    \binom{2r}{r}
    \left(
        \frac{t^2}{4}
    \right)^r
    =
    (1-t^2)^{-1/2}<\infty~.
    \label{eq: even Hermite series square norm}
\end{align}
Thus, \eqref{eq: even Hermite generating identity} holds in $L^2(\mathsf N(0,1))$.

Let $Z_1,\ldots,Z_m$ denote the independent $\mathsf N(0,I_q)$ vectors
under $P_0$. For fixed $\omega$, write
\begin{align}
    A_\omega
    &=
    U_\omega
    \operatorname{diag}\left(
        \lambda_{\omega,1},\ldots,\lambda_{\omega,q}
    \right)
    U_\omega^\top,
    \qquad
    Z_{k,r,\omega}
    =
    e_r^\top U_\omega^\top Z_k~.
    \label{eq: Gaussian covariance eigencoordinates}
\end{align}
The variables $Z_{k,r,\omega}$ are independent standard Gaussian
variables for fixed $\omega$, and
\begin{align}
    |ux\lambda_{\omega,r}|
    &\leq
    \rho<1,
    \qquad
    x\in[0,1]~.
    \label{eq: Gaussian covariance eigenvalue bound}
\end{align}
Applying \eqref{eq: even Hermite generating identity} to every pair
$(k,r)$ and collecting equal powers of $ux$ gives
\begin{align}
    \ell_{x,\omega}
    &=
    1+\sum_{j=1}^{\infty}(ux)^j\mathcal L_{j,\omega},
    \label{eq: Gaussian covariance likelihood expansion}
\end{align}
where
\begin{align}
    \mathcal L_{j,\omega}
    &=
    \sum_{\substack{
        h_{k,r}\geq0\\
        \sum_{k=1}^m\sum_{r=1}^q h_{k,r}=j
    }}
    \prod_{k=1}^m\prod_{r=1}^q
    \frac{
        \lambda_{\omega,r}^{h_{k,r}}\sqrt{(2h_{k,r})!}
    }{
        2^{h_{k,r}}h_{k,r}!
    }
    H_{2h_{k,r}}(Z_{k,r,\omega})~.
    \label{eq: Gaussian covariance likelihood coefficients explicit}
\end{align}
When expressed in any fixed orthonormal coordinates, every term in
$\mathcal L_{j,\omega}$ is a linear combination of products of Hermite
polynomials having total Hermite order $2j$. Therefore,
\eqref{eq: probabilists Hermite orthogonality} and independence give
\begin{align}
    \mathbb E_0[
        \mathcal L_{j,\omega}\mathcal L_{k,\omega'}
    ]
    &=0,
    \qquad j\neq k~,
    \label{eq: Gaussian covariance coefficient orthogonality}
\end{align}
for any $\omega$ and $\omega'$.

Choose $\alpha>1$ such that $\alpha\rho<1$. Orthogonality and \eqref{eq: even Hermite series square norm}, applied with $u$ replaced by $\alpha u$, give
\begin{align}
    \left\|
        \sum_{j>J}(ux)^j\mathcal L_{j,\omega}
    \right\|_{L^2(P_0)}^2
    &=
    \sum_{j>J}
    |ux|^{2j}
    \mathbb E_0[
        \mathcal L_{j,\omega}^2
    ]
    \nonumber\\
    &\leq
    \alpha^{-2J}
    \det(
        I_q-\alpha^2u^2x^2A_\omega^2
    )^{-m/2}
    \leq
    \alpha^{-2J}
    (1-\alpha^2\rho^2)^{-mq/2}~.
    \label{eq: Gaussian covariance uniform series tail}
\end{align}
Since the total variation of $\varpi_1-\varpi_0$ is at most two, \eqref{eq: Gaussian covariance uniform series tail} implies that
\begin{align}
    &\left\|
        \int
        \mathbb E_\omega\left[
            \sum_{j>J}(ux)^j\mathcal L_{j,\omega}
        \right]
        \,d(\varpi_1-\varpi_0)(x)
    \right\|_{L^2(P_0)}
    \leq
    2\alpha^{-J}
    (1-\alpha^2\rho^2)^{-mq/4}~.
    \label{eq: Gaussian covariance integrated series tail}
\end{align}
The right-hand side converges to zero. We may therefore average \eqref{eq: Gaussian covariance likelihood expansion} over $\omega$ and integrate it term by term. Since $\int d(\varpi_1-\varpi_0)=0$, we obtain
\begin{align}
    &\int
    \mathbb E_\omega[
        \ell_{x,\omega}
    ]
    \,d(\varpi_1-\varpi_0)(x)
    \nonumber\\
    &\qquad=
    \sum_{j=1}^{\infty}
    u^j
    \left(
        \int x^j\,d\varpi_1(x)
        -
        \int x^j\,d\varpi_0(x)
    \right)
    \mathbb E_\omega[
        \mathcal L_{j,\omega}
    ]
    \label{eq: Gaussian covariance integrated likelihood expansion}
\end{align}
in $L^2(P_0)$.

Now, define the sequence
\begin{align}
    a_j
    &=
    \mathbb E_0\left[
        \left(
            \mathbb E_\omega[
                \mathcal L_{j,\omega}
            ]
        \right)^2
    \right],
    \qquad j\geq1~.
    \label{eq: Gaussian covariance likelihood coefficients}
\end{align}
These numbers are nonnegative. Moreover, \eqref{eq: Gaussian covariance coefficient orthogonality} gives
\begin{align}
    \mathbb E_0\left[
        \mathbb E_\omega[
            \mathcal L_{j,\omega}
        ]
        \mathbb E_{\omega'}[
            \mathcal L_{k,\omega'}
        ]
    \right]
    &=
    \mathbb E_{\omega,\omega'}\left[
        \mathbb E_0[
            \mathcal L_{j,\omega}\mathcal L_{k,\omega'}
        ]
    \right]
    =0,
    \qquad j\neq k~.
    \label{eq: Gaussian covariance averaged coefficient orthogonality}
\end{align}
Taking the squared $L^2(P_0)$ norm in \eqref{eq: Gaussian covariance integrated likelihood expansion} now gives
\begin{align}
    &\left\|
        \int
        \mathbb E_\omega[
            \ell_{x,\omega}
        ]
        \,d(\varpi_1-\varpi_0)(x)
    \right\|_{L^2(P_0)}^2
    =
    \sum_{j=1}^{\infty}
    a_ju^{2j}
    \left(
        \int x^j\,d\varpi_1(x)
        -
        \int x^j\,d\varpi_0(x)
    \right)^2~.
    \label{eq: Gaussian covariance signed likelihood proof}
\end{align}
This proves
\eqref{eq: Gaussian covariance signed likelihood series}.

Taking $j=1$ in \eqref{eq: Gaussian covariance likelihood coefficients explicit} gives
\begin{align}
    H_{1,\omega}
    &=
    \frac12
    \sum_{k=1}^m
    \left(
        Z_k^\top A_\omega Z_k
        -
        \operatorname{tr}(A_\omega)
    \right)~.
    \label{eq: Gaussian covariance first coefficient}
\end{align}
Consequently, we have that
\begin{align}
    \mathbb E_\omega[
        H_{1,\omega}
    ]
    &=
    \frac12
    \sum_{k=1}^m
    \left(
        Z_k^\top
        \mathbb E_\omega[
            A_\omega
        ]
        Z_k
        -
        \operatorname{tr}\left(
            \mathbb E_\omega[
                A_\omega
            ]
        \right)
    \right)~.
    \label{eq: Gaussian covariance second order coefficient}
\end{align}
The summands are independent and centered. Hence, \cref{lem: wick} gives
\begin{align}
    a_1
    &=
    \frac14
    \sum_{k=1}^m
    \Var\left(
        Z_k^\top
        \mathbb E_\omega[
            A_\omega
        ]
        Z_k
    \right)
    =
    \frac14
    \sum_{k=1}^m
    2
    \left\|
        \mathbb E_\omega[
            A_\omega
        ]
    \right\|_{\mathrm F}^2
    =
    \frac m2
    \left\|
        \mathbb E_\omega[
            A_\omega
        ]
    \right\|_{\mathrm F}^2~.
    \label{eq: Gaussian covariance first coefficient proof}
\end{align}
This proves
\eqref{eq: Gaussian covariance first likelihood coefficient}.

It remains to calculate the cross likelihood. Let $g_r$ denote the
density of $\mathsf N(0,\Omega_r^{-1})$ for
$r\in\{0,1,2\}$. If
$\Omega_1+\Omega_2-\Omega_0$ is positive definite, then
\citet[Lemma~7.4]{cai-liu-zhou-2016} state that
\begin{align}
    \int_{\mathbb R^q}
    \frac{g_1(z)g_2(z)}{g_0(z)}
    \,dz
    &=
    \frac{
        \det(\Omega_1)^{1/2}
        \det(\Omega_2)^{1/2}
    }{
        \det(\Omega_0)^{1/2}
        \det(
            \Omega_1+\Omega_2-\Omega_0
        )^{1/2}
    }~.
    \label{eq: Gaussian density product identity}
\end{align}
Apply this identity with
\begin{align}
    \Omega_0
    &=
    I_q,
    \qquad
    \Omega_1
    =
    (I_q+uA_\omega)^{-1},
    \qquad
    \Omega_2
    =
    (I_q+uA_{\omega'})^{-1}~.
    \label{eq: Gaussian covariance precision identification}
\end{align}
The required positivity follows from
\begin{align}
    \Omega_1+\Omega_2-I_q
    &\succeq
    \frac{1-\rho}{1+\rho}I_q
    \succ0~.
    \label{eq: Gaussian covariance cross integral positivity}
\end{align}
Moreover, direct multiplication gives
\begin{align}
    \Omega_1+\Omega_2-I_q
    &=
    (I_q+uA_\omega)^{-1}
    (I_q-u^2A_\omega A_{\omega'})
    (I_q+uA_{\omega'})^{-1}~.
    \label{eq: Gaussian covariance determinant identity}
\end{align}
Therefore, \eqref{eq: Gaussian density product identity} and \eqref{eq: Gaussian covariance determinant identity} give, for one reference vector,
\begin{align}
    \mathbb E_0[
        \ell_{1,\omega}\ell_{1,\omega'}
    ]
    &=
    \det(
        I_q-u^2A_\omega A_{\omega'}
    )^{-1/2}~.
    \label{eq: Gaussian covariance one vector cross likelihood}
\end{align}
Independence of the $m$ reference vectors raises the right-hand side of \eqref{eq: Gaussian covariance one vector cross likelihood} to the $m$th power, which proves \eqref{eq: Gaussian covariance cross likelihood}, and thereby completes the proof.
\hfill\qed

\subsection{Proof of \cref{lem: alternating cycle identity}}

Regard each of the $N$ edges of $M$ as a segment with two endpoints.
The uniform matching $M'$ pairs these $2N$ endpoints uniformly. A
component of $M\cup M'$ containing $\ell$ edges from each matching is
therefore a loop containing $\ell$ of the original segments. Thus,
$J_\ell$ is the loop count denoted by $C_\ell(N)$ in
\citet[Sections~1.2--1.3]{arratia-barbour-ewens-tavare-2018}.

Equation~(3) of that reference gives the joint distribution of these
counts as the Ewens sampling formula with parameter $1/2$. In the
present notation,
\begin{align}
    P_{M'}\{J_\ell=j_\ell,\ \ell=1,\ldots,N\}
    &=
    \frac{
        N!
    }{
        \prod_{k=0}^{N-1}(k+1/2)
    }
    \prod_{\ell=1}^N
    \frac{1}{j_\ell!}
    \left(
        \frac{1}{2\ell}
    \right)^{j_\ell}
    \label{eq: alternating cycle Ewens formula}
\end{align}
for nonnegative integers $j_1,\ldots,j_N$ satisfying
\begin{align}
    \sum_{\ell=1}^N\ell j_\ell
    &=N~.
    \label{eq: alternating cycle count constraint}
\end{align}
The binomial series also gives
\begin{align}
    [u^N](1-u)^{-1/2}
    &=
    \frac{
        \prod_{k=0}^{N-1}(k+1/2)
    }{
        N!
    }~.
    \label{eq: alternating cycle normalizing coefficient}
\end{align}
Since $J_\ell=0$ for $\ell>N$, it is enough to take products over
$\ell=1,\ldots,N$. Multiplying
\eqref{eq: alternating cycle Ewens formula} by
$\prod_{\ell=1}^Nw_\ell^{j_\ell}$, summing over the count vectors that
satisfy \eqref{eq: alternating cycle count constraint}, and using
\eqref{eq: alternating cycle normalizing coefficient}, we obtain
\begin{align}
    \mathbb E_{M'}\left[
        \prod_{\ell=1}^Nw_\ell^{J_\ell}
    \right]
    &=
    \frac{1}{
        [u^N](1-u)^{-1/2}
    }
    \sum_{\substack{
        j_1,\ldots,j_N\geq0\\
        \sum_{\ell=1}^N\ell j_\ell=N
    }}
    \prod_{\ell=1}^N
    \frac{1}{j_\ell!}
    \left(
        \frac{w_\ell}{2\ell}
    \right)^{j_\ell}
    \nonumber\\
    &=
    \frac{
        [u^N]
        \prod_{\ell=1}^N
        \left\{
            \sum_{j=0}^{\infty}
            \frac{1}{j!}
            \left(
                \frac{w_\ell u^\ell}{2\ell}
            \right)^j
        \right\}
    }{
        [u^N](1-u)^{-1/2}
    }
    =
    \frac{
        [u^N]
        \exp\left(
            \frac12
            \sum_{\ell=1}^N
            \frac{w_\ell u^\ell}{\ell}
        \right)
    }{
        [u^N](1-u)^{-1/2}
    }~.
    \label{eq: alternating cycle coefficient calculation}
\end{align}
Terms with $\ell>N$ cannot contribute to the coefficient of $u^N$.
Thus, \eqref{eq: alternating cycle coefficient calculation} verifies
\eqref{eq: alternating cycle coefficient identity}.
\hfill\qed

\section{Proofs for Lemmas Supporting \cref{thm: relative bias regret}}

\subsection{Proof of \cref{lem: fixed multiplier calibration}}

For each $j\in\{0,1,\ldots,d\}$ and $\ell\in[g]$, let $\hat A_\ell^{(j)}$ be the matrix whose $(i,i')$ component is the indicator used to select the term $\hat\varepsilon_i^{(j)}\hat\varepsilon_{i'}^{(j)}$ in $\hat V_{n,d}^{(j)}(\underline\delta_{n,d}(\eta_\ell))$. That is, throughtout, we write
\begin{align}
    \hat V_{n,d}^{(j)}(\ell)
    &=\hat V_{n,d}^{(j)}(\underline\delta_{n,d}(\eta_\ell))
      =\frac{1}{n^2}(\hat\varepsilon^{(j)})^\top
      \hat A_\ell^{(j)}\hat\varepsilon^{(j)}
\end{align}
variously. Moreover, we have $\hat V_{n,d}(\ell)=\hat V_{n,d}^{(0)}(\ell)$. Define the oracle residuals 
\begin{align}
    \varepsilon_i^{(k;j)}
    &=Y_i^{(k)}-\tau^{(k)}
      -\gamma_{j,k}(Y_i^{(j)}-\tau^{(j)})
\end{align}
and variance estimator
\begin{align}
    \check V_{n,d}^{(j)}(\ell)
    &=\frac{1}{n^2}(\hat\varepsilon^{(j)})^\top
      \check A_\ell^{(j)}\hat\varepsilon^{(j)}~,
    \quad\text{where}\quad
    (\check A_\ell^{(j)})_{i,i'}
    =\mathbb I\left\{
      \left|
        \frac{1}{d}\sum_{\substack{k\ne j}}
        \varepsilon_i^{(k;j)}\varepsilon_{i'}^{(k;j)}
      \right|
      \geq\underline\delta_{n,d}(\eta_\ell)
    \right\}~.
\end{align}
In turn, define the random variables
\begin{align}
    \check\xi_{n,d}^2(\ell)
    &=\frac{1}{d}\sum_{j=1}^d
      \frac{(\check V_{n,d}^{(j)}(\ell)
        -\check V_{n,d}^{(0)}(\ell))^2}
      {1-\hat\omega_{0,j}^2}
    \quad\text{and}\quad
    S^2_{n,d}(\ell) = \frac{2}{n^4}\operatorname{tr}\left(
      \check A_\ell^{(0)}M_n\check A_\ell^{(0)}M_n \right)
\end{align}
as well as
\begin{align}
    \overline{\check V}_{n,d}(\ell)
    &=\frac{1}{d}\sum_{j=1}^d\check V_{n,d}^{(j)}(\ell)
    \quad\text{and}\quad
    \check D_{n,d}^2(\ell)
    =\frac{1}{d}\sum_{j=1}^d
      \left(
        \check V_{n,d}^{(j)}(\ell)
        -\overline{\check V}_{n,d}(\ell)
      \right)^2~.
\label{eq: fixed bank oracle calibration objects}
\end{align}
Both parts of the result are obtained from the application of the following two Lemmas.

\begin{lem}
\label{lem: fixed bank sample correlation normalization}$ $\\
Suppose that the conditions of \cref{thm: relative bias regret} hold. It holds that
\begin{align}
    P\left\{
        \min_{j\in[d]}(1-\hat\omega_{0,j}^2)>0
    \right\}
    &=1
\end{align}
for all sufficiently large $n$ and that
\begin{align}
    P\left\{
        \min_{j\in[d]}(1-\hat\omega_{0,j}^2)\geq c
    \right\}
    &\to1~,
    \label{eq: fixed bank sample correlation normalization}
\end{align}
as $n,d\to\infty$.
\end{lem}

\begin{lem}
\label{lem: fixed bank oracle and feasible calibration bounds}$ $\\
Suppose that the conditions of \cref{thm: relative bias regret} hold. The following conclusions hold uniformly over $\ell\in[g]$.

\noindent\textbf{(i)} There exist nonnegative random variables $\zeta_{n,d}^{1}(\ell)$ and $\zeta_{n,d}^{2}(\ell)$ satisfying
\begin{align}
    & \left|
        \frac{1}{n^2}
        \operatorname{tr}(\check A_\ell^{(0)}M_n)
        -\overline{\check V}_{n,d}(\ell)
    \right|
    +\left[
        cS_{n,d}^2(\ell)-\check D_{n,d}^2(\ell)
    \right]_+^{1/2}
    +
    \left|
        \frac{1}{n^2}
        \operatorname{tr}(\check A_\ell^{(0)}M_n)
        -\mathbb E[\check V_{n,d}^{(0)}(\ell)]
    \right| \nonumber \\
    &\leq
    \zeta_{n,d}^{1}(\ell)S_{n,d}(\ell)
    +\zeta_{n,d}^{2}(\ell)
    \label{eq: fixed bank oracle anchor approximation}
\end{align}
such that
\begin{align}
    \zeta_{n,d}^{1}(\ell)
    &=o_p(1)
    \quad\text{and}\quad
    \zeta_{n,d}^{2}(\ell)
    =O_p\left(
        V_n\frac{\log(V_n^{-1})}{\sqrt d}
        +V_n\sqrt{dV_n}
    \right)~.
    \label{eq: fixed bank oracle comparison remainder rates}
\end{align}

\noindent\textbf{(ii)} It holds that
\begin{align}
    \Var(\check V_{n,d}^{(0)}(\ell))
    &\lesssim
    V_n^2\left(
        V_n^{\eta_\ell}+\sqrt d\,V_n
    \right)\quad\text{and}
    \nonumber\\
    \frac{1}{d}\sum_{j=1}^d
    \mathbb E\left[
        \left(
            \check V_{n,d}^{(j)}(\ell)
            -\check V_{n,d}^{(0)}(\ell)
        \right)^2
    \right]
    &\lesssim
    V_n^2\left(
        V_n^{\eta_\ell}
        +dV_n
        +\frac{\log^2(V_n^{-1})}{d}
    \right)~.
    \label{eq: fixed bank oracle anchor moment bounds}
\end{align}

\noindent\textbf{(iii)} It holds that
\begin{align}
    & \left|
      \hat V_{n,d}(\ell)-\check V_{n,d}^{(0)}(\ell)
      -\mathbb E\left[
        \hat V_{n,d}(\ell)-\check V_{n,d}^{(0)}(\ell)
      \right]
    \right|
    +|\hat\xi_{n,d}(\ell)-\check\xi_{n,d}(\ell)|
    \nonumber \\
    &=
    O_p\left(
      V_n\left(
        \frac{\log(V_n^{-1})}{\sqrt d}
        +\sqrt{dV_n}\log^{3/2}(V_n^{-1})
        +\frac{s\sqrt d}{n}
      \right)
    \right)~.
    \label{eq: fixed bank feasible oracle transfer}
\end{align}
\end{lem}

\subsubsection{Part (i)}

Consider the decomposition
\begin{align}
    &\hat V_{n,d}(\ell)-\hat\xi_{n,d}(\ell)
      -\mathbb E[\hat V_{n,d}(\ell)]
    \nonumber\\
    &\quad=
      \left( \check V_{n,d}^{(0)}(\ell)-\check\xi_{n,d}(\ell)
      -\mathbb E[\check V_{n,d}^{(0)}(\ell)] \right)
    \nonumber\\
    &\quad+ \left(
      \hat V_{n,d}(\ell)-\check V_{n,d}^{(0)}(\ell)
      -\mathbb E[\hat V_{n,d}(\ell)-\check V_{n,d}^{(0)}(\ell)]
      -(\hat\xi_{n,d}(\ell)-\check\xi_{n,d}(\ell))\right)
    \label{eq: fixed bank feasible score decomposition}
\end{align}
Observe that we can write
\begin{align}
    \check V_{n,d}^{(0)}(\ell)-\check\xi_{n,d}(\ell)
      -\mathbb E[\check V_{n,d}^{(0)}(\ell)]
    &=
      \check V_{n,d}^{(0)}(\ell)
      -\frac{1}{n^2}\operatorname{tr}(\check A_\ell^{(0)}M_n)
      -\check\xi_{n,d}(\ell)
    \nonumber\\
    &\quad+
      \frac{1}{n^2}\operatorname{tr}(\check A_\ell^{(0)}M_n)
      -\mathbb E[\check V_{n,d}^{(0)}(\ell)]~.
    \label{eq: fixed bank oracle score decomposition}
\end{align}
On the almost-sure event in \cref{lem: fixed bank sample correlation normalization}, $0<1-\hat\omega_{0,j}^2\leq1$ for every $j\in[d]$. Thus, we have that
\begin{align}
    \check\xi_{n,d}^2(\ell)
    &\geq
    \frac{1}{d}\sum_{j=1}^d
    \left(
      \check V_{n,d}^{(j)}(\ell)
      -\check V_{n,d}^{(0)}(\ell)
    \right)^2
    =
    \left(
      \check V_{n,d}^{(0)}(\ell)
      -\overline{\check V}_{n,d}(\ell)
    \right)^2
    +\check D_{n,d}^2(\ell)
    \label{eq: fixed bank oracle anova bound}
\end{align}
almost surely for all sufficiently large $n$. In turn, the reverse triangle inequality and the elementary inequality $[\sqrt{u}-\sqrt{v}]_+\leq\sqrt{[u-v]_+}$ imply that 
\begin{align}
    \check\xi_{n,d}(\ell)
    &\geq
    \sqrt{
      \left(
        \check V_{n,d}^{(0)}(\ell)
        -\frac{1}{n^2}\operatorname{tr}(\check A_\ell^{(0)}M_n)
      \right)^2
      +cS_{n,d}^2(\ell)
    }
    \nonumber\\
    &\quad-
    \left|
      \frac{1}{n^2}\operatorname{tr}(\check A_\ell^{(0)}M_n)
      -\overline{\check V}_{n,d}(\ell)
    \right|
    -\left[
      cS_{n,d}^2(\ell)-\check D_{n,d}^2(\ell)
    \right]_+^{1/2}~.
    \label{eq: fixed bank oracle penalty lower bound}
\end{align}
By applying Part (i) of \cref{lem: fixed bank oracle and feasible calibration bounds} to \eqref{eq: fixed bank oracle penalty lower bound} and plugging the result into \eqref{eq: fixed bank oracle score decomposition}, we obtain
\begin{align}
    \check V_{n,d}^{(0)}(\ell)-\check\xi_{n,d}(\ell)
      -\mathbb E[\check V_{n,d}^{(0)}(\ell)]
    &\leq
      \check V_{n,d}^{(0)}(\ell)
      -\frac{1}{n^2}\operatorname{tr}(\check A_\ell^{(0)}M_n)
    \nonumber\\
    &\quad-
      \sqrt{
        \left(
          \check V_{n,d}^{(0)}(\ell)
          -\frac{1}{n^2}\operatorname{tr}(\check A_\ell^{(0)}M_n)
        \right)^2
        +cS_{n,d}^2(\ell)
      }
    \nonumber\\
    &\quad+
      \zeta_{n,d}^{1}(\ell)S_{n,d}(\ell)
      +\zeta_{n,d}^{2}(\ell)~.
    \label{eq: fixed bank oracle endpoint bound}
\end{align}
Hence, we have that
\begin{align}
    & \hat V_{n,d}(\ell)-\hat\xi_{n,d}(\ell)
      -\mathbb E[\hat V_{n,d}(\ell)] \nonumber \\
    &\leq
      \check V_{n,d}^{(0)}(\ell)
      -\frac{1}{n^2}\operatorname{tr}(\check A_\ell^{(0)}M_n)
    -
      \sqrt{
        \left(
          \check V_{n,d}^{(0)}(\ell)
          -\frac{1}{n^2}\operatorname{tr}(\check A_\ell^{(0)}M_n)
        \right)^2
        +cS_{n,d}^2(\ell)
      } \nonumber \\
    & +
      \zeta^1_{n,d}(\ell)S_{n,d}(\ell)
      +\zeta^2_{n,d}(\ell)
    +
      \left|
        \hat V_{n,d}(\ell)-\check V_{n,d}^{(0)}(\ell)
        -\mathbb E[
          \hat V_{n,d}(\ell)-\check V_{n,d}^{(0)}(\ell)
        ]
      \right|
    +
      |\hat\xi_{n,d}(\ell)-\check\xi_{n,d}(\ell)|~.
    \label{eq: fixed bank feasible endpoint bound}
\end{align}
by the triangle inequality.

Now, observe that \eqref{eq: population residual covariance and independence} implies that $\check A_\ell^{(0)}$ is independent of $\hat\varepsilon^{(0)}\sim\mathsf N(0,M_n)$. Thus, \cref{lem: wick} implies that the equality
\begin{align}
    \mathbb E\left[
      \left(
        \check V_{n,d}^{(0)}(\ell)
        -\frac{1}{n^2}\operatorname{tr}(\check A_\ell^{(0)}M_n)
      \right)^2
      \mid\check A_\ell^{(0)}
    \right]
    &=\frac{2}{n^4}\operatorname{tr}\left(
      \check A_\ell^{(0)}M_n\check A_\ell^{(0)}M_n
    \right)
    =S_{n,d}^2(\ell)
    \label{eq: fixed bank oracle conditional moments}
\end{align}
holds almost surely.

With this in place, choose a sequence $\epsilon_{n,d}\downarrow0$ such that
\begin{align}
    P\left\{
      \max_{\ell\in[g]}\zeta^1_{n,d}(\ell)>\epsilon_{n,d}
    \right\}
    &\to0~.
\end{align}

For every $x\in\mathbb R$ and $v\geq0$, and all sufficiently small
$\epsilon>0$, the inequality
\begin{align}
    x-\sqrt{x^2+cv^2}+\epsilon v>0
    &\quad\text{implies}\quad
    x^2>\frac{c^2}{16\epsilon^2}v^2~.
    \label{eq: fixed bank elementary absorption}
\end{align}

Consequently, Markov's inequality implies that
\begin{align}
    &P\Bigg\{
      \max_{\ell\in[g]}\Bigg(
      \check V_{n,d}^{(0)}(\ell)
      -\frac{1}{n^2}\operatorname{tr}(\check A_\ell^{(0)}M_n)
    \nonumber\\
    &\quad\quad\quad-
      \sqrt{
        \left(
          \check V_{n,d}^{(0)}(\ell)
          -\frac{1}{n^2}\operatorname{tr}(\check A_\ell^{(0)}M_n)
        \right)^2
        +cS_{n,d}^2(\ell)
      }
      +\zeta^1_{n,d}(\ell)S_{n,d}(\ell)
      \Bigg)>0
    \Bigg\}
    \nonumber\\
    &\leq
      P\left\{
        \max_{\ell\in[g]}\zeta^1_{n,d}(\ell)>\epsilon_{n,d}
      \right\}
      +\sum_{\ell=1}^g
      P\left\{
        \left(
          \check V_{n,d}^{(0)}(\ell)
          -\frac{1}{n^2}\operatorname{tr}(\check A_\ell^{(0)}M_n)
        \right)^2
        > \frac{c^2}{16\epsilon_{n,d}^2}S_{n,d}^2(\ell)
      \right\}
    \nonumber\\
    &\leq o(1)+ \frac{16g}{c^2}\epsilon_{n,d}^2
    =o(1)~.
    \label{eq: fixed bank endpoint absorption probability}
\end{align}
Thus, combining \eqref{eq: fixed bank feasible endpoint bound} and
\eqref{eq: fixed bank endpoint absorption probability}, we can conclude that
\begin{align}
    &\max_{\ell\in[g]}
      \frac{[
        \hat V_{n,d}(\ell)-\hat\xi_{n,d}(\ell)
        -\mathbb E[\hat V_{n,d}(\ell)]
      ]_+}{V_n}
    \nonumber\\
    &\leq
      \max_{\ell\in[g]}\frac{1}{V_n}\left\{
        \zeta^2_{n,d}(\ell)
        +\left|
          \hat V_{n,d}(\ell)-\check V_{n,d}^{(0)}(\ell)
          -\mathbb E[
            \hat V_{n,d}(\ell)-\check V_{n,d}^{(0)}(\ell)
          ]
        \right|+|\hat\xi_{n,d}(\ell)-\check\xi_{n,d}(\ell)|
      \right\}
    \label{eq: fixed bank feasible endpoint assembly}
\end{align}
with probability tending to one. By applying Parts (i) and (iii) of \cref{lem: fixed bank oracle and feasible calibration bounds} to \eqref{eq: fixed bank feasible endpoint assembly}, we find that
\begin{align}
    \max_{\ell\in[g]}
      \frac{[
        \hat V_{n,d}(\ell)-\hat\xi_{n,d}(\ell)
        -\mathbb E[\hat V_{n,d}(\ell)]
      ]_+}{V_n}
    &=O_p\left(
      \frac{\log(V_n^{-1})}{\sqrt d}
      +\sqrt{dV_n}\log^{3/2}(V_n^{-1})
      +\frac{s\sqrt d}{n}
    \right)~,
\end{align}
as required.\hfill\qed

\subsubsection{Part (ii)}

Consider the bound
\begin{align}
    &|\hat V_{n,d}(\ell)-\mathbb E[\hat V_{n,d}(\ell)]|
      +\hat\xi_{n,d}(\ell)
    \nonumber\\
    &\quad\leq
      |\check V_{n,d}^{(0)}(\ell)
        -\mathbb E[\check V_{n,d}^{(0)}(\ell)]|
      +\check\xi_{n,d}(\ell)
    \nonumber\\
    &\quad+
      \left|
        \hat V_{n,d}(\ell)-\check V_{n,d}^{(0)}(\ell)
        -\mathbb E[\hat V_{n,d}(\ell)-\check V_{n,d}^{(0)}(\ell)]
      \right|
      +|\hat\xi_{n,d}(\ell)-\check\xi_{n,d}(\ell)|~.
    \label{eq: fixed bank optimizer assembly}
\end{align}
Part (ii) of \cref{lem: fixed bank oracle and feasible calibration bounds} and Chebyshev's inequality imply that
\begin{align}
    \frac{
      |\check V_{n,d}^{(0)}(\ell)
        -\mathbb E[\check V_{n,d}^{(0)}(\ell)]|
    }{V_n}
    &=O_p\left(
      V_n^{\eta_\ell/2}+\sqrt{dV_n}
    \right)~.
    \label{eq: fixed bank oracle anchor rate}
\end{align}
In turn, \cref{lem: fixed bank sample correlation normalization} implies that 
\begin{align}
    \check\xi_{n,d}^2(\ell)
    &\lesssim
    \frac{1}{d}\sum_{j=1}^d
    \left(
      \check V_{n,d}^{(j)}(\ell)
      -\check V_{n,d}^{(0)}(\ell)
    \right)^2~.
\end{align}
holds with probability tending to one. Thus, Part (ii) of \cref{lem: fixed bank oracle and feasible calibration bounds} an Markov's inequality imply that
\begin{align}
    \frac{\check\xi_{n,d}(\ell)}{V_n}
    &=O_p\left(
      V_n^{\eta_\ell/2}
      +\frac{\log(V_n^{-1})}{\sqrt d}
      +\sqrt{dV_n}
    \right)~.\label{eq: check xi over V}
\end{align}
Combining \eqref{eq: fixed bank oracle anchor rate} and \eqref{eq: check xi over V}, we find that 
\begin{align}
    \frac{
      |\check V_{n,d}^{(0)}(\ell)
        -\mathbb E[\check V_{n,d}^{(0)}(\ell)]|
      +\check\xi_{n,d}(\ell)
    }{V_n}
    &=O_p\left(
      V_n^{\eta_\ell/2}
      +\frac{\log(V_n^{-1})}{\sqrt d}
      +\sqrt{dV_n}
    \right)~.
    \label{eq: fixed bank oracle optimizer rate}
\end{align}
Hence, by plugging Part (iii) of \cref{lem: fixed bank oracle and feasible calibration bounds} and \eqref{eq: fixed bank oracle optimizer rate} into \eqref{eq: fixed bank optimizer assembly}, we find that
\begin{align}
    \frac{
      |\hat V_{n,d}(\ell)-\mathbb E[\hat V_{n,d}(\ell)]|
      +\hat\xi_{n,d}(\ell)
    }{V_n}
    &=O_p\left(
      V_n^{\eta_\ell/2}
      +\frac{\log(V_n^{-1})}{\sqrt d}
      +\sqrt{dV_n}\log^{3/2}(V_n^{-1})
      +\frac{s\sqrt d}{n}
    \right)
\end{align}
as required.

\hfill\qed

\subsection{Proof of \cref{lem: risk selection bound}}

The identity \eqref{eq: fixed bank oracle conditional moments}, Markov's inequality, and the fact that $g$ is fixed imply that, for every $\epsilon>0$, there is a constant $C_\epsilon<\infty$ such that
\begin{align}
    \left|
      \check V_{n,d}^{(0)}(\ell)
      -\frac{1}{n^2}
       \operatorname{tr}(\check A_\ell^{(0)}M_n)
    \right|
    &\leq C_\epsilon S_{n,d}(\ell)
\end{align}
simultaneously over $\ell\in[g]$ with probability at least $1-\epsilon$ for all sufficiently large $n,d$. On this event,
\begin{align}
    &\sqrt{
      \left(
        \check V_{n,d}^{(0)}(\ell)
        -\frac{1}{n^2}
         \operatorname{tr}(\check A_\ell^{(0)}M_n)
      \right)^2
      +cS_{n,d}^2(\ell)
    }
    -
    \left(
      \check V_{n,d}^{(0)}(\ell)
      -\frac{1}{n^2}
       \operatorname{tr}(\check A_\ell^{(0)}M_n)
    \right)
    -\zeta^1_{n,d}(\ell)S_{n,d}(\ell)
    \nonumber\\
    &\quad\geq
      \left(
        \sqrt{C_\epsilon^2+c}
        -C_\epsilon
        -\zeta^1_{n,d}(\ell)
      \right)S_{n,d}(\ell)~.\label{eq: D11 first}
\end{align}
Since $\sqrt{C_\epsilon^2+c}-C_\epsilon>0$ and \eqref{eq: fixed bank oracle comparison remainder rates} implies that $\max_{\ell\in[g]}\zeta^1_{n,d}(\ell)=o_p(1)$, the right-hand side of \eqref{eq: D11 first} is bounded below by $c_\epsilon S_{n,d}(\ell)$ simultaneously over $\ell\in[g]$, with probability tending to one, for some constant $c_\epsilon>0$. Thus, \eqref{eq: fixed bank feasible endpoint bound}, \eqref{eq: fixed bank oracle comparison remainder rates}, and \eqref{eq: fixed bank feasible oracle transfer} imply that, after possibly enlarging $C_\epsilon$, there is a constant $c_\epsilon>0$ such that
\begin{align}
    \mathbb E[\hat V_{n,d}(\ell)]
      -\hat V_{n,d}(\ell)+\hat\xi_{n,d}(\ell)
    &\geq
      c_\epsilon S_{n,d}(\ell)
      -C_\epsilon V_n\mathcal{R}_{n,d}(s)
    \label{eq: risk selection score coercivity}
\end{align}
simultaneously over $\ell\in[g]$ with probability at least $1-\epsilon$ for all sufficiently large $n,d$.

Now, the definitions of $\hat k$ and $k^\star$, \eqref{eq: negative bias approx}, and \cref{lem: bias decomposition} imply that
\begin{align}
    &\mathbb E[\hat V_{n,d}(\hat k)]
      -\hat V_{n,d}(\hat k)+\hat\xi_{n,d}(\hat k)
    \nonumber\\
    &\quad\leq
      \mathbb E[\hat V_{n,d}(k^\star)]
      -\hat V_{n,d}(k^\star)+\hat\xi_{n,d}(k^\star)
      +C V_n\mathcal{R}_{n,d}(s)~.
    \label{eq: risk selection score comparison}
\end{align}
Combining \eqref{eq: risk selection score coercivity}, \eqref{eq: risk selection score comparison}, and \eqref{eq: optimizer side calibration}, and using the fact that $\epsilon>0$ is arbitrary, we obtain
\begin{align}
    \frac{S_{n,d}(\hat k)}{V_n}
    &=O_p\left(
        V_n^{\eta_{k^\star}/2}
        +\mathcal{R}_{n,d}(s)
    \right)~.
    \label{eq: risk selection selected conditional scale}
\end{align}
In turn, for an off-diagonal null pair $i,i'$, define
\begin{align}
    \pi_{n,d}(\ell)
    &=P\left\{(\check A_\ell^{(0)})_{i,i'}=1\right\}~.
\end{align}
The identity $M_n=H_n\Lambda_nH_n$ and \cref{assu: well conditioned} imply that $\|M_n\|_{\mathrm{op}}\lesssim1$. Thus, \eqref{eq: fixed bank oracle conditional moments}, \eqref{eq: fixed bank screen trace inequality}, and \eqref{eq: fixed bank full screen mean bounds} give
\begin{align}
    \left(\mathbb E[S_{n,d}^2(\ell)]\right)^{1/2}
    &\lesssim
      \frac{1}{n}\left(
        \sqrt{\pi_{n,d}(\ell)}
        +\sqrt{dV_n}
      \right)
    \label{eq: risk selection expected conditional scale}
\end{align}
uniformly over $\ell\in[g]$.

With this in place, consider the indices $\ell\in[g]$ satisfying
\begin{align}
    \pi_{n,d}(\ell)
    &>dV_n\log^2(V_n^{-1})~.
    \label{eq: risk selection frequent screening regime}
\end{align}
For these indices, $dV_n/\pi_{n,d}(\ell)\leq\log^{-2}(V_n^{-1})$. Hence, \eqref{eq: fixed bank full screen mean bounds}, \eqref{eq: fixed bank row count bound}, \eqref{eq: fixed bank full screen count variance}, \eqref{eq: fixed bank null screen probability}, \cref{lem: basic inequalities n}, display \eqref{eq: var bounds}, and \eqref{eq: regret rate condition}, together with Chebyshev's and Markov's inequalities, imply that
\begin{align}
    \frac{
      \sum_{i,i'=1}^n(\check A_\ell^{(0)})_{i,i'}
    }{n^2\pi_{n,d}(\ell)}
    &\overset{p}{\to}1
    \quad\text{and}\quad
    \frac{
      n^{-1}\sum_{i=1}^n
      \left(
        \sum_{i'=1}^n(\check A_\ell^{(0)})_{i,i'}
      \right)^2
    }{n^2\pi_{n,d}(\ell)}
    \overset{p}{\to}0
    \label{eq: risk selection screen count concentration}
\end{align}
simultaneously over these indices. Therefore, \eqref{eq: fixed bank oracle screen double centering}, \eqref{eq: fixed bank projected energy coercivity}, and \eqref{eq: risk selection screen count concentration} yield
\begin{align}
    S_{n,d}^2(\ell)
    &\gtrsim
      \frac{\pi_{n,d}(\ell)}{n^2}
    \label{eq: risk selection frequent screen energy}
\end{align}
with probability tending to one, simultaneously over these indices. On the other hand, for an index that does not satisfy \eqref{eq: risk selection frequent screening regime}, \eqref{eq: risk selection expected conditional scale} and \cref{lem: basic inequalities n}, display \eqref{eq: var bounds}, imply that
\begin{align}
    \left(\mathbb E[S_{n,d}^2(\ell)]\right)^{1/2}
    &\lesssim
      \frac{\sqrt{dV_n}}{n}
      \left(\log(V_n^{-1})+1\right)
    \lesssim
      V_n\mathcal R_{n,d}(s)~.\label{eq: above case}
\end{align}
For an index that satisfies \eqref{eq: risk selection frequent screening regime}, \eqref{eq: risk selection expected conditional scale}, \eqref{eq: risk selection frequent screen energy}, and \cref{lem: basic inequalities n}, display \eqref{eq: var bounds}, instead imply that
\begin{align}
    \left(\mathbb E[S_{n,d}^2(\ell)]\right)^{1/2}
    &\lesssim
      S_{n,d}(\ell)
      +\frac{\sqrt{dV_n}}{n}
    \lesssim
      S_{n,d}(\ell)+V_n\mathcal R_{n,d}(s)~.\label{eq: below case}
\end{align}
Consequently, we find that 
\begin{align}
    \max_{\ell\in[g]}
    \frac{
      \left(\mathbb E[S_{n,d}^2(\ell)]\right)^{1/2}
    }{
      S_{n,d}(\ell)+V_n\mathcal R_{n,d}(s)
    }
    &=O_p(1)
    \label{eq: risk selection scale comparison}
\end{align}
by applying \eqref{eq: above case} and \eqref{eq: below case}.

Finally, the law of total variance and
\eqref{eq: fixed bank oracle conditional moments} give the identity
\begin{align}
    \Var\left(\check V_{n,d}^{(0)}(\ell)\right)
    &=\mathbb E\left[S_{n,d}^2(\ell)\right]
      +\Var\left(
        \frac{1}{n^2}\operatorname{tr}(\check A_\ell^{(0)}M_n)
      \right)~.
\end{align}
Thus, Part (i) of \cref{lem: fixed bank oracle calibration aggregates}, \eqref{eq: fixed bank feasible oracle second moment}, and the triangle inequality imply that
\begin{align}
    \sqrt{
      \Var\left(
        \hat V_{n,d}(\underline\delta_{n,d}(\eta_\ell))
      \right)
    }
    \lesssim
      \left(\mathbb E[S_{n,d}^2(\ell)]\right)^{1/2}
      +V_n^{3/2}
      +V_n\mathcal{R}_{n,d}(s)
    \label{eq: risk selection variance assembly}
\end{align}
uniformly over $\ell\in[g]$. Hence, \eqref{eq: risk selection scale comparison} and \eqref{eq: risk selection variance assembly} imply that
\begin{align}
    \max_{\ell\in[g]}
    \frac{
      \sqrt{
        \Var\left(
          \hat V_{n,d}(\underline\delta_{n,d}(\eta_\ell))
        \right)
      }
    }{
      S_{n,d}(\ell)+V_n\mathcal{R}_{n,d}(s)
    }
    &=O_p(1)~.
    \label{eq: risk selection variance comparison}
\end{align}
Evaluating \eqref{eq: risk selection variance comparison} at $\ell=\hat k$ and applying \eqref{eq: risk selection selected conditional scale}, we obtain
\begin{align}
    \frac{
      \sqrt{
        \Var\left(
          \hat V_{n,d}(\underline\delta_{n,d}(\eta_\ell))
        \right)
      }
    }{V_n}
    \bigg|_{\ell=\hat k}
    &=O_p\left(
        V_n^{\eta_{k^\star}/2}
        +\mathcal{R}_{n,d}(s)
    \right)~,
\end{align}
as required.\hfill\qed

\subsection{Proof of \cref{lem: fixed bank sample correlation normalization}}

\cref{assu: well conditioned}, applied to $\Gamma_d$, and \cref{assu: homoskedasticity} imply that
\begin{align}
    1-|\gamma_{0,j}|
    &=\frac{1}{2}
      \min_{z\in\{-1,1\}}
      (e_0+ze_j)^\top\Gamma_d(e_0+ze_j)
    \gtrsim1
\end{align}
for every $j\in[d]$. Consequently, it holds that
\begin{align}
    \min_{j\in[d]}(1-\gamma_{0,j}^2)
    &\geq c~.
    \label{eq: fixed bank population squared correlation separation}
\end{align}
Part (ii) of \cref{lem: rotation and regression} implies that there exists a vector $\hat u^{(j)}\sim\mathsf N(0,M_n)$ such that
\begin{align}
    \hat\varepsilon^{(j)}
    &=\gamma_{0,j}\hat\varepsilon^{(0)}
      +\sqrt{1-\gamma_{0,j}^2}\,\hat u^{(j)}~,
    \quad\text{and}\quad
    \hat u^{(j)}
    \ci\hat\varepsilon^{(0)}~.
    \label{eq: fixed bank anchor innovation representation}
\end{align}
By \eqref{eq: fixed bank population squared correlation separation} and
\eqref{eq: fixed bank anchor innovation representation}, equality in
Cauchy--Schwarz would require the independent Gaussian vector
$\hat u^{(j)}$ to belong to the span of $\hat\varepsilon^{(0)}$. This
event has probability zero because $\operatorname{rank}(M_n)=n-1\geq2$
for all sufficiently large $n$. Hence, it holds that
\begin{align}
    |\hat\omega_{0,j}|
    &<1
    \label{eq: fixed bank strict sample correlation}
\end{align}
almost surely for every $j\in[d]$, as required.

Now, since $M_n$ is positive semi-definite, it holds that
\begin{align}
    \|M_n\|_{\mathrm F}^2
    &\leq\|M_n\|_{\mathrm{op}}\operatorname{tr}(M_n)~.
\end{align}
Thus, \cref{lem: basic inequalities n}, display \eqref{eq: n bound 4}, gives
\begin{align}
    \frac{\|M_n\|_{\mathrm F}}{\operatorname{tr}(M_n)}
    &\lesssim\sqrt{V_n}\quad\text{and}\quad
    \frac{\|M_n\|_{\mathrm{op}}}{\operatorname{tr}(M_n)}
    \lesssim V_n~.
    \label{eq: fixed bank normalized covariance bounds}
\end{align}
Each of $\hat\varepsilon^{(0)}$, $\hat u^{(j)}$, and
$(\hat\varepsilon^{(0)}+\hat u^{(j)})/\sqrt2$ is distributed as
$M_n^{1/2}Z$ for $Z\sim\mathsf N(0,I_n)$. Thus, the bounds \eqref{eq: fixed bank normalized covariance bounds} and \cref{lem: gaussian quadratic form tail}, display \eqref{eq: gaussian derivative quadratic tail} with $Q=M_n$, imply that
\begin{align}
    &\max_{j\in[d]}
    \Bigg\{
      \left|
        \frac{\|\hat\varepsilon^{(0)}\|_2^2}
        {\operatorname{tr}(M_n)}-1
      \right|
      +\left|
        \frac{\|\hat u^{(j)}\|_2^2}
        {\operatorname{tr}(M_n)}-1
      \right|
      +\left|
        \frac{(\hat\varepsilon^{(0)})^\top\hat u^{(j)}}
        {\operatorname{tr}(M_n)}
      \right|
    \Bigg\} \nonumber \\
    &=O_p\left(
      \sqrt{V_n\log(V_n^{-1})}
      +V_n\log(V_n^{-1})
    \right)
    =o_p(1)~.
    \label{eq: fixed bank normalized residual concentration}
\end{align}
The representation \eqref{eq: fixed bank anchor innovation representation} implies that
\begin{align}
    (\hat\varepsilon^{(0)})^\top\hat\varepsilon^{(j)}
    &=\gamma_{0,j}\|\hat\varepsilon^{(0)}\|_2^2
      +\sqrt{1-\gamma_{0,j}^2}\,
      (\hat\varepsilon^{(0)})^\top\hat u^{(j)}\quad\text{and}
    \nonumber\\
    \|\hat\varepsilon^{(j)}\|_2^2
    &=\gamma_{0,j}^2\|\hat\varepsilon^{(0)}\|_2^2
      +(1-\gamma_{0,j}^2)\|\hat u^{(j)}\|_2^2
      +2\gamma_{0,j}\sqrt{1-\gamma_{0,j}^2}\,
      (\hat\varepsilon^{(0)})^\top\hat u^{(j)}~.
    \label{eq: fixed bank exact sample correlation components}
\end{align}
By dividing \eqref{eq: fixed bank exact sample correlation components} by $\operatorname{tr}(M_n)$ and substituting the result into the definition of $\hat\omega_{0,j}$, we obtain
\begin{align}
    \hat\omega_{0,j}
    &=
      \frac{
        \displaystyle
        \gamma_{0,j}
        \frac{\|\hat\varepsilon^{(0)}\|_2^2}{\operatorname{tr}(M_n)}
        +\sqrt{1-\gamma_{0,j}^2}\,
        \frac{(\hat\varepsilon^{(0)})^\top\hat u^{(j)}}
        {\operatorname{tr}(M_n)}
      }{
        \displaystyle
        \left(
          \frac{\|\hat\varepsilon^{(0)}\|_2^2}
          {\operatorname{tr}(M_n)}
        \right)^{1/2}
      }
    \nonumber\\
    &\quad\cdot
      \Bigg(
        \gamma_{0,j}^2
        \frac{\|\hat\varepsilon^{(0)}\|_2^2}{\operatorname{tr}(M_n)}
        +(1-\gamma_{0,j}^2)
        \frac{\|\hat u^{(j)}\|_2^2}{\operatorname{tr}(M_n)}
        +2\gamma_{0,j}\sqrt{1-\gamma_{0,j}^2}\,
        \frac{(\hat\varepsilon^{(0)})^\top\hat u^{(j)}}
        {\operatorname{tr}(M_n)}
      \Bigg)^{-1/2}~.
    \label{eq: fixed bank exact sample correlation map}
\end{align}
The right-hand side of
\eqref{eq: fixed bank exact sample correlation map} equals
$\gamma_{0,j}$ when the two normalized squared norms equal one and the
normalized cross product equals zero. Its denominator is uniformly bounded
away from zero and its first derivatives are uniformly bounded on a fixed
neighborhood of these values by
\eqref{eq: fixed bank population squared correlation separation}. Thus, the
mean value theorem gives
\begin{align}
    |\hat\omega_{0,j}-\gamma_{0,j}|
    &\lesssim
      \left|
        \frac{\|\hat\varepsilon^{(0)}\|_2^2}
        {\operatorname{tr}(M_n)}-1
      \right|
      +\left|
        \frac{\|\hat u^{(j)}\|_2^2}
        {\operatorname{tr}(M_n)}-1
      \right|
      +
      \left|
        \frac{(\hat\varepsilon^{(0)})^\top\hat u^{(j)}}
        {\operatorname{tr}(M_n)}
      \right|
\end{align}
whenever the expression on the right-hand side is at most a sufficiently
small fixed constant. The bound \eqref{eq: fixed bank normalized residual concentration} therefore gives
\begin{align}
    \max_{j\in[d]}|\hat\omega^2_{0,j}-\gamma^2_{0,j}|
    \lesssim
    \max_{j\in[d]}|\hat\omega_{0,j}-\gamma_{0,j}|
    &=O_p\left(
      \sqrt{V_n\log(V_n^{-1})}
      +V_n\log(V_n^{-1})
    \right)
    =o_p(1)~,
    \label{eq: fixed bank uniform sample correlation approximation}
\end{align}
as required. \hfill\qed

\subsection{Proof of \cref{lem: fixed bank oracle and feasible calibration bounds}}

Let $\varepsilon^{(k;0,j)}$ denote the population residual from the regression of $Y^{(k)}$ on an intercept, $Y^{(0)}$, and $Y^{(j)}$. Define
\begin{align}
    (\check A_\ell^{(0,j)})_{i,i'}
    &=\mathbb I\left\{
      \left|
        \frac{1}{d-1}
        \sum_{k\in\{0,1,\ldots,d\}\setminus\{0,j\}}
        \varepsilon_{i}^{(k;0,j)}\varepsilon_{i'}^{(k;0,j)}
      \right|
      \geq\underline\delta_{n,d}(\eta_\ell)
    \right\}~.
\end{align}
Fix a pair $i\neq i'$ with $\lambda_{i,i'} = 0$. Define the sequence 
\begin{align}
    \pi_{n,d}(\ell)
    &=P\left\{
      (\check A_\ell^{(0)})_{i,i'}=1
    \right\}
    \quad\text{and observe that}\quad
    \pi_{n,d}(\ell)
    =H_{\underline\delta_{n,d}(\eta_\ell),d}(1,0)~.
\end{align}
Part (i) of \cref{lem: selected set characterization} and Part (i) of \cref{lem: rare null pair screening} imply that
\begin{align}
    \pi_{n,d}(\ell)
    &\lesssim V_n^{\eta_\ell}~.
    \label{eq: fixed bank null screen probability}
\end{align}
Fix the sequence
\begin{align}
    r_{n,d}(\ell)
    &=\frac{\pi_{n,d}(\ell)}{n^2}
      \frac{\log(V_n^{-1})}{\sqrt d}
      +V_n^2\left(
        \frac{\log^2(V_n^{-1})}{d}+dV_n
      \right)~.
    \label{eq: fixed bank oracle comparison rate}
\end{align}
The results are obtained from the applications of the following Lemmas.
\begin{lem}
\label{lem: fixed bank oracle calibration aggregates}$ $\\
Suppose that the conditions of \cref{thm: relative bias regret} hold.
The following conclusions hold uniformly over $\ell\in[g]$.

\noindent\textbf{(i)} It holds that
\begin{align}
    \Var\left(
      \frac{1}{n^2}\operatorname{tr}(\check A_\ell^{(0)}M_n)
    \right)
    &\lesssim V_n^3\quad\text{and}\quad
    \Var\left(\overline{\check V}_{n,d}(\ell)\right)
    \lesssim r_{n,d}(\ell)~.
    \label{eq: fixed bank oracle aggregate variance inputs}
\end{align}

\noindent\textbf{(ii)} It holds that
\begin{align}
    \left[
      cS_{n,d}^2(\ell)-\check D_{n,d}^2(\ell)
    \right]_+
    &=O_p\left( r_{n,d}(\ell)\right)~.
    \label{eq: fixed bank oracle dispersion input}
\end{align}

\noindent\textbf{(iii)} Simultaneously over the indices $\ell\in[g]$ satisfying
\begin{align}
    \frac{\pi_{n,d}(\ell)}{dV_n}
    &>\left(
      \frac{\log(V_n^{-1})}{\sqrt d}
    \right)^{-1/2}
    \label{eq: fixed bank frequent screen regime}
\end{align}
it holds that
\begin{align}
    S_{n,d}^2(\ell)
    &\geq c\frac{\pi_{n,d}(\ell)}{n^2}
    \label{eq: fixed bank frequent screen energy input}
\end{align}
with probability tending to one.

\end{lem}

\begin{lem}
\label{lem: fixed bank oracle screening replacement bounds}$ $\\
Suppose that the conditions of \cref{thm: relative bias regret} hold. The
following conclusions hold uniformly over $\ell\in[g]$.

\noindent\textbf{(i)} Uniformly over $i,i'\in[n]$ and $j\in[d]$, it holds that
\begin{align}
    P\left\{
      (\check A_\ell^{(0)})_{i,i'}=1
    \right\}
    +P\left\{
      (\check A_\ell^{(0,j)})_{i,i'}=1
    \right\}
    &\lesssim
      \pi_{n,d}(\ell)
      +\min\{1,d\lambda_{i,i'}^2\}~.
    \label{eq: fixed bank oracle screen probability inputs}
\end{align}

\noindent\textbf{(ii)} Uniformly over $j\in[d]$, it holds that
\begin{align}
    \mathbb E\left[
      \left(
        \frac{1}{n^2}(\hat\varepsilon^{(j)})^\top
        (\check A_\ell^{(j)}-\check A_\ell^{(0,j)})
        \hat\varepsilon^{(j)}
      \right)^2
    \right]
    &\lesssim r_{n,d}(\ell)\quad\text{and}
    \nonumber\\
    \mathbb E\left[
      \left(
        \frac{1}{n^2}(\hat\varepsilon^{(0)})^\top
        (\check A_\ell^{(0,j)}-\check A_\ell^{(0)})
        \hat\varepsilon^{(0)}
      \right)^2
    \right]
    &\lesssim r_{n,d}(\ell)~.
    \label{eq: fixed bank screen replacement moment inputs}
\end{align}
\end{lem}

\begin{lem}
\label{lem: fixed bank feasible oracle conditional moments}$ $\\
Suppose that the conditions of \cref{thm: relative bias regret} hold.
The following conclusions hold uniformly over $\ell\in[g]$.

\noindent\textbf{(i)} Uniformly over $j\in\{0,1,\ldots,d\}$, it holds that
\begin{align}
    \left\|
      \mathbb E\left[
        \hat V_{n,d}^{(j)}(\ell)-\check V_{n,d}^{(j)}(\ell)
        \mid\hat\varepsilon^{(j)}
      \right]
    \right\|_{L^2}
    \lesssim
      V_n\left(
        \frac{\log(V_n^{-1})}{\sqrt d}
        +\sqrt{dV_n}\log^{3/2}(V_n^{-1})
        +\frac{s\sqrt d}{n}
      \right)~.
    \label{eq: fixed bank feasible oracle conditional mean}
\end{align}

\noindent\textbf{(ii)} Uniformly over $j\in\{0,1,\ldots,d\}$, it holds that
\begin{align}
    \mathbb E\left[
      \Var\left(
        \hat V_{n,d}^{(j)}(\ell)-\check V_{n,d}^{(j)}(\ell)
        \mid\hat\varepsilon^{(j)}
      \right)
    \right]
    &\lesssim
      \frac{\pi_{n,d}(\ell)}{n^2}
      \left(
        \sqrt{V_n}\log(V_n^{-1})
        +V_n\sqrt{d\log(V_n^{-1})}
      \right)
    \nonumber\\
    &+
      V_n^2\left(
        \frac{\log^2(V_n^{-1})}{d}
        +dV_n\log^3(V_n^{-1})
      \right)~.
    \label{eq: fixed bank feasible oracle conditional variance}
\end{align}
\end{lem}

\subsubsection{Part (i)}

Observe that \eqref{eq: population residual covariance and independence} implies that we can evaluate
\begin{align}
    \mathbb E\left[
      \check V_{n,d}^{(0)}(\ell)
      \mid\check A_\ell^{(0)}
    \right]
    &=\frac{1}{n^2}
      \operatorname{tr}(\check A_\ell^{(0)}M_n)~.
    \label{eq: fixed bank target conditional mean}
\end{align}
Consider the inequality
\begin{align}
    &\left|
      \frac{1}{n^2}
      \operatorname{tr}(\check A_\ell^{(0)}M_n)
      -\overline{\check V}_{n,d}(\ell)
    \right|
      +\left[
        cS_{n,d}^2(\ell)-\check D_{n,d}^2(\ell)
      \right]_+^{1/2}
    +
    \left|
      \frac{1}{n^2}
      \operatorname{tr}(\check A_\ell^{(0)}M_n)
      -\mathbb E[\check V_{n,d}^{(0)}(\ell)]
    \right|
    \nonumber\\
    &\leq
      2\left|
        \frac{1}{n^2}
        \operatorname{tr}(\check A_\ell^{(0)}M_n)
        -\mathbb E[\check V_{n,d}^{(0)}(\ell)]
      \right|
    +
      \left|
        \overline{\check V}_{n,d}(\ell)
        -\mathbb E[\check V_{n,d}^{(0)}(\ell)]
      \right|
      +\left[
        cS_{n,d}^2(\ell)-\check D_{n,d}^2(\ell)
      \right]_+^{1/2}~.
    \label{eq: fixed bank oracle approximation assembly}
\end{align}
We bound each of the terms in \eqref{eq: fixed bank oracle approximation assembly} in succession. We begin with the first term. Part (i) of \cref{lem: fixed bank oracle calibration aggregates}, display \eqref{eq: fixed bank oracle aggregate variance inputs}, \eqref{eq: fixed bank target conditional mean}, and Chebyshev's inequality give
\begin{align}
    \left|
      \frac{1}{n^2}
      \operatorname{tr}(\check A_\ell^{(0)}M_n)
      -\mathbb E[\check V_{n,d}^{(0)}(\ell)]
    \right|
    &=O_p(V_n\sqrt{V_n})~.
    \label{eq: fixed bank target trace approximation}
\end{align}
Next, we handle the second term. \Cref{assu: gaussianity,assu: proportionality,assu: homoskedasticity} and Part (iii) of \cref{lem: rotation and regression}, applied to each $k\in\{0,1,\ldots,d\}\setminus\{0,j\}$, imply that
\begin{align}
    \check A_\ell^{(0,j)}
    &\ci
      (\hat\varepsilon^{(0)},\hat\varepsilon^{(j)})~.
    \label{eq: fixed bank common screen independence}
\end{align}
Thus, we can evaluate
\begin{align}
    \mathbb E\left[
      \frac{1}{n^2}
      \left(
        (\hat\varepsilon^{(j)})^\top
        \check A_\ell^{(0,j)}\hat\varepsilon^{(j)}
        -(\hat\varepsilon^{(0)})^\top
        \check A_\ell^{(0,j)}\hat\varepsilon^{(0)}
      \right)
      \mid\check A_\ell^{(0,j)}
    \right]
    &=0~.
    \label{eq: fixed bank common screen conditional centering}
\end{align}
Observe that we can write
\begin{align}
    \check V_{n,d}^{(j)}(\ell)-\check V_{n,d}^{(0)}(\ell)
    &=\frac{1}{n^2}(\hat\varepsilon^{(j)})^\top
      (\check A_\ell^{(j)}-\check A_\ell^{(0,j)})
      \hat\varepsilon^{(j)}
    \nonumber\\
    &\quad+\frac{1}{n^2}
      \left(
        (\hat\varepsilon^{(j)})^\top
        \check A_\ell^{(0,j)}\hat\varepsilon^{(j)}
        -(\hat\varepsilon^{(0)})^\top
        \check A_\ell^{(0,j)}\hat\varepsilon^{(0)}
      \right)
    \nonumber\\
    &\quad+\frac{1}{n^2}(\hat\varepsilon^{(0)})^\top
      (\check A_\ell^{(0,j)}-\check A_\ell^{(0)})
      \hat\varepsilon^{(0)}~.
    \label{eq: fixed bank common screen decomposition}
\end{align}
Taking expectations in \eqref{eq: fixed bank common screen decomposition}, and applying \eqref{eq: fixed bank common screen conditional centering}, Cauchy-Schwarz, and Part (ii) of
\cref{lem: fixed bank oracle screening replacement bounds}, gives
\begin{align}
    \max_{j\in[d]}
    \left|
      \mathbb E[\check V_{n,d}^{(j)}(\ell)]
      -\mathbb E[\check V_{n,d}^{(0)}(\ell)]
    \right|
    &+
    \left|
      \mathbb E[\overline{\check V}_{n,d}(\ell)]
      -\mathbb E[\check V_{n,d}^{(0)}(\ell)]
    \right|
    \lesssim\sqrt{r_{n,d}(\ell)}~.
    \label{eq: fixed bank auxiliary average expectation comparison}
\end{align}
Part (i) of \cref{lem: fixed bank oracle calibration aggregates}, display \eqref{eq: fixed bank oracle aggregate variance inputs}, \eqref{eq: fixed bank auxiliary average expectation comparison}, and Chebyshev's inequality therefore imply that
\begin{align}
    \left|
      \overline{\check V}_{n,d}(\ell)
      -\mathbb E[\check V_{n,d}^{(0)}(\ell)]
    \right|
    &=O_p\left(
      \sqrt{r_{n,d}(\ell)}
    \right)~.
    \label{eq: fixed bank auxiliary mean approximation}
\end{align}
To handle the third term, observe that the bound
\begin{align}
    \left[
      cS_{n,d}^2(\ell)-\check D_{n,d}^2(\ell)
    \right]_+^{1/2}
    &=O_p\left(
      \sqrt{r_{n,d}(\ell)}
    \right)~.
    \label{eq: fixed bank dispersion deficit bound}
\end{align}
follows from Part (ii) of \cref{lem: fixed bank oracle calibration aggregates}, display \eqref{eq: fixed bank oracle dispersion input}.

Now, since $\log(V_n^{-1})/\sqrt d\leq1$ eventually, if
\begin{align}
    \frac{\pi_{n,d}(\ell)}{dV_n}
    &\leq\left(
      \frac{\log(V_n^{-1})}{\sqrt d}
    \right)^{-1/2}~,
\end{align}
then \cref{lem: basic inequalities n}, display \eqref{eq: var bounds},
gives
\begin{align}
    \frac{\pi_{n,d}(\ell)}{n^2}
    \frac{\log(V_n^{-1})}{\sqrt d}
    &\lesssim
      V_n^2dV_n\left(
        \frac{\log(V_n^{-1})}{\sqrt d}
      \right)^{1/2}
    \leq V_n^2dV_n~.
\end{align}
In the complementary regime, Part (iii) of \cref{lem: fixed bank oracle calibration aggregates}, display \eqref{eq: fixed bank frequent screen energy input}, gives
\begin{align}
    \frac{\pi_{n,d}(\ell)}{n^2}
    \frac{\log(V_n^{-1})}{\sqrt d}
    &\lesssim
      \frac{\log(V_n^{-1})}{\sqrt d}S_{n,d}^2(\ell)~.
\end{align}
It follows that
\begin{align}
    \frac{\pi_{n,d}(\ell)}{n^2}
    \frac{\log(V_n^{-1})}{\sqrt d}
    &\lesssim
      \frac{\log(V_n^{-1})}{\sqrt d}S_{n,d}^2(\ell)
      +V_n^2dV_n~,
    \label{eq: fixed bank target screen energy absorption}
\end{align}
simultaneously over $\ell\in[g]$ with probability tending to one. Combining this bound with \eqref{eq: fixed bank oracle comparison rate} gives
\begin{align}
    \sqrt{r_{n,d}(\ell)}
    &\lesssim
      \left(
        \frac{\log(V_n^{-1})}{\sqrt d}
        S_{n,d}^2(\ell)
        +V_n^2\left(
          \frac{\log^2(V_n^{-1})}{d}+dV_n
        \right)
      \right)^{1/2}
    \nonumber\\
    &\leq
      C\left(
        \frac{\log(V_n^{-1})}{\sqrt d}
      \right)^{1/2}S_{n,d}(\ell)
      +CV_n\left(
        \frac{\log(V_n^{-1})}{\sqrt d}
        +\sqrt{dV_n}
      \right)~,
    \label{eq: fixed bank absorbed oracle comparison rate}
\end{align}
where we have used $\sqrt{x+y}\leq\sqrt x+\sqrt y$.

To conclude the proof, define the quantity
\begin{align}
    U_{n,d}(\ell)
    &=\Bigg(
      \left|
        \frac{1}{n^2}\operatorname{tr}(\check A_\ell^{(0)}M_n)
        -\overline{\check V}_{n,d}(\ell)
      \right|
      +\left[
        cS_{n,d}^2(\ell)-\check D_{n,d}^2(\ell)
      \right]_+^{1/2}
    +
      \left|
        \frac{1}{n^2}\operatorname{tr}(\check A_\ell^{(0)}M_n)
        -\mathbb E[\check V_{n,d}^{(0)}(\ell)]
      \right|
      \Bigg)
    \nonumber\\
    &\quad\cdot\Bigg(
      \left(
        \frac{\log(V_n^{-1})}{\sqrt d}
      \right)^{1/2}S_{n,d}(\ell)
      +V_n\left(
        \frac{\log(V_n^{-1})}{\sqrt d}
        +\sqrt{dV_n}
      \right)
      \Bigg)^{-1}~.
    \label{eq: fixed bank oracle multiplier definition}
\end{align}
The bounds \eqref{eq: fixed bank oracle approximation assembly}, \eqref{eq: fixed bank target trace approximation}, \eqref{eq: fixed bank auxiliary mean approximation}, \eqref{eq: fixed bank dispersion deficit bound}, and \eqref{eq: fixed bank absorbed oracle comparison rate}, together with $V_n\sqrt{V_n}\leq V_n\sqrt{dV_n}$ and the fact that $g$ is fixed, imply that
\begin{align}
    \max_{\ell\in[g]}U_{n,d}(\ell)
    &=O_p(1)~.
    \label{eq: fixed bank oracle multiplier bound}
\end{align}
Define the sequences
\begin{align}
    \zeta_{n,d}^{1}(\ell)
    =\left(
        \frac{\log(V_n^{-1})}{\sqrt d}
      \right)^{1/2}U_{n,d}(\ell)~,
    \quad\text{and}\quad
    \zeta_{n,d}^{2}(\ell)
    =V_n\left(
        \frac{\log(V_n^{-1})}{\sqrt d}
        +\sqrt{dV_n}
      \right)U_{n,d}(\ell)~.
    \label{eq: fixed bank oracle comparison sequences}
\end{align}
The bounds $\log(V_n^{-1})d^{-1/2}\to 0$ and \eqref{eq: fixed bank oracle multiplier bound} verify the rates \eqref{eq: fixed bank oracle comparison remainder rates}. By substituting \eqref{eq: fixed bank oracle comparison sequences} into \eqref{eq: fixed bank oracle multiplier definition}, the bound \eqref{eq: fixed bank oracle multiplier bound} implies the bound \eqref{eq: fixed bank oracle anchor approximation}.
\hfill\qed

\subsubsection{Part (ii)}

\cref{lem: fixed bank oracle screening replacement bounds}, display \eqref{eq: fixed bank oracle screen probability inputs}, and \cref{lem: basic inequalities n}, display \eqref{eq: n bound 1}, imply that
\begin{align}
    \frac{1}{n^2}\sum_{i,i'=1}^n
      P\left\{(\check A_\ell^{(0)})_{i,i'}=1\right\}
    +\max_{j\in[d]}\frac{1}{n^2}\sum_{i,i'=1}^n
      P\left\{(\check A_\ell^{(0,j)})_{i,i'}=1\right\}
    \lesssim
      \pi_{n,d}(\ell)+\sqrt d\,V_n~.
    \label{eq: fixed bank oracle screen probability aggregates}
\end{align}
Observe that, for any symmetric matrix $A$, the bound
\begin{align}
    \operatorname{tr}(AM_nAM_n)
    &=\left\|M_n^{1/2}AM_n^{1/2}\right\|_{\mathrm F}^2
    \leq\|M_n\|_{\mathrm{op}}^2\|A\|_{\mathrm F}^2~.
    \label{eq: fixed bank screen trace inequality}
\end{align}
holds. Thus, the bounds \eqref{eq: fixed bank oracle screen probability aggregates}, \eqref{eq: fixed bank screen trace inequality}, and \cref{lem: basic inequalities n}, display \eqref{eq: n bound 4}, imply
\begin{align}
    &
    \mathbb E\left[S_{n,d}^2(\ell)\right]
    +\max_{j\in[d]}
      \frac{2}{n^4}\mathbb E\left[
        \operatorname{tr}\left(
          \check A_\ell^{(0,j)}M_n
          \check A_\ell^{(0,j)}M_n
        \right)
      \right]
    \nonumber\\
    &\lesssim
      V_n^2\left(
        \pi_{n,d}(\ell)+\sqrt d\,V_n
      \right)
    \lesssim
      V_n^2\left(
        V_n^{\eta_\ell}+\sqrt d\,V_n
      \right)~,
    \label{eq: fixed bank target and common screen energies}
\end{align}
where the last inequality uses
\eqref{eq: fixed bank null screen probability}.

Recall that $\check A_\ell^{(0)}$ and $\hat\varepsilon^{(0)}$ are independent. Conditional on $\check A_\ell^{(0)}$, \cref{lem: wick} therefore gives
\begin{align}
    \Var\left(
      \check V_{n,d}^{(0)}(\ell)
      \mid\check A_\ell^{(0)}
    \right)
    &=\frac{2}{n^4}\operatorname{tr}\left(
      \check A_\ell^{(0)}M_n\check A_\ell^{(0)}M_n
    \right)
    =S_{n,d}^2(\ell)~.
\end{align}
The law of total variance, \eqref{eq: fixed bank target conditional mean},
Part (i) of \cref{lem: fixed bank oracle calibration aggregates}, display
\eqref{eq: fixed bank oracle aggregate variance inputs}, and
\eqref{eq: fixed bank target and common screen energies} imply that
\begin{align}
    \Var\left(\check V_{n,d}^{(0)}(\ell)\right)
    &=\mathbb E\left[S_{n,d}^2(\ell)\right]
      +\Var\left(
        \frac{1}{n^2}\operatorname{tr}(\check A_\ell^{(0)}M_n)
      \right)
    \nonumber\\
    &\lesssim
      V_n^2\left(
        V_n^{\eta_\ell}+\sqrt d\,V_n
      \right)~,
\end{align}
as required.

In turn, the identity \eqref{eq: residual covariance}, \eqref{eq: fixed bank common screen independence}, and \cref{lem: wick} imply that
\begin{align}
    \MoveEqLeft
    \mathbb E\left[
      \left\{
        \frac{1}{n^2}
        \left(
          (\hat\varepsilon^{(j)})^\top
          \check A_\ell^{(0,j)}\hat\varepsilon^{(j)}
          -(\hat\varepsilon^{(0)})^\top
          \check A_\ell^{(0,j)}\hat\varepsilon^{(0)}
        \right)
      \right\}^2
      \mid\check A_\ell^{(0,j)}
    \right]
    \nonumber\\
    &=\frac{4(1-\gamma_{0,j}^2)}{n^4}
      \operatorname{tr}\left(
        \check A_\ell^{(0,j)}M_n
        \check A_\ell^{(0,j)}M_n
      \right)~.
    \label{eq: fixed bank common screen conditional moment}
\end{align}
Taking expectations in
\eqref{eq: fixed bank common screen conditional moment} and applying
\eqref{eq: fixed bank target and common screen energies} yields
\begin{align}
    \mathbb E\left[
      \left\{
        \frac{1}{n^2}
        \left(
          (\hat\varepsilon^{(j)})^\top
          \check A_\ell^{(0,j)}\hat\varepsilon^{(j)}
          -(\hat\varepsilon^{(0)})^\top
          \check A_\ell^{(0,j)}\hat\varepsilon^{(0)}
        \right)
      \right\}^2
    \right]
    &\lesssim
      V_n^2\left(
        V_n^{\eta_\ell}+\sqrt d\,V_n
      \right)~.
    \label{eq: fixed bank common screen moment input}
\end{align}
Thus, by applying Cauchy-Schwarz to \eqref{eq: fixed bank common screen decomposition} and using Part (ii) of \cref{lem: fixed bank oracle screening replacement bounds}, display \eqref{eq: fixed bank screen replacement moment inputs}, we obtain
\begin{align}
    \mathbb E\left[
      \left(
        \check V_{n,d}^{(j)}(\ell)
        -\check V_{n,d}^{(0)}(\ell)
      \right)^2
    \right]
    &\lesssim
     r_{n,d}(\ell)
      +V_n^2\left(
        V_n^{\eta_\ell}+\sqrt d\,V_n
      \right)~.
    \label{eq: fixed bank anchor contrast intermediate bound}
\end{align}
Observe that \eqref{eq: fixed bank null screen probability} and \cref{lem: basic inequalities n}, display \eqref{eq: var bounds}, imply that
\begin{align}
    \frac{\pi_{n,d}(\ell)}{n^2}
    \frac{\log(V_n^{-1})}{\sqrt d}
    &\lesssim V_n^2V_n^{\eta_\ell}~.
\end{align}
Substituting this bound into \eqref{eq: fixed bank oracle comparison rate} gives
\begin{align}
    r_{n,d}(\ell)
    &\lesssim
      V_n^2\left(
        V_n^{\eta_\ell}
        +dV_n
        +\frac{\log^2(V_n^{-1})}{d}
      \right)~.
    \label{eq: fixed bank deterministic oracle comparison rate}
\end{align}
Thus, by combining \eqref{eq: fixed bank anchor contrast intermediate bound} and
\eqref{eq: fixed bank deterministic oracle comparison rate}, we can conclude that
\begin{align}
    \mathbb E\left[
      \left(
        \check V_{n,d}^{(j)}(\ell)
        -\check V_{n,d}^{(0)}(\ell)
      \right)^2
    \right]
    &\lesssim
      V_n^2\left(
        V_n^{\eta_\ell}
        +dV_n
        +\frac{\log^2(V_n^{-1})}{d}
      \right)~,
    \label{eq: fixed bank anchor contrast moment bound}
\end{align}
as required.
\hfill\qed

\subsubsection{Part (iii)}

The arithmetic--geometric mean inequality, $\pi_{n,d}(\ell)\leq1$, \cref{lem: basic inequalities n}, display \eqref{eq: var bounds}, and the law of total variance, applied to Parts (i) and (ii) of \cref{lem: fixed bank feasible oracle conditional moments}, give
\begin{align}
    \max_{j\in\{0,1,\ldots,d\}}
    \mathbb E\left[
      \left(
        \hat V_{n,d}^{(j)}(\ell)
        -\check V_{n,d}^{(j)}(\ell)
      \right)^2
    \right]
    \lesssim
      V_n^2\left(
        \frac{\log^2(V_n^{-1})}{d}
        +dV_n\log^3(V_n^{-1})
        +\frac{s^2d}{n^2}
      \right)~.
    \label{eq: fixed bank feasible oracle second moment}
\end{align}
In turn, Chebyshev's inequality and \eqref{eq: fixed bank feasible oracle second moment} imply that
\begin{align}
    & \left|
      \hat V_{n,d}(\ell)-\check V_{n,d}^{(0)}(\ell)
      -\mathbb E\left[
        \hat V_{n,d}(\ell)-\check V_{n,d}^{(0)}(\ell)
      \right]
    \right|
    \nonumber\\
    &=
      O_p\left(
        V_n\left(
          \frac{\log(V_n^{-1})}{\sqrt d}
          +\sqrt{dV_n}\log^{3/2}(V_n^{-1})
          +\frac{s\sqrt d}{n}
        \right)
      \right)~.
    \label{eq: fixed bank centered feasible oracle difference}
\end{align}
Likewise, we have that
\begin{align}
    \MoveEqLeft
    \left(
      \frac{1}{d}\sum_{j=1}^d
      \left(
        \hat V_{n,d}^{(j)}(\ell)-\check V_{n,d}^{(j)}(\ell)
        -\hat V_{n,d}(\ell)+\check V_{n,d}^{(0)}(\ell)
      \right)^2
    \right)^{1/2}
    \nonumber\\
    &=
      O_p\left(
        V_n\left(
          \frac{\log(V_n^{-1})}{\sqrt d}
          +\sqrt{dV_n}\log^{3/2}(V_n^{-1})
          +\frac{s\sqrt d}{n}
        \right)
      \right)~.
    \label{eq: fixed bank feasible oracle contrast average}
\end{align}
by Markov's inequality and Cauchy-Schwarz.

On the event in \cref{lem: fixed bank sample correlation normalization}, display \eqref{eq: fixed bank sample correlation normalization}, the reverse triangle inequality implies that
\begin{align}
    |\hat\xi_{n,d}(\ell)-\check\xi_{n,d}(\ell)|
    &\leq
      \left(
        \frac{1}{d}\sum_{j=1}^d
        \frac{
          \left(
            \hat V_{n,d}^{(j)}(\ell)-\check V_{n,d}^{(j)}(\ell)
            -\hat V_{n,d}(\ell)+\check V_{n,d}^{(0)}(\ell)
          \right)^2
        }{1-\hat\omega_{0,j}^2}
      \right)^{1/2}
    \nonumber\\
    &\lesssim
      \left(
        \frac{1}{d}\sum_{j=1}^d
        \left(
          \hat V_{n,d}^{(j)}(\ell)-\check V_{n,d}^{(j)}(\ell)
          -\hat V_{n,d}(\ell)+\check V_{n,d}^{(0)}(\ell)
        \right)^2
      \right)^{1/2}~.
    \label{eq: fixed bank calibration norm comparison}
\end{align}
This event has probability tending to one by \cref{lem: fixed bank sample correlation normalization}. Hence, by combining \eqref{eq: fixed bank centered feasible oracle difference}, \eqref{eq: fixed bank feasible oracle contrast average}, and \eqref{eq: fixed bank calibration norm comparison}, we can conclude that
\begin{align}
    \MoveEqLeft
    \left|
      \hat V_{n,d}(\ell)-\check V_{n,d}^{(0)}(\ell)
      -\mathbb E\left[
        \hat V_{n,d}(\ell)-\check V_{n,d}^{(0)}(\ell)
      \right]
    \right|
      +|\hat\xi_{n,d}(\ell)-\check\xi_{n,d}(\ell)|
    \nonumber\\
    &\lesssim
      \left|
        \hat V_{n,d}(\ell)-\check V_{n,d}^{(0)}(\ell)
        -\mathbb E\left[
          \hat V_{n,d}(\ell)-\check V_{n,d}^{(0)}(\ell)
        \right]
      \right|
    \nonumber\\
    &+
      \left(
        \frac{1}{d}\sum_{j=1}^d
        \left(
          \hat V_{n,d}^{(j)}(\ell)-\check V_{n,d}^{(j)}(\ell)
          -\hat V_{n,d}(\ell)+\check V_{n,d}^{(0)}(\ell)
        \right)^2
      \right)^{1/2}
    \nonumber\\
    &=
      O_p\left(
        V_n\left(
          \frac{\log(V_n^{-1})}{\sqrt d}
          +\sqrt{dV_n}\log^{3/2}(V_n^{-1})
          +\frac{s\sqrt d}{n}
        \right)
      \right)~,
\end{align}
as required.\hfill\qed

\subsection{Proof of  \cref{lem: fixed bank oracle calibration aggregates}}

For distinct $j,k\in[d]$, let $\check A_\ell^{(j,k)}$ be defined analogously to $\check A_\ell^{(0,j)}$, where instead we residualize the outcomes other than $j$ and $k$ on outcomes $j$ and $k$. In turn, for $B\subseteq[d]$ of fixed cardinality, let $\check A_\ell^{(B)}$ be defined analogously, by residualizing the outcomes outside $B$ on the outcomes in $B$. The result follows from the application of the following inequalities.

\begin{lem}
\label{lem: fixed bank selection matrix comparisons}$ $\\
Suppose that the conditions of \cref{thm: relative bias regret} hold. Fix an integer $m < C$. Uniformly over $\ell\in[g]$, distinct $j,k\in[d]$, and nonempty $B\subseteq[d]$ with $|B|\leq m$, the following conclusions
hold.

\noindent\textbf{(i)} For $j\in\{0,1,\ldots,d\}$, it holds that
\begin{align}
    \left|
      \Cov\left(
        (\check A_\ell^{(j)})_{i,i'},
        (\check A_\ell^{(j)})_{h,h'}
      \right)
    \right|
    &\lesssim
      \sum_{u\in\{i,i'\}}\sum_{v\in\{h,h'\}}\lambda_{u,v}^2~.
    \label{eq: fixed bank selection covariance comparison}
\end{align}

\noindent\textbf{(ii)} Take $j\in B$. It holds that
\begin{align}
    &\max\Bigg\{
      P\left\{
        (\check A_\ell^{(j)}-\check A_\ell^{(0,j)})_{i,i'}\neq0
      \right\},
      P\left\{
        (\check A_\ell^{(0,j)}-\check A_\ell^{(0)})_{i,i'}\neq0
      \right\},
    \nonumber\\
      &\quad \quad  \quad P\left\{
        (\check A_\ell^{(j)}-\check A_\ell^{(j,k)})_{i,i'}\neq0
      \right\},
      P\left\{
        (\check A_\ell^{(k)}-\check A_\ell^{(j,k)})_{i,i'}\neq0
      \right\},
    \nonumber\\
      & \quad \quad  \quad P\left\{
        (\check A_\ell^{(B)}-\check A_\ell^{(0)})_{i,i'}\neq0
      \right\},
      P\left\{
        (\check A_\ell^{(j)}-\check A_\ell^{(B)})_{i,i'}\neq0
      \right\}
    \Bigg\}
    \nonumber\\
    &\lesssim
      \pi_{n,d}(\ell)
      \frac{\log(V_n^{-1})}{\sqrt d}
      +\min\{1,d\lambda_{i,i'}^2\}~,
    \label{eq: fixed bank single screen crossing comparison}
\end{align}
and for distinct unordered pairs $\{i,i'\}$ and $\{h,h'\}$, that
\begin{align}
    & \max\Bigg\{
      \mathbb E\left[
        \left|
          (\check A_\ell^{(j)}-\check A_\ell^{(0,j)})_{i,i'}
          (\check A_\ell^{(j)}-\check A_\ell^{(0,j)})_{h,h'}
        \right|
      \right],
      \mathbb E\left[
        \left|
          (\check A_\ell^{(0,j)}-\check A_\ell^{(0)})_{i,i'}
          (\check A_\ell^{(0,j)}-\check A_\ell^{(0)})_{h,h'}
        \right|
      \right],
    \nonumber\\
      & \mathbb E\left[
        \left|
          (\check A_\ell^{(j)}-\check A_\ell^{(j,k)})_{i,i'}
          (\check A_\ell^{(j)}-\check A_\ell^{(j,k)})_{h,h'}
        \right|
      \right],
      \mathbb E\left[
        \left|
          (\check A_\ell^{(k)}-\check A_\ell^{(j,k)})_{i,i'}
          (\check A_\ell^{(k)}-\check A_\ell^{(j,k)})_{h,h'}
        \right|
      \right],
    \nonumber\\
      & \mathbb E\left[
        \left|
          (\check A_\ell^{(B)}-\check A_\ell^{(0)})_{i,i'}
          (\check A_\ell^{(B)}-\check A_\ell^{(0)})_{h,h'}
        \right|
      \right],
      \mathbb E\left[
        \left|
          (\check A_\ell^{(j)}-\check A_\ell^{(B)})_{i,i'}
          (\check A_\ell^{(j)}-\check A_\ell^{(B)})_{h,h'}
        \right|
      \right]
    \Bigg\}
    \lesssim\frac{1}{d}~.
    \label{eq: fixed bank two screen crossing comparison}
\end{align}

\noindent\textbf{(iii)} For $i\neq i'$, it holds that
\begin{align}
    \left|
      P\left\{(\check A_\ell^{(0)})_{i,i'}=1\right\}
      -\pi_{n,d}(\ell)
    \right|
    &\lesssim\min\{1,d\lambda_{i,i'}^2\}~.
    \label{eq: fixed bank baseline probability comparison}
\end{align}
For distinct $i,i',h\in[n]$, it holds that
\begin{align}
    P\left\{
      (\check A_\ell^{(0)})_{i,i'}=1,
      (\check A_\ell^{(0)})_{i,h}=1
    \right\}
    \lesssim
      \pi_{n,d}(\ell)^{3/2}
      +\min\left\{
        1,
        d\left(
          \lambda_{i,i'}^2+\lambda_{i,h}^2+\lambda_{i',h}^2
        \right)
      \right\}~.
    \label{eq: fixed bank shared row screening comparison}
\end{align}
Finally, for disjoint unordered pairs $\{i,i'\}$ and $\{h,h'\}$, it holds that
\begin{align}
    \MoveEqLeft
    \left|
      \Cov\left(
        (\check A_\ell^{(0)})_{i,i'},
        (\check A_\ell^{(0)})_{h,h'}
      \right)
    \right|
    \nonumber\\
    &\lesssim
      \left(
        P\left\{(\check A_\ell^{(0)})_{i,i'}=1\right\}
        +P\left\{(\check A_\ell^{(0)})_{h,h'}=1\right\}
      \right)
      \sum_{u\in\{i,i'\}}\sum_{v\in\{h,h'\}}
      \lambda_{u,v}^2~.
    \label{eq: fixed bank disjoint screen covariance comparison}
\end{align}
\end{lem}

\subsubsection{Part (i)}

By Part (i) of \cref{lem: fixed bank selection matrix comparisons}, we can evaluate
\begin{align}
    \Var\left(
      \frac{1}{n^2}\operatorname{tr}(\check A_\ell^{(j)}M_n)
    \right)
    &\leq\frac{1}{n^4}\sum_{i,i',h,h'=1}^n
      |\mu_{i,i'}\mu_{h,h'}|
      \left|
        \Cov\left(
          (\check A_\ell^{(j)})_{i,i'},
          (\check A_\ell^{(j)})_{h,h'}
        \right)
      \right|
    \lesssim V_n^3~,
    \label{eq: fixed bank oracle trace variance calculation}
\end{align}
where the final inequality follows from \cref{lem: basic inequalities n}, \eqref{eq: n bound 1} and \eqref{eq: n bound 3}, verifying the first bound in \eqref{eq: fixed bank oracle aggregate variance inputs}. 

Now, consider the decomposition
\begin{align}
    \overline{\check V}_{n,d}(\ell)
    &=\frac{1}{d}\sum_{j=1}^d
      \left(
        \check V_{n,d}^{(j)}(\ell)
        -\frac{1}{n^2}\operatorname{tr}(\check A_\ell^{(j)}M_n)
      \right)
    +
      \frac{1}{dn^2}\sum_{j=1}^d
      \operatorname{tr}(\check A_\ell^{(j)}M_n)~.
    \label{eq: fixed bank auxiliary average decomposition}
\end{align}
Observe that Minkowski's inequality and \eqref{eq: fixed bank oracle trace variance calculation} give
\begin{align}
    \Var\left(
      \frac{1}{dn^2}\sum_{j=1}^d
      \operatorname{tr}(\check A_\ell^{(j)}M_n)
    \right)^{1/2}
    &\leq
      \frac{1}{d}\sum_{j=1}^d
      \Var\left(
        \frac{1}{n^2}\operatorname{tr}(\check A_\ell^{(j)}M_n)
      \right)^{1/2}
    \lesssim V_n\sqrt{V_n}~.
    \label{eq: fixed bank auxiliary trace average variance}
\end{align}
To handle the first term in \eqref{eq: fixed bank auxiliary average decomposition}, observe that, as $\check A_\ell^{(j)}$ is independent of $\hat\varepsilon^{(j)}$, \cref{lem: wick} implies that
\begin{align}
    \Var\left(
      \check V_{n,d}^{(j)}(\ell)
      -\frac{1}{n^2}\operatorname{tr}(\check A_\ell^{(j)}M_n)
    \right)
    =\frac{2}{n^4}\mathbb E\left[
      \operatorname{tr}\left(
        \check A_\ell^{(j)}M_n
        \check A_\ell^{(j)}M_n
      \right)
    \right]
    \lesssim V_n^2~,
    \label{eq: fixed bank centered objective variance}
\end{align}
where we use the fact that 
\begin{align}
    \frac{1}{n^4}\operatorname{tr}\left(
      \check A_\ell^{(j)}M_n\check A_\ell^{(j)}M_n
    \right)
    &\leq\frac{1}{n^4}\|M_n\|_{\mathrm{op}}^2
      \|\check A_\ell^{(j)}\|_{\mathrm F}^2
    \lesssim V_n^2
\end{align}
follows from \cref{lem: basic inequalities n}, display \eqref{eq: n bound 4}, and the bound $\|\check A_\ell^{(j)}\|_{\mathrm F}^2\leq n^2$. Consider the decomposition
\begin{align}
    \check V_{n,d}^{(j)}(\ell)
      -\frac{1}{n^2}\operatorname{tr}(\check A_\ell^{(j)}M_n)
    &=\frac{1}{n^2}(\hat\varepsilon^{(j)})^\top
      \check A_\ell^{(j,k)}\hat\varepsilon^{(j)}
      -\frac{1}{n^2}\operatorname{tr}(\check A_\ell^{(j,k)}M_n)
    \nonumber\\
    &\quad+
      \frac{1}{n^2}(\hat\varepsilon^{(j)})^\top
      (\check A_\ell^{(j)}-\check A_\ell^{(j,k)})
      \hat\varepsilon^{(j)}
    -
      \frac{1}{n^2}\operatorname{tr}\left(
        (\check A_\ell^{(j)}-\check A_\ell^{(j,k)})M_n
      \right)~.
    \label{eq: fixed bank common selection decomposition}
\end{align}
Observe that \Cref{lem: wick} and Part (ii) of \cref{lem: fixed bank selection matrix comparisons} imply that
\begin{align}
    & \mathbb E\Bigg[
      \Bigg(
        \frac{1}{n^2}(\hat\varepsilon^{(j)})^\top
        (\check A_\ell^{(j)}-\check A_\ell^{(j,k)})
        \hat\varepsilon^{(j)}
        -\frac{1}{n^2}\operatorname{tr}\left(
          (\check A_\ell^{(j)}-\check A_\ell^{(j,k)})M_n
        \right)
      \Bigg)^2
    \Bigg]
    \nonumber\\
    &\leq\frac{C}{n^4}\sum_{i,i',h,h'=1}^n
      \left(
        |\mu_{i',h}\mu_{h',i}|+|\mu_{i,i'}\mu_{h,h'}|
      \right)
      \mathbb I\left\{\{i,i'\}=\{h,h'\}\right\}
    \cdot \Big(
        \pi_{n,d}(\ell)\frac{\log(V_n^{-1})}{\sqrt d}
        +\min\{1,d\lambda_{i,i'}^2\}
      \Big)
    \nonumber\\
    &\quad+
      \frac{C}{n^4}\sum_{i,i',h,h'=1}^n
      \left(
        |\mu_{i',h}\mu_{h',i}|+|\mu_{i,i'}\mu_{h,h'}|
      \right)
      \mathbb I\left\{\{i,i'\}\neq\{h,h'\}\right\}
      \frac{1}{d}
    \lesssim r_{n,d}(\ell)~,
    \label{eq: fixed bank common screen replacement bound}
\end{align}
where for the last inequality, we have used the fact that \cref{lem: basic inequalities n}, displays \eqref{eq: var bounds} and \eqref{eq: n bound 1}--\eqref{eq: n bound 3}, gives
\begin{align}
    \sum_{i,i',h,h'=1}^n
      \left(
        |\mu_{i',h}\mu_{h',i}|+|\mu_{i,i'}\mu_{h,h'}|
      \right)
      \mathbb I\left\{\{i,i'\}=\{h,h'\}\right\}
    &\lesssim n^2~,
    \nonumber\\
    \sum_{i,i',h,h'=1}^n
      \left(
        |\mu_{i',h}\mu_{h',i}|+|\mu_{i,i'}\mu_{h,h'}|
      \right)
      \mathbb I\left\{\{i,i'\}\neq\{h,h'\}\right\}
    &\lesssim n^4V_n^2~,\quad\text{and}
    \nonumber\\
    \frac{1}{n^4}\sum_{i,i'=1}^n
      \min\{1,d\lambda_{i,i'}^2\}
    &\lesssim dV_n^3~,
    \label{eq: fixed bank common screen coefficient bounds}
\end{align}
respectively. In turn, as the matrix $\check A_\ell^{(j,k)}$ is independent of outcomes $j$ and $k$, \cref{lem: wick} implies that
\begin{align}
    & \Cov\Bigg(
      \frac{1}{n^2}(\hat\varepsilon^{(j)})^\top
        \check A_\ell^{(j,k)}\hat\varepsilon^{(j)}
      -\frac{1}{n^2}\operatorname{tr}(\check A_\ell^{(j,k)}M_n),
      \frac{1}{n^2}(\hat\varepsilon^{(k)})^\top
        \check A_\ell^{(j,k)}\hat\varepsilon^{(k)}
      -\frac{1}{n^2}\operatorname{tr}(\check A_\ell^{(j,k)}M_n)
    \Bigg)
    \nonumber\\
    &=\frac{2\gamma_{j,k}^2}{n^4}\mathbb E\left[
      \operatorname{tr}\left(
        \check A_\ell^{(j,k)}M_n
        \check A_\ell^{(j,k)}M_n
      \right)
    \right]
    \lesssim\gamma_{j,k}^2V_n^2~.
    \label{eq: fixed bank common screen covariance}
\end{align}
Likewise, by applying Part (ii) of \cref{lem: rotation and regression} and \cref{lem: wick} in a similar way, we can evaluate
\begin{align}
    &\Cov\Bigg(
      \frac{1}{n^2}(\hat\varepsilon^{(j)})^\top
        (\check A_\ell^{(j)}-\check A_\ell^{(j,k)})
        \hat\varepsilon^{(j)}
      -\frac{1}{n^2}\operatorname{tr}\left(
        (\check A_\ell^{(j)}-\check A_\ell^{(j,k)})M_n
      \right), \nonumber \\
      &\quad \quad \quad \quad \quad \quad 
      \frac{1}{n^2}(\hat\varepsilon^{(k)})^\top
        \check A_\ell^{(j,k)}\hat\varepsilon^{(k)}
      -\frac{1}{n^2}\operatorname{tr}(\check A_\ell^{(j,k)}M_n)
    \Bigg)
    \nonumber\\
    &=\frac{2\gamma_{j,k}^2}{n^4}\mathbb E\left[
      \operatorname{tr}\left(
        (\check A_\ell^{(j)}-\check A_\ell^{(j,k)})M_n
        \check A_\ell^{(j,k)}M_n
      \right)
    \right]
    \lesssim\gamma_{j,k}^2V_n\sqrt{r_{n,d}(\ell)}~,
    \label{eq: fixed bank mixed common screen bound}
\end{align}
where the final inequality follows from Cauchy-Schwarz, \eqref{eq: fixed bank common screen replacement bound}, and the bound used in \eqref{eq: fixed bank common screen covariance}. In turn, Cauchy-Schwarz and \eqref{eq: fixed bank common screen replacement bound} give
\begin{align}
    \MoveEqLeft
    \Bigg|
      \Cov\Bigg(
        \frac{1}{n^2}(\hat\varepsilon^{(j)})^\top
          (\check A_\ell^{(j)}-\check A_\ell^{(j,k)})
          \hat\varepsilon^{(j)}
        -\frac{1}{n^2}\operatorname{tr}\left(
          (\check A_\ell^{(j)}-\check A_\ell^{(j,k)})M_n
        \right),
    \nonumber\\
    &\hspace{23mm}
        \frac{1}{n^2}(\hat\varepsilon^{(k)})^\top
          (\check A_\ell^{(k)}-\check A_\ell^{(j,k)})
          \hat\varepsilon^{(k)}
        -\frac{1}{n^2}\operatorname{tr}\left(
          (\check A_\ell^{(k)}-\check A_\ell^{(j,k)})M_n
        \right)
      \Bigg)
    \Bigg|
    \lesssim r_{n,d}(\ell)~.
    \label{eq: fixed bank two replacement covariance}
\end{align}
Thus, by applying the decomposition \eqref{eq: fixed bank common selection decomposition} and the bounds  \eqref{eq: fixed bank common screen covariance}, \eqref{eq: fixed bank mixed common screen bound}, and
\eqref{eq: fixed bank two replacement covariance}, we find that
\begin{align}
    & \left|
      \Cov\left(
        \check V_{n,d}^{(j)}(\ell)
        -\frac{1}{n^2}\operatorname{tr}(\check A_\ell^{(j)}M_n),
        \check V_{n,d}^{(k)}(\ell)
        -\frac{1}{n^2}\operatorname{tr}(\check A_\ell^{(k)}M_n)
      \right)
    \right|
    \nonumber\\
    &\lesssim
      \gamma_{j,k}^2V_n^2
      +r_{n,d}(\ell)
      +\gamma_{j,k}^2V_n\sqrt{r_{n,d}(\ell)}~.
    \label{eq: fixed bank centered objective covariance}
\end{align}
By combining \eqref{eq: fixed bank centered objective variance} and \eqref{eq: fixed bank centered objective covariance}, and using \cref{lem: auxiliary outcome covariance bounds}, we obtain
\begin{align}
    \MoveEqLeft
    \Var\left(
      \frac{1}{d}\sum_{j=1}^d
      \left\{
        \check V_{n,d}^{(j)}(\ell)
        -\frac{1}{n^2}\operatorname{tr}(\check A_\ell^{(j)}M_n)
      \right\}
    \right)
    \lesssim
      \frac{V_n^2}{d}
      +r_{n,d}(\ell)
      +\frac{V_n\sqrt{r_{n,d}(\ell)}}{d}
    \lesssim r_{n,d}(\ell)~,
    \label{eq: fixed bank centered oracle average variance}
\end{align}
where we have used \cref{assu: homoskedasticity} and Cauchy-Schwarz to absorb the covariance terms. Hence, Minkowski's inequality, \eqref{eq: fixed bank auxiliary average decomposition}, \eqref{eq: fixed bank centered oracle average variance}, and \eqref{eq: fixed bank auxiliary trace average variance} imply that 
\begin{align}
    \Var\left(\overline{\check V}_{n,d}(\ell)\right)^{1/2}
    &\lesssim\sqrt{r_{n,d}(\ell)}+V_n\sqrt{V_n}
    \lesssim\sqrt{r_{n,d}(\ell)}~,
\end{align}
where the final inequality uses $V_n^3\lesssim r_{n,d}(\ell)$, as required.
\hfill\qed

\subsubsection{Part (ii)}

Fix a sufficiently large integer $m$ and partition all but at most $m-1$
auxiliary outcomes into a collection $\mathcal B$ of size-$m$ blocks. For
$B\in\mathcal B$ and $j\in B$, define the quantities
\begin{align}
    \check Q_{n,d,B}^{(j)}(\ell)
    &=\frac{1}{n^2}(\hat\varepsilon^{(j)})^\top
      \check A_\ell^{(B)}\hat\varepsilon^{(j)}~,
    \quad
    \overline{\check Q}_{n,d,B}(\ell)
    =\frac{1}{m}\sum_{j\in B}\check Q_{n,d,B}^{(j)}(\ell)~,\quad\text{and}
    \nonumber\\
    S_{n,d,B}^2(\ell)& =\frac{2}{n^4}\operatorname{tr}\left(
      \check A_\ell^{(B)}M_n\check A_\ell^{(B)}M_n
    \right)~.
    \label{eq: fixed bank block proof objects}
\end{align}
As the matrix $\check A_\ell^{(B)}$ is independent of the outcomes in $B$,  \cref{lem: wick}, \cref{lem: even gaussian covariance contraction}, and \cref{lem: auxiliary outcome covariance bounds} together with the bound $\sum_{j,k\in B}\gamma_{j,k}^2\lesssim m$ imply that 
\begin{align}
    \mathbb E\Bigg[
      \left(
        \check Q_{n,d,B}^{(j)}(\ell)
        -\frac{1}{n^2}\operatorname{tr}(\check A_\ell^{(B)}M_n)
      \right)
      \left(
        \check Q_{n,d,B}^{(k)}(\ell)
        -\frac{1}{n^2}\operatorname{tr}(\check A_\ell^{(B)}M_n)
      \right)
      \mathrel{\Big|}\check A_\ell^{(B)}
    \Bigg]
    & =\gamma_{j,k}^2S_{n,d,B}^2(\ell)~,
    \nonumber\\
    \mathbb E\Bigg[
      \left(
        \check Q_{n,d,B}^{(j)}(\ell)
        -\frac{1}{n^2}\operatorname{tr}(\check A_\ell^{(B)}M_n)
      \right)^4
      \mathrel{\Big|}\check A_\ell^{(B)}
    \Bigg]
    & \lesssim S_{n,d,B}^4(\ell)~,
    \nonumber\\
    \mathbb E\left[
      \left(
        \frac{1}{m}\sum_{j\in B}
        \left(
          \check Q_{n,d,B}^{(j)}(\ell)
          -\frac{1}{n^2}\operatorname{tr}(\check A_\ell^{(B)}M_n)
        \right)
      \right)^4
      \mathrel{\Big|}\check A_\ell^{(B)}
    \right]
    & \lesssim\frac{S_{n,d,B}^4(\ell)}{m^2}~,
    \label{eq: fixed bank block conditional moments}
\end{align}
and 
\begin{align}
    & \left|
      \Cov\Bigg(
        \left(
          \check Q_{n,d,B}^{(j)}(\ell)
          -\frac{1}{n^2}\operatorname{tr}(\check A_\ell^{(B)}M_n)
        \right)^2,
        \left(
          \check Q_{n,d,B}^{(k)}(\ell)
          -\frac{1}{n^2}\operatorname{tr}(\check A_\ell^{(B)}M_n)
        \right)^2
        \mathrel{\Big|}\check A_\ell^{(B)}
      \Bigg)
    \right| \nonumber \\
    & \lesssim\gamma_{j,k}^2S_{n,d,B}^4(\ell) \label{eq: fixed bank block conditional moments cov}
\end{align}
hold almost surely. 

Now, consider the decomposition
\begin{align}
    & \frac{1}{m}\sum_{j\in B}
    \left(
      \check Q_{n,d,B}^{(j)}(\ell)
      -\overline{\check Q}_{n,d,B}(\ell) 
    \right)^2\nonumber \\
    &=\frac{1}{m}\sum_{j\in B}
      \left(
        \check Q_{n,d,B}^{(j)}(\ell)
        -\frac{1}{n^2}\operatorname{tr}(\check A_\ell^{(B)}M_n)
      \right)^2
    -
      \left(
        \frac{1}{m}\sum_{j\in B}
        \left(
          \check Q_{n,d,B}^{(j)}(\ell)
          -\frac{1}{n^2}\operatorname{tr}(\check A_\ell^{(B)}M_n)
        \right)
      \right)^2~.
    \label{eq: fixed bank block dispersion identity}
\end{align}
Observe that the inequalities \eqref{eq: fixed bank block conditional moments} and \eqref{eq: fixed bank block conditional moments cov} imply that 
\begin{align}
    \mathbb E\left[
      \frac{1}{m}\sum_{j\in B}
      \left(
        \check Q_{n,d,B}^{(j)}(\ell)
        -\overline{\check Q}_{n,d,B}(\ell)
      \right)^2
      \mathrel{\Big|}\check A_\ell^{(B)}
    \right]
    & \geq\frac{3}{4}S_{n,d,B}^2(\ell)\quad\text{and}
    \nonumber\\
    \Var\left(
      \frac{1}{m}\sum_{j\in B}
      \left(
        \check Q_{n,d,B}^{(j)}(\ell)
        -\overline{\check Q}_{n,d,B}(\ell)
      \right)^2
      \mathrel{\Big|}\check A_\ell^{(B)}
    \right)
    &\lesssim\frac{S_{n,d,B}^4(\ell)}{m}~,
    \label{eq: fixed bank block dispersion moments}
\end{align}
respectively. Thus, we find that
\begin{align}
    \MoveEqLeft
    P\left\{
      \frac{1}{m}\sum_{j\in B}
      \left(
        \check Q_{n,d,B}^{(j)}(\ell)
        -\overline{\check Q}_{n,d,B}(\ell)
      \right)^2
      <\frac{1}{4}S_{n,d,B}^2(\ell)
      \mathrel{\Big|}\check A_\ell^{(B)}
    \right\}
    \leq\frac{C}{m}~.
    \label{eq: fixed bank block dispersion probability}
\end{align}
by Chebyshev's inequality.

Observe that the reverse triangle inequality gives
\begin{align}
    \mathbb{E}\left[ \left(S_{n,d,B}(\ell)-S_{n,d}(\ell)\right)^2 \right]
    &\leq
    \mathbb{E}\left[ \frac{2}{n^4}\operatorname{tr}\left(
      (\check A_\ell^{(B)}-\check A_\ell^{(0)})M_n
      (\check A_\ell^{(B)}-\check A_\ell^{(0)})M_n
    \right)\right]
    ~.
    \label{eq: fixed bank block comparisons}
\end{align}
As the vector $\hat\varepsilon^{(j)}$ is independent of $(\check A_\ell^{(j)},\check A_\ell^{(B)})$, \Cref{lem: wick} implies that we can evaluate
\begin{align}
    & \mathbb E\left[
      \left(
        \check V_{n,d}^{(j)}(\ell)
        -\check Q_{n,d,B}^{(j)}(\ell)
      \right)^2
    \right]
    \nonumber\\
    & =\frac{2}{n^4}\mathbb E\left[
      \operatorname{tr}\left(
        (\check A_\ell^{(j)}-\check A_\ell^{(B)})M_n
        (\check A_\ell^{(j)}-\check A_\ell^{(B)})M_n
      \right)
    \right]
    +
      \mathbb E\left[
        \left(
          \frac{1}{n^2}\operatorname{tr}\left(
            (\check A_\ell^{(j)}-\check A_\ell^{(B)})M_n
          \right)
        \right)^2
      \right]~.
    \label{eq: fixed bank block objective wick identity}
\end{align}
Part (ii) of \cref{lem: fixed bank selection matrix comparisons} and the bound \eqref{eq: fixed bank common screen coefficient bounds} imply that we can bound \eqref{eq: fixed bank block comparisons} and the two terms in
\eqref{eq: fixed bank block objective wick identity} by a constant
multiple of $r_{n,d}(\ell)$. Thus, we can conclude that
\begin{align}
    \mathbb E\Bigg[
      \frac{m}{d}\sum_{B\in\mathcal B}
      \left(S_{n,d,B}(\ell)-S_{n,d}(\ell)\right)^2
    +
      \frac{1}{d}\sum_{B\in\mathcal B}\sum_{j\in B}
      \left(
        \check V_{n,d}^{(j)}(\ell)
        -\check Q_{n,d,B}^{(j)}(\ell)
      \right)^2
    \Bigg]
    \lesssim r_{n,d}(\ell)
    \label{eq: fixed bank block comparison bound}
\end{align}
by averaging these bounds.

Define the event
\begin{align}
    \mathcal{E}(B) = \left\{ \frac{1}{m}\sum_{j\in B}
      \left(
        \check Q_{n,d,B}^{(j)}(\ell)
        -\overline{\check Q}_{n,d,B}(\ell)
      \right)^2
    \geq\frac{1}{4}S_{n,d,B}^2(\ell) \right\}~.
    \label{eq: fixed bank good block event}
\end{align}
Observe that \eqref{eq: fixed bank block dispersion probability} and Markov's inequality imply that
\begin{align}
    P\left\{
    \frac{m}{d} \sum_{B \in \mathcal{B}}
    \mathbb{I}\{ \mathcal{E}(B)\} \geq  1/2
    \right\} \geq 1 - \frac{C}{m}\label{eq: good events}
\end{align}
Thus, on the event in \eqref{eq: good events}, we can evaluate
\begin{align}
    \check D_{n,d}^2(\ell)
    &=\frac{1}{d}\min_{t\in\mathbb R}
      \sum_{j=1}^d\left(\check V_{n,d}^{(j)}(\ell)-t\right)^2
    \nonumber\\
    &\geq\frac{1}{d}\sum_{B\in\mathcal B}
      \min_{t\in\mathbb R}\sum_{j\in B}
      \left(\check V_{n,d}^{(j)}(\ell)-t\right)^2
    \nonumber\\
    &\geq\frac{1}{2d}\sum_{B\in\mathcal B}\sum_{j\in B}
      \left(
        \check Q_{n,d,B}^{(j)}(\ell)
        -\overline{\check Q}_{n,d,B}(\ell)
      \right)^2
    -
      \frac{1}{d}\sum_{B\in\mathcal B}\sum_{j\in B}
      \left(
        \check V_{n,d}^{(j)}(\ell)
        -\check Q_{n,d,B}^{(j)}(\ell)
      \right)^2
    \nonumber\\
    &\geq\frac{m}{8d}\sum_{B\in\mathcal B}
      \mathbb I\{\mathcal{E}(B)\} S_{n,d,B}^2(\ell)
    -
      \frac{1}{d}\sum_{B\in\mathcal B}\sum_{j\in B}
      \left(
        \check V_{n,d}^{(j)}(\ell)
        -\check Q_{n,d,B}^{(j)}(\ell)
      \right)^2
    \nonumber\\
    &\geq cS_{n,d}^2(\ell)
      -C\frac{m}{d}\sum_{B\in\mathcal B}
      \left(S_{n,d,B}(\ell)-S_{n,d}(\ell)\right)^2
    -
      \frac{1}{d}\sum_{B\in\mathcal B}\sum_{j\in B}
      \left(
        \check V_{n,d}^{(j)}(\ell)
        -\check Q_{n,d,B}^{(j)}(\ell)
      \right)^2~.
    \label{eq: fixed bank block dispersion assembly}
\end{align}
Thus, to conclude the proof, fix $\epsilon>0$, choose $m$ so that $C/m<\epsilon/2$. By combining \eqref{eq: fixed bank block comparison bound}, \eqref{eq: good events}, \eqref{eq: fixed bank block dispersion assembly}, and Markov's inequality, we find that
\begin{align}
    \left[
      cS_{n,d}^2(\ell)-\check D_{n,d}^2(\ell)
    \right]_+
    &=O_p\left(r_{n,d}(\ell)\right)~,
\end{align}
as required.\hfill\qed

\subsubsection{Part (iii)}

Part (iii) of \cref{lem: fixed bank selection matrix comparisons} and \cref{lem: basic inequalities n}, display \eqref{eq: n bound 1}, imply that
\begin{align}
    n^2\left(\pi_{n,d}(\ell)-CdV_n\right)
    &\leq
      \mathbb E\left[
        \sum_{i,i'=1}^n(\check A_\ell^{(0)})_{i,i'}
      \right]
    \leq n^2\left(\pi_{n,d}(\ell)+CdV_n\right)~.
    \label{eq: fixed bank full screen mean bounds}
\end{align}
Thus, Part (iii) of \cref{lem: fixed bank selection matrix comparisons}, displays \eqref{eq: fixed bank baseline probability comparison} and
\eqref{eq: fixed bank shared row screening comparison}, give
\begin{align}
    \mathbb E\left[
      \sum_{i=1}^n
      \left(
        \sum_{i'=1}^n(\check A_\ell^{(0)})_{i,i'}
      \right)^2
    \right]
    &\lesssim
      n^2\left(\pi_{n,d}(\ell)+dV_n\right)
      +n^3\pi_{n,d}(\ell)^{3/2}
      +dn^3V_n~,
    \label{eq: fixed bank row count bound}
\end{align}
where we have used the fact that \cref{lem: basic inequalities n} implies that
\begin{align}
    & \sum_{i,i',h,h'=1}^n
      \left|
        \Cov\left(
          (\check A_\ell^{(0)})_{i,i'},
          (\check A_\ell^{(0)})_{h,h'}
        \right)
      \right|
      \mathbb I\left\{|\{i,i',h,h'\}|=4\right\}
    \nonumber\\
    &\lesssim
    \sum_{i,i',h,h'=1}^n
      \left(
        P\left\{(\check A_\ell^{(0)})_{i,i'}=1\right\}
        +P\left\{(\check A_\ell^{(0)})_{h,h'}=1\right\}
      \right)
      \sum_{u\in\{i,i'\}}\sum_{v\in\{h,h'\}}
      \lambda_{u,v}^2
    \nonumber\\
    &\lesssim
      n^4V_n\left(\pi_{n,d}(\ell)+dV_n\right)~.
    \label{eq: fixed bank disjoint covariance sum}
\end{align}
As a consequence, we can evaluate
\begin{align}
    \Var\left(
      \sum_{i,i'=1}^n(\check A_\ell^{(0)})_{i,i'}
    \right)
    &\lesssim
      n^2\left(\pi_{n,d}(\ell)+dV_n\right)
      +n^3\pi_{n,d}(\ell)^{3/2}
      +dn^3V_n \nonumber \\
    & +
      n^4V_n\left(\pi_{n,d}(\ell)+dV_n\right)~,
    \label{eq: fixed bank full screen count variance}
\end{align}
by combining \eqref{eq: fixed bank full screen mean bounds} and \eqref{eq: fixed bank row count bound}.

Now observe that, under the condition \eqref{eq: fixed bank frequent screen regime}, the bounds \eqref{eq: fixed bank full screen mean bounds}--\eqref{eq: fixed bank full screen count variance} imply that
\begin{align}
    \frac{
      \Var\left(
        \sum_{i,i'=1}^n(\check A_\ell^{(0)})_{i,i'}
      \right)
    }{n^4\pi_{n,d}(\ell)^2}
    &\lesssim
      \frac{\pi_{n,d}(\ell)+dV_n}
        {n^2\pi_{n,d}(\ell)^2}
      +\frac{1}{n\sqrt{\pi_{n,d}(\ell)}} \nonumber \\
      & +\frac{dV_n}{n\pi_{n,d}(\ell)^2}
      +\frac{V_n}{\pi_{n,d}(\ell)}
      +\frac{dV_n^2}{\pi_{n,d}(\ell)^2}
    \to 0~,\quad\text{and}
    \nonumber\\
    \frac{
      \mathbb E\left[
        \sum_{i=1}^n
        \left(
          \sum_{i'=1}^n(\check A_\ell^{(0)})_{i,i'}
        \right)^2
      \right]
    }{n^3\pi_{n,d}(\ell)}
    &\lesssim
      \frac{1}{n}
      +\frac{dV_n}{n\pi_{n,d}(\ell)}
      +\sqrt{\pi_{n,d}(\ell)}
      +\frac{dV_n}{\pi_{n,d}(\ell)}
    \to0~.
    \label{eq: fixed bank normalized screen count bounds}
\end{align}
Chebyshev's and Markov's inequalities therefore imply that
\begin{align}
    \frac{
      \sum_{i,i'=1}^n(\check A_\ell^{(0)})_{i,i'}
    }{n^2\pi_{n,d}(\ell)}
    &\overset p \to 1
    \quad\text{and}\quad
    \frac{
      n^{-1}\sum_{i=1}^n
      \left(
        \sum_{i'=1}^n(\check A_\ell^{(0)})_{i,i'}
      \right)^2
    }{n^2\pi_{n,d}(\ell)}
    \overset p\to0
    \label{eq: fixed bank screen count concentration}
\end{align}
respectively. 

Recall that we can write
\begin{align}
    \left\|H_n\check A_\ell^{(0)}H_n\right\|_{\mathrm F}^2
    &=\sum_{i,i'=1}^n(\check A_\ell^{(0)})_{i,i'}
      -\frac{2}{n}\sum_{i=1}^n
      \left(
        \sum_{i'=1}^n(\check A_\ell^{(0)})_{i,i'}
      \right)^2
    +
      \frac{1}{n^2}
      \left(
        \sum_{i,i'=1}^n(\check A_\ell^{(0)})_{i,i'}
      \right)^2~.
    \label{eq: fixed bank oracle screen double centering}
\end{align}
The conditions \eqref{eq: fixed bank screen count concentration} imply that
\begin{align}
    \left\|H_n\check A_\ell^{(0)}H_n\right\|_{\mathrm F}^2
    &\geq cn^2\pi_{n,d}(\ell)
    \label{eq: fixed bank oracle projected screen energy}
\end{align}
with probability tending to one, simultaneously over the stated indices.
\cref{assu: well conditioned}, applied to $\Lambda_n$, gives
\begin{align}
    \operatorname{tr}(AM_nAM_n)
    &\gtrsim\|H_nAH_n\|_{\mathrm F}^2
    \label{eq: fixed bank projected energy coercivity}
\end{align}
for every symmetric matrix $A$. By applying this inequality with $A=\check A_\ell^{(0)}$ and using the definition of $S_{n,d}^2(\ell)$, we can conclude that
\begin{align}
    S_{n,d}^2(\ell)
    &\geq c\frac{\pi_{n,d}(\ell)}{n^2}~,
\end{align}
as required. \hfill\qed

\subsection{Proof of \cref{lem: fixed bank oracle screening replacement bounds}}

\subsubsection{Part (i)}

Fix $i \neq i'$. Observe that 
\begin{align}
    \MoveEqLeft
    P\left\{(\check A_\ell^{(0)})_{i,i'}=1\right\}
    +P\left\{(\check A_\ell^{(0,j)})_{i,i'}=1\right\}
    \nonumber\\
    &\leq
      2P\left\{(\check A_\ell^{(0)})_{i,i'}=1\right\}
      +P\left\{
        (\check A_\ell^{(0,j)}-\check A_\ell^{(0)})_{i,i'}
        \neq0
      \right\}
    \label{eq: fixed bank screen probability triangle}
\end{align}
Parts (ii) and (iii) of \cref{lem: fixed bank selection matrix comparisons} imply that
\begin{align}
    P\left\{(\check A_\ell^{(0)})_{i,i'}=1\right\}
    +P\left\{(\check A_\ell^{(0,j)})_{i,i'}=1\right\}
    \lesssim
      \pi_{n,d}(\ell)+\min\{1,d\lambda_{i,i'}^2\}~.
\end{align}
For $i=i'$, the same conclusion is immediate from $\lambda_{i,i}=1$.
\hfill\qed

\subsubsection{Part (ii)}

Recall that the vector $\hat\varepsilon^{(j)}$ is independent of
$(\check A_\ell^{(j)},\check A_\ell^{(0,j)})$. Thus, by conditioning on
these matrices, \cref{lem: wick} implies that we can evaluate
\begin{align}
    &\mathbb E\left[
      \left(
        \frac{1}{n^2}(\hat\varepsilon^{(j)})^\top
        (\check A_\ell^{(j)}-\check A_\ell^{(0,j)})
        \hat\varepsilon^{(j)}
      \right)^2
    \right]
    \nonumber\\
    &=\frac{2}{n^4}\mathbb E\left[
      \operatorname{tr}\left(
        (\check A_\ell^{(j)}-\check A_\ell^{(0,j)})M_n
        (\check A_\ell^{(j)}-\check A_\ell^{(0,j)})M_n
      \right)
    \right]
    +
      \mathbb E\left[
        \left(
          \frac{1}{n^2}\operatorname{tr}\left(
            (\check A_\ell^{(j)}-\check A_\ell^{(0,j)})M_n
          \right)
        \right)^2
      \right]
    \nonumber\\
    &=\frac{2}{n^4}
      \sum_{i,i',h,h'=1}^n
      \mu_{i',h}\mu_{h',i}
      \mathbb E\left[
        (\check A_\ell^{(j)}-\check A_\ell^{(0,j)})_{i,i'}
        (\check A_\ell^{(j)}-\check A_\ell^{(0,j)})_{h,h'}
      \right]
    \nonumber\\
    &+
      \frac{1}{n^4}
      \sum_{i,i',h,h'=1}^n
      \mu_{i,i'}\mu_{h,h'}
      \mathbb E\left[
        (\check A_\ell^{(j)}-\check A_\ell^{(0,j)})_{i,i'}
        (\check A_\ell^{(j)}-\check A_\ell^{(0,j)})_{h,h'}
      \right]~.
    \label{eq: fixed bank oracle replacement wick expansion}
\end{align}
Every entry of $\check A_\ell^{(j)}-\check A_\ell^{(0,j)}$ belongs to $\{-1,0,1\}$. Part (ii) of \cref{lem: fixed bank selection matrix comparisons} therefore implies that
\begin{align}
    &\mathbb E\left[
      \left|
        (\check A_\ell^{(j)}-\check A_\ell^{(0,j)})_{i,i'}
        (\check A_\ell^{(j)}-\check A_\ell^{(0,j)})_{h,h'}
      \right|
    \right]
    \nonumber\\
    &\lesssim
      \mathbb I\left\{\{i,i'\}=\{h,h'\}\right\}
      \left(
        \pi_{n,d}(\ell)\frac{\log(V_n^{-1})}{\sqrt d}
        +\min\{1,d\lambda_{i,i'}^2\}
      \right)
      +
      \mathbb I\left\{\{i,i'\}\neq\{h,h'\}\right\}
      \frac{1}{d}~.
    \label{eq: fixed bank oracle replacement entry bound}
\end{align}
For each $i,i'\in[n]$, there are at most two ordered pairs
$(h,h')\in[n]^2$ satisfying
$\{i,i'\}=\{h,h'\}$. Hence,
\cref{lem: basic inequalities n}, display \eqref{eq: n bound 2},
implies that
\begin{align}
    &\sum_{i,i',h,h'=1}^n
      \left(
        |\mu_{i',h}\mu_{h',i}|
        +|\mu_{i,i'}\mu_{h,h'}|
      \right)
      \mathbb I\left\{\{i,i'\}=\{h,h'\}\right\}
    \lesssim n^2
    \quad\text{and}
    \nonumber\\
    &\sum_{i,i',h,h'=1}^n
      \left(
        |\mu_{i',h}\mu_{h',i}|
        +|\mu_{i,i'}\mu_{h,h'}|
      \right)
      \mathbb I\left\{\{i,i'\}=\{h,h'\}\right\}
      \min\{1,d\lambda_{i,i'}^2\}
    \lesssim
      \sum_{i,i'=1}^n
      \min\{1,d\lambda_{i,i'}^2\}~.
    \label{eq: fixed bank oracle replacement equal pair coefficient sums}
\end{align}
Moreover, \cref{lem: basic inequalities n}, display
\eqref{eq: n bound 3}, gives
\begin{align}
    &\sum_{i,i',h,h'=1}^n
      \left(
        |\mu_{i',h}\mu_{h',i}|
        +|\mu_{i,i'}\mu_{h,h'}|
      \right)
      \mathbb I\left\{\{i,i'\}\neq\{h,h'\}\right\}
    \leq
      2\left(
        \sum_{i,i'=1}^n|\mu_{i,i'}|
      \right)^2
    \lesssim n^4V_n^2~.
    \label{eq: fixed bank oracle replacement distinct pair coefficient sum}
\end{align}
Taking absolute values in the final two lines of \eqref{eq: fixed bank oracle replacement wick expansion} and applying \eqref{eq: fixed bank oracle replacement entry bound}, \eqref{eq: fixed bank oracle replacement equal pair coefficient sums}, and \eqref{eq: fixed bank oracle replacement distinct pair coefficient sum} gives
\begin{align}
    &\mathbb E\left[
      \left(
        \frac{1}{n^2}(\hat\varepsilon^{(j)})^\top
        (\check A_\ell^{(j)}-\check A_\ell^{(0,j)})
        \hat\varepsilon^{(j)}
      \right)^2
    \right]
    \nonumber\\
    &\lesssim
      \frac{\pi_{n,d}(\ell)}{n^2}
      \frac{\log(V_n^{-1})}{\sqrt d}
      +\frac{1}{n^4}\sum_{i,i'=1}^n
      \min\{1,d\lambda_{i,i'}^2\}
      +\frac{V_n^2}{d}~.
    \label{eq: fixed bank oracle replacement moment reduction}
\end{align}
\cref{lem: basic inequalities n}, displays \eqref{eq: var bounds} and \eqref{eq: n bound 1}, implies that
\begin{align}
    \frac{1}{n^4}\sum_{i,i'=1}^n
      \min\{1,d\lambda_{i,i'}^2\}
    &\leq\frac{\sqrt d}{n^4}
      \sum_{i,i'=1}^n|\lambda_{i,i'}|
    \lesssim dV_n^3~.
    \label{eq: fixed bank oracle replacement signal sum}
\end{align}
Hence, the bounds \eqref{eq: fixed bank oracle replacement moment reduction} and \eqref{eq: fixed bank oracle replacement signal sum}, together with $\log(V_n^{-1})\geq1$, verify the first bound in \eqref{eq: fixed bank screen replacement moment inputs}.

As the vector $\hat\varepsilon^{(0)}$ is independent of $(\check A_\ell^{(0,j)},\check A_\ell^{(0)})$, replacing $\hat\varepsilon^{(j)}$ by $\hat\varepsilon^{(0)}$ and $\check A_\ell^{(j)}-\check A_\ell^{(0,j)}$ by $\check A_\ell^{(0,j)}-\check A_\ell^{(0)}$ in the same calculation
verifies the second bound. \hfill\qed

\subsection{Proof of
\cref{lem: fixed bank feasible oracle conditional moments}}

\subsubsection{Part (i)}

The condition \eqref{eq: regret rate condition} and \cref{lem: basic inequalities n}, display \eqref{eq: var bounds}, imply that \eqref{eq: variance component local condition} holds for all sufficiently large $n,d$.
The proof of Part (ii) of \cref{lem: variance approximation components}, display \eqref{eq: conditional mean variance component}, implies that
\begin{align}
    &
    \left\|
      \mathbb E\left[
        \hat V_{n,d}^{(j)}(\ell)-\check V_{n,d}^{(j)}(\ell)
        \mathrel{\Big|}\hat\varepsilon^{(j)}
      \right]
    \right\|_{L^2}
    \nonumber\\
    &\lesssim V_n\Bigg(
      \frac{1}{\sqrt d}
      +\frac{s\sqrt d}{n}
      +V_n\sqrt d
      +V_n^{1+\eta_\ell}
      \left(
        d\log(V_n^{-1})+\log^2(V_n^{-1})
      \right)
    \Bigg)~.
    \label{eq: fixed bank feasible oracle raw conditional mean}
\end{align}
for each outcome $j$ in $0,\ldots,d$. The condition \eqref{eq: regret rate condition} and \cref{lem: basic inequalities n}, display \eqref{eq: var bounds}, imply that
\begin{align}
    &\frac{1}{\sqrt d}
      +\frac{s\sqrt d}{n}
      +V_n\sqrt d
      +V_n^{1+\eta_\ell}
      \left(
        d\log(V_n^{-1})+\log^2(V_n^{-1})
      \right)
    \nonumber\\
    &\lesssim
      \frac{\log(V_n^{-1})}{\sqrt d}
      +\sqrt{dV_n}\log^{3/2}(V_n^{-1})
      +\frac{s\sqrt d}{n}~.
    \label{eq: fixed bank feasible oracle conditional mean absorption}
\end{align}
The proof is complete by combining \eqref{eq: fixed bank feasible oracle raw conditional mean} and \eqref{eq: fixed bank feasible oracle conditional mean absorption}.
\hfill\qed

\subsubsection{Part (ii)}

Recall that we can write
\begin{align}
    \hat V_{n,d}^{(j)}(\ell)-\check V_{n,d}^{(j)}(\ell)
    &=\frac{1}{n^2}\sum_{i,i'=1}^n
      \hat\varepsilon_i^{(j)}\hat\varepsilon_{i'}^{(j)}
      (\hat A_\ell^{(j)}-\check A_\ell^{(j)})_{i,i'}~.
    \label{eq: fixed bank feasible oracle exact difference}
\end{align}
Consequently, we can evaluate
\begin{align}
    \MoveEqLeft
    \mathbb E\left[
      \Var\left(
        \hat V_{n,d}^{(j)}(\ell)-\check V_{n,d}^{(j)}(\ell)
        \mathrel{\Big|}\hat\varepsilon^{(j)}
      \right)
    \right]
    \nonumber\\
    &=\frac{1}{n^4}\sum_{i,i',h,h'=1}^n
      \mathbb E\Bigg[
        \hat\varepsilon_i^{(j)}\hat\varepsilon_{i'}^{(j)}
        \hat\varepsilon_h^{(j)}\hat\varepsilon_{h'}^{(j)}
        \Cov\left(
          (\hat A_\ell^{(j)}-\check A_\ell^{(j)})_{i,i'},
          (\hat A_\ell^{(j)}-\check A_\ell^{(j)})_{h,h'}
          \mathrel{\Big|}\hat\varepsilon^{(j)}
        \right)
      \Bigg]~.
    \label{eq: fixed bank feasible oracle exact conditional variance}
\end{align}
The result follows from the application of the following inequalities.

\begin{lem}
\label{lem: fixed bank feasible oracle covariance inputs}$ $\\
Suppose that the conditions of \cref{thm: relative bias regret} hold.  The
following conclusions hold uniformly over $\ell\in[g]$.

\noindent\textbf{(i)} Uniformly over $j\in\{0,1,\ldots,d\}$, it holds that
\begin{align}
    &\frac{1}{n^4}\sum_{i,i'=1}^n
      \mathbb E\left[
        \left(
          \hat\varepsilon_i^{(j)}\hat\varepsilon_{i'}^{(j)}
        \right)^2
        \left(
          (\hat A_\ell^{(j)}-\check A_\ell^{(j)})_{i,i'}
        \right)^2
      \right]
    \nonumber\\
    &\lesssim
      \frac{\pi_{n,d}(\ell)}{n^2}
      \left(
        \sqrt{V_n}\log(V_n^{-1})
        +V_n\sqrt{d\log(V_n^{-1})}
      \right)
      +dV_n^3\log^3(V_n^{-1})~.
    \label{eq: fixed bank feasible oracle equal pair input}
\end{align}

\noindent\textbf{(ii)} Uniformly over $j\in\{0,1,\ldots,d\}$, it holds that
\begin{align}
    &\Bigg|
      \frac{1}{n^4}\sum_{i,i',h,h'=1}^n
      \mathbb E\Bigg[
        \hat\varepsilon_i^{(j)}\hat\varepsilon_{i'}^{(j)}
        \hat\varepsilon_h^{(j)}\hat\varepsilon_{h'}^{(j)}
    \cdot
        \Cov\left(
          (\hat A_\ell^{(j)}-\check A_\ell^{(j)})_{i,i'},
          (\hat A_\ell^{(j)}-\check A_\ell^{(j)})_{h,h'}
          \mathrel{\Big|}\hat\varepsilon^{(j)}
        \right)
    \nonumber\\
    &\quad\quad\quad\quad\quad\quad\quad\quad\quad\quad\quad\quad\quad\quad\quad\quad
    \cdot\mathbb I\left\{
        \{i,i'\}\neq\{h,h'\},
        \ i=i'\text{ or }h=h'
      \right\}
      \Bigg]
    \Bigg|
    \nonumber\\
    &\lesssim
      V_n^2\left(
        \frac{\log^2(V_n^{-1})}{d}
        +dV_n\log^3(V_n^{-1})
      \right)~.
    \label{eq: fixed bank feasible oracle diagonal covariance input}
\end{align}

\noindent\textbf{(iii)} Uniformly over $j\in\{0,1,\ldots,d\}$, it holds that
\begin{align}
    & \Bigg|
      \frac{1}{n^4}\sum_{i,i',h,h'=1}^n
      \mathbb E\Bigg[
        \hat\varepsilon_i^{(j)}\hat\varepsilon_{i'}^{(j)}
        \hat\varepsilon_h^{(j)}\hat\varepsilon_{h'}^{(j)}
    \cdot
        \Cov\left(
          (\hat A_\ell^{(j)}-\check A_\ell^{(j)})_{i,i'},
          (\hat A_\ell^{(j)}-\check A_\ell^{(j)})_{h,h'}
          \mathrel{\Big|}\hat\varepsilon^{(j)}
        \right)
    \nonumber\\
    & 
    \quad\quad\quad\quad\quad\quad\quad\quad\quad\quad\quad\quad\quad\quad\quad\quad\cdot\mathbb I\left\{
        i\neq i',\ h\neq h',\ |\{i,i',h,h'\}|=3
      \right\}
      \Bigg]
    \Bigg|
    \lesssim dV_n^3~.
    \label{eq: fixed bank feasible oracle shared index covariance input}
\end{align}

\noindent\textbf{(iv)} Uniformly over $j\in\{0,1,\ldots,d\}$, it holds that
\begin{align}
    & \Bigg|
      \frac{1}{n^4}\sum_{i,i',h,h'=1}^n
      \mathbb E\Bigg[
        \hat\varepsilon_i^{(j)}\hat\varepsilon_{i'}^{(j)}
        \hat\varepsilon_h^{(j)}\hat\varepsilon_{h'}^{(j)}
    \cdot
        \Cov\left(
          (\hat A_\ell^{(j)}-\check A_\ell^{(j)})_{i,i'},
          (\hat A_\ell^{(j)}-\check A_\ell^{(j)})_{h,h'}
          \mathrel{\Big|}\hat\varepsilon^{(j)}
        \right)
    \nonumber\\
    &\quad\quad\quad\quad\quad\quad\quad\quad\quad\quad\quad\quad\quad\quad\quad\quad
    \cdot\mathbb I\left\{|\{i,i',h,h'\}|=4\right\}
      \Bigg]
    \Bigg|
    \lesssim dV_n^3~.
    \label{eq: fixed bank feasible oracle disjoint covariance input}
\end{align}
\end{lem}

\noindent Observe that $(\hat A_\ell^{(j)}-\check A_\ell^{(j)})_{i,i'} =(\hat A_\ell^{(j)}-\check A_\ell^{(j)})_{i',i}$ for all indices. Hence each ordered pair $(i,i')$ has at most two partners $(h,h')$ with $\{i,i'\}=\{h,h'\}$. Cauchy-Schwarz therefore implies that
\begin{align}
    & \Bigg|
      \frac{1}{n^4}\sum_{i,i',h,h'=1}^n
      \mathbb E\Bigg[
        \hat\varepsilon_i^{(j)}\hat\varepsilon_{i'}^{(j)}
        \hat\varepsilon_h^{(j)}\hat\varepsilon_{h'}^{(j)}
        \Cov\left(
          (\hat A_\ell^{(j)}-\check A_\ell^{(j)})_{i,i'},
          (\hat A_\ell^{(j)}-\check A_\ell^{(j)})_{h,h'}
          \mathrel{\Big|}\hat\varepsilon^{(j)}
        \right)
    \mathbb I\left\{\{i,i'\}=\{h,h'\}\right\}
      \Bigg]
    \Bigg|
    \nonumber\\
    &\leq\frac{2}{n^4}\sum_{i,i'=1}^n
      \mathbb E\left[
        \left(
          \hat\varepsilon_i^{(j)}\hat\varepsilon_{i'}^{(j)}
        \right)^2
        \left(
          (\hat A_\ell^{(j)}-\check A_\ell^{(j)})_{i,i'}
        \right)^2
      \right]~.
    \label{eq: fixed bank feasible oracle equal pair reduction}
\end{align}
Part (i) of \cref{lem: fixed bank feasible oracle covariance inputs} therefore bounds the equal-pair contribution by the right-hand side of \eqref{eq: fixed bank feasible oracle equal pair input}. Thus, as the equal unordered pairs and the three classes in Parts (ii)-(iv) of \cref{lem: fixed bank feasible oracle covariance inputs} partition the sum in \eqref{eq: fixed bank feasible oracle exact conditional variance}, by combining Parts (i)-(iv) of \cref{lem: fixed bank feasible oracle covariance inputs}, and using $\log(V_n^{-1})\geq1$, we obtain
\begin{align}
&
    \mathbb E\left[
      \Var\left(
        \hat V_{n,d}^{(j)}(\ell)-\check V_{n,d}^{(j)}(\ell)
        \mathrel{\Big|}\hat\varepsilon^{(j)}
      \right)
    \right]
    \nonumber\\
    &\lesssim
      \frac{\pi_{n,d}(\ell)}{n^2}
      \left(
        \sqrt{V_n}\log(V_n^{-1})
        +V_n\sqrt{d\log(V_n^{-1})}
      \right)
    +
      V_n^2\left(
        \frac{\log^2(V_n^{-1})}{d}
        +dV_n\log^3(V_n^{-1})
      \right)~,
\end{align}
as required.\hfill\qed

\section{Proofs for Lemmas Stated in \cref{app: examples}}

\subsection{Proof of \cref{lem: bias irreducible}}

Each row of $\Lambda_n$ has $k^2$ nonzero off-diagonal entries, each equal to $a/k^2$, and has row sum $1+a$. Moreover, the eigenvalues of each block of $\Lambda_n$ are $1-a/k^2$ and $1+a$. Thus, for all sufficiently large $k$, the covariance matrices satisfy \cref{assu: homoskedasticity,assu: sparsity,assu: well conditioned}, and
\begin{align}
    V_n
    &=\frac{1+a}{n}~.
\end{align}
For every fixed $\eta\in[0,1]$, we also have
\begin{align}
    dV_n^{1+\eta}\log(V_n^{-1})
    +\frac{s\sqrt d}{n}
    &\lesssim
      k^{-3-4\eta}\log k+k^{-3/2}
      \to0~.
\end{align}
Hence, the conditions of \cref{lem: bias decomposition} hold.

Since $\Gamma_d=I_{d+1}$ and $d=k$, it holds that $v_0^{(d)}=1/k$. Let $Q_{D,k}(\delta)$ denote $Q_{i,i}(\delta)$ and let $Q_{O,k}(\delta)$ denote $Q_{i,i'}(\delta)$ when $i\neq i'$ and $\lambda_{i,i'}=a/k^2$. Counting the diagonal and nonzero off-diagonal terms in \eqref{eq: B def} gives
\begin{align}
    \frac{B_{n,d}(\delta)}{V_n}+1
    &=\frac{Q_{D,k}(\delta)+aQ_{O,k}(\delta)}{1+a}~.
    \label{eq: example exact bias}
\end{align}
The two probabilities defining $Q_{O,k}(\delta)$ are evaluated under the distributions
\begin{align}
    \mathsf N\left(
      \frac{a}{k^2},
      \frac{1}{k}\left(1+\frac{a^2}{k^4}\right)
    \right)
    \quad\text{and}\quad
    \mathsf N\left(0,\frac{1}{k}\right)~.
\end{align}
The Kullback--Leibler divergence of the first distribution from the second is
\begin{align}
    \frac{1}{2}\left(
      \frac{a^2}{k^3}
      +\frac{a^2}{k^4}
      -\log\left(1+\frac{a^2}{k^4}\right)
    \right)
    &=O(k^{-3})~.
\end{align}
Therefore, \cref{lem: standard information distance bounds}, display
\eqref{eq: cited Pinsker inequality}, implies that
\begin{align}
    \sup_{\delta\geq0}|Q_{O,k}(\delta)|
    &=O(k^{-3/2})~.
\end{align}
Since $Q_{D,k}(\delta)\leq1$, \eqref{eq: example exact bias} yields
\begin{align}
    \sup_{\delta\geq0}\frac{B_{n,d}(\delta)}{V_n}
    &\leq
      -\frac{a}{1+a}+O(k^{-3/2})~.
\end{align}
The right-hand side is negative for all sufficiently large $k$. Hence, for every sequence $\delta_k$, it holds that
\begin{align}
    \liminf_{k\to\infty}
    \frac{|B_{n,d}(\delta_k)|}{V_n}
    &\geq\frac{a}{1+a}~,
\end{align}
as required.\hfill\qed

\subsection{Proof of \cref{lem: variance sharpness}}

In this example, it holds that $V_n=1/n$ and $v_0^{(d)}=1/d$. By the representation \eqref{eq: endpoint asymptotic represent}, it holds that
\begin{align}
    \underline\delta_{n,d}(\eta)\sqrt d
    &=\sqrt{2\eta\log n}\{1+o(1)\}~.
    \label{eq: variance example endpoint}
\end{align}
For any $i\neq i'$, define
\begin{align}
    \pi_{n,d}
    &=
    P\left\{
        \left|
            \frac{1}{d}\sum_{j=1}^d
            (Y_i^{(j)}-\tau^{(j)})
            (Y_{i'}^{(j)}-\tau^{(j)})
        \right|
        \geq
        \underline\delta_{n,d}(\eta)
    \right\}~.
\end{align}
Since $\Lambda_n=I_n$ and $\Gamma_d=I_{d+1}$, the summands are independent products of independent standard Gaussian variables. Therefore, \cref{lem: weighted gaussian product moderate deviation} and \eqref{eq: variance example endpoint} imply that
\begin{align}
    \pi_{n,d}
    &=n^{-\eta+o(1)}
      =V_n^{\eta+o(1)}~.
    \label{eq: variance example screen probability}
\end{align}
The restrictions on $\alpha$ imply that
\begin{align}
    \frac{\pi_{n,d}}
    {dV_n\log^2(V_n^{-1})}
    &=\frac{n^{1-\alpha-\eta+o(1)}}{\log^2n}
      \to\infty~.
\end{align}
Consequently, it holds that
\begin{align}
    \pi_{n,d}
    &>
    dV_n\log^2(V_n^{-1})
\end{align}
for all sufficiently large $n$. Thus, the bound \eqref{eq: risk selection frequent screen energy} implies that
\begin{align}
    S_{n,d}^2(1)
    &\gtrsim
    \frac{\pi_{n,d}}{n^2}
\end{align}
with probability tending to one, Since $S_{n,d}^2(1)$ is nonnegative, this then implies that
\begin{align}
    \mathbb E[S_{n,d}^2(1)]
    &\gtrsim\frac{\pi_{n,d}}{n^2}~.
\end{align}
Hence, we find that
\begin{align}
    \Var\left(
      \check V_{n,d}^{(0)}(1)
    \right)
    &\gtrsim V_n^2\pi_{n,d}
    \label{eq: variance example oracle lower bound}
\end{align}
by the identity \eqref{eq: fixed bank oracle conditional moments} and the law of total variance.

It remains to transfer this bound to the feasible estimator. By
\eqref{eq: fixed bank feasible oracle second moment}, it holds that
\begin{align}
    &\frac{
      \mathbb E\left[
        \left(
          \hat V_{n,d}(\underline\delta_{n,d}(\eta))
          -\check V_{n,d}^{(0)}(1)
        \right)^2
      \right]
    }{V_n^2\pi_{n,d}}
    \nonumber\\
    &\quad\lesssim
      n^{\eta-\alpha+o(1)}\log^2n
      +n^{\eta+\alpha-1+o(1)}\log^3n
      +n^{\eta+\alpha-2+o(1)}
      \to0~,
    \label{eq: variance example feasible transfer}
\end{align}
where the convergence follows from $\eta<\alpha<1-\eta$. The reverse triangle inequality, applied after centering, and \eqref{eq: variance example oracle lower bound} therefore give
\begin{align}
    \frac{
      \sqrt{\Var(
        \hat V_{n,d}(\underline\delta_{n,d}(\eta))
      )}
    }{V_n}
    &\gtrsim V_n^{\eta/2+o(1)}~.
\end{align}
On the other hand, \cref{lem: variance bound} gives the matching upper bound, because
\begin{align}
    \sqrt{dV_n^{1+\eta}\log(V_n^{-1})}
    +\frac{1}{\sqrt d}
    +\left(\frac{\sqrt d}{n}\right)^{1/2}
    &=o(V_n^{\eta/2})~.
\end{align}
Hence, it follows that
\begin{align}
    \frac{
      \sqrt{\Var(
        \hat V_{n,d}(\underline\delta_{n,d}(\eta))
      )}
    }{V_n}
    &=V_n^{\eta/2+o(1)}~,
\end{align}
as required.\hfill\qed

\subsection{Proof of \cref{lem: pathology}}

Recall that, for every $x>0$, the
standard normal distribution satisfies
\begin{align}
    \frac{x}{1+x^2}
    \frac{\exp(-x^2/2)}{\sqrt{2\pi}}
    &\leq
    \Phi(-x)
    \leq
    \frac{1}{x}
    \frac{\exp(-x^2/2)}{\sqrt{2\pi}}~.
    \label{eq: pathology normal tail inequalities}
\end{align}
Now, in this example, it holds $V_n=1/n$ and $v_0^{(d)}=1/d$.  Thus, \eqref{eq: endpoint asymptotic represent} and \eqref{eq: pathology normal tail inequalities} imply that
\begin{align}
    \underline\delta_{n,d}(\eta_j)\sqrt d
    &=\sqrt{2\eta_j\log n}(1+o(1))
    \quad\text{and}\quad
    F^+(
        \underline\delta_{n,d}(\eta_j),0,\frac{1}{d}
    )
    =V_n^{\eta_j+o(1)}~.
    \label{eq: pathology null tail probabilities}
\end{align}
for each for $j\in\{1,2\}$.

With this in place, since only the diagonal entries of $\Lambda_n$ are nonzero, it holds that
\begin{align}
    \frac{
        B_{n,d}(\underline\delta_{n,d}(\eta_j))
    }{V_n}
    &=
    F^+\left(
        \underline\delta_{n,d}(\eta_j),1,\frac{2}{d}
    \right)
    -F^+\left(
        \underline\delta_{n,d}(\eta_j),0,\frac{1}{d}
    \right)
    -1~.
    \label{eq: pathology exact bias approximation}
\end{align}
Moreover, for all sufficiently large $n$, it holds that
$\underline\delta_{n,d}(\eta_j)<1/2$. Consequently, we obtain
\begin{align}
    0
    &\leq
    1-F^+\left(
        \underline\delta_{n,d}(\eta_j),1,\frac{2}{d}
    \right)
    \leq
    \Phi\left(-\sqrt{d/8}\right)
    \leq
    \exp\left(-\frac{d}{16}\right)~,
    \label{eq: pathology diagonal Gaussian probability}
\end{align}
where the middle inequality follows because $|Z_{1,2/d}|<\underline\delta_{n,d}(\eta_j)$ implies that $Z_{1,2/d}<1/2$, and the final inequality follows from \eqref{eq: pathology normal tail inequalities}. Hence, we find that
\begin{align}
    \frac{
        B_{n,d}(\underline\delta_{n,d}(\eta_j))
    }{V_n}
    &=-V_n^{\eta_j+o(1)}~,
    \label{eq: pathology approximate bias}
\end{align}
by combining \eqref{eq: pathology null tail probabilities}, \eqref{eq: pathology exact bias approximation}, and \eqref{eq: pathology diagonal Gaussian probability}.

The error in \cref{lem: bias decomposition} is negligible relative to \eqref{eq: pathology approximate bias}. In particular, after division by $V_n^{\eta_j}$, the bound \eqref{eq: bias approximation} implies that
\begin{align}
    \frac{
        dV_n^{1+\eta_j}\log(V_n^{-1})
        +d^{-1/2}
        +\sqrt d/n
    }{
        V_n^{\eta_j}
    }
    \lesssim
    n^{\alpha-1}\log n
    +n^{\eta_j-\alpha/2}
    +n^{\eta_j+\alpha/2-1}
    \to0~,
    \label{eq: pathology bias approximation error}
\end{align}
uniformly over $j\in\{1,2\}$, where the convergence follows from $2\eta_2<\alpha<1-2\eta_1$ and $\eta_2<1/4$. Hence, we obtain the approximation
\begin{align}
    \frac{\operatorname{Bias}_{n,d}(j)}{V_n}
    &=-V_n^{\eta_j+o(1)}~,
    \qquad j\in\{1,2\}~.
    \label{eq: pathology bias conclusion}
\end{align}
Since $\eta_1<\eta_2$, the representation \eqref{eq: pathology bias conclusion} implies that $k^\star=2$ for all sufficiently large $n$.

Now, in this example, the entries of the matrices $\check A_j^{(0)}$ satisfy
\begin{align}
    (\check A_j^{(0)})_{i,i'}
    &=\mathbb I\left\{
        \left|
            \frac{1}{d}\sum_{r=1}^d
            (Y_i^{(r)}-\tau^{(r)})
            (Y_{i'}^{(r)}-\tau^{(r)})
        \right|
        \geq
        \underline\delta_{n,d}(\eta_j)
    \right\}~.
    \label{eq: pathology selection matrices}
\end{align}
Recall that, for any $i\neq i'$, the probability of retaining the pair at the $j$th threshold is
\begin{align}
    \pi_{n,d}(j)
    &=P\left\{(\check A_j^{(0)})_{i,i'}=1\right\}~.
\end{align}
By \cref{lem: weighted gaussian product moderate deviation} and the representations \eqref{eq: pathology null tail probabilities}, we have that 
\begin{align}
    \pi_{n,d}(j)
    &=n^{-\eta_j+o(1)}~,
    \qquad j\in\{1,2\}~.
    \label{eq: pathology selection probabilities}
\end{align}
It follows that
\begin{align}
    \pi_{n,d}(2)
    &=o(\pi_{n,d}(1))~.
    \label{eq: pathology probability comparison}
\end{align}
The restrictions on $\alpha$ and \eqref{eq: pathology selection probabilities} imply that
\begin{align}
    \frac{\pi_{n,d}(1)}
    {dV_n\log^2(V_n^{-1})}
    &=\frac{n^{1-\alpha-\eta_1+o(1)}}{\log^2n}
    \to\infty~.
\end{align}
Consequently, the limits \eqref{eq: risk selection screen count concentration}, applied at the threshold $\underline\delta_{n,d}(\eta_1)$, gives
\begin{align}
    \frac{
        \sum_{i,i'=1}^n(\check A_1^{(0)})_{i,i'}
    }{
        n^2\pi_{n,d}(1)
    }
    &\overset p\to1~.
    \label{eq: pathology first selection count}
\end{align}
In turn, symmetry across off-diagonal pairs gives
\begin{align}
    \mathbb E\left[
        \sum_{i,i'=1}^n(\check A_2^{(0)})_{i,i'}
    \right]
    &\leq
    n^2\pi_{n,d}(2)+n~.
\end{align}
Thus, Markov's inequality, \eqref{eq: pathology selection probabilities}, and \eqref{eq: pathology probability comparison} imply that
\begin{align}
    \frac{
        \sum_{i,i'=1}^n(\check A_2^{(0)})_{i,i'}
    }{
        n^2\pi_{n,d}(1)
    }
    &\overset p\longrightarrow0~.
    \label{eq: pathology second selection count}
\end{align}
Hence, we can conclude that
\begin{align}
    \sum_{i,i'=1}^n
    (\check A_1^{(0)}-\check A_2^{(0)})_{i,i'}
    &=n^2\pi_{n,d}(1)(1+o_p(1))~.
    \label{eq: pathology selection difference count}
\end{align}
by combining \eqref{eq: pathology first selection count} and \eqref{eq: pathology second selection count}.

We next bound the largest row sum of $\check A_1^{(0)}-\check A_2^{(0)}$. Exponential Markov's inequality and the moment-generating function of a chi-square variable give
\begin{align}
    P\left\{
        \left|
            \frac{1}{d}\sum_{r=1}^d
            (Y_i^{(r)}-\tau^{(r)})^2-1
        \right|>x
    \right\}
    &\leq
    2\exp\left(-\frac{dx^2}{8}\right)~,
    \qquad 0<x<1~.
    \label{eq: pathology row norm probability}
\end{align}
Taking $x=4\sqrt{\log(n)/d}$ in \eqref{eq: pathology row norm probability} and applying the union bound give
\begin{align}
    P\left\{
        \max_{i\in[n]}
        \left|
            \frac{1}{d}\sum_{r=1}^d
            (Y_i^{(r)}-\tau^{(r)})^2-1
        \right|
        >4\sqrt{\frac{\log n}{d}}
    \right\}
    &\leq\frac{2}{n}
    \to0~.
    \label{eq: pathology uniform row norm}
\end{align}
\begin{samepage}
Conditional on $(Y_i^{(r)})_{r=1}^d$, the off-diagonal entries in row $i$
of $\check A_1^{(0)}$ are independent and identically distributed Bernoulli
variables. Their success probability is
\begin{align}
    2\Phi\left(
        -\frac{
            d\underline\delta_{n,d}(\eta_1)
        }{
            (\sum_{r=1}^d(Y_i^{(r)}-\tau^{(r)})^2)^{1/2}
        }
    \right)~.
    \label{eq: pathology conditional row probability}
\end{align}
\end{samepage}
On the event in \eqref{eq: pathology uniform row norm}, the representation \eqref{eq: endpoint asymptotic represent} implies that
\begin{align}
    \frac{
        d^2\underline\delta_{n,d}^2(\eta_1)
    }{
        2\sum_{r=1}^d(Y_i^{(r)}-\tau^{(r)})^2
    }
    &=\eta_1\log n(1+o(1))
    \label{eq: pathology conditional row exponent}
\end{align}
uniformly over $i\in[n]$. Thus, it follows from \eqref{eq: pathology normal tail inequalities} that the probability in \eqref{eq: pathology conditional row probability} is at most $n^{-7\eta_1/8}$, uniformly over $i\in[n]$, for all sufficiently large $n$.

If $X$ is a binomial variable with parameters $m$ and $p$, its moment-generating function and the exponential Markov's inequality give
\begin{align}
    P\{X\geq x\}
    &\leq
    \left(\frac{\exp(1)mp}{x}\right)^x~,
    \qquad x\geq mp~.
    \label{eq: pathology binomial probability}
\end{align}
Applying \eqref{eq: pathology binomial probability} conditionally to each
row, with $x=n^{1-3\eta_1/4}$, and then applying the union bound gives
\begin{align}
    & P\left\{
        \max_{i\in[n]}
        \sum_{\substack{i'=1\\i'\neq i}}^n
        (\check A_1^{(0)})_{i,i'}
        \geq n^{1-3\eta_1/4}
    \right\}
    \nonumber\\
    &\quad\leq
    P\left\{
        \max_{i\in[n]}
        \left|
            \frac{1}{d}\sum_{r=1}^d
            (Y_i^{(r)}-\tau^{(r)})^2-1
        \right|
        >4\sqrt{\frac{\log n}{d}}
    \right\}
    +n\left(\exp(1)n^{-\eta_1/8}\right)^{n^{1-3\eta_1/4}}
    \to0~.
\end{align}
The diagonal contributes at most one to each row. Hence, we obtain
\begin{align}
    \max_{i\in[n]}
    \sum_{i'=1}^n
    (\check A_1^{(0)}-\check A_2^{(0)})_{i,i'}
    &=O_p\left(n^{1-3\eta_1/4}\right)
    \label{eq: pathology maximum row sum}
\end{align}
for the fact that $\check A_1^{(0)}-\check A_2^{(0)}$ are nonnegative and bounded above by the corresponding entries of $\check A_1^{(0)}$.

Since the entries of $\check A_1^{(0)}-\check A_2^{(0)}$ are zero or one, we can write
\begin{align}
    & \left\|
        H_n(\check A_1^{(0)}-\check A_2^{(0)})H_n
    \right\|_{\mathrm F}^2 \nonumber \\
    & =\sum_{i,i'=1}^n
        (\check A_1^{(0)}-\check A_2^{(0)})_{i,i'}
    -
    \frac{2}{n}\sum_{i=1}^n
    \left(
        \sum_{i'=1}^n
        (\check A_1^{(0)}-\check A_2^{(0)})_{i,i'}
    \right)^2
    +
    \frac{1}{n^2}
    \left(
        \sum_{i,i'=1}^n
        (\check A_1^{(0)}-\check A_2^{(0)})_{i,i'}
    \right)^2~.
    \label{eq: pathology projected matrix identity}
\end{align}
The second term on the right-hand side of \eqref{eq: pathology projected matrix identity} satisfies
\begin{align}
    &\frac{1}{n}\sum_{i=1}^n
    \left(
        \sum_{i'=1}^n
        (\check A_1^{(0)}-\check A_2^{(0)})_{i,i'}
    \right)^2 \nonumber \\
    & \leq
    \frac{1}{n}
    \left(
        \max_{i\in[n]}
        \sum_{i'=1}^n
        (\check A_1^{(0)}-\check A_2^{(0)})_{i,i'}
    \right)
    \sum_{i,i'=1}^n
    (\check A_1^{(0)}-\check A_2^{(0)})_{i,i'}
    =
    o_p\left(n^2\pi_{n,d}(1)\right)~,
\end{align}
by \eqref{eq: pathology selection difference count} and \eqref{eq: pathology maximum row sum}. The final term in \eqref{eq: pathology projected matrix identity} is also $o_p(n^2\pi_{n,d}(1))$, because \eqref{eq: pathology selection difference count} and \eqref{eq: pathology selection probabilities} imply that it is of order $O_p(n^2\pi_{n,d}^2(1))$. Substituting these two bounds into \eqref{eq: pathology projected matrix identity} yields
\begin{align}
    \left\|
        H_n(\check A_1^{(0)}-\check A_2^{(0)})H_n
    \right\|_{\mathrm F}^2
    =n^2\pi_{n,d}(1)(1+o_p(1))~.
    \label{eq: pathology projected matrix energy}
\end{align}
Moreover, $\|H_n\|_{\mathrm{op}}=1$, while symmetry and non-negativity imply
that
\begin{align}
    \left\|
        H_n(\check A_1^{(0)}-\check A_2^{(0)})H_n
    \right\|_{\mathrm{op}}
    &\leq
    \left\|
        \check A_1^{(0)}-\check A_2^{(0)}
    \right\|_{\mathrm{op}}
    \leq
    \max_{i\in[n]}
    \sum_{i'=1}^n
    (\check A_1^{(0)}-\check A_2^{(0)})_{i,i'}~.
\end{align}
Thus, we obtain
\begin{align}
    \frac{
        \left\|
            H_n(\check A_1^{(0)}-\check A_2^{(0)})H_n
        \right\|_{\mathrm{op}}
    }{
        \left\|
            H_n(\check A_1^{(0)}-\check A_2^{(0)})H_n
        \right\|_{\mathrm F}
    }
    &=O_p\left(n^{-\eta_1/4+o(1)}\right)
    \overset p\to0~.
    \label{eq: pathology projected matrix ratio}
\end{align}
by applying \eqref{eq: pathology maximum row sum} and \eqref{eq: pathology projected matrix energy}.

We next obtain the limiting distribution of $\check V_{n,d}^{(0)}(1)-\check V_{n,d}^{(0)}(2)$. Observe that we can write
\begin{align}
    \check V_{n,d}^{(0)}(1)-\check V_{n,d}^{(0)}(2)
    &=\frac{1}{n^2}
    (\varepsilon^{(0)})^\top
    H_n(\check A_1^{(0)}-\check A_2^{(0)})H_n
    \varepsilon^{(0)}~.
    \label{eq: pathology objective difference}
\end{align}
Since $\Lambda_n=I_n$ and $\Gamma_d=I_{d+1}$, the vector $\varepsilon^{(0)}$ is distributed as $\mathsf N(0,I_n)$ and is independent of $(\check A_1^{(0)},\check A_2^{(0)})$. Conditional on the two selection matrices, diagonalization and \cref{lem: berry esseen} imply that
\begin{align}
    \MoveEqLeft\sup_{x\in\mathbb R}
    \left|
    P\left\{
        \frac{
            (\varepsilon^{(0)})^\top
            H_n(\check A_1^{(0)}-\check A_2^{(0)})H_n
            \varepsilon^{(0)}
            -\operatorname{tr}\left(
                H_n(\check A_1^{(0)}-\check A_2^{(0)})H_n
            \right)
        }{
            \sqrt{2}
            \left\|
                H_n(\check A_1^{(0)}-\check A_2^{(0)})H_n
            \right\|_{\mathrm F}
        }
        \leq x
        \,\middle|\,
        \check A_1^{(0)},\check A_2^{(0)}
    \right\}
    -\Phi(x)
    \right|
    \nonumber\\
    &\quad\lesssim
    \frac{
        \left\|
            H_n(\check A_1^{(0)}-\check A_2^{(0)})H_n
        \right\|_{\mathrm{op}}
    }{
        \left\|
            H_n(\check A_1^{(0)}-\check A_2^{(0)})H_n
        \right\|_{\mathrm F}
    }
    \overset p\longrightarrow0~,
    \label{eq: pathology conditional normal approximation}
\end{align}
Since the conditional approximation error is bounded by one, taking
expectations over $\check A_1^{(0)}$ and $\check A_2^{(0)}$ gives the
corresponding unconditional convergence.

It remains to evaluate the centering and scaling in \eqref{eq: pathology conditional normal approximation}. Since $\underline\delta_{n,d}(\eta_2)<1/2$ for all sufficiently large $n$, the terms $(\check A_1^{(0)})_{i,i}$ and $(\check A_2^{(0)})_{i,i}$ can differ only if
\begin{align}
    \frac{1}{d}\sum_{r=1}^d
    (Y_i^{(r)}-\tau^{(r)})^2
    &<\underline\delta_{n,d}(\eta_2)<\frac{1}{2}~.\label{eq: sum of squares}
\end{align}
The term on the left-hand side of \eqref{eq: sum of squares} has a chi-square distribution with $d$ degrees of freedom. Exponential Markov's inequality therefore gives
\begin{align}
    P\left\{
        (\check A_1^{(0)})_{i,i}
        \neq(\check A_2^{(0)})_{i,i}
        \text{ for some }i\in[n]
    \right\}
    \leq
    nP\left\{\chi_d^2<\frac{d}{2}\right\}
    \leq
    n\exp\left(\frac{d}{4}\right)2^{-d/2}
    \to0~.
    \label{eq: pathology diagonal agreement}
\end{align}
On the event in \eqref{eq: pathology diagonal agreement}, the diagonal of
$\check A_1^{(0)}-\check A_2^{(0)}$ is zero. Thus,
\eqref{eq: pathology selection difference count} implies that
\begin{align}
    \operatorname{tr}\left(
        H_n(\check A_1^{(0)}-\check A_2^{(0)})H_n
    \right)
    &=-\frac{1}{n}
    \sum_{i,i'=1}^n
    (\check A_1^{(0)}-\check A_2^{(0)})_{i,i'}
    \nonumber\\
    &=-n\pi_{n,d}(1)(1+o_p(1))~.
\end{align}
Since $V_n=1/n$, it follows that
\begin{align}
    \frac{
        n^{-2}\operatorname{tr}\left(
            H_n(\check A_1^{(0)}-\check A_2^{(0)})H_n
        \right)
    }{
        V_n\sqrt{2\pi_{n,d}(1)}
    }
    &=-\sqrt{\frac{\pi_{n,d}(1)}{2}}(1+o_p(1))
    \overset p\longrightarrow0~,
\end{align}
while \eqref{eq: pathology projected matrix energy} gives
\begin{align}
    \frac{
        n^{-2}\sqrt{2}
        \left\|
            H_n(\check A_1^{(0)}-\check A_2^{(0)})H_n
        \right\|_{\mathrm F}
    }{
        V_n\sqrt{2\pi_{n,d}(1)}
    }
    &\overset p\to1~.
\end{align}
Combining these two displays with \eqref{eq: pathology objective difference}
and \eqref{eq: pathology conditional normal approximation}, we obtain
\begin{align}
    \frac{
        \check V_{n,d}^{(0)}(1)-\check V_{n,d}^{(0)}(2)
    }{
        V_n\sqrt{2\pi_{n,d}(1)}
    }
    &\overset d\to\mathsf N(0,1)~.
    \label{eq: pathology oracle objective limit}
\end{align}
We finally transfer this limit to the feasible estimators. Applying
\eqref{eq: fixed bank feasible oracle second moment} to both thresholds, we obtain
\begin{align}
    \MoveEqLeft\frac{
        \mathbb E\left[
            \left(
                \hat V_{n,d}(\underline\delta_{n,d}(\eta_1))
                -\hat V_{n,d}(\underline\delta_{n,d}(\eta_2))
                -\check V_{n,d}^{(0)}(1)
                +\check V_{n,d}^{(0)}(2)
            \right)^2
        \right]
    }{
        V_n^2\pi_{n,d}(1)
    }
    \nonumber\\
    &\quad\lesssim
    n^{\eta_1-\alpha+o(1)}\log^2n
    +n^{\eta_1+\alpha-1+o(1)}\log^3n
    +n^{\eta_1+\alpha-2+o(1)}
    \to0~,
    \label{eq: pathology feasible objective approximation}
\end{align}
where the convergence follows from $2\eta_2<\alpha<1-2\eta_1$. Hence, \eqref{eq: pathology oracle objective limit} and \eqref{eq: pathology feasible objective approximation} imply that
\begin{align}
    \frac{
        \hat V_{n,d}(\underline\delta_{n,d}(\eta_1))
        -\hat V_{n,d}(\underline\delta_{n,d}(\eta_2))
    }{
        V_n\sqrt{2\pi_{n,d}(1)}
    }
    &\overset d\longrightarrow\mathsf N(0,1)~.
\end{align}
The limiting distribution is continuous at zero. Therefore, we can conclude that
\begin{align}
    P\{\check k=1\}
    &\to\frac{1}{2}~.
    \label{eq: pathology first candidate probability}
\end{align}
by the definition of $\check k$.

It remains to establish \eqref{eq: relative bias regret failure}. By \eqref{eq: pathology bias conclusion} and the fact that $k^\star=2$, we obtain
\begin{align}
    \frac{
        |\operatorname{Bias}_{n,d}(\check k)|
        -|\operatorname{Bias}_{n,d}(k^\star)|
    }{
        V_n
    }
    &=\mathbb I\{\check k=1\}n^{-\eta_1+o(1)}~.
    \label{eq: pathology realized regret}
\end{align}
The restrictions on $\eta_1$, $\eta_2$, and $\alpha$ imply that
\begin{align}
    \min\left(
        \frac{\eta_2}{2},
        \frac{1-\alpha}{2},
        \frac{\alpha}{2},
        \frac{2-\alpha}{4}
    \right)
    &>\eta_1~.
\end{align}
Consequently, for some $\epsilon>0$,
\begin{align}
    V_n^{\eta_{k^\star}/2}+\mathcal R_{n,d}(1)
    &\lesssim
    n^{-\eta_2/2}
    +n^{-(1-\alpha)/2}\log^{3/2}n
    +n^{-\alpha/2}\log n
    +n^{-(2-\alpha)/4}
    =o\left(n^{-\eta_1-\epsilon}\right)~.
    \label{eq: pathology regret rate}
\end{align}
\enlargethispage{2\baselineskip}
For every fixed $C>0$, equations \eqref{eq: pathology realized regret} and \eqref{eq: pathology regret rate} imply that the event in \eqref{eq: relative bias regret failure} agrees with the event $\{\check k=1\}$ for all sufficiently large $n$. Hence, the conclusion follows from \eqref{eq: pathology first candidate probability}. \hfill\qed

\section{Gaussian Calculus\label{app: gaussian calculus}}

\subsection{Auxiliary Lemmas}

In this subsection, we collect several generic definitions and Lemmas that will be applied repeatedly throughout the proofs given in this Appendix. For a positive-definite $2\times2$ matrix $A$, let $X_1,\ldots,X_d$ be independent with distribution $\mathsf N(0,A)$ and define the function
\begin{align}
    p_{\delta,d}(A)
    &=P\left\{
        \left|
            \frac{1}{d}\sum_{j=1}^d
            \theta_{d,j}X_{j,1}X_{j,2}
        \right|\geq\delta
    \right\}~.
\end{align}
For positive-definite $A$, $B$ and $R\in\mathbb R^{2\times2}$, let $(X_j,Y_j)_{j=1}^d$ be independent centered Gaussian vectors with covariance
\begin{align}
    \Cov\left(
        \begin{pmatrix}X_j\\Y_j\end{pmatrix}
    \right)
    &=
    \begin{pmatrix}A&R\\R^\top&B\end{pmatrix}~,
\end{align}
and define the quantity
\begin{align}
    \Psi_{\delta,d}(A,B,R)
    &=\Cov\Bigg(
        \mathbb{I}\left\{
            \left|
                \frac{1}{d}\sum_{j=1}^d
                \theta_{d,j}X_{j,1}X_{j,2}
            \right|\geq\delta
        \right\},
        \mathbb{I}\left\{
            \left|
                \frac{1}{d}\sum_{j=1}^d
                \theta_{d,j}Y_{j,1}Y_{j,2}
            \right|\geq\delta
        \right\}
    \Bigg)~.
\end{align}
For a scalar map $F$ on an open subset of a finite-dimensional matrix space, its Fr\'echet derivative at $X$ is the linear map $\mathrm DF(X)$ satisfying
\begin{align}
    \lim_{\|H\|_{\mathrm F}\to0}
    \frac{|F(X+H)-F(X)-\mathrm DF(X)[H]|}{\|H\|_{\mathrm F}}
    &=0~.
\end{align}
Let the operator $\mathrm D_R$ differentiate with respect to the cross-covariance matrix $R$, the operator $\mathrm D_{A,B}$ differentiate jointly with respect to the two marginal covariance matrices, and a superscript $2$ denote the corresponding bilinear second derivative. At various points in the argument, we make use of the following bounds.

\begin{lem}[\citealp{veraar2009}, Corollary 3.5]
\label{lem: even gaussian covariance contraction}$ $\\
Let $U\in\mathbb R^p$ and $W\in\mathbb R^q$ be jointly centered Gaussian vectors such that
\begin{align}
    \Var(U)&=I_p,
    \qquad
    \Var(W)=I_q,
    \qquad
    \Cov(U,W)=R~.
\end{align}
If $f(U)$ and $g(W)$ are centered and square-integrable and at least one of
$f$ and $g$ is even, then
\begin{align}
    |\Cov(f(U),g(W))|
    &\leq
    \|R\|_{\mathrm{op}}^2
    \sqrt{\Var(f(U))\Var(g(W))}~.
\end{align}
\end{lem}

\begin{lem}
\label{lem: gaussian threshold covariance bounds}$ $\\
Write $R=\sum_{q,r=1}^2r_{q,r}e_qe_r^\top$. Suppose that, for a fixed $\kappa\in(0,1)$, the bound
\begin{align}
    \kappa I_4
    &\preceq
    \begin{pmatrix}A&R\\R^\top&B\end{pmatrix}
    \preceq\kappa^{-1}I_4~,
    \label{eq: gaussian covariance eigenvalue bounds}
\end{align}
hold and let $p\in(0,1]$ satisfy $\max\{p_{\delta,d}(A),p_{\delta,d}(B)\}\leq p$.

\noindent\textbf{(i)} For $G,H\in\mathbb R^{2\times2}$, it holds that
\begin{align}
    |\mathrm D_R\Psi_{\delta,d}(A,B,R)[G]|
    &\leq Cp\|R\|_{\mathrm F}\|G\|_{\mathrm F}\quad\text{and}
    \nonumber\\
    |\mathrm D_R^2\Psi_{\delta,d}(A,B,R)[G,H]|
    &\leq Cp\|G\|_{\mathrm F}\|H\|_{\mathrm F}~.
    \label{eq: gaussian cross block derivative bounds}
\end{align}

\noindent\textbf{(ii)} For symmetric $2\times2$ matrices
$H_A,H_B,K_A,K_B$, it holds that
\begin{align}
    &|\mathrm D_{A,B}\Psi_{\delta,d}(A,B,R)[H_A,H_B]|
    \nonumber\\
    &\leq
    Cp\left(
        \sqrt{d(1+\log(p^{-1}))}+1+\log(p^{-1})
    \right)
    \|R\|_{\mathrm F}^2
    (\|H_A\|_{\mathrm F}+\|H_B\|_{\mathrm F})
    \label{eq: gaussian marginal first derivative bound}
\end{align}
and
\begin{align}
    & |\mathrm D_{A,B}^2\Psi_{\delta,d}(A,B,R)
        [(H_A,H_B),(K_A,K_B)]| \nonumber \\
    &\leq
    Cp\left(
        d(1+\log(p^{-1}))+(1+\log(p^{-1}))^2
    \right)
    \|R\|_{\mathrm F}^2
    (\|H_A\|_{\mathrm F}+\|H_B\|_{\mathrm F})
    (\|K_A\|_{\mathrm F}+\|K_B\|_{\mathrm F})~.
    \label{eq: gaussian marginal second derivative bound}
\end{align}

\noindent\textbf{(iii)} If \eqref{eq: gaussian covariance eigenvalue bounds} holds after $R$ is replaced by every element of
\begin{equation}
    \operatorname{conv}\left(
        \{0,R\}\cup
        \{r_{q,r}e_qe_r^\top:q,r\in[2]\}
    \right),
    \label{eq: gaussian radial interaction domain}
\end{equation}
then
\begin{align}
    &\left|
        \Psi_{\delta,d}(A,B,R)
        -\sum_{q,r=1}^2
        \Psi_{\delta,d}(A,B,r_{q,r}e_qe_r^\top)
    \right|
    \leq
    Cp\sum_{q,r=1}^2|r_{q,r}|
    \left(
        \sum_{q',r'=1}^2|r_{q',r'}|-|r_{q,r}|
    \right)~.
    \label{eq: gaussian radial interaction bound}
\end{align}

\noindent\textbf{(iv)} For each $q,r\in[2]$, $x\in\mathbb R$ such that \eqref{eq: gaussian covariance eigenvalue bounds} holds with $A=I_2$, $B=I_2$, and $R =xe_qe_r^\top$, it holds that
\begin{align}
    \mathrm D_{A,B}\Psi_{\delta,d}
    (I_2,I_2,xe_qe_r^\top)
    [e_1e_2^\top+e_2e_1^\top,0]
    &=0~,
    \quad\text{and}\quad
    \nonumber\\
    \mathrm D_{A,B}\Psi_{\delta,d}
    (I_2,I_2,xe_qe_r^\top)
    [0,e_1e_2^\top+e_2e_1^\top]
    &=0~.
    \label{eq: gaussian canonical off diagonal derivatives}
\end{align}
Moreover, for arbitrary symmetric $H_A,H_B$, it holds that
\begin{align}
    &|\mathrm D_{A,B}\Psi_{\delta,d}
    (I_2,I_2,xe_qe_r^\top)[H_A,H_B]|
    \leq
    C\sqrt d\,x^2
    (\|H_A\|_{\mathrm F}+\|H_B\|_{\mathrm F})~.
    \label{eq: gaussian canonical marginal derivative bound}
\end{align}
\end{lem}

\begin{lem}\label{lem: Gaussian three row} $ $\\
Let $K$ be a positive-definite $3\times3$ matrix. Suppose that $\kappa\in(0,1)$ is fixed, that
\begin{align}
    \kappa I_3
    &\preceq I_3+t(K-I_3)\preceq\kappa^{-1}I_3,
    \qquad t\in[0,1]~,
    \label{eq: gaussian three row covariance path}
\end{align}
and, for some $p\in(0,1]$,
\begin{align}
    &\max\Big\{
        p_{\delta,d}\big(
            (I_3+t(K-I_3))_{\{1,2\},\{1,2\}}
        \big),
        p_{\delta,d}\big(
            (I_3+t(K-I_3))_{\{1,3\},\{1,3\}}
        \big)
    \Big\}
    \leq p,
    \qquad t\in[0,1]~.
    \label{eq: gaussian three row probability envelope}
\end{align}
It holds that
\begin{align}
    &\left|
        \Psi_{(1,2),(1,3)}(K,\delta)
        -\Psi_{(1,2),(1,3)}(I_3,\delta)
        \right.
    \nonumber\\
    &\hspace{28mm}\left.
        -\mathrm D\Psi_{(1,2),(1,3)}(I_3,\delta)
        [\mathsf{diag}(K_{11}-1,K_{22}-1,K_{33}-1)]
    \right|
    \nonumber\\
    &\quad\leq
    Cp\left(
        d(1+\log(p^{-1}))
        +(1+\log(p^{-1}))^2
    \right)\|K-I_3\|_{\mathrm F}^2
    \label{eq: gaussian three row expansion bound}
\end{align}
and that
\begin{align}
    \mathrm D\Psi_{(1,2),(1,3)}(I_3,\delta)[H]
    &=\mathrm D\Psi_{(1,2),(1,3)}(I_3,\delta)
        [\mathsf{diag}(H_{11},H_{22},H_{33})]~\quad\text{and}
    \nonumber\\
    \left|
        \mathrm D\Psi_{(1,2),(1,3)}(I_3,\delta)[H]
    \right|
    &\leq C\sqrt d\,\|H\|_{\mathrm F}
    \label{eq: gaussian three row linear term}
\end{align}
for every symmetric $3\times3$ matrix $H$.

\end{lem}

\noindent The following two Lemmas are used in proofs for Lemmas supporting \cref{thm: relative bias regret}.

\begin{lem}
\label{lem: fixed bank sign comparison}$ $\\
Fix integers $r\geq2$ and $m\geq1$. Let $Z_1,\ldots,Z_m$ be independent centered Gaussian vectors in $\mathbb R^r$ with common correlation matrix $K$ satisfying
\begin{align}
    cI_r
    &\preceq K\preceq CI_r~.
    \label{eq: fixed bank sign covariance condition}
\end{align}
Write $P_K$ for their joint law and $P_{I_r}$ for the corresponding law with identity covariance. If an event $E\subseteq(\mathbb R^r)^m$ is unchanged when any one of the $r$ coordinates is multiplied by $-1$ in every vector, then
\begin{align}
    \left|P_K\{E\}-P_{I_r}\{E\}\right|
    &\lesssim
      \min\left\{1,m\|K-I_r\|_{\mathrm F}^2\right\}~,
    \label{eq: fixed bank sign comparison}
\end{align}
where the implicit constant depends on the dimension $r$.
\end{lem}

\begin{lem}
\label{lem: high rank relative crossing}$ $\\
Let $G$ be a standard Gaussian vector in $\mathbb R^n$, let $A_0$ and $B$ be symmetric $n\times n$ matrices, and, for $u\in[0,1]$, define
\begin{align}
    A_u
    =A_0+uB~,
    \quad\text{and}\quad
    Q_u
    =G^\top A_uG-\operatorname{tr}(A_0)~.
\end{align}
Let $\alpha_{u,1},\ldots,\alpha_{u,n}$ denote the eigenvalues of $A_u$, counted with multiplicity. Suppose that the bounds
\begin{align}
    \|A_u\|_{\mathrm F}^2
    &\asymp1~,
    \quad
    \|A_u\|_{\mathrm{op}}
    \lesssim\frac{1}{\sqrt d}~,
    \quad\text{and}\quad
    \left|
      \left\{q\in[n]:
        |\alpha_{u,q}|\geq\frac{c}{\sqrt d}
      \right\}
    \right|
    \geq cd
    \label{eq: fixed bank high rank spectral conditions}
\end{align}
hold uniformly over $u\in[0,1]$.

\noindent\textbf{(i)} Suppose that $x=x(d)$ satisfies $1\leq x=o(\sqrt d)$ and that the bound
\begin{align}
    (1+x)^2\|B\|_{\mathrm F}
      +(1+x)|\operatorname{tr}(B)|
    &\leq c
    \label{eq: fixed bank high rank local condition}
\end{align}
holds. Then, uniformly over $\epsilon\in\{-1,1\}$ and $y\in[x/2,2x]$, the bound
\begin{align}
    P\{\epsilon Q_0\geq y\}
    &\geq\exp(-Cx^2)
    \label{eq: fixed bank high rank shifted tail lower bound}
\end{align}
holds once $d$ is sufficiently large. Uniformly over $\epsilon\in\{-1,1\}$ and $y\in[x/2,2x]$, the bound
\begin{align}
    P\left\{
      \mathbb I\{\epsilon Q_1\geq y\}
      \neq\mathbb I\{\epsilon Q_0\geq y\}
    \right\}
    &\leq
      C\left(
        (1+x)^2\|B\|_{\mathrm F}
        +(1+x)|\operatorname{tr}(B)|
      \right)
      P\{\epsilon Q_0\geq y\}
\end{align}
also holds.

\noindent\textbf{(ii)} The bound
\begin{align}
    \sup_{y\in\mathbb R}
    P\left\{
      \mathbb I\{Q_1\geq y\}
      \neq\mathbb I\{Q_0\geq y\}
    \right\}
    &\lesssim
      \|B\|_{\mathrm F}+|\operatorname{tr}(B)|
    \label{eq: fixed bank high rank global crossing}
\end{align}
holds when $d$ is sufficiently large.
\end{lem}

\begin{lem}
\label{lem: gaussian threshold endpoint covariance bounds}$ $\\
Let $\vartheta_1,\ldots,\vartheta_d$ be nonnegative sequence satisfying
\begin{align}
    \sum_{k=1}^d\vartheta_k^2
    &\asymp d~,
    &\max_{k\in[d]}\vartheta_k
    &\lesssim1~,\quad\text{and}
    &\left|
      \left\{k\in[d]:\vartheta_k\geq c\right\}
    \right|
    &\geq cd~.
    \label{eq: gaussian endpoint weight conditions}
\end{align}
Let $\Psi_{\delta,d}^{\vartheta}$ be defined analogously to
$\Psi_{\delta,d}$, with $\vartheta_k$ in place of $\theta_{d,k}$. Let $K_{ZZ}$ be a correlation matrix satisfying
\begin{align}
    cI_m
    &\preceq K_{ZZ}\preceq CI_m
    \quad\text{and}\quad
    \Delta_K
    =
      \|K_{ZU}\|_{\mathrm F}
      +\|K_{UU}\|_{\mathrm F}
      \leq c~.
    \label{eq: gaussian endpoint covariance conditions}
\end{align}
Let $(Z_k,U_k)_{k=1}^d$ be independent copies of a centered Gaussian vector in $\mathbb R^m\times\mathbb R^m$ satisfying
\begin{align}
    \Cov\left(
      \begin{pmatrix}
        Z_k\\
        U_k
      \end{pmatrix}
    \right)
    &=
    \begin{pmatrix}
      K_{ZZ}&K_{ZU}\\
      K_{ZU}^\top&K_{UU}
    \end{pmatrix}~.
    \label{eq: gaussian endpoint covariance blocks}
\end{align}
For $a,b\in[m]$ and $t,w\in\{0,1\}$, define
\begin{align}
    S_{a,b}(t)
    &=\mathbb I\left\{
      \left|
        \frac{1}{d}\sum_{k=1}^d
        \vartheta_k
        (Z_{k,a}-tU_{k,a})
        (Z_{k,b}-tU_{k,b})
      \right|
      \geq\delta
    \right\}~,\nonumber \\
    \rho_{a,b}(t,q)
    &=
      (K_{ZZ})_{a,b+2}
      -q(K_{ZU})_{a,b+2}
      -t(K_{ZU})_{b+2,a}
      +tq(K_{UU})_{a,b+2}~,\quad\text{and}\nonumber\\
\mathcal J_{a,b}
    &=
      \sum_{t,q\in\{0,1\}}(-1)^{t+q}
      \Psi_{\delta,d}^{\vartheta}\left(
        I_2,
        I_2,
        \rho_{a,b}(t,q)e_ae_b^\top
      \right)~.
    \label{eq: gaussian single entry endpoint term}
\end{align}
The following conclusions hold for $d$ sufficiently large.

\noindent\textbf{(i)} If $m=3$, there is a deterministic quantity $\chi_{\delta,d}^{\vartheta}$, that can depend on $\delta$ and $\vartheta_1,\ldots,\vartheta_d$, such that
\begin{align}
|\chi_{\delta,d}^{\vartheta}| \lesssim\sqrt d\quad\text{and}\quad
    &\left|
      \Cov\left(
        S_{1,2}(1)-S_{1,2}(0),
        S_{1,3}(1)-S_{1,3}(0)
      \right)
      -\chi_{\delta,d}^{\vartheta}(K_{UU})_{1,1}
    \right|
    \nonumber\\
    &\quad\lesssim
      d\left(
        \|K_{ZZ}-I_3\|_{\mathrm F}\Delta_K
        +\Delta_K^2
      \right)~.
    \label{eq: gaussian shared coordinate endpoint expansion}
\end{align}

\noindent\textbf{(ii)} If $m = 4$ and $|(K_{ZZ})_{1,2}|+|(K_{ZZ})_{3,4}|\leq c$, then
\begin{align}
    &\left|
      \Cov\left(
        S_{1,2}(1)-S_{1,2}(0),
        S_{3,4}(1)-S_{3,4}(0)
      \right)
      -\sum_{a,b=1}^2\mathcal J_{a,b}
    \right|
    \nonumber\\
    &\lesssim
      d\Delta_K^2
      \left(
        \|(K_{ZZ})_{[2],\{3,4\}}\|_{\mathrm F}
        +\Delta_K
      \right)^2
      +
      \Delta_K^2
      \sum_{a,b=1}^2
      \left(
        |(K_{ZZ})_{a,b+2}|+\Delta_K
      \right)^2
    \nonumber\\
    &+
      \sqrt d\,\Delta_K
      \left(
        |(K_{ZZ})_{1,2}|+|(K_{ZZ})_{3,4}|+\Delta_K
      \right)
      \left(
        \|(K_{ZZ})_{[2],\{3,4\}}\|_{\mathrm F}
        +\Delta_K
      \right)
    \nonumber\\
    &+
      \Delta_K
      \sum_{a,b,a',b'=1}^2
      \left(
        |(K_{ZZ})_{a,b+2}|+\Delta_K
      \right)
      \left(
        |(K_{ZZ})_{a',b'+2}|+\Delta_K
      \right)
      \mathbb I\{(a,b)\neq(a',b')\}~.
    \label{eq: gaussian four coordinate refined endpoint bound}
\end{align}

\noindent\textbf{(iii)} If $m=4$ and $a,b\in[2]$, it holds that
\begin{align}
    |\mathcal J_{a,b}|
    &\lesssim
      \left(
        |(K_{ZU})_{a,b+2}|
        +|(K_{ZU})_{b+2,a}|
        +|(K_{UU})_{a,b+2}|
      \right)
    \nonumber\\
    &\quad\cdot
      \left(
        |(K_{ZZ})_{a,b+2}|
        +|(K_{ZU})_{a,b+2}|
        +|(K_{ZU})_{b+2,a}|
        +|(K_{UU})_{a,b+2}|
      \right)~.
    \label{eq: gaussian single entry endpoint bound}
\end{align}

\noindent\textbf{(iv)} If $m=4$, it holds that
\begin{align}
    &\left|
      \Cov\left(
        S_{1,2}(1)-S_{1,2}(0),
        S_{3,4}(1)-S_{3,4}(0)
      \right)
      -\sum_{a,b=1}^2\mathcal J_{a,b}
    \right|
    \lesssim
      d\Delta_K^2
      +\Delta_K
      \sum_{a,b=1}^2|(K_{ZZ})_{a,b+2}|~.
    \label{eq: gaussian four coordinate general endpoint bound}
\end{align}
\end{lem}

\subsection{Proof of \cref{lem: feasible product screening covariance classes}}

We begin by collecting several bounds and definitions that will be applied repeatedly throughout the proof. For every nonempty $T\subseteq[n]$ with $|T|\leq4$, define
\begin{align}
    \Delta_T
    &=\max_{u,v\in T}|\bar\mu_{u,v}-\mu_{u,v}|~.
    \label{eq: feasible class local perturbation}
\end{align}
Whenever $T$ is the set of distinct indices among $i,i',k,k'$, H\"older's inequality, \cref{lem: wick}, and \cref{lem: projection residualization bounds}, display \eqref{eq: feasible projection higher moments}, give
\begin{align}
    &\mathbb E\Bigg[
        \left|
            \hat\varepsilon_i^{(0)}
            \hat\varepsilon_{i'}^{(0)}
            \hat\varepsilon_k^{(0)}
            \hat\varepsilon_{k'}^{(0)}
        \right|
        \Delta_T^q
        \exp\Bigg(
            C\left(
                \sqrt{
                    d\log\left(
                        V_n^{-1}
                    \right)
                }
                +\log\left(
                    V_n^{-1}
                \right)
            \right)\Delta_T
        \Bigg)
    \Bigg]
    \nonumber\\
    &\lesssim V_n^q
    \sum_{m=0}^\infty
    \left(
        CV_n
        \left(
            \sqrt{
                d\log\left(
                    V_n^{-1}
                \right)
            }
            +\log\left(
                V_n^{-1}
            \right)
        \right)
    \right)^m
    (m+1)^q
    \lesssim V_n^q~,
    \label{eq: feasible class weighted local perturbation moments}
\end{align}
uniformly over the indices under consideration.

In turn, fix $(u,v)\in T\times T$ such that $(u,v)\notin\mathcal S_n$. Fix
\begin{align}
    A_0
    &=
    (\mu_{r,s})_{r,s\in\{u,v\}},
    \qquad
    A_1
    =
    (\bar\mu_{r,s})_{r,s\in\{u,v\}},
\end{align}
and  $A\in\operatorname{conv}\{I_2,A_0,A_1\}$. Convexity of the Frobenius norm, \cref{lem: basic inequalities n}, display \eqref{eq: n bound 3}, and \eqref{eq: feasible class local perturbation} give
\begin{align}
    \|A-I_2\|_{\mathrm F}
    &\leq
    C\left(
        V_n+\Delta_T
    \right).
    \label{eq: feasible class marginal matrix distance}
\end{align}
Applying Part (i) of \cref{lem: rare null pair screening}, displays
\eqref{eq: rare null probability envelope} and
\eqref{eq: rare screening probability growth}, therefore gives
\begin{align}
    p_{\delta,d}(A)
    &\leq
    CV_n^\eta
    \exp\Bigg(
        C\left(
            \sqrt{
                d\log\left(
                    V_n^{-1}
                \right)
            }
            +
            \log\left(
                V_n^{-1}
            \right)
        \right)\Delta_T
    \Bigg)
    \label{eq: feasible class transported marginal probability}
\end{align}
uniformly over $\delta\in\Delta_{n,d}(\eta)$.

Finally, observe that \cref{assu: well conditioned}, applied to $\Lambda_n$, \cref{lem: basic inequalities n}, display \eqref{eq: n bound 2}, and \eqref{eq: feasible class local perturbation} imply that
\begin{align}
    \kappa I_{|T|}
    &\preceq
    ((1-t)M_n+t\bar M_n)_{T,T}
    \preceq\kappa^{-1}I_{|T|},
    \qquad t\in[0,1]~.
    \label{eq: feasible class local covariance geometry}
\end{align}
Moreover, \cref{lem: basic inequalities n}, displays \eqref{eq: n bound 1} and \eqref{eq: n bound 3}, imply that
\begin{align}
    \max_{i\in[n]}\sum_{i'=1}^n|\mu_{i,i'}|
    &\leq
    \max_{i\in[n]}\sum_{i'=1}^n|\lambda_{i,i'}|
    +n\max_{i,i'\in[n]}|\mu_{i,i'}-\lambda_{i,i'}|
    \lesssim nV_n~.
    \label{eq: feasible class centered row sum}
\end{align}

\subsubsection{Part (i)}

Fix $a=(i,i')$ and $b=(k,k')$, and let $T=\{i,i',k,k'\}$. The arrays $\widetilde M_n$ and $\bar M_n$ agree on $T\times T$ unless at least one entry is truncated. Thus, uniformly over $\delta\geq0$, we can evaluate
\begin{align}
    \left|
        \Psi_{a,b}(\widetilde M_n,\delta)
        -\Psi_{a,b}(\bar M_n,\delta)
    \right|
    &\leq
    \mathbb I\left\{
        \max_{u,v\in T}
        |\tilde\mu_{u,v}-\mu_{u,v}|>c
    \right\}
    \nonumber\\
    &\leq
    \sum_{u,v\in T}
    \mathbb I\left\{
        |\tilde\mu_{u,v}-\mu_{u,v}|>c
    \right\}
    \leq
    c^{-3}
    \sum_{u,v\in T}
    |\tilde\mu_{u,v}-\mu_{u,v}|^3~.
    \label{eq: feasible class truncation reduction}
\end{align}
Now, for every $u,v\in T$, Cauchy--Schwarz, \cref{assu: gaussianity,lem: wick}, \cref{lem: basic inequalities n}, display \eqref{eq: n bound 2}, and \cref{lem: projection residualization bounds}, display \eqref{eq: feasible projection higher moments}, with $m=3$, give
\begin{align}
    &\mathbb E\left[
        \left|
            \hat\varepsilon_i^{(0)}
            \hat\varepsilon_{i'}^{(0)}
            \hat\varepsilon_k^{(0)}
            \hat\varepsilon_{k'}^{(0)}
        \right|
        |\tilde\mu_{u,v}-\mu_{u,v}|^3
    \right] \nonumber \\
    & \leq
    \mathbb E\left[
        \left(
            \hat\varepsilon_i^{(0)}
            \hat\varepsilon_{i'}^{(0)}
            \hat\varepsilon_k^{(0)}
            \hat\varepsilon_{k'}^{(0)}
        \right)^2
    \right]^{1/2}
    \|\tilde\mu_{u,v}-\mu_{u,v}\|_6^3
    \leq
    C V_n^3~.
    \label{eq: feasible class weighted raw perturbation third moment}
\end{align}
Thus, the bounds \eqref{eq: feasible class truncation reduction} and \eqref{eq: feasible class weighted raw perturbation third moment} imply that
\begin{align}
    \sup_{\delta\geq0}
    |\mathcal R_{1,n,d}(\delta)|
    &\leq
    \frac{1}{n^4}
    \sum_{i,i',k,k'=1}^n
    \mathbb E\Bigg[
        \left|
            \hat\varepsilon_i^{(0)}
            \hat\varepsilon_{i'}^{(0)}
            \hat\varepsilon_k^{(0)}
            \hat\varepsilon_{k'}^{(0)}
        \right|
        \nonumber\\
    &\hspace{35mm}\cdot
        \sup_{\delta\geq0}
        \left|
            \Psi_{(i,i'),(k,k')}(\widetilde M_n,\delta)
            -\Psi_{(i,i'),(k,k')}(\bar M_n,\delta)
        \right|
    \Bigg]
    \nonumber\\
    &\leq
    \frac{c^{-3}}{n^4}
    \sum_{i,i',k,k'=1}^n
    \sum_{u,v\in T}
    \mathbb E\left[
        \left|
            \hat\varepsilon_i^{(0)}
            \hat\varepsilon_{i'}^{(0)}
            \hat\varepsilon_k^{(0)}
            \hat\varepsilon_{k'}^{(0)}
        \right|
        |\tilde\mu_{u,v}-\mu_{u,v}|^3
    \right]
    \nonumber\\
    &\leq
    \frac{C}{n^4}
    \sum_{i,i',k,k'=1}^n
    |T|^2V_n^3
    \leq
    CV_n^3~,
\end{align}
as required. 
\hfill\qed

\subsubsection{Part (ii)}

The bound is obtained by considering the terms in the decomposition
\begin{align}
    \mathcal R_{2,n,d}(\delta)
    &=\mathcal R_{2,1,n,d}(\delta)
    +\mathcal R_{2,2,n,d}(\delta)
    +\mathcal R_{2,3,n,d}(\delta)~.
    \label{eq: repeated index exact three contribution decomposition}
\end{align}
where
\begin{align}
    \begin{pmatrix}
        \mathcal R_{2,1,n,d}(\delta)\\
        \mathcal R_{2,2,n,d}(\delta)\\
        \mathcal R_{2,3,n,d}(\delta)
    \end{pmatrix}
    &\quad=\frac{1}{n^4}\sum_{i,i',k,k'=1}^n
    \mathbb E\Bigg[
        \hat\varepsilon_i^{(0)}\hat\varepsilon_{i'}^{(0)}
        \hat\varepsilon_k^{(0)}\hat\varepsilon_{k'}^{(0)}
        \left(
            \Psi_{(i,i'),(k,k')}(\bar M_n,\delta)
            -\Psi_{(i,i'),(k,k')}(M_n,\delta)
        \right)
        \nonumber\\
    &\hspace{40mm}\cdot
        \begin{pmatrix}
            \mathbb{I}\{i=i'\ \mathrm{or}\ k=k'\}\\[1mm]
            \mathbb{I}\{i\neq i',\ k\neq k',\
                (k,k')\in\{(i,i'),(i',i)\}\}\\[1mm]
            \mathbb{I}\{i\neq i',\ k\neq k',\
                |\{i,i',k,k'\}|=3\}
        \end{pmatrix}
    \Bigg]~.
    \label{eq: repeated index three contribution definitions}
\end{align}
We use \cref{lem: even gaussian covariance contraction} to bound
$\mathcal R_{2,1,n,d}(\delta)$. Fix $\Xi\in\{M_n,\bar M_n\}$ and define
\begin{align}
    U
    &=\left(\Xi_{i,i}^{-1/2}G_i^{(j)}\right)_{j=1}^d~,
    \quad
    W
    &= 
    \begin{cases}
    \left(\Xi_{k,k}^{-1/2}G_k^{(j)}\right)_{j=1}^d~,&k=k'~,\\
    \left(
        \left((\Xi_{u,v})_{u,v\in\{k,k'\}}\right)^{-1/2}
        (G_k^{(j)}, G_{k'}^{(j)})^{\top}
    \right)_{j=1}^d~,&k\neq k^\prime~.
    \end{cases}
\end{align}
Observe that $U$ and $W$ wach have  identify covariance matrices and that \eqref{eq: feasible class local covariance geometry} gives
\begin{align}
    \|\Cov(U,W)\|_{\mathrm{op}}^2
    &\lesssim
    \begin{cases}
        \Xi_{i,k}^2~,&k= k'~,\\
    \Xi_{i,k}^2+\Xi_{i,k'}^2~,&k\neq k'~.
    \end{cases}
    \label{eq: repeated index diagonal standardized cross covariance}
\end{align}
Thus, as the first indicator in \eqref{eq: generic product screening covariance} is an even function of $U$, we can evaluate
\begin{align}
    |\Psi_{(i,i),(k,k')}(\Xi,\delta)|
    &\lesssim
    \begin{cases}
        \Xi_{i,k}^2~,&k=k'~,\\
        \left(\Xi_{i,k}^2+\Xi_{i,k'}^2\right)
    p_{\delta,d}\left((\Xi_{u,v})_{u,v\in\{k,k'\}}\right)^{1/2}~,&k\neq k'~.
    \end{cases}
    \label{eq: repeated index diagonal pointwise contraction}
\end{align}
by applying \cref{lem: even gaussian covariance contraction} and the bound $\Var(\mathbb I\{E\})\leq P\{E\}$.

We consider three cases. First, consider the case $(k,k')\in\mathcal S_n$. Here, the bounds \eqref{eq: feasible class weighted local perturbation moments} with $q\in\{0,2\}$, \eqref{eq: repeated index diagonal pointwise contraction}, and $\bar\mu_{i,k}^2+\bar\mu_{i,k'}^2 \lesssim\mu_{i,k}^2+\mu_{i,k'}^2+\Delta_{\{i,k,k'\}}^2$ imply that 
\begin{align}
    \sup_{\delta\geq0}
    \mathbb E\Bigg[
        \left|
            (\hat\varepsilon_i^{(0)})^2
            \hat\varepsilon_k^{(0)}
            \hat\varepsilon_{k'}^{(0)}
        \right|
        \left|
            \bar\Psi_{(i,i),(k,k')}(\delta)
        \right|
    \Bigg]
    &\lesssim
    \mu_{i,k}^2+\mu_{i,k'}^2+V_n^2~.
\end{align}
Moreover, \cref{lem: basic inequalities n}, display \eqref{eq: n bound 4}, implies that
\begin{align}
    \|M_n\|_{\mathrm F}^2
    &=\operatorname{tr}(M_n^2)
    \leq\|M_n\|_{\mathrm{op}}\operatorname{tr}(M_n)
    \leq Cn^2V_n~.
    \label{eq: repeated index centered frobenius bound}
\end{align}
Now, by \cref{assu: sparsity}, for each $k$ there are at most $s+1$ values of $k'$ such that $(k,k')\in\mathcal S_n$. Thus, \eqref{eq: repeated index centered frobenius bound} and \cref{lem: basic inequalities n}, display \eqref{eq: var bounds}, give
\begin{align}
    \sum_{i=1}^n
    \sum_{(k,k')\in\mathcal S_n}
    \left(
        \mu_{i,k}^2+\mu_{i,k'}^2+V_n^2
    \right)
    &\leq
    2(s+1)\|M_n\|_{\mathrm F}^2
    +n|\mathcal S_n|V_n^2
    \lesssim n^2(s+1)V_n~.
    \label{eq: repeated index diagonal coefficient sums 1}
\end{align}
Consequently, we find that 
\begin{align}
    \frac{C}{n^4}
    \sum_{i=1}^n
    \sum_{(k,k')\in\mathcal S_n}
    \left(
        \mu_{i,k}^2+\mu_{i,k'}^2+V_n^2
    \right)
    &\lesssim \frac{(s+1)V_n}{n^2}
    \lesssim V_n^2\frac{s\sqrt d}{n}~.
    \label{eq: repeated index diagonal supported contribution}
\end{align}
We next consider pairs $(k,k')\notin\mathcal S_n$ such that $\lambda_{i,k}\neq0$ or $\lambda_{i,k'}\neq0$. The principal blocks of $M_n$ and $\bar M_n$ indexed by $\{i,k,k'\}$ satisfy \eqref{eq: finite gaussian comparison domain} by \eqref{eq: feasible class local covariance geometry}. Hence, \cref{lem: finite gaussian covariance comparison}, display \eqref{eq: finite gaussian covariance transport}, and \eqref{eq: feasible class local perturbation} give
\begin{align}
    \sup_{\delta\geq0}
    \left|
        \bar\Psi_{(i,i),(k,k')}(\delta)
    \right|
    &\lesssim \sqrt d\,\Delta_{\{i,k,k'\}}~.
    \label{eq: repeated index diagonal correlated comparison}
\end{align}
\Cref{assu: sparsity} gives
\begin{align}
    \sum_{i,k,k'=1}^n
    \mathbb I\left\{
        (k,k')\notin\mathcal S_n,\
        \lambda_{i,k}\neq0\ \mathrm{or}\ \lambda_{i,k'}\neq0
    \right\}
    &\leq
    2n^2(s+1)~.
\end{align}
Therefore, \eqref{eq: feasible class weighted local perturbation moments} with $q=1$ gives
\begin{align}
    \begin{multlined}[b]
        \sup_{\delta\geq0}
        \frac{C}{n^4}
        \sum_{i,k,k'=1}^n
        \mathbb I\left\{
            (k,k')\notin\mathcal S_n,\
            \lambda_{i,k}\neq0\ \mathrm{or}\ \lambda_{i,k'}\neq0
        \right\}\\
        {}\cdot
        \mathbb E\Bigg[
            \left|
                (\hat\varepsilon_i^{(0)})^2
                \hat\varepsilon_k^{(0)}
                \hat\varepsilon_{k'}^{(0)}
            \right|
            \left|
                \bar\Psi_{(i,i),(k,k')}(\delta)
            \right|
        \Bigg]
    \end{multlined}
    &\lesssim \frac{n^2(s+1)\sqrt d\,V_n}{n^4}
    \lesssim 
    V_n^2\frac{s\sqrt d}{n}~,
    \label{eq: repeated index diagonal correlated contribution}
\end{align}
where the final inequality uses \cref{lem: basic inequalities n}, display \eqref{eq: var bounds}. Finally, we consider the case that $\lambda_{k,k'}=\lambda_{i,k}=\lambda_{i,k'}=0$. \Cref{lem: basic inequalities n}, display \eqref{eq: n bound 3}, gives $|\mu_{i,k}|+|\mu_{i,k'}|\lesssim V_n$. Thus, \eqref{eq: repeated index diagonal pointwise contraction} and \eqref{eq: feasible class transported marginal probability} give
\begin{align}
    & \sup_{\delta\in\Delta_{n,d}(\eta)}
    \left|
        \bar\Psi_{(i,i),(k,k')}(\delta)
    \right| \nonumber \\
    &\lesssim 
    \left(
        V_n^2+\Delta_{\{i,k,k'\}}^2
    \right)
    V_n^{\eta/2}
    \exp\Bigg(
        C\left(
            \sqrt{d\log(V_n^{-1})}
            +\log(V_n^{-1})
        \right)
        \Delta_{\{i,k,k'\}}
    \Bigg)~.
\end{align}
Multiplying this bound by the absolute residual product and applying \eqref{eq: feasible class weighted local perturbation moments} with $q\in\{0,2\}$ yields
\begin{align}
    \sup_{\delta\in\Delta_{n,d}(\eta)}
    \mathbb E\Bigg[
        \left|
            (\hat\varepsilon_i^{(0)})^2
            \hat\varepsilon_k^{(0)}
            \hat\varepsilon_{k'}^{(0)}
        \right|
        \left|
            \bar\Psi_{(i,i),(k,k')}(\delta)
        \right|
    \Bigg]
    &\lesssim V_n^{2+\eta/2}~.\label{eq: repeated index diagonal null contribution}
\end{align}
Observe that \cref{lem: basic inequalities n}, display \eqref{eq: var bounds} implies that
\begin{align}
    \frac{V_n^{2+\eta/2}}{n}
    &\lesssim
    V_n^{3+\eta/2}
    \lesssim
    V_n^3
    \lesssim
    V_n^2\frac{s\sqrt d}{n}~.
    \label{eq: repeated index diagonal null contribution bound}
\end{align}
Hence, we can conclude that
\begin{align}
    \sup_{\delta\in\Delta_{n,d}(\eta)}
    |\mathcal R_{2,1,n,d}(\delta)|
    &\lesssim V_n^2\frac{s\sqrt d}{n}
    \label{eq: repeated index diagonal pair bound}
\end{align}
by combining the bounds \eqref{eq: repeated index diagonal supported contribution}, \eqref{eq: repeated index diagonal correlated contribution}, and \eqref{eq: repeated index diagonal null contribution} 

Next, we give a bound for the term $\mathcal R_{2,2,n,d}(\delta)$. Fix $i\neq i'$ and $\Xi\in\{M_n,\bar M_n\}$. Observe that
\begin{align}
    \Psi_{(i,i'),(i,i')}(\Xi,\delta)
    &=\Psi_{(i,i'),(i',i)}(\Xi,\delta)
    \nonumber\\
    &=p_{\delta,d}\left((\Xi_{u,v})_{u,v\in\{i,i'\}}\right)
        \left(
            1-p_{\delta,d}\left(
                (\Xi_{u,v})_{u,v\in\{i,i'\}}
            \right)
        \right)
    \leq
    p_{\delta,d}\left((\Xi_{u,v})_{u,v\in\{i,i'\}}\right)~.
    \label{eq: repeated index equal pair pointwise bound}
\end{align}
Thus, by the first equality in \eqref{eq: repeated index equal pair pointwise bound}, we can write
\begin{align}
    \mathcal R_{2,2,n,d}(\delta)
    &=\frac{2}{n^4}\sum_{i,i'=1}^n
    \mathbb I\{i\neq i'\}
    \mathbb E\left[
        (\hat\varepsilon_i^{(0)}\hat\varepsilon_{i'}^{(0)})^2
        \bar\Psi_{(i,i'),(i,i')}(\delta)
    \right]~.
    \label{eq: repeated index equal pair canonical sum}
\end{align}
Consequently, by the second equality in \eqref{eq: repeated index equal pair pointwise bound}, \eqref{eq: feasible class transported marginal probability}, and the bound \eqref{eq: feasible class weighted local perturbation moments} with $q=0$ we obtain
\begin{align}
    \sup_{\delta\in\Delta_{n,d}(\eta)}
    |\mathcal R_{2,2,n,d}(\delta)|
    &\lesssim
    \frac{1}{n^4}\sum_{i,i'=1}^n
    \mathbb I\{i\neq i'\}
    \sup_{\delta\in\Delta_{n,d}(\eta)}
    \mathbb E\Bigg[
        \left|(
            \hat\varepsilon_i^{(0)}
            \hat\varepsilon_{i'}^{(0)}
        )^2\right|
        \max_{\Xi\in\{M_n,\bar M_n\}}
        p_{\delta,d}\left(
            (\Xi_{u,v})_{u,v\in\{i,i'\}}
        \right)
    \Bigg]
    \nonumber\\
    &\lesssim 
    \frac{1}{n^4}\sum_{i,i'=1}^n
    \mathbb I\{i\neq i'\}
    \left(
        \mathbb I\{(i,i')\in\mathcal S_n\}
        +V_n^\eta
        \mathbb I\{(i,i')\notin\mathcal S_n\}
    \right)
    \nonumber\\
    &\lesssim 
    \left(
        \frac{s}{n^3}
        +\frac{V_n^\eta}{n^2}
    \right)
    \lesssim 
    V_n^2\left(
        \frac{s\sqrt d}{n}
        +V_n^{\eta}
    \right)~,
    \label{eq: repeated index equal pair bound}
\end{align}
where the final two inequalities use $|\{(i,i')\in\mathcal S_n:i\neq i'\}|\leq ns$ and \cref{lem: basic inequalities n}, display \eqref{eq: var bounds}.

Finally, we give a bound for the term $\mathcal R_{2,3,n,d}(\delta)$. Observe that we can write
\begin{align}
    \mathcal R_{2,3,n,d}(\delta)
    &=\frac{4}{n^4}\sum_{i,j,k=1}^n
    \mathbb I\{|\{i,j,k\}|=3\}
    \mathbb E\left[
        (\hat\varepsilon_i^{(0)})^2
        \hat\varepsilon_j^{(0)}\hat\varepsilon_k^{(0)}
        \bar\Psi_{(i,j),(i,k)}(\delta)
    \right]~.
    \label{eq: repeated index three row canonical sum}
\end{align}
by symmetry. We split the sum in \eqref{eq: repeated index three row canonical sum} using the identity
\begin{align}
    \mathbb I\{|\{i,j,k\}|=3\}
    &=\mathbb I\Big\{
        |\{i,j,k\}|=3,
        \max\{|\lambda_{i,j}|,|\lambda_{i,k}|,|\lambda_{j,k}|\}>0
    \Big\}
    \nonumber\\
    &\quad+
    \mathbb I\Big\{
        |\{i,j,k\}|=3,
        \lambda_{i,j}=\lambda_{i,k}=\lambda_{j,k}=0
    \Big\}~.
    \label{eq: repeated index exact correlation split}
\end{align}
For the first indicator in \eqref{eq: repeated index exact correlation split}, \cref{assu: sparsity} gives
\begin{align}
    &\sum_{i,j,k=1}^n
    \mathbb I\Big\{
        |\{i,j,k\}|=3,
        \max\{|\lambda_{i,j}|,|\lambda_{i,k}|,|\lambda_{j,k}|\}>0
    \Big\}
    \nonumber\\
    &\leq
    n\sum_{i,j=1}^n
        \mathbb I\{i\neq j,\ \lambda_{i,j}\neq0\}
    +n\sum_{i,k=1}^n
        \mathbb I\{i\neq k,\ \lambda_{i,k}\neq0\}
    +
    n\sum_{j,k=1}^n
        \mathbb I\{j\neq k,\ \lambda_{j,k}\neq0\}
    \leq3n^2s~.
    \label{eq: repeated index correlated triple count}
\end{align}
The principal blocks of $M_n$ and $\bar M_n$, ordered as $(i,j,k)$, satisfy \eqref{eq: finite gaussian comparison domain} by \eqref{eq: feasible class local covariance geometry}. Hence, \cref{lem: finite gaussian covariance comparison}, display \eqref{eq: finite gaussian covariance transport}, implies that
\begin{align}
    \sup_{\delta\geq0}
    |\bar\Psi_{(i,j),(i,k)}(\delta)|
    &\leq C\sqrt d\,\Delta_{\{i,j,k\}}~.
    \label{eq: repeated index correlated triple comparison}
\end{align}
By combining \eqref{eq: repeated index correlated triple count}, \eqref{eq: repeated index correlated triple comparison}, and \eqref{eq: feasible class weighted local perturbation moments} with $q=1$, we obtain
\begin{align}
    &\sup_{\delta\in\Delta_{n,d}(\eta)}
    \frac{4}{n^4}
    \Bigg|
    \sum_{i,j,k=1}^n
    \mathbb I\Big\{
        |\{i,j,k\}|=3,
        \max\{|\lambda_{i,j}|,|\lambda_{i,k}|,|\lambda_{j,k}|\}>0
    \Big\}
    \mathbb E\left[
        (\hat\varepsilon_i^{(0)})^2
        \hat\varepsilon_j^{(0)}\hat\varepsilon_k^{(0)}
        \bar\Psi_{(i,j),(i,k)}(\delta)
    \right]
    \Bigg|
    \nonumber\\
    &\lesssim
    \frac{\sqrt d}{n^4}
    \sum_{i,j,k=1}^n
    \mathbb I\Big\{
        |\{i,j,k\}|=3,
        \max\{|\lambda_{i,j}|,|\lambda_{i,k}|,|\lambda_{j,k}|\}>0
    \Big\}
    \mathbb E\left[
        \left|
            (\hat\varepsilon_i^{(0)})^2
            \hat\varepsilon_j^{(0)}\hat\varepsilon_k^{(0)}
        \right|
        \Delta_{\{i,j,k\}}
    \right]
    \nonumber\\
    &\lesssim
    \frac{n^2s\sqrt d\,V_n}{n^4}
    \lesssim
    V_n^2\frac{s\sqrt d}{n}~,
    \label{eq: repeated index nonzero correlation aggregate}
\end{align}
where the final inequality uses \cref{lem: basic inequalities n}, display \eqref{eq: var bounds}.

We next treat the second indicator in \eqref{eq: repeated index exact correlation split}. Fix distinct $i,j,k$ such that $\lambda_{i,j}=\lambda_{i,k}=\lambda_{j,k}=0$, and order every $3\times3$ block below as $(i,j,k)$. Then \cref{lem: basic inequalities n}, display \eqref{eq: n bound 3}, gives
\begin{align}
    \left\|
        (\mu_{u,v})_{u,v\in\{i,j,k\}}-I_3
    \right\|_{\mathrm F}
    &\leq CV_n\quad\text{and}
    \nonumber\\
    \left\|
        (\bar\mu_{u,v})_{u,v\in\{i,j,k\}}-I_3
    \right\|_{\mathrm F}
    &\leq C\left(
        V_n+\Delta_{\{i,j,k\}}
    \right)~.
    \label{eq: repeated index null three row distances}
\end{align}
Observe that \eqref{eq: repeated index null three row distances} implies that, for every $\Xi\in\{M_n,\bar M_n\}$ and $t\in[0,1]$, the bound
\begin{align}
    \kappa I_3
    &\preceq
    I_3+t\left(
        (\Xi_{u,v})_{u,v\in\{i,j,k\}}-I_3
    \right)
    \preceq
    \kappa^{-1}I_3
    \label{eq: repeated index null three row covariance paths}
\end{align}
holds.  Moreover, the two marginal blocks along paths of the form satisfy
\begin{align}
    \left(
        I_3+t\left(
            (\Xi_{u,v})_{u,v\in\{i,j,k\}}-I_3
        \right)
    \right)_{\{1,2\},\{1,2\}}
    &\in
    \operatorname{conv}\left\{
        I_2,
        (\mu_{u,v})_{u,v\in\{i,j\}},
        (\bar\mu_{u,v})_{u,v\in\{i,j\}}
    \right\}\quad\text{and}
    \nonumber\\
    \left(
        I_3+t\left(
            (\Xi_{u,v})_{u,v\in\{i,j,k\}}-I_3
        \right)
    \right)_{\{1,3\},\{1,3\}}
    &\in
    \operatorname{conv}\left\{
        I_2,
        (\mu_{u,v})_{u,v\in\{i,k\}},
        (\bar\mu_{u,v})_{u,v\in\{i,k\}}
    \right\}
    \label{eq: repeated index null three row marginal paths}
\end{align}
for each $\Xi\in\{M_n,\bar M_n\}$ and $t\in[0,1]$. For fixed $\delta\in\Delta_{n,d}(\eta)$, define
\begin{align}
    p
    =\max\Bigg\{
    &V_n^\eta, \nonumber\\
    &\sup_{\substack{\Xi\in\{M_n,\bar M_n\}\\t\in[0,1]}}
        \max\Bigg\{
            p_{\delta,d}\left(
                \left(
                    I_3+t\left(
                        (\Xi_{u,v})_{u,v\in\{i,j,k\}}-I_3
                    \right)
                \right)_{\{1,2\},\{1,2\}}
            \right),
    \nonumber\\
    &
    \quad\quad\quad\quad\quad\quad\quad\quad
    p_{\delta,d}\left(
                \left(
                    I_3+t\left(
                        (\Xi_{u,v})_{u,v\in\{i,j,k\}}-I_3
                    \right)
                \right)_{\{1,3\},\{1,3\}}
            \right)
        \Bigg\}
    \Bigg\}
    \label{eq: repeated index null three row probability envelope}
\end{align}
and observe that $p\in(0,1]$. Because $(i,j),(i,k)\notin\mathcal S_n$, \eqref{eq: feasible class transported marginal probability} and \eqref{eq: repeated index null three row marginal paths} imply that
\begin{align}
    p
    &\leq
    CV_n^\eta
    \exp\Bigg(
        C\left(
            \sqrt{
                d\log\left(
                    V_n^{-1}
                \right)
            }
            +\log\left(
                V_n^{-1}
            \right)
        \right)
        \Delta_{\{i,j,k\}}
    \Bigg)~.
    \label{eq: repeated index null three row probability bounds}
\end{align}
As a consequence, we have that
\begin{align}
    &p\left(
        d(1+\log(p^{-1}))
        +(1+\log(p^{-1}))^2
    \right)
    \nonumber\\
    &\leq
    CV_n^\eta
    \left(
        d\log(V_n^{-1})
        +\log^2\left(V_n^{-1}\right)
    \right)
    \exp\Bigg(
        C\left(
            \sqrt{
                d\log\left(
                    V_n^{-1}
                \right)
            }
            +\log\left(
                V_n^{-1}
            \right)
        \right)
        \Delta_{\{i,j,k\}}
    \Bigg)~.
    \label{eq: repeated index null three row logarithmic envelope}
\end{align}
The bounds \eqref{eq: repeated index null three row covariance paths} and \eqref{eq: repeated index null three row probability envelope} verify the hypotheses of \cref{lem: Gaussian three row}. Applying that lemma separately to the $M_n$ and $\bar M_n$ blocks and using \eqref{eq: repeated index null three row distances} and \eqref{eq: repeated index null three row logarithmic envelope} yields the bound
\begin{align}
    &\max_{\Xi\in\{M_n,\bar M_n\}}\left|
        \Psi_{(i,j),(i,k)}(\Xi,\delta)
        -
        \Psi_{(1,2),(1,3)}(I_3,\delta) \right.\nonumber \\
    & \left. \quad\quad\quad\quad\quad\quad
    -\mathrm D\Psi_{(1,2),(1,3)}(I_3,\delta)
        \left[
            \mathsf{Diag}\left(
                \Xi_{i,i}-1,
                \Xi_{j,j}-1,
                \Xi_{k,k}-1
            \right)
        \right]
    \right|
    \nonumber\\
    &
    \lesssim V_n^\eta
    \left(
        d\log(V_n^{-1})
        +\log^2\left(V_n^{-1}\right)
    \right)
    \nonumber\\
    &\quad
    \cdot
    \exp\Bigg(
        C\left(
            \sqrt{
                d\log\left(
                    V_n^{-1}
                \right)
            }
            +\log\left(
                V_n^{-1}
            \right)
        \right)
        \Delta_{\{i,j,k\}}
    \Bigg)
    \left(
        V_n^2
        +\Delta_{\{i,j,k\}}^2
    \right)~.
    \label{eq: repeated index null three row endpoint approximations}
\end{align}
Subtracting the approximation for the $M_n$ block from the approximation for the $\bar M_n$ block gives
\begin{align}
    &\left|
        \bar\Psi_{(i,j),(i,k)}(\delta) - 
        -\mathrm D\Psi_{(1,2),(1,3)}(I_3,\delta)
        \left[
            \mathsf{Diag}\left(
                \bar\mu_{i,i}-\mu_{i,i},
                \bar\mu_{j,j}-\mu_{j,j},
                \bar\mu_{k,k}-\mu_{k,k}
            \right)
        \right]
    \right|
    \nonumber\\
    &\lesssim 
    V_n^\eta
    \left(
        d\log(V_n^{-1})
        +\log^2\left(V_n^{-1}\right)
    \right)
    \nonumber\\
    &\quad\cdot
    \exp\Bigg(
        C\left(
            \sqrt{
                d\log\left(
                    V_n^{-1}
                \right)
            }
            +\log\left(
                V_n^{-1}
            \right)
        \right)
        \Delta_{\{i,j,k\}}
    \Bigg)
    \left(
        V_n^2
        +\Delta_{\{i,j,k\}}^2
    \right)~.
    \label{eq: repeated index null three row endpoint expansion}
\end{align}
Now observe that
\begin{align}
    &\sum_{i,j,k=1}^n
    (\hat\varepsilon_i^{(0)})^2
    \hat\varepsilon_j^{(0)}\hat\varepsilon_k^{(0)}
    \mathrm D\Psi_{(1,2),(1,3)}(I_3,\delta)
    \left[
        \mathsf{Diag}\left(
            \bar\mu_{i,i}-\mu_{i,i},
            \bar\mu_{j,j}-\mu_{j,j},
            \bar\mu_{k,k}-\mu_{k,k}
        \right)
    \right]
    \nonumber\\
    &=\mathrm D\Psi_{(1,2),(1,3)}(I_3,\delta) \nonumber \\
    &\quad
    \Bigg[
        \mathsf{Diag}\Bigg(
                \sum_{i=1}^n
                (\hat\varepsilon_i^{(0)})^2
                (\bar\mu_{i,i}-\mu_{i,i})
            \sum_{j=1}^n\hat\varepsilon_j^{(0)}
            \sum_{k=1}^n\hat\varepsilon_k^{(0)},
            \sum_{i=1}^n(\hat\varepsilon_i^{(0)})^2
                \sum_{j=1}^n
                \hat\varepsilon_j^{(0)}
                (\bar\mu_{j,j}-\mu_{j,j})
            \sum_{k=1}^n\hat\varepsilon_k^{(0)},
            \nonumber\\
    &\quad\quad
            \sum_{i=1}^n(\hat\varepsilon_i^{(0)})^2
            \sum_{j=1}^n\hat\varepsilon_j^{(0)}
                \sum_{k=1}^n
                \hat\varepsilon_k^{(0)}
                (\bar\mu_{k,k}-\mu_{k,k})
        \Bigg)
    \Bigg]
    =0~,
    \label{eq: repeated index null three row cancellation}
\end{align}
by the equality $\sum_{i=1}^n\hat\varepsilon_i^{(0)}=0$. Thus, we find that
\begin{align}
    &\sum_{i,j,k=1}^n
    \mathbb I\Big\{
        |\{i,j,k\}|=3,
        \lambda_{i,j}=\lambda_{i,k}=\lambda_{j,k}=0
    \Big\}
    (\hat\varepsilon_i^{(0)})^2
    \hat\varepsilon_j^{(0)}\hat\varepsilon_k^{(0)}
    \nonumber\\
    &\hspace{12mm}\cdot
    \mathrm D\Psi_{(1,2),(1,3)}(I_3,\delta)
    \left[
        \mathsf{diag}\left(
            \bar\mu_{i,i}-\mu_{i,i},
            \bar\mu_{j,j}-\mu_{j,j},
            \bar\mu_{k,k}-\mu_{k,k}
        \right)
    \right]
    \nonumber\\
    &=-\sum_{i,j,k=1}^n
    \left(
        1-
        \mathbb I\Big\{
            |\{i,j,k\}|=3,
            \lambda_{i,j}=\lambda_{i,k}=\lambda_{j,k}=0
        \Big\}
    \right)
    (\hat\varepsilon_i^{(0)})^2
    \hat\varepsilon_j^{(0)}\hat\varepsilon_k^{(0)}
    \nonumber\\
    &\hspace{12mm}\cdot
    \mathrm D\Psi_{(1,2),(1,3)}(I_3,\delta)
    \left[
        \mathsf{diag}\left(
            \bar\mu_{i,i}-\mu_{i,i},
            \bar\mu_{j,j}-\mu_{j,j},
            \bar\mu_{k,k}-\mu_{k,k}
        \right)
    \right]~.
    \label{eq: repeated index null three row restricted cancellation}
\end{align}
The number of terms on the right-hand side of \eqref{eq: repeated index null three row restricted cancellation} satisfies
\begin{align}
    &\sum_{i,j,k=1}^n
    \left(
        1-
        \mathbb I\Big\{
            |\{i,j,k\}|=3,
            \lambda_{i,j}=\lambda_{i,k}=\lambda_{j,k}=0
        \Big\}
    \right) \nonumber \\
    &\leq
    3n^2
    +n\sum_{i,j=1}^n
        \mathbb I\{i\neq j,\ \lambda_{i,j}\neq0\}
    +n\sum_{i,k=1}^n
        \mathbb I\{i\neq k,\ \lambda_{i,k}\neq0\}
    +
    n\sum_{j,k=1}^n
        \mathbb I\{j\neq k,\ \lambda_{j,k}\neq0\}\nonumber\\
    &
    \leq Cn^2(s+1)~.
    \label{eq: repeated index null three row complement count}
\end{align}
Hence, by applying the derivative bound in \cref{lem: Gaussian three row}, the bound \eqref{eq: feasible class weighted local perturbation moments} with $q=1$, the bounds \eqref{eq: repeated index null three row restricted cancellation} and \eqref{eq: repeated index null three row complement count}, and \cref{lem: basic inequalities n}, display \eqref{eq: var bounds}, we obtain
\begin{align}
    &\sup_{\delta\in\Delta_{n,d}(\eta)}
    \frac{4}{n^4}
    \Bigg|
    \sum_{i,j,k=1}^n
    \mathbb I\Big\{
        |\{i,j,k\}|=3,
        \lambda_{i,j}=\lambda_{i,k}=\lambda_{j,k}=0
    \Big\}
    \nonumber\\
    &\hspace{22mm}\cdot
    \mathbb E\Bigg[
        (\hat\varepsilon_i^{(0)})^2
        \hat\varepsilon_j^{(0)}\hat\varepsilon_k^{(0)}
        \mathrm D\Psi_{(1,2),(1,3)}(I_3,\delta)
        \left[
            \mathsf{diag}\left(
                \bar\mu_{i,i}-\mu_{i,i},
                \bar\mu_{j,j}-\mu_{j,j},
                \bar\mu_{k,k}-\mu_{k,k}
            \right)
        \right]
    \Bigg]
    \Bigg|
    \nonumber\\
    &\lesssim 
    \frac{n^2(s+1)\sqrt d\,V_n}{n^4}
    \lesssim CV_n^2\frac{s\sqrt d}{n}~.
    \label{eq: repeated index null three row linear aggregate}
\end{align}
Now, observe that the bound \eqref{eq: feasible class weighted local perturbation moments} with $q\in\{0,2\}$ gives
\begin{align}
    &\mathbb E\Bigg[
        \left|
            (\hat\varepsilon_i^{(0)})^2
            \hat\varepsilon_j^{(0)}\hat\varepsilon_k^{(0)}
        \right|
        \nonumber\\
    &\quad\quad
    \cdot \exp\Bigg(
            C\left(
                \sqrt{
                    d\log\left(
                        V_n^{-1}
                    \right)
                }
                +\log\left(
                    V_n^{-1}
                \right)
            \right)
            \Delta_{\{i,j,k\}}
        \Bigg)
        \left(
            V_n^2
            +\Delta_{\{i,j,k\}}^2
        \right)
    \Bigg]
    \lesssim V_n^2~.
    \label{eq: repeated index null three row weighted remainder}
\end{align}
Thus, we find that 
\begin{align}
        &\sup_{\delta\in\Delta_{n,d}(\eta)}
        \frac{4}{n^4}
        \Bigg|
        \sum_{i,j,k=1}^n
        \mathbb I\{|\{i,j,k\}|=3\}\cdot
        \mathbb I\{
            \lambda_{i,j}=\lambda_{i,k}=\lambda_{j,k}=0
        \} \nonumber \\
        &\quad\quad
        \cdot \mathbb E\Bigg[
            (\hat\varepsilon_i^{(0)})^2
            \hat\varepsilon_j^{(0)}
            \hat\varepsilon_k^{(0)} 
            \Bigg(\bar\Psi_{(i,j),(i,k)}(\delta)\nonumber \\
            & \quad\quad\quad\quad\quad\quad\quad\quad\quad\quad
            -\mathrm D\Psi_{(1,2),(1,3)}(I_3,\delta)
            \quad\Big[
                \mathsf{diag}\Big(
                    \bar\mu_{i,i}-\mu_{i,i},
                    \bar\mu_{j,j}-\mu_{j,j},
                    \bar\mu_{k,k}-\mu_{k,k}
                \Big)
            \Big]
            \Bigg)
        \Bigg]
        \Bigg| \nonumber \\
    &\lesssim 
    \frac{V_n^\eta}{n^4}
    \left(
        d\log(V_n^{-1})
        +\log^2(V_n^{-1})
    \right) \nonumber \\
    & \quad\quad \sum_{i,j,k=1}^n
    \mathbb E\Bigg[
        \left|
            (\hat\varepsilon_i^{(0)})^2
            \hat\varepsilon_j^{(0)}
            \hat\varepsilon_k^{(0)}
        \right|
        \exp\Bigg(
            C\left(
                \sqrt{d\log(V_n^{-1})}
                +\log(V_n^{-1})
            \right)
            \Delta_{\{i,j,k\}}
        \Bigg)
        \left(
            V_n^2+\Delta_{\{i,j,k\}}^2
        \right)
    \Bigg]
    \nonumber\\
    &\lesssim 
    \frac{V_n^{2+\eta}}{n}
    \left(
        d\log(V_n^{-1})
        +\log^2(V_n^{-1})
    \right)
    \lesssim 
    V_n^2\left(
        dV_n^{1+\eta}
        \log(V_n^{-1})
        +V_n^\eta
    \right)~,
    \label{eq: repeated index null three row remainder aggregate}
\end{align}
where the first inequality uses \eqref{eq: repeated index null three row endpoint expansion}, the second uses \eqref{eq: repeated index null three row weighted remainder}
and the bound of $n^3$ on the number of triples, and the final inequality uses \cref{lem: basic inequalities n}, display \eqref{eq: var bounds}, and the boundedness of $x\mapsto x\log^2(x^{-1})$ on $(0,1]$. Consequently, we find that
\begin{align}
    \sup_{\delta\in\Delta_{n,d}(\eta)}
    |\mathcal R_{2,3,n,d}(\delta)|
    \lesssim 
    V_n^2\left(
        V_n^\eta
        +\frac{s\sqrt d}{n}
        +dV_n^{1+\eta}
        \log(V_n^{-1})
    \right)~,
    \label{eq: repeated index three row bound}
\end{align}
by substituting \eqref{eq: repeated index nonzero correlation aggregate}, \eqref{eq: repeated index null three row linear aggregate}, and \eqref{eq: repeated index null three row remainder aggregate} into \eqref{eq: repeated index three row canonical sum}, according to the split \eqref{eq: repeated index exact correlation split}.

To conclude the proof, observe that
\begin{align}
    &\sup_{\delta\in\Delta_{n,d}(\eta)}
    |\mathcal R_{2,n,d}(\delta)|
    \lesssim V_n^2\Bigg(
        V_n^{\eta}
        +\frac{s\sqrt d}{n}
        +dV_n^{1+\eta}
        \log(V_n^{-1})
    \Bigg)~.
\end{align}
follows by \eqref{eq: repeated index diagonal pair bound}, \eqref{eq: repeated index equal pair bound}, and \eqref{eq: repeated index three row bound}, as required.\hfill\qed

\subsubsection{Part (iii)}

Fix four distinct indices $i,i',k,k'$ such that $(i,i')\in\mathcal S_n$ or $(k,k')\in\mathcal S_n$, and let $T=\{i,i',k,k'\}$. Order the corresponding principal blocks by the pairs $(i,i')$ and $(k,k')$, and write
\begin{align}
    (M_n)_{T,T}
    &=\begin{pmatrix}A_0&R_0\\R_0^\top&B_0\end{pmatrix},
    \qquad
    (\bar M_n)_{T,T}
    =\begin{pmatrix}A_1&R_1\\R_1^\top&B_1\end{pmatrix}~.
\end{align}
For $t\in[0,1]$, define the paths
\begin{align}
    A_t&=(1-t)A_0+tA_1,
    &B_t&=(1-t)B_0+tB_1,
    &R_t&=(1-t)R_0+tR_1~.
\end{align}
The fundamental theorem of calculus implies that
\begin{align}
    \bar\Psi_{(i,i'),(k,k')}(\delta)
    &=\int_0^1
    \mathrm D_{A,B}\Psi_{\delta,d}(A_t,B_t,R_t)
    [A_1-A_0,B_1-B_0]\,\mathrm dt
    \nonumber\\
    &+
    \int_0^1
    \mathrm D_R\Psi_{\delta,d}(A_t,B_t,R_t)
    [R_1-R_0]\,\mathrm dt~.
    \label{eq: feasible class third contribution integral}
\end{align}
Observe that we can evaluate
\begin{align}
    &\|A_1-A_0\|_{\mathrm F}
    +\|B_1-B_0\|_{\mathrm F}
    +\|R_1-R_0\|_{\mathrm F}
    \leq C\Delta_T,\quad\text{and}
    \nonumber\\
    &\|R_t\|_{\mathrm F}
    \leq
    \left(
        \sum_{u\in\{i,i'\}}
        \sum_{v\in\{k,k'\}}
        \mu_{u,v}^2
    \right)^{1/2}
    +C\Delta_T~.
    \label{eq: feasible class third contribution inputs}
\end{align}
The covariance matrices in \eqref{eq: feasible class third contribution integral} satisfy \eqref{eq: gaussian covariance eigenvalue bounds} by \eqref{eq: feasible class local covariance geometry}. Applying Parts (i) and (ii) of \cref{lem: gaussian threshold covariance bounds} with $p=1$, and then substituting \eqref{eq: feasible class third contribution inputs}, gives
\begin{align}
    |\bar\Psi_{(i,i'),(k,k')}(\delta)|
    &\leq
    C\sum_{u\in\{i,i'\}}
    \sum_{v\in\{k,k'\}}
    \Big(
        |\mu_{u,v}|\Delta_T+\Delta_T^2
        +\sqrt d\left(
            \mu_{u,v}^2\Delta_T+\Delta_T^3
        \right)
    \Big)~.
    \label{eq: feasible class third contribution pointwise bound}
\end{align}
Multiplying \eqref{eq: feasible class third contribution pointwise bound} by the absolute residual product and applying \eqref{eq: feasible class weighted local perturbation moments} with $q\in\{1,2,3\}$ yields
\begin{align}
    &\mathbb E\left[
        \left|
            \hat\varepsilon_i^{(0)}\hat\varepsilon_{i'}^{(0)}
            \hat\varepsilon_k^{(0)}\hat\varepsilon_{k'}^{(0)}
        \right|
        |\bar\Psi_{(i,i'),(k,k')}(\delta)|
    \right]
    \nonumber\\
    &\leq
    C\sum_{u\in\{i,i'\}}
    \sum_{v\in\{k,k'\}}
    \Bigg(
        |\mu_{u,v}|V_n
        +V_n^2
        +\sqrt d\left(
            \mu_{u,v}^2V_n
            +V_n^3
        \right)
    \Bigg)~.
    \label{eq: feasible class third contribution weighted bound}
\end{align}
Observe that
\begin{align}
    &\sum_{(i,i')\in\mathcal S_n}
    \sum_{k,k'=1}^n
    \sum_{u\in\{i,i'\}}
    \sum_{v\in\{k,k'\}}1
    \leq4|\mathcal S_n|n^2
    \leq Cn^3(s+1)\quad\text{and}
    \nonumber\\
    &\sum_{(i,i')\in\mathcal S_n}
    \sum_{k,k'=1}^n
    \sum_{u\in\{i,i'\}}
    \sum_{v\in\{k,k'\}}
    \left(|\mu_{u,v}|+\mu_{u,v}^2\right)
    \nonumber\\
    &\lesssim n|\mathcal S_n|
    \max_{u\in[n]}
    \sum_{v=1}^n
    \left(|\mu_{u,v}|+\mu_{u,v}^2\right)
    \leq Cn^3(s+1)V_n~,
    \label{eq: feasible class third contribution coefficient sums}
\end{align}
where the second line uses \eqref{eq: feasible class centered row sum} and \cref{lem: basic inequalities n}, display \eqref{eq: n bound 2}. Bounding the union in \eqref{eq: feasible covariance third aggregate} by the two symmetric cases gives
\begin{align}
    \sup_{\delta\geq0}|\mathcal R_{3,n,d}(\delta)|
    &\leq
    \frac{2C}{n^4}
    \sum_{(i,i')\in\mathcal S_n}
    \sum_{k,k'=1}^n
    \sum_{u\in\{i,i'\}}
    \sum_{v\in\{k,k'\}}
    \Bigg(
        |\mu_{u,v}|V_n
        +V_n^2 +\sqrt d\left(
            \mu_{u,v}^2V_n
            +V_n^3
        \right)
    \Bigg)
    \nonumber\\
    &\lesssim
    \frac{n^3(s+1)}{n^4}
    \left(
        V_n^2
        +\sqrt d\,V_n^2
    \right)
    \lesssim
    V_n^2\frac{s\sqrt d}{n}~,
\end{align}
as required.\hfill\qed

\subsubsection{Part (iv)}

Fix $(i,i',k,k')\in\mathcal Q_n$ such that $(i,i')\notin\mathcal S_n$ and $(k,k')\notin\mathcal S_n$. Put $i_1=i$, $i_2=i'$, $k_1=k$, and $k_2=k'$, and let $T=\{i_1,i_2,k_1,k_2\}$. Order the corresponding principal blocks by the pairs $(i_1,i_2)$ and $(k_1,k_2)$, and write
\begin{align}
    (M_n)_{T,T}
    &=\begin{pmatrix}A_0&R_0\\R_0^\top&B_0\end{pmatrix},
    &
    (\bar M_n)_{T,T}
    &=\begin{pmatrix}A_1&R_1\\R_1^\top&B_1\end{pmatrix},
\end{align}
where
\begin{align}
    R_0
    &=\begin{pmatrix}
        \mu_{i,k}&\mu_{i,k'}\\
        \mu_{i',k}&\mu_{i',k'}
    \end{pmatrix},
    &
    R_1
    &=\begin{pmatrix}
        \bar\mu_{i,k}&\bar\mu_{i,k'}\\
        \bar\mu_{i',k}&\bar\mu_{i',k'}
    \end{pmatrix}~.
    \label{eq: feasible class fourth endpoint blocks}
\end{align}
For $\ell\in\{0,1\}$ and $q,r\in[2]$, define
\begin{align}
    R_{\ell,qr}
    &=(R_\ell)_{q,r}e_qe_r^\top,
    &
    R_\ell
    &=\sum_{a,b=1}^2R_{\ell,ab}~.
    \label{eq: feasible class positional cross components}
\end{align}
Consider the representation
\begin{align}
    \bar\Psi_{(i,i'),(k,k')}(\delta)
    &=\left(
        \Psi_{\delta,d}(A_1,B_1,R_1)
        -\sum_{q,r=1}^2
        \Psi_{\delta,d}(A_1,B_1,R_{1,qr})
    \right)
    \nonumber\\
    &\quad-
    \left(
        \Psi_{\delta,d}(A_0,B_0,R_0)
        -\sum_{q,r=1}^2
        \Psi_{\delta,d}(A_0,B_0,R_{0,qr})
    \right)
    \nonumber\\
    &\quad+
    \sum_{q,r=1}^2
    \left(
        \Psi_{\delta,d}(A_1,B_1,R_{1,qr})
        -\Psi_{\delta,d}(A_1,B_1,R_{0,qr})
    \right)
    \nonumber\\
    &\quad+
    \sum_{q,r=1}^2
    \left(
        \Psi_{\delta,d}(A_1,B_1,R_{0,qr})
        -\Psi_{\delta,d}(A_0,B_0,R_{0,qr})
    \right)~.
    \label{eq: feasible class exact endpoint decomposition}
\end{align}
The result follows by developing bounds for each of the four terms in \eqref{eq: feasible class exact endpoint decomposition}.

We begin by establishing a set of bounds that will be used to handle more than one term. Observe that, for every $q,r\in[2]$ and $\ell\in\{0,1\}$, we have
\begin{align}
    \|R_{1,qr}-R_{0,qr}\|_{\mathrm F}
    &\leq\Delta_T~,
    \nonumber\\
    \|R_{\ell,qr}\|_{\mathrm F}
    &\leq|\mu_{i_q,k_r}|+\Delta_T
    \leq
    |\lambda_{i_q,k_r}|+CV_n+\Delta_T~,
    \nonumber\\
    \|A_\ell-I_2\|_{\mathrm F}
    +\|B_\ell-I_2\|_{\mathrm F}
    &\leq C\left(
        V_n+\Delta_T
    \right)~.
    \label{eq: feasible class fourth endpoint inputs}
\end{align}
\cref{assu: well conditioned}, applied to $\Lambda_n$, and \cref{assu: homoskedasticity} give $1-|\lambda_{i_q,k_r}|\geq c$. Thus, the bounds \eqref{eq: feasible class fourth endpoint inputs} give that
\begin{align}
    \kappa I_4
    &\preceq
    \begin{pmatrix}A&R\\R^\top&B\end{pmatrix}
    \preceq
    \kappa^{-1}I_4
    \label{eq: feasible class one entry covariance geometry}
\end{align}
whenever
\begin{align}
    A&\in\operatorname{conv}\{I_2,A_0,A_1\}~,
    &B&\in\operatorname{conv}\{I_2,B_0,B_1\}~,
    \quad\text{and}\quad
    R\in\operatorname{conv}\{0,R_{0,qr},R_{1,qr}\}~.
\end{align}
The bound \eqref{eq: feasible class transported marginal probability} implies, for $h\in\{0,1,2\}$ and $G\in\operatorname{conv}\{I_2,A_0,A_1\}
\cup\operatorname{conv}\{I_2,B_0,B_1\}$, that
\begin{align}
    &p_{\delta,d}(G)
    \left(
        1+\log(p_{\delta,d}(G)^{-1})
    \right)^h
    \nonumber\\
    &\leq
    CV_n^\eta
    \log^h\left(V_n^{-1}\right)
    \exp\Bigg(
        C\left(
            \sqrt{
                d\log\left(
                    V_n^{-1}
                \right)
            }
            +\log\left(
                V_n^{-1}
            \right)
        \right)\Delta_T
    \Bigg)~,
    \label{eq: feasible class fourth marginal probability bound}
\end{align}
uniformly over $\delta\in\Delta_{n,d}(\eta)$. Moreover, observe that we have that
\begin{align}
    \sum_{i,i',k,k'=1}^n
    \sum_{q,r=1}^2
    \left(|\mu_{i_q,k_r}|+V_n\right)
    & \leq Cn^4V_n~,
    \nonumber\\
    \sum_{i,i',k,k'=1}^n
    \sum_{q,r=1}^2
    \mu_{i_q,k_r}^2
    & \leq Cn^4V_n~,\quad\text{and}
    \nonumber\\
    \sum_{i,i',k,k'=1}^n
    \sum_{q,r=1}^2
    \sum_{\tilde q,\tilde r=1}^2
    \mathbb I\{(q,r)\neq(\tilde q,\tilde r)\}
    \left(|\mu_{i_q,k_r}|+V_n\right)
    \left(|\mu_{i_{\tilde q},k_{\tilde r}}|+V_n\right)
    & \leq Cn^4V_n^2
    \label{eq: feasible class fourth coefficient sums}
\end{align}
by the bound \eqref{eq: feasible class centered row sum} and \cref{lem: basic inequalities n}, display \eqref{eq: n bound 2}.

We now bound the first two terms of \eqref{eq: feasible class exact endpoint decomposition}. Part (iii) of \cref{lem: gaussian threshold covariance bounds}, \eqref{eq: feasible class fourth endpoint inputs}, and \eqref{eq: feasible class one entry covariance geometry} give
\begin{align}
    &\left|
        \Psi_{\delta,d}(A_\ell,B_\ell,R_\ell)
        -\sum_{q,r=1}^2
        \Psi_{\delta,d}(A_\ell,B_\ell,R_{\ell,qr})
    \right|
    \nonumber\\
    &\leq
    C\max\{p_{\delta,d}(A_\ell),p_{\delta,d}(B_\ell)\}
    \sum_{q,r=1}^2
    \sum_{\tilde q,\tilde r=1}^2
    \mathbb I\{(q,r)\neq(\tilde q,\tilde r)\}
    \left(|\mu_{i_q,k_r}|+\Delta_T\right)
    \left(|\mu_{i_{\tilde q},k_{\tilde r}}|+\Delta_T\right)
    \label{eq: feasible class interaction pointwise bound}
\end{align}
for $\ell\in\{0,1\}$. By multiplying \eqref{eq: feasible class interaction pointwise bound} by the absolute residual product, applying \eqref{eq: feasible class fourth marginal probability bound} with $h=0$ and \eqref{eq: feasible class weighted local perturbation moments} for moment powers $0$, $1$, and $2$, we obtain
\begin{align}
    &\sup_{\delta\in\Delta_{n,d}(\eta)}
    \frac{1}{n^4}
    \sum_{(i,i',k,k')\in\mathcal Q_n}
    \mathbb I\{(i,i')\notin\mathcal S_n\}
    \mathbb I\{(k,k')\notin\mathcal S_n\}
    \nonumber\\
    &\quad\cdot
    \mathbb E\Bigg[
        \left|
            \hat\varepsilon_i^{(0)}\hat\varepsilon_{i'}^{(0)}
            \hat\varepsilon_k^{(0)}\hat\varepsilon_{k'}^{(0)}
        \right|
        \sum_{\ell=0}^1
        \left|
            \Psi_{\delta,d}(A_\ell,B_\ell,R_\ell)
            -\sum_{q,r=1}^2
            \Psi_{\delta,d}(A_\ell,B_\ell,R_{\ell,qr})
        \right|
    \Bigg]
    \nonumber\\
    &\lesssim
    \frac{CV_n^\eta}{n^4}
    \sum_{i,i',k,k'=1}^n
    \sum_{q,r=1}^2
    \sum_{\tilde q,\tilde r=1}^2
    \mathbb I\{(q,r)\neq(\tilde q,\tilde r)\}
    \cdot
    \left(|\mu_{i_q,k_r}|+V_n\right)
    \left(|\mu_{i_{\tilde q},k_{\tilde r}}|+V_n\right)
    \lesssim CV_n^{2+\eta}~.
    \label{eq: feasible class interaction aggregate}
\end{align}
For the third line in \eqref{eq: feasible class exact endpoint decomposition}, the fundamental theorem of calculus gives, for $q,r\in[2]$,
\begin{align}
    &\Psi_{\delta,d}(A_1,B_1,R_{1,qr})
    -\Psi_{\delta,d}(A_1,B_1,R_{0,qr})
    \nonumber\\
    &=\int_0^1
    \mathrm D_R\Psi_{\delta,d}\left(
        A_1,B_1,
        R_{0,qr}+t(R_{1,qr}-R_{0,qr})
    \right)
    [R_{1,qr}-R_{0,qr}]\,\mathrm dt~.
    \label{eq: feasible class cross entry integral}
\end{align}
Part (i) of \cref{lem: gaussian threshold covariance bounds} and \eqref{eq: feasible class fourth endpoint inputs} therefore imply that
\begin{align}
    &\left|
        \Psi_{\delta,d}(A_1,B_1,R_{1,qr})
        -\Psi_{\delta,d}(A_1,B_1,R_{0,qr})
    \right|
    \lesssim \max\{p_{\delta,d}(A_1),p_{\delta,d}(B_1)\}
    \left(|\mu_{i_q,k_r}|+\Delta_T\right)\Delta_T~.
    \label{eq: feasible class cross entry pointwise bound}
\end{align}
Thus, \eqref{eq: feasible class fourth marginal probability bound} with $h=0$, \eqref{eq: feasible class weighted local perturbation moments} for moment powers $1$ and $2$, and \eqref{eq: feasible class fourth coefficient sums} give
\begin{align}
    &\sup_{\delta\in\Delta_{n,d}(\eta)}
    \frac{1}{n^4}
    \sum_{(i,i',k,k')\in\mathcal Q_n}
    \mathbb I\{(i,i')\notin\mathcal S_n\}
    \mathbb I\{(k,k')\notin\mathcal S_n\}
    \nonumber\\
    &\quad\cdot
    \mathbb E\Bigg[
        \left|
            \hat\varepsilon_i^{(0)}\hat\varepsilon_{i'}^{(0)}
            \hat\varepsilon_k^{(0)}\hat\varepsilon_{k'}^{(0)}
        \right|
        \sum_{q,r=1}^2
        \left|
            \Psi_{\delta,d}(A_1,B_1,R_{1,qr})
            -\Psi_{\delta,d}(A_1,B_1,R_{0,qr})
        \right|
    \Bigg]
    \nonumber\\
    &\lesssim
    \frac{V_n^\eta}{n^4}
    \sum_{i,i',k,k'=1}^n
    \sum_{q,r=1}^2
    \left(
        |\mu_{i_q,k_r}|V_n
        +V_n^2
    \right)
    \lesssim V_n^{2+\eta}~.
    \label{eq: feasible class cross entry aggregate}
\end{align}
It remains to bound the fourth line in \eqref{eq: feasible class exact endpoint decomposition}. Put
\begin{align}
    H_A&=A_1-A_0,
    &H_B&=B_1-B_0,
    &A_t&=A_0+tH_A,
    &B_t&=B_0+tH_B~.
\end{align}
Two applications of the fundamental theorem of calculus give, for every $q,r\in[2]$,
\begin{align}
    &\Psi_{\delta,d}(A_1,B_1,R_{0,qr})
    -\Psi_{\delta,d}(A_0,B_0,R_{0,qr})
    \nonumber\\
    &=\mathrm D_{A,B}\Psi_{\delta,d}
        (I_2,I_2,R_{0,qr})[H_A,H_B]
    \nonumber\\
    &\quad+
    \int_0^1\int_0^1
    \mathrm D_{A,B}^2\Psi_{\delta,d}\Big(
        I_2+w(A_t-I_2),
        I_2+w(B_t-I_2),
        R_{0,qr}
    \Big)
    \nonumber\\
    &\hspace{25mm}\Big[
        (A_t-I_2,B_t-I_2),(H_A,H_B)
    \Big]\,\mathrm dw\,\mathrm dt~.
    \label{eq: feasible class marginal expansion}
\end{align}
Part (iv) of \cref{lem: gaussian threshold covariance bounds}, display \eqref{eq: gaussian canonical off diagonal derivatives}, eliminates the off-diagonal entries of $H_A$ and $H_B$. Hence, the linear term in \eqref{eq: feasible class marginal expansion} can be expressed
\begin{align}
    &\mathrm D_{A,B}\Psi_{\delta,d}
        (I_2,I_2,R_{0,qr})[H_A,H_B]
    \nonumber\\
    &=\sum_{a=1}^2
    (\bar\mu_{i_a,i_a}-\mu_{i_a,i_a})
    \mathrm D_{A,B}\Psi_{\delta,d}
        (I_2,I_2,R_{0,qr})[e_ae_a^\top,0]
    \nonumber\\
    &\quad+
    \sum_{b=1}^2
    (\bar\mu_{k_b,k_b}-\mu_{k_b,k_b})
    \mathrm D_{A,B}\Psi_{\delta,d}
        (I_2,I_2,R_{0,qr})[0,e_be_b^\top]~.
    \label{eq: feasible class marginal linear components}
\end{align}
On the unrestricted set $[n]^4$, retain  $i_1=i$, $i_2=i'$, $k_1=k$, and $k_2=k'$. Define $H_A=(\bar\mu_{i_a,i_b}-\mu_{i_a,i_b})_{a,b=1}^2$ and $H_B=(\bar\mu_{k_a,k_b}-\mu_{k_a,k_b})_{a,b=1}^2$. For each $q,r\in[2]$, put $R_{0,qr}=\mu_{i_q,k_r}e_qe_r^\top$ when $i_q\neq k_r$, and set the associated linear summand to zero when $i_q=k_r$. For every $a\in[2]$, the coordinate $k_{3-r}$ is then free, so the corresponding $A$-direction sum satisfies
\begin{align}
    &\sum_{i_1,i_2,k_1,k_2=1}^n
    \left(\prod_{c=1}^2
        \hat\varepsilon_{i_c}^{(0)}
        \hat\varepsilon_{k_c}^{(0)}
    \right)
    \mathbb I\{i_q\neq k_r\}
    (\bar\mu_{i_a,i_a}-\mu_{i_a,i_a})
    \mathrm D_{A,B}\Psi_{\delta,d}
        (I_2,I_2,\mu_{i_q,k_r}e_qe_r^\top)
        [e_ae_a^\top,0]
    \nonumber\\
    &=\left(
        \sum_{k_{3-r}=1}^n
        \hat\varepsilon_{k_{3-r}}^{(0)}
    \right)
    \sum_{i_1,i_2,k_r=1}^n
    \left(\prod_{c=1}^2\hat\varepsilon_{i_c}^{(0)}\right)
    \hat\varepsilon_{k_r}^{(0)}
    \mathbb I\{i_q\neq k_r\}
    (\bar\mu_{i_a,i_a}-\mu_{i_a,i_a})
    \nonumber\\
    &\quad\cdot
    \mathrm D_{A,B}\Psi_{\delta,d}
        (I_2,I_2,\mu_{i_q,k_r}e_qe_r^\top)
        [e_ae_a^\top,0]
    =0~,
    \label{eq: feasible class unrestricted A direction cancellation}
\end{align}
by $\sum_{j=1}^n\hat\varepsilon_j^{(0)}=0$. Similarly, for every $b\in[2]$, the coordinate $i_{3-q}$ is free, and the corresponding $B$-direction sum is
\begin{align}
    &\sum_{i_1,i_2,k_1,k_2=1}^n
    \left(\prod_{c=1}^2
        \hat\varepsilon_{i_c}^{(0)}
        \hat\varepsilon_{k_c}^{(0)}
    \right)
    \mathbb I\{i_q\neq k_r\}
    (\bar\mu_{k_b,k_b}-\mu_{k_b,k_b})
    \mathrm D_{A,B}\Psi_{\delta,d}
        (I_2,I_2,\mu_{i_q,k_r}e_qe_r^\top)
        [0,e_be_b^\top]
    \nonumber\\
    &=\left(
        \sum_{i_{3-q}=1}^n
        \hat\varepsilon_{i_{3-q}}^{(0)}
    \right)
    \sum_{i_q,k_1,k_2=1}^n
    \hat\varepsilon_{i_q}^{(0)}
    \left(\prod_{c=1}^2\hat\varepsilon_{k_c}^{(0)}\right)
    \mathbb I\{i_q\neq k_r\}
    (\bar\mu_{k_b,k_b}-\mu_{k_b,k_b})
    \nonumber\\
    &\quad\cdot
    \mathrm D_{A,B}\Psi_{\delta,d}
        (I_2,I_2,\mu_{i_q,k_r}e_qe_r^\top)
        [0,e_be_b^\top]
    =0
    \label{eq: feasible class unrestricted B direction cancellation}
\end{align}
by $\sum_{j=1}^n\hat\varepsilon_j^{(0)}=0$. The off-diagonal identities in \eqref{eq: gaussian canonical off diagonal derivatives} eliminate the remaining entries of $H_A$ and $H_B$. Hence, summing \eqref{eq: feasible class unrestricted A direction cancellation} and \eqref{eq: feasible class unrestricted B direction cancellation} over $a,b,q,r\in[2]$ gives
\begin{align}
    &\sum_{i_1,i_2,k_1,k_2=1}^n
    \left(\prod_{c=1}^2
        \hat\varepsilon_{i_c}^{(0)}
        \hat\varepsilon_{k_c}^{(0)}
    \right)
    \sum_{q,r=1}^2
    \mathrm D_{A,B}\Psi_{\delta,d}
        (I_2,I_2,R_{0,qr})[H_A,H_B]
    =0~,
    \label{eq: feasible class complete linear cancellation}
\end{align}
where each positional summand is understood to be zero when $i_q=k_r$. Consequently, the sum restricted to the indices in \eqref{eq: feasible covariance fourth aggregate} equals
\begin{align}
    &\sum_{(i,i',k,k')\in\mathcal Q_n}
    \mathbb I\{(i,i')\notin\mathcal S_n\}
    \mathbb I\{(k,k')\notin\mathcal S_n\}
    \hat\varepsilon_i^{(0)}\hat\varepsilon_{i'}^{(0)}
    \hat\varepsilon_k^{(0)}\hat\varepsilon_{k'}^{(0)}
    \sum_{q,r=1}^2
    \mathrm D_{A,B}\Psi_{\delta,d}
        (I_2,I_2,R_{0,qr})[H_A,H_B]
    \nonumber\\
    &=-\sum_{i,i',k,k'=1}^n
    \mathbb I\Big\{
        |\{i,i',k,k'\}|<4
        \ \mathrm{or}\ (i,i')\in\mathcal S_n
        \ \mathrm{or}\ (k,k')\in\mathcal S_n
    \Big\}
    \nonumber\\
    &\quad\cdot
    \hat\varepsilon_i^{(0)}\hat\varepsilon_{i'}^{(0)}
    \hat\varepsilon_k^{(0)}\hat\varepsilon_{k'}^{(0)}
    \sum_{q,r=1}^2
    \mathrm D_{A,B}\Psi_{\delta,d}
        (I_2,I_2,R_{0,qr})[H_A,H_B]~.
    \label{eq: feasible class restricted linear cancellation}
\end{align}
Fix $q,r\in[2]$ and distinct values $u,v$. The number of terms on the right-hand side of \eqref{eq: feasible class restricted linear cancellation} with $(i_q,k_r)=(u,v)$ is bounded by
\begin{align}
    &\#\Big\{
        (i_1,i_2,k_1,k_2)\in[n]^4:
        (i_q,k_r)=(u,v),
        \ |\{i_1,i_2,k_1,k_2\}|<4
        \nonumber\\
    &\hspace{37mm}
        \mathrm{or}\ (i_1,i_2)\in\mathcal S_n
        \ \mathrm{or}\ (k_1,k_2)\in\mathcal S_n
    \Big\}
    \leq5n+2n(s+1)
    \leq Cn(s+1)~.
    \label{eq: feasible class excluded completion count}
\end{align}
Observe that
\begin{align}
    \sum_{\substack{u,v=1\\u\neq v}}^n\mu_{u,v}^2
    &\leq Cn^2V_n~.
    \label{eq: feasible class direct squared cross sum}
\end{align}
by displays \eqref{eq: n bound 2} and \eqref{eq: n bound 3} in \cref{lem: basic inequalities n}. Thus, Part (iv) of \cref{lem: gaussian threshold covariance bounds}, display \eqref{eq: gaussian canonical marginal derivative bound}, \eqref{eq: feasible class weighted local perturbation moments} for moment power $1$, and the bounds \eqref{eq: feasible class restricted linear cancellation} through \eqref{eq: feasible class direct squared cross sum} imply that
\begin{align}
    &\sup_{\delta\geq0}
    \frac{1}{n^4}
    \Bigg|
    \sum_{(i,i',k,k')\in\mathcal Q_n}
    \mathbb I\{(i,i')\notin\mathcal S_n\}
    \mathbb I\{(k,k')\notin\mathcal S_n\}
    \nonumber\\
    &\quad\quad\quad\quad\quad\cdot
    \mathbb E\Bigg[
        \hat\varepsilon_i^{(0)}\hat\varepsilon_{i'}^{(0)}
        \hat\varepsilon_k^{(0)}\hat\varepsilon_{k'}^{(0)}
        \sum_{q,r=1}^2
        \mathrm D_{A,B}\Psi_{\delta,d}
        (I_2,I_2,R_{0,qr})[H_A,H_B]
    \Bigg]
    \Bigg|
    \nonumber\\
    &\lesssim 
    \frac{\sqrt d\,V_n}{n^4}
    \left(n(s+1)\right)
    \sum_{\substack{u,v=1\\u\neq v}}^n\mu_{u,v}^2
    \lesssim V_n^2
    \frac{(s+1)\sqrt d}{n}~.
    \label{eq: feasible class marginal linear aggregate}
\end{align}
Finally, we bound the double integral in \eqref{eq: feasible class marginal expansion}. Its directions satisfy
\begin{align}
    \|A_t-I_2\|_{\mathrm F}
    +\|B_t-I_2\|_{\mathrm F}
    &\leq C\left(
        V_n+\Delta_T
    \right),
    \nonumber\\
    \|H_A\|_{\mathrm F}+\|H_B\|_{\mathrm F}
    &\leq C\Delta_T~.
    \label{eq: feasible class marginal path distances}
\end{align}
Part (ii) of \cref{lem: gaussian threshold covariance bounds}, \eqref{eq: feasible class fourth marginal probability bound} with $h\in\{1,2\}$, and \eqref{eq: feasible class marginal path distances} bound the absolute value of this integral by
\begin{align}
    &CV_n^\eta
    \left(
        d\log(V_n^{-1})
        +\log^2\left(V_n^{-1}\right)
    \right)
    \nonumber\\
    &\quad\cdot
    \exp\Bigg(
        C\left(
            \sqrt{
                d\log\left(
                    V_n^{-1}
                \right)
            }
            +\log\left(
                V_n^{-1}
            \right)
        \right)\Delta_T
    \Bigg)
    \mu_{i_q,k_r}^2
    \left(V_n+\Delta_T\right)\Delta_T~.
    \label{eq: feasible class marginal remainder pointwise bound}
\end{align}
Multiplying \eqref{eq: feasible class marginal remainder pointwise bound} by the absolute residual product, applying \eqref{eq: feasible class weighted local perturbation moments} for moment powers $1$ and $2$, and using \eqref{eq: feasible class fourth coefficient sums}, we obtain
\begin{align}
    &\sup_{\delta\in\Delta_{n,d}(\eta)}
    \frac{1}{n^4}
    \sum_{(i,i',k,k')\in\mathcal Q_n}
    \mathbb I\{(i,i')\notin\mathcal S_n\}
    \mathbb I\{(k,k')\notin\mathcal S_n\}
    \nonumber\\
    &\quad\cdot
    \mathbb E\Bigg[
        \left|
            \hat\varepsilon_i^{(0)}\hat\varepsilon_{i'}^{(0)}
            \hat\varepsilon_k^{(0)}\hat\varepsilon_{k'}^{(0)}
        \right|
        \sum_{q,r=1}^2
        \Bigg|
            \int_0^1\int_0^1
            \mathrm D_{A,B}^2\Psi_{\delta,d}\Big(
                I_2+w(A_t-I_2),
                I_2+w(B_t-I_2),
                R_{0,qr}
            \Big)
            \nonumber\\
    &\hspace{65mm}\Big[
                (A_t-I_2,B_t-I_2),(H_A,H_B)
            \Big]\,\mathrm dw\,\mathrm dt
        \Bigg|
    \Bigg]
    \nonumber\\
    &\lesssim 
    V_n^{3+\eta}
    \left(
        d\log(V_n^{-1})
        +\log^2\left(V_n^{-1}\right)
    \right)~.
    \label{eq: feasible class marginal remainder aggregate}
\end{align}
Substituting \eqref{eq: feasible class interaction aggregate}, \eqref{eq: feasible class cross entry aggregate}, \eqref{eq: feasible class marginal linear aggregate}, and \eqref{eq: feasible class marginal remainder aggregate} into \eqref{eq: feasible class exact endpoint decomposition} and then into \eqref{eq: feasible covariance fourth aggregate}, we find that
\begin{align}
    \sup_{\delta\in\Delta_{n,d}(\eta)}
    |\mathcal R_{4,n,d}(\delta)|
    &\lesssim
    V_n^{2+\eta}
    +V_n^2
        \frac{(s+1)\sqrt d}{n} 
    +
    V_n^{3+\eta}
    \left(
        d\log(V_n^{-1})
        +\log^2\left(V_n^{-1}\right)
    \right)
    \nonumber\\
    &\lesssim
    V_n^2\Bigg(
        V_n^{\eta}
        +\frac{s\sqrt d}{n}
        +dV_n^{1+\eta}
        \log(V_n^{-1})
    \Bigg)~,
\end{align}
as required, where the final inequality uses \cref{lem: basic inequalities n}, display \eqref{eq: var bounds}, and the boundedness of $x\mapsto x\log^2(x^{-1})$ on $(0,1]$.\hfill\qed

\subsection{Proof of \cref{lem: fixed bank selection matrix comparisons}}

We begin by introducing some notation, and giving some bounds, that will be used throughout the course of the proof. Let $J$ and $K$ denote any of the sets $\{0\}$, $\{j\}$, $\{k\}$, $\{0,j\}$, $\{j,k\}$, or $B$. Let $h_J=d+1-|J|$, let $E_J$ select the coordinates in $J$, and let $J^c$ denote the complement of $J$. Define the matrices
\begin{align}
    \Omega_d^{(J)}
    &=\Gamma_{J^c,J^c}
      -\Gamma_{J^c,J}\Gamma_{J,J}^{-1}\Gamma_{J,J^c}
    ~,\quad
    \Pi_J
    =\Gamma_d^{1/2}E_J\Gamma_{J,J}^{-1}
      E_J^\top\Gamma_d^{1/2}~,\quad\text{and}
    \nonumber\\
    V_J
    &=\frac{1}{d+1}
      (I_{d+1}-\Pi_J)\Gamma_d(I_{d+1}-\Pi_J)~.
    \label{eq: fixed bank anchor matrices}
\end{align}
The Schur-complement formula, \cref{assu: well conditioned}, and
\cref{assu: homoskedasticity} give
\begin{align}
    cI_{h_J}
    &\preceq\Omega_d^{(J)}
    \preceq CI_{h_J}~,
    &
    \|\Omega_d^{(J)}\|_{\mathrm F}^2
    &=\|\bar\Gamma_d\|_{\mathrm F}^2+O(1)\asymp d~,
    \nonumber\\
    \operatorname{tr}(\Omega_d^{(J)})
    &=h_J+O(1)~,
    &
    \frac{\|\Omega_d^{(J)}\|_{\mathrm F}}{h_J}
    &=\sqrt{v_0^{(d)}}\left(1+O(d^{-1})\right)~,
    \label{eq: fixed bank anchor covariance geometry}
\end{align}
Moreover, it holds that
\begin{align}
    \operatorname{rank}(V_J-V_K)
    &\leq2(|J|+|K|)~,
    &
    \|V_J-V_K\|_{\mathrm{op}}
      +\|V_J-V_K\|_{\mathrm F}
      +|\operatorname{tr}(V_J-V_K)|
    &\lesssim\frac{1}{d}~,
    \nonumber\\
    \|V_J\|_{\mathrm{op}}
    &\lesssim\frac{1}{d}~,\quad\text{and}
    &\|V_J\|_{\mathrm F}^2
    &\asymp\frac{1}{d}~.
    \label{eq: fixed bank anchor matrix geometry}
\end{align}
Finally, define
\begin{align}
    Z_i
    &=\Gamma_d^{-1/2}
      \left(Y_i-\mathbb E[Y_i]\right)^\top~,
    \qquad i\in[n]~.
\end{align}
By \cref{assu: gaussianity,assu: proportionality}, the vectors $Z_1,\ldots,Z_n$ are jointly centered Gaussian and satisfy
\begin{align}
    \Cov(Z_i,Z_{i'})
    &=\lambda_{i,i'}I_{d+1}~.
    \label{eq: fixed bank common gaussian rows}
\end{align}
The $k$th component of the vector $\Gamma_d^{1/2}(I_{d+1}-\Pi_J)Z_i$ gives the population residual from regressing $Y_i^{(k)}$ on an intercept and the outcomes indexed by $J$. Consequently, we can write
\begin{align}
    (\check A_\ell^{(J)})_{i,i'}
    &=\mathbb I\left\{
      |Z_i^\top V_JZ_{i'}|
      \geq\frac{h_J}{d+1}
      \underline\delta_{n,d}(\eta_\ell)
    \right\}~.
    \label{eq: fixed bank common gaussian representation}
\end{align}

\subsubsection{Part (i)}

If the unordered pairs $\{i,i'\}$ and $\{h,h'\}$ share an index, the
right-hand side of \eqref{eq: fixed bank selection covariance comparison}
contains $\lambda_{u,u}^2=1$ for that index. The conclusion then follows
from the fact that the covariance of two indicator variables is bounded by one. Suppose, then, that the pairs $\{i,i'\}$ and $\{h,h'\}$ are disjoint. First suppose that $i\neq i'$ and $h\neq h'$. Consider the random variables
\begin{align}
    U
    &=
      \left(
        \begin{pmatrix}
          1&\lambda_{i,i'}\\
          \lambda_{i,i'}&1
        \end{pmatrix}^{-1/2}
        \begin{pmatrix}
          Z_{i,q}\\
          Z_{i',q}
        \end{pmatrix}
      \right)_{q=1}^{d+1}\quad\text{and}\quad
    W
    =
      \left(
        \begin{pmatrix}
          1&\lambda_{h,h'}\\
          \lambda_{h,h'}&1
        \end{pmatrix}^{-1/2}
        \begin{pmatrix}
          Z_{h,q}\\
          Z_{h',q}
        \end{pmatrix}
      \right)_{q=1}^{d+1}~.
\end{align}
It follows from \eqref{eq: fixed bank common gaussian rows} that $\Var(U)
=I_{2(d+1)}$, $\Var(W)=I_{2(d+1)}$, and
\begin{align}
    \Cov(U,W)
    &=
      I_{d+1}\otimes\left[
        \begin{pmatrix}
          1&\lambda_{i,i'}\\
          \lambda_{i,i'}&1
        \end{pmatrix}^{-1/2}
        (\lambda_{u,v})_{
          u\in\{i,i'\},\ v\in\{h,h'\}}
        \begin{pmatrix}
          1&\lambda_{h,h'}\\
          \lambda_{h,h'}&1
        \end{pmatrix}^{-1/2}
      \right]~.
    \label{eq: fixed bank standardized cross covariance}
\end{align}
By the representation \eqref{eq: fixed bank common gaussian representation}, we can write the indicators $(\check A_\ell^{(j)})_{i,i'}$ and $(\check A_\ell^{(j)})_{h,h'}$ as functions $f(U)$ and $g(W)$, respectively. Observe that, replacing $U$ by $-U$ replaces both $Z_i$ and $Z_{i'}$ by their negatives and leaves $Z_i^\top V_{\{j\}}Z_{i'}$ unchanged. Hence, $f$ and $g$ are even. Subtracting their expectations preserves this property.

Now, \cref{assu: well conditioned}, applied to $\Lambda_n$, uniformly bounds the
operator norms of the two inverse square roots in \eqref{eq: fixed bank standardized cross covariance}. Consequently, it holds that
\begin{align}
    \|\Cov(U,W)\|_{\mathrm{op}}^2
    &\lesssim
      \left\|
        (\lambda_{u,v})_{
          u\in\{i,i'\},\ v\in\{h,h'\}}
      \right\|_{\mathrm{op}}^2
    \nonumber\\
    &\leq
      \left\|
        (\lambda_{u,v})_{
          u\in\{i,i'\},\ v\in\{h,h'\}}
      \right\|_{\mathrm F}^2
    =
      \sum_{u\in\{i,i'\}}\sum_{v\in\{h,h'\}}
      \lambda_{u,v}^2~.
\end{align}
Therefore, \cref{lem: even gaussian covariance contraction} and the
fact that the variance of an indicator is at most one give
\begin{align}
    \left|
      \Cov\left(
        (\check A_\ell^{(j)})_{i,i'},
        (\check A_\ell^{(j)})_{h,h'}
      \right)
    \right|
    &\leq
      \|\Cov(U,W)\|_{\mathrm{op}}^2
      \sqrt{
        \Var\left((\check A_\ell^{(j)})_{i,i'}\right)
        \Var\left((\check A_\ell^{(j)})_{h,h'}\right)
      }
    \nonumber\\
    &\lesssim
      \sum_{u\in\{i,i'\}}\sum_{v\in\{h,h'\}}
      \lambda_{u,v}^2~,
\end{align}
as required. If $i=i'$, take $U=Z_i$ in the argument above. If $h=h'$, then take $W=Z_h$. The same calculation then applies with the corresponding one-row or one-column cross-covariance matrix and gives the same bound. \hfill\qed

\subsubsection{Part (ii)}

We first prove \eqref{eq: fixed bank single screen crossing comparison}.
Fix $i\neq i'$ and suppose that $Z_i$ and $Z_{i'}$ are independent. Write
\begin{align}
    G
    &=
      (Z_{i,0},Z_{i',0},\ldots,Z_{i,d},Z_{i',d})^\top
    \quad\text{and}\quad
    S
    =
      \frac12(e_1e_2^\top+e_2e_1^\top)~.
\end{align}
Under $P_{I_2}$, the vector $G$ is standard Gaussian and satisfies
\begin{align}
    G^\top(V_L\otimes S)G
    &=Z_i^\top V_LZ_{i'}~,
    \qquad L\in\{0,J\}~.
\end{align}
Define the matrices
\begin{align}
    A_0
    &=
      \frac{V_0\otimes S}
      {\sqrt2\|V_0\otimes S\|_{\mathrm F}}
    \quad\text{and}\quad
    A_J
    =
      \frac{(d/h_J)V_J\otimes S}
      {\sqrt2\|V_0\otimes S\|_{\mathrm F}}~.
    \label{eq: fixed bank one pair matrices}
\end{align}
Observe that we can write
\begin{align}
    \sqrt2\|V_0\otimes S\|_{\mathrm F}
    &=\|V_0\|_{\mathrm F}
    =\frac{\|\Omega_d\|_{\mathrm F}}{d+1}~.
\end{align}
Since $h_0=d$, the representation \eqref{eq: fixed bank common gaussian representation} implies that we can write
\begin{align}
    (\check A_\ell^{(0)})_{i,i'}
    &=
      \mathbb I\left\{
        |G^\top A_0G|
        \geq
        \frac{
          d\,\underline\delta_{n,d}(\eta_\ell)
        }{\|\Omega_d\|_{\mathrm F}}
      \right\}
    \quad\text{and}\quad
    (\check A_\ell^{(J)})_{i,i'}
    &=
      \mathbb I\left\{
        |G^\top A_JG|
        \geq
        \frac{
          d\,\underline\delta_{n,d}(\eta_\ell)
        }{\|\Omega_d\|_{\mathrm F}}
      \right\}~.
    \label{eq: fixed bank one pair quadratic representation}
\end{align}
The difference between the matrices in \eqref{eq: fixed bank one pair matrices} can be expressed
\begin{align}
    A_J-A_0
    &=
      \left(\frac{d}{h_J}-1\right)A_0
      +
      \frac{d}{h_J}
      \frac{(V_J-V_0)\otimes S}
      {\sqrt2\|V_0\otimes S\|_{\mathrm F}}~.
    \label{eq: fixed bank one pair perturbation}
\end{align}
The bounds \eqref{eq: fixed bank anchor covariance geometry} and \eqref{eq: fixed bank anchor matrix geometry} imply that
\begin{align}
    2\|A_0\|_{\mathrm F}^2
    &=1~,
    &
    \|A_0\|_{\mathrm{op}}
    &\lesssim d^{-1/2}~,
    &
    \|A_J-A_0\|_{\mathrm F}
      +\|A_J-A_0\|_{\mathrm{op}}
    &\lesssim d^{-1/2}~,
    \nonumber\\
    \operatorname{tr}(A_0)
    &=\operatorname{tr}(A_J-A_0)=0~,
    &
    \left|\frac{d}{h_J}-1\right|
    &\lesssim d^{-1}~,
    &
    \operatorname{rank}\left((V_J-V_0)\otimes S\right)
    &\lesssim1~.
    \label{eq: fixed bank one pair matrix bounds}
\end{align}
Let $\alpha_1(u),\ldots,\alpha_{2(d+1)}(u)$ denote the eigenvalues of
$A_0+u(A_J-A_0)$. The eigenvalue bounds for $V_0$, the two nonzero
eigenvalues of $S$, \eqref{eq: fixed bank one pair perturbation}, and
rank interlacing give, uniformly over $u\in[0,1]$, that 
\begin{align}
    \|A_0+u(A_J-A_0)\|_{\mathrm F}^2
    &\asymp1~,
    &
    \|A_0+u(A_J-A_0)\|_{\mathrm{op}}
    &\lesssim d^{-1/2}~,\quad\text{and}
    \nonumber \\
    \left|
      \left\{
        r:
        |\alpha_r(u)|\geq\frac{c}{\sqrt d}
      \right\}
    \right|
    &\geq cd~.
    \label{eq: fixed bank one pair path geometry}
\end{align}
Now, Part (ii) of \cref{lem: selected set characterization}, \eqref{eq: auxiliary outcome weight and scale bounds}, and \eqref{eq: regret rate condition} give
\begin{align}
    \left(
      \frac{
        d\,\underline\delta_{n,d}(\eta_\ell)
      }{\|\Omega_d\|_{\mathrm F}}
    \right)^2
    &=
      2\eta_\ell\log(V_n^{-1})(1+o(1))~,
    &
    \frac{
      d\,\underline\delta_{n,d}(\eta_\ell)
    }{\|\Omega_d\|_{\mathrm F}}
    &=o(\sqrt d)~,\quad\text{and}
    \nonumber\\
    \left(
      1+
      \frac{
        d\,\underline\delta_{n,d}(\eta_\ell)
      }{\|\Omega_d\|_{\mathrm F}}
    \right)^2
    \|A_J-A_0\|_{\mathrm F}
    &\lesssim
      \frac{\log(V_n^{-1})}{\sqrt d}
    =o(1)~.
    \label{eq: fixed bank one pair crossing conditions}
\end{align}
Thus, by applying Part (i) of \cref{lem: high rank relative crossing} to the two inequalities defining the absolute values in \eqref{eq: fixed bank one pair quadratic representation}, we find that 
\begin{align}
    &P_{I_2}\left\{
      (\check A_\ell^{(J)})_{i,i'}
      \neq
      (\check A_\ell^{(0)})_{i,i'}
    \right\}
    \nonumber\\
    &\leq
      \sum_{a\in\{-1,1\}}
      P\Bigg\{
        \mathbb I\left\{
          aG^\top A_JG
          \geq
          \frac{
            d\,\underline\delta_{n,d}(\eta_\ell)
          }{\|\Omega_d\|_{\mathrm F}}
        \right\}
        \neq
        \mathbb I\left\{
          aG^\top A_0G
          \geq
          \frac{
            d\,\underline\delta_{n,d}(\eta_\ell)
          }{\|\Omega_d\|_{\mathrm F}}
        \right\}
      \Bigg\}
    \nonumber\\
    &\lesssim
      \frac{\log(V_n^{-1})}{\sqrt d}
      \sum_{a\in\{-1,1\}}
      P\left\{
        aG^\top A_0G
        \geq
        \frac{
          d\,\underline\delta_{n,d}(\eta_\ell)
        }{\|\Omega_d\|_{\mathrm F}}
      \right\}
    =
      \pi_{n,d}(\ell)
      \frac{\log(V_n^{-1})}{\sqrt d}~.
    \label{eq: fixed bank independent anchor crossing}
\end{align}
Now, with this in place, for two conditioning sets $J$ and $K$, the inequality
\begin{align}
    \mathbb I\left\{
      (\check A_\ell^{(J)})_{i,i'}
      \neq
      (\check A_\ell^{(K)})_{i,i'}
    \right\}
    &\leq
      \mathbb I\left\{
        (\check A_\ell^{(J)})_{i,i'}
        \neq
        (\check A_\ell^{(0)})_{i,i'}
      \right\}
    +
      \mathbb I\left\{
        (\check A_\ell^{(K)})_{i,i'}
        \neq
        (\check A_\ell^{(0)})_{i,i'}
      \right\}
\end{align}
and the bound \eqref{eq: fixed bank independent anchor crossing} imply that
\begin{align}
    P_{I_2}\left\{
      (\check A_\ell^{(J)})_{i,i'}
      \neq
      (\check A_\ell^{(K)})_{i,i'}
    \right\}
    &\lesssim
      \pi_{n,d}(\ell)
      \frac{\log(V_n^{-1})}{\sqrt d}~.
\end{align}
Replacing either $Z_i$ or $Z_{i'}$ by its negative leaves both indicators in the preceding display unchanged. Moreover, it holds that
\begin{align}
    \left\|
      \begin{pmatrix}
        1&\lambda_{i,i'}\\
        \lambda_{i,i'}&1
      \end{pmatrix}
      -I_2
    \right\|_{\mathrm F}^2
    &=2\lambda_{i,i'}^2~.
\end{align}
Thus, \cref{lem: fixed bank sign comparison}, applied to
$Z_i$ and $Z_{i'}$ with $m=d+1$, gives
\begin{align}
    P\left\{
      (\check A_\ell^{(J)})_{i,i'}
      \neq
      (\check A_\ell^{(K)})_{i,i'}
    \right\}
    &\lesssim
      \pi_{n,d}(\ell)
      \frac{\log(V_n^{-1})}{\sqrt d}
      +\min\{1,d\lambda_{i,i'}^2\}~,
    \label{eq: fixed bank general anchor crossing}
\end{align}
For $i=i'$, this bound follows from $\min\{1,d\lambda_{i,i}^2\}=1$, which verifies \eqref{eq: fixed bank single screen crossing comparison}.

Next, we verify \eqref{eq: fixed bank two screen crossing comparison}. Fix two distinct unordered pairs $\{i,i'\}$ and $\{h,h'\}$, and let $T$ contain the distinct indices among $i,i',h,h'$. Define symmetric matrices $S_1$ and $S_2$, indexed by $T$, by
\begin{align}
    z^\top S_1z
    &=z_iz_{i'}~,
    \quad
    z^\top S_2z
    &=z_hz_{h'}~,
    \qquad z=(z_u)_{u\in T}\in\mathbb R^{|T|}~.
    \label{eq: fixed bank two pair spatial matrices}
\end{align}
Let $W_0,\ldots,W_d$ be independent standard Gaussian vectors in
$\mathbb R^{|T|}$. By \eqref{eq: fixed bank common gaussian rows}, we
may write
\begin{align}
    (Z_{u,r})_{u\in T}
    &=\Lambda_{T,T}^{1/2}W_r~,
    \qquad r\in\{0,1,\ldots,d\}~.
\end{align}
For $W=(W_0^\top,\ldots,W_d^\top)^\top$, it follows that
\begin{align}
    Z_i^\top V_LZ_{i'}
    &=
      W^\top\left(
        V_L\otimes
        \Lambda_{T,T}^{1/2}S_1\Lambda_{T,T}^{1/2}
      \right)W~,
    \nonumber\\
    Z_h^\top V_LZ_{h'}
    &=
      W^\top\left(
        V_L\otimes
        \Lambda_{T,T}^{1/2}S_2\Lambda_{T,T}^{1/2}
      \right)W~,
    \qquad L\in\{J,K\}~.
    \label{eq: fixed bank two pair quadratic coupling}
\end{align}
For $q\in\{1,2\}$, define
\begin{align}
    s_q^2
    &=
      2\left\|
        V_K\otimes
        \Lambda_{T,T}^{1/2}S_q\Lambda_{T,T}^{1/2}
      \right\|_{\mathrm F}^2~,
    &
    A_q
    &=
      \frac{
        V_K\otimes
        \Lambda_{T,T}^{1/2}S_q\Lambda_{T,T}^{1/2}
      }{s_q}~,\quad\text{and}
    \nonumber\\
    B_q
    &=
      \frac{
        (V_J-V_K)\otimes
        \Lambda_{T,T}^{1/2}S_q\Lambda_{T,T}^{1/2}
      }{s_q}~.
    \label{eq: fixed bank two pair matrices}
\end{align}
Since the two unordered pairs are distinct, $S_1$ and $S_2$ are
linearly independent. Part (ii) of
\cref{assu: well conditioned} and
\eqref{eq: fixed bank anchor covariance geometry} therefore imply that
\begin{align}
    s_q
    &\asymp d^{-1/2}~,
    \qquad q\in\{1,2\}~,
    &
    \|t_1A_1+t_2A_2\|_{\mathrm F}^2
    &\asymp\|t\|_2^2~,
    &
    \|t_1A_1+t_2A_2\|_{\mathrm{op}}
    &\lesssim
      \frac{\|t\|_2}{\sqrt d}~
    \label{eq: fixed bank two pair pencil}
\end{align}
for every $t=(t_1,t_2)\in\mathbb R^2$. In turn, the bounds \eqref{eq: fixed bank anchor matrix geometry} imply that
\begin{align}
    \operatorname{rank}(B_1)+\operatorname{rank}(B_2)
    &\lesssim1
    \quad\text{and}\quad
    \sum_{q=1}^2
      \left(
        \|B_q\|_{\mathrm F}
        +\|B_q\|_{\mathrm{op}}
        +|\operatorname{tr}(B_q)|
      \right)
    \lesssim d^{-1/2}
    \label{eq: fixed bank two pair perturbations}
\end{align}
respectively.

Now, let $\mathcal P$ be the orthogonal projection onto $\operatorname{range}(B_1)+\operatorname{range}(B_2)$. It holds that
\begin{align}
    \operatorname{rank}(\mathcal P)
    &\lesssim1
    \quad\text{and}\quad
    B_q
    =\mathcal P B_q\mathcal P~,
    \qquad q\in\{1,2\}~.
    \label{eq: fixed bank perturbation projection}
\end{align}
We now bound the conditional density of
\begin{align}
    \left(
      W^\top A_1W,
      W^\top A_2W
    \right)
\end{align}
given $\mathcal PW$. For $t=(t_1,t_2)\in\mathbb R^2$, define
\begin{align}
    A(t)
    &=t_1A_1+t_2A_2~,
    \quad\text{and}\quad
    \widetilde A(t)
    =
      (I-\mathcal P)A(t)(I-\mathcal P)~.
\end{align}
The identity
\begin{align}
    A(t)-\widetilde A(t)
    &=
      \mathcal PA(t)+A(t)\mathcal P
      -\mathcal PA(t)\mathcal P
\end{align}
and \eqref{eq: fixed bank two pair pencil} imply that
\begin{align}
    \operatorname{rank}\left(A(t)-\widetilde A(t)\right)
    &\lesssim1
    \quad\text{and}\quad
    \|A(t)-\widetilde A(t)\|_{\mathrm F}
    \lesssim\frac{\|t\|_2}{\sqrt d}~.
\end{align}
Consequently, it holds that
\begin{align}
    \|\widetilde A(t)\|_{\mathrm F}^2
    &\asymp\|t\|_2^2
    \quad\text{and}\quad
    \|\widetilde A(t)\|_{\mathrm{op}}
    \lesssim\frac{\|t\|_2}{\sqrt d}~.
    \label{eq: fixed bank compressed pencil norms}
\end{align}
Since $W$ has dimension at most $4(d+1)$,
\eqref{eq: fixed bank compressed pencil norms} implies that at least $cd$
eigenvalues of $\widetilde A(t)$ have absolute value at least
$c\|t\|_2/\sqrt d$. Define
\begin{align}
    \varphi_{\mathcal PW}(t)
    &=
      \mathbb E\left[
        \exp\left(
          \mathrm i
          \sum_{q=1}^2t_qW^\top A_qW
        \right)
        \mathrel{\Big|}\mathcal PW
      \right]~.
\end{align}
Conditional on $\mathcal PW$, the vector $(I-\mathcal P)W$ is standard Gaussian on the range of $I-\mathcal P$ and is independent of $\mathcal PW$. The Gaussian quadratic-form transform in \citealp{muirhead1982}, Proof of Theorem~1.4.2, together with completion of the square, gives
\begin{align}
    |\varphi_{\mathcal PW}(t)|
    &\leq
      \det\left(
        I+4\widetilde A(t)^2
      \right)^{-1/4}
    \leq
      \left(
        1+c\frac{\|t\|_2^2}{d}
      \right)^{-cd}~.
    \label{eq: fixed bank conditional characteristic function}
\end{align}
The right-hand side has integral bounded by a constant over $\mathbb R^2$.The inversion theorem for integrable characteristic functions
(\citealp{feller1991introduction}, Chapter~XV.3) therefore gives a
conditional density $f_{\mathcal PW}$ satisfying
\begin{align}
    \sup_{z\in\mathbb R^2}
      f_{\mathcal PW}(z)
    &\leq
      \frac{1}{(2\pi)^2}
      \int_{\mathbb R^2}
      |\varphi_{\mathcal PW}(t)|\,\mathrm dt
    \lesssim1
    \label{eq: fixed bank conditional quadratic density}
\end{align}
almost surely.

With this in place, recall that we can write
\begin{align}
    (\check A_\ell^{(K)})_{i,i'}
    &=
      \mathbb I\left\{
        |W^\top A_1W|
        \geq
        \frac{
          h_K\underline\delta_{n,d}(\eta_\ell)
        }{(d+1)s_1}
      \right\}~,
    \nonumber\\
    (\check A_\ell^{(J)})_{i,i'}
    &=
      \mathbb I\left\{
        |W^\top(A_1+B_1)W|
        \geq
        \frac{
          h_J\underline\delta_{n,d}(\eta_\ell)
        }{(d+1)s_1}
      \right\}~,
    \nonumber\\
    (\check A_\ell^{(K)})_{h,h'}
    &=
      \mathbb I\left\{
        |W^\top A_2W|
        \geq
        \frac{
          h_K\underline\delta_{n,d}(\eta_\ell)
        }{(d+1)s_2}
      \right\}~,\quad\text{and}
    \nonumber\\
    (\check A_\ell^{(J)})_{h,h'}
    &=
      \mathbb I\left\{
        |W^\top(A_2+B_2)W|
        \geq
        \frac{
          h_J\underline\delta_{n,d}(\eta_\ell)
        }{(d+1)s_2}
      \right\}~.
    \label{eq: fixed bank two pair indicator representation}
\end{align}
The reverse triangle inequality applied to the first two identities in
\eqref{eq: fixed bank two pair indicator representation} gives
\begin{align}
    \left\{
      (\check A_\ell^{(J)}-\check A_\ell^{(K)})_{i,i'}
      \neq0
    \right\}
    &\subseteq
      \bigcup_{a\in\{-1,1\}}
      \Bigg\{
        \left|
          W^\top A_1W
          -
          \frac{
            ah_K\underline\delta_{n,d}(\eta_\ell)
          }{(d+1)s_1}
        \right|
    \nonumber\\
    &\hspace{38mm}\leq
        |W^\top B_1W|
        +
        \frac{
          |h_J-h_K|\underline\delta_{n,d}(\eta_\ell)
        }{(d+1)s_1}
      \Bigg\}~.
    \label{eq: fixed bank first pair boundary}
\end{align}
The final two identities give
\begin{align}
    \left\{
      (\check A_\ell^{(J)}-\check A_\ell^{(K)})_{h,h'}
      \neq0
    \right\}
    &\subseteq
      \bigcup_{a\in\{-1,1\}}
      \Bigg\{
        \left|
          W^\top A_2W
          -
          \frac{
            ah_K\underline\delta_{n,d}(\eta_\ell)
          }{(d+1)s_2}
        \right|
    \nonumber\\
    &\hspace{38mm}\leq
        |W^\top B_2W|
        +
        \frac{
          |h_J-h_K|\underline\delta_{n,d}(\eta_\ell)
        }{(d+1)s_2}
      \Bigg\}~.
    \label{eq: fixed bank second pair boundary}
\end{align}
By \eqref{eq: fixed bank perturbation projection}, we can write
\begin{align}
    W^\top B_qW
    &=
      (\mathcal PW)^\top B_q(\mathcal PW)~,
    \qquad q\in\{1,2\}~.
\end{align}
Thus, the two quantities on the right-hand sides of
\eqref{eq: fixed bank first pair boundary} and
\eqref{eq: fixed bank second pair boundary} are measurable with respect
to $\mathcal PW$. Moreover, \cref{lem: wick} and
\eqref{eq: fixed bank two pair perturbations} give
\begin{align}
    \mathbb E\left[(W^\top B_qW)^2\right]
    &=2\|B_q\|_{\mathrm F}^2
      +\operatorname{tr}(B_q)^2
    \lesssim\frac{1}{d}~,
    \qquad q\in\{1,2\}~.
\end{align}
Part (ii) of \cref{lem: selected set characterization} and
\eqref{eq: auxiliary outcome weight and scale bounds} also give
\begin{align}
    \underline\delta_{n,d}(\eta_\ell)^2
    &\lesssim\frac{\log(V_n^{-1})}{d}~,\quad\text{and}
    \nonumber\\
    \left(
      \frac{
        |h_J-h_K|\underline\delta_{n,d}(\eta_\ell)
      }{(d+1)s_q}
    \right)^2
    &\lesssim
      \frac{\log(V_n^{-1})}{d^2}
    =o(d^{-1})~,
    \qquad q\in\{1,2\}~.
\end{align}
Consequently,
\begin{align}
    \mathbb E\left[
      \left(
        |W^\top B_qW|
        +
        \frac{
          |h_J-h_K|\underline\delta_{n,d}(\eta_\ell)
        }{(d+1)s_q}
      \right)^2
    \right]
    &\lesssim\frac{1}{d}~,
    \qquad q\in\{1,2\}~.
    \label{eq: fixed bank two pair width}
\end{align}
By combining \eqref{eq: fixed bank first pair boundary} and \eqref{eq: fixed bank second pair boundary} with the conditional density bound \eqref{eq: fixed bank conditional quadratic density}, we obtain
\begin{align}
    &P\left\{
      (\check A_\ell^{(J)}-\check A_\ell^{(K)})_{i,i'}\neq0,
      (\check A_\ell^{(J)}-\check A_\ell^{(K)})_{h,h'}\neq0
      \mathrel{\Big|}\mathcal PW
    \right\}
    \nonumber\\
    &\lesssim
      \prod_{q=1}^2
      \left(
        |W^\top B_qW|
        +
        \frac{
          |h_J-h_K|\underline\delta_{n,d}(\eta_\ell)
        }{(d+1)s_q}
      \right)~.
\end{align}
Taking expectations and applying Cauchy-Schwarz and \eqref{eq: fixed bank two pair width}, we then find that
\begin{align}
    &P\left\{
      (\check A_\ell^{(J)}-\check A_\ell^{(K)})_{i,i'}\neq0,
      (\check A_\ell^{(J)}-\check A_\ell^{(K)})_{h,h'}\neq0
    \right\}
    \nonumber\\
    &\lesssim
      \prod_{q=1}^2
      \mathbb E\left[
        \left(
          |W^\top B_qW|
          +
          \frac{
            |h_J-h_K|\underline\delta_{n,d}(\eta_\ell)
          }{(d+1)s_q}
        \right)^2
      \right]^{1/2}
    \lesssim\frac{1}{d}~.
    \label{eq: fixed bank simultaneous anchor crossing}
\end{align}
Finally, each entry of
$\check A_\ell^{(J)}-\check A_\ell^{(K)}$ belongs to
$\{-1,0,1\}$. Hence, we can conclude that
\begin{align}
    &\mathbb E\left[
      \left|
        (\check A_\ell^{(J)}-\check A_\ell^{(K)})_{i,i'}
        (\check A_\ell^{(J)}-\check A_\ell^{(K)})_{h,h'}
      \right|
    \right]
    \nonumber\\
    &=
      P\left\{
        (\check A_\ell^{(J)}-\check A_\ell^{(K)})_{i,i'}\neq0,
        (\check A_\ell^{(J)}-\check A_\ell^{(K)})_{h,h'}\neq0
      \right\}
    \lesssim\frac{1}{d}~.
\end{align}
as required.
\hfill\qed

\subsubsection{Part (iii)}

For $i\neq i'$, the representation \eqref{eq: fixed bank common gaussian representation} gives
\begin{align}
    P_{I_2}\left\{
      (\check A_\ell^{(0)})_{i,i'}=1
    \right\}
    &=
      P_{I_2}\left\{
        |Z_i^\top V_0Z_{i'}|
        \geq\frac{h_0}{d+1}
        \underline\delta_{n,d}(\eta_\ell)
      \right\}
    =\pi_{n,d}(\ell)~.
    \label{eq: fixed bank common gaussian representation iii}
\end{align}
The event in \eqref{eq: fixed bank common gaussian representation iii}
is unchanged if either $Z_i$ or $Z_{i'}$ is multiplied by $-1$.
Therefore, \cref{lem: fixed bank sign comparison}, applied with $r=2$,
$m=d+1$, and correlation matrix
\begin{align}
    \begin{pmatrix}
      1&\lambda_{i,i'}\\
      \lambda_{i,i'}&1
    \end{pmatrix}~,
\end{align}
gives
\begin{align}
    \left|
      P\left\{(\check A_\ell^{(0)})_{i,i'}=1\right\}
      -\pi_{n,d}(\ell)
    \right|
    &\lesssim
      \min\{1,(d+1)\lambda_{i,i'}^2\}
    \lesssim
      \min\{1,d\lambda_{i,i'}^2\}~,
\end{align}
as required.

We next consider two pairs sharing one index. Under $P_{I_3}$, let $Z_1,Z_2,Z_3$ denote the three independent standard Gaussian vectors in \eqref{eq: fixed bank common gaussian representation}. Choose an orthogonal matrix $Q_d$ satisfying
\begin{align}
    Q_d^\top V_0Q_d
    &=
      \frac{1}{d+1}
      \operatorname{diag}\left(
        \theta_{d,1},\ldots,\theta_{d,d},0
      \right)~.
\end{align}
For $a\in[3]$, let $Z_{a,q}$ denote the $q$th coordinate of $Q_d^\top Z_a$. These variables are independent and standard Gaussian. Since $h_0=d$, if both indicators equal one, there are $a_1,a_2\in\{-1,1\}$ for which the corresponding signed sum is at least twice the threshold. Thus, it holds that
\begin{align}
    &P_{I_3}\left\{
      (\check A_\ell^{(0)})_{1,2}=1,
      (\check A_\ell^{(0)})_{1,3}=1
    \right\}
    \nonumber\\
    &\leq
      \sum_{a_1,a_2\in\{-1,1\}}
      P\left\{
        \sum_{q=1}^d
        \frac{\theta_{d,q}}{\|\Omega_d\|_{\mathrm F}}
        Z_{1,q}
        \frac{a_1Z_{2,q}+a_2Z_{3,q}}{\sqrt 2}
        \geq
        \frac{
          \sqrt 2d\,\underline\delta_{n,d}(\eta_\ell)
        }{\|\Omega_d\|_{\mathrm F}}
      \right\}~.
    \label{eq: fixed bank independent shared index probability}
\end{align}
For every $a_1,a_2\in\{-1,1\}$, the two Gaussian factors in each summand of \eqref{eq: fixed bank independent shared index probability} are independent and standard, and the pairs are independent across $q$. Hence, \cref{lem: weighted gaussian product moderate deviation} gives, uniformly over $a_1,a_2\in\{-1,1\}$ and $\ell\in[g]$, that
\begin{align}
    \log\pi_{n,d}(\ell)
    &=
      \left(-\eta_\ell+o(1)\right)
      \log(V_n^{-1})~,
    \nonumber\\
    \log P\left\{
      \sum_{q=1}^d
      \frac{\theta_{d,q}}{\|\Omega_d\|_{\mathrm F}}
      Z_{1,q}
      \frac{a_1Z_{2,q}+a_2Z_{3,q}}{\sqrt 2}
      \geq
      \frac{
        \sqrt 2d\,\underline\delta_{n,d}(\eta_\ell)
      }{\|\Omega_d\|_{\mathrm F}}
    \right\}
    &=
      \left(-2\eta_\ell+o(1)\right)
      \log(V_n^{-1})~.
    \label{eq: fixed bank shared index moderate deviations}
\end{align}
Hence, we can evaluate
\begin{align}
    &\log\left( \pi_{n,d}(\ell)^{-3/2}
        P\left\{
          \displaystyle
          \sum_{q=1}^d
          \frac{\theta_{d,q}}{\|\Omega_d\|_{\mathrm F}}
          Z_{1,q}
          \frac{a_1Z_{2,q}+a_2Z_{3,q}}{\sqrt 2}
          \geq
          \displaystyle
          \frac{
            \sqrt 2d\,\underline\delta_{n,d}(\eta_\ell)
          }{\|\Omega_d\|_{\mathrm F}}
        \right\}
    \right) \nonumber \\
    &=
      \left(
        -2\eta_\ell+\frac{3}{2}\eta_\ell+o(1)
      \right)
      \log(V_n^{-1})
    =
      \left(-\frac{\eta_\ell}{2}+o(1)\right)
      \log(V_n^{-1})
    \to-\infty~.
    \label{eq: fixed bank shared index exponent comparison}
\end{align}
Therefore, since $g$ is fixed and $\eta_1>0$, \eqref{eq: fixed bank independent shared index probability} implies that
\begin{align}
    P_{I_3}\left\{
      (\check A_\ell^{(0)})_{1,2}=1,
      (\check A_\ell^{(0)})_{1,3}=1
    \right\}
    &\lesssim
      \pi_{n,d}(\ell)^{3/2}~.
    \label{eq: fixed bank independent shared index bound}
\end{align}
In turn, for distinct $i,i',h$, replacing any one of $Z_i$, $Z_{i'}$, or $Z_h$ by its negative leaves both $(\check A_\ell^{(0)})_{i,i'}$ and $(\check A_\ell^{(0)})_{i,h}$ unchanged. Moreover,
\begin{align}
    \left\|
      (\lambda_{u,v})_{u,v\in\{i,i',h\}}-I_3
    \right\|_{\mathrm F}^2
    &=
      2\left(
        \lambda_{i,i'}^2+\lambda_{i,h}^2+\lambda_{i',h}^2
      \right)~.
\end{align}
Hence, \cref{lem: fixed bank sign comparison}, applied with $r=3$ and $m=d$, gives
\begin{align}
    &P\left\{
      (\check A_\ell^{(0)})_{i,i'}=1,
      (\check A_\ell^{(0)})_{i,h}=1
    \right\}
    \lesssim
      \pi_{n,d}(\ell)^{3/2}
      +\min\left\{
        1,
        d\left(
          \lambda_{i,i'}^2+\lambda_{i,h}^2+\lambda_{i',h}^2
        \right)
      \right\}~,
\end{align}
as required.

Finally, suppose that $\{i,i'\}$ and $\{h,h'\}$ are disjoint.
Applying \cref{lem: even gaussian covariance contraction} to the
centered indicators and using the bound established in Part (i), we
obtain
\begin{align}
    & \left|
      \Cov\left(
        (\check A_\ell^{(0)})_{i,i'},
        (\check A_\ell^{(0)})_{h,h'}
      \right)
    \right| \nonumber \\
    &\lesssim
      \left(
        \sum_{u\in\{i,i'\}}\sum_{v\in\{h,h'\}}
        \lambda_{u,v}^2
      \right)
      \left(
        P\left\{(\check A_\ell^{(0)})_{i,i'}=1\right\}
        P\left\{(\check A_\ell^{(0)})_{h,h'}=1\right\}
      \right)^{1/2}
    \nonumber\\
    &\lesssim
      \left(
        P\left\{(\check A_\ell^{(0)})_{i,i'}=1\right\}
        +P\left\{(\check A_\ell^{(0)})_{h,h'}=1\right\}
      \right)
      \sum_{u\in\{i,i'\}}\sum_{v\in\{h,h'\}}
      \lambda_{u,v}^2~,
\end{align}
as required.
\hfill\qed

\subsection{Proof of \cref{lem: fixed bank feasible oracle covariance inputs}}

For each $j$, let $\theta_{d,j,1},\ldots,\theta_{d,j,d}$ be the eigenvalues of $\Omega_d^{(\{j\})}$ and define
\begin{align}
    H_{n,j}
    &=H_n-
      \frac{
        \hat\varepsilon^{(j)}(\hat\varepsilon^{(j)})^\top
      }{
        \|\hat\varepsilon^{(j)}\|_2^2
      }~.
    \label{eq: fixed bank auxiliary projection}
\end{align}
For every nonempty $T\subseteq[n]$ with $|T|\leq4$, define
\begin{align}
    L_{j,T}
    &=\left\|
      \left(\Lambda_n(I_n-H_{n,j})\right)_{T,T}
      \right\|_{\mathrm F}
    +
      \left\|
      \left(
        (I_n-H_{n,j})\Lambda_n(I_n-H_{n,j})
      \right)_{T,T}
      \right\|_{\mathrm F}~.
    \label{eq: fixed bank local projection size}
\end{align}
Observe that \Cref{lem: basic inequalities n}, display \eqref{eq: n bound 1}, and \eqref{eq: projection gaussian direction product} therefore give
\begin{align}
    \|L_{j,T}\|_{2m}
    &\lesssim mV_n
    \label{eq: fixed bank local projection moments}
\end{align}
uniformly over $j$, $T$, and the positive integers $m$. Thus, \cref{lem: wick} and H\"older's inequality imply that
\begin{align}
    &\mathbb E\left[
      \left|
        \hat\varepsilon_i^{(j)}
        \hat\varepsilon_{i'}^{(j)}
        \hat\varepsilon_h^{(j)}
        \hat\varepsilon_{h'}^{(j)}
      \right|
      L_{j,T}^q
    \right]
    \lesssim V_n^q~,
    \qquad
    q\in\left\{\frac12,1,2,3\right\}~,
    \label{eq: fixed bank weighted local projection moments}
\end{align}
uniformly over $i,i',h,h'$, $j$, and the nonempty sets $T$ with $|T|\leq4$.

\subsubsection{Part (i)}

Fix an off-diagonal pair $(i,i')$ and set $T=\{i,i'\}$. Let $Z_1,\ldots,Z_d$ be independent standard Gaussian vectors in $\mathbb R^n$, independent of $\hat\varepsilon^{(j)}$, and write $Z=(Z_1^\top,\ldots,Z_d^\top)^\top$. Define
\begin{align}
    A(0)
    &=\operatorname{\mathsf{diag}}_{k\in[d]}\left\{
      \frac{\theta_{d,j,k}}
      {2\|\Omega_d^{(\{j\})}\|_{\mathrm F}}
      \Lambda_n^{1/2}
      (e_ie_{i'}^\top+e_{i'}e_i^\top)
      \Lambda_n^{1/2}
      \right\}\quad\text{and}
    \nonumber\\
    A(1)
    &=\operatorname{\mathsf{diag}}_{k\in[d]}\left\{
      \frac{\theta_{d,j,k}}
      {2\|\Omega_d^{(\{j\})}\|_{\mathrm F}}
      \Lambda_n^{1/2}H_{n,j}
      (e_ie_{i'}^\top+e_{i'}e_i^\top)
      H_{n,j}\Lambda_n^{1/2}
      \right\}
    \label{eq: fixed bank equal pair coefficient endpoints}
\end{align}
and set
\begin{align}
    A(t)
    &=(1-t)A(0)+tA(1)
    \quad\text{and}\quad
    B
    =A(1)-A(0)~.
    \label{eq: fixed bank equal pair coefficient path}
\end{align}
Conditionally on $\hat\varepsilon^{(j)}$, \cref{lem: rotation and regression} implies that we can write
\begin{align}
    (\check A_\ell^{(j)})_{i,i'}
    &=\mathbb I\left\{
      \left|
        \frac{\|\Omega_d^{(\{j\})}\|_{\mathrm F}}{d}
        Z^\top A(0)Z
      \right|
      \geq\underline\delta_{n,d}(\eta_\ell)
    \right\}\quad\text{and}
    \nonumber\\
    (\hat A_\ell^{(j)})_{i,i'}
    &=\mathbb I\left\{
      \left|
        \frac{\|\Omega_d^{(\{j\})}\|_{\mathrm F}}{d}
        Z^\top A(1)Z
      \right|
      \geq\underline\delta_{n,d}(\eta_\ell)
    \right\}~.
    \label{eq: fixed bank equal pair gaussian representation}
\end{align}
Now, observe that we can evaluate
\begin{align}
    \operatorname{tr}(A(0))
    &=\lambda_{i,i'}
      \frac{\operatorname{tr}(\Omega_d^{(\{j\})})}
      {\|\Omega_d^{(\{j\})}\|_{\mathrm F}}
      \quad\text{and}\quad
    2\|A(0)\|_{\mathrm F}^2
    =1+\lambda_{i,i'}^2~.
    \label{eq: fixed bank equal pair oracle geometry}
\end{align}
By applying  applying \eqref{eq: fixed bank anchor covariance geometry} gives
\begin{align}
    \|B\|_{\mathrm F}^2
    &\lesssim L_{j,T}+L_{j,T}^2~,
    &
    \|B\|_{\mathrm{op}}
    &\lesssim
      \frac{\sqrt{L_{j,T}}+L_{j,T}}{\sqrt d}~,
    &
    |\operatorname{tr}(B)|
    &\lesssim\sqrt d\,L_{j,T}~.
    \label{eq: fixed bank equal pair perturbation geometry}
\end{align}
The bounds \eqref{eq: fixed bank anchor covariance geometry}, \eqref{eq: fixed bank equal pair oracle geometry}, and \eqref{eq: fixed bank equal pair perturbation geometry} imply that, on
$\{L_{j,T}\leq c\}$, it holds that
\begin{align}
    \|A(t)\|_{\mathrm F}^2
    &\asymp1~,
    &
    \|A(t)\|_{\mathrm{op}}
    &\lesssim d^{-1/2}~,
    &
    \left|
      \left\{q\in[nd]:
        |\alpha_q(A(t))|\geq c d^{-1/2}
      \right\}
    \right|
    &\geq cd
    \label{eq: fixed bank equal pair path geometry}
\end{align}
uniformly over $t\in[0,1]$.

With this in place, we first consider the pairs satisfying
\begin{align}
    \sqrt d\,|\lambda_{i,i'}|
    &\leq
      c\frac{\underline\delta_{n,d}(\eta_\ell)}
      {\|\Omega_d^{(\{j\})}\|_{\mathrm F}/d}~.
    \label{eq: fixed bank equal pair first class}
\end{align}
Observe that
\begin{align}
    \frac{\underline\delta_{n,d}(\eta_\ell)}
      {\|\Omega_d^{(\{j\})}\|_{\mathrm F}/d}
    &\asymp\sqrt{\log(V_n^{-1})}
    \label{eq: fixed bank equal pair threshold scale}
\end{align}
by \eqref{eq: fixed bank anchor covariance geometry}. Moreover, after decreasing the constant in \eqref{eq: fixed bank equal pair first class}, \eqref{eq: fixed bank equal pair oracle geometry} also gives
\begin{align}
    &\frac{\underline\delta_{n,d}(\eta_\ell)}
      {\|\Omega_d^{(\{j\})}\|_{\mathrm F}/d}
      \mathbin{\pm}\operatorname{tr}(A(0))
    \in
      \left[
        \frac{1}{2}
        \frac{\underline\delta_{n,d}(\eta_\ell)}
          {\|\Omega_d^{(\{j\})}\|_{\mathrm F}/d},
        2
        \frac{\underline\delta_{n,d}(\eta_\ell)}
          {\|\Omega_d^{(\{j\})}\|_{\mathrm F}/d}
      \right]~.
    \label{eq: fixed bank equal pair centered thresholds}
\end{align}
On the event
\begin{align}
    \log(V_n^{-1})\sqrt{L_{j,T}}
      +\sqrt{d\log(V_n^{-1})}\,L_{j,T}
    &\leq c~,
    \label{eq: fixed bank equal pair crossing event}
\end{align}
the bounds \eqref{eq: fixed bank equal pair perturbation geometry} and \eqref{eq: fixed bank equal pair path geometry} allow us to apply Part (i) of \cref{lem: high rank relative crossing} with $A_0=A(0)$, $B=A(1)-A(0)$, and $G=Z$. Thus, by applying Part (i) of \cref{lem: high rank relative crossing} to the upper and lower boundaries in \eqref{eq: fixed bank equal pair centered thresholds}, and adding the two bounds, we obtain
\begin{align}
    &P\left\{
      (\hat A_\ell^{(j)})_{i,i'}
      \neq(\check A_\ell^{(j)})_{i,i'}
      \mathrel{\Big|}\hat\varepsilon^{(j)}
    \right\}
    \nonumber\\
    &\quad\lesssim
      P\left\{(\check A_\ell^{(j)})_{i,i'}=1\right\}
      \left(
        \log(V_n^{-1})\sqrt{L_{j,T}}
        +\sqrt{d\log(V_n^{-1})}\,L_{j,T}
      \right)~.
    \label{eq: fixed bank equal pair conditional crossing}
\end{align}
The moment bound \eqref{eq: fixed bank local projection moments} implies that, for every fixed $M>0$,
\begin{align}
    P\left\{
      \log(V_n^{-1})\sqrt{L_{j,T}}
      +\sqrt{d\log(V_n^{-1})}\,L_{j,T}>c
    \right\}
    &=o(V_n^M)~.
    \label{eq: fixed bank equal pair crossing complement}
\end{align}
The first conclusion in Part (i) of
\cref{lem: high rank relative crossing}, applied with $B=0$, also gives
\begin{align}
    P\left\{(\check A_\ell^{(j)})_{i,i'}=1\right\}
    &\geq\exp\{-C\log(V_n^{-1})\}~.
    \label{eq: fixed bank equal pair oracle probability lower bound}
\end{align}
Consequently, Cauchy-Schwarz, \eqref{eq: fixed bank equal pair crossing complement}, and \eqref{eq: fixed bank equal pair oracle probability lower bound} give
\begin{align}
    &\mathbb E\left[
      \left(
        \hat\varepsilon_i^{(j)}\hat\varepsilon_{i'}^{(j)}
      \right)^2
      \mathbb I\left\{
        \log(V_n^{-1})\sqrt{L_{j,T}}
        +\sqrt{d\log(V_n^{-1})}\,L_{j,T}>c
      \right\}
    \right]
    \nonumber\\
    &\quad=o\left(
      P\left\{(\check A_\ell^{(j)})_{i,i'}=1\right\}
      \sqrt{V_n}\log(V_n^{-1})
    \right)~.
    \label{eq: fixed bank equal pair crossing complement contribution}
\end{align}
On the complement of the event in \eqref{eq: fixed bank equal pair crossing complement}, take expectations in \eqref{eq: fixed bank equal pair conditional crossing} and apply \eqref{eq: fixed bank weighted local projection moments}. Combining the result with
\eqref{eq: fixed bank equal pair crossing complement contribution} gives
\begin{align}
    &\mathbb E\left[
      \left(
        \hat\varepsilon_i^{(j)}
        \hat\varepsilon_{i'}^{(j)}
      \right)^2
      \left(
        (\hat A_\ell^{(j)})_{i,i'}
        -(\check A_\ell^{(j)})_{i,i'}
      \right)^2
    \right]
    \nonumber\\
    &\quad\lesssim
      P\left\{(\check A_\ell^{(j)})_{i,i'}=1\right\}
      \left(
        \sqrt{V_n}\log(V_n^{-1})
        +V_n\sqrt{d\log(V_n^{-1})}
      \right)~.
    \label{eq: fixed bank equal pair first class pointwise bound}
\end{align}
Parts (ii) and (iii) of
\cref{lem: fixed bank selection matrix comparisons} give
\begin{align}
    P\left\{(\check A_\ell^{(j)})_{i,i'}=1\right\}
    &\lesssim
      \pi_{n,d}(\ell)+\min\{1,d\lambda_{i,i'}^2\}~.
    \label{eq: fixed bank equal pair oracle probability upper bound}
\end{align}
Consequently, we find that
\begin{align}
    &\sum_{i,i'=1}^n
      P\left\{(\check A_\ell^{(j)})_{i,i'}=1\right\}
      \mathbb I\left\{
        i\neq i',
        \ \sqrt d\,|\lambda_{i,i'}|
        \leq
        c\frac{\underline\delta_{n,d}(\eta_\ell)}
        {\|\Omega_d^{(\{j\})}\|_{\mathrm F}/d}
      \right\}
    \nonumber\\
    &\quad\lesssim
      n^2\left(\pi_{n,d}(\ell)+\sqrt d\,V_n\right)~.
    \label{eq: fixed bank equal pair first class probability sum}
\end{align}
Thus, we find that
\begin{align}
    &\frac{1}{n^4}\sum_{i,i'=1}^n
      \mathbb E\Bigg[
        \left(
          \hat\varepsilon_i^{(j)}
          \hat\varepsilon_{i'}^{(j)}
        \right)^2
        \left(
          (\hat A_\ell^{(j)})_{i,i'}
          -(\check A_\ell^{(j)})_{i,i'}
        \right)^2
    \nonumber\\
    &\hspace{39mm}{}
      \mathbb I\left\{
        i\neq i',
        \ \sqrt d\,|\lambda_{i,i'}|
        \leq
        c\frac{\underline\delta_{n,d}(\eta_\ell)}
        {\|\Omega_d^{(\{j\})}\|_{\mathrm F}/d}
      \right\}
      \Bigg]
    \nonumber\\
    &\quad\lesssim
      \frac{\pi_{n,d}(\ell)}{n^2}
      \left(
        \sqrt{V_n}\log(V_n^{-1})
        +V_n\sqrt{d\log(V_n^{-1})}
      \right)
      +dV_n^3\log^3(V_n^{-1})~.
    \label{eq: fixed bank equal pair first class aggregate}
\end{align}
by multiplying \eqref{eq: fixed bank equal pair first class pointwise bound} by the indicator in \eqref{eq: fixed bank equal pair first class probability sum} and summing.

For the remaining off-diagonal pairs, \cref{lem: basic inequalities n}, display \eqref{eq: n bound 1}, and \eqref{eq: fixed bank equal pair threshold scale} give
\begin{align}
    &\sum_{i,i'=1}^n
      \mathbb I\left\{
        i\neq i',
        \ \sqrt d\,|\lambda_{i,i'}|
        >
        c\frac{\underline\delta_{n,d}(\eta_\ell)}
        {\|\Omega_d^{(\{j\})}\|_{\mathrm F}/d}
      \right\}
    \lesssim
      \frac{n^2\sqrt d\,V_n}{\sqrt{\log(V_n^{-1})}}~.
    \label{eq: fixed bank equal pair second class count}
\end{align}
On $\{L_{j,T}\leq c\}$, apply Part (ii) of \cref{lem: high rank relative crossing} to the upper boundary, and then to the lower boundary after replacing $(A(0),B)$ by $(-A(0),-B)$. Adding the two bounds and applying \eqref{eq: fixed bank equal pair perturbation geometry} gives
\begin{align}
    &P\left\{
      (\hat A_\ell^{(j)})_{i,i'}
      \neq(\check A_\ell^{(j)})_{i,i'}
      \mathrel{\Big|}\hat\varepsilon^{(j)}
    \right\}
    \lesssim
      \sqrt{L_{j,T}}+\sqrt d\,L_{j,T}~.
    \label{eq: fixed bank equal pair global conditional crossing}
\end{align}
By multiplying \eqref{eq: fixed bank equal pair global conditional crossing} by the squared residual product, applying \eqref{eq: fixed bank weighted local projection moments}, and bounding the event $\{L_{j,T}>c\}$ by $L_{j,T}^2$, we obtain
\begin{align}
    &\mathbb E\left[
      \left(
        \hat\varepsilon_i^{(j)}\hat\varepsilon_{i'}^{(j)}
      \right)^2
      \left(
        (\hat A_\ell^{(j)})_{i,i'}
        -(\check A_\ell^{(j)})_{i,i'}
      \right)^2
    \right]
    \lesssim
      \sqrt{V_n}+\sqrt d\,V_n+V_n^2~.
    \label{eq: fixed bank equal pair second class pointwise bound}
\end{align}
Hence, the bound \eqref{eq: fixed bank equal pair second class count} and \eqref{eq: fixed bank equal pair second class pointwise bound} give
\begin{align}
    &\frac{1}{n^4}\sum_{i,i'=1}^n
      \mathbb E\Bigg[
        \left(
          \hat\varepsilon_i^{(j)}
          \hat\varepsilon_{i'}^{(j)}
        \right)^2
        \left(
          (\hat A_\ell^{(j)})_{i,i'}
          -(\check A_\ell^{(j)})_{i,i'}
        \right)^2
    \nonumber\\
    &\hspace{39mm}{}
      \mathbb I\left\{
        i\neq i',
        \ \sqrt d\,|\lambda_{i,i'}|
        >
        c\frac{\underline\delta_{n,d}(\eta_\ell)}
        {\|\Omega_d^{(\{j\})}\|_{\mathrm F}/d}
      \right\}
      \Bigg]
    \lesssim dV_n^3\log^3(V_n^{-1})~.
    \label{eq: fixed bank equal pair second class aggregate}
\end{align}
Finally, Gaussian fourth moments give
\begin{align}
    \frac{1}{n^4}\sum_{i=1}^n
      \mathbb E\left[
        (\hat\varepsilon_i^{(j)})^4
        \left(
          (\hat A_\ell^{(j)})_{i,i}
          -(\check A_\ell^{(j)})_{i,i}
        \right)^2
      \right]
    &\lesssim n^{-3}
    \lesssim dV_n^3\log^3(V_n^{-1})~.
    \label{eq: fixed bank equal pair diagonal aggregate}
\end{align}
Combining \eqref{eq: fixed bank equal pair first class aggregate}, \eqref{eq: fixed bank equal pair second class aggregate}, and \eqref{eq: fixed bank equal pair diagonal aggregate} proves \eqref{eq: fixed bank feasible oracle equal pair input}. \hfill\qed

\subsubsection{Part (ii)}

Take $i=i'$ and set $T=\{i,h,h'\}$. Let $G$ be a standard Gaussian vector in $\mathbb R^d$, independent of $\hat\varepsilon^{(j)}$. \cref{lem: rotation and regression}
implies that
\begin{align}
    P\left\{
      (\check A_\ell^{(j)})_{i,i}=0
      \mathrel{\Big|}\hat\varepsilon^{(j)}
    \right\}
    &=
    P\left\{
      \left|G^\top\frac{\Omega_d^{(\{j\})}}{d}G\right|
      <\underline\delta_{n,d}(\eta_\ell)
    \right\}~,\quad\text{and}
    \nonumber\\
    P\left\{
      (\hat A_\ell^{(j)})_{i,i}=0
      \mathrel{\Big|}\hat\varepsilon^{(j)}
    \right\}
    &=
    P\left\{
      \left|
        G^\top
        \frac{
          e_i^\top H_{n,j}\Lambda_nH_{n,j}e_i
        }{d}
        \Omega_d^{(\{j\})}G
      \right|
      <\underline\delta_{n,d}(\eta_\ell)
      \mathrel{\Big|}\hat\varepsilon^{(j)}
    \right\}~.
    \label{eq: fixed bank diagonal gaussian representation}
\end{align}
On $\{L_{j,T}\leq c\}$, \cref{assu: well conditioned} and \eqref{eq: fixed bank anchor covariance geometry} imply that
\begin{align}
    \min\left\{
      \operatorname{tr}\left(
        \frac{1}{d}\Omega_d^{(\{j\})}
      \right),
      \operatorname{tr}\left(
        \frac{e_i^\top H_{n,j}\Lambda_nH_{n,j}e_i}{d}
        \Omega_d^{(\{j\})}
      \right)
    \right\}
    &\geq c~,
    \nonumber\\
    \max\left\{
      \left\|\frac{1}{d}\Omega_d^{(\{j\})}\right\|_{\mathrm F},
      \left\|
        \frac{e_i^\top H_{n,j}\Lambda_nH_{n,j}e_i}{d}
        \Omega_d^{(\{j\})}
      \right\|_{\mathrm F}
    \right\}
    &\lesssim d^{-1/2}~,\quad\text{and}
    \nonumber\\
    \max\left\{
      \left\|\frac{1}{d}\Omega_d^{(\{j\})}\right\|_{\mathrm{op}},
      \left\|
        \frac{e_i^\top H_{n,j}\Lambda_nH_{n,j}e_i}{d}
        \Omega_d^{(\{j\})}
      \right\|_{\mathrm{op}}
    \right\}
    &\lesssim d^{-1}~.
    \label{eq: fixed bank diagonal quadratic geometry}
\end{align}
Moreover, \cref{lem: selected set characterization}, \cref{lem: auxiliary outcome covariance bounds}, and \eqref{eq: regret rate condition} give
\begin{align}
    \max_{\ell\in[g]}
    \underline\delta_{n,d}(\eta_\ell)
    &\lesssim
      \sqrt{\frac{\log(V_n^{-1})}{d}}
      \to0~.
\end{align}
Thus, for all sufficiently large $n,d$, on $\{L_{j,T}\leq c\}$, the threshold $\underline\delta_{n,d}(\eta_\ell)$ is smaller than half of each of the two traces in \eqref{eq: fixed bank diagonal quadratic geometry}, uniformly over $\ell\in[g]$. By applying \cref{lem: gaussian quadratic form tail}, display
\eqref{eq: gaussian derivative quadratic tail}, separately to the two probabilities in \eqref{eq: fixed bank diagonal gaussian representation} we obtain
\begin{align}
    &P\left\{
      (\check A_\ell^{(j)})_{i,i}=0
      \mathrel{\Big|}\hat\varepsilon^{(j)}
    \right\}
    +
    P\left\{
      (\hat A_\ell^{(j)})_{i,i}=0
      \mathrel{\Big|}\hat\varepsilon^{(j)}
    \right\}
    \lesssim e^{-cd}~.
\end{align}
Hence, the union bound, and the fact that if two indicators differ, at least one is equal zero, imply that, on
$\{L_{j,T}\leq c\}$, the bound
\begin{align}
    &P\left\{
      (\hat A_\ell^{(j)}-\check A_\ell^{(j)})_{i,i}\neq0
      \mathrel{\Big|}\hat\varepsilon^{(j)}
    \right\}
    \nonumber\\
    &\quad\leq
    P\left\{
      (\check A_\ell^{(j)})_{i,i}=0
      \mathrel{\Big|}\hat\varepsilon^{(j)}
    \right\}
    +
    P\left\{
      (\hat A_\ell^{(j)})_{i,i}=0
      \mathrel{\Big|}\hat\varepsilon^{(j)}
    \right\}
    \lesssim e^{-cd}
\end{align}
holds. Since this probability is at most one on
$\{L_{j,T}>c\}$, it follows that
\begin{align}
    &P\left\{
      (\hat A_\ell^{(j)}-\check A_\ell^{(j)})_{i,i}\neq0
      \mathrel{\Big|}\hat\varepsilon^{(j)}
    \right\}
    \lesssim
      e^{-cd}+\mathbb I\{L_{j,T}>c\}~.
    \label{eq: fixed bank diagonal disagreement probability}
\end{align}
Since every entry of $\hat A_\ell^{(j)}-\check A_\ell^{(j)}$ is bounded by one, \eqref{eq: fixed bank diagonal disagreement probability} gives
\begin{align}
    \left|
      \Cov\left(
        (\hat A_\ell^{(j)}-\check A_\ell^{(j)})_{i,i},
        (\hat A_\ell^{(j)}-\check A_\ell^{(j)})_{h,h'}
        \mathrel{\Big|}\hat\varepsilon^{(j)}
      \right)
    \right|
    &\lesssim
      e^{-cd}+\mathbb I\{L_{j,T}>c\}~.
    \label{eq: fixed bank diagonal pointwise covariance}
\end{align}
By symmetry, \eqref{eq: fixed bank diagonal pointwise covariance} bounds every covariance in Part (ii). There are at most $2n^3$ ordered unequal pairs in this part. Hence, Gaussian eighth moments, \eqref{eq: fixed bank weighted local projection moments} with $q=3$, and \eqref{eq: regret rate condition} give
\begin{align}
    &\Bigg|
      \frac{1}{n^4}\sum_{i,i',h,h'=1}^n
      \mathbb E\Bigg[
        \hat\varepsilon_i^{(j)}
        \hat\varepsilon_{i'}^{(j)}
        \hat\varepsilon_h^{(j)}
        \hat\varepsilon_{h'}^{(j)}
    \cdot
        \Cov\left(
          (\hat A_\ell^{(j)}-\check A_\ell^{(j)})_{i,i'},
          (\hat A_\ell^{(j)}-\check A_\ell^{(j)})_{h,h'}
          \mathrel{\Big|}\hat\varepsilon^{(j)}
        \right)
    \nonumber\\
    &\hspace{26mm}{}\cdot
        \mathbb I\left\{
          \{i,i'\}\neq\{h,h'\},
          \ i=i'\text{ or }h=h'
        \right\}
      \Bigg]
    \Bigg|
    \nonumber\\
    &\quad\lesssim
      \frac{e^{-cd}+V_n^3}{n}
    \lesssim
      V_n^2\left(
        \frac{\log^2(V_n^{-1})}{d}
        +dV_n\log^3(V_n^{-1})
      \right)~,
\end{align}
as required.
\hfill\qed

\subsubsection{Part (iii)}

Define the matrix
\begin{align}
    H_{n,j}(t)
    &=I_n-t(I_n-H_{n,j})~,
    \qquad t\in[0,1]~.
    \label{eq: fixed bank target projection path}
\end{align}
For $r,s\in[n]$ and $t\in[0,1]$, define
\begin{align}
    S_{r,s}(t)
    &=\mathbb I\left\{
      \left|
        \frac{1}{d}
        \sum_{k=0}^d
        (H_{n,j}(t)\varepsilon^{(k;j)})_r
        (H_{n,j}(t)\varepsilon^{(k;j)})_s
        \mathbb I\{k\neq j\}
      \right|
      \geq\underline\delta_{n,d}(\eta_\ell)
    \right\}~.
    \label{eq: fixed bank screening projection path}
\end{align}
Observe that we can write
\begin{align}
    (\hat A_\ell^{(j)}-\check A_\ell^{(j)})_{r,s}
    &=S_{r,s}(1)-S_{r,s}(0)~.
    \label{eq: fixed bank screening endpoint difference}
\end{align}
Now, fix distinct $u,v,w$, and set $T=\{u,v,w\}$, ordered as $(u,v,w)$.
Conditional on $\hat\varepsilon^{(j)}$, the spectral decomposition of
$\Omega_d^{(\{j\})}$ and \cref{lem: rotation and regression} imply that
\begin{align}
    \left(
      S_{u,v}(0),S_{u,v}(1),S_{u,w}(0),S_{u,w}(1)
    \right)
    &\overset{d}{=}
    \left(
      S_{1,2}(0),S_{1,2}(1),S_{1,3}(0),S_{1,3}(1)
    \right)~,
    \label{eq: fixed bank shared index endpoint distribution}
\end{align}
where the variables on the right are those in \cref{lem: gaussian threshold endpoint covariance bounds} with
\begin{align}
    K_{ZZ}
    &=(\Lambda_n)_{T,T}~,
    &K_{ZU}
    &=\left(\Lambda_n(I_n-H_{n,j})\right)_{T,T}~,
    \nonumber\\
    K_{UU}
    &=\left(
      (I_n-H_{n,j})\Lambda_n(I_n-H_{n,j})
      \right)_{T,T}~,
    &\vartheta_k
    &=\theta_{d,j,k}~.
    \label{eq: fixed bank shared index endpoint identification}
\end{align}
On $\{L_{j,T}\leq c\}$, \eqref{eq: fixed bank anchor covariance geometry} and \cref{assu: well conditioned}, applied to $\Lambda_n$, permit application of Part (i) of \cref{lem: gaussian threshold endpoint covariance bounds}, which gives a deterministic quantity $\chi_{\ell,j}$, independent of $u,v,w$, such that
\begin{align}
    |\chi_{\ell,j}|
    &\lesssim\sqrt d
    \label{eq: fixed bank shared index coefficient bound}
\end{align}
and
\begin{align}
    &\left|
      \Cov\left(
        (\hat A_\ell^{(j)}-\check A_\ell^{(j)})_{u,v},
        (\hat A_\ell^{(j)}-\check A_\ell^{(j)})_{u,w}
        \mathrel{\Big|}\hat\varepsilon^{(j)}
      \right)
    -\chi_{\ell,j}e_u^\top(I_n-H_{n,j})\Lambda_n
        (I_n-H_{n,j})e_u
    \right|
    \nonumber\\
    &\quad\lesssim
      d\left(
        \|(\Lambda_n)_{T,T}-I_3\|_{\mathrm F}L_{j,T}
        +L_{j,T}^2
      \right)~.
    \label{eq: fixed bank shared index covariance expansion}
\end{align}
As differences of indicator variables are bounded by one in absolute value, the bounds \eqref{eq: fixed bank shared index coefficient bound} and $\mathbb I\{L_{j,T}>c\}\lesssim L_{j,T}^2$ imply that \eqref{eq: fixed bank shared index covariance expansion} holds for all values of $L_{j,T}$.

The unrestricted sum of the linear term in \eqref{eq: fixed bank shared index covariance expansion} satisfies
\begin{align}
    &\sum_{u,v,w=1}^n
      (\hat\varepsilon_u^{(j)})^2
      \hat\varepsilon_v^{(j)}
      \hat\varepsilon_w^{(j)}
      e_u^\top(I_n-H_{n,j})\Lambda_n
      (I_n-H_{n,j})e_u
    \nonumber\\
    &\quad=
      \left(
        \sum_{u=1}^n
        (\hat\varepsilon_u^{(j)})^2
        e_u^\top(I_n-H_{n,j})\Lambda_n
        (I_n-H_{n,j})e_u
      \right)
      \left(\sum_{v=1}^n\hat\varepsilon_v^{(j)}\right)
      \left(\sum_{w=1}^n\hat\varepsilon_w^{(j)}\right)
    =0~.
    \label{eq: fixed bank shared index linear cancellation}
\end{align}
Consequently, we have that 
\begin{align}
    &\sum_{u,v,w=1}^n
      (\hat\varepsilon_u^{(j)})^2
      \hat\varepsilon_v^{(j)}
      \hat\varepsilon_w^{(j)}
      e_u^\top(I_n-H_{n,j})\Lambda_n
      (I_n-H_{n,j})e_u
      \mathbb I\{|\{u,v,w\}|=3\}
    \nonumber\\
    &\quad=
      -\sum_{u,v,w=1}^n
      (\hat\varepsilon_u^{(j)})^2
      \hat\varepsilon_v^{(j)}
      \hat\varepsilon_w^{(j)}
      e_u^\top(I_n-H_{n,j})\Lambda_n
      (I_n-H_{n,j})e_u
      \mathbb I\{|\{u,v,w\}|<3\}~.
    \label{eq: fixed bank shared index restricted cancellation}
\end{align}
Since there are at most $Cn^2$ triples satisfying
$|\{u,v,w\}|<3$, \eqref{eq: fixed bank shared index coefficient bound},
\eqref{eq: fixed bank shared index restricted cancellation}, and
\eqref{eq: fixed bank weighted local projection moments} give
\begin{align}
    &\left|
      \chi_{\ell,j}
      \sum_{u,v,w=1}^n
      \mathbb E\Big[
        (\hat\varepsilon_u^{(j)})^2
        \hat\varepsilon_v^{(j)}
        \hat\varepsilon_w^{(j)}
        e_u^\top(I_n-H_{n,j})\Lambda_n
        (I_n-H_{n,j})e_u
      \Big]
      \mathbb I\{|\{u,v,w\}|=3\}
    \right|
    \nonumber\\
    &\quad=
    \left|
      \chi_{\ell,j}
      \sum_{u,v,w=1}^n
      \mathbb E\Big[
        (\hat\varepsilon_u^{(j)})^2
        \hat\varepsilon_v^{(j)}
        \hat\varepsilon_w^{(j)}
        e_u^\top(I_n-H_{n,j})\Lambda_n
        (I_n-H_{n,j})e_u
      \Big]
      \mathbb I\{|\{u,v,w\}|<3\}
    \right|
    \nonumber\\
    &\quad\leq
      |\chi_{\ell,j}|
      \sum_{u,v,w=1}^n
      \mathbb E\left[
        \left|
          (\hat\varepsilon_u^{(j)})^2
          \hat\varepsilon_v^{(j)}
          \hat\varepsilon_w^{(j)}
        \right|
        L_{j,\{u,v,w\}}
      \right]
      \mathbb I\{|\{u,v,w\}|<3\}
    \lesssim
      \sqrt d\,n^2V_n~.
    \label{eq: fixed bank shared index linear aggregate}
\end{align} Moreover, for distinct $u,v,w$, \cref{lem: basic inequalities n},
display \eqref{eq: n bound 1}, gives
\begin{align}
    \|(\Lambda_n)_{\{u,v,w\},\{u,v,w\}}-I_3\|_{\mathrm F}
    &\lesssim
      |\lambda_{u,v}|+|\lambda_{u,w}|+|\lambda_{v,w}|~,\quad\text{and}
    \nonumber\\
    \sum_{u,v,w=1}^n
      \left(
        |\lambda_{u,v}|+|\lambda_{u,w}|+|\lambda_{v,w}|
      \right)
      \mathbb I\{|\{u,v,w\}|=3\}
    &\lesssim n^3V_n~.
    \label{eq: fixed bank shared index population sum}
\end{align}
For distinct $u,v,w$, applying \eqref{eq: fixed bank weighted local projection moments} with $q=1$ and $q=2$ gives
\begin{align}
    \mathbb E\left[
        |
            (\hat\varepsilon_u^{(j)})^2
            \hat\varepsilon_v^{(j)}
            \hat\varepsilon_w^{(j)}
        |
        L_{j,\{u,v,w\}}
    \right]
    &\lesssim V_n~
    \quad\text{and}\quad
    \mathbb E\left[
        |
            (\hat\varepsilon_u^{(j)})^2
            \hat\varepsilon_v^{(j)}
            \hat\varepsilon_w^{(j)}
        |
        L_{j,\{u,v,w\}}^2
    \right]
    \lesssim V_n^2~.\label{eq: two conditional residual bounds}
\end{align}
By multiplying both sides of \eqref{eq: fixed bank shared index covariance expansion} by $|(\hat\varepsilon_u^{(j)})^2 \hat\varepsilon_v^{(j)}\hat\varepsilon_w^{(j)}|$ and taking expectations, the bounds \eqref{eq: two conditional residual bounds} imply
\begin{align}
    &\mathbb E\Bigg[
        \left|
            (\hat\varepsilon_u^{(j)})^2
            \hat\varepsilon_v^{(j)}
            \hat\varepsilon_w^{(j)}
        \right|
        \Bigg|
            \Cov\left(
                (\hat A_\ell^{(j)}-\check A_\ell^{(j)})_{u,v},
                (\hat A_\ell^{(j)}-\check A_\ell^{(j)})_{u,w}
                \mathrel{\Big|}\hat\varepsilon^{(j)}
            \right)
    \nonumber\\
    &\hspace{40mm}
            {}-\chi_{\ell,j}e_u^\top(I_n-H_{n,j})\Lambda_n
                (I_n-H_{n,j})e_u
        \Bigg|
    \Bigg]
    \nonumber\\
    &\quad\lesssim
        d\left(
            V_n
            \left\|
                (\Lambda_n)_{\{u,v,w\},\{u,v,w\}}-I_3
            \right\|_{\mathrm F}
            +V_n^2
        \right)~.
\end{align}
Consequently, we find that 
\begin{align}
    &\sum_{u,v,w=1}^n
      \mathbb E\Bigg[
        \left|
          (\hat\varepsilon_u^{(j)})^2
          \hat\varepsilon_v^{(j)}
          \hat\varepsilon_w^{(j)} \right| 
          \bigg\vert
          \Cov\left(
            (\hat A_\ell^{(j)}-\check A_\ell^{(j)})_{u,v},
            (\hat A_\ell^{(j)}-\check A_\ell^{(j)})_{u,w}
            \mathrel{\Big|}\hat\varepsilon^{(j)}
          \right)
    \nonumber\\
    &\hspace{20mm}
          {}-\chi_{\ell,j}e_u^\top(I_n-H_{n,j})\Lambda_n
            (I_n-H_{n,j})e_u
        \bigg\vert
        \mathbb I\{|\{u,v,w\}|=3\}
      \Bigg] \nonumber\\
    & 
    \lesssim
    dV_n
    \sum_{\substack{u,v,w=1\\|\{u,v,w\}|=3}}^n
        \left\|
            (\Lambda_n)_{\{u,v,w\},\{u,v,w\}}-I_3
        \right\|_{\mathrm F}
    +dn^3V_n^2
    \lesssim dn^3V_n^2~,
    \label{eq: fixed bank shared index remainder aggregate}
\end{align}
where the last inequality follows from \eqref{eq: fixed bank shared index population sum}. By symmetry, the same bounds hold when the common index occupies any of the other three positions. Combining \eqref{eq: fixed bank shared index linear aggregate} and \eqref{eq: fixed bank shared index remainder aggregate} gives
\begin{align}
    &\Bigg|
      \frac{1}{n^4}\sum_{i,i',h,h'=1}^n
      \mathbb E\Bigg[
        \hat\varepsilon_i^{(j)}
        \hat\varepsilon_{i'}^{(j)}
        \hat\varepsilon_h^{(j)}
        \hat\varepsilon_{h'}^{(j)}
    \Cov\left(
          (\hat A_\ell^{(j)}-\check A_\ell^{(j)})_{i,i'},
          (\hat A_\ell^{(j)}-\check A_\ell^{(j)})_{h,h'}
          \mathrel{\Big|}\hat\varepsilon^{(j)}
        \right)
    \nonumber\\
    &\hspace{60mm}{}\cdot
        \mathbb I\left\{
          i\neq i',
          \ h\neq h',
          \ |\{i,i',h,h'\}|=3
        \right\}
      \Bigg]
    \Bigg|
    \nonumber\\
    &\quad\lesssim
      \frac{\sqrt d\,n^2V_n+dn^3V_n^2}{n^4}
    \lesssim dV_n^3~,
\end{align}
as required.
\hfill\qed

\subsubsection{Part (iv)}

Let $\Psi_{\delta,d}^{(j)}$ be defined analogously to $\Psi_{\delta,d}$, with $\theta_{d,j,k}$ in place of $\theta_{d,k}$. For $a,b\in[2]$ and distinct $r,s\in[n]$ satisfying $L_{j,\{r,s\}}\leq c$, define
\begin{align}
    \Delta_{a,b}(r,s)
    &=\sum_{t,q\in\{0,1\}}(-1)^{t+q}
      \Psi_{\underline\delta_{n,d}(\eta_\ell),d}^{(j)}
      \left(
        I_2,
        I_2,
        e_r^\top H_{n,j}(t)\Lambda_nH_{n,j}(q)e_s
        e_ae_b^\top
      \right)~.
    \label{eq: fixed bank four index single entry rectangle}
\end{align}
Set $\Delta_{a,b}(r,s)=0$ for the remaining pairs. To bound the nonzero terms, identify coordinates $a$ and $b+2$ in \cref{lem: gaussian threshold endpoint covariance bounds} with $r$ and $s$, respectively, using the two-coordinate blocks
\begin{align}
    (K_{ZZ})_{\{a,b+2\},\{a,b+2\}}
    &=(\Lambda_n)_{\{r,s\},\{r,s\}}~,
    \nonumber\\
    (K_{ZU})_{\{a,b+2\},\{a,b+2\}}
    &=\left(\Lambda_n(I_n-H_{n,j})\right)_{\{r,s\},\{r,s\}}~,\quad\text{and}
    \nonumber\\
    (K_{UU})_{\{a,b+2\},\{a,b+2\}}
    &=\left(
      (I_n-H_{n,j})\Lambda_n(I_n-H_{n,j})
      \right)_{\{r,s\},\{r,s\}}~,
    \label{eq: fixed bank single entry endpoint identification}
\end{align}
For each coordinate in $[4]\setminus\{a,b+2\}$, take the corresponding $Z$-coordinate to be independent standard Gaussian and set the corresponding $U$-coordinate equal to zero. Then $\Delta_K\lesssim L_{j,\{r,s\}}$, and the scalar covariance in \eqref{eq: gaussian single entry endpoint term} is $e_r^\top H_{n,j}(t)\Lambda_nH_{n,j}(q)e_s$. Part (iii) of \cref{lem: gaussian threshold endpoint covariance bounds} therefore gives
\begin{align}
    |\Delta_{a,b}(r,s)|
    &\lesssim
      L_{j,\{r,s\}}
      \left(
        |\lambda_{r,s}|+L_{j,\{r,s\}}
      \right)~.
    \label{eq: fixed bank four index single entry bound}
\end{align}
The same bound holds for the remaining pairs because $\Delta_{a,b}(r,s)=0$.

Fix four distinct indices $i_1,i_2,h_1,h_2$, and order $T=\{i_1,i_2,h_1,h_2\}$ as $(i_1,i_2,h_1,h_2)$. Conditional on $\hat\varepsilon^{(j)}$, the spectral decomposition of $\Omega_d^{(\{j\})}$ and \cref{lem: rotation and regression} imply that
\begin{align}
    &\left(
      S_{i_1,i_2}(0),S_{i_1,i_2}(1),
      S_{h_1,h_2}(0),S_{h_1,h_2}(1)
    \right)
    \overset{d}{=}
    \left(
      S_{1,2}(0),S_{1,2}(1),
      S_{3,4}(0),S_{3,4}(1)
    \right)~,
    \label{eq: fixed bank four index endpoint distribution}
\end{align}
where the variables on the right are those in \cref{lem: gaussian threshold endpoint covariance bounds} with
\begin{align}
    K_{ZZ}
    &=(\Lambda_n)_{T,T}~,
    &K_{ZU}
    &=\left(\Lambda_n(I_n-H_{n,j})\right)_{T,T}~,
    \nonumber\\
    K_{UU}
    &=\left(
      (I_n-H_{n,j})\Lambda_n(I_n-H_{n,j})
      \right)_{T,T}~,
    &\vartheta_k
    &=\theta_{d,j,k}~.
    \label{eq: fixed bank four index endpoint identification}
\end{align}
Define the approximation error
\begin{align}
    \mathcal E_{\ell,j}(i_1,i_2,h_1,h_2)
    &=\Cov\left(
        (\hat A_\ell^{(j)}-\check A_\ell^{(j)})_{i_1,i_2},
        (\hat A_\ell^{(j)}-\check A_\ell^{(j)})_{h_1,h_2}
        \mathrel{\Big|}\hat\varepsilon^{(j)}
      \right)
      -\sum_{a,b=1}^2\Delta_{a,b}(i_a,h_b)~.
    \label{eq: fixed bank four index endpoint decomposition}
\end{align}
For every quadruple containing fewer than four distinct indices, set $\mathcal E_{\ell,j}(i_1,i_2,h_1,h_2)=0$. The identities \eqref{eq: fixed bank screening endpoint difference} and \eqref{eq: fixed bank four index endpoint distribution}, together with \eqref{eq: fixed bank anchor covariance geometry} and \cref{assu: well conditioned}, applied to $\Lambda_n$, permit application of Parts (ii) and (iv) of \cref{lem: gaussian threshold endpoint covariance bounds} on $\{L_{j,T}\leq c\}$. Since $L_{j,\{i_a,h_b\}}\leq L_{j,T}$, each $\Delta_{a,b}(i_a,h_b)$ equals the corresponding term $\mathcal J_{a,b}$ in \cref{lem: gaussian threshold endpoint covariance bounds}.

On $\{L_{j,T}\leq c\}$, Part (ii) of \cref{lem: gaussian threshold endpoint covariance bounds} gives, whenever $|\lambda_{i_1,i_2}|+|\lambda_{h_1,h_2}|\leq c$, that
\begin{align}
    |\mathcal E_{\ell,j}(i_1,i_2,h_1,h_2)|
    &\lesssim
      dL_{j,T}^2
      \left(
        \|(\Lambda_n)_{\{i_1,i_2\},\{h_1,h_2\}}\|_{\mathrm F}
        +L_{j,T}
      \right)^2
    \nonumber\\
    &\quad+
      \sqrt d\,L_{j,T}
      \left(
        |\lambda_{i_1,i_2}|+|\lambda_{h_1,h_2}|+L_{j,T}
      \right)
      \left(
        \|(\Lambda_n)_{\{i_1,i_2\},\{h_1,h_2\}}\|_{\mathrm F}
        +L_{j,T}
      \right)
    \nonumber\\
    &\quad+
      L_{j,T}^2\sum_{a,b=1}^2
      \left(
        |\lambda_{i_a,h_b}|+L_{j,T}
      \right)^2
    \nonumber\\
    &\quad+
      L_{j,T}\sum_{a,b,a',b'=1}^2
      \left(
        |\lambda_{i_a,h_b}|+L_{j,T}
      \right)
      \left(
        |\lambda_{i_{a'},h_{b'}}|+L_{j,T}
      \right)
      \mathbb I\{(a,b)\neq(a',b')\}~.
    \label{eq: fixed bank four index refined pointwise bound}
\end{align}
On the same event, Part (iv) of \cref{lem: gaussian threshold endpoint covariance bounds} gives, without the preceding restriction, that
\begin{align}
    |\mathcal E_{\ell,j}(i_1,i_2,h_1,h_2)|
    &\lesssim
      dL_{j,T}^2
      +L_{j,T}\sum_{a,b=1}^2|\lambda_{i_a,h_b}|~.
    \label{eq: fixed bank four index general pointwise bound}
\end{align}
Now, observe that we can evaluate
\begin{align}
    &\sum_{i_1,i_2,h_1,h_2=1}^n
      \left(
        \sum_{a,b=1}^2|\lambda_{i_a,h_b}|
        +\sum_{a,b=1}^2\lambda_{i_a,h_b}^2
      \right)
      \mathbb I\{|\{i_1,i_2,h_1,h_2\}|=4\}
    \lesssim n^4V_n\quad\text{and}
    \nonumber\\
    &\sum_{i_1,i_2,h_1,h_2=1}^n
    \mathbb I\{|\{i_1,i_2,h_1,h_2\}|=4\}
      \Bigg(
        \left(
          |\lambda_{i_1,i_2}|+|\lambda_{h_1,h_2}|
        \right)
        \sum_{a,b=1}^2|\lambda_{i_a,h_b}|
    \nonumber\\
    &\hspace{15mm}+
        \sum_{a,b,a',b'=1}^2
        |\lambda_{i_a,h_b}\lambda_{i_{a'},h_{b'}}|
        \mathbb I\{(a,b)\neq(a',b')\}
      \Bigg)
    \lesssim n^4V_n^2~.
    \label{eq: fixed bank four index coefficient sums}
\end{align}
by applying \cref{lem: basic inequalities n}, display \eqref{eq: n bound 1}. Thus, \eqref{eq: fixed bank four index refined pointwise bound}, \eqref{eq: fixed bank weighted local projection moments}, and \eqref{eq: fixed bank four index coefficient sums} imply that
\begin{align}
    &\frac{1}{n^4}
      \sum_{i_1,i_2,h_1,h_2=1}^n
      \mathbb E\left[
        \left|
          \hat\varepsilon_{i_1}^{(j)}
          \hat\varepsilon_{i_2}^{(j)}
          \hat\varepsilon_{h_1}^{(j)}
          \hat\varepsilon_{h_2}^{(j)}
        \right|
        |\mathcal E_{\ell,j}(i_1,i_2,h_1,h_2)|
    \right.
    \nonumber\\
    &\hspace{30mm}\left.
        {}\cdot\mathbb I\left\{
          |\lambda_{i_1,i_2}|+|\lambda_{h_1,h_2}|\leq c,
          \ L_{j,T}\leq c,
          \ |\{i_1,i_2,h_1,h_2\}|=4
        \right\}
      \right]
    \lesssim dV_n^3~.
    \label{eq: fixed bank four index first aggregate}
\end{align}
On the other hand, the indicator $\mathbb I\{|\lambda_{i_1,i_2}|+|\lambda_{h_1,h_2}|>c\}$ is bounded by a constant multiple of $|\lambda_{i_1,i_2}|+|\lambda_{h_1,h_2}|$. Thus, the bound \eqref{eq: fixed bank four index coefficient sums} gives
\begin{align}
    \sum_{i_1,i_2,h_1,h_2=1}^n
      \mathbb I\left\{
        |\lambda_{i_1,i_2}|+|\lambda_{h_1,h_2}|>c
      \right\}
    &\lesssim n^4V_n\quad\text{and}
    \nonumber\\
    \sum_{i_1,i_2,h_1,h_2=1}^n
      \sum_{a,b=1}^2|\lambda_{i_a,h_b}|
      \mathbb I\left\{
        |\lambda_{i_1,i_2}|+|\lambda_{h_1,h_2}|>c
      \right\}
    &\lesssim n^4V_n^2~.
    \label{eq: fixed bank four index complementary coefficient sums}
\end{align}
Consequently, we find that 
\begin{align}
    &\frac{1}{n^4}
      \sum_{i_1,i_2,h_1,h_2=1}^n
      \mathbb E\left[
        \left|
          \hat\varepsilon_{i_1}^{(j)}
          \hat\varepsilon_{i_2}^{(j)}
          \hat\varepsilon_{h_1}^{(j)}
          \hat\varepsilon_{h_2}^{(j)}
        \right|
        |\mathcal E_{\ell,j}(i_1,i_2,h_1,h_2)|
    \right.
    \nonumber\\
    &\hspace{30mm}\left.
        {}\cdot\mathbb I\left\{
          |\lambda_{i_1,i_2}|+|\lambda_{h_1,h_2}|>c,
          \ L_{j,T}\leq c,
          \ |\{i_1,i_2,h_1,h_2\}|=4
        \right\}
      \right]
    \lesssim dV_n^3~.
    \label{eq: fixed bank four index second aggregate}
\end{align}
by applying the bounds \eqref{eq: fixed bank four index general pointwise bound}, \eqref{eq: fixed bank weighted local projection moments}, and \eqref{eq: fixed bank four index complementary coefficient sums}.

For each $a,b\in[2]$, observe that we can evaluate
\begin{align}
    &\sum_{i_1,i_2,h_1,h_2=1}^n
      \hat\varepsilon_{i_1}^{(j)}
      \hat\varepsilon_{i_2}^{(j)}
      \hat\varepsilon_{h_1}^{(j)}
      \hat\varepsilon_{h_2}^{(j)}
      \Delta_{a,b}(i_a,h_b)
    =0~,
    \label{eq: fixed bank four index positional cancellation}
\end{align}
because the two residuals whose indices do not appear in
$\Delta_{a,b}(i_a,h_b)$ each sum to zero. For each fixed pair
$(i_a,h_b)$, at most $5n$ choices of the remaining indices contain fewer
than four distinct values. Therefore,
\eqref{eq: fixed bank four index single entry bound},
\eqref{eq: fixed bank weighted local projection moments}, and
\cref{lem: basic inequalities n}, display \eqref{eq: n bound 1}, imply
that
\begin{align}
    &\frac{1}{n^4}\left|
      \sum_{i_1,i_2,h_1,h_2=1}^n
      \mathbb E\left[
        \hat\varepsilon_{i_1}^{(j)}
        \hat\varepsilon_{i_2}^{(j)}
        \hat\varepsilon_{h_1}^{(j)}
        \hat\varepsilon_{h_2}^{(j)}
        \sum_{a,b=1}^2\Delta_{a,b}(i_a,h_b)
        \mathbb I\{|\{i_1,i_2,h_1,h_2\}|=4\}
      \right]
    \right|
    \lesssim\frac{V_n^2}{n}~.
    \label{eq: fixed bank four index positional completion}
\end{align}
Finally, a covariance of two indicator differences and each $\Delta_{a,b}(r,s)$ are bounded by a constant. Hence, $|\mathcal E_{\ell,j}(i_1,i_2,h_1,h_2)|\lesssim1$ for four distinct indices. Thus, $\mathbb I\{L_{j,T}>c\}\lesssim L_{j,T}^3$ and \eqref{eq: fixed bank weighted local projection moments} give
\begin{align}
    &\frac{1}{n^4}
      \sum_{i_1,i_2,h_1,h_2=1}^n
      \mathbb E\left[
        \left|
          \hat\varepsilon_{i_1}^{(j)}
          \hat\varepsilon_{i_2}^{(j)}
          \hat\varepsilon_{h_1}^{(j)}
          \hat\varepsilon_{h_2}^{(j)}
          |\mathcal E_{\ell,j}(i_1,i_2,h_1,h_2)|
        \right| \nonumber \right.\\
    & \quad\quad\quad\quad\quad\quad\quad\quad
    \left.
        \mathbb I\left\{
          L_{j,T}>c,
          \ |\{i_1,i_2,h_1,h_2\}|=4
        \right\}
      \right]
    \lesssim V_n^3~.
    \label{eq: fixed bank four index exceptional aggregate}
\end{align}
Combining \eqref{eq: fixed bank four index endpoint decomposition}, \eqref{eq: fixed bank four index first aggregate}, \eqref{eq: fixed bank four index second aggregate}, \eqref{eq: fixed bank four index positional completion}, and \eqref{eq: fixed bank four index exceptional aggregate} we can conclude that
\begin{align}
    &\Bigg|
      \frac{1}{n^4}
      \sum_{i_1,i_2,h_1,h_2=1}^n
      \mathbb E\Bigg[
        \hat\varepsilon_{i_1}^{(j)}
        \hat\varepsilon_{i_2}^{(j)}
        \hat\varepsilon_{h_1}^{(j)}
        \hat\varepsilon_{h_2}^{(j)}
    \Cov\left(
          (\hat A_\ell^{(j)}-\check A_\ell^{(j)})_{i_1,i_2},
          (\hat A_\ell^{(j)}-\check A_\ell^{(j)})_{h_1,h_2}
          \mathrel{\Big|}\hat\varepsilon^{(j)}
        \right)
    \nonumber\\
    &\hspace{40mm}{}\cdot
        \mathbb I\{|\{i_1,i_2,h_1,h_2\}|=4\}
      \Bigg]
    \Bigg|
    \lesssim
      \frac{V_n^2}{n}+V_n^3+dV_n^3
    \lesssim dV_n^3~,
\end{align}
as required. \hfill\qed

\subsection{Proof of \cref{lem: gaussian threshold covariance bounds}}

Throughout this proof, the constant $C$ may depend on the fixed value of $\kappa$. For a multi-index $\alpha=(\alpha_1,\ldots,\alpha_{2d})\in\mathbb N_0^{2d}$, write $|\alpha|=\sum_{l=1}^{2d}\alpha_l$, $\alpha!=\prod_{l=1}^{2d}\alpha_l!$, and $t^\alpha=\prod_{l=1}^{2d}t_l^{\alpha_l}$. Let $H_\alpha$ denote the product of the normalized probabilists' Hermite polynomials determined by
\begin{align}
    \exp\left(t^\top z-\frac{\|t\|_2^2}{2}\right)
    &=
    \sum_{\alpha\in\mathbb N_0^{2d}}
        H_\alpha(z)\frac{t^\alpha}{\sqrt{\alpha!}}~,
    \label{eq: gaussian derivative hermite generating function}
\end{align}
for every $z,t\in\mathbb R^{2d}$.

We use the following standard facts concerning finite-dimensional representations of Wiener-chaos.
\begin{lem}[\citealp{janson1997}, Theorems~2.6, 4.5, and~4.9]
\label{lem: gaussian chaos transfer}$ $\\
Let $(Z,W)$ be centered jointly Gaussian vectors in $\mathbb R^{2d}$ with standard marginal covariances and cross-covariance $S$. For $m\geq1$, let $\mathcal T_m(S)$ be the matrix $(\mathbb E[H_\alpha(Z)H_\beta(W)])_{|\alpha|=|\beta|=m}$. Then the degree-$m$ chaos is identified with $\operatorname{Sym}^m(\mathbb R^{2d})$, the matrix $\mathcal T_m(S)$ is the restriction of $S^{\otimes m}$ to that space, and chaos components of different degrees are orthogonal. Moreover, it holds that
\begin{align}
    \|\mathcal T_m(S)\|_{\mathrm{op}}
    &\leq
    \|S\|_{\mathrm{op}}^m~,
    \nonumber\\
    \|\mathrm D\mathcal T_m(S)[G]\|_{\mathrm{op}}
    &\leq
    m\|S\|_{\mathrm{op}}^{m-1}\|G\|_{\mathrm{op}}~,\quad\text{and}
    \nonumber\\
    \|\mathrm D^2\mathcal T_m(S)[G,H]\|_{\mathrm{op}}
    &\leq
    m(m-1)\|S\|_{\mathrm{op}}^{m-2}
    \|G\|_{\mathrm{op}}\|H\|_{\mathrm{op}}~,
    \label{eq: gaussian derivative transfer bounds}
\end{align}
for every $m\geq2$ and all $2d\times2d$ matrices $G,H$.
\end{lem}

Define the centered threshold functions
\begin{align}
    f_A(z)
    &=
    \mathbb I\left\{
        \left|
            \frac{1}{d}\sum_{j=1}^d\theta_{d,j}
            (A^{1/2}z_j)_1(A^{1/2}z_j)_2
        \right|\geq\delta
    \right\}
    -p_{\delta,d}(A)
    \quad\text{and}
    \nonumber\\
    f_B(w)
    &=
    \mathbb I\left\{
        \left|
            \frac{1}{d}\sum_{j=1}^d\theta_{d,j}
            (B^{1/2}w_j)_1(B^{1/2}w_j)_2
        \right|\geq\delta
    \right\}
    -p_{\delta,d}(B)~,
    \label{eq: gaussian derivative centered threshold functions}
\end{align}
for $z,w\in\mathbb R^{2d}$. For $m\geq1$, let $f_{A,m} =(\mathbb E[f_A(Z)H_\alpha(Z)])_{|\alpha|=m}$ and $f_{B,m} =(\mathbb E[f_B(W)H_\alpha(W)])_{|\alpha|=m}$. Parseval's identity applied to the two functions in \eqref{eq: gaussian derivative centered threshold functions} gives
\begin{align}
    \sum_{m\geq1}\|f_{A,m}\|_{\ell^2}^2
    &=
    \mathbb E[f_A(Z)^2]
    \leq p_{\delta,d}(A)
    \quad\text{and}\quad
    \sum_{m\geq1}\|f_{B,m}\|_{\ell^2}^2
    =
    \mathbb E[f_B(W)^2]
    \leq p_{\delta,d}(B)~,
    \label{eq: gaussian derivative hermite coefficient bounds}
\end{align}
for every positive-definite $A$ and $B$. The degree-zero coefficients vanish because the functions are centered. The odd-degree coefficients vanish because both functions are even and $H_\alpha(-z)=(-1)^{|\alpha|}H_\alpha(z)$. Applying the cross-moment pairing in \cref{lem: gaussian chaos transfer} with $S=I_d\otimes A^{-1/2}RB^{-1/2}$ to the $L^2$ chaos expansions of the two functions in \eqref{eq: gaussian derivative centered threshold functions} gives the exact representation
\begin{align}
    \Psi_{\delta,d}(A,B,R)
    &=
    \sum_{\substack{m\geq2\\m\ \mathrm{even}}}
    \left\langle
        f_{A,m},
        \mathcal T_m(I_d\otimes A^{-1/2}RB^{-1/2})
        f_{B,m}
    \right\rangle~,
    \label{eq: gaussian derivative even hermite representation}
\end{align}
where the series converges by Cauchy--Schwarz under the joint Gaussian law and
\eqref{eq: gaussian derivative hermite coefficient bounds}.

The covariance bounds in \eqref{eq: gaussian covariance eigenvalue bounds} imply that $\kappa I_2\preceq A,B\preceq\kappa^{-1}I_2$. Congruence by $\operatorname{diag}(A^{-1/2},B^{-1/2})$ and the singular-value decomposition therefore give
\begin{align}
    \begin{pmatrix}
        I_2&A^{-1/2}RB^{-1/2}\\
        B^{-1/2}R^\top A^{-1/2}&I_2
    \end{pmatrix}
    &=
    \begin{pmatrix}A^{-1/2}&0\\0&B^{-1/2}\end{pmatrix}
    \begin{pmatrix}A&R\\R^\top&B\end{pmatrix}
    \begin{pmatrix}A^{-1/2}&0\\0&B^{-1/2}\end{pmatrix}
    \succeq\kappa^2I_4~,
    \nonumber\\
    \|A^{-1/2}RB^{-1/2}\|_{\mathrm{op}}
    &\leq1-\kappa^2~, \quad\text{and}
    \nonumber\\
    \|A^{-1/2}RB^{-1/2}\|_{\mathrm F}
    &\leq
    \|A^{-1/2}\|_{\mathrm{op}}
    \|R\|_{\mathrm F}
    \|B^{-1/2}\|_{\mathrm{op}}
    \leq\kappa^{-1}\|R\|_{\mathrm F}~,
    \label{eq: gaussian derivative standardized covariance bounds}
\end{align}
whenever \eqref{eq: gaussian covariance eigenvalue bounds} holds.

Finally, we record two fixed-degree estimates used in Part (ii). Let $\mathcal H_m=\operatorname{span}\{H_\alpha:|\alpha|=m\}$ and let $\Pi_m$ denote the orthogonal projection onto $\mathcal H_m$. We interpret $H_\beta=0$ if a coordinate of $\beta$ is negative. For $2d\times2d$ matrices $M$ and $N$, define
\begin{align}
    (L_Mg)(z)
    &=
    \sum_{a,b=1}^{2d}M_{a,b}z_b
    \frac{\mathrm d g(z)}{\mathrm d z_a}
    \quad\text{and}\quad
    (\mathcal Q_{M,N}g)(z)
    =
    \sum_{a,b,c,e=1}^{2d}
        M_{a,b}N_{c,e}z_bz_e
        \frac{\mathrm d^2g(z)}{\mathrm d z_a\,\mathrm d z_c}~,
    \label{eq: gaussian derivative chaos operators}
\end{align}
for every polynomial $g$. Differentiating \eqref{eq: gaussian derivative hermite generating function} with respect to $z_a$ and $t_b$ gives
\begin{align}
    \frac{\mathrm d H_\alpha}{\mathrm d z_a}
    &=
    \sqrt{\alpha_a}\,
    H_{(\alpha_1,\ldots,\alpha_a-1,\ldots,\alpha_{2d})}\quad\text{and}
    \nonumber\\
    z_bH_\alpha
    &=
    \sqrt{\alpha_b+1}\,
    H_{(\alpha_1,\ldots,\alpha_b+1,\ldots,\alpha_{2d})}
    +\sqrt{\alpha_b}\,
    H_{(\alpha_1,\ldots,\alpha_b-1,\ldots,\alpha_{2d})}~,
    \label{eq: gaussian derivative hermite recurrences}
\end{align}
for every multi-index $\alpha$. Under the symmetric-tensor identification in \cref{lem: gaussian chaos transfer}, the recurrences in \eqref{eq: gaussian derivative hermite recurrences} and Cauchy-Schwarz give
\begin{align}
    \|L_Mg\|_{L^2}
    &\leq
    C(m+1)
    (\|M\|_{\mathrm{op}}+\|M\|_{\mathrm F})
    \|g\|_{L^2}~,
    \label{eq: gaussian derivative fixed degree operator bound}
\end{align}
for every $m\geq0$ and $g\in\mathcal H_m$. The product rule and two applications of \eqref{eq: gaussian derivative fixed degree operator bound}
give
\begin{align}
    \mathcal Q_{M,N}
    &=L_ML_N-L_{NM}\quad\text{and}
    \nonumber\\
    \|\mathcal Q_{M,N}g\|_{L^2}
    &\leq
    C(m+1)^2
    (\|M\|_{\mathrm{op}}+\|M\|_{\mathrm F})
    (\|N\|_{\mathrm{op}}+\|N\|_{\mathrm F})
    \|g\|_{L^2}~,
    \label{eq: gaussian derivative second operator bound}
\end{align}
for every $m\geq0$ and $g\in\mathcal H_m$.

\subsubsection{Part (i)}

At fixed $A$ and $B$, the standardized cross-covariance in \eqref{eq: gaussian derivative even hermite representation} is linear in $R$. Thus, the bounds \eqref{eq: gaussian derivative transfer bounds} and \eqref{eq: gaussian derivative standardized covariance bounds} imply that
\begin{align}
    &\left|
        \mathrm D_R^2\Psi_{\delta,d}(A,B,R)[G,H]
    \right|
    \nonumber\\
    &\quad\leq
    C\|G\|_{\mathrm F}\|H\|_{\mathrm F}
    \sum_{\substack{m\geq2\\m\ \mathrm{even}}}
    m(m-1)(1-\kappa^2)^{m-2}
    \|f_{A,m}\|_{\ell^2}\|f_{B,m}\|_{\ell^2}
    \leq
    Cp\|G\|_{\mathrm F}\|H\|_{\mathrm F}~,
    \label{eq: gaussian derivative cross hessian bound}
\end{align}
where the final inequality follows from Cauchy--Schwarz, \eqref{eq: gaussian derivative hermite coefficient bounds}, and the boundedness of $m(m-1)(1-\kappa^2)^{m-2}$. The series in \eqref{eq: gaussian derivative even hermite representation} starts at degree two, so $\mathrm D_R\Psi_{\delta,d}(A,B,0)[G]=0$. The covariance matrices on the segment from $0$ to $R$ satisfy \eqref{eq: gaussian covariance eigenvalue bounds} by convexity. Thus, we obtain the bound
\begin{align}
    |\mathrm D_R\Psi_{\delta,d}(A,B,R)[G]|
    &=\left|
        \int_0^1
        \mathrm D_R^2\Psi_{\delta,d}(A,B,tR)[R,G]
        \,\mathrm dt
    \right|
    \leq Cp\|R\|_{\mathrm F}\|G\|_{\mathrm F}~,
\end{align}
which completes the proof.\hfill\qed

\subsubsection{Part (ii)}

Let $\Omega$ satisfy $\kappa I_2/2\preceq\Omega\preceq2\kappa^{-1}I_2$, let $X_1,\ldots,X_d$ be independent $\mathsf N(0,\Omega)$ vectors, and let $P_\Omega$ and $\mathbb E_\Omega$ denote their joint law and expectation. Write $\mathcal A_{\delta,d} =\{x:|d^{-1}\sum_{j=1}^d\theta_{d,j}x_{j,1}x_{j,2}|\geq\delta\}$. For every multi-index $\alpha$, write $H_{\alpha,\Omega}(x)=H_\alpha((\Omega^{-1/2}x_j)_{j=1}^d)$. For $m\geq1$, write $f_{\Omega,m} =(\mathbb E_\Omega[ \mathbb I\{\mathcal A_{\delta,d}\}H_{\alpha,\Omega}])_{|\alpha|=m}$. The equality $\mathbb E_\Omega[H_{\alpha,\Omega}]=0$ for $|\alpha|=m\geq1$ gives $f_{\Omega,m}=f_{A,m}$ when $\Omega=A$ and $f_{\Omega,m}=f_{B,m}$ when $\Omega=B$.

Differentiating the Gaussian log density once and twice, while holding
$X_1,\ldots,X_d$ fixed, gives
\begin{align}
    \mathcal S_\Omega(H)
    &=
    \frac{1}{2}\sum_{j=1}^d
    \left(
        X_j^\top\Omega^{-1}H\Omega^{-1}X_j
        -\operatorname{tr}(\Omega^{-1}H)
    \right)~,
    \nonumber\\
    \mathcal J_\Omega(H,K)
    &=
    \mathrm D_\Omega\mathcal S_\Omega(H)[K]
    \nonumber\\
    &=
    \frac{1}{2}\sum_{j=1}^d
    \left(
        -X_j^\top\Omega^{-1}K\Omega^{-1}H\Omega^{-1}X_j \right. \nonumber\\ 
    & \left.
    \quad\quad\quad\quad\quad\quad
    -X_j^\top\Omega^{-1}H\Omega^{-1}K\Omega^{-1}X_j
        +\operatorname{tr}(\Omega^{-1}K\Omega^{-1}H)
    \right)~,
    \label{eq: gaussian derivative marginal score identities}
\end{align}
for all symmetric $2\times2$ matrices $H$ and $K$.

The resolvent representation for the inverse square root and differentiation
under its integral give
\begin{align}
    \Omega^{-1/2}
    &=
    \frac{1}{\pi}\int_0^\infty
        t^{-1/2}(\Omega+tI_2)^{-1}\,\mathrm dt~,
    \nonumber\\
    \mathrm D(\Omega^{-1/2})[H]
    &=
    -\frac{1}{\pi}\int_0^\infty
        t^{-1/2}(\Omega+tI_2)^{-1}
        H(\Omega+tI_2)^{-1}\,\mathrm dt~,
    \nonumber\\
    \mathrm D^2(\Omega^{-1/2})[H,K]
    &=
    \frac{1}{\pi}\int_0^\infty t^{-1/2}
    \left(
        (\Omega+tI_2)^{-1}H(\Omega+tI_2)^{-1}
        K(\Omega+tI_2)^{-1}
    \right.
    \nonumber\\
    &\hspace{32mm}\left.
        {}+
        (\Omega+tI_2)^{-1}K(\Omega+tI_2)^{-1}
        H(\Omega+tI_2)^{-1}
    \right)\,\mathrm dt~,
    \nonumber\\
    \|\mathrm D(\Omega^{-1/2})[H]\|_{\mathrm F}
    &\leq C\|H\|_{\mathrm F}
    \quad\text{and}\quad
    \|\mathrm D^2(\Omega^{-1/2})[H,K]\|_{\mathrm F}
    \leq C\|H\|_{\mathrm F}\|K\|_{\mathrm F}~,
    \label{eq: gaussian derivative inverse square root bounds}
\end{align}
for all symmetric $2\times2$ matrices $H$ and $K$.
The final line follows by bounding
$\|(\Omega+tI_2)^{-1}\|_{\mathrm{op}}$ by
$(\kappa/2+t)^{-1}$ in the two integrals in
\eqref{eq: gaussian derivative inverse square root bounds}. The derivatives of the Gaussian density and of $H_{\alpha,\Omega}$ are locally bounded by an integrable polynomial times a Gaussian density. Dominated convergence and the product rule, together with \eqref{eq: gaussian derivative marginal score identities}, therefore give
\begin{align}
    \bigl(\mathrm Df_{\Omega,m}[H]\bigr)_\alpha
    &=
    \mathbb E_\Omega\left[
        \mathbb I\{\mathcal A_{\delta,d}\}
        \left(
            \mathrm DH_{\alpha,\Omega}[H]
            +H_{\alpha,\Omega}\mathcal S_\Omega(H)
        \right)
    \right]\quad\text{and}
    \nonumber\\
    \bigl(\mathrm D^2f_{\Omega,m}[H,K]\bigr)_\alpha
    &=
    \mathbb E_\Omega\left[
        \mathbb I\{\mathcal A_{\delta,d}\}
        \left(
            \mathrm D^2H_{\alpha,\Omega}[H,K]
            +\mathrm DH_{\alpha,\Omega}[H]\mathcal S_\Omega(K)
        \right.
    \right.
    \nonumber\\
    &\hspace{15mm}\left.\left.
            {}+
            \mathrm DH_{\alpha,\Omega}[K]\mathcal S_\Omega(H)
            +H_{\alpha,\Omega}
            \left(
                \mathcal S_\Omega(H)\mathcal S_\Omega(K)
                +\mathcal J_\Omega(H,K)
            \right)
        \right)
    \right]~,
    \label{eq: gaussian derivative marginal coefficient identities}
\end{align}
for every $|\alpha|=m$ and all symmetric $2\times2$ matrices $H$ and $K$.

We use the following lemma to control the score terms appearing in \eqref{eq: gaussian derivative marginal coefficient identities}.

\begin{lem}[\citealp{rudelson-vershynin-2013}, Theorem~1.1]
\label{lem: gaussian quadratic form tail}$ $\\
Let $Z\sim\mathsf N(0,I_m)$ and let $Q$ be a symmetric $m\times m$ matrix. There is a universal constant $c>0$ such that
\begin{align}
    P\left\{
        |Z^\top QZ-\operatorname{tr}(Q)|>x
    \right\}
    &\leq
    2\exp\left(
        -c\min\left\{
            \frac{x^2}{\|Q\|_{\mathrm F}^2},
            \frac{x}{\|Q\|_{\mathrm{op}}}
        \right\}
    \right)
    \label{eq: gaussian derivative quadratic tail}
\end{align}
holds for every $x>0$.
\end{lem}
\noindent Let $\mathcal A$ be an event with probability at most $q\in(0,1]$. For $r\in\{1,2\}$, H\"older's inequality and the tail bound in \cref{lem: gaussian quadratic form tail}, display \eqref{eq: gaussian derivative quadratic tail}, give
\begin{align}
    \left\|
        \mathbb I\{\mathcal A\}
        (Z^\top QZ-\operatorname{tr}(Q))
    \right\|_{L^{2r}}
    &\leq
    q^{1/(2r)-1/(2r(1+\log(q^{-1})))}
    \|Z^\top QZ-\operatorname{tr}(Q)\|_{
        L^{2r(1+\log(q^{-1}))}}
    \nonumber\\
    &\leq
    Cq^{1/(2r)}
    \left(
        \sqrt{1+\log(q^{-1})}\,\|Q\|_{\mathrm F}
        +(1+\log(q^{-1}))\|Q\|_{\mathrm{op}}
    \right)~,
    \label{eq: gaussian derivative restricted quadratic moments}
\end{align}
where the second inequality uses the consequence $\|Z^\top QZ-\operatorname{tr}(Q)\|_{L^u} \leq C(\sqrt u\|Q\|_{\mathrm F}+u\|Q\|_{\mathrm{op}})$ for $u\geq1$ of \eqref{eq: gaussian derivative quadratic tail}, together with the fact that $q^{-1/(2r(1+\log(q^{-1})))}$ is uniformly bounded.

Write $X_j=\Omega^{1/2}Z_j$, where $Z=(Z_1^\top,\ldots,Z_d^\top)^\top\sim\mathsf N(0,I_{2d})$. Substitution into \eqref{eq: gaussian derivative marginal score identities}
gives
\begin{align}
    \mathcal S_\Omega(H)
    &=
    Z^\top
    \left[
        \frac{1}{2}I_d\otimes
        (\Omega^{-1/2}H\Omega^{-1/2})
    \right]Z
    -
    \operatorname{tr}\left[
        \frac{1}{2}I_d\otimes
        (\Omega^{-1/2}H\Omega^{-1/2})
    \right]~,
    \nonumber\\
    \mathcal J_\Omega(H,K)
        -\mathbb E_\Omega[\mathcal J_\Omega(H,K)]
    &=
    Z^\top
    \left[
        -\frac{1}{2}I_d\otimes
        \left(
            \begin{gathered}
            \Omega^{-1/2}K\Omega^{-1}H\Omega^{-1/2}
            \\
            {}+
            \Omega^{-1/2}H\Omega^{-1}K\Omega^{-1/2}
            \end{gathered}
        \right)
    \right]Z
    \nonumber\\
    &\quad-
    \operatorname{tr}
    \left[
        -\frac{1}{2}I_d\otimes
        \left(
            \begin{gathered}
            \Omega^{-1/2}K\Omega^{-1}H\Omega^{-1/2}
            \\
            {}+
            \Omega^{-1/2}H\Omega^{-1}K\Omega^{-1/2}
            \end{gathered}
        \right)
    \right]~,
    \nonumber\\
    \mathbb E_\Omega[\mathcal J_\Omega(H,K)]
    &=
    -\frac{d}{2}
    \operatorname{tr}(\Omega^{-1}K\Omega^{-1}H)~,
    \label{eq: gaussian derivative marginal score quadratic forms}
\end{align}
for all symmetric $2\times2$ matrices $H$ and $K$. Thus, we find that
\begin{align}
    \left\|
        \frac{1}{2}I_d\otimes
        (\Omega^{-1/2}H\Omega^{-1/2})
    \right\|_{\mathrm F}
    &\leq C\sqrt d\,\|H\|_{\mathrm F}~,
    \nonumber\\
    \left\|
        \frac{1}{2}I_d\otimes
        (\Omega^{-1/2}H\Omega^{-1/2})
    \right\|_{\mathrm{op}}
    &\leq C\|H\|_{\mathrm F}~,
    \nonumber\\
    \left\|
        -\frac{1}{2}I_d\otimes
        \left(
            \begin{gathered}
            \Omega^{-1/2}K\Omega^{-1}H\Omega^{-1/2}
            \\
            {}+\Omega^{-1/2}H\Omega^{-1}K\Omega^{-1/2}
            \end{gathered}
        \right)
    \right\|_{\mathrm F}
    &\leq
    C\sqrt d\,\|H\|_{\mathrm F}\|K\|_{\mathrm F}~,
    \nonumber\\
    \left\|
        -\frac{1}{2}I_d\otimes
        \left(
            \begin{gathered}
            \Omega^{-1/2}K\Omega^{-1}H\Omega^{-1/2}
            \\
            {}+\Omega^{-1/2}H\Omega^{-1}K\Omega^{-1/2}
            \end{gathered}
        \right)
    \right\|_{\mathrm{op}}
    &\leq
    C\|H\|_{\mathrm F}\|K\|_{\mathrm F}~,\quad\text{and}
    \nonumber\\
    d^{-1}|\mathbb E_\Omega[\mathcal J_\Omega(H,K)]|
    &\leq
    C\|H\|_{\mathrm F}\|K\|_{\mathrm F}~,
    \label{eq: gaussian derivative marginal score quadratic bounds}
\end{align}
for all symmetric $2\times2$ matrices $H$ and $K$. Apply \eqref{eq: gaussian derivative restricted quadratic moments} to the two centered quadratic forms in \eqref{eq: gaussian derivative marginal score quadratic forms}, using the matrix norm bounds in \eqref{eq: gaussian derivative marginal score quadratic bounds}. For the product of two scores, apply \eqref{eq: gaussian derivative restricted quadratic moments} with $r=2$ and then Cauchy--Schwarz. Since
$2\sqrt{d(1+\log(q^{-1}))}(1+\log(q^{-1}))
\leq d(1+\log(q^{-1}))+(1+\log(q^{-1}))^2$, we obtain
\begin{align}
    \|\mathbb I\{\mathcal A\}\mathcal S_\Omega(H)\|_{L^2(P_\Omega)}
    &\leq
    C\sqrt q
    \left(
        \sqrt{d(1+\log(q^{-1}))}
        +1+\log(q^{-1})
    \right)\|H\|_{\mathrm F}~,
    \nonumber\\
    \|\mathbb I\{\mathcal A\}
        \mathcal S_\Omega(H)\mathcal S_\Omega(K)\|_{L^2(P_\Omega)}
    &\leq
    C\sqrt q
    \left(
        d(1+\log(q^{-1}))
        +(1+\log(q^{-1}))^2
    \right)
    \|H\|_{\mathrm F}\|K\|_{\mathrm F}~,
    \nonumber\\
    \|\mathbb I\{\mathcal A\}\mathcal J_\Omega(H,K)\|_{L^2(P_\Omega)}
    &\leq
    C\sqrt q
    \left(
        d(1+\log(q^{-1}))
        +(1+\log(q^{-1}))^2
    \right)
    \|H\|_{\mathrm F}\|K\|_{\mathrm F}~,
    \label{eq: gaussian derivative restricted marginal scores}
\end{align}
for every event $\mathcal A$ satisfying $P_\Omega\{\mathcal A\}\leq q$.

With this in place, we now control the derivatives of the Hermite basis that appear in \eqref{eq: gaussian derivative marginal coefficient identities}. Differentiating \eqref{eq: gaussian derivative hermite generating function} and then setting $x_j=\Omega^{1/2}z_j$ gives
\begin{align}
    \left.
        \mathrm D_\Omega H_{\alpha,\Omega}(x)[H]
    \right|_{x_j=\Omega^{1/2}z_j}
    &=
    L_{I_d\otimes(
        \mathrm D(\Omega^{-1/2})[H]\Omega^{1/2})}
    H_\alpha(z)~,
    \nonumber\\
    \left.
        \mathrm D_\Omega^2H_{\alpha,\Omega}(x)[H,K]
    \right|_{x_j=\Omega^{1/2}z_j}
    &=
    \left(
        L_{I_d\otimes(
            \mathrm D^2(\Omega^{-1/2})[H,K]\Omega^{1/2})}
    \right.
    \nonumber\\
    &\hspace{5mm}\left.
        {}+
        \mathcal Q_{
            I_d\otimes(
                \mathrm D(\Omega^{-1/2})[H]\Omega^{1/2}),
            I_d\otimes(
                \mathrm D(\Omega^{-1/2})[K]\Omega^{1/2})
        }
    \right)H_\alpha(z)~,
    \label{eq: gaussian derivative hermite basis identities}
\end{align}
for every multi-index $\alpha$ and all symmetric $2\times2$ matrices $H,K$. For $Y\in L^2(P_\Omega)$, change variables by $x_j=\Omega^{1/2}z_j$ and take adjoints in \eqref{eq: gaussian derivative hermite basis identities}. Applying \eqref{eq: gaussian derivative inverse square root bounds}, \eqref{eq: gaussian derivative fixed degree operator bound}, \eqref{eq: gaussian derivative second operator bound}, $\|I_d\otimes E\|_{\mathrm{op}}=\|E\|_{\mathrm{op}}$, and $\|I_d\otimes E\|_{\mathrm F}=\sqrt d\|E\|_{\mathrm F}$, gives
\begin{align}
    \left(
        \sum_{|\alpha|=m}
        \left|
            \mathbb E_\Omega[
                Y\,\mathrm DH_{\alpha,\Omega}[H]]
        \right|^2
    \right)^{1/2}
    &\leq
    C\sqrt d\,(m+1)\|H\|_{\mathrm F}
        \|Y\|_{L^2(P_\Omega)}~,
    \nonumber\\
    \left(
        \sum_{|\alpha|=m}
        \left|
            \mathbb E_\Omega[
                Y\,\mathrm D^2H_{\alpha,\Omega}[H,K]]
        \right|^2
    \right)^{1/2}
    &\leq
    Cd(m+1)^2\|H\|_{\mathrm F}\|K\|_{\mathrm F}
        \|Y\|_{L^2(P_\Omega)}~,
    \label{eq: gaussian derivative fixed degree basis bounds}
\end{align}
for every $m\geq1$.

For $q\in(0,1]$, suppose that $p_{\delta,d}(\Omega)\leq q$. Bessel's inequality, \eqref{eq: gaussian derivative marginal coefficient identities}, \eqref{eq: gaussian derivative restricted marginal scores}, and \eqref{eq: gaussian derivative fixed degree basis bounds}, with $\mathcal A=\mathcal A_{\delta,d}$, give
\begin{align}
    \|f_{\Omega,m}\|_{\ell^2}
    &\leq\sqrt q~,
    \nonumber\\
    \|\mathrm Df_{\Omega,m}[H]\|_{\ell^2}
    &\leq
    C\sqrt q\,(m+1)
    \left(
        \sqrt{d(1+\log(q^{-1}))}
        +1+\log(q^{-1})
    \right)\|H\|_{\mathrm F}~\quad\text{and}
    \nonumber\\
    \|\mathrm D^2f_{\Omega,m}[H,K]\|_{\ell^2}
    &\leq
    C\sqrt q\,(m+1)^2
    \left(
        d(1+\log(q^{-1}))
        +(1+\log(q^{-1}))^2
    \right)
    \|H\|_{\mathrm F}\|K\|_{\mathrm F}~,
    \label{eq: gaussian derivative fixed degree coefficient bounds}
\end{align}
for every $m\geq1$ and all symmetric $2\times2$ matrices $H,K$. The dominated-convergence argument used above shows that $\Omega\mapsto p_{\delta,d}(\Omega)$ is continuous. Hence, there is a neighborhood $\mathcal N$ of $(A,B)$ on which the joint covariance remains uniformly well conditioned, both marginal probabilities are at most $\min\{1,2p\}$, and the standardized cross-covariance has operator norm at most $1-c$. For each of the finitely many powers used below,
\begin{align}
    q\left(1+\log(q^{-1})\right)^r
    &\lesssim
      p\left(1+\log(p^{-1})\right)^r,
    \qquad 0<q\leq\min\{1,2p\}~.
\end{align}
Thus, $q$ equal to $p_{\delta,d}(A')$ and $p_{\delta,d}(B')$ may be replaced by $p$ in \eqref{eq: gaussian derivative fixed degree coefficient bounds}.

The product rule and \eqref{eq: gaussian derivative inverse square root bounds} imply that, uniformly on $\mathcal N$,
\begin{align}
    \|I_d\otimes A'^{-1/2}RB'^{-1/2}\|_{\mathrm{op}}
    &\leq
    1-\frac{3\kappa^2}{4}~,
    \nonumber\\
    \|I_d\otimes A'^{-1/2}RB'^{-1/2}\|_{\mathrm{op}}
    & \leq C\|R\|_{\mathrm F}~,
    \nonumber\\
    \left\|
        \mathrm D_{A,B}
        (I_d\otimes A'^{-1/2}RB'^{-1/2})
        [H_A,H_B]
    \right\|_{\mathrm{op}}
    &\leq
    C\|R\|_{\mathrm F}
    (\|H_A\|_{\mathrm F}+\|H_B\|_{\mathrm F})~,
    \nonumber\\
    \left\|
        \mathrm D_{A,B}^2
        (I_d\otimes A'^{-1/2}RB'^{-1/2})
        [(H_A,H_B),(K_A,K_B)]
    \right\|_{\mathrm{op}}
    &\leq
    C\|R\|_{\mathrm F}
    (\|H_A\|_{\mathrm F}+\|H_B\|_{\mathrm F})
    \nonumber\\
    &\quad\cdot
    (\|K_A\|_{\mathrm F}+\|K_B\|_{\mathrm F})~,
    \label{eq: gaussian derivative standardized marginal derivatives}
\end{align}
for all symmetric direction matrices $H_A,H_B,K_A,K_B$. Applying \eqref{eq: gaussian derivative transfer bounds} to \eqref{eq: gaussian derivative standardized marginal derivatives} gives, for $m\geq2$,
\begin{align}
    \left\|
        \mathcal T_m(I_d\otimes A'^{-1/2}RB'^{-1/2})
    \right\|_{\mathrm{op}}
    &\leq
    C\|R\|_{\mathrm F}^2
    e^{-cm}~,
    \nonumber\\
    \left\|
        \mathrm D_{A,B}
        \mathcal T_m(I_d\otimes A'^{-1/2}RB'^{-1/2})
        [H_A,H_B]
    \right\|_{\mathrm{op}}
    &\leq
    Cm\|R\|_{\mathrm F}^2
    e^{-cm}(\|H_A\|_{\mathrm F}+\|H_B\|_{\mathrm F})~,
    \label{eq: gaussian derivative marginal transfer bounds}
\end{align}
and
\begin{align}
    &\left\|
        \mathrm D_{A,B}^2
        \mathcal T_m(I_d\otimes A'^{-1/2}RB'^{-1/2})
        [(H_A,H_B),(K_A,K_B)]
    \right\|_{\mathrm{op}} \nonumber \\
    &\leq
    Cm^2\|R\|_{\mathrm F}^2
    e^{-cm}
    (\|H_A\|_{\mathrm F}+\|H_B\|_{\mathrm F})
    (\|K_A\|_{\mathrm F}+\|K_B\|_{\mathrm F})~,
    \label{eq: gaussian derivative marginal second transfer bound}
\end{align}
uniformly for $(A',B')\in\mathcal N$.

To complete the proof, observe that the first-order product rule, \eqref{eq: gaussian derivative fixed degree coefficient bounds}, and \eqref{eq: gaussian derivative marginal transfer bounds} give
\begin{align}
    & \sup_{(A',B')\in\mathcal N}
    \left|
            \mathrm D_{A,B}\langle f_{A',m},
            \mathcal T_m(I_d\otimes A'^{-1/2}RB'^{-1/2})
            f_{B',m}\rangle[H_A,H_B]
    \right| \nonumber \\
    &\leq
    Cp
    \left(
        \sqrt{d(1+\log(p^{-1}))}
        +1+\log(p^{-1})
    \right)
    \|R\|_{\mathrm F}^2 \nonumber \\
    & \quad \quad \cdot (m+1)
    e^{-cm}
    (\|H_A\|_{\mathrm F}+\|H_B\|_{\mathrm F})~,
    \label{eq: gaussian derivative marginal first summand derivative bound}
\end{align}
for every even $m\geq2$. The second-order product rule and the same bounds give
\begin{align}
    & \sup_{(A',B')\in\mathcal N}
    \left|
            \mathrm D_{A,B}^2\langle f_{A',m},
            \mathcal T_m(I_d\otimes A'^{-1/2}RB'^{-1/2})
            f_{B',m}\rangle
            [(H_A,H_B),(K_A,K_B)]
    \right| \nonumber \\
    &\leq
    Cp
    \left(
        d(1+\log(p^{-1}))
        +(1+\log(p^{-1}))^2
    \right) \nonumber \\
    & \quad\quad\|R\|_{\mathrm F}^2
    (m+1)^2
    e^{-cm}
    (\|H_A\|_{\mathrm F}+\|H_B\|_{\mathrm F})
    (\|K_A\|_{\mathrm F}+\|K_B\|_{\mathrm F})~,
    \label{eq: gaussian derivative marginal second summand derivative bound}
\end{align}
for every even $m\geq2$. The bounds in \eqref{eq: gaussian derivative marginal first summand derivative bound}-\eqref{eq: gaussian derivative marginal second summand derivative bound} are summable uniformly on every closed product neighborhood of $(A,B)$ contained in $\mathcal N$. Hence, \eqref{eq: gaussian derivative even hermite representation} may be differentiated term by term twice. Summing the resulting bounds gives
\begin{align}
    |\mathrm D_{A,B}\Psi_{\delta,d}(A,B,R)[H_A,H_B]|
    &\leq
    Cp
    \left(
        \sqrt{d(1+\log(p^{-1}))}
        +1+\log(p^{-1})
    \right)
    \|R\|_{\mathrm F}^2
    \nonumber\\
    &\quad\cdot
    (\|H_A\|_{\mathrm F}+\|H_B\|_{\mathrm F})~,\quad\text{and}
    \nonumber\\
    \left|
            \mathrm D_{A,B}^2\Psi_{\delta,d}(A,B,R)
            [(H_A,H_B),(K_A,K_B)]
    \right| 
    &\leq
    Cp
    \left(
        d(1+\log(p^{-1}))
        +(1+\log(p^{-1}))^2
    \right)
    \|R\|_{\mathrm F}^2
    \nonumber\\
    &\quad\cdot
    (\|H_A\|_{\mathrm F}+\|H_B\|_{\mathrm F})
    (\|K_A\|_{\mathrm F}+\|K_B\|_{\mathrm F})~,
    \label{eq: gaussian derivative marginal final bounds}
\end{align}
as required.

\subsubsection{Part (iii)}

Write $R^{q,r}=r_{q,r}e_qe_r^\top$. Since $\Psi_{\delta,d}(A,B,0)=0$ and $R=\sum_{q,r=1}^2R^{q,r}$, two applications of the fundamental theorem of calculus give
\begin{align}
    &\Psi_{\delta,d}(A,B,R)
    -\sum_{q,r=1}^2\Psi_{\delta,d}(A,B,R^{q,r})
    \nonumber\\
    &=\sum_{q,r=1}^2\int_0^1\int_0^1
        t\,\mathrm D_R^2\Psi_{\delta,d}
        \left(A,B,t((1-u)R^{q,r}+uR)\right)
        [R-R^{q,r},R^{q,r}]
        \,\mathrm du\,\mathrm dt~.
    \label{eq: gaussian derivative radial exact integral}
\end{align}
Every covariance matrix in \eqref{eq: gaussian derivative radial exact integral} belongs to \eqref{eq: gaussian radial interaction domain}. Part (i), display \eqref{eq: gaussian cross block derivative bounds}, therefore gives
\begin{align}
    &\left|
        \Psi_{\delta,d}(A,B,R)
        -\sum_{q,r=1}^2\Psi_{\delta,d}(A,B,R^{q,r})
    \right|
    \nonumber\\
    &\lesssim 
    p\sum_{q,r=1}^2
        \|R-R^{q,r}\|_{\mathrm F}\|R^{q,r}\|_{\mathrm F}
    \lesssim
    p\sum_{q,r=1}^2|r_{q,r}|
    \left(
        \sum_{q',r'=1}^2|r_{q',r'}|-|r_{q,r}|
    \right)~,
\end{align}
as required.
\hfill\qed

\subsubsection{Part (iv)}

Put $F=e_1e_2^\top+e_2e_1^\top$ and $J_q=2e_qe_q^\top-I_2$. Changing the sign of the Gaussian coordinate not indexed by $q$ in the first margin gives
\begin{align}
    \Psi_{\delta,d}(I_2+tF,I_2,xe_qe_r^\top)
    &=\Psi_{\delta,d}
        (J_q(I_2+tF)J_q,I_2,J_qxe_qe_r^\top)
    \nonumber\\
    &=\Psi_{\delta,d}(I_2-tF,I_2,xe_qe_r^\top)~.
\end{align}
The corresponding sign change by $J_r$ in the second margin gives
\begin{align}
    \Psi_{\delta,d}(I_2,I_2+tF,xe_qe_r^\top)
    &=\Psi_{\delta,d}
        (I_2,J_r(I_2+tF)J_r,xe_qe_r^\top J_r)
    \nonumber\\
    &=\Psi_{\delta,d}(I_2,I_2-tF,xe_qe_r^\top)~.
\end{align}
Both functions of $t$ are even. Differentiating them at $t=0$ proves the two identities in \eqref{eq: gaussian canonical off diagonal derivatives}. Moreover, Part (ii), display \eqref{eq: gaussian marginal first derivative bound}, with $p=1$ gives
\begin{align}
    &|\mathrm D_{A,B}\Psi_{\delta,d}
        (I_2,I_2,xe_qe_r^\top)[H_A,H_B]|
    \nonumber\\
    &\quad\leq
    C(\sqrt d+1)x^2
    (\|H_A\|_{\mathrm F}+\|H_B\|_{\mathrm F})
    \leq
    C\sqrt d\,x^2
    (\|H_A\|_{\mathrm F}+\|H_B\|_{\mathrm F})~,
\end{align}
as required.
\hfill\qed

\subsection{Proof of \cref{lem: Gaussian three row}}

We continue to use the notation introduced in \cref{lem: gaussian threshold covariance bounds}. Let $K_t=I_3+t(K-I_3)$. For each $t\in[0,1]$, let $P_t$ denote the joint law of independent vectors $Z_1,\ldots,Z_d\sim\mathsf N(0,K_t)$, and let $\mathbb E_t$ denote expectation under $P_t$. Differentiating the logarithm of the joint Lebesgue density of $Z_1,\ldots,Z_d$ with respect to $t$, while holding their observed values fixed, gives
\begin{align}
    \mathcal S_t
    &=\frac{1}{2}\sum_{j=1}^d
    \left(
        Z_j^\top K_t^{-1}(K-I_3)K_t^{-1}Z_j
        -\operatorname{tr}(K_t^{-1}(K-I_3))
    \right)~.
    \label{eq: gaussian derivative three row score}
\end{align}
Differentiating $\mathcal S_t$ with respect to $t$, again holding $Z_1,\ldots,Z_d$ fixed, gives
\begin{align}
    \frac{\mathrm d\mathcal S_t}{\mathrm d t}
    &=-\sum_{j=1}^d
        Z_j^\top K_t^{-1}(K-I_3)K_t^{-1}(K-I_3)K_t^{-1}Z_j
        +
        \frac{d}{2}\operatorname{tr}\left(
            K_t^{-1}(K-I_3)K_t^{-1}(K-I_3)
        \right)~.
    \label{eq: gaussian derivative three row score derivative}
\end{align}
Under $P_t$, write $Z_j=K_t^{1/2}W_j$, where $W_1,\ldots,W_d$ are independent $\mathsf N(0,I_3)$ vectors, and set $W=(W_1^\top,\ldots,W_d^\top)^\top$. Define
\begin{align}
    Q_{\mathcal S_t}
    &=\frac{1}{2}I_d\otimes
        \left(K_t^{-1/2}(K-I_3)K_t^{-1/2}\right)\quad\text{and}
    \nonumber\\
    \widetilde Q_{\mathcal S_t}
    &=-I_d\otimes\left(
        K_t^{-1/2}(K-I_3)K_t^{-1}(K-I_3)K_t^{-1/2}
    \right)~.
    \label{eq: gaussian derivative three row quadratic matrices}
\end{align}
Substitution into \eqref{eq: gaussian derivative three row score} and \eqref{eq: gaussian derivative three row score derivative} gives
\begin{align}
    \mathcal S_t
    &=W^\top Q_{\mathcal S_t}W
        -\operatorname{tr}(Q_{\mathcal S_t})~,
    \nonumber\\
    \frac{\mathrm d\mathcal S_t}{\mathrm d t}
        -\mathbb E_t[\frac{\mathrm d\mathcal S_t}{\mathrm d t}]
    &=W^\top \widetilde Q_{\mathcal S_t}W
        -\operatorname{tr}(\widetilde Q_{\mathcal S_t})~,\quad\text{and}
    \nonumber\\
    \mathbb E_t[\frac{\mathrm d\mathcal S_t}{\mathrm d t}]
    &=-\frac{d}{2}\operatorname{tr}\left(
        K_t^{-1}(K-I_3)K_t^{-1}(K-I_3)
    \right)~.
    \label{eq: gaussian derivative three row centered quadratics}
\end{align}
The eigenvalue bounds for $K_t$ in \eqref{eq: gaussian three row covariance path} imply
\begin{align}
    \|Q_{\mathcal S_t}\|_{\mathrm{op}}
    &\leq C\|K-I_3\|_{\mathrm F}~,
    \qquad
    \|Q_{\mathcal S_t}\|_{\mathrm F}
    \leq C\sqrt d\,\|K-I_3\|_{\mathrm F}~,
    \nonumber\\
    \|\widetilde Q_{\mathcal S_t}\|_{\mathrm{op}}
    &\leq C\|K-I_3\|_{\mathrm F}^2~,
    \qquad
    \|\widetilde Q_{\mathcal S_t}\|_{\mathrm F}
    \leq C\sqrt d\,\|K-I_3\|_{\mathrm F}^2~,\quad\text{and}
    \nonumber\\
    |\mathbb E_t[\frac{\mathrm d\mathcal S_t}{\mathrm d t}]|
    &\leq Cd\|K-I_3\|_{\mathrm F}^2~.
    \label{eq: gaussian derivative three row score bounds}
\end{align}
Fix $t\in[0,1]$ and a Borel event $E$ satisfying $P_t\{E\}\leq p$. Cauchy--Schwarz and the triangle inequality give
\begin{align}
    \mathbb E_t[\mathbb I\{E\}|\mathcal S_t|]
    &\leq
    \sqrt p\,
    \mathbb E_t[\mathbb I\{E\}\mathcal S_t^2]^{1/2}\quad\text{and}
    \nonumber\\
    \mathbb E_t[\mathbb I\{E\}|\frac{\mathrm d\mathcal S_t}{\mathrm d t}|]
    &\leq
    \sqrt p\,
    \mathbb E_t\left[
        \mathbb I\{E\}
        |\frac{\mathrm d\mathcal S_t}{\mathrm d t}
            -\mathbb E_t[\frac{\mathrm d\mathcal S_t}{\mathrm d t}]|^2
    \right]^{1/2}
    +p|\mathbb E_t[\frac{\mathrm d\mathcal S_t}{\mathrm d t}]|~.
    \label{eq: gaussian derivative three row moment reduction}
\end{align}
Apply \eqref{eq: gaussian derivative restricted quadratic moments} to the two centered quadratic forms, and use \eqref{eq: gaussian derivative three row score bounds} in the resulting bounds. Together with \eqref{eq: gaussian derivative three row moment reduction}, we obtain
\begin{align}
    \mathbb E_t[\mathbb I\{E\}|\mathcal S_t|]
    &\leq
    Cp\left(
        \sqrt{d(1+\log(p^{-1}))}+1+\log(p^{-1})
    \right)\|K-I_3\|_{\mathrm F}\quad\text{and}
    \nonumber\\
    \mathbb E_t\left[
        \mathbb I\{E\}
        (\mathcal S_t^2+|\frac{\mathrm d\mathcal S_t}{\mathrm d t}|)
    \right]
    &\leq
    Cp\left(
        d(1+\log(p^{-1}))+(1+\log(p^{-1}))^2
    \right)\|K-I_3\|_{\mathrm F}^2~.
    \label{eq: gaussian derivative three row restricted scores}
\end{align}
respectively. 

Now, for $z=(z_1^\top,\ldots,z_d^\top)^\top\in\mathbb R^{3d}$ with $z_j\in\mathbb R^3$ and $r\in\{2,3\}$, define the event
\begin{align}
    E_{1,r}
    &=\left\{z:
        \left|
            \frac{1}{d}\sum_{j=1}^d
            \theta_{d,j}z_{j,1}z_{j,r}
        \right|\geq\delta
    \right\}~.
    \label{eq: gaussian derivative three row threshold events}
\end{align}
The eigenvalue bounds in \eqref{eq: gaussian three row covariance path}
imply that the joint Gaussian density and its first two derivatives with
respect to $t$ are dominated, uniformly over $t\in[0,1]$, by an integrable
function. Differentiating under the integral sign and applying
\eqref{eq: gaussian derivative three row score}--\eqref{eq: gaussian derivative three row score derivative}
therefore give
\begin{align}
    \frac{\mathrm d}{\mathrm d t}P_t\{E\}
    &=\mathbb E_t[\mathbb I\{E\}\mathcal S_t],
    \quad\text{and}\quad
    \frac{\mathrm d^2}{\mathrm d t^2}P_t\{E\}
    =\mathbb E_t\left[
        \mathbb I\{E\}(\mathcal S_t^2+\frac{\mathrm d\mathcal S_t}{\mathrm d t})
    \right]
    \label{eq: gaussian derivative three row probability derivatives}
\end{align}
for any Borel event $E$. The two events in \eqref{eq: gaussian derivative three row threshold events} satisfy
\begin{align}
    \Psi_{(1,2),(1,3)}(K_t,\delta)
    &=P_t\{E_{1,2}\cap E_{1,3}\}
        -P_t\{E_{1,2}\}P_t\{E_{1,3}\}~.
\end{align}
Differentiating this identity twice gives
\begin{align}
    \frac{\mathrm d^2}{\mathrm d t^2}
    \Psi_{(1,2),(1,3)}(K_t,\delta)
    &=\frac{\mathrm d^2}{\mathrm d t^2}P_t\{E_{1,2}\cap E_{1,3}\}
    -\left(\frac{\mathrm d^2}{\mathrm d t^2}P_t\{E_{1,2}\}\right)
        P_t\{E_{1,3}\}
    \nonumber\\
    &\quad-2\left(\frac{\mathrm d}{\mathrm d t}P_t\{E_{1,2}\}\right)
        \left(\frac{\mathrm d}{\mathrm d t}P_t\{E_{1,3}\}\right)
    -P_t\{E_{1,2}\}
        \left(\frac{\mathrm d^2}{\mathrm d t^2}P_t\{E_{1,3}\}\right)~.
    \label{eq: gaussian derivative three row covariance derivative}
\end{align}
The bound \eqref{eq: gaussian three row probability envelope} gives $P_t\{E_{1,2}\}\leq p$ and $P_t\{E_{1,3}\}\leq p$, and that thereby $P_t\{E_{1,2}\cap E_{1,3}\}\leq p$. Apply \eqref{eq: gaussian derivative three row restricted scores} and \eqref{eq: gaussian derivative three row probability derivatives} to each of these three events. The two first-derivative factors in the third term of \eqref{eq: gaussian derivative three row covariance derivative} have product bounded by
\begin{align}
    p^2\left(
        \sqrt{d(1+\log(p^{-1}))}+1+\log(p^{-1})
    \right)^2
    &\leq
    Cp\left(
        d(1+\log(p^{-1}))+(1+\log(p^{-1}))^2
    \right)~.\label{eq: gaussian derivative three row second derivative bound}
\end{align}
Thus, we find that
\begin{align}
    \left|
        \frac{\mathrm d^2}{\mathrm d t^2}
        \Psi_{(1,2),(1,3)}(K_t,\delta)
    \right|
    &\leq
    Cp\left(
        d(1+\log(p^{-1}))+(1+\log(p^{-1}))^2
    \right)\|K-I_3\|_{\mathrm F}^2
    \label{eq: gaussian derivative three row second derivative}
\end{align}
by multiplying the bound \eqref{eq: gaussian derivative three row second derivative bound} by $\|K-I_3\|_{\mathrm F}^2$ and substituting into \eqref{eq: gaussian derivative three row covariance derivative}

The same density-domination argument, applied to every matrix entry on a neighborhood of $I_3$, shows that $K\mapsto\Psi_{(1,2),(1,3)}(K,\delta)$ is Fr\'echet differentiable at $I_3$ and that its derivative in direction $H$ is the corresponding path derivative. Fix $1\leq q<r\leq3$ and set $H=e_qe_r^\top+e_re_q^\top$. Choose a diagonal matrix $J$ whose diagonal entries belong to $\{-1,1\}$ such that $JHJ=-H$. The transformation $Z_j\mapsto JZ_j$ leaves both threshold events unchanged and replaces the covariance matrix $I_3+tH$ by $J(I_3+tH)J=I_3-tH$. Thus, we have
\begin{align}
    \Psi_{(1,2),(1,3)}(I_3-tH,\delta)
    &=\Psi_{(1,2),(1,3)}(I_3+tH,\delta)~.
    \label{eq: gaussian derivative three row sign symmetry}
\end{align}
Differentiating this identity at $t=0$ gives
\begin{align}
    \mathrm D\Psi_{(1,2),(1,3)}(I_3,\delta)[H]
    &=0~.
\end{align}
The Fr\'echet derivative is linear in its direction, so it vanishes on every symmetric matrix with zero diagonal. Decomposing an arbitrary symmetric matrix $H$ into its diagonal and zero-diagonal parts gives
\begin{align}
    \mathrm D\Psi_{(1,2),(1,3)}(I_3,\delta)[H]
    &=\mathrm D\Psi_{(1,2),(1,3)}(I_3,\delta)
        [\mathsf{diag}(H_{11},H_{22},H_{33})]~.
\end{align}
For an arbitrary symmetric $H$, apply \cref{lem: finite gaussian covariance comparison} with $m=3$, $K_0=I_3$, $K_1=I_3+tH$, $E=E_{1,2}$, and $F=E_{1,3}$. For sufficiently small $|t|$, the covariance matrices satisfy \eqref{eq: finite gaussian comparison domain}. Hence, the bound \eqref{eq: finite gaussian covariance transport} gives
\begin{align}
    \left|
        \Psi_{(1,2),(1,3)}(I_3+tH,\delta)
        -\Psi_{(1,2),(1,3)}(I_3,\delta)
    \right|
    &\leq C\sqrt d\,|t|\|H\|_{\mathrm F}~.
\end{align}
Dividing by $|t|$ and letting $t\to0$ gives
\begin{align}
    \left|
        \mathrm D\Psi_{(1,2),(1,3)}(I_3,\delta)[H]
    \right|
    &\leq C\sqrt d\,\|H\|_{\mathrm F}~,
\end{align}
as required. 

Taylor's formula with integral remainder gives the exact identity
\begin{align}
    &\Psi_{(1,2),(1,3)}(K,\delta)
        -\Psi_{(1,2),(1,3)}(I_3,\delta)
        -\mathrm D\Psi_{(1,2),(1,3)}(I_3,\delta)[K-I_3] \nonumber \\
    &=\int_0^1(1-t)
        \frac{\mathrm d^2}{\mathrm d t^2}
        \Psi_{(1,2),(1,3)}(K_t,\delta)\,\mathrm dt~.
\end{align}
The first identity in \eqref{eq: gaussian three row linear term} replaces the derivative on the left-hand side by its value in the direction $\mathsf{diag}(K_{11}-1,K_{22}-1,K_{33}-1)$. Taking absolute values, applying \eqref{eq: gaussian derivative three row second derivative} with the same $p$ for every $t\in[0,1]$, and using $\int_0^1(1-t)\,\mathrm dt=1/2$ give
\begin{align}
    & \Bigg|
            \Psi_{(1,2),(1,3)}(K,\delta)
            -\Psi_{(1,2),(1,3)}(I_3,\delta) -\mathrm D\Psi_{(1,2),(1,3)}(I_3,\delta)
            [\mathsf{Diag}(K_{11}-1,K_{22}-1,K_{33}-1)]
        \Bigg| \nonumber \\
    &\leq
    Cp\left(
        d(1+\log(p^{-1}))+(1+\log(p^{-1}))^2
    \right)
    \|K-I_3\|_{\mathrm F}^2~,
\end{align}
as required.\hfill\qed

\subsection{Proof of \cref{lem: fixed bank sign comparison}}

Write $H=K-I_r$, and let $\mathcal J_r$ be the set of diagonal matrices
whose diagonal entries belong to $\{-1,1\}$. For $J\in\mathcal J_r$
and $t\in[0,1]$, define
\begin{align}
    K_{J}(t)
    &=I_r+tJHJ
    =(1-t)I_r+tJKJ~.
    \label{eq: fixed bank sign covariance path}
\end{align}
By \eqref{eq: fixed bank sign covariance condition}, uniformly over
$J\in\mathcal J_r$ and $t\in[0,1]$, it holds that
\begin{align}
    cI_r
    &\preceq K_{J}(t)\preceq c^{-1}I_r
    \quad\text{and}\quad
    \|JHJ\|_{\mathrm F}
    =\|H\|_{\mathrm F}~.
    \label{eq: fixed bank sign covariance path bounds}
\end{align}
Consequently, it holds that
\begin{align}
    \operatorname{tr}\left(
      (K_{J}(t)^{-1}JHJ)^2
    \right)
    &=
      \left\|
        K_{J}(t)^{-1/2}JHJK_{J}(t)^{-1/2}
      \right\|_{\mathrm F}^2
    \lesssim\|H\|_{\mathrm F}^2~.
    \label{eq: fixed bank sign trace bound}
\end{align}
Now, let $P_0=P_{I_r}$, and let $L_{J,t}$ be the likelihood ratio of $P_{K_{J}(t)}$ relative to $P_0$, given by
\begin{align}
    L_{J,t}(z_1,\ldots,z_m)
    &=\det(K_{J}(t))^{-m/2}
      \exp\left(
        -\frac12\sum_{l=1}^m
        z_l^\top(K_{J}(t)^{-1}-I_r)z_l
      \right)~.
    \label{eq: fixed bank sign likelihood}
\end{align}
Write $\ell_{J,t}=\log L_{J,t}$. Differentiating
\eqref{eq: fixed bank sign likelihood} gives
\begin{align}
    \frac{\mathrm d\ell_{J,t}}{\mathrm d t}
    &=\frac12\sum_{l=1}^m
      \left(
        z_l^\top K_{J}(t)^{-1}JHJK_{J}(t)^{-1}z_l
        -\operatorname{tr}(K_{J}(t)^{-1}JHJ)
      \right)~,
    \nonumber\\
    \frac{\mathrm d^2\ell_{J,t}}{\mathrm d t^2}
    &=\frac m2\operatorname{tr}\left(
        (K_{J}(t)^{-1}JHJ)^2
      \right)
    \nonumber\\
    &\quad-
      \sum_{l=1}^m z_l^\top
      K_{J}(t)^{-1}JHJK_{J}(t)^{-1}JHJK_{J}(t)^{-1}z_l~,\quad\text{and}
    \nonumber\\
    \frac{\mathrm d^2L_{J,t}}{\mathrm d t^2}
    &=L_{J,t}\left(
      \left(\frac{\mathrm d\ell_{J,t}}{\mathrm d t}\right)^2
      +\frac{\mathrm d^2\ell_{J,t}}{\mathrm d t^2}
      \right)~.
    \label{eq: fixed bank sign likelihood derivatives}
\end{align}
Under $P_{K_{J}(t)}$, each summand in the first line of \eqref{eq: fixed bank sign likelihood derivatives} has mean zero.
Moreover, it holds that
\begin{align}
    &K_{J}(t)^{-1}JHJK_{J}(t)^{-1}JHJK_{J}(t)^{-1}
    \nonumber\\
    &\quad=
      K_{J}(t)^{-1/2}
      \left(
        K_{J}(t)^{-1/2}JHJK_{J}(t)^{-1/2}
      \right)^2
      K_{J}(t)^{-1/2}
    \succeq0~.
    \label{eq: fixed bank sign second derivative positivity}
\end{align}
Thus, \cref{lem: wick} and \eqref{eq: fixed bank sign trace bound} imply that
\begin{align}
    \left\|
      \frac{\mathrm d\ell_{J,t}}{\mathrm d t}
    \right\|_{L^2(P_{K_{J}(t)})}^2
    &=
      \frac m2\operatorname{tr}\left(
        (K_{J}(t)^{-1}JHJ)^2
      \right)
    \lesssim m\|H\|_{\mathrm F}^2\quad\text{and}
    \nonumber\\
    \left\|
      \frac{\mathrm d^2\ell_{J,t}}{\mathrm d t^2}
    \right\|_{L^1(P_{K_{J}(t)})}
    &\leq
      \frac{3m}{2}\operatorname{tr}\left(
        (K_{J}(t)^{-1}JHJ)^2
      \right)
    \lesssim m\|H\|_{\mathrm F}^2~.
    \label{eq: fixed bank sign logarithm derivative moments}
\end{align}
Since $L_{J,t}\,\mathrm dP_0=\mathrm dP_{K_{J}(t)}$, the last line of \eqref{eq: fixed bank sign likelihood derivatives}
and \eqref{eq: fixed bank sign logarithm derivative moments} imply that
\begin{align}
    & \sup_{J\in\mathcal J_r}\sup_{t\in[0,1]}
    \left\|
      \frac{\mathrm d^2L_{J,t}}{\mathrm d t^2}
    \right\|_{L^1(P_0)} \nonumber \\
    &\leq
      \sup_{J\in\mathcal J_r}\sup_{t\in[0,1]}
      \left(
        \left\|
          \frac{\mathrm d\ell_{J,t}}{\mathrm d t}
        \right\|_{L^2(P_{K_{J}(t)})}^2
        +
        \left\|
          \frac{\mathrm d^2\ell_{J,t}}{\mathrm d t^2}
        \right\|_{L^1(P_{K_{J}(t)})}
      \right)
    \lesssim m\|H\|_{\mathrm F}^2~.
    \label{eq: fixed bank sign likelihood derivative bound}
\end{align}
Because $K$ is a correlation matrix, $H$ has zero diagonal. Therefore, it holds that
\begin{align}
    \operatorname{tr}(JHJ)
    &=0
    \quad\text{and}\quad
    \frac{1}{2^r}\sum_{J\in\mathcal J_r}JHJ
    =0~.
\end{align}
Since $L_{J,0}=1$, the first line of \eqref{eq: fixed bank sign likelihood derivatives} now gives
\begin{align}
    \frac{1}{2^r}\sum_{J\in\mathcal J_r}
    \left.
      \frac{\mathrm d L_{J,t}}{\mathrm d t}
    \right|_{t=0}
    &=\frac12\sum_{l=1}^m
      z_l^\top
      \left(
        \frac{1}{2^r}\sum_{J\in\mathcal J_r}JHJ
      \right)z_l
    =0~.
    \label{eq: fixed bank sign first derivative cancellation}
\end{align}
For each $J\in\mathcal J_r$, Taylor's formula with integral remainder gives
\begin{align}
    L_{J,1}
    &=1+
      \left.
        \frac{\mathrm d L_{J,t}}{\mathrm d t}
      \right|_{t=0}
      +
      \int_0^1(1-t)
      \frac{\mathrm d^2L_{J,t}}{\mathrm d t^2}\,\mathrm dt~.
    \label{eq: fixed bank sign taylor identity}
\end{align}
Averaging \eqref{eq: fixed bank sign taylor identity} over $J\in\mathcal J_r$, applying \eqref{eq: fixed bank sign first derivative cancellation}, and using \eqref{eq: fixed bank sign likelihood derivative bound} yield
\begin{align}
    \left\|
      \frac{1}{2^r}\sum_{J\in\mathcal J_r}L_{J,1}-1
    \right\|_{L^1(P_0)}
    &\leq
      \frac{1}{2^r}\sum_{J\in\mathcal J_r}
      \int_0^1(1-t)
      \left\|
        \frac{\mathrm d^2L_{J,t}}{\mathrm d t^2}
      \right\|_{L^1(P_0)}\,\mathrm dt
    \nonumber\\
    &\lesssim m\|H\|_{\mathrm F}^2~.
    \label{eq: fixed bank sign mixture bound}
\end{align}
Since $K_{J}(1)=JKJ$ and $E$ is invariant under the corresponding coordinate sign changes,
\begin{align}
    P_K\{E\}
    &=\frac{1}{2^r}\sum_{J\in\mathcal J_r}P_{JKJ}\{E\}
    =\int_E
      \frac{1}{2^r}\sum_{J\in\mathcal J_r}L_{J,1}\,\mathrm dP_0~.
\end{align}
Therefore, \eqref{eq: fixed bank sign mixture bound} implies that
\begin{align}
    \left|P_K\{E\}-P_{I_r}\{E\}\right|
    &\leq
      \left\|
        \frac{1}{2^r}\sum_{J\in\mathcal J_r}L_{J,1}-1
      \right\|_{L^1(P_0)}
    \lesssim m\|K-I_r\|_{\mathrm F}^2~.
\end{align}
Combining this result with the bound $\left|P_K\{E\}-P_{I_r}\{E\}\right|\leq1$ completes the proof.\hfill\qed

\subsection{Proof of \cref{lem: high rank relative crossing}}

We first establish a density bound used in both parts of the Lemma. Fix symmetric matrices $A$ and $B$, where $A$ satisfies \eqref{eq: fixed bank high rank spectral conditions}, and let $Z$ be a standard Gaussian vector in $\mathbb R^n$. For every Borel set $S\subseteq\mathbb R$, define
\begin{align}
    \nu_0(S)
    &=P\left\{
      Z^\top AZ-\operatorname{tr}(A)\in S
    \right\}~,
    \nonumber\\
    \nu_k(S)
    &=\mathbb E\left[
      |Z^\top BZ|^k
      \mathbb I\left\{
        Z^\top AZ-\operatorname{tr}(A)\in S
      \right\}
    \right]~,
    \qquad k\in\{1,2\}~.
    \label{eq: fixed bank high rank weighted measures}
\end{align}
For $z\in\mathbb R$ and $\zeta$ in a neighborhood of zero, define
\begin{align}
    \varphi(z,\zeta)
    &=\mathbb E\left[
      \exp\left(
        iz\left(
          Z^\top AZ-\operatorname{tr}(A)
        \right)
        +\zeta Z^\top BZ
      \right)
    \right]
    \nonumber\\
    &=\exp\left(-iz\operatorname{tr}(A)\right)
      \det(I_n-2izA-2\zeta B)^{-1/2}~,
    \label{eq: fixed bank high rank gaussian transform}
\end{align}
where the second equality follows from the Gaussian quadratic-form transform used in \citealp{muirhead1982}, Proof of Theorem 1.4.2. By writing
\begin{align}
    \widehat{\nu}_k(z)
    &=\int_{\mathbb R}e^{izw}\,\mathrm d\nu_k(w)~,
\end{align}
we have
\begin{align}
    \widehat{\nu}_0(z)
    &=\varphi(z,0)
    \quad\text{and}\quad
    \widehat{\nu}_2(z)
    =\left.
      \frac{\mathrm d^2\varphi(z,\zeta)}
      {\mathrm d\zeta^2}
      \right|_{\zeta=0}~.
\end{align}
Differentiating \eqref{eq: fixed bank high rank gaussian transform}
gives
\begin{align}
    \widehat{\nu}_2(z)
    &=\widehat{\nu}_0(z)\Bigg(
      2\operatorname{tr}\left(
        B(I_n-2izA)^{-1}B(I_n-2izA)^{-1}
      \right)
    +
      \operatorname{tr}\left(
        B(I_n-2izA)^{-1}
      \right)^2
      \Bigg)~.
    \label{eq: fixed bank high rank weighted transform}
\end{align}
Let $\alpha_1,\ldots,\alpha_n$ denote the eigenvalues of $A$. The bound \eqref{eq: fixed bank high rank spectral conditions} implies that
\begin{align}
    |\widehat{\nu}_0(z)|
    &=\prod_{q=1}^n
      \left(1+4z^2\alpha_q^2\right)^{-1/4}
    \leq
      \left(
        1+c\frac{z^2}{d}
      \right)^{-cd}~.
    \label{eq: fixed bank high rank transform bound}
\end{align}
For real $z$, it holds that
\begin{align}
    \|(I_n-2izA)^{-1}\|_{\mathrm{op}}
    &\leq1
    \quad\text{and}\quad
    \|A(I_n-2izA)^{-1}\|_{\mathrm F}
    \lesssim1~.
\end{align}
Moreover, it holds that
\begin{align}
    \left|
      \operatorname{tr}\left(
        B(I_n-2izA)^{-1}
      \right)
    \right|
    &=\left|
      \operatorname{tr}(B)
      +2iz\operatorname{tr}\left(
        BA(I_n-2izA)^{-1}
      \right)
    \right|
    \nonumber\\
    &\lesssim
      |\operatorname{tr}(B)|+|z|\|B\|_{\mathrm F}~,\quad\text{and}
    \nonumber\\
    \left|
      \operatorname{tr}\left(
        B(I_n-2izA)^{-1}B(I_n-2izA)^{-1}
      \right)
    \right|
    &\leq\|B\|_{\mathrm F}^2~.
\end{align}
Combining these bounds with \eqref{eq: fixed bank high rank weighted transform} yields
\begin{align}
    |\widehat{\nu}_2(z)|
    &\lesssim
      \left(
        (1+z^2)\|B\|_{\mathrm F}^2
        +|\operatorname{tr}(B)|^2
      \right)
      \left(
        1+c\frac{z^2}{d}
      \right)^{-cd}~.
    \label{eq: fixed bank high rank weighted transform bound}
\end{align}
By splitting the integral at $\sqrt d$, we can evaluate
\begin{align}
    &\int_{\mathbb R}(1+z^2)
      \left(
        1+c\frac{z^2}{d}
      \right)^{-cd}\,\mathrm dz
    \nonumber\\
    &\quad\lesssim
      \int_{|z|\leq\sqrt d}
      (1+z^2)e^{-cz^2}\,\mathrm dz
      +
      d^{3/2}e^{-cd}
      \int_1^\infty s^{2-cd}\,\mathrm ds
    \lesssim1~.
    \label{eq: fixed bank high rank transform integral}
\end{align}
Therefore, we have that
\begin{align}
    \int_{\mathbb R}|\widehat{\nu}_0(z)|\,\mathrm dz
    &\lesssim1
    \quad\text{and}\quad
    \int_{\mathbb R}|\widehat{\nu}_2(z)|\,\mathrm dz
    \lesssim
      \|B\|_{\mathrm F}^2
      +|\operatorname{tr}(B)|^2~.
    \label{eq: fixed bank high rank transform integrability}
\end{align}
The inversion theorem for finite measures with integrable Fourier transforms (\citealp{feller1991introduction}, Chapter XV.3) implies that $\nu_0$ and $\nu_2$ have continuous densities $f_0$ and $f_2$, satisfying
\begin{align}
    \|f_0\|_\infty
    &\lesssim1
    \quad\text{and}\quad
    \|f_2\|_\infty
    \lesssim
      \|B\|_{\mathrm F}^2
      +|\operatorname{tr}(B)|^2~.
    \label{eq: fixed bank high rank squared density bounds}
\end{align}
Since $\nu_1$ is absolutely continuous with respect to $\nu_0$, it also has a Lebesgue density, denoted by $f_1$. Conditional Cauchy-Schwarz applied to
\eqref{eq: fixed bank high rank weighted measures} gives
\begin{align}
    f_1(w)^2
    &\leq f_0(w)f_2(w)
\end{align}
for almost every $w$. Hence, it holds that
\begin{align}
    \|f_1\|_\infty
    &\lesssim
      \|B\|_{\mathrm F}
      +|\operatorname{tr}(B)|~.
    \label{eq: fixed bank high rank common weighted density bound}
\end{align}

\subsubsection{Part (i)}

Fix $y\in[x/2,2x]$. First, consider the upper tail. Define
\begin{align}
    H_u
    &=G^\top A_uG-\operatorname{tr}(A_u)~,
    &
    y_u
    &=y-u\operatorname{tr}(B)~,
    &
    K_u(t)
    &=\log\mathbb E[\exp(tH_u)]~.
\end{align}
Then $Q_u-y=H_u-y_u$, and \eqref{eq: fixed bank high rank local condition} gives $y_u\asymp x$ uniformly over $u\in[0,1]$. For $0\leq t\leq C(1+x)$, \eqref{eq: fixed bank high rank spectral conditions} and $x=o(\sqrt d)$ also give $t\|A_u\|_{\mathrm{op}}=o(1)$ uniformly over $u\in[0,1]$. Orthogonal invariance of the standard Gaussian distribution implies that
\begin{align}
    H_u
    &\overset{d}{=}
      \sum_{q=1}^n
      \alpha_{u,q}(Z_q^2-1)~,
\end{align}
where $Z_1,\ldots,Z_n$ are independent standard Gaussian variables. It follows that we can write
\begin{align}
    K_u(t)
    &=-t\operatorname{tr}(A_u)
      -\frac12\sum_{q=1}^n
      \log(1-2t\alpha_{u,q})~,
    \nonumber\\
    K_u'(t)
    &=2t\sum_{q=1}^n
      \frac{\alpha_{u,q}^2}
      {1-2t\alpha_{u,q}}~,\quad\text{and}
    &
    K_u''(t)
    &=2\sum_{q=1}^n
      \frac{\alpha_{u,q}^2}
      {(1-2t\alpha_{u,q})^2}~.
\end{align}
The Frobenius and operator-norm bounds in \eqref{eq: fixed bank high rank spectral conditions} therefore imply that
\begin{align}
    K_u''(t)
    &\asymp1~,
    \qquad
    0\leq t\leq C(1+x)~.
    \label{eq: fixed bank high rank second derivative bound}
\end{align}
Since $K_u'(0)=0$, integrating \eqref{eq: fixed bank high rank second derivative bound} gives $K_u'(t)\asymp t$. Thus, there is a unique $t_u>0$ satisfying
\begin{align}
    K_u'(t_u)
    &=y_u~,\quad
    t_u
    \asymp x~.
    \label{eq: fixed bank high rank tilting parameter}
\end{align}
Define the probability measure $P_{u,t_u}$ by
\begin{align}
    \frac{\mathrm dP_{u,t_u}}{\mathrm dP}
    &=\exp\left(
      t_uH_u-K_u(t_u)
    \right)~.
    \label{eq: fixed bank high rank change of measure}
\end{align}
Under $P_{u,t_u}$, the vector $G$ is centered Gaussian with covariance
$(I_n-2t_uA_u)^{-1}$. Define
\begin{align}
    T_u
    &=(I_n-2t_uA_u)^{-1/2}
      A_u(I_n-2t_uA_u)^{-1/2}\quad\text{and}
    \nonumber\\
    B_u
    &=(I_n-2t_uA_u)^{-1/2}
      B(I_n-2t_uA_u)^{-1/2}~.
    \label{eq: fixed bank high rank transformed matrices}
\end{align}
Observe that
\begin{align}
    \operatorname{tr}(T_u)-\operatorname{tr}(A_u)
    &=2t_u\sum_{q=1}^n
      \frac{\alpha_{u,q}^2}
      {1-2t_u\alpha_{u,q}}
    =K_u'(t_u)
    =y_u~.
    \label{eq: fixed bank high rank transformed trace}
\end{align}
Thus, we have, under $P_{u,t_u}$, that
\begin{align}
    \left(
      H_u-y_u,
      G^\top BG
    \right)
    &\overset{d}{=}
      \left(
        Z^\top T_uZ-\operatorname{tr}(T_u),
        Z^\top B_uZ
      \right)~,
    \label{eq: fixed bank high rank transformed representation}
\end{align}
where $Z$ has the standard $n$-dimensional Gaussian distribution.

The matrices $A_u$ and $T_u$ have the same eigenvectors, and the
eigenvalues of $T_u$ are
\begin{align}
    \frac{\alpha_{u,q}}
    {1-2t_u\alpha_{u,q}}~,
    \qquad q\in[n]~.
\end{align}
Thus,
\eqref{eq: fixed bank high rank spectral conditions} and
$t_u\|A_u\|_{\mathrm{op}}=o(1)$ imply that
\begin{align}
    \|T_u\|_{\mathrm F}^2
    &\asymp1~,
    &
    \|T_u\|_{\mathrm{op}}
    &\lesssim\frac{1}{\sqrt d}~,\quad\text{and}
    &
    \left|
      \left\{q\in[n]:
        \left|
          \frac{\alpha_{u,q}}
          {1-2t_u\alpha_{u,q}}
        \right|
        \geq\frac{c}{\sqrt d}
      \right\}
    \right|
    &\geq cd~.
    \label{eq: fixed bank high rank tilted spectral conditions}
\end{align}
Now, observe that the first coordinate in \eqref{eq: fixed bank high rank transformed representation} satisfies
\begin{align}
    H_u-y_u
    &\overset{d}{=}
      \sum_{q=1}^n
      \frac{\alpha_{u,q}}
      {1-2t_u\alpha_{u,q}}
      (Z_q^2-1)
\end{align}
under $P_{u,t_u}$. Consequently, we have that
\begin{align}
    \Var_{u,t_u}(H_u-y_u)
    &=2\sum_{q=1}^n
      \frac{\alpha_{u,q}^2}
      {(1-2t_u\alpha_{u,q})^2}
    \asymp1\quad\text{and}
    \nonumber\\
    \sum_{q=1}^n
      \mathbb E\left[
        \left|
          \frac{\alpha_{u,q}}
          {1-2t_u\alpha_{u,q}}
          (Z_q^2-1)
        \right|^3
      \right]
    &\lesssim
      \|T_u\|_{\mathrm{op}}\|T_u\|_{\mathrm F}^2
    \lesssim\frac{1}{\sqrt d}~.
    \label{eq: fixed bank high rank tilted moments}
\end{align}
A centered Gaussian variable whose variance is bounded above and away from zero assigns probability of order $(1+t_u)^{-1}$ to $[0,(1+t_u)^{-1}]$. Therefore, \cref{lem: berry esseen} and \eqref{eq: fixed bank high rank tilted moments} imply that
\begin{align}
    P_{u,t_u}\left\{
      0\leq H_u-y_u\leq(1+t_u)^{-1}
    \right\}
    &\geq\frac{c}{1+t_u}~.
    \label{eq: fixed bank high rank tilted interval probability}
\end{align}
Using \eqref{eq: fixed bank high rank change of measure} and restricting the expectation to the event in \eqref{eq: fixed bank high rank tilted interval probability}, we obtain
\begin{align}
    P\{Q_u\geq y\}
    &=P\{H_u\geq y_u\}
    \nonumber\\
    &=\exp\left(
        K_u(t_u)-t_uy_u
      \right)
      \mathbb E_{u,t_u}\left[
        \exp\left(
          -t_u(H_u-y_u)
        \right)
        \mathbb I\{H_u-y_u\geq0\}
      \right]
    \nonumber\\
    &\geq
      \frac{c}{1+t_u}
      \exp\left(
        K_u(t_u)-t_uy_u
        -\frac{t_u}{1+t_u}
      \right)
    \geq
      \frac{c}{1+t_u}
      \exp\left(
        K_u(t_u)-t_uy_u
      \right)~.
    \label{eq: fixed bank high rank tail comparison}
\end{align}
Moreover, \eqref{eq: fixed bank high rank tilting parameter} and \eqref{eq: fixed bank high rank second derivative bound} give
\begin{align}
    K_u(t_u)-t_uy_u
    &=K_u(t_u)-t_uK_u'(t_u)
    =-\int_0^{t_u}sK_u''(s)\,\mathrm ds
    \geq-Cx^2~.
\end{align}
Since $t_u\asymp x$ and $x\geq1$, it follows that
\begin{align}
    \inf_{u\in[0,1]}P\{Q_u\geq y\}
    &\geq\exp(-Cx^2)~.
    \label{eq: fixed bank high rank tail along path}
\end{align}

With these bounds in place,  observe that \eqref{eq: fixed bank high rank transformed matrices} implies that 
\begin{align}
    \|B_u\|_{\mathrm F}
    &\lesssim\|B\|_{\mathrm F}~,\quad\text{and}
    \nonumber\\
    \operatorname{tr}(B_u)
    &=\operatorname{tr}\left(
        B(I_n-2t_uA_u)^{-1}
      \right)
    =\operatorname{tr}(B)
      +2t_u\operatorname{tr}\left(
        BA_u(I_n-2t_uA_u)^{-1}
      \right)~.
\end{align}
Hence, Cauchy-Schwarz and \eqref{eq: fixed bank high rank tilting parameter} imply that
\begin{align}
    |\operatorname{tr}(B_u)|
    &\lesssim
      |\operatorname{tr}(B)|
      +x\|B\|_{\mathrm F}~.
    \label{eq: fixed bank high rank transformed trace bound}
\end{align}
Applying \eqref{eq: fixed bank high rank common weighted density bound} with $A=T_u$ and $B=B_u$, and using \eqref{eq: fixed bank high rank tilted spectral conditions} and \eqref{eq: fixed bank high rank transformed trace bound}, gives a function $\widetilde a_u$ such that, for every Borel set $S\subseteq\mathbb R$,
\begin{align}
    \mathbb E_{u,t_u}\left[
      |G^\top BG|
      \mathbb I\{H_u-y_u\in S\}
    \right]
    &=
      \int_S\widetilde a_u(w)\,\mathrm dw
    \quad\text{and}\nonumber \\
    \|\widetilde a_u\|_\infty
    &\lesssim
      (1+x)\|B\|_{\mathrm F}
      +|\operatorname{tr}(B)|~.
    \label{eq: fixed bank high rank tilted weighted density}
\end{align}
The density $f_0$ constructed above, applied with $A=A_u$, also shows
that $Q_u$ has a continuous density.

Choose continuously differentiable, non-decreasing functions $F_\tau$
such that $0\leq F_\tau\leq1$,
$F_\tau(w)=0$ for $w\leq-\tau$,
$F_\tau(w)=1$ for $w\geq\tau$, and
\begin{align}
    \int_{\mathbb R}F_\tau'(w)\,\mathrm dw
    &=1~.
\end{align}
By \eqref{eq: fixed bank high rank change of measure}, \eqref{eq: fixed bank high rank transformed representation}, and \eqref{eq: fixed bank high rank tilted weighted density}, it holds that
\begin{align}
    &\mathbb E\left[
      |G^\top BG|F_\tau'(Q_u-y)
    \right]
    \nonumber\\
    &\quad=
      \exp\left(
        K_u(t_u)-t_uy_u
      \right)
      \int_{\mathbb R}
      e^{-t_uw}F_\tau'(w)
      \widetilde a_u(w)\,\mathrm dw
    \leq
      \exp\left(
        K_u(t_u)-t_uy_u+t_u\tau
      \right)
      \|\widetilde a_u\|_\infty~.
\end{align}
Rearranging \eqref{eq: fixed bank high rank tail comparison} gives
\begin{align}
    \exp\left(
      K_u(t_u)-t_uy_u
    \right)
    &\leq
      C(1+t_u)P\{Q_u\geq y\}~.
\end{align}
Therefore, we have that
\begin{align}
    &\mathbb E\left[
      |G^\top BG|F_\tau'(Q_u-y)
    \right]
    \nonumber\\
    &\quad\leq
      Ce^{Cx\tau}
      \left(
        (1+x)^2\|B\|_{\mathrm F}
        +(1+x)|\operatorname{tr}(B)|
      \right)
      P\{Q_u\geq y\}~.
    \label{eq: fixed bank high rank smooth derivative bound}
\end{align}
For $0\leq v\leq u\leq1$, the chain rule and
\begin{align}
    \frac{\mathrm d Q_s}{\mathrm d s}
    &=G^\top BG
\end{align}
give
\begin{align}
    &\left|
      \mathbb E[F_\tau(Q_u-y)]
      -\mathbb E[F_\tau(Q_v-y)]
    \right|
    \quad\leq
      \int_v^u
      \mathbb E\left[
        |G^\top BG|F_\tau'(Q_s-y)
      \right]\,\mathrm ds~.
    \label{eq: fixed bank high rank smooth tail path}
\end{align}
Since each $Q_s$ has a continuous density, dominated convergence in \eqref{eq: fixed bank high rank smooth tail path} and \eqref{eq: fixed bank high rank smooth derivative bound} imply that
\begin{align}
    &\left|
      P\{Q_u\geq y\}
      -P\{Q_v\geq y\}
    \right|
    \nonumber\\
    &\quad\leq
      C\left(
        (1+x)^2\|B\|_{\mathrm F}
        +(1+x)|\operatorname{tr}(B)|
      \right)
      \int_v^uP\{Q_s\geq y\}\,\mathrm ds~.
    \label{eq: fixed bank high rank tail path bound}
\end{align}
Gronwall's inequality \cite[Chapter~III, \S1, Theorem~1.1, p.~24]{hartman2002ordinary}, applied to \eqref{eq: fixed bank high rank tail path bound}, gives
\begin{align}
    P\{Q_u\geq y\}
    &\leq
      P\{Q_0\geq y\}
      \exp\left(
        Cu\left(
          (1+x)^2\|B\|_{\mathrm F}
          +(1+x)|\operatorname{tr}(B)|
        \right)
      \right)
    \lesssim P\{Q_0\geq y\}~,
    \label{eq: fixed bank high rank tail path comparison}
\end{align}
where the last inequality follows from \eqref{eq: fixed bank high rank local condition}.

Finally, because $u\mapsto Q_u$ is affine and $F_\tau$ is
nondecreasing, we have that
\begin{align}
    |F_\tau(Q_1-y)-F_\tau(Q_0-y)|
    &=\int_0^1
      |G^\top BG|F_\tau'(Q_u-y)\,\mathrm du~.
\end{align}
Taking expectations and applying \eqref{eq: fixed bank high rank smooth derivative bound} and \eqref{eq: fixed bank high rank tail path comparison} give
\begin{align}
    &\mathbb E\left[
      |F_\tau(Q_1-y)-F_\tau(Q_0-y)|
    \right]
    \nonumber\\
    &\quad\leq
      Ce^{Cx\tau}
      \left(
        (1+x)^2\|B\|_{\mathrm F}
        +(1+x)|\operatorname{tr}(B)|
      \right)
      P\{Q_0\geq y\}~.
\end{align}
Since $Q_0$ and $Q_1$ have continuous densities, dominated convergence as $\tau$ decreases to zero yields
\begin{align}
    &P\left\{
      \mathbb I\{Q_1\geq y\}
      \neq
      \mathbb I\{Q_0\geq y\}
    \right\}
    \nonumber\\
    &\quad\leq
      C\left(
        (1+x)^2\|B\|_{\mathrm F}
        +(1+x)|\operatorname{tr}(B)|
      \right)
      P\{Q_0\geq y\}~.
    \label{eq: fixed bank high rank positive crossing}
\end{align}
The bound \eqref{eq: fixed bank high rank tail along path} at $u=0$ gives the upper-tail probability bound, while \eqref{eq: fixed bank high rank positive crossing} gives the upper-tail comparison. Replacing $(A_0,B,Q_u)$ by $(-A_0,-B,-Q_u)$ proves both conclusions for the lower tail.
\hfill\qed

\subsubsection{Part (ii)}

Applying \eqref{eq: fixed bank high rank common weighted density bound} with $A=A_u$ and $B=B$, and translating by $u\operatorname{tr}(B)$, gives a function $a_u$ such that, for every Borel set $S\subseteq\mathbb R$,
\begin{align}
    \mathbb E\left[
      |G^\top BG|
      \mathbb I\{Q_u\in S\}
    \right]
    &=\int_Sa_u(w)\,\mathrm dw
    \quad\text{and}\quad
    \sup_{u\in[0,1]}\|a_u\|_\infty
    \lesssim
      \|B\|_{\mathrm F}
      +|\operatorname{tr}(B)|~.
    \label{eq: fixed bank high rank global weighted density}
\end{align}
The density $f_0$ constructed above also shows that every $Q_u$ has a continuous density. Fix $y\in\mathbb R$ and choose $F_\tau$ as in Part (i). Since $u\mapsto Q_u$ is affine, we obtain
\begin{align}
    &\mathbb E\left[
      |F_\tau(Q_1-y)-F_\tau(Q_0-y)|
    \right]
    \nonumber\\
    &\quad=
      \int_0^1
      \mathbb E\left[
        |G^\top BG|F_\tau'(Q_u-y)
      \right]\,\mathrm du
    =
      \int_0^1\int_{\mathbb R}
      F_\tau'(w-y)a_u(w)\,\mathrm dw\,\mathrm du
    \lesssim
      \|B\|_{\mathrm F}
      +|\operatorname{tr}(B)|~.
\end{align}
As $\tau$ decreases to zero, it holds that
\begin{align}
    |F_\tau(Q_1-y)-F_\tau(Q_0-y)|
    &\to
      \mathbb I\left\{
        \mathbb I\{Q_1\geq y\}
        \neq
        \mathbb I\{Q_0\geq y\}
      \right\}
\end{align}
almost surely. The proof is then complete by applying dominated convergence and taking the supremum over $y\in\mathbb R$.

\subsection{Proof of
\cref{lem: gaussian threshold endpoint covariance bounds}}

\subsubsection{Part (i)}

For every positive semidefinite $6\times6$ matrix $\Sigma$, define
\begin{align}
    \Psi_{\delta,d}^{\vartheta}(\Sigma)
    &=
    \Cov\Bigg(
      \mathbb I\left\{
        \left|
          \frac{1}{d}\sum_{k=1}^d
          \vartheta_kX_{k,1}X_{k,2}
        \right|
        \geq\delta
      \right\},
      \mathbb I\left\{
        \left|
          \frac{1}{d}\sum_{k=1}^d
          \vartheta_kY_{k,1}Y_{k,3}
        \right|
        \geq\delta
      \right\}
    \Bigg)~,
    \label{eq: gaussian endpoint covariance functional}
\end{align}
where $(X_k,Y_k)_{k=1}^d$ are independent centered Gaussian vectors
with covariance $\Sigma$. For $r,s\in\{0,1\}$, let $\Sigma_{r,s}$ be the covariance matrix of $(Z_k-rU_k,Z_k-sU_k)$. We can write
\begin{align}
    \Sigma_{r,s}
    &=
    \begin{pmatrix}
      K_{ZZ}-r(K_{ZU}+K_{ZU}^\top)+rK_{UU}
      &
      K_{ZZ}-sK_{ZU}-rK_{ZU}^\top+rsK_{UU}
      \\
      K_{ZZ}-rK_{ZU}-sK_{ZU}^\top+rsK_{UU}
      &
      K_{ZZ}-s(K_{ZU}+K_{ZU}^\top)+sK_{UU}
    \end{pmatrix}~.
    \label{eq: gaussian endpoint covariance matrices}
\end{align}
Consequently, it holds that
\begin{align}
    \Cov\left(
      S_{1,2}(1)-S_{1,2}(0),
      S_{1,3}(1)-S_{1,3}(0)
    \right)
    =
      \sum_{r,s\in\{0,1\}}(-1)^{r+s}
      \Psi_{\delta,d}^{\vartheta}(\Sigma_{r,s})~.
    \label{eq: gaussian endpoint exact rectangle}
\end{align}
Moreover, \eqref{eq: gaussian endpoint covariance matrices} and
\eqref{eq: gaussian endpoint covariance conditions} imply that
\begin{align}
    \max_{r,s\in\{0,1\}}
      \|\Sigma_{r,s}-\Sigma_{0,0}\|_{\mathrm F}
    &\lesssim\Delta_K
    \quad\text{and}\quad
    \nonumber\\
    \sum_{r,s\in\{0,1\}}(-1)^{r+s}
      (\Sigma_{r,s}-\Sigma_{0,0})
    &=
      \begin{pmatrix}
        0&K_{UU}\\
        K_{UU}&0
      \end{pmatrix}~.
    \label{eq: gaussian endpoint alternating direction}
\end{align}
We will expand the four terms in
\eqref{eq: gaussian endpoint exact rectangle} about their common
covariance $\Sigma_{0,0}$.

We first establish some necessary derivative bounds. For $h\in[0,1]$ and $\tau\in[0,c]$, define
\begin{align}
    &\Sigma_{0,0}
      +h(\Sigma_{r,s}-\Sigma_{0,0})
      +\tau I_6
    \quad\text{and}\quad
    &\begin{pmatrix}
      I_3+h(K_{ZZ}-I_3)&I_3+h(K_{ZZ}-I_3)\\
      I_3+h(K_{ZZ}-I_3)&I_3+h(K_{ZZ}-I_3)
    \end{pmatrix}
    +\tau I_6~.
    \label{eq: gaussian endpoint covariance paths}
\end{align}
Stack $(X_k^\top,Y_k^\top)^\top$, $k\in[d]$, into $W$. For $a=(a_1,a_2)\in\mathbb R^2$, let $A(a)$ be the symmetric matrix satisfying
\begin{align}
    W^\top A(a)W
    &=
      \frac{
        \sum_{k=1}^d\vartheta_k
        \left(
          a_1X_{k,1}X_{k,2}
          +a_2Y_{k,1}Y_{k,3}
        \right)
      }{
        \left(\sum_{k=1}^d\vartheta_k^2\right)^{1/2}
      }~.
    \label{eq: gaussian endpoint quadratic matrices}
\end{align}
For any covariance matrix $\Sigma$ appearing in
\eqref{eq: gaussian endpoint covariance paths}, let
$\alpha_1(a,\Sigma),\ldots,\alpha_{6d}(a,\Sigma)$ be the eigenvalues of
\begin{align}
    (I_d\otimes\Sigma)^{1/2}
    A(a)
    (I_d\otimes\Sigma)^{1/2}~.
\end{align}
The bound \eqref{eq: gaussian endpoint weight conditions}, \eqref{eq: gaussian endpoint covariance conditions}, and \eqref{eq: gaussian endpoint covariance matrices} imply that
\begin{align}
    c\|a\|_2^2
    \leq
      \left\|
        (I_d\otimes\Sigma)^{1/2}
        A(a)
        (I_d\otimes\Sigma)^{1/2}
      \right\|_{\mathrm F}^2
    &\leq C\|a\|_2^2~,
    \nonumber\\
    \left\|
      (I_d\otimes\Sigma)^{1/2}
      A(a)
      (I_d\otimes\Sigma)^{1/2}
    \right\|_{\mathrm{op}}
    &\lesssim\frac{\|a\|_2}{\sqrt d}~,
    \quad\text{and}\quad
    \nonumber\\
    \left|
      \left\{
        l\in[6d]:
        |\alpha_l(a,\Sigma)|
        \geq\frac{c\|a\|_2}{\sqrt d}
      \right\}
    \right|
    &\geq cd~.
    \label{eq: gaussian endpoint pencil bounds}
\end{align}
For positive-definite $\Sigma$, let $\varphi_\Sigma$ be the characteristic function of the centered quadratic-form pair in \eqref{eq: gaussian endpoint quadratic matrices}. The Gaussian quadratic-form transform in \citealp{muirhead1982}, Proof of Theorem~1.4.2, gives
\begin{align}
    \varphi_\Sigma(a)
    &=
      \exp\left(
        -i\operatorname{tr}((I_d\otimes\Sigma)A(a))
      \right)
      \det\left(
        I_{6d}-2i(I_d\otimes\Sigma)A(a)
      \right)^{-1/2}~.
    \label{eq: gaussian endpoint characteristic function}
\end{align}
It follows from \eqref{eq: gaussian endpoint pencil bounds} that
\begin{align}
    |\varphi_\Sigma(a)|
    &\leq
      \left(
        1+c\frac{\|a\|_2^2}{d}
      \right)^{-cd}~.
    \label{eq: gaussian endpoint determinant bound}
\end{align}
Jacobi's formula applied to
\eqref{eq: gaussian endpoint characteristic function} gives
\begin{align}
    \frac{
      \mathrm D_\Sigma\varphi_\Sigma(a)[H]
    }{
      \varphi_\Sigma(a)
    }
    &=
      i\operatorname{tr}\Bigg(
        \left[
          \left(
            I_{6d}-2i(I_d\otimes\Sigma)A(a)
          \right)^{-1}
          -I_{6d}
        \right]
        (I_d\otimes H)A(a)
      \Bigg)~.
    \label{eq: gaussian endpoint characteristic derivative identity}
\end{align}
The identity
\begin{align}
    &\left(
      I_{6d}-2i(I_d\otimes\Sigma)A(a)
    \right)^{-1}
    -I_{6d}
    \nonumber\\
    &\quad=
      2i(I_d\otimes\Sigma)^{1/2}
      \left(
        I_{6d}
        -2i(I_d\otimes\Sigma)^{1/2}
        A(a)
        (I_d\otimes\Sigma)^{1/2}
      \right)^{-1}
      (I_d\otimes\Sigma)^{1/2}A(a)
    \label{eq: gaussian endpoint inverse identity}
\end{align}
shows that the right-hand side of
\eqref{eq: gaussian endpoint characteristic derivative identity}
contains two factors of $A(a)$. Since the middle inverse in
\eqref{eq: gaussian endpoint inverse identity} has operator norm at
most one, differentiating
\eqref{eq: gaussian endpoint characteristic function} with respect to
$\Sigma$ and $a$ gives
\begin{align}
    \left|\frac{
      \mathrm d^{\beta_1+\beta_2}\varphi_\Sigma(a)
    }{
      \mathrm d a_1^{\beta_1}\,\mathrm d a_2^{\beta_2}
    }\right|
    &\lesssim
      (1+\|a\|_2)^C
      \left(
        1+c\frac{\|a\|_2^2}{d}
      \right)^{-cd}~,
    \nonumber\\
    \left|
      \frac{
        \mathrm d^{\beta_1+\beta_2}
      }{
        \mathrm d a_1^{\beta_1}\,\mathrm d a_2^{\beta_2}
      }
      \mathrm D_\Sigma\varphi_\Sigma(a)[H]
    \right|
    &\lesssim
      \|H\|_{\mathrm F}
      (1+\|a\|_2)^C
      \left(
        1+c\frac{\|a\|_2^2}{d}
      \right)^{-cd}~,\quad\text{and}
    \nonumber\\
    \left|
      \frac{
        \mathrm d^{\beta_1+\beta_2}
      }{
        \mathrm d a_1^{\beta_1}\,\mathrm d a_2^{\beta_2}
      }
      \mathrm D_\Sigma^2\varphi_\Sigma(a)[H,G]
    \right|
    &\lesssim
      \|H\|_{\mathrm F}\|G\|_{\mathrm F}
      (1+\|a\|_2)^C
      \left(
        1+c\frac{\|a\|_2^2}{d}
      \right)^{-cd}
    \label{eq: gaussian endpoint characteristic derivative bounds}
\end{align}
for $\beta\in\{0,1,2\}^2$. The right-hand sides are integrable uniformly in $d$ by the calculation in \eqref{eq: fixed bank high rank transform integral}.

Let $g_\Sigma$ be the density of the centered pair in \eqref{eq: gaussian endpoint quadratic matrices}. The Fourier inversion theorem for integrable characteristic functions (\citealp{feller1991introduction}, Chapter~XV.3), followed by two integrations by parts in each coordinate, gives
\begin{align}
    \|q\|_{L^1(\mathbb R^2)}
    &\lesssim
      \sum_{\beta_1,\beta_2\in\{0,2\}}
      \left\|
        \frac{
          \mathrm d^{\beta_1+\beta_2}\widehat q
        }{
          \mathrm d a_1^{\beta_1}\,\mathrm d a_2^{\beta_2}
        }
      \right\|_{L^1(\mathbb R^2)}
    \label{eq: gaussian endpoint weighted inversion bound}
\end{align}
whenever the terms on the right are integrable. Applying this bound to \eqref{eq: gaussian endpoint characteristic derivative bounds} yields
\begin{align}
    \sum_{|\gamma|\leq2}
      \left\|\frac{
        \mathrm d^{\gamma_1+\gamma_2}g_\Sigma
      }{
        \mathrm d x_1^{\gamma_1}\,\mathrm d x_2^{\gamma_2}
      }\right\|_{L^1(\mathbb R^2)}
    &\lesssim1~,
    \nonumber\\
    \sum_{|\gamma|\leq1}
      \left\|
        \frac{
          \mathrm d^{\gamma_1+\gamma_2}
        }{
          \mathrm d x_1^{\gamma_1}\,\mathrm d x_2^{\gamma_2}
        }
        \mathrm D_\Sigma g_\Sigma[H]
      \right\|_{L^1(\mathbb R^2)}
    &\lesssim\|H\|_{\mathrm F}~,
    \quad\text{and}\quad
    \left\|
      \mathrm D_\Sigma^2g_\Sigma[H,G]
    \right\|_{L^1(\mathbb R^2)}
    \lesssim\|H\|_{\mathrm F}\|G\|_{\mathrm F}~.
    \label{eq: gaussian endpoint centered density bounds}
\end{align}
The representation \eqref{eq: gaussian endpoint quadratic matrices} and Cauchy-Schwarz imply that
\begin{align}
    &\left\|
      \mathrm D_\Sigma
      \begin{pmatrix}
        \operatorname{tr}((I_d\otimes\Sigma)A(e_1))\\
        \operatorname{tr}((I_d\otimes\Sigma)A(e_2))
      \end{pmatrix}
      [H]
    \right\|_2
    \lesssim\sqrt d\,\|H\|_{\mathrm F}~.
    \label{eq: gaussian endpoint quadratic mean bounds}
\end{align}
The joint and marginal probabilities defining $\Psi_{\delta,d}^{\vartheta}(\Sigma)$ are integrals of the translated density over sets that do not depend on $\Sigma$. Therefore, \eqref{eq: gaussian endpoint centered density bounds}, \eqref{eq: gaussian endpoint quadratic mean bounds}, and the chain rule
give
\begin{align}
    \left|
      \mathrm D_\Sigma
      \Psi_{\delta,d}^{\vartheta}(\Sigma)[H]
    \right|
    &\lesssim
      \sqrt d\,\|H\|_{\mathrm F}
    \quad\text{and}\quad
    \left|
      \mathrm D_\Sigma^2
      \Psi_{\delta,d}^{\vartheta}(\Sigma)[H,G]
    \right|
    \lesssim
      d\|H\|_{\mathrm F}\|G\|_{\mathrm F}~.
    \label{eq: gaussian endpoint covariance derivative bounds}
\end{align}
The bounds in \eqref{eq: gaussian endpoint characteristic derivative bounds} are uniform in $\tau$. Dominated convergence in the inverse Fourier integrals therefore extends \eqref{eq: gaussian endpoint covariance derivative bounds} to $\tau=0$ along the covariance paths in \eqref{eq: gaussian endpoint covariance paths}.

For a positive-definite $3\times3$ matrix $K$, define
\begin{align}
    \mathrm D_K\Psi_{\delta,d}^{\vartheta}[H]
    &=
      \lim_{\tau\downarrow0}
      \mathrm D_\Sigma
      \Psi_{\delta,d}^{\vartheta}
      \left(
        \begin{pmatrix}
          K&K\\
          K&K
        \end{pmatrix}
        +\tau I_6
      \right)[H]~.
    \label{eq: gaussian endpoint regularized derivative}
\end{align}
The first bound in
\eqref{eq: gaussian endpoint covariance derivative bounds} gives
\begin{align}
    \left|
      \mathrm D_K\Psi_{\delta,d}^{\vartheta}[H]
    \right|
    &\lesssim\sqrt d\,\|H\|_{\mathrm F}~.
    \label{eq: gaussian endpoint regularized derivative bound}
\end{align}
The second bound in
\eqref{eq: gaussian endpoint covariance derivative bounds}, applied
along the second path in
\eqref{eq: gaussian endpoint covariance paths}, gives
\begin{align}
    &\left|
      \mathrm D_{K_{ZZ}}\Psi_{\delta,d}^{\vartheta}[H]
      -\mathrm D_{I_3}\Psi_{\delta,d}^{\vartheta}[H]
    \right|
    \lesssim
      d\|K_{ZZ}-I_3\|_{\mathrm F}\|H\|_{\mathrm F}~.
    \label{eq: gaussian endpoint derivative stability}
\end{align}
Moreover, for every diagonal matrix $J$ whose diagonal entries belong to $\{-1,1\}$, the transformation $(X_k,Y_k)\mapsto(JX_k,JY_k)$ leaves both indicators in \eqref{eq: gaussian endpoint covariance functional} unchanged. Averaging over the eight choices of $J$ eliminates the off-diagonal entries of $G$. The first indicator does not depend on $X_{k,3}$, and the second does not depend on $Y_{k,2}$. Hence, there is a deterministic quantity $\chi_{\delta,d}^{\vartheta}$ such that
\begin{align}
    \mathrm D_{I_3}\Psi_{\delta,d}^{\vartheta}
    \left[
      \begin{pmatrix}
        0&G\\
        G^\top&0
      \end{pmatrix}
    \right]
    &=
      \chi_{\delta,d}^{\vartheta}G_{1,1}
    \quad\text{and}\quad
    |\chi_{\delta,d}^{\vartheta}|
    \lesssim\sqrt d~,
    \label{eq: gaussian endpoint canonical derivative}
\end{align}
where the second conclusion follows from \eqref{eq: gaussian endpoint regularized derivative bound}.

Now, Taylor's formula along the first path in \eqref{eq: gaussian endpoint covariance paths}, applied first with $\tau>0$ and then with $\tau\downarrow0$, gives
\begin{align}
    &\left|
      \Psi_{\delta,d}^{\vartheta}(\Sigma_{r,s})
      -\Psi_{\delta,d}^{\vartheta}(\Sigma_{0,0})
      -\mathrm D_{K_{ZZ}}\Psi_{\delta,d}^{\vartheta}
        [\Sigma_{r,s}-\Sigma_{0,0}]
    \right|
    \lesssim d\Delta_K^2~.
    \label{eq: gaussian endpoint covariance taylor bound}
\end{align}
Consequently, \eqref{eq: gaussian endpoint exact rectangle} and \eqref{eq: gaussian endpoint alternating direction} imply that
\begin{align}
    &\left|
      \Cov\left(
        S_{1,2}(1)-S_{1,2}(0),
        S_{1,3}(1)-S_{1,3}(0)
      \right)
      -
      \mathrm D_{K_{ZZ}}\Psi_{\delta,d}^{\vartheta}
      \left[
        \begin{pmatrix}
          0&K_{UU}\\
          K_{UU}&0
        \end{pmatrix}
      \right]
    \right|
    \lesssim d\Delta_K^2~.
    \label{eq: gaussian endpoint first order expansion}
\end{align}
Finally, \eqref{eq: gaussian endpoint derivative stability}, \eqref{eq: gaussian endpoint canonical derivative}, and \eqref{eq: gaussian endpoint covariance conditions} give
\begin{align}
    &\left|
      \mathrm D_{K_{ZZ}}\Psi_{\delta,d}^{\vartheta}
      \left[
        \begin{pmatrix}
          0&K_{UU}\\
          K_{UU}&0
        \end{pmatrix}
      \right]
      -\chi_{\delta,d}^{\vartheta}(K_{UU})_{1,1}
    \right|
    \nonumber\\
    &\lesssim
      d\|K_{ZZ}-I_3\|_{\mathrm F}\|K_{UU}\|_{\mathrm F}
    \lesssim
      d\|K_{ZZ}-I_3\|_{\mathrm F}\Delta_K~.
    \label{eq: gaussian endpoint canonical comparison}
\end{align}
Combining \eqref{eq: gaussian endpoint first order expansion} and \eqref{eq: gaussian endpoint canonical comparison} gives
\begin{align}
    &\left|
      \Cov\left(
        S_{1,2}(1)-S_{1,2}(0),
        S_{1,3}(1)-S_{1,3}(0)
      \right)
      -\chi_{\delta,d}^{\vartheta}(K_{UU})_{1,1}
    \right|
    \lesssim
      d\left(
        \|K_{ZZ}-I_3\|_{\mathrm F}\Delta_K
        +\Delta_K^2
      \right)~.
\end{align}
Together with the second bound in \eqref{eq: gaussian endpoint canonical derivative}, this proves \eqref{eq: gaussian shared coordinate endpoint expansion}.
\hfill\qed

\subsubsection{Part (ii)}

Extend the definition of $\rho_{a,b}(t,q)$ to $(t,q)\in[0,1]^2$ and define
\begin{align}
    A(t)
    &=\left(
      (K_{ZZ})_{a,b}
      -t(K_{ZU})_{a,b}
      -t(K_{ZU})_{b,a}
      +t^2(K_{UU})_{a,b}
    \right)_{a,b=1}^2~,
    \nonumber\\
    B(q)
    &=\left(
      (K_{ZZ})_{a+2,b+2}
      -q(K_{ZU})_{a+2,b+2}
      -q(K_{ZU})_{b+2,a+2}
      +q^2(K_{UU})_{a+2,b+2}
    \right)_{a,b=1}^2~,\quad\text{and}
    \nonumber\\
    R(t,q)
    &=\left(\rho_{a,b}(t,q)\right)_{a,b=1}^2~.
    \label{eq: gaussian endpoint four coordinate path}
\end{align}
Applying the fundamental theorem of calculus in $t$ and $q$ gives
\begin{align}
    \Cov\left(
      S_{1,2}(1)-S_{1,2}(0),
      S_{3,4}(1)-S_{3,4}(0)
    \right)
    =
      \int_0^1\int_0^1
      \frac{\mathrm d^2}{\mathrm d t\,\mathrm d q}
      \Psi_{\delta,d}^{\vartheta}
      \left(A(t),B(q),R(t,q)\right)
      \,\mathrm dt\,\mathrm dq~.
    \label{eq: gaussian endpoint four coordinate integral}
\end{align}
We compare the integrand with the four terms defining $\mathcal J_{a,b}$ by writing
\begin{align}
    &\frac{\mathrm d^2}{\mathrm d t\,\mathrm d q}
      \Psi_{\delta,d}^{\vartheta}(A(t),B(q),R(t,q))
      -\sum_{a,b=1}^2
      \frac{\mathrm d^2}{\mathrm d t\,\mathrm d q}
      \Psi_{\delta,d}^{\vartheta}
      \left(I_2,I_2,\rho_{a,b}(t,q)e_ae_b^\top\right)
    \nonumber\\
    &=
      \frac{\mathrm d^2}{\mathrm d t\,\mathrm d q}
      \left(
        \Psi_{\delta,d}^{\vartheta}(A(t),B(q),R(t,q))
        -\Psi_{\delta,d}^{\vartheta}(I_2,I_2,R(t,q))
      \right)
    \nonumber\\
    &+
      \frac{\mathrm d^2}{\mathrm d t\,\mathrm d q}
      \left(
        \Psi_{\delta,d}^{\vartheta}(I_2,I_2,R(t,q))
        -\sum_{a,b=1}^2
        \Psi_{\delta,d}^{\vartheta}
        \left(I_2,I_2,\rho_{a,b}(t,q)e_ae_b^\top\right)
      \right)~.
    \label{eq: gaussian endpoint derivative decomposition}
\end{align}
Direct differentiation of \eqref{eq: gaussian endpoint four coordinate path} gives
\begin{align}
    &\sup_{t,q\in[0,1]}
      \left(
        \left\|\frac{\mathrm d A(t)}{\mathrm d t}\right\|_{\mathrm F}
        +\left\|\frac{\mathrm d B(q)}{\mathrm d q}\right\|_{\mathrm F}
        +\left\|\frac{\mathrm d R(t,q)}{\mathrm d t}\right\|_{\mathrm F}
        +\left\|\frac{\mathrm d R(t,q)}{\mathrm d q}\right\|_{\mathrm F}
        +\left\|\frac{\mathrm d^2R(t,q)}{\mathrm d t\,\mathrm d q}\right\|_{\mathrm F}
      \right)
    \lesssim\Delta_K~,
    \nonumber\\
    &\sup_{t,q\in[0,1]}\|R(t,q)\|_{\mathrm F}
    \lesssim
      \|(K_{ZZ})_{[2],\{3,4\}}\|_{\mathrm F}+\Delta_K~,\quad\text{and}
    \nonumber\\
    &\sup_{t,q\in[0,1]}
      \left(
        \|A(t)-I_2\|_{\mathrm F}
        +\|B(q)-I_2\|_{\mathrm F}
      \right)
    \lesssim
      |(K_{ZZ})_{1,2}|+|(K_{ZZ})_{3,4}|+\Delta_K~.
    \label{eq: gaussian endpoint four coordinate path bounds}
\end{align}
Observe that by \eqref{eq: gaussian endpoint covariance conditions} and \eqref{eq: gaussian endpoint four coordinate path bounds}, the covariance matrices with blocks $(A(t),B(q),R(t,q))$, $(I_2,I_2,R(t,q))$, and $(I_2,I_2,\rho_{a,b}(t,q)e_ae_b^\top)$ satisfy \eqref{eq: gaussian covariance eigenvalue bounds}. By convexity, the same is true for the covariance matrix with marginal blocks $I_2+v(A(t)-I_2)$ and $I_2+v(B(q)-I_2)$ and cross-covariance $R(t,q)$, for every $v\in[0,1]$. In particular, \eqref{eq: gaussian derivative standardized covariance bounds} gives $\|R(t,q)\|_{\mathrm{op}}\leq1-c$ uniformly over the square. Under \eqref{eq: gaussian endpoint weight conditions}, Parts (i) and (ii) of \cref{lem: gaussian threshold covariance bounds} also apply to $\Psi_{\delta,d}^{\vartheta}$ with $p=1$.

We first bound the first term in \eqref{eq: gaussian endpoint derivative decomposition}. The representation in \cref{lem: gaussian chaos transfer} gives, in addition to \eqref{eq: gaussian derivative transfer bounds},
\begin{align}
    \left\|
      \mathrm D^3\mathcal T_s(S)[G_1,G_2,G_3]
    \right\|_{\mathrm{op}}
    &\leq
      s(s-1)(s-2)\|S\|_{\mathrm{op}}^{s-3}
      \prod_{v=1}^3\|G_v\|_{\mathrm{op}}~,
      \qquad s\geq3~.
\end{align}
Using this bound in the degree-$s$ expansion \eqref{eq: gaussian derivative even hermite representation}, together with \eqref{eq: gaussian derivative fixed degree coefficient bounds} and \eqref{eq: gaussian derivative inverse square root bounds}, gives
\begin{align}
    &\left|
      \mathrm D_{A,B}\mathrm D_R^r
      \Psi_{\delta,d}^{\vartheta}(A,B,R)
      [(H_A,H_B),G_1,\ldots,G_r]
    \right|
    \nonumber\\
    &\quad\lesssim
      \sqrt d\,\|R\|_{\mathrm F}^{2-r}
      (\|H_A\|_{\mathrm F}+\|H_B\|_{\mathrm F})
      \prod_{v=1}^r\|G_v\|_{\mathrm F}~,
      \qquad r\in\{1,2\}~.
    \label{eq: gaussian endpoint mixed derivative bound}
\end{align}
The chain rule implies that
\begin{align}
    &\frac{\mathrm d^2}{\mathrm d t\,\mathrm d q}
      \left[
        \Psi_{\delta,d}^{\vartheta}(A(t),B(q),R(t,q))
        -\Psi_{\delta,d}^{\vartheta}(I_2,I_2,R(t,q))
      \right]
    \nonumber\\
    &\quad=
      \mathrm D_{A,B}^2\Psi_{\delta,d}^{\vartheta}
      (A(t),B(q),R(t,q))
      \left[
        \left(\frac{\mathrm d A(t)}{\mathrm d t},0\right),
        \left(0,\frac{\mathrm d B(q)}{\mathrm d q}\right)
      \right]
    \nonumber\\
    &\qquad+
      \mathrm D_{A,B}\mathrm D_R\Psi_{\delta,d}^{\vartheta}
      (A(t),B(q),R(t,q))
      \left[
        \left(\frac{\mathrm d A(t)}{\mathrm d t},0\right),
        \frac{\mathrm d R(t,q)}{\mathrm d q}
      \right]
    \nonumber\\
    &\qquad+
      \mathrm D_{A,B}\mathrm D_R\Psi_{\delta,d}^{\vartheta}
      (A(t),B(q),R(t,q))
      \left[
        \left(0,\frac{\mathrm d B(q)}{\mathrm d q}\right),
        \frac{\mathrm d R(t,q)}{\mathrm d t}
      \right]
    \nonumber\\
    &\qquad+
      \left[
        \mathrm D_R^2\Psi_{\delta,d}^{\vartheta}(A(t),B(q),R(t,q))
        -\mathrm D_R^2\Psi_{\delta,d}^{\vartheta}(I_2,I_2,R(t,q))
      \right]
      \left[
        \frac{\mathrm d R(t,q)}{\mathrm d t},
        \frac{\mathrm d R(t,q)}{\mathrm d q}
      \right]
    \nonumber\\
    &\qquad+
      \left[
        \mathrm D_R\Psi_{\delta,d}^{\vartheta}(A(t),B(q),R(t,q))
        -\mathrm D_R\Psi_{\delta,d}^{\vartheta}(I_2,I_2,R(t,q))
      \right]
      \left[
        \frac{\mathrm d^2R(t,q)}{\mathrm d t\,\mathrm d q}
      \right]~.
    \label{eq: gaussian endpoint marginal chain rule}
\end{align}
For $r\in\{1,2\}$, the differences in the final two terms of \eqref{eq: gaussian endpoint marginal chain rule} satisfy
\begin{align}
    &\Big[
      \mathrm D_R^r\Psi_{\delta,d}^{\vartheta}(A(t),B(q),R(t,q))
      -\mathrm D_R^r\Psi_{\delta,d}^{\vartheta}(I_2,I_2,R(t,q))
    \Big][G_1,\ldots,G_r]
    \nonumber\\
    &\quad=
      \int_0^1
      \mathrm D_{A,B}\mathrm D_R^r\Psi_{\delta,d}^{\vartheta}
      \left(
        I_2+v(A(t)-I_2),
        I_2+v(B(q)-I_2),
        R(t,q)
      \right)
    \nonumber\\
    &\hspace{22mm}\left[
        (A(t)-I_2,B(q)-I_2),
        G_1,\ldots,G_r
      \right]\,\mathrm dv~.
    \label{eq: gaussian endpoint marginal derivative path}
\end{align}
Thus, we find that 
\begin{align}
    &\left|
      \frac{\mathrm d^2}{\mathrm d t\,\mathrm d q}
      \left[
        \Psi_{\delta,d}^{\vartheta}(A(t),B(q),R(t,q))
        -\Psi_{\delta,d}^{\vartheta}(I_2,I_2,R(t,q))
      \right]
    \right|
    \nonumber\\
    &\quad\lesssim
      d\|R(t,q)\|_{\mathrm F}^2
      \left\|\frac{\mathrm d A(t)}{\mathrm d t}\right\|_{\mathrm F}
      \left\|\frac{\mathrm d B(q)}{\mathrm d q}\right\|_{\mathrm F}
    \nonumber\\
    &\qquad+
      \sqrt d\,\|R(t,q)\|_{\mathrm F}
      \left(
        \left\|\frac{\mathrm d A(t)}{\mathrm d t}\right\|_{\mathrm F}
        \left\|\frac{\mathrm d R(t,q)}{\mathrm d q}\right\|_{\mathrm F}
        +\left\|\frac{\mathrm d B(q)}{\mathrm d q}\right\|_{\mathrm F}
        \left\|\frac{\mathrm d R(t,q)}{\mathrm d t}\right\|_{\mathrm F}
      \right)
    \nonumber\\
    &\qquad+
      \sqrt d
      \left(
        \|A(t)-I_2\|_{\mathrm F}+\|B(q)-I_2\|_{\mathrm F}
      \right)
      \left(
        \left\|\frac{\mathrm d R(t,q)}{\mathrm d t}\right\|_{\mathrm F}
        \left\|\frac{\mathrm d R(t,q)}{\mathrm d q}\right\|_{\mathrm F}
        +\|R(t,q)\|_{\mathrm F}
        \left\|\frac{\mathrm d^2R(t,q)}{\mathrm d t\,\mathrm d q}\right\|_{\mathrm F}
      \right)
    \nonumber\\
    &\quad\lesssim
      d\Delta_K^2
      \left(
        \|(K_{ZZ})_{[2],\{3,4\}}\|_{\mathrm F}+\Delta_K
      \right)^2
    \nonumber\\
    &\qquad+
      \sqrt d\,\Delta_K
      \left(
        |(K_{ZZ})_{1,2}|+|(K_{ZZ})_{3,4}|+\Delta_K
      \right)
      \left(
        \|(K_{ZZ})_{[2],\{3,4\}}\|_{\mathrm F}+\Delta_K
      \right)
    \label{eq: gaussian endpoint marginal derivative difference}
\end{align}
by Part (ii) of \cref{lem: gaussian threshold covariance bounds}, \eqref{eq: gaussian endpoint mixed derivative bound}, and \eqref{eq: gaussian endpoint marginal derivative path}, applied to \eqref{eq: gaussian endpoint marginal chain rule}.

Next, we bound the second term in \eqref{eq: gaussian endpoint derivative decomposition}. Let $\Phi_s^{\vartheta}(A,B,R)$ denote the degree-$s$ term in \eqref{eq: gaussian derivative even hermite representation}. Define the function
\begin{align}
    \widetilde\Phi_s^{\vartheta}(R)
    &=
      \Phi_s^{\vartheta}(I_2,I_2,R)
      -\sum_{a,b=1}^2
      \Phi_s^{\vartheta}
      \left(
        I_2,I_2,R_{a,b}e_ae_b^\top
      \right)
    \label{eq: gaussian endpoint degree interaction term}
\end{align}
and observe that we can write
\begin{align}
    &\Psi_{\delta,d}^{\vartheta}(I_2,I_2,R)
      -\sum_{a,b=1}^2
      \Psi_{\delta,d}^{\vartheta}
      \left(
        I_2,I_2,R_{a,b}e_ae_b^\top
      \right)
    =
      \sum_{\substack{s\geq2\\s\ \mathrm{even}}}
      \widetilde\Phi_s^{\vartheta}(R)~.
    \label{eq: gaussian endpoint degree interaction expansion}
\end{align}
At $A=B=I_2$, every monomial in
$\Phi_s^{\vartheta}(I_2,I_2,R)$ is a constant multiple of
\begin{align}
    \prod_{v=1}^s R_{a_v,b_v}~,
    \qquad
    a_v,b_v
    &\in[2]~,
\end{align}
where $a_v$ identifies the coordinate associated with the first
indicator and $b_v$ identifies the coordinate associated with the
second indicator. Changing the signs of $(X_{k,1})_{k=1}^d$ or
$(X_{k,2})_{k=1}^d$ does not change the first indicator defining
$\Psi_{\delta,d}^{\vartheta}(I_2,I_2,R)$. The corresponding statement
holds for $(Y_{k,1})_{k=1}^d$ and $(Y_{k,2})_{k=1}^d$ in the second
indicator. By the definition \eqref{eq: gaussian derivative hermite generating function}, such a monomial can have a nonzero coefficient only if
\begin{align}
    &\left|\{v\in[s]:a_v=1\}\right|,
    \quad
    \left|\{v\in[s]:a_v=2\}\right|,
    \quad
    \left|\{v\in[s]:b_v=1\}\right|,
    \quad
    \left|\{v\in[s]:b_v=2\}\right|
    \label{eq: gaussian endpoint canonical coefficient parity}
\end{align}
are all even When $s=2$, these restrictions require $a_1=a_2$ and $b_1=b_2$. Therefore, every degree-two monomial is the square of a single entry of $R$. Thus, we find
\begin{align}
    \Phi_2^{\vartheta}(I_2,I_2,R)
    &=\sum_{a,b=1}^2
      \Phi_2^{\vartheta}
      \left(I_2,I_2,R_{a,b}e_ae_b^\top\right)~,
    \quad\text{and}\quad
    \widetilde\Phi_2^{\vartheta}(R)
    =0~.\label{eq: gaussian endpoint degree two cancellation}
\end{align}
Now fix an even $s\geq4$. The subtraction in \eqref{eq: gaussian endpoint degree interaction term} removes every monomial having only one positive exponent. For each remaining monomial, and each choice of indices for which the corresponding derivative is nonzero, \eqref{eq: gaussian endpoint canonical coefficient parity} gives
\begin{align}
    \left|
      \left\{(c,e)\in[2]^2:
        \nu_{c,e}
        -\mathbb I\{(c,e)=(a,b)\}>0
      \right\}
    \right|
    &\geq2~,
    \nonumber\\
    \sum_{c,e=1}^2
      \left(
        \nu_{c,e}
        -\mathbb I\{(c,e)=(a,b)\}
        -\mathbb I\{(c,e)=(a',b')\}
      \right)
    &=s-2
    \geq2~.
    \label{eq: gaussian endpoint remaining factor counts}
\end{align}
The first line of \eqref{eq: gaussian endpoint remaining factor counts} allows us to
retain two distinct factors after one differentiation. The second allows us to retain two factors after two differentiations. Applying \cref{lem: gaussian chaos transfer} to the remaining factors, Parseval's identity and Cauchy--Schwarz to the degree-$s$ Hermite coefficient vectors, we obtain the bounds
\begin{align}
    \left|
      \mathrm D_R\widetilde\Phi_s^{\vartheta}(R)[G]
    \right|
    &\lesssim
      s\binom{s-1}{2}
      \|R\|_{\mathrm{op}}^{(s-1)-2}\|G\|_{\mathrm F}
      \sum_{a,b,a',b'=1}^2
      |R_{a,b}R_{a',b'}|
      \mathbb I\{(a,b)\neq(a',b')\}
    \nonumber\\
    &\lesssim
      (s+1)^4\|R\|_{\mathrm{op}}^{s-3}\|G\|_{\mathrm F}
      \sum_{a,b,a',b'=1}^2
      |R_{a,b}R_{a',b'}|
      \mathbb I\{(a,b)\neq(a',b')\}~,\quad\text{and}
    \nonumber\\
    \left|
      \mathrm D_R^2\widetilde\Phi_s^{\vartheta}(R)[G,H]
    \right|
    &\lesssim
      s(s-1)\binom{s-2}{2}
      \|R\|_{\mathrm{op}}^{(s-2)-2}
      \|G\|_{\mathrm F}\|H\|_{\mathrm F}
      \sum_{a,b=1}^2|R_{a,b}|^2
    \nonumber\\
    &\lesssim
      (s+1)^4\|R\|_{\mathrm{op}}^{s-4}
      \|G\|_{\mathrm F}\|H\|_{\mathrm F}
      \sum_{a,b=1}^2|R_{a,b}|^2~.
    \label{eq: gaussian endpoint degree derivative bounds}
\end{align}
The chain rule gives
\begin{align}
    &\frac{\mathrm d^2}{\mathrm d t\,\mathrm d q}
      \widetilde\Phi_s^{\vartheta}(R(t,q))
    \nonumber\\
    &\quad=
      \mathrm D_R^2\widetilde\Phi_s^{\vartheta}(R(t,q))
      \left[
        \frac{\mathrm d R(t,q)}{\mathrm d t},
        \frac{\mathrm d R(t,q)}{\mathrm d q}
      \right]
      +
      \mathrm D_R\widetilde\Phi_s^{\vartheta}(R(t,q))
      \left[
        \frac{\mathrm d^2R(t,q)}{\mathrm d t\,\mathrm d q}
      \right]~.
    \label{eq: gaussian endpoint degree interaction chain rule}
\end{align}
Since $\|R(t,q)\|_{\mathrm{op}}\leq1-c$, \eqref{eq: gaussian endpoint four coordinate path bounds} and \eqref{eq: gaussian endpoint degree derivative bounds} imply that
\begin{align}
    &\left|
      \mathrm D_R^2\widetilde\Phi_s^{\vartheta}(R(t,q))
      \left[
        \frac{\mathrm d R(t,q)}{\mathrm d t},
        \frac{\mathrm d R(t,q)}{\mathrm d q}
      \right]
    \right|
    \lesssim
      (s+1)^4e^{-cs}
      \Delta_K^2\sum_{a,b=1}^2|R_{a,b}(t,q)|^2~,
    \nonumber\\
    &\left|
      \mathrm D_R\widetilde\Phi_s^{\vartheta}(R(t,q))
      \left[
        \frac{\mathrm d^2R(t,q)}{\mathrm d t\,\mathrm d q}
      \right]
    \right|
    \lesssim
      (s+1)^4e^{-cs}\Delta_K \nonumber \\
    &\quad\quad \quad\quad \quad\quad \quad\quad \quad\quad \quad\quad \quad\quad 
    \cdot \sum_{a,b,a',b'=1}^2
      |R_{a,b}(t,q)R_{a',b'}(t,q)|
      \mathbb I\{(a,b)\neq(a',b')\}~.
    \label{eq: gaussian endpoint degree interaction bound}
\end{align}
Since $(s+1)^4e^{-cs}$ is summable and
\begin{align}
    |R_{a,b}(t,q)|
    &\lesssim
      |(K_{ZZ})_{a,b+2}|+\Delta_K~,
\end{align}
we find that
\begin{align}
    &\left|
      \frac{\mathrm d^2}{\mathrm d t\,\mathrm d q}
      \left[
        \Psi_{\delta,d}^{\vartheta}(I_2,I_2,R(t,q))
        -\sum_{a,b=1}^2
        \Psi_{\delta,d}^{\vartheta}
        \left(
          I_2,I_2,\rho_{a,b}(t,q)e_ae_b^\top
        \right)
      \right]
    \right|
    \nonumber\\
    &\quad\lesssim
      \Delta_K^2
      \sum_{a,b=1}^2
      \left(
        |(K_{ZZ})_{a,b+2}|+\Delta_K
      \right)^2
      +
      \Delta_K
      \sum_{a,b,a',b'=1}^2
      \left(
        |(K_{ZZ})_{a,b+2}|+\Delta_K
      \right)
    \nonumber\\
    &\hspace{38mm}\cdot
      \left(
        |(K_{ZZ})_{a',b'+2}|+\Delta_K
      \right)
      \mathbb I\{(a,b)\neq(a',b')\}~.
    \label{eq: gaussian endpoint canonical interaction bound}
\end{align}
by combining the bounds bounds \eqref{eq: gaussian endpoint degree interaction expansion}, \eqref{eq: gaussian endpoint degree two cancellation}, and \eqref{eq: gaussian endpoint degree interaction chain rule}--\eqref{eq: gaussian endpoint degree interaction bound}.

To complete the proof, observe that \eqref{eq: gaussian endpoint four coordinate integral}, \eqref{eq: gaussian single entry endpoint term}, and the fundamental theorem of calculus, imply
\begin{align}
    &\Cov\left(
      S_{1,2}(1)-S_{1,2}(0),
      S_{3,4}(1)-S_{3,4}(0)
    \right)
      -
      \sum_{a,b=1}^2\mathcal J_{a,b}
    \nonumber\\
    &\quad=
      \int_0^1\int_0^1
      \frac{\mathrm d^2}{\mathrm d t\,\mathrm d q}
      \Bigg[
        \Psi_{\delta,d}^{\vartheta}
        \left(A(t),B(q),R(t,q)\right)
        -
        \sum_{a,b=1}^2
        \Psi_{\delta,d}^{\vartheta}
        \left(
          I_2,I_2,\rho_{a,b}(t,q)e_ae_b^\top
        \right)
      \Bigg]
      \,\mathrm dt\,\mathrm dq~.
    \label{eq: gaussian endpoint integrated decomposition}
\end{align}
Substituting \eqref{eq: gaussian endpoint derivative decomposition} into \eqref{eq: gaussian endpoint integrated decomposition} and applying the triangle inequality gives
\begin{align}
    &\left|
      \Cov\left(
        S_{1,2}(1)-S_{1,2}(0),
        S_{3,4}(1)-S_{3,4}(0)
      \right)
      -
      \sum_{a,b=1}^2\mathcal J_{a,b}
    \right|
    \nonumber\\
    &\quad\leq
      \int_0^1\int_0^1
      \left|
        \frac{\mathrm d^2}{\mathrm d t\,\mathrm d q}
        \left[
          \Psi_{\delta,d}^{\vartheta}
          \left(A(t),B(q),R(t,q)\right)
          -
          \Psi_{\delta,d}^{\vartheta}
          \left(I_2,I_2,R(t,q)\right)
        \right]
      \right|
      \,\mathrm dt\,\mathrm dq
    \nonumber\\
    &\qquad+
      \int_0^1\int_0^1
      \left|
        \frac{\mathrm d^2}{\mathrm d t\,\mathrm d q}
        \left[
          \Psi_{\delta,d}^{\vartheta}
          \left(I_2,I_2,R(t,q)\right)
          -
          \sum_{a,b=1}^2
          \Psi_{\delta,d}^{\vartheta}
          \left(
            I_2,I_2,\rho_{a,b}(t,q)e_ae_b^\top
          \right)
        \right]
      \right|
      \,\mathrm dt\,\mathrm dq~.
\end{align}
The first integral is bounded by
\eqref{eq: gaussian endpoint marginal derivative difference}, and the
second is bounded by
\eqref{eq: gaussian endpoint canonical interaction bound}. Since both
integrals are over $[0,1]^2$, their sum gives
\eqref{eq: gaussian four coordinate refined endpoint bound}.
\hfill\qed

\subsubsection{Part (iii)}

For fixed $a,b\in[2]$, direct differentiation gives
\begin{align}
    \frac{\mathrm d\rho_{a,b}(t,q)}{\mathrm d t}
    &=- (K_{ZU})_{b+2,a}+q(K_{UU})_{a,b+2}~,
    \nonumber\\
    \frac{\mathrm d\rho_{a,b}(t,q)}{\mathrm d q}
    &=- (K_{ZU})_{a,b+2}+t(K_{UU})_{a,b+2}~,\quad\text{and}
    \nonumber\\
    \frac{\mathrm d^2\rho_{a,b}(t,q)}{\mathrm d t\,\mathrm d q}
    &=(K_{UU})_{a,b+2}~.
    \label{eq: gaussian endpoint single entry derivatives}
\end{align}
The chain rule, Part (i) of
\cref{lem: gaussian threshold covariance bounds} with $p=1$, and
\eqref{eq: gaussian endpoint single entry derivatives} imply that
\begin{align}
    &\left|
      \frac{\mathrm d^2}{\mathrm d t\,\mathrm d q}
      \Psi_{\delta,d}^{\vartheta}
      \left(I_2,I_2,\rho_{a,b}(t,q)e_ae_b^\top\right)
    \right|
    \nonumber\\
    &\quad\lesssim
      \left(
      |(K_{ZU})_{a,b+2}|
      +|(K_{ZU})_{b+2,a}|
      +|(K_{UU})_{a,b+2}|
    \right)
    \nonumber\\
    &\quad\cdot
    \left(
      |(K_{ZZ})_{a,b+2}|
      +|(K_{ZU})_{a,b+2}|
      +|(K_{ZU})_{b+2,a}|
      +|(K_{UU})_{a,b+2}|
      \right)~.
    \label{eq: gaussian endpoint single entry derivative bound}
\end{align}
Integrating \eqref{eq: gaussian endpoint single entry derivative bound} over $[0,1]^2$ verifies \eqref{eq: gaussian single entry endpoint bound}. \hfill\qed

\subsubsection{Part (iv)}

By applying the chain rule to $(t,q)\mapsto\Psi_{\delta,d}^{\vartheta}(A(t),B(q),R(t,q))$, Parts (i) and (ii) of \cref{lem: gaussian threshold covariance bounds}, with $p=1$, and \eqref{eq: gaussian endpoint mixed derivative bound} give
\begin{align}
    &\left|
      \frac{\mathrm d^2}{\mathrm d t\,\mathrm d q}
      \Psi_{\delta,d}^{\vartheta}(A(t),B(q),R(t,q))
    \right|
    \lesssim
      d\Delta_K^2
      +\Delta_K
      \sum_{a,b=1}^2|(K_{ZZ})_{a,b+2}|~.
    \label{eq: gaussian endpoint general derivative bound}
\end{align}
Part (iii) and the definition of $\Delta_K$ imply that
\begin{align}
    \sum_{a,b=1}^2|\mathcal J_{a,b}|
    &\lesssim
      \Delta_K^2
      +\Delta_K\sum_{a,b=1}^2|(K_{ZZ})_{a,b+2}|~.
\end{align}
The proof is complete by integrating \eqref{eq: gaussian endpoint general derivative bound} in \eqref{eq: gaussian endpoint four coordinate integral}, verifying \eqref{eq: gaussian four coordinate general endpoint bound}.\hfill\qed

\end{spacing}
\end{appendix}
\end{document}